\documentclass[12pt,a4paper,oneside,openright]{book}
\usepackage [english] {babel}
\usepackage{slashed}
\usepackage [T1]{fontenc}
\usepackage{graphicx}
\usepackage{dcolumn}
\usepackage{float}
\usepackage{amssymb}
\usepackage{tensor}
\usepackage{amsmath}
\usepackage[font=small,labelfont=bf]{caption}
\usepackage{hyperref}
\usepackage{ragged2e}
\usepackage[normalem]{ulem}
\usepackage{tocbibind}
\usepackage{cancel}
\usepackage{epigraph}
\usepackage{enumerate}
\usepackage{footnote}
\input epsf
\usepackage{fancyhdr}
\usepackage{cite}

\hypersetup{
    colorlinks=false, linktocpage=true,
    citecolor=red,
    filecolor=black,
    linkcolor=blue,
    urlcolor=cyan
}

\def\simge{\mathrel{\rlap{\raise 0.511ex \hbox{$>$}}{\lower 0.511ex \hbox{$\sim$}}}}
\def\simle{\mathrel{\rlap{\raise 0.511ex \hbox{$<$}}{\lower 0.511ex \hbox{$\sim$}}}} 
\def\lsim{\mathrel{\rlap{\lower3pt\hbox{\hskip0pt$\sim$}}
    \raise1pt\hbox{$<$}}}         %less than or approx. symbol
\def\gsim{\mathrel{\rlap{\lower4pt\hbox{\hskip1pt$\sim$}}
    \raise1pt\hbox{$>$}}}         %greater than or approx. symbol

\def\slash#1{\setbox0=\hbox{$#1$}\dimen0=\wd0                    
      \setbox1=\hbox{/} \dimen1=\wd1 \ifdim\dimen0>\dimen1
      \rlap{\hbox to \dimen0{\hfil/\hfil}} #1                        \else                                       
      \rlap{\hbox to \dimen1{\hfil$#1$\hfil}}  
      /   \fi}                                         

\newcommand{\be}{\begin{equation}}
\newcommand{\ee}{\end{equation}}
\newcommand{\bea}{\begin{eqnarray}}
\newcommand{\eea}{\end{eqnarray}}
\newcommand{\pp}{\prime\prime}
\newcommand{\p}{\prime}
\newcommand{\mH}{\mathcal{H}}

\newcommand{\Om}{\Omega_m}
\newcommand{\Omo}{\Omega_m^0}

\newcommand{\OMo}{\Omega_{M}^0}
\newcommand{\Oro}{\Omega_{r}^0}
\newcommand{\OL}{\Omega_{\Lambda}}

\newcommand{\OLo}{\Omega_{\Lambda}^0}

\newcommand{\OD}{\Omega_{D}}
\newcommand{\ODo}{\Omega_{D}^0}

\newcommand{\rc}{\rho_c}
\newcommand{\rco}{\rho_{c}^0}
\newcommand{\rmo}{\rho_{m}^0}
\newcommand{\rro}{\rho_{r}^0}

\newcommand{\rmr}{\rho_m}

\newcommand{\rR}{\rho_r}
\newcommand{\rD}{\rho_D}

\newcommand{\wCC}{\omega_\CC}

\newcommand{\rL}{\rho_{\CC}}

\newcommand{\rLo}{\rho_{\CC}^0}

\newcommand{\wD}{\omega_D}

\newcommand{\CC}{\Lambda}

\newcommand{\xiR}{\xi'}
\newcommand{\CH}{C_H}
\newcommand{\CHd}{C_{\dot{H}}}

\newcommand{\nueff}{\nu_{\rm eff}}
\newcommand{\zeff}{\zeta_{\rm eff}}

\newcommand{\rRo}{\rho_r^0}

\newcommand{\Hd}{\dot{H}}

\newcommand{\prm}{\delta\rho_m}
\newcommand{\prD}{\delta\rho_D}
\newcommand{\poD}{\delta\omega_D}
\newcommand{\wDo}{\omega_D^{(0)}}

\newcommand{\DA}{${\cal D}$A}
\newcommand{\DAU}{${\cal D}$A1}
\newcommand{\DAD}{${\cal D}$A2}
\newcommand{\DAT}{${\cal D}$A3}
\newcommand{\DCU}{${\cal D}$C1}
\newcommand{\DCD}{${\cal D}$C2}
\newcommand{\DC}{${\cal D}$C}
\newcommand{\DHlin}{${\cal D}$H}

\newcommand{\rma}{\rho_m}
\newcommand{\rr}{\rho_r}
\newcommand{\dG}{\dot{G}}

\newcommand{\drL}{\dot{\rho}_\Lambda}
\newcommand{\dH}{\dot{H}}
\newcommand{\nueffp}{\nu_{\rm eff}'}
\begin{document}

\frontmatter

\thispagestyle{empty}

\mbox{}
\begin{center}

\vskip 0.6cm

%________________________________________________________________
\vspace{-0.2cm}

\vskip 2.5cm
{\huge {\bf VACUUM ENERGY IN QUANTUM FIELD THEORY AND COSMOLOGY}}\\
\vskip 3.2cm
{\Large \sc Adrià Gómez-Valent}
\vskip 0.6cm
{\Large Departament de Física Quàntica i Astrofísica}\\
\vskip 0.25cm
{\Large Universitat de Barcelona}\\
\vskip 2.5cm
{\Large September 2017}
\vskip 4cm
\begin{figure}[H]
 \centering
 \includegraphics[width=400pt]{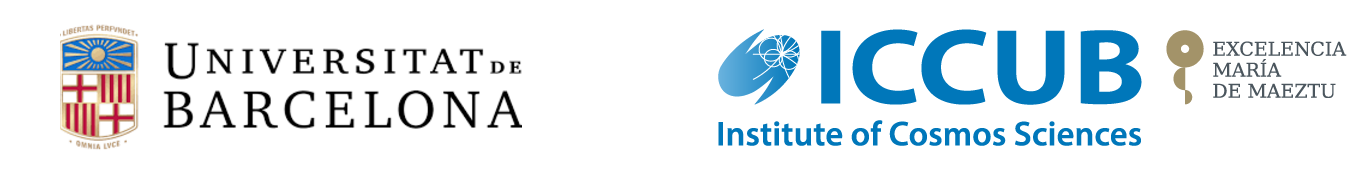}
\end{figure}

\end{center}

% pagina bianca dopo il titolo
%\thispagestyle{empty}
%\null
\newpage

\thispagestyle{empty}
\pagestyle{plain}
\vspace*{\fill}
\begin{center}
\textcircled{c} 2017
\vskip 0.4cm
Adri\`a G\'omez-Valent
\vskip 0.4cm
ALL RIGHTS RESERVED
\end{center}
\thispagestyle{empty}
\null
\newpage

\pagestyle{empty}

\begin{flushleft}
Programa de Doctorat en Física
\newline
\line(1,0){460}
\end{flushleft}
\vskip 2cm
\begin{center}
{\Large \bf Vacuum energy in Quantum Field Theory and Cosmology}
\vskip 1.0cm
by
\vskip 1.0cm
{\bf \Large Adrià Gómez-Valent}
\end{center}
\vskip 1.5cm
A dissertation submitted to the Faculty of Physics of the University of Barcelona in partial fulfillment of the requirements for the degree of {\it Doctor per la Universitat de Barcelona}. This thesis has been carried out under the supervision of
\vskip 1.5cm
\begin{center}
{\bf \Large Dr. Joan Solà Peracaula}\,,
\end{center}
\vskip 1.5cm
and under the tutoring of {\it Dr. Domènec Espriu}, both full professors of Theoretical Physics of the Department of Quantum Physics and Astrophysics of the University of Barcelona.

\vskip 1.5cm
\begin{flushright}
Barcelona, September 2017
\end{flushright}

\thispagestyle{empty}
\null
\newpage

\pagestyle{empty}
%\thispagestyle{empty}
% Copertina

\mbox{}

\vskip 0.6cm

%________________________________________________________________
\vspace{-0.2cm}

\vskip 2.5cm

\vskip 4cm
\begin{center}
{\it To my family, because they have always illuminated \\my path with their unconditional love. \\
And specially, to my parents, \\I owe everything to them.}
\end{center}

\thispagestyle{empty}
\null
\newpage

\pagestyle{plain}

\chapter{Acknowledgements}

\epigraph{<<Topant de cap en una i altra soca,\\avan\c{c}ant d'esma pel cam\'i de l'aigua,\\
se'n ve la vaca tota sola. \'Es cega.>>}{\itshape - Joan Maragall, extract from the poem ``La vaca cega'' (1895)}
\justify
This thesis is the culmination of many years of effort, illusion and sacrifices that go far beyond the PhD research period. I would like to firstly thank in broad everybody who has taught me something about life in this long way, even those who did it before the arrival of Physics to my life, during my infancy and early adolescence, contributing to forge the man I am today. 
\newline
\newline
I deem fair to acknowledge the task of those teachers of the primary and high school who live their profession with vocation, knowing how to exploit the capabilities of each student. I particularly thank those teachers from the high school IES Cambrils that taught me during my time there. I would like to make special mention of Pilar P\'erez, who has always been so close to the students, and Lloren\c{c} Porquer. I will always remember his first class on the free fall.  
\newline
\newline
I would also like to recognize the work carried out by all those who honestly dedicate their work to push the frontiers of Science a bit farther. This thesis would have not been possible without the previous knowledge in the field. Hopefully, this dissertation will also contribute to push the existing borders a bit away. 
\newline
\newline
I thank my PhD supervisor, Joan Sol\`a, for guiding me during these years and for giving me the opportunity of working on the amazing topic this thesis is about. He has been for me an example of commitment and passion for our profession. I also want to thank him the many meetings we have shared and his understanding of my personal circumstances. 
\newline
\newline
I also want to express my gratitude to Prof. Dom\`enec Espriu for agreeing to be the tutor of my thesis, and Prof. Federico Mescia for telling me about the possibility of doing the PhD with Joan when I finished the master course.
\newline
\newline
I thank the professors Harald Fritzsch, Antonio L\'opez Maroto and Spyros Basilakos for having accepted to be part of the PhD Thesis Committee; and also the professors Enric Verdaguer, Federico Mescia and Jordi Miralda-Escud\'e for having accepted to be supply members of such Committee. It has been a huge honor for me. 
\newline
\newline
I would also want to sincerely thank the professors Jaume Garriga, Tomeu Fiol, Emilio Elizalde and Jaume de Haro for their amiability and for accepting to take part of the original configuration of the Committee.
\newline
\newline
Of course, I would also like to thank the University of Barcelona and the Institute of Cosmos Sciences for funding the major part of my PhD research with an APIF grant, including my 3-month stay in the University Federico II of Napoli. I would also like to explicitly thank the Department of Quantum Physics and Astrophysics of the University of Barcelona, together with the disappeared Department of Structure and Constituents of Matter, for funding my participation in: the High Energy Workshop held in Benasque in September 2015; the CPAN conferences celebrated in Segovia in December of the same year; the 51st conference in the Rencontres
de Moriond on Cosmology held in the beautiful Italian town of La Thuille in March 2016; and the amazing International School of Subnuclear Physics held in Erice (Sicily) in June 2016. I enjoyed a lot each of these experiences.
\newline
\newline
I thank Prof. Salvatore Capozziello for reading this thesis, for being international referee of it, and also for allowing me to carry out a 3-month stay in the Dipartimento di Fisica ``Ettore Pancini'' of the University Federico II di Napoli. During my stay, I carried out a project about the possible explanation of the rotational curves of galaxies in the context of the running vacuum models and without making use of any dark matter component. The project is currently in its last stages, but still work in progress. The output of this research has not been included in this dissertation, which is about the cosmological implications of the running vacuum models, not about the ones at the astrophysical domain. I am also very grateful for the hospitality and friendliness offered by the colleagues of the Astrophysics office, in Napoli. I will never forget it.
\newline
\newline
I would also like to extend my deepest gratitude to Prof. Nick Mavromatos for his kind interest in reading this dissertation, and for being international referee of it. It has been a great honor for me.
\newline
\newline
And to Prof. Spyros Basilakos, the student Javier de Cruz P\'erez and Elahe Karimkhani, PhD, for their collaboration in some of the works presented in this thesis.
\newline
\newline
As a physicist, I strongly believe that everything should ultimately be explainable in terms of physical laws, even feelings. I do not know what rules our destinies. I do not comprehend at all which is the role played by quantum mechanics in our brains, the determinism in our lives, but it seems to me that there is not too much room for freedom in the choice of the most important aspects of our lives. Even those that we believe that are the product of our will, might not be that. In this sense, I do not differentiate too well what is merit, and what is the result of a pile of unwanted decisions thought of as wanted. Probably the phrase ``I am me and my circumstance'' of the Spanish philosopher Ortega y Gasset suits well in between these lines. I do not know who I have to thank the fact of having been born in this ``here'' and this ``now'', in the bosom of the beautiful family I belong to. Let me thank God for this, understanding here God as the set of primordial conditions in the Universe and physical laws that have given rise to my particular circumstances.
\newline
\newline
I would like to thank my family for teaching me the meaning of love. Probably, the most important lesson one can learn in life. This thesis is dedicated to them.
\newline
\newline
I will always thank my sister M\`onica for having allowed me to live with her during my Physics degree and the realization of the master course. She probably will never understand the scope of this fact and the impact it had in my life. I also thank her for having summed to my family her husband, Jordi, and Marcel, my beloved godson. 
\newline
\newline
I also want to thank my other sister, Mireia, for being an example of meticulousness in her work. I strongly admire her. And my niece, Nadia, who is for me like a little sister, and has always encouraged me with her optimism and joyfulness. 
\newline
\newline
But if I am what I am now is mainly due to my parents, who have always believed in me much more than what I believe in myself. I thank them for having considered my education as a priority and not as something secondary, and for having fought hard in life for giving me more opportunities than those they and my grandparents had. 
\newline
\newline
Agraeixo de tot cor als meus pares que m'hagin inculcat el seu esperit de superaci\'o i lluita a la vida, i que m'hagin ensenyat que la bona educaci\'o i l'\`etica no s\'on q\"{u}estions de classe.
\newline
\newline
A la meva mare, qui sempre ha mirat pels fills per sobre de tot. Gr\`acies per ensenyar-me a ser exigent amb m\'i mateix i a lluitar per all\`o que volem i estimem.
\newline
\newline
A mi padre, mi mejor amigo y compa\~nero de vida. Gracias por tus consejos y tus sonrisas. Gracias por darme la oportunidad de conocer el deporte desde peque\~no, a trav\'es del cual he aprendido m\'as sobre las personas, y tambi\'en a saber que el mundo va mucho m\'as all\'a de los libros. 
\newline
\newline
A totes aquelles persones que m'estimen, gr\`acies, gr\`acies, i mil cops gr\`acies.

\thispagestyle{empty}
\null
\newpage

\chapter{Preface}

\epigraph{<<And it does me good to do what's difficult. That doesn't stop me having a tremendous need for, shall I say the word - for religion - so I go outside at night to paint the stars, and I always dream a painting like that.>>}{\itshape - Vincent Van Gogh, in a letter to his brother (1888)}

\justify

In the centenary of the introduction of the cosmological constant by Albert Einstein in his gravitational field equations, we now know that the Universe is expanding at an increasing rate. The most ``simple'' explanation of such positive acceleration is precisely given by the cosmological term, but we are still facing the question whether the ultimate cause of it
is a rigid (i.e. estrictly constant) $\Lambda$-term, some sort of mildly evolving dynamical vacuum energy, or maybe some unknown form of dark energy (DE)
different from the vacuum, or even a modification of General Relativity at cosmological scales.
\newline
\newline
We are currently living in a very privileged and exciting epoch, in the beginning of the precision Cosmology era. The exploration of the Cosmos at large scales is now a reality. We can look out there with our sophisticated instruments and try to guess how mother Nature gave rise to the special Universe in which we live in, which is the mechanism that rules the accelerated expansion. We can scrutinize the most recondite corners of the Cosmos and try to extract valuable information from observations. But many questions are still pending to be answered, namely the issues concerning the matter-antimatter asymmetry in the Universe, the existence of primordial non-gaussianities, neutrino's masses and hierarchy, the nature of dark energy and dark matter, etc. Hopefully, with the aid of all the experimental and technological advances that are to come in the near future we will be in position to find some clues about these (still open) mysteries.
\newline
\newline
The presence of dark energy in our Universe seems to be a well established empirical reality in modern Cosmology. The facts supporting its existence are robust, innumerable and overwhelming. They are compellingly expressed through the observational body of evidence confirming the accelerated expansion of the Universe, evidence that has been piling up in the last two decades from a large variety of cosmological sources. On these grounds it all looks as if the historical investigation on the DE has been a research path full of color and successes. It is so, in part, but not exactly so. For one thing, real knowledge and understanding of the physical quantities being measured seems to lag ``a bit'' behind the outstanding record achieved by the modern observations. It is disquieting to recognize that our theoretical conceptions, bold as they are in some cases, did not live up to our expectations. In fact, they did not follow a pace comparable to the astounding developments undergone by the observational tools and state-of-the-art instruments that have been (and are being) used to explore our patch of surrounding Cosmos. That this dual pace between empirical and theoretical knowledge is a factual reality in Cosmology becomes apparent when we try to answer the next obvious scientific question at stake: what is the dark energy that we ``observe''? Is it a substance, is it a field, may be an effect associated to something else  more fundamental than matter and fields? Or perhaps it is the quantum vacuum in action, maybe behaving in a way that we are not sufficiently prepared to fully understand it yet?  Cosmologists have worked hard to decipher the dark energy code, but the bare facts tell us that despite the staggering technological victories in exploring the Universe in the last twenty years, we are still facing a longstanding, stubbornly persistent, almost hopeless ignorance about the physical nature of the DE and hence on the ultimate cause of the observed acceleration of the Universe. This profound and distressing conflict is, of course, at the root of the Cosmological Constant Problem (CCP), most likely the biggest theoretical conundrum of Fundamental Physics of all times.  
\newline
\newline 
In this thesis I describe the CCP in the Introduction (Chapter \ref{chap:Introduction}), and I make a short (but exhaustive) historical review of the cosmological term from its birth, a hundred years ago, up to the present. I also introduce the standard cosmological model, the $\Lambda$CDM, together with its associated theoretical and observational problems. This leads to the study of alternative scenarios that are potentially able to alleviate the aforementioned drawbacks. I review in a quite general way some of them, just to locate the reader in the right context. The last two sections of the Introduction are mainly focused to motivate a particular class of models, the so-called running vacuum models (RVM's), which arise from the renormalization group formalism of Quantum Field Theory in curved spacetime applied to Cosmology. These models are the main object of study of this thesis. In them, the vacuum energy density evolves with the expansion, having a direct dependence on the Hubble rate. The latter somehow parametrizes the dynamics of the cosmological vacuum. This time evolution of $\rho_\Lambda$ can only respect the fulfillment of the local energy conservation equation if the Newtonian coupling acquires some dynamics too, and/or if the laws that describe the matter-radiation energy densities acquire some anomalous behavior. All these scenarios give rise to a very rich phenomenological palette of different RVM's, which is amply studied in the subsequent chapters, in the main body of the thesis.
\newline
\newline 
Let me summarize the main content of this dissertation. In Chapter \ref{chap:Atype} we analyse some RVM's in which the variation of $\Lambda$ is due to an energy exchange between vacuum and matter sectors. We investigate in detail the background solutions, together with the linear and nonlinear perturbations. The latter is carried out with an improved version of the Press \& Schechter formalism. In Chapter \ref{chap:DynamicalDE} we explore a DE class (denoted as $\mathcal{D}$-class) of models in which the DE is self-conserved and its energy density is described by the same law encountered in the RVM's. In this case, though, the dynamical behavior of the DE is possible thanks to a time-dependent deviation of the DE equation of state (EoS) parameter from the vacuum one. We also study the linear structure formation for these models and analyze the effect of the DE perturbations at deep subhorizon scales. In Chapter \ref{chap:Gtype} we explore the possibility that both, $\Lambda$ and the Newton coupling $G$ vary with the expansion in the so-called G-type models, being in this case both, matter and radiation, self-conserved. We analyze the phenomenological implications in this framework. Some small hints in favor of vacuum dynamics are found. This is in short the content of the first part of this thesis. In the second one, we refine the observational data sets, we improve the statistical analyses carried out in the first part, and enlarge the fitting parameter space. These improvements are firstly presented in Chapter \ref{chap:AandGRevisited}, where again we study the G-type models, together with the A-type ones, a subclass of RVM's which has also been studied in detail in Chapter \ref{chap:Atype}. Here, the statistical signal in favor of the vacuum dynamics reaches the $4\sigma$ c.l., something that has not been previously found in the literature. In Chapter \ref{chap:MPLAbased} we focus our attention to a concrete quintessence model, the original Peebles \& Ratra (PR) model. Interestingly enough, we find that the important evidences in favor of dynamical DE are also traced by this model, in which one starts from a well-defined scalar field Lagrangian. We also find a better agreement with the data than the one offered by the $\Lambda$CDM if we use some simple DE parametrizations, as the XCDM and the CPL ones. This is also quite remarkable, since it seems to indicate that the aforementioned signal in favor of the DE dynamics can be mimicked not only through a particular model, but by an ample spectrum of them, although, of course, not all of them perform equally good. All these results are also unprecedented in the literature. In Chapter \ref{chap:PRDbased} we analyze in detail a RVM in which vacuum interacts with dark matter, together with two more phenomenological dynamical vacuum models that one can also find in the market. We also reanalyze the DE parametrizations mentioned before and the PR model in the light of the most updated data sets. We confirm the presence of a $\sim 4\sigma$ c.l. signal sitting on the top of the data in favor of an evolving-in-time DE component. We present a dedicated study on the cosmological data used in the analysis and discuss the existing correlations among them. In addition, we also provide a thorough explanation about why important collaborations as Planck or BOSS have proved incapable of detecting such positive signal. Finally, in Chapter \ref{chap:H0tension} we explore the $H_0$ tension between the measurements of Planck and the Hubble Space Telescope (HST) in light of vacuum dynamics in the Universe. We prove that in the context of the $\Lambda$CDM the inclusion of the large scale structure (LSS) data in the fitting analysis also drags the preferred value of the Hubble parameter far away from the Planck's best-fit value, reaching a $4-5\sigma$ tension with it, being the tension with the HST value greater than $5\sigma$. The tension between the value of $H_0$ derived from baryon acoustic oscillations and cosmic microwave background data and the one inferred from LSS formation data is removed in the context of some of the dynamical vacuum models under consideration, but the tension with the HST value is kept at the level of $>5\sigma$. We also show that the complete set of data used in our analysis seems to prefer vacuum dynamics rather than DE dynamics, i.e. the vacuum EoS $w=-1$ is preferred over an EoS parameter different from $-1$.
\newline
\newline
I have also included five appendices after Chapter \ref{chap:H0tension}. In Appendix \ref{ch:appZPE} I regularize the zero-point energy of a free scalar field in Minkowski spacetime using the momentum cutoff and dimensional regularization methods, and renormalize the regularized expression using two minimal subtraction schemes. It is a very standard calculation, but I deem useful to provide the explicit derivation here together with the most important formulas, which are often called from the main body of the thesis. In Appendix \ref{ch:appZPEcurved} I obtain the effective (vacuum) action in curved spacetime using the canonical quantization procedure. In the literature one usually finds an alternative way of computing this object, through the path integral formalism and making use of e.g. the adiabatic expansion of the Green functions. The method used in this appendix leads to the same expression of the effective vacuum action in the low-energy Universe that the one obtained using the more involved procedure of the path integral formalism. It is, though, less general. In Appendix \ref{ch:appPert} I derive the differential equations that govern the evolution of matter density perturbations in the RVM's that are studied in the main body of this thesis. In appendix \ref{ch:appCollapse} I describe the procedure to compute the collapse density threshold, which is a quantity that plays a crucial role in the Press \& Schechter formalism. This appendix is added in support of the analysis of nonlinear perturbations carried out in Chapter \ref{chap:Atype}. Finally, in Appendix \ref{chap:App5} I provide some technical details on the cosmological observables used and statistical analyses carried out in the second part of this thesis, together with some comments on the fitting procedure and the Fisher matrix formalism.
\newline
\newline
I hope that the reader enjoys reading these pages as much as the author has enjoyed writing them. The path just starts here. But where does it lead? Let us see.

\pagestyle{plain}
\chapter{List of publications}
\label{chap:PublicationList}
\justify
The research carried out during this PhD thesis has given rise to the following list of papers, most of them already published in top international journals: 

\begin{enumerate}

\item {\it Dynamical vacuum energy in the expanding Universe confronted with observations: a dedicated study.}\\
A. Gómez-Valent, J. Solà, and S. Basilakos\\
\href{http://iopscience.iop.org/article/10.1088/1475-7516/2015/01/004/meta;jsessionid=0FFE952C483D87B81C214D5B73E6A404.c2.iopscience.cld.iop.org}{JCAP {\bf 1501}, 004 (2015)} ;  [\href{https://arxiv.org/abs/1409.7048v3}{arXiv:1409.7048}]

\item {\it Vacuum models with a linear and a quadratic term in H: structure formation and number counts analysis.}\\
A. Gómez-Valent and J. Solà\\
\href{http://mnras.oxfordjournals.org/content/448/3/2810}{Mon. Not. R. Astron. Soc. {\bf 448}, 2810 (2015)} ;  [\href{https://arxiv.org/abs/1412.3785v3}{arXiv:1412.3785}]
 
\item {\it The $\bar{\Lambda}$CDM cosmology: From inflation to dark energy through running $\Lambda$.}\\
J. Solà and A. Gómez-Valent\\
\href{http://www.worldscientific.com/doi/abs/10.1142/S0218271815410035}{Int. J. Mod. Phys. D{\bf 24}, 1541003 (2015)} ;  [\href{https://arxiv.org/abs/1501.03832v1}{arXiv:1501.03832}]

\item {\it Hints of dynamical vacuum energy in the expanding Universe.}\\
J. Solà, A. Gómez-Valent, and J. de Cruz Pérez\\
\href{http://iopscience.iop.org/article/10.1088/2041-8205/811/1/L14/meta}{Astrophys. J. Lett. {\bf 811}, L14 (2015)} ; [\href{https://arxiv.org/abs/1506.05793v2}{arXiv:1506.05793}]

\item {\it Background history and cosmic perturbations for a general system of self-conserved dynamical dark energy and matter.}\\
A. Gómez-Valent, J. Solà, and E. Karimkhani\\
\href{http://iopscience.iop.org/article/10.1088/1475-7516/2015/12/048/meta}{JCAP {\bf 1512}, 048 (2015)} ; [\href{https://arxiv.org/abs/1509.03298v3}{arXiv:1509.03298}]

\item {\it First evidence of running cosmic vacuum: challenging the concordance model.}\\
J. Solà, A. Gómez-Valent, and J. de Cruz Pérez\\
\href{http://iopscience.iop.org/article/10.3847/1538-4357/836/1/43/meta}{Astrophys. J. {\bf 836}, 43 (2017)} ; [\href{https://arxiv.org/abs/1602.02103}{arXiv:1602.02103}]

\item {\it Dynamical dark energy: scalar fields and running vacuum.}\\
J. Solà, A. Gómez-Valent, and J. de Cruz Pérez\\
\href{http://www.worldscientific.com/doi/abs/10.1142/S0217732317500547}{Mod. Phys. Lett. A{\bf32}, 1750054 (2017)} ; [\href{https://arxiv.org/abs/1610.08965v5}{arXiv:1610.08965}]

\item{\it Dynamical Vacuum against a rigid Cosmological Constant.}\\
J. Solà, J. de Cruz Pérez, and A. Gómez-Valent\\
Submitted for publication in Phys. Rev. Lett. ; [\href{https://www.arxiv.org/abs/1606.00450v3}{arXiv:1606.00450}]

\item{\it Towards the firsts compelling signs of vacuum dynamics in modern cosmological observations.}\\
J. Solà, J. de Cruz Pérez, and A. Gómez-Valent\\
Submitted for publication in Phys. Rev. D ; [\href{https://arxiv.org/abs/1703.08218}{arXiv:1703.08218}]

\item{\it The $H_0$ tension in light of vacuum dynamics in the Universe.}\\
J. Solà, A. Gómez-Valent, and J. de Cruz Pérez\\
\href{http://www.sciencedirect.com/science/article/pii/S0370269317307852?via%3Dihub}{Phys. Lett. B{\bf 774}, 317 (2017)} ; [\href{https://arxiv.org/abs/1705.06723}{arXiv:1705.06723}]

\item{\it Vacuum dynamics in the Universe versus a rigid $\Lambda=$const.}\\
J. Solà, A. Gómez-Valent, and J. de Cruz Pérez\\
\href{http://www.worldscientific.com/doi/abs/10.1142/S0217751X17300149}{Int. J. Mod. Phys. A{\bf32}, 1730014 (2017)} ; [\href{https://arxiv.org/abs/1709.07451}{arXiv:1709.07451}]

\end{enumerate}

\noindent
It is also worthwhile to mention the following contribution to the peer-reviewed proceedings of the 51st conference in the Rencontres de Moriond series:

\begin{enumerate}
\item{\it Running vacuum versus the $\Lambda$CDM.}\\
A. Gómez-Valent, J. Solà, and J. de Cruz Pérez\\
Published in Proceedings of the 51st Rencontres de Moriond, La Thuile, Italy, March 2016, eds. E. Augé, J. Dumarchez, and J. Tran Thanh Van, p. 251 ; [\href{https://arxiv.org/abs/1605.06448}{arXiv:1605.06448}]

\end{enumerate}

%%%%%%%%%%%%%%%%%%%%%%%%%%%%%%%%%%%%%%%%%%%%%%%%%%%%%%%%%%%%%%%%%%%%%%%%%%%%%%
\tableofcontents              % this may be commented out in the proof stage
%%%%%%%%%%%%%%%%%%%%%%%%%%%%%%%%%%%%%%%%%%%%%%%%%%%%%%%%%%%%%%%%%%%%%%%%%%%%%%

\mainmatter

\newpage
\thispagestyle{empty}
\null

\pagestyle{fancy}
\fancyhf{}
\fancyhead[CO]{\nouppercase{\leftmark}}
\fancyhead[CE]{\nouppercase{\rightmark}}
\cfoot{\thepage}

\chapter{Introduction}
\label{chap:Introduction}

%%%%%%%%%%%%%%%%%%%%%%%%%%%%%%%%%
% CITES 
%%%%%%%%%%%%%%%%%%%%%%%%%%%%%%%%%

\epigraph{<<If I have seen further, it is by standing on the shoulders of giants.>>}{\itshape - Sir Isaac Newton, in a letter to Robert Hooke (1675)}

\epigraph{<<Even thus by the great sages 'tis confessed\\The phoenix dies, and then is born again.>>}{\itshape - Dante Alighieri, from the ``Divine Comedy'' (1320)}

%%%%%%%%%%%%%%%%%%%%%%%%%%%%%%%%%
% Introducció de la introducció
%%%%%%%%%%%%%%%%%%%%%%%%%%%%%%%%%

\vskip 0.5cm

%\newpage 

\section{Brief history of the Cosmological Constant}
\label{chap:historyCC}

I begin the introductory chapter of my thesis with this short historical review of the not that short history of the Cosmological Constant (CC). I have divided it in two subsections. In the first one, I explain the historical remarks from the introduction of the CC in the General Relativity (GR) field equations by A. Einstein up to the present, without entering the details about the possible theoretical connection between the CC and the vacuum energy. Thus, I deal with the CC just as a pure ``geometrical'' object, and basically explain which has been its role in Cosmology. In the second one I cover the summarized history of the vacuum energy from the beginning of the quantum revolution up to the discovery of the Higgs boson in the LHC, including the modern formulation of the CC problem, which was carried out for the first time by Y.B. Zel'dovich in 1967. This will allow me to introduce the important concept of zero-point energy, together with other possible vacuum energy contributions to the CC. 

Hopefully, this historical review will enlighten the subsequent parts of this dissertation, clearly showing where we come from and drawing a general picture of the current status of the problems affecting the CC.
   
%\vskip 0.5cm

%%%%%%%%%%%%%%%%%%%%%%%%%%%%%%%%%
% Historical review Part I
%%%%%%%%%%%%%%%%%%%%%%%%%%%%%%%%%

\subsection{Cosmology and the CC}
\label{subsec:HistPartI}

Albert Einstein completed the General theory of Relativity in 1915, after finding the field equations \cite{EinsteinFieldEqGR},
\be\label{eq:EinsteinOrFielEq}
G_{\mu\nu}\equiv R_{\mu\nu}-\frac{1}{2}g_{\mu\nu}R=8\pi G T_{\mu\nu}\,,\qquad\footnote{In this thesis I partially adopt the natural units convention. I use $c=1$ and $\hbar=1$, but I keep, e.g. $G\ne 1$. Another important issue: Greek letters in superscripts and subscripts will take values between $0$ and $4$, being $0$ the temporal component; Latin Letters will only run between $1$ and $3$, i.e. they will denote spatial components.}
\ee  
where $G_{\mu\nu}$ is the so-called Einstein tensor, $g_{\mu\nu}$ and $R_{\mu\nu}$ are the metric and Ricci tensors, respectively, $R\equiv g^{\mu\nu}R_{\mu\nu}$ is the Ricci scalar, and $T_{\mu\nu}$ is the energy-momentum tensor. This compact set of equations establishes the interplay between the matter-energy content of the Universe and the geometry of the spacetime as a whole. As it is usually said, matter tells geometry how to curve and geometry tells matter how to move. 

Shortly after his fascinating discovery, and fully aware of the potential power of his theory, Einstein focused his efforts in studying the implications that the newborn theoretical framework could have in the cosmological arena. As starting point, he considered a hypothetical homogeneous and globally static Universe. The latter assumption was quite reasonable at the time, due to the existing empirical evidence supporting the low relative velocities of the observed stars\footnote{The author is not sure whether Einstein ignored or not the studies carried out by Vesto Slipher between 1912 and 1917, which led him to the experimental discovery of the Doppler shifts in spiral nebulae. But even if Einstein was aware of this discovery, there was no consensus in that time about whether spiral nebulae were independent star systems outside the Milky Way or not (the famous Shapley-Curtis Great Debate was held in 1920 and Hubble's determinant work appeared in 1929). Thus, in 1917 there was no strong enough reason to think about a dynamical Universe.}. The question was whether GR could give rise to this kind of Universe or not. Einstein knew, of course, that the Newtonian Poisson's equation,
\be\label{eq:PoissonOr}
\nabla^2\phi=4\pi G\rho_m\,,
\ee
is immediately recovered from \eqref{eq:EinsteinOrFielEq} in the low-energy limit. Actually, it was a kind of needed requisite for GR that Einstein demanded himself to its theory. He was also aware that Newtonian gravity does not allow the kind of Universe he had in mind. It does not matter whether we consider an Euclidean spatially finite or an infinite Universe. In the former case, it is obvious that gravitational attraction makes the Universe to collapse and, in the latter, it can be proved that a non-null matter energy density uniformly distributed in the infinite space cannot fulfill Poisson's equation if the Universe under consideration is supposed to be static. Thus, he needed to change his original field equations \eqref{eq:EinsteinOrFielEq} in order to obtain a modified form of Poisson's equation in the low-energy regime, since \eqref{eq:EinsteinOrFielEq} naturally produces a dynamical Universe. The outcome of his research was presented in his memorable paper, ``Cosmological considerations on the general theory of relativity'' \cite{EinsteinCC1917}, submitted to the Royal Prussian Academy of Sciences in Berlin\footnote{In German: {\it K{\"o}niglich Preussischen Akademie der Wissenschaften zu Berlin}.} on February 8th of 1917, almost exactly one century ago at the time of writing this dissertation. In his paper, after recalling to the reader the inherent problems associated to \eqref{eq:PoissonOr} in the cosmological context, Einstein pointed out that the following modified form of Poisson's equation can eliminate the existing drawback,
\be\label{eq:PoissonMod}
\nabla^2\phi{\bf -\lambda\phi}=4\pi G\rho_m\,,\qquad \footnote{{\it Historical curiosity:} The German physicists Hugo von Seeliger and Carl Neumann had proposed this modification of the original Poisson's equation in the mid-1890's in order to avoid the existing cosmological conundrums. As a consequence, the gravitational potential generated by a point-like massive particle departs from the Newtonian expression, it is Yukawa-like and reads $\phi=-\frac{mG}{r}e^{-r\sqrt{\lambda}}$, with a very small $\lambda$. Asaph Hall demonstrated in 1894 that this kind of potential could solve the precession of the perihelion of Mercury problem. Nevertheless it could not solve other existing open problems of that epoch affecting various Solar System bodies.}
\ee
where $\lambda$ is a universal constant\footnote{At this stage, I keep the notation for the CC that was originally used by Einstein in \cite{EinsteinCC1917}. From \eqref{eq:EinsteinModFielEq} on I will denote the CC with $\Lambda$ (instead of $\lambda$), since the former is the Greek letter currently mostly used in the literature.}. According to this equation, a static and homogeneous Universe becomes possible if the gravitational potential takes the constant value $\phi=-4\pi G\rho_m/\lambda$. Analogously to the addition of this new term containing the constant $\lambda$ in the original Poisson's equation, Einstein modified his original field equations \eqref{eq:EinsteinOrFielEq} by also adding a constant so as to be able to explain in a consistent way his static Universe,
\be\label{eq:EinsteinModFielEq}
R_{\mu\nu}-\frac{1}{2}g_{\mu\nu}R-\Lambda g_{\mu\nu}=8\pi G T_{\mu\nu}\,.
\ee  
Einstein conceived $\Lambda$ (the CC) as a new constant of Nature\footnote{Actually the non-relativistic limit of the 00 component of \eqref{eq:EinsteinModFielEq} does not lead to an equation like \eqref{eq:PoissonMod}, but to $\nabla^2\phi=4\pi G\rho_m-\Lambda$. Notice that the constant $\lambda$ used in \eqref{eq:PoissonMod} cannot be directly identified with $\Lambda$. Despite being both constant and playing a similar role, they are {\it not} equal.}. This new term preserves the general covariance of the theory and, if it is small enough, it can pass all the Solar Sytem tests. He showed that if $\Lambda=4\pi G \rho_m$, it is able to counteract the attractive gravitational force generated by the uniformly distributed mass that fills the Universe, making possible a homogeneous, static and spherically closed Universe with curvature radius $R_E=\Lambda^{-1/2}$\footnote{It is usually called Einstein's radius.}. 
It is a well-known fact that \cite{EinsteinCC1917} represents the birth of modern Cosmology, and it goes hand in hand with the birth of the Cosmological Constant. As we are about to see, the CC has had (an, indeed, still has) a very convulsed life. It has been killed many times, but like the phoenix, it has always managed to rise from the ashes, becoming again and again required to match theory with observations. Unfortunately for Einstein, his Universe had severe stability problems, as was much later effectively discovered by Lema\^itre in \cite{Lemaitre1927} (see also the work of Sir A.S. Eddington, \cite{EddingtonInestab1930}). This means that a small disturbance in the matter distribution makes his Universe to collapse or expand forever. But in the late 1920's other models had already gained leadership in front of the static one, due to the found experimental evidence in favor of a dynamical Universe. 

In the same year of the publication of Einstein's paper, i.e. in 1917, Willem de Sitter presented his own model in \cite{DeSitter1917}. Therein he showed that, contrary to Mach's principle and Einstein's ideas, matter is not needed to produce inertia. He was the first theoretician who paid close attention to Slipher's work. He considered a Universe dominated by $\Lambda$ and completely absent of matter, in which the Universe expands and therefore, is dynamical. It is considered the first and more simple expanding Universe model, but despite its simplicity it is of great importance, since it predicted for the first time the recession of extragalactic nebulae from us with a huge velocity, as a simple consequence of the properties of the gravitational field, without having to assume that we are located at a privileged point of the Universe. This is known as ``de Sitter effect''. The problem with the de Sitter's metric as it was originally presented is that there does not exist any preferred rest frame. It was not written in proper Robertson-Walker form, and in particular does not have a universal cosmic time. Thus, in order to derive a velocity-distance relation it was needed to make an extra assumption and different assumptions led to different relations. This feature made the de Sitter effect quite difficult to interpret in absolute terms\footnote{Some of the physicists contributing to the discussion on the interpretation of the de Sitter effect were Weyl, Eddington, Lanczos, Silberstein and Tolman. I omit the details in the current historical review, which does not aim to be that exhaustive. A full study of all the historical details would deserve a separate PhD dissertation.}.

Also the two posterior seminal papers of A. Friedmann \cite{Friedmann1922,Friedmann1924} must be mentioned. In \cite{Friedmann1922} he analyzed from a very mathematical perspective the various solutions to GR field equations, including the possibility of a zero or negative CC. He obtained the ``periodic world'' solution. In the second reference, he also explored the effect of alternative spatial curvatures. Curiously, in 1923 Einstein believed he had found an error in \cite{Friedmann1922} and he published a brief note claiming this, but Friedmann showed that his calculations were indeed correct. Consequently, Einstein withdrew his objection to the result by publishing a retraction of his comment, with a sentence that was finally deleted before the publication:

\begin{center}
\parbox{13cm}{{\it I consider that Mr Friedmann's results are correct and shed new light. They show that the field equations admit, for the structure of spherically symmetric space, in addition to static solutions, dynamical solutions, \sout{but a physical significance can hardly be ascribed to them}.}} 
\end{center}

\vskip 0.3cm
But also in 1923, Einstein wrote in a postcard to Weyl:

\begin{center}
\parbox{13cm}{{\it If there is no quasi-static world, then away with the cosmological term.}} 
\end{center}

\vskip 0.3cm
In light of these events, I am not able to determine up to what extent was Einstein obdurate with the idea of a static Universe. In that epoch there was no crucial experimental evidence against his cosmological model, but neither a crucial evidence in favor of it. According to the first statement, it seems that he had some kind of prejudices against dynamical cosmologies, but according to the second one, he did not discard the latter possibility either. Of course, he was not aware of the pathological stability problems affecting his static Universe and this did not allow him to discard his model from a pure theoretical point of view in that time. 

In 1924, A. Eddington pointed out in reference \cite{EddingtonSliph1924} that $36$ out of the $41$ spectral shifts of galaxies measured by V. Slipher were redshifted, what he thought was a hint in favor of the de Sitter effect\footnote{In \cite{EddingtonSliph1924} one can find a list of radial velocities of spiral nebulae measured by Vesto Slipher in the Lowell Observatory, some of them unpublished in that time. It is curious the comment of Eddington about the blueshifted spectrum of Andromeda (which was measured with very high precision). He noticed that this data point was difficult to be explained with the de Sitter model. Of course, the distances of the spiral nebulae to us were still unknown and this made impossible to ensure the existence of some kind of correlation between the spectral shifts and the distances to the sources.}. 

Some years later, in 1927, the Belgian astronomer and Catholic Jesuit priest Georges Lema\^itre published his cosmological model in \cite{Lemaitre1927}\footnote{Here, it is worth to mention another example of the apparent ``fight'' of Einstein against the non-static cosmological models. In the Solvay conference of 1927, he told Lema\^itre: {\it Your calculations are correct, but your physical insight is ``tout a fait abominable''.}}. He was surely inspired by Hubble's work\footnote{Lema\^itre had attended a meeting of the American Astronomical Society in 1924-1925 in which E. Hubble presented some results, probably the first steps towards his posterior extraordinary work of 1929. It is important to notice that in 1925 Hubble had already proved the extragalactic nature of the spiral nebula NGC 6822 \cite{Hubble1925}.}. In his paper, he considered again the presence of a non-null $\Lambda$. Two of his main conclusions were: 1) The radius of the Universe increases without limits from an asymptotic value $R_E$ for $t = -\infty$, where $R_E$ is Einstein's radius, i.e. the Universe evolves from an asymptotically static state which, of course, is unstable; 2) The recession velocities of extragalactic nebulae are a cosmical effect of the expansion of the Universe. He estimated for the very first time the value of the Hubble constant by using the velocities of the galaxies as measured by Slipher and published by Gustaf Str\"{o}mberg in \cite{Stromberg1925}, and the distances obtained from brightness measurements by Hubble in 1926 \cite{Hubble1926}. Lema\^itre determined that its value was around $625\,{\rm km/s/Mpc}$. Sadly for him, \cite{Lemaitre1927} had a very little impact, since the journal in which it was published was not well-known outside Belgium. A. Eddington\footnote{In 1930 Eddington wrote a long comment in \cite{EddingtonInestab1930} on Lema\^itre's work, in which he described it as a {\it brilliant solution} to the outstanding cosmological problems of that time.} helped to translate the paper into English and this allowed it to be published in a high-impact journal in 1931 \cite{Lemaitre1927}, but the part containing the estimation of the Hubble constant was not included. It is known that Lema\^itre recommended not to include this point in the translated version. In a letter to the editor of The Monthly Notices of the Royal Astronomical Society, he wrote (see \cite{MarioLivio2011}, and references therein):

\begin{center}
\parbox{13cm}{{\it I did not find advisable to reprint the provisional discussion of radial velocities which is clearly of no actual interest, and also the geometrical note, which could be replaced by a small bibliography of ancient and new papers on the subject.}} 
\end{center}

\vskip 0.3cm
In the same year of his paper republication he presented a modified version of his original model, the ``loitering'' or ``hesitating'' model, which did not start from an unstable Einstein's Universe, but with an initial singularity (called by Lema\^itre {\it primeval atom}, {\it a unique atom the atomic weight of which is the total mass of the Universe} \cite{Lemaitre1931}). With positive curvature, positive CC and a total mass very near the one found in the Einstein static Universe, Lema\^itre's one emerges from a very high energy density state and evolves through a decelerated phase until the matter density drops to near the Einstein value. In that stage, the Universe behaves almost as a static Einstein Universe, but then the instability does his job and the expansion starts speeding up. The CC dominates over the matter energy density and the Universe evolves towards a pure de Sitter phase. This model is currently considered as the first seed of the celebrated Big Bang theory.    

\begin{figure}
\centering
\includegraphics[angle=0,width=0.7\linewidth]{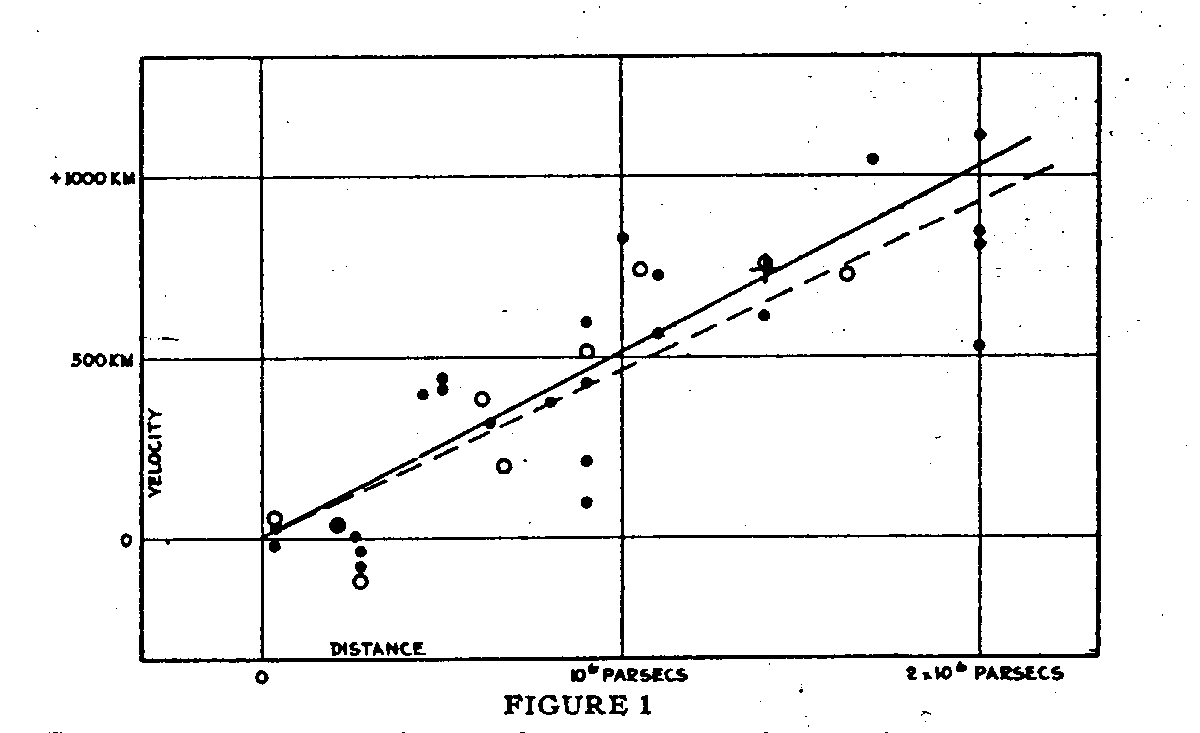}
\caption[Original velocity-distance plot from Hubble's 1929 paper \cite{Hubble1929}]{\label{fig:HubbleOriPlot}%
\scriptsize {Original plot with the velocity-distance relation measured by Hubble in 1929. For further details see \cite{Hubble1929}.}}
\end{figure}

In 1929 Edwin Hubble published in \cite{Hubble1929} the results of the investigations carried out during the preceding years. He experimentally confirmed what from then on became known as Hubble's law, the linear relation between the distance of the extragalactic sources and their recession velocity, i.e. $v=H_0\cdot d$, with $H_0=525\,{\rm km/s/Mpc}$ (not that far from Lema\^itre's predicted value, although certainly quite far from the current accepted one, which is around $70\,{\rm km/s/Mpc}$). Two important astronomers had a deep influence on Hubble's work, Henrietta S. Leavitt and V. Slipher. I have talked about the latter before. If he could not find out Hubble's law before Hubble is because he did not make use of Leavitt's important discovery\footnote{Slipher noticed in 1917 that most of the nebulae with positive velocities were close to one direction in the sky, whilst the few of the nebulae in the opposite direction had negative velocities. This anisotropy, which turned out to be accidental, did not allow him to even suspect the expansion of the Universe. Slipher suggested that the redshifts could be due to the Milky Way's motion relative to the system of spiral nebulae.}, a relation between the luminosity period of the Cepheid variable stars and their absolute luminosity. This relation allowed Hubble to determine the distance of the extragalactic nebulae to us, what finally allowed to close the Great Debate \cite{GreatDebate} too, in favor of H.D. Curtis, and confirming the old Kant's concept of ``island Universes''. It was the revolutionary confirmation of the fact that the Universe is not only composed by the Milky Way. Two years later, Hubble and Humason published a new paper, \cite{HubbleHumason1931}, estimating $H_0=558\,{\rm km/s/Mpc}$ and reconfirming all the previous results.

It is curious because Hubble's investigations had as main aim the confirmation/rejection of the de Sitter effect, not Lema\^itre's estimation of $H_0$! This can be clearly seen in the conclusions of \cite{Hubble1929}, in which he states: {\it The outstanding feature, however, is the possibility that the velocity-distance relation may represent the de Sitter effect}. But in January 1930, Eddington and de Sitter publicly rejected the latter's theory as inadequate to explain the Hubble law and, almost immediately, Lema\^itre reminded Eddington of his 1927 paper, which was then acclaimed by the latter. This points out that, probably, it would be fairer to name Hubble's law as Lema\^itre-Hubble's law (or even Lema\^itre's law!).

With these observational evidences, the dynamical nature of the Universe became generally accepted. Even by Einstein, who then thought that there was no special reason to keep the CC in his field equations. Applying the principle of economy of thought (Occam's razor criterion), he argued that it had to be {\it rejected from the point of view of logical economy}, see e.g. \cite{EinsteinBook1946}. In this reference he also ensured that

\begin{center}
\parbox{13cm}{{\it If Hubble's expansion had been discovered at the time of the creation of the general theory of relativity, the cosmological member would never have been introduced.}} 
\end{center}

In 1931, shortly after coming back from a three-months stay in the United States during which he had the opportunity of meeting Hubble in Mount Wilson Observatory, Einstein published his new cosmological model in \cite{FriedEinsteinUniv1931}, based on the previous calculations of A. Friedmann \cite{Friedmann1922}. The result was later known as ``Friedmann-Einstein'' Universe. This is the first scientific publication in which Einstein abandoned the notion of a static Universe and explicitly stated that his solution of 1917 is not stable. So his new cosmological view was motivated from both, observational and theoretical reasons. Thus, he set $\Lambda=0$ in Friedmann's equations, and considered positive curvature. The resulting Universe undergoes an expansion followed by a contraction, and ends with a singularity in which (according to Einstein) GR breaks down.

In the paper \cite{EinsteinDeSitter1932} (1932), coauthored by Einstein and de Sitter, they explored the possibility of a flat Universe with no CC, the so-called Einstein-de Sitter model. Experimentally there was no evidence favoring a curved Universe, so it was also a matter of simplicity to dispense with the curvature. It is an interesting model, since it analyzes the special case in which the total matter energy density is equal to the critical one. A lower mass density would imply a hyperbolic spatial geometry, which would produce an ever increasing expansion rate, while a cosmos of higher mass density would be spherical and would eventually collapse. 

Although Einstein renounced to the use of the CC, it never disappeared completely from the scene. Some cosmologists retained this term\footnote{Two examples: A. Eddington wrote in \cite{Eddington1933}, in connection with the Universe age problem: {\it We have no reason to think that $\Lambda$ is not so small as to be entirely beyond observation. [...] I would as soon think of reverting to Newtonian theory as of dropping the cosmical constant}; And Tolman, in \cite{Tolman1934}: {\it For regions of great size, it can be shown that effects could result from a very small value of $\Lambda$. Hence, for cosmological considerations we shall retain the possibility that the quantity $\Lambda$, customarily known as the cosmological constant, may not necessarily be exactly equal to zero}.}, since it was not excluded from observations and made the field equations more general. Despite this, the CC remained in the shadows during many years, waiting for new opportunities to revive the interest of the scientific community. And they indeed arrived in some sporadic occasions. In the early 1960's, A. Sandage reinvoked $\Lambda$ in \cite{Sandage1961} in order to cure some problems related with the age of the Universe, which theoretical prediction turned out to be lower than the age of some stellar populations if the CC was not taken into account. Then, the estimates of the age of these stars decreased and $\Lambda$ became unnecessary again. Later on, in the late 60's, it was noticed that the number counts of quasars as a function of redshift showed a peak at $z\sim 2$ \cite{QSusa,QSurss1,QSurss2}. As a possible explanation, there were some proposals claiming that this could probably be a positive indication in favor of Lema\^itre's loitering model\footnote{It is in this time when Y.B. Zel'dovich wrote his striking paper \cite{Zeldovich1967}, trying to estimate $\Lambda$ by using elements from Particle Physics Theory. See Sect. \ref{subsec:HistPartII}.}. Nowadays it is well-known that the peak is centered at $z\simeq 2.5$, but it is thought to have an explanation of astrophysical origin rather than cosmological. This excess of quasars at this particular redshift is usually interpreted as the evolution in the rate of violent activity in the nuclei of galaxies. Also some early applications of the redshift-magnitude relation to giant elliptical galaxies set the stage for an ephemeral return of the CC. 

Between the mid-60's and the mid-80's two outstanding hits should be remarked: the detection of the Cosmic Microwave Background (CMB) by A.A. Penzias and R.W. Wilson in 1964 \cite{PenziasWilson1965}, which definitively confirmed the correctness of the Big Bang theory\footnote{The existence of this background radiation had been theoretically predicted for the first time by G. Gamow in 1946 \cite{Gamow1946} (see also \cite{Gamow1948}) and two years later by Alpher and Herman \cite{AlpherHerman1948}.}; and the theoretical construction of the inflationary paradigm \cite{Starobinsky1980,Guth1981,Linde1982,PaulSteinhardt1982}\footnote{Some remarkable works which were, indeed, the precursors of such inflationary models appeared between the late 1960's and the beginning of the 1980's. Just to cite some of them: \cite{Gliner1966,Gliner1970,Sakharov1966,Misner1969,Kazanas1980,Sato1981}.}, which constitutes a solid viable solution to the flatness, horizon and monopole problems, and provides a mechanism for the origin of the large-scale structure in the Universe. A non-zero CC was invoked again in \cite{TurnerEtAl1984} and \cite{Peebles1984} in order to reconcile the flat Universe predicted by Inflation and the apparent age of globular clusters with the fact that many estimates for the clustered mass density on large scales suggested that there was an insufficient amount of non-relativistic matter\footnote{See also the corresponding chapter on the $\Omega$ problem in \cite{KolbTurnerBook}}. Although the status of the cosmological constant was good during the 80's and 90's, there were also some studies that seemed to put it on doubt, e.g. the first generation of distant SNIa observations in the mid-90's \cite{SNIaAgainstCC}, in which the authors measured $\Omega^{(0)}_\Lambda<0.51$\footnote{The superscript (0) will refer from now on to the present value of the quantity under consideration. In some occasions also 0 (without parenthesis) is used.}  at the $95\%$ c.l.

\begin{figure}
\centering
\includegraphics[angle=0,width=0.7\linewidth]{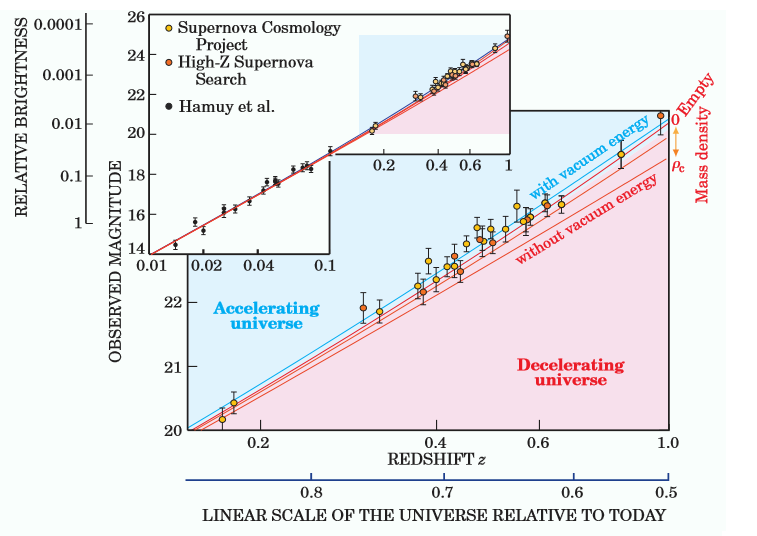}
\caption[Observed magnitude-redshift for the SNIa used to detect $\Lambda>0$ in 1998]{\label{fig:RiessPerlmutter}%
\scriptsize{The observed magnitude-redshift points for well-measured SNIa that led to the detection of a non-null positive $\Lambda$, \cite{SNIaRiess,SNIaPerl}. Also shown the data points (for closer SNIa, at $z<0.1$) measured by the Cal\'an/Tololo Project \cite{Hamuy1,Hamuy2} a few years before the celebrated discovery.}}
\end{figure}

The final confirmation of the non-zero (positive) value of the CC and the current (positive) acceleration of the Universe came precisely by the hand of the accurate measurement of SNIa luminosity distances as a function of the redshift carried out by two independent groups in 1998-1999, the High-Z Supernova Search team \cite{SNIaRiess}, led by Adam Riess and Brian P. Schmidt, and the Supernova Cosmology Project \cite{SNIaPerl}, led by Saul Perlmutter (cf. Fig. \ref{fig:RiessPerlmutter})\footnote{S. Perlmutter, B.P. Schmidt and A. Riess were awarded with the Nobel Prize in 2011 ``for the discovery of the accelerating expansion of the Universe through observations of distant supernovae''.}. These authors measured $\Lambda>0$ with a confidence level of more than $3\sigma$. In the subsequent years these evidences were reinforced with alternative probes, as CMB, Baryon Acoustic Oscillations (BAO), etc., and now this fact is known with even a greater accuracy. For instance, the TT,TE,EE+lowP+lensing analysis carried out by the Planck Collaboration \cite{Planck2015} shows that $\Lambda>0$ with a confidence level of $\sim 79\sigma$, which is a quite remarkable level of statistical significance. Of course, this number can change depending on the data set used to confront the model with observations, but the important point is that there currently exists a huge evidence for the positive acceleration of the Universe's expansion, which remains above the $50\sigma$ c.l. when Planck's CMB data is taken into account (even when only the CMB TT+lowP Planck's data are used). This allows us to assure with high degree of confidence that the CC accounts for roughly the $70\%$ of the critical energy density of the Universe, which translates to the following value of the CC energy density, $\rho_\Lambda\sim 10^{-29}$ g/cm$^3$. One could think that this quantity is a misery (compared, of course, with Earth's standards), but even being a misery, it turns out that this small amount of energy density associated to $\Lambda$ dominates the current expansion of our vast Universe!\footnote{As a simple exercise one can compare the measured value of $\rho_\Lambda$ with the energy density that one would naively ascribe to a hydrogen atom by dividing the proton mass with the third power of the Bohr radius (I do not take into account the negative binding energy between the electron and the proton). The result gives the following density, $\rho_H\sim 11.3\,{\rm g/cm^3}$, which is $30$ orders of magnitude larger than the CC energy density. If $\Lambda$ dominates the current expansion despite being so tiny is because the Universe is extremely empty. The mean matter energy density is approximately equal to one proton per cubic meter.} 

So the statement of the visionary G. Lema\^itre in \cite{Lemaitre1946} (1946), 

\begin{center}
\parbox{13cm}{{\it The history of science provides many instances of discoveries which have been made for reasons which are no longer considered satisfactory. It may be that the discovery of the cosmological constant is such a case,}} 
\end{center}

\vskip 0.3cm
\noindent is very suitable. After one hundred years from the birth of $\Lambda$, we can state that now it is more alive than ever. Although it was introduced by Einstein to obtain a static Cosmos, the CC currently constitutes the ``easiest'' explanation of the positive acceleration of the Universe (what a paradox!). But many questions concerning the ultimate nature of the CC are still unanswered. In the standard cosmological model, the $\Lambda$ - Cold Dark Matter ($\Lambda$CDM) model, $\Lambda$ is usually associated with some form of vacuum energy. Let me give a very naive explanation about the possible link between the cosmological constant and the concept of vacuum energy, since at first glance it is not obvious at all that such a connection exists. As it has been shown in \eqref{eq:EinsteinModFielEq}, Einstein introduced the cosmological term in the left hand side ({\it l.h.s.}) of his field equations, interpreting it as a {\it Universellen Konstante}, as a pure geometrical contribution, without pointing out any fundamental origin. We can easily move this term to the right hand side ({\it r.h.s.}) of \eqref{eq:EinsteinModFielEq}, obtaining
\be\label{eq:EinsteinModFielEq2}
G_{\mu\nu}=8\pi G (T_{\mu\nu}+T_{\mu\nu}^\Lambda)\quad {\rm with}\quad T_{\mu\nu}^\Lambda=\frac{\Lambda}{8\pi G} g_{\mu\nu}\,.
\ee   
Let us consider the energy-momentum tensor of a perfect fluid,
\be\label{eq:EMTPF}
T_{\mu\nu}=(\rho+p)u_\mu u_\nu-p g_{\mu\nu}\,,
\ee  
where $u_{\mu}$ is the $4$-velocity that the fluid has with respect to the observer, and $\rho$ and $p$ are the energy density and isotropic pressure, respectively, measured in the rest frame of the fluid. Now, a quick comparison of the last expression with the CC term of the {\it l.h.s.} of \eqref{eq:EinsteinModFielEq} allows us to determine that $p_\Lambda=-\rho_\Lambda$ (and $\Lambda=8\pi G\rho_\Lambda$). That this is also the EoS that characterizes the vacuum if its associated energy density is constant can be proved in a straightforward way, just by making use of the first law of Thermodynamics applied to the whole Universe. Let us consider a pure de Sitter Universe, only ``filled'' with vacuum. As no form of energy can enter or exit the Universe (by definition the Universe is everything, so it makes no sense to talk about the exterior of it), the variation of internal energy, i.e. $dU$, can only be due to the work exerted by the internal pressure, so we find,
\be\label{eq:dU1}
dU = -p_{\rm vac} dV\,,
\ee  
where $dV$ is a differential increase/decrease of the Universe's volume. On the other hand, we have
\be
dU =d\left(\rho_{\rm vac} V\right)\,.
\ee
Now, if we consider a constant vacuum energy density we finally obtain $dU=\rho_{\rm vac} dV$. By matching the last expression with \eqref{eq:dU1} we find that the EoS for the vacuum is $p_{\rm vac}=-\rho_{\rm vac}$ if the vacuum energy density is constant, as we aimed to prove. This explicitly shows us that any form of vacuum contribution with constant energy density will strictly behave as a CC (and vice versa), just because they enter the field equations with the same mathematical structure.

Another way of reaching the same conclusion is the following. If we locate ourselves in a free-falling inertial frame, we will locally describe the spacetime with the Minkowskian metric $\eta_{\mu\nu}$. Now, we can perform a Lorentz transformation. Vacuum is Lorentz invariant in flat spacetime (actually, it is invariant under the whole Poincar\'e group, which also includes translations). Thus, it must not be sensitive to this transformation, that is, we must measure the same vacuum energy as before. This means that the vacuum energy-momentum tensor $T^{\rm vac}_{\mu\nu}$ must be invariant too, and the only way to achieve this is by demanding $T_{\mu\nu}^{\rm vac}\propto \eta_{\mu\nu}$. Finally, using the principle of general covariance one obtains $T_{\mu\nu}^{\rm vac}\propto g_{\mu\nu}$. And this is precisely the form that takes the CC term in the {\it l.h.s.} of \eqref{eq:EinsteinModFielEq}. One could argue that this reasoning is not very satisfactory, in the sense that, in principle, if we did not stick {\it stricto sensu} to the usual minimal substitution recipe consisting in changing $\eta_{\mu\nu}$ by $g_{\mu\nu}$ when going from Minkowski to a curved spacetime, we would have to consider the possibility of having a vacuum energy momentum-tensor with the general form $T_{\mu\nu}^{\rm vac}=a_0g_{\mu\nu}+a_1R_{\mu\nu}+(...)$ with the $a_i$'s being functions built up from invariants, i.e. a linear combination of a constant, $R$, $R^2$, $R^{\mu\nu}R_{\mu\nu}$, $R^{\mu\nu\beta\alpha}R_{\mu\nu\beta\alpha}$, etc. It would be allowed by the general covariance principle too. Also in this case local Lorentz invariance is recovered, since locally $T_{\mu\nu}^{\rm vac}\propto \eta_{\mu\nu}$. Curiously, this more complicated form of the vacuum energy-momentum tensor is approximately the form that is found from the effective action of Quantum Field Theory (QFT) in curved spacetime (cf. Sect. \ref{subsec:RVMintro}). But the contribution we are interested in at this stage is basically the first one, $T_{\mu\nu}^{\rm vac}=\rho_{\rm vac} g_{\mu\nu}$. This is directly considered a vacuum contribution, whilst the other ones are usually considered as higher derivative corrections to the original Einstein-Hilbert action \eqref{eq:EHaction} or terms that modify the effective value of the Newtonian coupling $G$. 

These two alternative arguments (the thermodynamical and the invariance one) show that, apart from a pure geometrical term in the {\it l.h.s.} of Einstein's equations, we could also receive, in principle, several vacuum contributions in the {\it r.h.s.} with the same mathematical structure of the former. In practice, we cannot disentangle the phenomenological effect of these terms, simply because we only have experimental access to their sum. This is the effective quantity that enters the cosmological formulas that we put under test through the various observational probes,
\be
\rho_{\Lambda,{\rm eff}} = \frac{\Lambda}{8\pi G}+\rho_{\rm ind}\,,
\ee
where the subdindex ``eff'' refers to the effective (measured, physical) value of the cosmological term, namely the sum of the CC term, which is usually considered as the pure vacuum one, what remains after the removal of any other form of energy (from now on I will denote it as $\rho_{\rm vac}\equiv \Lambda /8\pi G$), and the (induced, $\rho_{\rm ind}$) vacuum contributions coming from the various fields existing in Nature. But how many contributions should we include in $\rho_{\rm ind}$? In the next subsection I will try to answer this question. This will allow us to talk about some key concepts, as the one of zero-point energy (ZPE) and the so-called ``old cosmological constant problem'', probably the most notorious conundrum in current theoretical physics. I will explain them together with some historical remarks related with the quantum vacuum.

%%%%%%%%%%%%%%%%%%%%%%%%%%%%%%%%%
% Historical review Part II
%%%%%%%%%%%%%%%%%%%%%%%%%%%%%%%%%

\subsection{Vacuum energy and the CC}
\label{subsec:HistPartII}

Quantum Field Theory tells us that empty space is far from being empty. Vacuum-to-vacuum fluctuations of the matter fields are present even in the absence of interactions and are a pure quantum effect. They are represented by Feynman closed loop diagrams without external legs (vacuum blob diagrams), and account for the contribution of the infinite number of oscillators which can vibrate in all possible frequencies $\omega_{\vec{k}}$. This is a direct consequence of the Heisenberg uncertainty principle. These diagrams add an infinite quantity to the total energy density of the quantum system under consideration. This infinite contribution is equal to the theoretically predicted energy of the system when it is in the ground state. For instance, the Hamiltonian operator corresponding to a free scalar field takes the following form,
\be
\hat{\mathcal{H}}=\int\frac{d^3k}{(2\pi)^3}\frac{\omega_{\vec{k}}}{2}\left(a_{\vec{k}}a_{\vec{k}}^\dag+a_{\vec{k}}^\dag a_{\vec{k}}\right)\,,
\ee
where $a_{\vec{k}}$ and $a_{\vec{k}}^\dag$ are the creation and annihilation operators, respectively, which obey the usual commutation relation $[a_{\vec{p}},a^{\dag}_{\vec{k}}]=(2\pi)^3\delta(\vec{p}-\vec{k})$, and $\omega_{\vec{k}}=(|\vec{k}|^2+m^2)^{1/2}$ is the energy of the mode with wavenumber $|\vec{k}|$ if the scalar particle has mass $m$. The ground state (with $0$ real particles) is then characterized by the following energy,
\be
E=<0|\hat{\mathcal{H}}|0>=\int d^3k\frac{\omega_{\vec{k}}}{2}\delta(0)=V\times\int \frac{d^3k}{(2\pi)^3}\frac{\omega_{\vec{k}}}{2}\,.
\ee
In the second equality I have used the relation $\delta(0)=V/(2\pi)^3$, where $V$ is the (infinite) space volum. Thus, the energy density reads
\be
\rho=\int \frac{d^3k}{(2\pi)^3}\frac{\omega_{\vec{k}}}{2}\,.
\ee
This integral diverges if arbitrarily short modes are taken into account. This is the so-called zero-point energy density of the corresponding quantum field\footnote{The concept of ZPE is not exclusively linked to QFT, but also to more basic Quantum Mechanics. It already appears in e.g. the quantization of the quantum harmonic oscillator.}. This first ZPE contribution (higher order ones are obtained by switching on the interactions) is usually eliminated from the field theoretical calculations by hand using the normal (or Wick) ordered product ansatz, since it is often argued that we cannot measure this infinite amount of energy, but only energy differences with respect to the ground state. This simple argumentation might not work when trying to analyze the gravitational implications. According to GR, in principle one has to take into account all the energy contributions and include them in the {\it r.h.s.} of Einstein's equations. {\it A priori}, all forms of energy are expected to back-react and curve the spacetime in which they ``live'' in, and these effects should be somehow quantifiable\footnote{There are some works in the literature arguing that only vacuum energy effects induced by the curvature of spacetime are, indeed, measurable. This is equivalent to say that the vacuum energy computed in Minkowski spacetime does not gravitate. I will deal with this point in Sect. \ref{subsec:RVMintro}. In connection with it, see also the historical remark on the Casimir effect, in p. 15.}. It is obvious that a big conflict arises then, since the ZPE seems to acquire a huge (technically infinite) value, whilst the measured CC is derisory in comparison. But we will come back to this point later on. Now, let me briefly review some remarkable historical events related with the ZPE. We will see that some aspects of its problematic nature were already hinted in the early days of Quantum Mechanics. 

The concept of ZPE emerged a few years after the birth of Planck's first theory for the black-body radiation, when Planck presented in \cite{Planck2T} his second quantum theory, in which the black-body emitted quanta discretely, but absorbed radiation in a continuous way. He derived the following average energy of an oscillator of frequency $\omega$ in equilibrium with radiation at temperature $T$ (here I restore $\hbar\ne 1$ so as to explicitly show the quantum nature of these energy), 
\be\label{PlanckLaw}
E_\omega=\frac{\hbar\omega}{e^{\hbar \omega/kT}-1}+\frac{\hbar \omega}{2}\,.
\ee
In the limit $T\to 0$ the above formula yields a residual non-null energy, $E_\omega\to\frac{\hbar \omega}{2}$, and as was later noticed by Einstein and Stern in \cite{EinsteinStern1913}, this term is crucial to retrieve the classical result (when $kT\gg \hbar\omega$), $E_\omega\to kT$. Thus, this seemed to indicate that the ZPE could have an important physical role. 

In 1916 W. Nernst \cite{Nernst1916} proposed the idea of an ``empty'' space filled with electromagnetic radiation, what he called ``Licht\"ather'' (light aether). It is probably closer to the field theoretical picture of the vacuum that we have nowadays. He proposed it as a possible mechanism to prevent the heat death of the Universe, by letting the appearance of atoms and chemical elements from the ether fluctuations. He found that this aether seemed to contain an insurmountable amount of energy. But the first one in studying the possible cosmological consequences of these enormous zero-point electromagnetic energies was probably Wilhelm Lenz, in 1926 \cite{Lenz1926}, who analyzed their effect in the framework of Einstein's cosmological model of 1917. He noted that due to the large value of such energy density, its gravitational effects must be somehow neutralized in order to produce a reasonable Cosmos. Afterwards, Wolfang Pauli estimated in the late 20's and in a more formal way in 1933\footnote{Pauli did not publish these calculations, but a reference to his work can be found in \cite{EnzThellung1960}.} which would be the curvature radius in Einstein's static Universe if the CC energy density was given by the electromagnetic zero-point energy. It consists in a quite straightforward calculation. By considering that the quantum soup filling the Universe is formed by photons and using an ultraviolet momentum cutoff $\Lambda_c$ he obtained,
\be\label{eq:PauliEst}
\rho_{em}=<0|\int\frac{1}{2}(\hat{E}^2+\hat{B}^2)d^3x|0>=\frac{1}{2}\int_0^{\Lambda_c}\frac{d^3k}{(2\pi)^3}|\vec{k}|=\frac{\Lambda_c^4}{16\pi^2}\,.
\ee
For the momentum cutoff he took the inverse of the classical electron radius $r_e$, i.e. $\Lambda_c=2\pi/r_e=2\pi m_e/\alpha$ (where $\alpha$ is the fine-structure constant). The resulting Universe's radius reads,
\be
R_E=\Lambda^{-1/2}=\frac{1}{\sqrt{8\pi G\rho_{em}}}=\frac{\alpha^2}{(2\pi)^{3/2}G^{1/2}m_e^2}\sim 32.7\,{\rm km}\,,
\ee
which is extremely small as a direct consequence of the large vacuum energy density computed from \eqref{eq:PauliEst}. In order to counteract the huge repulsive force generated by the CC, all the matter of the Universe must be concentrated in a very tiny volume. It is worth to point out that \eqref{eq:PauliEst} corresponds to the regularized expression for the energy density (it has not been renormalized yet!). After renormalization, the result must be independent of the cutoff regulator $\Lambda_c$. For photons, and after proper renormalization, one obtains $\rho_{em}=0$ because of the massless nature of these particles. But notice that one could have considered vacuum fluctuations of massive particles (of mass $m$), instead of photons, and in this case a quartic term $\rho_{em}\sim m^4$ remains after the corresponding renormalization, i.e. (see Appendix \ref{ch:appZPE} for details). Pauli's calculation is evidently too short in scope, but serves us to see in a quantitative way the intrinsic problems associated with the ZPE and the meaning of the so-called CC problem, i.e. the extraordinary mismatch between the theoretical expectation of $\rho_{\rm ind}$ and the value inferred from observations.

After Pauli's investigations, I would like to remark Lema\^itre's quote in \cite{Lemaitre1934},

\begin{center}
\parbox{13cm}{{\it Everything happens as though the energy in vacuo would be different from zero. In order that absolute motion, i.e. motion relative to vacuum, may not be detected, we must associate a pressure $p =- \rho c^2$ to the energy of vacuum. This is essentially the meaning of the cosmological constant [...]}}
\end{center}

\vskip 0.3cm
\noindent where he explicitly links the CC with the energy inherent to the vacuum, although he does not point out its possible quantum mechanical origin. In that time also Matvei Bronstein explored the idea of $\Lambda$ as a form of cosmic energy in \cite{Bronstein1933}, and he even went a step further, being the first to suggest a time-varying $\Lambda=\Lambda(t)$. While he formerly did it {\it merely for the sake of generality}, he realized that the dynamical evolution of $\Lambda$ could be related with Bohr's suggestion \cite{Bohr1932} that energy might not be conserved in some nuclear processes. He thought that there could be an exchange of energy between the cosmological term and ordinary matter. These revolutionary considerations, though, did not attract a lot of attention in the subsequent years\footnote{As an exception, see the paper of William McCrea \cite{McCrea1951}, published in 1951, where the author considered the creation of matter out of the cosmological term, in the context of the steady-state model.}.

In 1948 H. Casimir predicted in \cite{Casimir1948} what was later on called the ``Casimir effect'', the attraction force between two parallel conducting plates due to changes in the quantum electromagnetic zero-point energy. The first measurement of this phenomenon was performed ten years later by M. Sparnaay \cite{Sparnaay1958}, but still with very large uncertainties. In fact, they were of the same order of the measurement itself. The experimental confirmation of Casimir's theoretical prediction did not arrive until the late 90's \cite{CasimirExp1,CasimirExp2}, when it became evident that the Casimir effect is indeed real and measurable\footnote{Whether Casimir's effect is ultimately explained in terms of vacuum fluctuations or not is still something under discussion, since it is possible to derive such effect without referring to the ZPE of quantum fields \cite{Rugh1999}.}. The uncertainties of these measurements were of the order of the $5\%$ and $1\%$, respectively, and were even reduced during the first years of the 2000's. 

But the first published discussion on the problematic impact of the quantum effects on the value of the CC came in 1967. In his seminal work \cite{Zeldovich1967} (and in more detail, a year later in \cite{Zeldovich1968}), Yakov B. Zel'dovich reformulated the CC problem in the way we know it today. In view of the detected excess of number counts of quasars at $z\sim 2$ several authors \cite{QSusa,QSurss1,QSurss2} proposed to recover a non-null $\Lambda$ together with Lema\^itre's loitering model. According to it, this excess could be caused by the extreme slowing down of the Universe. A plateau in the scale factor $a(t)$ plot (with $t$ being the cosmic time) at $a\sim 1/3$ could be the responsible of this excess, since the emitted light by all the quasars during this period of time (in which $a(t)$ remains almost constant) would be equally redshifted, generating the observed number count's peak at $z\sim 2$. Zel'dovich attempted to give a fundamental explanation to the needed value of $\Lambda$. He used the mass of the proton as the typical mass of particle physics and noticed that the leading ZPE contribution $\rho_{\rm ZPE}\sim m_p^4\sim 1\,{\rm GeV^4}$ is many orders of magnitude larger than the current critical energy density of the Universe, which reads,
\be
\rho_c^{(0)}=\frac{3H_0^2}{8\pi G}=\frac{3}{8\pi}(H_0M_P)^2\sim 10^{-47} {\rm GeV^4}\,,
\ee   
where $H_0\sim 10^{-42}\,{\rm GeV}$ and the Planck's mass $M_P\sim 10^{19}\,{\rm GeV}$. In his own words, {\it it is clear that such an estimate} for $\rho_\Lambda$ {\it has nothing in common with reality}. He was aware that these $47$ orders of magnitude difference between the predicted value of $\rho_\Lambda$ and $\rho_c^{(0)}$ posed a huge problem, of course. In order to alleviate it, he assumed (without many explanations) that the zero-point energies, as well as higher order electromagnetic corrections to them, are effectively canceled to zero in the theory. And then, inspired by Dirac's Large Numbers Hypothesis (LNH) \cite{DiracLNH}\footnote{Dirac pointed out in \cite{DiracLNH} (following the research line initiated by Weyl and Eddington some years before) that the ratio of some physical quantities involving characteristic fundamental constants of the micro and macro-cosmos, as for instance the ratio between the Universe's age $T_u$ and the atomic unit of time, i.e. $\tau\equiv e^2/4\pi\epsilon_0 m_ec^3$, and the ratio of the gravitational force and the electric one between the electron and the proton inside a hydrogen atom, give rise to very similar (and extremely tiny) quantities. For the former, and taking into account that according to Hubble's measurements $T_u\sim 2\,{\rm Gyrs}$, he obtained $\tau/T_u\sim 10^{-40}$, the same as for the latter, $4\pi \epsilon_0 Gm_pm_e/e^2\sim 10^{-40}$. He thought that these coincidences were far from accidental and believed that it was a hint in favor of and underlying deep connection between Quantum Mechanics and the macroscopic world physics. Assuming that atomic parameters cannot change with time and that $\tau/T_u\sim F_{g}/F_{e}$ throughout all the cosmic history, Dirac predicted the variation of Newton's coupling throughout the cosmic history, i.e. $G\sim 1/T_u$.} he extended it by including $\Lambda$, and estimated a more reasonable order of magnitude for the latter. By using the dimensionless quantity characterizing the gravitational interaction for a proton $Gm_p^2/\hbar c\sim 10^{-38}$ and matching it with the ratio between the Compton length of the proton and Einstein's radius he obtained, in natural units: 
\be
\Lambda\sim G^2 m_p^6\longrightarrow \rho_\Lambda\sim G m_p^6\sim 10^{-38}\,{\rm GeV^4}\,.
\ee
This result can be interpreted as the energy density that is due to the gravitational attraction between pairs of virtual protons when the number density of these excitations is equal to $1/\lambda_p^3$, with $\lambda_p=1/m_p$ the Compton length of the proton. Despite the problem was considerably alleviated by proceeding in this way, it still remained, since $\rho_\Lambda$ was roughly $9$ orders of magnitude larger than the critical one! Of course, one could ask: why do we have to use the mass of the proton as the typical characteristic mass of particle physics? Why not, for instance, the mass of the pion, i.e. $m_\pi\sim 0.1$ GeV? In this case, $\rho_\Lambda\sim 10^{-44}\,{\rm GeV^4}$, which is ``only'' three orders of magnitude larger that the critical energy density of the Universe. But still, why should we only take into account the quantum excitations of one type of particle? In principle, this is totally unjustified. All the matter fields of the Standard Model (SM) of particle physics (even those coming from a hypothetical extension of it) contribute with their associated ZPE, which would be different from zero even in the free field theory, as it has been mentioned before. As Zel'dovich wrote in \cite{Zeldovich1968}, {\it the genie has been let out of the bottle, and it is no longer easy to force it back in}! From Zel'dovich's work on, many other authors speculated about the idea of constructing $\rho_\Lambda$ by combining different fundamental constants. But despite it is certainly possible to achieve this (although by using quite feeble and {\it ad hoc} arguments), one must still explain why the ``first order'' ZPE contributions do not gravitate or, at least, why its effects are so weakened with respect what is predicted from theory. 

The realization that the problem becomes further aggravated by the spontaneous symmetry breaking mechanism (or Higgs mechanism) \cite{EnglertSSB,HiggsSSB} in modern gauge theories, a key ingredient of the successful standard model of electroweak interactions, came only a few years later \cite{Linde1974,Veltman1975,Dreitlein1974}. The problem has become even more real after the discovery of the Higgs boson in the Large Hadron Collider (LHC) in 2012 \cite{HiggsDiscovery1,HiggsDiscovery2}. In the SM, the spontaneous symmetry breaking (SSB) mechanism  is used to generate the weak bosons and fermions masses through the use of a complex doublet of scalar fields. This mechanism is the only known way to generate these masses by preserving at the same time the underlying gauge invariance of the theory, which is crucial to make the theory renormalizable. In order to show the role played by the SSB mechanism, let me consider a more simplified framework, in which I only take one real scalar field. The SSB mechanism requires the introduction of a scalar field potential in the SM Lagrangian. If we want to keep the theory renormalizable, then we must omit the use of operators with mass dimensions greater than four. In addition, we do not use odd powers of $\phi$ so as to obtain a symmetric $V(\phi)$. Thus, the classical scalar field potential reads, 
\be\label{eq:HiggsPotential}
V(\phi)=\frac{1}{2}m^2\phi^2+\frac{1}{4!}\lambda\phi^4\,,
\ee
The SSB phenomenon consists in the spontaneous breaking of the symmetry of the vacuum. It is easy to see that if $m^2>0$, then the vacuum state is symmetric\footnote{This simplified Higgs potential is symmetric under a reflection transformation, i.e. $\phi\leftrightarrow -\phi$. Notice that if $m^2>0$, an excitation of the field from the vacuum expectation value can be written as $\phi=\delta\phi$, and $\delta\phi^n=(-\delta\phi)^n$ because $n$ is an even power (see the form of \eqref{eq:HiggsPotential}).}, the vacuum expectation value (VEV) is $v\equiv<0|\phi|0>=0$ and the ground state reads $<V(\phi)>=0$. This absolutely changes if the SSB mechanism comes into play, i.e. when $m^2$ becomes negative\footnote{It is believed that this happened during the electroweak phase transition, $10^{-11}\,{\rm s}$ after Inflation, when the temperature of the Universe cooled down below the electroweak energy scale, i.e. $T_{EW}\sim v\sim 246\,{\rm GeV}$. See \cite{BookGorbunovRubakovHBBT} and references therein.}. In this case, the VEV is non-null, 
\be\label{eq:vevSSB}
v=\sqrt{-\frac{6m^2}{\lambda}}\,,
\ee 
and the symmetry gets broken. It can be seen in a manifest way by substituting $\phi=v+\delta\phi$ in \eqref{eq:HiggsPotential}, where $\delta\phi$ is a small excitation of the field. By doing so we obtain, 
\be
V(\delta\phi)=\frac{m^2}{2}\left(m^2+\delta\phi^2+2v\delta\phi\right)+\frac{\lambda}{4!}\left(v^4+\delta\phi^4+4v^3\delta\phi+4v\delta\phi^3+6v^2\delta\phi^2\right)\,,
\ee
which is clearly not invariant under the mirror transformation, i.e. $V(\delta\phi)\ne V(-\delta\phi)$, due to the presence of the odd powers of $\delta\phi$ in the last expression. Thus, after the SSB the original symmetry disappears, and the potential acquires a non-zero (negative) value, which represents the classical contribution of the Higgs potential to the overall vacuum energy density,
\be\label{eq:VSSB}
V^{clas}_{Higgs}=-\frac{3}{2}\frac{m^4}{\lambda}=\frac{1}{4}m^2v^2<0\,.
\ee
We can write the last result in terms of the mass of the Higgs boson $m_H$ and Fermi's scale $M_F$, by taking into account 
\be
m_H^2=\frac{\partial^2 V}{\partial\phi^2}\bigg\rvert_{\phi=v}=m^2+\frac{1}{2}\lambda v^2=-2m^2\,,
\ee
and
\be
M_F=G_F^{-1/2} \qquad \frac{G_F}{\sqrt{2}}=\frac{g^2}{8M_W^2}=\frac{1}{2v^2}\,,
\ee  
with $G_F$ being Fermi's constant, $g$ the weak coupling and $M_W$ the mass of the charged weak bosons W$^{\pm}$, with $M_W=\frac{1}{2}gv$. Plugging these expressions in \eqref{eq:VSSB} one obtains:
\be\label{eq:VhiggsClassical}
V^{clas}_{Higgs}=-\frac{1}{8\sqrt{2}}m_H^2M_F^2\sim -1.19\cdot 10^{8}\,{\rm GeV^4}\,,
\ee
where I have used $m_H\sim 125\,{\rm GeV}$ and $M_F\sim 293.132\,{\rm GeV}$. This contribution to the cosmic vacuum energy density is, again, quite impressive. It surmounts the measured value of the CC energy density, i.e. $\rho_\Lambda\sim 10^{-47}\,{\rm GeV^4}$, in $55$ orders of magnitude, a huge quantity. Thus, it worsens even more the cosmological constant problem, since the Higgs particle has been detected in the lab and, therefore, its existence is already an observational fact. The associated Higgs potential must be there, just because it is inherent to the Higgs particle itself. And not just the classical part of the Higgs potential, but also its associated quantum corrections, entering at all orders in perturbation theory, i.e.
\be
V_{Higgs}=V^{clas}_{Higgs}+\hbar V^{(1)}_{Higgs}+\hbar^2 V^{(2)}_{Higgs}+...\,,
\ee
where I explicitly write the $\hbar$ dependence in order to show up the quantum nature of the corrections to the classical term $V_{Higgs}^{clas}$ computed in \eqref{eq:VhiggsClassical}.

On top of the zero-point energies of the matter fields and the induced contribution of the Higgs potential to the CC, also some other additional vacuum terms could come into play. For instance, those predicted in Quantum Chromodynamics (QCD). Due to the asymptotic freedom property of its gauge coupling, which makes QCD non-perturbative at low energies, the vacuum structure of this SM sector is extremely involved, and it is still an area under intense study. Despite this, it is commonly asserted that low-energy QCD is responsible for the existence of quark and gluon condensates which make the vacuum expectation values of the quark and gluon fields to be different from zero. Taking into account that the QCD energy scale is around $\Lambda_{QCD}\sim 0.2\,{\rm GeV}$, one expects to receive contributions of order $\rho_{QCD}\sim \Lambda_{QCD}^4/16\pi^2\sim 10^{-5}\,{\rm GeV^4}$ to the overall vacuum energy, which is again totally incompatible with the observed value of $\rho_\Lambda$, since their ratio yields an exorbitant number, i.e. $\rho_{QCD}/\rho_{\Lambda}\sim 10^{42}$.

Let me summarize the situation. The measured value of the CC should be the result of the following sum:
\begin{eqnarray}
\rho_{\Lambda, {\rm eff}} &=& \frac{\Lambda}{8\pi G} + V^{clas}_{Higgs}+\hbar V^{(1)}_{Higgs}+\hbar^2 V^{(2)}_{Higgs}+...+\nonumber\\
&& +\rho_{QCD}+\hbar V^{(1)}_{ZPE}+\hbar^2 V^{(2)}_{ZPE}+...
\end{eqnarray}
where $V^{(i)}_{ZPE}$ accounts for the contributions of all the matter fields to the ZPE at ith order, and each term $V^{(i)}_{Higgs}$ accounts for the quantum corrections to the classical Higgs potential at ith order too. $\Lambda$ is the bare cosmological constant, i.e. the pure geometrical term appearing in Einstein's field equations. Substituting some of these terms by their approximate numerical values we find,
\begin{eqnarray}\label{eq:sumVac}
10^{-47}\,{\rm GeV^4}&\approx& \frac{\Lambda}{8\pi G} -10^{8} \,{\rm GeV^4}+\hbar V^{(1)}_{Higgs}+\hbar^2 V^{(2)}_{Higgs}+...+\nonumber\label{eq:FineTun}\\
&& +10^{-5}\,{\rm GeV^4}+\hbar V^{(1)}_{ZPE}+\hbar^2 V^{(2)}_{ZPE}+...
\end{eqnarray}
where we have seen before that the individual contributions of the various fields to the ZPE at the lowest order, i.e. $V^{(1)}_{ZPE}$, depend on the masses of the particles, but generally give rise to a number which exceeds in many orders of magnitude the measured value $\rho_{\Lambda,{\rm eff}}$\footnote{Only a neutrino mass of order of the meV seems to give an acceptable estimate, since in this case $\rho_\nu\sim(1\,{\rm meV})^4\sim 10^{-48}\,{\rm GeV^4}$.}. In addition, it is well-known that the quantum corrections of the Higgs potential are also huge. Actually, they are sizable to $\rho_{\Lambda,{\rm eff}}$ even at $21$st order of perturbation theory, in which there are thousands of loop diagrams involved in the calculation\footnote{See \cite{SolaReview2013} for a very detailed discussion on this point and the vacuum energy in general.}! In addition, we must bear in mind that the Universe has undergone different phase transitions, what means that some of the contributions appearing in \eqref{eq:sumVac}, e.g. the electroweak one, have not remained constant throughout the entire cosmic history\footnote{See \cite{ShapiroSola1999}, where a first theoretical attempt to overcome such difficulties is presented.}. This complicates even more the situation and shows in a quite obvious way the intricate status of the ``old'' cosmological constant problem, which basically consists in an incredible fine-tuning puzzle. If one wants to compute $\rho_{\Lambda,{\rm eff}}$ by working at a certain order of perturbations one has to split the bare CC $\Lambda$ in order to produce a counterterm $\delta\Lambda$ capable of absorbing all these huge quantities. This is the usual renormalization procedure amply used in QFT to remove the infinities coming from the divergent loop diagrams (after applying, of course, the usual regularization procedure in a concrete scheme). But this positive counterterm must be very precisely fine-tuned in (at least) $55$ digits so as to cancel the leading electroweak contribution, and this does not seem very natural. Moreover, it is worth to notice that when we change the order of perturbation theory in which we are working, we are forced to readjust this counterterm several orders of magnitude in order to keep $\rho_{\Lambda,{\rm eff}}$ equal to the observed one, which is also a clear sign of the rawness of the problem we are facing.

One of the possible shortcuts to alleviate the CC problem would be the existence of a symmetry in the theory that allowed the exact cancellation of the vacuum energy. Pauli pointed out that the total ZPE density could be set to zero by imposing a carefully fine-tuned cancellation between boson and fermion contributions \cite{Pauli1951}. This could be achieved in the context of non-broken Supersymmetry (SUSY), as was shown in \cite{WessZumino1974} more than $40$ years ago. But we know that SUSY cannot be perfectly realized in Nature (it must be, at least, softly broken), since the theoretically predicted superpartners of the known particles of the SM in non-broken SUSY have not been detected yet in any accelerator (not even in the LHC). This points out that these superpartners must have a larger masses, i.e. $>\mathcal{O}(1\,\rm TeV)$. Thus, theories incorporating SUSY are not able to sufficiently improve the status of the problem with respect to the SM case that has been discussed above \cite{WeinbergReview1989}. Moreover, we must bear in mind that even if this theory was able to produce an exact cancellation of the vacuum energy, we would still have to explain the small measured value of $\rho_{\Lambda,{\rm eff}}$. Neither string theory manage to solve the problem, with its ``string landscape'' with more than $10^{500}$ possible vacua, together with the anthropic arguments employed \cite{SusskindString}.

Without any shadow of doubt, revealing the ultimate origin of the measured cosmological term has become one of the greatest motivations of the present-day physics community. The CC, what according to G. Gamow was considered by Einstein himself his ``greatest blunder'' \cite{GamowEinsteinBlunder}, has turned out to be one of the cornerstones of the standard cosmological model \footnote{Right next, after quoting Einstein's words, Gamow wrote in \cite{GamowEinsteinBlunder} the following
prescient words, which still resonate in our ears with undamped strength: ``But this `blunder', rejected by Einstein, is still sometimes used by cosmologists even today, and the cosmological constant denoted by the Greek letter $\Lambda$ rears its ugly head again and again and again''.}. Since the end of the last century we can assure that $\Lambda$ or some other sort of dark energy (DE)\footnote{The term ``dark energy'' was coined in \cite{HutererTurner1999} to refer in a generic way all the possible forms of unclustered energy able to cause the current positive acceleration of the Universe, due to their negative pressure.} behaving in a very similar way is there, pervading the entire Universe, and dominating its expansion. Whether it is possible to solve the fundamental cosmological constant problem with the current available theoretical tools or not is something debatable, but in view of the failure of the not small amount of different approaches undertaken in the last years to explain the observed tiny value of the (effective) CC, it is worth to go on with the research, trying to characterize $\Lambda$ as best as possible in order to obtain more hints that help to ease the construction of the appropriate theoretical framework that finally manage to explain in a self-consistent way what the hell is triggering the positive acceleration of the Universe. Moreover, we should welcome those frameworks that, despite not solving completely the CC problem, are able to shed some light on it and to explain the current experimental data with proficiency. 

This thesis focuses on one of these interesting frameworks, which rises high as a viable alternative to the $\Lambda$CDM. Motivated from QFT in curved spacetime, the class of Running Vacuum Models (RVM's) that is studied in the main body of this manuscript consider a time-evolving cosmological term, together with a possible time evolution of the Newtonian coupling and/or the anomalous conservation law for the matter and/or radiation components that fill the Universe. This could be potentially linked to the decay of the vacuum through some unknown mechanism and/or to the variation of other fundamental constants (as the QCD scale or the fine structure constant) throughout the cosmic history. Quite remarkably, these models are capable of fitting the available observational data with astonishing accuracy, even much better than the $\Lambda$CDM. Thus, after many years working towards the improvement of the characterization of $\Lambda$, we are now probably starting to be in a position to answer one of the most ``basic'' (and relevant!) questions affecting its nature: is $\Lambda$ (or more generically, the DE) dynamical? One of the main outcomes of this work is that we now have important evidences in favor of such dynamical character of the DE\footnote{A dynamical form of DE is theoretically very welcomed because: 1) it leaves the door opened to the existence of a possible link between the late-time acceleration of the Universe and the mechanism that triggered Inflation, which is characterized by a much larger energy scale; 2) it is something needed to solve the so-called ``coincidence'' problem associated to a rigid CC. See \ref{subsec:TPLCDM} for a detailed discussion on this point.}! 

But before introducing the RVM's from a theoretical perspective and, of course, before talking about their phenomenological performance, it is highly convenient to explain the most relevant features of the standard cosmological model, together with its theoretical and experimental status, and the possible alternatives that have been proposed along the years to cure or try to alleviate some of the problems related with it. It will turn out to be very practical to have the $\Lambda$CDM formulas handy, since it will ease the future comparisons with the expressions of the RVM's, and also because it will serve to set the notation convention that will be used throughout the entire dissertation. Hence, I hope this quick review will be extremely useful, especially for those readers that are not familiar with the $\Lambda$CDM.

%%%%%%%%%%%%%%%%%%%%%%%%%%%%%%%%%%%%%%%%%%%%%%%%%%%%%%%%%%%%%%%
%%%%%%%%%%%%%%%%%%%%%%%%%%%%%%%%%%%%%%%%%%%%%%%%%%%%%%%%%%%%%%%
%%%%%%%%%%%%%%%%%%%%%%%%%%%%%%%%%%%%%%%%%%%%%%%%%%%%%%%%%%%%%%% 

\section{The standard cosmological model}
\label{sec:LambdaCDM}

The standard cosmological model, also known as $\Lambda$CDM, vanilla, or concordance model, explains the expansion of the Universe from a very small fraction of a second after the Bang up to the undergoing accelerated phase, going of course by the various fundamental processes occurred in the meanwhile, as e.g. the nucleosynthesis of light elements \cite{AlpherBetheGamow1948}, the decoupling of matter and radiation (the origin of the CMB, see p. 9 for references), or the formation of the large-scale structures in the Universe. The $\Lambda$CDM incorporates elements of a very diverse spectrum of physical branches, ranging from astronomy to nuclear and particle physics. Even computing has been important to establish its foundations. It makes use of quantum mechanics and general relativity, and some of its extensions (as e.g. Inflation) are also a good playground for those theories that try to unify the four forces of Nature. The success of the model lies in its ability to properly fit a large variety of observational data. But there are still open questions affecting not only theoretical crucial points (as we have seen in the previous section), but also experimental ones. A walk through this section will enable us to pinpoint the strength of the model and also its weak side.

The layout of this part of the thesis dedicated to the standard cosmological model is the following. I review the basic background formulas of the model in \ref{subsec:BasicsLCDM}. Next, I comment on its current theoretical and observational status in \ref{subsec:TPLCDM} and \ref{subsec:OPLCDM}, respectively. This will motivate the introduction of some alternative models in Sect. \ref{subsec:AlternativesLCDM}. Their main ideas will be presented in a very concise way, together with their possible associated drawbacks. I just aim to offer the reader a panoramic vision of the possible options one can find in the literature, all of them trying to shed some light in the various problems affecting the standard cosmological model with a rigid $\Lambda$ term.  

\subsection{Some basics}
\label{subsec:BasicsLCDM}

In the standard big bang cosmological model the cosmic dynamics is described by GR and its field equations \eqref{eq:EinsteinModFielEq}. Two key features determine the form of the metric tensor: the large-scale isotropy and homogeneity of the Universe. These properties, which evidently can only be checked in the region to which we have observational access, are postulated in the Cosmological Principle. In fact, we cannot directly test homogeneity, since we cannot observe the large-scale structure on 3D spatial hypersurfaces. We observe on the past light cone at effectively a single cosmological time, which means that we can only directly test isotropy about our world line. Isotropy about all galaxy world lines implies homogeneity, so if our world line is not special, i.e. if we adopt the Copernican Principle, then we can deduce the Cosmological Principle on the basis of isotropy. Isotropy is exceedingly sustained by radiation backgrounds as CMB observations \cite{COBE,WMAP9,Planck2015}\footnote{Of course, the CMB does not have a perfect black-body spectrum. It shows very faint anisotropies of the order of $\Delta T/T\sim 10^{-4}-10^{-5}$, once the dipole contribution is removed (it is generated by our peculiar motion with respect to the CMB rest frame). These tiny fluctuations do not alter the cosmology at the background level, in which no perturbations are considered. They can be neglected in first approximation without giving rise to sizable errors affecting the background expansion.} and counts of sources observed at wavelengths ranging from radio to gamma rays; and homogeneity in 2D spatial (angular) hypersurfaces by the measurements and statistical analysis carried out by several galaxy surveys, e.g. the WiggleZ Dark Energy Survey \cite{WiggleZsurvey} or the Baryon Oscillation Spectroscopic Survey (BOSS) \cite{BOSSsurvey}. See the recent paper \cite{HomogeneityBOSSDR12}, in which it is concluded that the matter distribution in our Universe becomes homogenous at scales larger than $(63.3\pm 0.7)h^{-1}$ Mpc. These observational facts motivate the use of the so-called Friedmann-Lema\^itre-Robertson-Walker (FLRW) metric, through which the Cosmological Principle is explicitly implemented\footnote{Apart from the works of Friedmann and Lema\^itre discussed in Sect. \ref{subsec:HistPartI}, see also \cite{Robertson,Walker}.}, 
\be\label{eq:FLRW}
ds^2=dt^2-a^2(t)\left[\frac{dr^2}{1-k\frac{r^2}{r_0^2}}+r^2 (d\theta^2+sin^2\theta\,d\varphi^2)\right]\,,
\ee
where $t$ is the cosmic time, $(0\leq r<r_0,0\leq \theta\leq\pi,0\leq \varphi\leq 2\pi)$ are the comoving coordinates and $k$ is the curvature parameter, which can take the values $+1$, $-1$ or $0$ depending on whether the space geometry is closed, open or flat, respectively. The scale factor $a(t)$ gives us important information, since it allows us to translate radial comoving distances into physical ones at a given cosmic time, just by computing the product $d_{\rm phys}=a(t)\Delta r$. In addition, $a(t)$ allows us to perform the link between observations and the theoretical predictions, in the sense that we cannot directly measure the cosmic time at which a given cosmological event occurred, but only how the wavelength $\lambda_e$ of the light emitted by this object has been lengthened through its cosmic walk until reaching our position, where we measure $\lambda_o$. This is possible thanks to the available spectroscopic techniques. Usually, this information is encoded in the redshift $z$. If we choose the scale factor to be $a_0=1$ at present, the following relations apply,
\be\label{eq:reshift}
\lambda_e=a\lambda_o\qquad a\equiv\frac{1}{1+z}\qquad z=\frac{\lambda_o}{\lambda_e}-1\,.
\ee
Now, let me set the notation regarding the elements of differential geometry entering the field equations. The sign convention that will be used throughout this thesis is the timelike one, also denoted as $(-,+,+)$ in the table of sign conventions of Ref. \cite{MisnerThorneWheeler}. That is, apart from the signature $(+,-,-,-)$ for the metric and the $+$ sign in the {\it r.h.s.} of \eqref{eq:EinsteinModFielEq}, I make use of 
\be\label{eq:RiemannTensor}
R^{\alpha}{}_{\beta\gamma\delta}=\Gamma^{\alpha}{}_{\beta\delta,\gamma}-\Gamma^{\alpha}{}_{\beta\gamma,\delta}+\Gamma^{\lambda}_{\beta\delta}\Gamma^{\alpha}{}_{\lambda\gamma}-\Gamma^{\lambda}_{\beta\gamma}\Gamma^{\alpha}{}_{\lambda\delta}\,,
\ee
\be\label{eq:RicciTensor}
R_{\beta\delta}=R^{\alpha}{}_{\beta\alpha\delta}\,,
\ee
for the Riemann and Ricci tensors, respectively. The Chrystoffel symbols and the Ricci (or curvature) scalar are computed as usual,
\be\label{eq:ChrystoffelSymbols}
\Gamma^{\mu}{}_{\nu\alpha}=\frac{1}{2}g^{\mu\beta}(g_{\nu\beta,\alpha}+g_{\alpha\beta,\nu}-g_{\nu\alpha,\beta})\,,
\ee
\be\label{eq:RicciScalar}
R=g^{\mu\nu}R_{\mu\nu}\,.
\ee
At the background level, one can treat the different components filling the Universe (radiation, non-relativistic matter and $\Lambda_{\rm eff}$) as perfect fluids uniformly distributed in space, so they can be described by the energy-momentum tensor \eqref{eq:EMTPF}. This is totally consistent with the Cosmological Principle introduced before. Plugging in \eqref{eq:EinsteinModFielEq} the FLRW metric and the geometrical expressions presented above, together with the aforementioned energy-momentum tensors, and choosing the $00$ and ij components, one obtains the Friedmann and pressure equations, respectively\footnote{See Appendix  \ref{ch:appPert} for the explicit calculation of these objects and equations with the FLRW metric at both, background and perturbed levels.}. They read,
\be\label{eq:FriedmannLCDM}
H^2=\frac{8\pi G}{3} (\rho_M+\rho_{\Lambda,{\rm eff}})-\frac{k}{a^2}\,,
\ee  
\be\label{eq:PressureLCDM}
3H^2+2\dot{H}=-8\pi G (p_M+p_{\Lambda,{\rm eff}})-\frac{k}{a^2}\,,
\ee
where $\rho_M=\rho_m+\rho_r$, being $\rho_m$ and $\rho_r$ the matter and radiation energy densities, respectively, and $H=\dot{a}/a$ the Hubble function, with the dot denoting a derivative with respect to the cosmic time. These two independent equations govern the expansion of the Universe. It is important to notice that we ``only'' deal with two equations and, therefore, we can only have two unknown functions, since otherwise the system is not determined and one obtains at most a degenerate solution. This situation is more relaxed than it seems {\it prima facie} because all the perfect fluids can be characterized by their EoS at the background level, $p_i=\omega_i\rho_i$, and this reduces the number of unknowns. For non-relativistic matter $\omega_m=0$, for radiation $\omega_r=1/3$, whereas for the effective cosmological term $\omega_{\Lambda,{\rm eff}}=-1$, since we treat it as pure vacuum. The combination of \eqref{eq:FriedmannLCDM} and \eqref{eq:PressureLCDM} yields the acceleration equation,
\be\label{eq:acceEq}
\frac{\ddot{a}}{a}=-\frac{4\pi G}{3}\sum_{i=1}^{N}\left(\rho_i+3p_i\right)=\frac{4\pi G}{3}\left(2\rho_{\Lambda,{\rm eff}}-2\rho_r-\rho_m\right)\,,
\ee
where in the second equality I have already used the EoS parameters for the various components. From here it becomes clear that $\rho_m$ and $\rho_r$ slow down the expansion, whilst $\rho_{\Lambda,{\rm eff}}$ speeds it up, since all the energy densities are positive. In general, a fluid contributes positively to the acceleration if $\rho_i+3p_i<0$, as it is plain from \eqref{eq:acceEq}. It is also convenient to define the density parameters,
\be\label{eq:DensityPara}
\Omega_i(a)=\frac{\rho_i(a)}{\rho_c(a)}\qquad \Omega_k(a)=-\frac{k}{[aH(a)]^2}\,,
\ee
where the subscript $i$ labels the various components entering the model, and 
\be
\rho_c(a)=\frac{3H^2(a)}{8\pi G}
\ee
is the critical energy density of the Universe. Let me now comment on a couple of aspects related with the curvature. The analysis TT+lowP+lensing+BAO carried out by the Planck Collaboration in \cite{Planck2015} shows that $\Omega_k^{(0)}=0.000\pm0.005$ at $2\sigma$ c.l. Taking this into account together with the second expression of \eqref{eq:DensityPara}, this means that we can ensure with a $\sim 95\%$ of c.l. that the radius of the Universe (in case it is closed) is at least one order of magnitude larger than the radius of the visible Universe\footnote{See also Ref. \cite{RatraSugiyama2017}, where the authors find $\Omega_k^{(0)}=-0.008\pm0.002$ at $1\sigma$ c.l. upon applying the same data set, but using a modified shape of the primordial power spectrum which according to them is more consistent with the starting hypothesis of a non-flat Universe.}. The measured low value of $\Omega_k^{(0)}$ tells us that the current total energy density of the Universe $\rho^{(0)}_{\rm tot}$ is very near (if not exact to) the critical one, i.e. $\Omega^{(0)}_{\rm tot}\equiv\Omega^{(0)}_m+\Omega^{(0)}_r+\Omega^{(0)}_{\Lambda,{\rm eff}}\approx 1$. This poses a huge problem of fine-tuning in the initial conditions of the radiation-dominated expansion, i.e. let us say at $a_i$. It is the following. By means of the Friedmann equation \eqref{eq:FriedmannLCDM} it is possible to see that $[\Omega_{\rm tot}^{-1}(a)-1]\rho_{\rm tot}(a) a^2=-3k/8\pi G$, the {\it r.h.s.} being obviously a constant. Thus, one finds $\rho_{\rm tot}(a_i)a_i^2/\rho^{(0)}_{\rm tot}a_0^2\sim 10^{53-59}$ \cite{Guth1981}, depending on the energy scale at which inflation ends, typically at some point in between of the GUT scale M$_{\rm GUT}\sim 10^{16-17}$ GeV and Planck's one. Here $a_0$ is again the current value of the scale factor. This and the fact that $\Omega_{\rm tot}^{(0)}\approx 1$ forces $\Omega_{\rm tot}(a_i)$ to be extremely fine-tuned to $1$, concretely in more than these $53$-$59$ digits! This problem\footnote{Indeed, it is only a problem if we consider that $k\ne0$. If the Universe was perfectly flat, i.e. if $k=0$, then no fine-tuning would be needed because in this case $\Omega_{\rm tot}(a)=1,\,\forall a$. But having a perfect flat Universe seems quite unnatural, despite being in fact compatible with current observations.} is cured by the inflationary scenario, which strictly speaking is not part of the $\Lambda$CDM model. In this framework, a quasi-de Sitter phase with exponential growth triggered by the presence of a very energetic component (or components, in plural) of associated negative pressure allows to flatten the space and the needed value of $\Omega_{\rm tot}(a_i)$ is obtained in a natural way, i.e. without fine-tuning\footnote{Inflation is also able to cure other important problems, see Sect. \ref{subsec:TPLCDM} for details.}. Summing up, a very flat Universe is favored by current observations and this can be reconciled with the standard cosmological model if Inflation is incorporated to the latter. From now on we will neglect the curvature term in all the cosmological formulas, by effectively taking $k=0$. The reader must keep in mind, though, that current observations do not rule out the $k\ne 0$ scenario \cite{Planck2015,RatraSugiyama2017}, and this is an important point worth to recall for future investigations.

Another key relation can be found thanks to the Bianchi identities, 
\be\label{eq:BianchiIndentity}
\nabla^\mu G_{\mu\nu}=0\,, 
\ee
which by construction of the Einstein tensor are automatically satisfied. Upon the application of the covariant derivative to \eqref{eq:EinsteinModFielEq}, and picking the temporal component we get:
\be\label{eq:consBianchi}
\nabla^\mu T_{\mu 0}=0\longrightarrow \sum_{i=1}^N\left[\dot{\rho}_i+3H(\rho_i+p_i)\right]=0\,,
\ee
where $N$ is the total number of perfect fluid components. Using the corresponding EoS parameter for each of them one can use this equation to obtain the background evolution of all the energy densities. This equation can be also obtained as a first integral of the system \eqref{eq:FriedmannLCDM}-\eqref{eq:PressureLCDM}. It is very useful, since for non-coupled components one finds some extra relations that enlarge the two initial independent equations. In terms of the scale factor they read,
\be\label{eq:rhoConsEq}
\rho^\prime_i+3\frac{\rho_i}{a}(1+\omega_i)=0\qquad i=1,...,N\,,
\ee
with the prime denoting, of course, a derivative with respect to the scale factor. A simple integration yields, 
\be\label{eq:rhoaw}
\rho_i(a)=\rho_i^{(0)}a^{-3(1+\omega_i)}\,,
\ee
and therefore,
\begin{center}
\begin{equation}
\begin{array}{rrcll} 
{\rm Radiation}:& \quad\rho_{r}(a)&=&\rho_r^{(0)}a^{-4} &\,,\\
{\rm Matter}:& \quad\rho_{m}(a)&=&\rho_m^{(0)}a^{-3}&\,,\\
{\rm \Lambda_{\rm eff}}:& \quad\rho_{\Lambda,{\rm eff}}(a)&=&\rho^{(0)}_{\Lambda,{\rm eff}}&\,.\\
\end{array}
\label{eq:DensitiesLCDM}
\end{equation}
\end{center}
This allows us to completely close the system, by computing the Hubble function with the use of e.g. \eqref{eq:FriedmannLCDM},
\be\label{eq:HubbleLCDM}
H^2(a)=H_0^2\left[1+\Omega_m^{(0)}(a^{-3}-1)+\Omega_r^{(0)}(a^{-4}-1)\right]\,,
\ee
where $H_0\equiv H(a=1)$ is the Hubble constant, which is fixed by observations. It is usually expressed as $H_0=100h\,{\rm km\,s^{-1}Mpc^{-1}}$, with $h$ being the dimensionless Hubble parameter. The pressureless non-relativistic component results from the sum of baryonic and cold dark matter (CDM)\footnote{We must differentiate CDM from hot dark matter (HDM). The former was non-relativistic when it decoupled from photons, whilst the latter was still relativistic in that time. The $\Lambda$CDM relies on CDM to explain the amount of structure observed in the Universe \cite{FrenkWhite2012}. Although the existence of HDM is not ruled out, the CMB puts severe constraints on it. In mixed models of CDM+HDM the amount of HDM is limited to be at most of a few percent.}, i.e. $\rho_m(a)=\rho_b(a)+\rho_{dm}(a)$. Baryons accounts for the $\sim5\%$ energy content of the Universe, whereas DM for the $\sim25\%$. The latter interacts gravitationally with the former, but not electromagnetically. In principle, a very weak interaction of non-gravitational nature with standard particles is not excluded by experiments, although despite the big efforts made so far there has not been any direct detection neither. There are also some hints of non-null very weak interactions between DM particles coming from astronomical observations. The existence of dark matter is needed to properly fit the cosmological data, in which it plays a crucial role e.g. in the large-scale structure formation process. Moreover, it is also required to explain several phenomena at lower scales, as the flat rotation curves of spiral galaxies, the orbital peculiar motions of individual galaxies inside clusters, or the results from the weak lensing mass reconstruction of the interacting Bullet cluster. It is possible that a very small fraction of DM is explained by baryonic objects like the so-called Massive Astrophysical Compact Halo Objects (MACHO's) or cold molecular gas clouds, which are both very difficult to detect, but still, the biggest part of the DM component must be described by some kind of non-baryonic particle, as for instance the axion or the Weakly Interacting Massive Particle (WIMP), which is usually predicted in supersymmetric theories\footnote{For a short review on DM, see \cite{ReviewDM}.}.

Radiation contributes nowadays with less than the $0.01\%$ of the total energy content of the Universe. It is only composed by photons and neutrinos, since in the standard $\Lambda$CDM model no HDM is considered. Both components are relativistic and its energy densities read, 
\be\label{eq:PhotonAndNeutrinoDens}
\rho_\gamma=\frac{\pi^2}{15}T_\gamma^4 \qquad \rho_\nu=N_{\rm eff}\frac{7\pi^2}{120}T_\nu^4\,,
\ee 
where $N_{\rm eff}=3.046$ \cite{Mangano2005} is the effective number of neutrino species, and $T_\gamma$ and $T_\nu$ are the photon and neutrino temperatures, respectively. They are linked via the relation $T_\nu/T_\gamma=(4/11)^{1/3}$, which is obtained from the conservation of entropy before and after the $e^{-}/e^{+}$ annihilation, which happened before the BBN. Using these expressions we can write the total radiation energy density as follows, 
\be\label{eq:TotalRadDens}
\rho_r = \rho_\gamma+\rho_\nu=\frac{\pi^2}{15}T_\gamma^4\left[1+\frac{7}{8}\left(\frac{4}{11}\right)^{4/3}N_{\rm eff}\right]\,.
\ee 
The present value of $T_\gamma$ corresponds to the measured temperature of the CMB photons, $T_\gamma=2.72548\pm 0.00057$ \cite{Fixsen2009}. Using this we can express the current radiation density parameter only as a function of the dimensionless Hubble parameter,
\be\label{eq:RadFit}
\Omega_r^{(0)}=4.18343\cdot10^{-5}\,h^{-2}\,.
\ee
After Inflation, radiation dominated the dynamics of the Universe. According as it got redshifted by the expansion ($\nu\propto 1/a$, being here $\nu$ the frequency of the photon), its energy density decayed faster than matter's one, so there was a moment (usually referred to as matter-radiation equality time) in which $\rho_m$ surpassed $\rho_r$. In the $\Lambda$CDM model this is predicted to happen at  
\be
z_{\rm eq}=\frac{\Omega^{(0)}_m}{\Omega^{(0)}_r}-1\sim 3500\,.
\ee 
From this moment on non-relativistic matter became dominant and DM structures began to form by gravitational attraction. This process was enhanced after recombination, when baryons released from photons and became susceptible to the DM gravitational potential wells that had been formed before. It occurred at $z_{*}$ (in the so-called last scattering surface, where the CMB photons come from). It takes a value close to $z_{*}\approx 1090$, but its precise value depends weakly on the parameters, and it can be obtained from the fitted formula\,\cite{HuSugiyama},
\begin{equation}\label{zlastscatt}
z_*=1048\,\left[1+0.00124\,\left(\Omega_b^{(0)}
h^2\right)^{-0.738}\right]\left[1+g_1 \left(\Omega_m^{(0)}
h^2\right)^{g_2}\right]\,,
\end{equation}
with
\begin{equation}\label{zlastscatt2}
g_1=\frac{0.0783\left(\Omega_b^{(0)}
h^2\right)^{-0.238}}{1+39.5\,\left(\Omega_b^{(0)}
h^2\right)^{0.763}}\ \   \qquad
g_2=\frac{0.560}{1+21.1\left(\Omega_b^{(0)} h^2\right)^{1.81}}\,.
\end{equation}
Before concluding this section, let me show the relation between the scale factor and the cosmic time in those epochs in which the Universe is dominated by a single component. In this case, one can neglect the other density contributions and take only into account the energy density of the dominant component in \eqref{eq:HubbleLCDM}. After integrating the resulting expression, one obtains $a(t)$. For a radiation or matter dominated (RD or MD, respectively) Universe we find,
\be
a(t)\sim t^{\frac{2}{3(1+\omega_i)}}\,.
\ee
This is an important analytical result that tells us that in a RD Universe $a(t)\sim t^{1/2}$, whereas in a MD one $a(t)\sim t^{2/3}$. In case $\Lambda$ is the dominant component, $a(t)\sim e^{\sqrt{\frac{\Lambda}{3}}t}$, and the expansion becomes de Sitter-like. It will probably be the case in the remote future, when the matter fluid becomes extremely diluted, and its contribution to the overall energy density becomes negligible.

\subsection{Theoretical problems associated to $\Lambda$}
\label{subsec:TPLCDM}

In Sect. \ref{subsec:HistPartII} I have amply discussed the main theoretical problem affecting the cosmological term, the old CC problem. We have seen that it arises when we try to explain the tiny measured value of $\rho_{\Lambda,{\rm eff}}$ by using the standard QFT formalism. Although the renormalization of the vacuum energy is in principle feasible, it must be performed in a very unnatural way so as to cancel the various vacuum contributions in an extremely fine-tuned form. In other words, the counterterm coming from the bare $\Lambda$ must acquire a value which is orders of magnitude larger than the measured one. In this section I will not explain further about the CC problem, which I consider has been analyzed in detail in Sect. \ref{subsec:HistPartII}, but I will introduce another problem that a rigid CC carries around or, at least, this is what is usually stated in the literature.

It is the so-called cosmic coincidence problem, and it is somewhat more controversial than the CC one. In fact, there is no clear consensus on the relevance of this problem among the experts in the field. In the $\Lambda$CDM model $\Lambda_{\rm eff}$ remains constant during all the cosmic history, and so is the associated energy density. Conversely, the matter dilutes with the expansion and therefore $\rho_m\sim a^{-3}$. It is an observational fact that both energy densities are of the same order of magnitude at present, i.e. $r(z=0)\equiv\rho^{(0)}_{\Lambda,{\rm eff}}/\rho^{(0)}_m\approx 7/3=\mathcal{O}(1)$. The ratio $r(z)$ tends to $0$ in the past and to infinity in the future. But why is this ratio precisely nowadays of order unity? Is this just a coincidence? And if so, how can it be explained in physical grounds?

Before going on with the analysis of the problem, we should firstly understand what is a coincidence or how can we define it in more concrete terms. The cosmic coincidence can be considered a real coincidence only if:

\begin{enumerate}
\item The evolution of $\rho_m$ is not linked to the one of $\rho_{\Lambda,{\rm eff}}$, i.e. if both energy densities do not talk to each other, just evolve in an independent way.
\item The equality of both energy densities could have happened at any stage of the Universe's expansion with equal probability, but happens precisely today or in a span of time small enough around the current time. 
\end{enumerate}

In the $\Lambda$CDM model the first requisite is clearly fulfilled, since there is no linkage between $\rho_m$ and $\rho_{\Lambda,{\rm eff}}$. The latter is constant, independently of the value that the matter energy density takes. The fulfillment of the second condition is not that obvious and must be studied with more detail. First of all it is important to notice that probability arguments can sometimes be a bit fuzzy. If we considered the period of the expansion of the Universe in which $r\sim\mathcal{O}(1)$ and we compared it with the entire cosmic history, which is in principle infinitely large, then we could conclude that the aforementioned period is infinitely tiny and, therefore, the fact of living in the epoch in which this happens would be regarded as a huge coincidence. One could use in this case anthropic arguments in order to explain the observed coincidence\footnote{The first modern application of the anthropic principle was probably carried out by R.H. Dicke in \cite{Dicke1961}. Therein, he argued that Dirac's relation $G\sim 1/T_{u}$ \cite{DiracLNH} could simply be a selection effect: fundamental physical laws connect Newton's coupling to the lifetime of the main sequence stars, and these stars are necessary for the existence of life. At any other epoch, when the equality did not hold, there would be no intelligent life around to notice the discrepancy.}. But one must keep in mind that these anthropic arguments do not solve the problem at all. Obviously, if we measure the observed value of the CC is because we exist and if we exist is because the CC acquires this precise value, but this argument does not explain why the CC takes the value we measure! It is fairer to compare, though, the period in which $r\sim\mathcal{O}(1)$ with the current Universe's age.  But still, we can do it in several ways, e.g. by using the cosmic time scale or the redshift scale. We can distinguish two characteristic events: the coincidence redshift at which $\rho_m(z_{\rm coinc})=\rho_{\Lambda,{\rm eff}}(z_{\rm coinc})$,
\be
z_{\rm coinc}=\left(\frac{\Omega^{(0)}_{\Lambda,{\rm eff}}}{\Omega_m^{(0)}}\right)^{1/3}-1\approx 0.33\,,
\ee
and the redshift at which the acceleration of the Universe becomes positive, i.e. $\ddot{a}>0$. We will refer to the latter as transition redshift. It can be computed from \eqref{eq:acceEq} by neglecting the contribution of the relativistic component,  
\be
z_{tr}=\left(\frac{2\Omega^{(0)}_{\Lambda,{\rm eff}}}{\Omega_m^{(0)}}\right)^{1/3}-1\approx 0.67\,.
\ee
The fact that $z_{tr}>z_{\rm coinc}$ shows us that the Universe started its positive acceleration before the domination of $\Lambda_{\rm eff}$, and this is simply due to the extra factor $2$ accompanying $\rho_{\Lambda,{\rm eff}}$ in \eqref{eq:acceEq}. Taking into account that $\infty>z>0$ from the far past up to now, it seems an incredible coincidence to find these two redshifts so close to $0$. But let us translate $z_{tr}$ and $z_{\rm coinc}$ to cosmic time. Considering that the age of the Universe at redshift $z$ is given by 
\be
t(z)=\int_{z}^{\infty}\frac{dz^\p}{(1+z^\p)H(z^\p)}\,,
\ee
we find that $t_{\rm coinc}\sim 10$ Gyrs and $t_{tr}\sim 7.5$ Gyrs. The expression for the present age of the Universe reads,
\be
t(z=0)\equiv t_0=\frac{H_0^{-1}}{3\sqrt{\Omega_{\Lambda,{\rm eff}}^{(0)}}}\ln\left(\frac{1+\sqrt{\Omega_{\Lambda,{\rm eff}}^{(0)}}}{1-\sqrt{\Omega_{\Lambda,{\rm eff}}^{(0)}}}\right)\sim 13.7\,{\rm Gyrs}\,.
\ee 
By computing the ratios $t_{tr}/t_0\approx 0.55$ and $t_{\rm coinc}/t_0\approx 0.73$, we see that in terms of the cosmic time the coincidence of the cosmic coincidence problem seems to be totally non-existent. Notice that $r\sim\mathcal{O}(1)$ almost during $\sim 1/4-1/2$ of the current Universe's age! This idea is remarked in \cite{BianchiRovelli2010}, and it certainly lowers the significance of the coincidence problem. Nevertheless, it is very well-known that a slight variation of the value of $\Lambda_{\rm eff}$ would have had catastrophic consequences. For instance, a larger value would have inhibited the formation of galaxies due to the increase of the vacuum repulsive effects, and probably life would not have existed either. Despite this, one should not consider the value of $\Lambda_{\rm eff}$ as a mere coincidence, but as something that is potentially explainable. Thus, understanding such ``coincidence'' basically consists in solving the old CC problem\footnote{See Ref. \cite{LXCDM} for the study of the cosmic coincidence from the perspective of a model with a DE mixture formed by a variable cosmological term, $\Lambda$, and a dynamical ``cosmon'' possibly interacting with $\Lambda$ but not with matter.}.

Another issue that must also be tackled has to do with the generation of the inflationary phase in the very early Universe. I have mentioned before that Inflation allows to solve some of the problems that affect the standard cosmological model. For instance, it is able to give rise to a low curvature Universe without fine-tuning the initial conditions at the beginning of the radiation-dominated era. It also manages to explain why angular patches of the sky that in the context of the standard $\Lambda$CDM model have always been causally disconnected show the observed high degree of homogeneity. This is the ``horizon'' problem, which becomes evident from the CMB map, where only very little deviations from the mean relic photon's temperature are observed in different directions. The causal connection is realized in the inflationary paradigm thanks to the initial exponential expansion of the Universe, which succeeds in stretching the small density perturbations of the quantum fields up to scales larger than the Hubble horizon of that epoch. In addition, these stretched perturbations put the seeds of the cosmic structures that would be formed later on. Thus, Inflation has become a key piece of the cosmological puzzle. The well-behaved picture of the Cosmos provided by the $\Lambda$CDM model would not be that successful without this inflationary phase. So now it takes part of the natural extension of the concordance model, giving rise to the Lambda Cold Dark Matter Inflationary Big Bang Theory. But how can we generate Inflation? Is there a link between it and the current observed positive accelerated expansion of the Universe? Notice that both phases are characterized by two very different energy scales,
\be                                       
\frac{\rho_{\rm inf}}{\rho_c^{(0)}}\sim\frac{\rho_{\rm GUT}}{\rho_c^{(0)}}\sim 10^{111-115}\,,
\ee
so this connection cannot be made in the context of a rigid cosmological term, precisely because no evolution of $\Lambda_{\rm eff}$ is allowed. It is just a constant, so its energy density becomes negligible at $z\gg z_{tr}$, when matter, radiation and then the component (or components) that trigger the Inflation rule the expansion. In order to establish this connection one is enforced to endow $\Lambda_{\rm eff}$ with some dynamics. By doing so one breaks the standard $\Lambda$CDM picture. These dynamics can be implemented through the use of scalar fields, for example. And then, these evolving fields could potentially link both energy scales, explaining the current positive accelerated phase in the context of quintessence or alternatives the like (cf. Sect. \ref{subsec:AlternativesLCDM}). These alternative theoretical constructions are not problem-free, of course, and it is not clear either that this connection does indeed exist. It is, though, a beautiful idea, which would make the cosmic coincidence problem exposed before even much weaker. We will see in the subsequent sections that the RVM's integrate all these ideas in the context of a varying (or running) vacuum energy density, in which no auxiliary scalar fields are needed. The idea of the aforementioned link between the two positively accelerated phases of the Universe in such models become even much appealing, since in this case they are explained with an evolving vacuum energy in a kind of unified framework, which is based on the formalism of QFT in curved spacetime. As it will be seen later on, a growing-in-the-past vacuum energy is favored by observations, and it is able to eventually surpass the radiation energy density and trigger Inflation. All these concepts will be developed in the next sections. Now, I am only limiting myself to present the theoretical problems that affect $\Lambda_{\rm eff}$ and motivate the study of the aforementioned alternatives. 

%%%%%%%%%%%%%%%%%%%%%%%%%%%%%%%%%%%%%%%%%%%%%%%%%%%%%%%%%%%%%%%%%%%%%%
%%%%%%%%%%%%%%%%%%%%%%%%%%%%%%%%%%%%%%%%%%%%%%%%%%%%%%%%%%%%%%%%%%%%%%
%%%%%%%%%%%%%%%%%%%%%%%%%%%%%%%%%%%%%%%%%%%%%%%%%%%%%%%%%%%%%%%%%%%%%%
%%%%%%%%%%%%%%%%%%%%%%%%%%%%%%%%%%%%%%%%%%%%%%%%%%%%%%%%%%%%%%%%%%%%%%

\subsection{$\Lambda$CDM. Current observational status}
\label{subsec:OPLCDM}

The success of the $\Lambda$CDM model basically lies in its capability of explaining the large variety of available observational data. This is not an easy task at all taking into account that we live in the precision Cosmology era, in which more and more stringent tests are used to constrain the models. Despite this, the $\Lambda$CDM is still there, resisting the new generation of galaxy surveys and CMB probes. But, according to the current observational data, can the $\Lambda$CDM be considered the final concordance model?

Planck's measurements of the CMB temperature and E-mode polarization auto- and cross-spectra (TT, TE, EE) \cite{Planck2015} are in very good agreement with the vanilla 6-parameter $\Lambda$CDM model. The cosmological parameters inferred from these spectra are also highly consistent, regardless of which combination of spectra is used for the parameter estimation. Apart from the analysis of CMB anisotropies, which give us information about how was the Universe in the pre-recombination era, i.e. at $z>z_{*}$\footnote{Of course, CMB anisotropies also contain distinct signatures of the post-recombination Universe as well, since angles between two points in the CMB map depend on the background cosmology that has ruled the expansion from $z_*$ up to now. Also the reionization process comes into play at $z_{re}\sim 6-9$, when the first stars formed. In addition, some DE signatures were imprinted by the process of gravitational lensing and the integrated Sachs-Wolfe effect. See \cite{HuThesis} for further details.}, there are also a bunch of observables of low and intermediate redshifts that also contribute to break degeneracies and constrain the parameter space of the model. Some examples are: i) Supernovae of type Ia, which are used as standardizable candles and provide us a list of luminosity distances as a function of the redshift (at $z<2$) \cite{BetouleJLA}; ii) Measurements of the Hubble function at redshifts lower than $3$ obtained from BAO measurements and differential-age techniques applied to passively evolving galaxies (see Table \ref{compilationH} for some references); iii) Other cosmological distances extracted from the analysis of the BAO peak in the density correlation function (or, equivalently, from the matter power-spectrum), obtained by several galaxy surveys; iv) Large-scale structure (LSS) observables, as number counts or redshift-space distortions (RSD) induced by the peculiar velocities of galaxies falling under the action of the gravitational wells; v) gamma-ray burst data \cite{GammaRay1,GammaRay2,GammaRay3}, vi) strong and weak lensing data \cite{LensingData}; vii) HII galaxy data \cite{HIIdata}; viii) cluster gas mass fraction data \cite{GasMassFraction1,GasMassFraction2}.

But there are also potential evidences of substantial tension between the vanilla $\Lambda$CDM model and some data. For instance, the tension observed in the constraints in the $\Omega_m^{(0)}-\sigma_8$ plane derived from the primary CMB temperature and polarization measurements of the Planck satellite \cite{Planck2015}, those obtained by the CFHTLenS weak lensing survey \cite{Heymans2013} and the Kilo Degree Survey (KiDS) \cite{KIDS450}, and the Planck analysis on Sunyaev-Zel'dovich cluster counts \cite{PlanckXXIV} (see also the analysis of KIDS+2dFLenS \cite{KIDS2dFLenS}). The latter four sources favor significantly lower values of the root mean square matter fluctuation amplitude on $R=8h^{-1}$ Mpc spheres, i.e. $\sigma_8$. There is also an apparent tension between measurements of the growth rate extracted from RSD and the value predicted by the Planck best-fit parameters \cite{Macaulay2013}. It seems that cosmic structures prefer a weaker growth than the one predicted in the $\Lambda$CDM model (cf. Fig. 1 of \cite{BernalVerdeCuesta2016}, which is quite illustrative). 

There also exists some tension between $\Lambda$CDM parameters inferred from the Planck temperature fluctuations spectrum at the multipoles accessible to WMAP ($l\leq 1000$) and at higher multipoles ($l\geq 1000$), see the thorough analysis presented in \cite{Addison2016}.

It is also found a significant $3.4\sigma$ tension between the non-local measurements of $H_0$, e.g. \cite{Planck2015}, and local ones, e.g. \cite{RiessH0}, see Chapter \ref{chap:H0tension} for more references and details. Also a $2.5\sigma$ tension between the determination of $H(z=2.34)$ and the angular distance $D_A(z=2.34)$ by the BOSS collaboration in \cite{Delubac2015} and the $\Lambda$CDM with the best-fit parameters obtained from Planck in combination with the Wilkinson Microwave Anisotropy Probe (WMAP) \cite{WMAP9} polarization data. This tension is reduced to a $2.2\sigma$ one when the best-fit values are inferred from WMAP9 in combination with the Atacama Cosmology Telescope (ACT) \cite{ACTSievers} and the South Pole Telescope (SPT) \cite{SPTStory} data. In relation with this tension, in 2014 V. Sahni, A. Shafieloo and A.A. Starobinsky proposed in \cite{SahniShafielooStarobinsky} a model-independent test based on the following diagnostic,
\be\label{eq:Omh2Diagnostic}
Omh^2(z_i;z_j)=\frac{h^2(z_i)-h^2(z_j)}{(1+z_i)^3-(1+z_j)^3}\,,
\ee
where $h(z)=H(z)/100\,{\rm km/s/Mpc}$. For the $\Lambda$CDM, the two-point diagnostic boils down to $Omh^2(z_i;z_j)=\Omega_m^{(0)}h^2$, which is constant for any pair $z_i$, $z_j$. Using this testing tool and the known observational information on $H(z)$ at the three redshift values $z=0$ \cite{Efstathiou2014}, $0.57$ \cite{Samushia2013}, $2.34$ \cite{Delubac2015}, the aforementioned authors observed that the average result is: $Omh^2_{\rm mean,obs} = 0.122 \pm 0.010$, with very little variation from any pair of points taken and almost independent of the chosen value of $H(z=0)$, see \cite{SahniShafielooStarobinsky} for details. The obtained result is significantly smaller than the corresponding Planck's value of the two-point diagnostic, which is constant and given by $Omh^2_{\Lambda}=0.1415\pm 0.0019$ \cite{Planck2015}. According to the authors of \cite{SahniShafielooStarobinsky}, this departure might reveal (in the absence of systematics in the CMB and SDSS data sets) a hint in favor of the dynamical nature of the DE. Results in the same line have been obtained more recently by using an enlarged list of $H(z)$ data points in \cite{Ding2015,Zheng2016}. See also \cite{Ferreira2017}. 

Some people argue, though, that these discrepancies among data sets could be due to the existence of some kind of systematics in the data, see e.g. \cite{Addison2016}\footnote{The same authors from \cite{Addison2016} state in the Discussion section: {\it Cosmology beyond standard $\Lambda$CDM cannot be ruled
out as the dominant cause of tension}. Thus, it seems that they do not exclude the possibility of New Physics either.}, but the real truth is that the possibility of these tensions to be pointing out the incompleteness of the concordance model is not excluded. In this case, New Physics would be required to explain the observations. Thus, the door is opened for those models able to reconcile theory and experiment. 

See \cite{Raveri2016} for an exhaustive analysis of the concordance level between the $\Lambda$CDM and different data sets, and also \cite{BeyondLambda} for a more complete review of the current observational status of the standard model. 

%%%%%%%%%%%%%%%%%%%%%%%%%%%%%%%%%%%%%%%%%%%%%%%%%%%%%%%%%%%%%%%%%%%%%%
%%%%%%%%%%%%%%%%%%%%%%%%%%%%%%%%%%%%%%%%%%%%%%%%%%%%%%%%%%%%%%%%%%%%%%
%%%%%%%%%%%%%%%%%%%%%%%%%%%%%%%%%%%%%%%%%%%%%%%%%%%%%%%%%%%%%%%%%%%%%%
%%%%%%%%%%%%%%%%%%%%%%%%%%%%%%%%%%%%%%%%%%%%%%%%%%%%%%%%%%%%%%%%%%%%%%

\subsection{Alternatives to the $\Lambda$CDM}
\label{subsec:AlternativesLCDM}

The problems affecting the CC have motivated cosmologists to search for alternative approaches to explain the current positive acceleration of the Universe. They range from modified gravity models, in which the {\it l.h.s.} of Einstein's equations is modified due to the changes introduced in the original Einstein-Hilbert action with CC \eqref{eq:EHaction}, to models in which some kind of modified form of matter is used as a source of DE. In the latter case, such modifications are implemented in the {\it r.h.s.} of the gravitational field equations. In principle, one can treat gravity modifications like a pure change in the matter sector, just by moving the modified terms arising in the {\it l.h.s.} to the {\it r.h.s.} and considering them as part of the total energy-momentum tensor. The resulting framework cannot be distinguished from the starting one by only analyzing gravitational effects. Of course, from the QFT perspective, there is a clear difference between both scenarios, since in the second one there is an obvious change in the field content of the theory, contrary to what happens in the modified gravity case. Now, I review some of these alternative models in a rather general way, but trying to pinpoint the most important aspects of each of them.

\subsubsection{Scalar fields in the late-time Cosmology}

Scalar fields, $\phi$, have been used in Cosmology since long ago, most conspicuously in the context of Brans-Dicke theories \cite{BD61}, where $G\propto 1/\phi(t)$, and subsequently in general scalar-tensor theories. The general scalar-tensor action reads,
\be
S=\int d^4x\sqrt{-g}\left[\frac{1}{2}f(R,\phi)+\frac{1}{2}h(\phi)(\nabla\phi)^2\right]+S_m(g_{\mu\nu},\Psi_m)\,,
\ee
where $f$ is a function of the scalar field $\phi$ and the Ricci scalar $R$, $h$ depends also on $\phi$, and $S_m$ is the action that governs the behavior of the matter fields $\Psi_m$, which can be minimally or non-minimally coupled to gravity. The Brans-Dicke theory is just a particular case, in which $f(R,\phi)=\phi R$ and $h(\phi)=\omega_{BD}/\phi$, $\omega_{BD}$ being the (constant) Brans-Dicke parameter. These theories are also considered to be inside the large group of modified gravity theories for obvious reasons.

Soon after, scalar fields also played a role as a strategy to endow the vacuum and the cosmological term with some time dependence in a QFT context, $\Lambda=\Lambda(\phi(t))$, and in some cases with the purpose to adjust dynamically its value. Some of the old approaches to the CC problem from the scalar field perspective can be found in \cite{EndoFukui1977and1982,Fujii1982,Dolgov1983,Abbott1985,Barr1987and1988,Ford1987,Weiss87}. Among the
proposed dynamical mechanisms, let me mention the cosmon model \cite{PecceiSolaWetterich1987}, which was subsequently discussed in detail in \cite{WeinbergReview1989}. In all cases, a more or less obvious form of fine tuning underlies the adjusting mechanisms. For this reason scalar fields were later used mostly to ascribe a possible evolution to the vacuum energy with the hope to explain
the cosmic coincidence problem, giving rise to the notion of quintessence and the like, cf. \cite{PeeblesRatra88b,PeeblesRatra88a,Wetterich88,Wetterich95,Caldwell98,ZlatevWangSteinhardt99a,ZlatevWangSteinhardt99b,Amendola2000}, among many other alternatives. See e.g. the reviews \cite{Padmanabhan2003,PeeblesRatra2003,Copeland2006,Tsujikawa2013} and the book \cite{BookAmendolaTsujikawa}. Quintessence and phantom dark energy models are constructed by introducing a scalar field minimally coupled to gravity. The action for this scalar field takes the following form,
\be\label{eq:QPDE}
S_\phi=\int d^4x\sqrt{-g}\left[\frac{\xi}{2}g^{\mu\nu}\partial_\mu\phi\partial_\nu\phi-V(\phi)\right]\,,
\ee
where $\xi=+1$ for quintessential models, and $\xi=-1$ for the phantom DE ones. $V(\phi)$ is the scalar field potential. An adequate shape is crucial to ensure a correct phenomenological behavior of the model. The (light) scalar field is taken to be classical and homogeneous so as to fulfill the Cosmological Principle discussed in Sect. \ref{subsec:BasicsLCDM}. Thus, it can only depend on the cosmic time, i.e. $\phi=\phi(t)$. The associated sound speed must be high enough in order not to produce an excess of structure in the Universe and suppress the DE density perturbations at small scales. In the non-interacting versions of these models, the scalar field does not exchange energy with other sectors. By varying the action \eqref{eq:QPDE} with respect to the scalar field and using the FLRW metric one obtains the modified Klein-Gordon equation,
\be
\ddot{\phi}+3H\dot{\phi}+\xi\frac{\partial V}{\partial\phi}=0\,.
\ee   
On the other hand, the variation of \eqref{eq:QPDE} with respect to the metric yields the energy-momentum tensor of the scalar field,
\be
T^{(\phi)}_{\mu\nu}=\xi\partial_\mu\phi\partial_\nu\phi-g_{\mu\nu}\left[\frac{\xi}{2}g^{\mu\nu}\partial_\mu\phi\partial_\nu\phi-V(\phi)\right]\,,
\ee
from where one can derive the expressions for the energy density and pressure associated to the scalar field,
\be                                             
\rho_\phi=\xi\frac{\dot{\phi}^2}{2}+V(\phi)\,,
\ee
\be
p_\phi=\xi\frac{\dot{\phi}^2}{2}-V(\phi)\,.
\ee
The EoS parameter varies with the expansion according to the following formula,
\be
\omega_\phi=\frac{p_\phi}{\rho_\phi}=\frac{\xi\frac{\dot{\phi}^2}{2}-V(\phi)}{\xi\frac{\dot{\phi}^2}{2}+V(\phi)}\,.
\ee
Quintessence DE gives $\omega_\phi>-1$ during all the cosmic history, whereas phantom DE $\omega_\phi<-1$ when $V(\phi)>\dot{\phi}^2/2$. A value $\omega_\phi(z=0)<-1$ in the current time is not excluded by the constraints provided by some observational teams. For instance, Planck Collaboration \cite{Planck2015} gives $\omega_{\rm DE}=-1.019^{+0.075}_{-0.080}$, whereas the BOSS one \cite{Aubourg2015} $\omega_{\rm DE}=-0.97\pm 0.05$\footnote{We will see in Sects. \ref{chap:MPLAbased} and \ref{chap:PRDbased} that the use of a data set which includes the state-of-the-art RSD+BAO+CMB measurements allows us to determine a value of $\omega_{\rm DE}$ which is in the quintessence region with $\sim 4\sigma$ c.l. See also the results presented in \cite{ApJnostre,MPLAnostre,PRLnostre,PRDnostre}.}. The key point, though, is that in both kind of DE models $\omega_\phi\simeq -1$ and $\rho_\phi\simeq{\rm const}$ when $V(\phi)\gg\dot{\phi}^2$. Thus, they are in principle able to mimic the behavior of the CC at present, but this is something that ultimately depends on the exact form of $V(\phi)$. 

Notice that in the phantom DE scenario, the energy density is not bounded by below because of the negative kinetic term. This causes the problematic instability of the vacuum state \cite{BookAmendolaTsujikawa,CarrHofTrod2003}. Another characteristic feature of these models is that the energy density of a phantom field increases with the expansion [cf. \eqref{eq:rhoaw}]. Not only that, the scale factor goes to infinity in a finite amount of time thanks to the superaccelerated phase that the Universe undergoes, which lasts until a future singularity is reached. It is the so-called Big Rip, since all bounded objects (even atoms) rip apart \cite{CaldwellPDE2003,NesserisPerivo2004}. The reason why this happens can be easily understood. For instance, a body orbiting around a massive object of mass $M$ in an orbit of radius $R$ will become unbounded when $-(4\pi/3)(\rho+3p)R^3\sim M$, where $\rho$ and $p$ are the total energy density and pressure of the system, respectively, so they take into account the contribution of the DE too. If $\omega_\phi>-1$ then $-(\rho+3p)$ decreases with time and, therefore, a system which is currently bounded remains bounded forever. In other words, the cosmic expansion is not able to tear the system up. In contradistinction, a phantom (or ghost) scalar field makes $-(\rho+3p)$ to increase in the phantom-dominated era, so there is a moment in which the bounded system just breaks. But, of course, this (sort of catastrophic) feature of the phantom DE models is not a drawback for the model itself, since {\it a priori} nothing prevents the Universe to have such dramatic end.

The weakest points of these scalar field models are probably the following: 

\begin{itemize}
\item They do not even try to solve or alleviate the old CC problem. They just presume that the CC vanishes due to some unspecified mechanism (maybe some unknown symmetry) and use these scalar fields in order to provide some dynamics to the DE. It is important to remark here that these new scalar fields do not only contribute at the classical level to the effective cosmological term with their associated potential energy in the ground state, but also ``carry'' their own ZPE! Therefore, they keep alive the unwanted CC problem despite the {\it ad hoc} removal of the other vacuum contributions. 
\item It is sometimes difficult to establish a clear connection between these scalar fields and a fundamental theory, although some quintessence potentials can be well motivated from a theoretical point of view, e.g. from supergravity, see \cite{BookAmendolaTsujikawa} and references therein.
\item The energy density associated to the scalar field $\rho_\phi$ equals the critical one when $m_\phi\sim H$. This means that $m_\phi\sim H_0\sim 10^{-33}$ eV. One could consider quite unnatural that a particle with such a tiny mass could explain the physics of the current Universe, regarding that the energy scale of the latter is $\rho_c^{(0)1/4}\sim 10^{-3}$ eV, i.e. $30$ orders of magnitude larger than $m_\phi$. And what probably is the most shocking feature, the mass of the scalar turns out to be many orders of magnitude smaller than any of the masses of the scalar fields usually encountered in particle physics. 
\end{itemize}

\begin{table}[t]
\centering
    \begin{tabular}{ | c | c |}
    \hline
    {\bf Scalar field Potential} & {\bf Expression} \\\hline
		 Linear \cite{Garriga04,Perivolaropoulos05} & $V(\phi)=\lambda\phi$ \\\hline
		 Inverse power-law \cite{PeeblesRatra88b,PeeblesRatra88a} & $V(\phi)=\frac{1}{2}\kappa M_{P}^2\phi^{-\alpha}$ \\ \hline
    Exponential \cite{Wetterich88,Wetterich95,CopelandLiddleWands98,DoubleExpBarreiro} & $V(\phi)=V_0 e^{-\kappa\lambda\phi}$ \\\hline
		Double exponential \cite{DoubleExpBarreiro} & $V(\phi)=V_0(e^{-\kappa\lambda_1\phi}+e^{-\kappa\lambda_2\phi})$\\ \hline
		Pseudo-Nambu-Goldstone \cite{Frieman95,Choi2000} & $V(\phi)=V_0\left[C+\cos\left(\frac{\phi}{f}\right)\right]$ \\\hline
		Hyperbolic cosine \cite{SahniWang2000} & $V(\phi)=V_0\left[\cosh(\kappa\lambda\phi)-1\right]$ \\\hline
		Exponential + power-law \cite{AlbrechtSkordis2000} & $V(\phi)=V_0e^{-\kappa\lambda\phi}\left[A+(\kappa\phi-B)^2\right]$\\\hline
		SUGRA motivated \cite{BraxMartinSUGRA1,BraxMartinSUGRA2}& $V(\phi)=V_0^{4+\alpha}\phi^{-\alpha}e^{\frac{\kappa^2 n^2\phi}{2}}$ \\\hline
    \end{tabular}
		\caption[List of some quintessence potentials found in the literature]{{\scriptsize List of some quintessence potentials that can be found in the literature. Here, $\kappa^2=8\pi/M_P^2$.}}
		\label{tab:TabQuintessence}
\end{table}

Despite being affected by these problems, some of these models are able to explain the data with great efficiency and much better than the $\Lambda$CDM. This is the case, for instance, of the Peebles-Ratra (PR) quintessence model \cite{PeeblesRatra88b,PeeblesRatra88a}, which takes an inverse power-law potential, i.e. $V(\phi)\sim \phi^{-\alpha}$. I analyze this model in detail from the theoretical and phenomenological point of view in Chapter \ref{chap:MPLAbased}, which is basically based on the dedicated study \cite{MPLAnostre}. In the same section I also make some brief comments on quintessence models with both, single \cite{Wetterich88,CopelandLiddleWands98} and double exponential potentials \cite{DoubleExpBarreiro}. We will see that the former type of potential, i.e. $V(\phi)\sim e^{-\lambda\phi}$, is incapable of fulfilling the BBN bounds and driving at the same time the current positive accelerated phase, whereas the double exponential potential does not suffer from this problem. Exponential quintessence models, together with the PR one, are called freezing models. They generate the so-called tracker solutions \cite{ZlatevWangSteinhardt99b}, in which the DE keeps track of the matter component and finally dominates the expansion, when the movement of the scalar field is slowed down enough and the potential energy becomes much greater than the kinetic one. Tracker solutions are interesting because are able to alleviate the initial fine-tuning problem. Different initial scalar field configurations in the post-inflationary stage give rise to the same late-time expansion once the parameters of the potential are fixed.

There are also thawing models, in which the scalar field is ``frozen'' by Hubble friction and $\omega_\phi\simeq -1$ until recently, and then $\phi$ begins to evolve and $\omega_\phi$ starts to grow. As an example, we can mention the model with scalar field potential $V(\phi)=M^4 \left[1+\cos(\phi/f)\right]$ \cite{Frieman95}. Actually, the list of alternative non-interactying quintessence models is almost endless. I refer the reader to Table \ref{tab:TabQuintessence} for a more complete list and to the book \cite{BookAmendolaTsujikawa} for more details and references. But I would also like to mention the existence of interacting (or coupled) quintessence models. In them, the scalar field exchanges energy and momentum with non-relativistic matter. This interaction is implemented through the source vector $Q_\mu=QT_m\nabla_\mu\phi$, where $T_m$ is the trace of the matter energy-momentum tensor and $Q$ is the coupling strength, which in principle can be different for baryons and DM. In this way, the covariant energy-momentum conservation equation can be split as follows, 
\be
\nabla^\mu T^{(m)}_{\mu\nu}=-Q_\nu\qquad  \nabla^\mu T^{(\phi)}_{\mu\nu}=+Q_\nu    \,.
\ee
As an example of these kind of models with exponential potential for the scalar field and a linear coupling see \cite{Amendola2000}. The main aim of the study of coupled quintessence is to alleviate the cosmic coincidence problem through the explicit linkage between DE and matter. 

To end this brief summary on the existing scalar field proposals in connection with the late-time Cosmology, let me also mention the K-essence models. They result from a direct generalization of the action \eqref{eq:QPDE},
\be
S_\phi=\int d^4x\sqrt{-g}\,P(\phi,X)\,,
\ee
where $X\equiv (\nabla\phi)^2/2$. Different kind of K-essence models are obtained from the low-energy effective string theory, ghost condensate models, and others \cite{BookAmendolaTsujikawa}. They can also lead to tracking solutions and, therefore, they are able to  soften the fine-tuning issue. 

\subsubsection{Chaplygin gas}

\begin{figure}
\centering
\includegraphics[angle=0,width=0.5\linewidth]{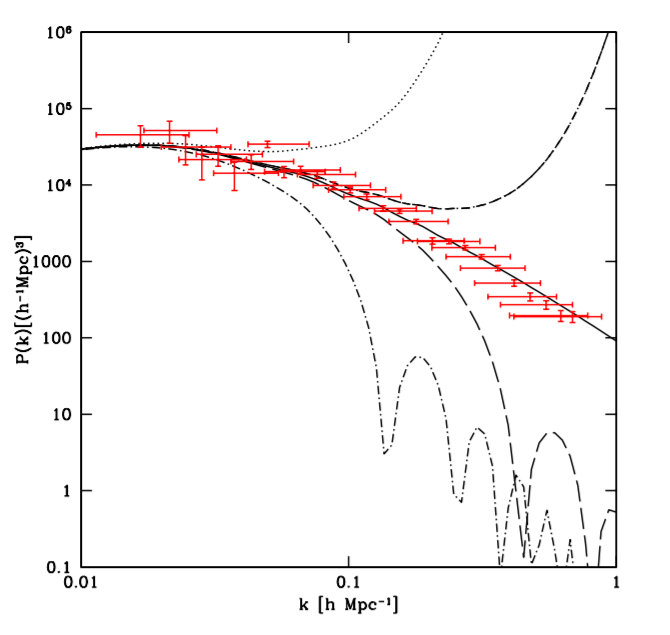}
\caption[P(k) in the generalized Chaplygin gas model.]{\label{fig:chapylginGasPk}%
\scriptsize {Predicted power spectrum $P(k)$ in the GCG model. From top to bottom, the curves correspond to $\alpha=-10^{-4},\,-10^{-5},\,0\,(\Lambda{\rm CDM}),\,10^{-5},\,10^{-4}$, respectively [cf. \eqref{eq:GCGeos}]. The data points are those from the 2dF Galaxy Redshift Survey \cite{Tegmark04}. From \cite{Sandvik2004}.}}
\end{figure}

The Chaplygin gas hydrodynamic model was presented in \cite{chaplyginGas1}. It unifies dark energy and dark matter in one single (exotic) component, which is usually referred to as unified dark matter or ``quartessence''. The EoS of this fluid in the original model takes a very simple form, $p=-A/\rho$, where $A$ is a positive constant. It can be extended to a more general one \cite{chaplyginGas2}, 
\be\label{eq:GCGeos}
p=-A\rho^{-\alpha}\,.
\ee
This is the EoS of the generalized Chaplygin gas (GCG), which can be obtained from the parametrization invariant Nambu-Goto d-brane action in a $(d + 1,1)$ spacetime. Notice that if $\alpha>1$, pressure decreases in the past, and this is precisely what we are looking for, since the fluid must behave as non-relativistic matter in that limit. At late times, the negative pressure becomes important and the Chaplygin gas can drive the positive cosmic acceleration. Introducing \eqref{eq:GCGeos} in \eqref{eq:rhoConsEq} one obtains,
\be
\rho(a)=\left[A+\frac{B}{a^{3(1+\alpha)}}\right]^{1/(1+\alpha)}\,,
\ee
where $B$ is an integration constant. It is clear that $\rho(a)\sim a^{-3}$ for $a\ll 1$ and $\rho(a)={\rm const}$ for $a\gg 1$. Equivalently, we find that the EoS parameter tends to $\omega\to 0$ and $\omega\to -1$, respectively. This model seems to be at least potentially able to fit the cosmological data. Indeed, it is at the background level \cite{Makler03}. Problems can arise, though, at the linear perturbation regime, since values $|\alpha|>10^{-5}$ modify excessively the matter power spectrum with respect to the $\Lambda$CDM one. Fig. \ref{fig:chapylginGasPk} shows that these values of $\alpha$ cause large fluctuations or blowup in the matter power spectrum that are not observed. This basically rules out the original model with $\alpha=1$ and put severe constraints to the $\alpha$ parameter of the GCG (see e.g. Fig. 3 of \cite{Sandvik2004}).   

\subsubsection{Some phenomenological proposals}

Some of the old cosmological kinematic models based on attributing a phenomenological direct time-dependence to the CC term, $\Lambda=\Lambda(t)$, without an obvious relation to scalar fields can be found in \cite{OzerTaha8687,Bertolami86,Freese87,Carvalho92,Waga93,LimaMaia93,ArcuriWaga94,Arbab97}. Many other works are available in the literature, the reader can also consult the reviews e.g. \cite{Overduin98,Vishwakarma01,SolaReview2013} and references therein. 

One should not consider these phenomenological approaches as something useless or with lack of interest. Quite the opposite, it is important to keep an open mind because we can learn many things from the analysis of these models, which are usually easier to deal with from a mathematical point of view. They could be able to mimic some of the key features of the underlying (and probably more fundamental) theory. 

\subsubsection{Modified gravity models}

In modified gravity models the geometrical part of Einstein's equations themselves is changed so as to explain the late-time accelerated expansion without the need of any DE component. Among these models one finds $f(R)$ gravity, Gauss-Bonnet gravity, or braneworld models. Also the scalar-tensor theories mentioned before are inside this large group. 

$f(R)$ theories are described by the following action,
\be\label{eq:f(R)}
S=\frac{1}{16\pi G}\int d^4x\,\sqrt{-g} f(R)+S_m(g_{\mu\nu},\Psi_m)\,.
\ee
In order to derive the field equations there are two different approaches which are coincident if $f(R)$ is linear in $R$. They are the metric and Palatini formalisms \cite{BookAmendolaTsujikawa}. In the first case, the connections are computed using \eqref{eq:ChrystoffelSymbols} and the field equations are obtained upon varying the action \eqref{eq:f(R)} with respect to the metric $g_{\mu\nu}$. In the second case, the Chrystoffel symbols are treated as independent degrees of freedom together with the metric tensor, so one must also vary \eqref{eq:f(R)} with respect to $\Gamma^{\mu}{}_{\nu\alpha}$ under the important assumption that the matter action does not depend on the connection. In this way one obtains the equation for the Christoffel symbols. There is actually even a third version of $f(R)$ gravity, the so-called metric-affine $f(R)$ gravity, in which one uses the Palatini variation but abandons the
assumption that the matter action is independent of the connection. This is the most general $f(R)$ formalism. For a review of these theories see e.g. Refs. \cite{SotiriouFaraoni2008,CapozzielloMariafelicia2011,ModifiedGRClifton}.

A more general modification of GR can be implemented by adding to the Einstein-Hilbert Lagrangian \eqref{eq:EHaction} a function of geometrical invariants $f(R,R^{\mu\nu}R_{\mu\nu},R_{\mu\nu\alpha\delta}R^{\mu\nu\alpha\delta},...)$. As we will see later in Sect. \ref{subsec:RVMintro}, these kind of Lagrangians with higher order curvature terms arise in the effective low-energy gravitational action when quantum corrections are taken into account \cite{BirrellDavies,ParkerToms}. They have been investigated since the 60's, see e.g. \cite{Sakharov1968}.

As an example of theories containing this more general kind of terms in the Lagrangian, we find the Gauss-Bonnet dark energy models, in which one adds to the quintessence action the following term
\be\label{eq:ActionGB}
S_{GB}=\int d^4x\,\sqrt{-g}\, b(\phi)R_{GB}^2\,,
\ee 
where $b(\phi)$ is a function of the scalar field $\phi$ and
\be\label{eq:GaussBonnet}
R^2_{GB}\equiv R^2-4R_{\mu\nu}R^{\mu\nu}+R_{\mu\nu\alpha\delta}R^{\mu\nu\alpha\delta}
\ee
is the Gauss-Bonnet term. It is a topological invariant, which means that the variation of \eqref{eq:ActionGB} with respect to the metric is zero and, therefore, this term does not add any additional contribution to the usual gravitational field equations. Despite this, the Gauss-Bonnet term intervenes in a physical way, due to its coupling with the scalar field, which modifies the equation of motion for $\phi$. 

Some modified gravity models also provide a way to suppress the effect of the large vacuum energy density predicted in QFT giving rise to the current effective (and extremely tiny as compared with particle physics standards) value of the cosmological term without requiring the fine-tuning issue that has been discussed in Sect. \ref{subsec:HistPartII}. This is quite remarkable, and can be done through the so-called relaxing mechanism, which is implemented with an effective action that takes the form $\mathcal{F}(R,R^2_{GB})$, see \cite{BauerSolaStefancic} and references therein. These models have proven to be technically able to solve the hard part of the old CC problem by making use of a direct modification of the gravitational sector. Nevertheless, these action functionals are not derived from any fundamental theory, and this fact keeps the CC problem still open.

For more details on modified gravity and braneworld DE models, see \cite{BookAmendolaTsujikawa} and references therein. 

\subsubsection{Void models}

Here I would just like to mention an alternative to the $\Lambda$CDM model that also departs from those that have been introduced before. It is the void model, in which the current cosmic positive acceleration is not explained in terms of the presence of a DE component nor a direct modification of Einstein's GR, but as a local inhomogeneity in the matter content of the Universe. See e.g. \cite{Tomita2000,Sarkar2007}. 

%%%%%%%%%%%%%%%%%%%%%%%%%%%%%%%%%%%%%%%%%%%%%%%%%%%%%
%%%%%%%%%%%%%%%%%%%%%%%%%%%%%%%%%%%%%%%%%%%%%%%%%%%%%
%%%%%%%%%%%%%%%%%%%%%%%%%%%%%%%%%%%%%%%%%%%%%%%%%%%%%

\section{More about vacuum energy and QFT}
\label{sec:MoreVacuumQFT}

In this section I want to go a step further in the characterization of the vacuum energy in Minkowski spacetime. This will be done in \ref{subsec:EffectivePot}. In \ref{subsec:RVMintro} I will introduce the running vacuum models, motivating them from the renormalization group formalism in QFT in curved spacetime.

\subsection{Effective potential in Minkowski spacetime}
\label{subsec:EffectivePot}
In Sect. \ref{subsec:HistPartII} I have discussed which is the contribution of the electroweak (EW) vacuum to the total vacuum energy, but this calculation has been performed in a rather coarse way. Notice that I have only worked at the classical level, i.e. with the classical potential \eqref{eq:HiggsPotential}. I have searched for the minimum of such potential in order to compute the classical value of the scalar field and I have substituted it in $V(\phi)$ so as to evaluate the leading EW contribution. On the other hand, in Appendix \ref{ch:appZPE} I have computed the renormalized ZPE of a massive scalar field in flat spacetime without taking into account the self-interaction term, i.e. $\lambda\phi^2/4!$. In this section I want to cure these deficiencies and draw a broader picture of the vacuum energy in flat spacetime, working in a more formal framework. I want to show how to compute the quantum corrections of the classical potential \eqref{eq:HiggsPotential} in Minkowski spacetime, together with the ZPE of the scalar field. 

The starting point is the generating functional of the correlation functions, which is defined as follows,
\be\label{eq:GeneratingZ}
Z[J]=\mathcal{N}\int \mathcal{D}\phi\,exp\left[\frac{i}{\hbar}\left\{S_m[\phi]+\int d^4x J(x)\phi(x)\right\}\right]\,,
\ee  
where $J(x)$ is an external classical source, $S_m[\phi]$ is the classical action of the scalar field, and $\mathcal{N}$ is a normalization factor chosen such that $Z[0]=1$. $Z[J]$ gives us the transition amplitude between vacuum states in the presence of a non-null $J(x)$. This is a very well-known result, which is obtained from the path-integral formalism of QFT. It is also convenient to define $W[J]\equiv -i\hbar\ln Z[J]$, from which we obtain,
\be
\frac{\delta W[J]}{\delta J(x)}=\frac{1}{Z[J]}\int \mathcal{D}\phi\,\phi\,e^{\frac{i}{\hbar}S_m[\phi]+\frac{i}{\hbar}\int d^4x J(x)\phi(x)}=<\Omega|\hat{\phi}(x,J)|\Omega>\equiv\phi_{cl,J}(x)\,,
\ee
where $\phi_{cl,J}(x)$ is the mean field, i.e. the expectation value of the field in the ground state when there is a source. Notice that here I use $|\Omega>$ to denote the vacuum state instead of $|0>$ in order to remark that $|\Omega>\ne|0>$ when $J\ne 0 $.

In the computation of the EW classical contribution to the vacuum energy performed in Sect. \ref{subsec:HistPartII} I have implicitly used the fact that in the ground state we expect the classical field $\phi_{cl}$ to be invariant under spacetime translations. Assuming this, we obtain $S[\phi_{cl}]=-\int d^4x V(\phi_{cl})=-\mathcal{V}_4\cdot V(\phi_{cl})$, with $\mathcal{V}_4$ being the spacetime volume. This means that in this case the minimization of the action is equivalent to the minimization of the classical potential. But as I have mentioned before, this is a pure classical object. One would like to have an effective action $\Gamma$ which incorporated the quantum corrections up to the desired order in perturbation theory, such that  
\be
\Gamma[\phi_{cl,J}]=-\int d^4x V_{eff}(\phi_{cl,J})=-\mathcal{V}_4\cdot V_{eff}(\phi_{cl,J})\,.
\ee 
This is achieved by Legendre-transforming $W[J]$,
\be\label{eq:LegendreGamma}
\Gamma[\phi_{cl,J}]=W[J]-\int d^4y J(y)\phi_{cl,J}(y)\,.
\ee
This object is called effective action. One can easily show that 
\be
\frac{\delta \Gamma[\phi_{cl,J}]}{\delta\phi_{cl,J}(x)}=-J(x)\,,
\ee
and this is telling us that in absence of sources the following equation must be fulfilled,                                                                       
\be\label{eq:EqClassicalFi}
\frac{\delta \Gamma[\phi_{cl,J}]}{\delta\phi_{cl,J}(x)}\Bigr|_{J=0}=0\,.
\ee
Therefore, if one is able to build the effective action, one can also compute the ``classical'' scalar field\footnote{Actually, it is not purely classical, since it minimizes not the classical potential, but the effective one, which incorporates the quantum corrections up to the chosen order.} in absence of sources by solving \eqref{eq:EqClassicalFi}, and the minimum of the effective potential (the vacuum energy, which is what we are ultimately interested in) by substituting it in \eqref{eq:LegendreGamma}. But \eqref{eq:LegendreGamma} is, in fact, a very crude expression. In order to deal with it, one has to express the {\it r.h.s.} of this relation in terms of $\phi_{cl,J}$ instead of $J$. This can be done, see e.g. \cite{Peskin,WeinbergBook}. The result at one-loop order reads, 
\be\label{eq:EffAction}
\Gamma[\phi_{cl}]=S_1[\phi_{cl}]+\frac{i\hbar}{2}\ln\,{\rm det}\left[-\frac{\delta^2S_1[\phi]}{\delta\phi(x)\delta\phi(y)}\Bigr|_{\phi=\phi_{cl}}\right]+\delta S[\phi_{cl}]+\mathcal{O}(\hbar^2)\,,
\ee  
where here I have identified $\phi_{cl}\equiv\phi_{cl,J=0}$, and the bare action has been split into a piece depending on renormalized parameters $S_1$ and one containing the counterterms $\delta S$,
\be
S_m[\phi_{cl}]=S_1[\phi_{cl}]+\delta S[\phi_{cl}]\,.
\ee
Let me apply \eqref{eq:EffAction} in the case of a scalar field with classical action
\be
S_m[\phi]=\int d^4{x}\,\left(\frac{1}{2}\partial_\mu\phi\partial^\mu\phi-V(\phi)\right)\,.
\ee
It is easy to find,
\be\label{eq:OperatorFunctional}
\frac{\delta^2S_1[\phi]}{\delta\phi(x)\delta\phi(y)}\Bigr|_{\phi=\phi_{cl}}=-\left[\Box_x+V^{\pp}(\phi_{cl})\right]\delta(x-y)\,.
\ee
where $V^{\pp}\equiv\frac{\partial^2 V}{\partial\phi^2}$. Moreover,
\be
\ln\,{\rm det}\left[-\frac{\delta^2S_1[\phi]}{\delta\phi(x)\delta\phi(y)}\Bigr|_{\phi=\phi_{cl}}\right]=\int d^4y\,\lim_{x\to y}\sum_i\ln\,[-\lambda_i(x,y)]\,.
\ee
The subscript $i$ labels the eigenvalues $\lambda_i$ of the operator \eqref{eq:OperatorFunctional}. One can find them by diagonalizing \eqref{eq:OperatorFunctional}. It turns out that $i$ is a continuous label and can be directly identified with the momentum associated to a given mode, $\lambda_k=-k^2+V^{\pp}(\phi_{cl})$. Thus, we find,
\be
\ln\,{\rm det}\left[-\frac{\delta^2S_1[\phi]}{\delta\phi(x)\delta\phi(y)}\Bigr|_{\phi=\phi_{cl}}\right]=\mathcal{V}_4\int\frac{d^4k}{(2\pi)^4}\ln[-k^2+V^{\pp}(\phi_{cl})]\,,
\ee
so the one-loop effective potential takes the following form,
\be\label{eq:EffectivePotential}
V_{eff}(\phi_{cl})=V_1(\phi_{cl})+\delta V(\phi_{cl})-\frac{i\hbar}{2}\int\frac{d^4k}{(2\pi)^4}\ln[-k^2+V_1^{\pp}(\phi_{cl})]+\mathcal{O}(\hbar^2)\,.
\ee
Note that the integral appearing in the expression above is clearly divergent. We need to regularize it, and then renormalize the result in order to obtain a physical output under the appropriate renormalization conditions. But before doing this, I would like to remark that this integral contains the one-loop ZPE of the scalar field and the one-loop quantum correction of the potential. As in \cite{SolaReview2013}, we can split the integral in two parts by doing
\be\label{eq:potentialSplit}
\ln[-k^2+V_1^{\pp}(\phi_{cl})]= \ln\left[\frac{-k^2+V_1^{\pp}(\phi_{cl})}{-k^2+m^2}\right]+\ln[-k^2+m^2]\,,          
\ee
where $m$ is the renormalized mass (not the bare one). The first part in the {\it r.h.s.} contains the aforementioned one-loop correction of the classical potential and in general depends on the scalar field, whilst the second one is the one-loop field-independent ZPE. Notice that in the free-field case $V_1^{\pp}(\phi_{cl})=m^2$ and, therefore, after renormalizing the potential with the cosmological constant we obtain $\rho_{\Lambda,{\rm eff}}=\frac{1}{2}m^2\phi_{cl}^2+{\rm ZPE_{ren}}$, with ZPE$_{ren}$ denoting the renormalized ZPE. The minimum of the resulting effective potential is trivially found to be at $\phi_{cl}=0$, so $\rho_{\Lambda,{\rm eff}}={\rm ZPE_{ren}}$. In this case we can take the result derived in Appendix \ref{ch:appZPE} as the total vacuum energy density associated to the scalar field. But let me use now \eqref{eq:EffectivePotential} to compute the vacuum energy in a more general case than the free scalar one, when  $V_1^{\pp}(\phi_{cl})\ne m^2$. Let us firstly regularize the integral appearing in the {\it r.h.s.} of \eqref{eq:EffectivePotential}. I use the dimensional regularization technique. The first step consists in performing a Wick rotation, i.e. $k^\mu=(k^{0},\vec{k})\to k_E^\mu=(ik^{0}_E,\vec{k})$, in order to Euclideanize the integral,
\be
\mu^{4-d}\int\frac{d^dk}{(2\pi)^d}\ln[-k^2+V_1^{\pp}(\phi_{cl})]=i\mu^{4-d}\int\frac{d^dk_E}{(2\pi)^d}\ln[k_E^2+V_1^{\pp}(\phi_{cl})]=(*)
\ee
By doing the following mathematical trick
\be
(*)=-i\mu^{4-d}\frac{\partial}{\partial\alpha}\int\frac{d^dk_E}{(2\pi)^d}\frac{1}{[k_E^2+V_1^{\pp}(\phi_{cl})]^\alpha}\Bigr|_{\alpha=0}\,,
\ee
and applying the formalism of Sect. \ref{sec:DimReg} one finally obtains,
\be\label{eq:interm}
\mu^{4-d}\int\frac{d^dk}{(2\pi)^d}\ln[-k^2+V_1^{\pp}(\phi_{cl})]=-i\mu^{4-d}\frac{\Gamma\left(-\frac{d}{2}\right)}{(4\pi)^{d/2}}[V_1^{\pp}(\phi_{cl})]^{d/2}\,.
\ee
$\mu$ is an arbitrary parameter with dimensions of mass that must be introduced in order to keep the result with dimensions of energy to the quartic power. Logically, the result is still divergent in the limit $d\to 4$, but we will perform this limit only after the corresponding renormalization. Let me assume now that the only contributions to the effective energy density $\rho_{\Lambda,{\rm eff}}$ are given by the effective potential ($\rho_{\rm ind}=V_{eff}(\phi_{cl})$) of the scalar field presented in \eqref{eq:EffectivePotential} and the cosmological term ($\rho_{\rm vac}$),
\be
\rho_{\Lambda,{\rm eff}}=\rho_{\rm vac}+V_{eff}(\phi_{cl})\,.
\ee 
I remark that at this stage we are dealing with the bare CC and bare effective potential. Plugging \eqref{eq:interm} in \eqref{eq:EffectivePotential}, and after some algebra (see Appendix \ref{sec:DimReg}) one finds,
\be
\rho_{\Lambda,{\rm eff}}=\rho_{\rm vac}+V(\phi_{cl})-\frac{\hbar}{64\pi^2}[V_1^{\pp}(\phi_{cl})]^2\left[\frac{1}{\epsilon}-\gamma+\frac{3}{2}+\ln\left(\frac{4\pi\mu^2}{V_1^{\pp}(\phi_{cl})}\right)+\mathcal{O}(\epsilon^2)\right]+\mathcal{O}(\hbar^2)\,,
\ee
where I have defined $d\equiv 4-2\epsilon$. We must do the limit $\epsilon\to 0$ in order to obtain the final result, but before this we have to absorb the problematic pole $1/\epsilon$ of the last expression by using the counterterms contained in $\rho_{\rm vac}$ and $\delta V(\phi_{cl})$ (recall that $V=V_1+\delta V$). Obviously, this will only be possible if the power of the scalar fields contained in the potential are lower or equal than $4$. Only in this case the scalar field theory is renormalizable. Let us work out the result in the particular case in which we have the Higgs-like potential \eqref{eq:HiggsPotential}, which do satisfy this condition. In this case,
\begin{eqnarray}
\rho_{\Lambda, {\rm eff}} &=& \rho_\Lambda+\frac{m^2}{2}\phi_{cl}^2+\frac{\lambda}{4!}\phi_{cl}^4+\delta\rho_{\Lambda}+\frac{\delta_m}{2}\phi_{cl}^2+\frac{\delta_\lambda}{4!}\phi_{cl}^4-\\
&&-\frac{\hbar}{64\pi^2}\left(m^2+\frac{\lambda}{2}\phi_{cl}^2\right)^2\left[\frac{1}{\epsilon}-\gamma+\frac{3}{2}+\ln\left(\frac{4\pi\mu^2}{m^2+\frac{\lambda}{2}\phi_{cl}^2}\right)+\mathcal{O}(\epsilon^2)\right]+\mathcal{O}(\hbar^2)\,,\nonumber
\end{eqnarray}
where $m$ and $\lambda$ are the renormalized mass and coupling constant, respectively, and $\delta_m$ and $\delta_\lambda$ are the associated counterterms. I have also split $\rho_{\rm vac}$ in a finite part $\rho_\Lambda$ plus a counterterm $\delta\rho_\Lambda$. The three aforementioned counterterms are sufficient to absorb the divergences appearing in the last expression. By adopting the $\overline{MS}$ renormalization scheme (see Sect. \ref{sec:MSappendix} in the Appendix) we find, 
\begin{eqnarray}
\delta_m &=& \frac{\hbar}{32\pi^2}\lambda m^2\left[\frac{1}{\epsilon}+\frac{3}{2}-\gamma\right]\,,\nonumber\\
\delta_\lambda &=& \frac{3\hbar}{32\pi^2}\lambda^2\left[\frac{1}{\epsilon}+\frac{3}{2}-\gamma\right]\,,\label{eq:counterterms}\\
\delta\rho_\Lambda &=& \frac{\hbar}{64\pi^2}m^4\left[\frac{1}{\epsilon}+\frac{3}{2}-\gamma\right]\,,\nonumber\\
\end{eqnarray}
so I finally obtain,
\begin{eqnarray}\label{eq:EffEDHiggsFlat}
\rho_{\Lambda, {\rm eff}} &=& \rho_\Lambda+\frac{m^2}{2}\phi_{cl}^2+\frac{\lambda}{4!}\phi_{cl}^4-\nonumber\\
&&-\frac{\hbar}{64\pi^2}\left(m^2+\frac{\lambda}{2}\phi_{cl}^2\right)^2\ln\left(\frac{4\pi\mu^2}{m^2+\frac{\lambda}{2}\phi_{cl}^2}\right)+\mathcal{O}(\hbar^2)\,,
\end{eqnarray}
which is known as Coleman-Weinberg potential \cite{ColeWeinPot}. Notice that the counterterm coming from the CC renormalizes the ZPE, whilst the counterterms coming from the bare potential renormalize the potential itself. The effective energy density must be renormalization scale invariant. This means that it must not change under a shift of the value of $\mu$, since the latter is by definition an arbitrary parameter (we have only demanded it to have mass dimensions),
\be\label{eq:RGInvariance}
\frac{d\rho_{\Lambda,{\rm eff}}}{d\mu}=0\,.
\ee 
In order to fulfill this relation, the quantities $\rho_\Lambda$, $m$, $\lambda$ and $\phi_{cl}$ must acquire, in principle, some implicit dependence on $\mu$. This is something totally needed if one wants to cancel the explicit dependence on this parameter. Moreover, it is important to note the following point. According to \eqref{eq:potentialSplit}, the effective potential can be split as 
\be
V_{eff}(m(\mu),\lambda(\mu),\phi_{cl}(\mu),\mu)=V_{ZPE}(m(\mu),\lambda(\mu),\mu)+V_{scalar}(m(\mu),\lambda(\mu),\phi(\mu),\mu)\,,
\ee
so we can rewrite \eqref{eq:RGInvariance} as follows,
\be
\frac{d}{d\mu}\left[\rho_\Lambda(\mu)+V_{ZPE}(m,\lambda,\mu)+V_{scalar}(m,\lambda,\phi,\mu)\right]=0\,.
\ee  
And not only that, taking into account that variations induced in $V_{scalar}$ due to variations of $\mu$ can only be compensated by changes in $V_{scalar}$ itself, and that equivalently, that variations induced in $V_{ZPE}$ due to variations of $\mu$ can only be compensated by changes in $\rho_\Lambda$, we find that the last relation can be separated into two renormalization group equations (RGE's),
\be\label{eq:RGE1}
\frac{d}{d\mu}\left[\rho_\Lambda(\mu)+V_{ZPE}(m,\lambda,\mu)\right]=0\,,
\ee
\be\label{eq:RGE2}
\frac{d}{d\mu}\left[V_{scalar}(m,\lambda,\phi,\mu)\right]=0\,.
\ee
In the non-interacting theory, i.e. when $\lambda=0$, we retrieve the well-known result \eqref{eq:renZPE} and the beta function for the cosmological term \eqref{eq:betaFunc}, 
\be\label{eq:betaFuncMain}
\beta_\Lambda^{(1)}\equiv\mu\frac{d\rho_\Lambda(\mu)}{d\mu}=\frac{\hbar m^4}{32\pi^2}\,,
\ee
In fact, it turns out that this is also obtained in the interacting case at one-loop order. But which is the role of $\mu$? Is it relevant? Can we endow it with some kind of physical meaning? Can we learn something about the running of the physical constants appearing in the vacuum effective action? Is, indeed, this running a real running? 

In order to tackle these questions, let me first talk about an example coming from massless Quantum Electrodynamics (QED), the effective theory describing the interactions of photons and U(1)$_{em}$-charged particles when the energy of the latter are much larger than their masses, i.e. $|q^2|\gg m^2$. The electromagnetic part of the one-loop renormalized effective action takes the form
\be\label{eq:EMGamma}
\Gamma^{(1)}_{em}=-\frac{1}{4e^2(\mu)}\int d^4x F_{\mu\nu}\left[1-\frac{\hbar e^2(\mu)}{12\pi^2}\ln\left(-\frac{\Box}{\mu^2}\right)\right]F^{\mu\nu}\,,
\ee
where $e(\mu)$ is the renormalized QED charge in the $\overline{MS}$ scheme. The effective action is, of course, $\mu$-independent. This allows us obtain the one-loop $\beta$-function for $e$,
\be\label{eq:betaQEDUV}
\beta_e^{\overline{MS}(1)}=\mu\frac{de(\mu)}{d\mu}=\frac{\hbar e^3}{12\pi^2}\,.
\ee
From this RGE or directly from \eqref{eq:EMGamma} it seems that the running QED charge satisfies (in the high energy limit)
\be
\frac{1}{e^2(|q^2|)}=\frac{1}{e^2(\mu^2)}-\frac{\hbar}{12\pi^2}\ln\left(\frac{|q^2|}{\mu^2}\right)\,,
\ee
which shows that the QED charge increases with the energy, something that has been very well-tested by experiments. Thus, the $\beta$-function for $e$ obtained in the $\overline{MS}$ renormalization scheme is able to reproduce correctly the predictions of the theory in the high energy-limit. It is important to remark that the RG-invariance of the effective action is kept untouched. In this case, though, the energy scale $\mu$ can be directly identified with a physical quantity, the energy of the electron. This is a crucial aspect of the RG-method in QFT: being the combination $|q^2|/\mu^2$ the natural variable in the renormalized scattering amplitudes, the RG helps us to find out physical quantum effects that ``run'' with the energy or some external field by just inspecting the $\mu$-dependence of the various parts of the S-matrix and effective action. But the $MS$ scheme is efficient only in the UV limit. If one wants to obtain the physical $\beta$-function in the IR limit, one has to make use of a mass-dependent renormalization scheme, as the momentum subtraction one. In this case, 
\be                                       
\beta^{(1)}_{e}=\frac{\hbar e^3}{2\pi^2}\int_{0}^{1}dx\,x(1-x)\frac{M^2x(1-x)}{m_e^2+M^2 x(1-x)}\,,
\ee
where $m_e$ is the mass of the electron and the momentum subtraction has been performed at $q^2=M^2$. From this expression one recovers \eqref{eq:betaQEDUV} in the UV (when $M\gg m_e$), and obtains
\be\label{eq:QEDbetaLOW}
\beta^{(1)}_{e,IR}=\frac{\hbar e^3}{60\pi^2}\cdot\frac{M^2}{m_e^2}+\mathcal{O}\left(\frac{M^4}{m_e^4}\right)
\ee
in the IR (when $M\ll m_e$). This quadratic decoupling fulfills the Appelquist-Carazzone theorem \cite{AppelquistCarazzoneTheorem}. Contrary to what one could naively think, particles with a mass larger than the typical energy scale of the process under consideration do not decouple completely from the theory. In other words, the decoupling of massive particles does not have a ``sharp cut-off'' behavior. The quantum effects of those particles with masses $m\gg M$ can have a non-null impact on the physical running of the parameter under study.

Now, let me go back to the cosmological scenario and consider again the $\beta$-function for the CC \eqref{eq:betaFuncMain}, which has been computed by applying the $\overline{MS}$ renormalization scheme. The first problem that we encounter when we try to make use of the RG method in the cosmological context is that in this case the energy scale $\mu$ is something difficult to interpret. The RGE for the CC can only acquire a physical meaning after setting the link between $\mu$ and a physical (directly or indirectly measurable) energy scale of the problem. This is not an easy task. In QED this identification ($\mu^2\to|q^2|$) is immediate, by direct comparison of the $\beta$-function obtained with the (mass-dependent) momentum-subtraction scheme in the high energy limit with the results obtained with the (mass-independent) $\overline{MS}$ scheme. In the QCD case, it is not possible to apply at low energies a mass-dependent scheme because of the quark-confinement. Despite this, the high energy limit of the theory is correctly reproduced and, again, one can identify the scale $\mu$ with the typical energy of the scattering process. 

I am interested in the potential running of $\rho_\Lambda$ in the cosmological context, but I have computed \eqref{eq:betaFuncMain} in flat spacetime! Thus, {\it a priori} we cannot rely on this result, since the expansion of the Universe is described by a dynamical metric. But still, could we endow $\mu$ with a physical meaning in Minkowski? The first important thing to notice is that, obviously, Minkowski spacetime is static and, therefore, the CC must be also time-independent in this case. This means that $\mu$ cannot induce a dynamical evolution of the CC. Moreover, vacuum blobs cannot interact with the gravitational field, since the mere presence of such field would render the metric deviate automatically from $\eta_{\mu\nu}$. Thus, in this case $\mu$ can only behave as a quantity that parametrizes the sequence of possible renormalization conditions that lead to $\rho_{\Lambda,{\rm eff}}=0$ \footnote{We must demand $\rho_{\Lambda,{\rm eff}}=0$ in order Minkowski's spacetime to be a global solution of Einstein's equations in vacuum, which is a kind of minimal mathematical requirement for ensuring the consistency of our calculations, since up to now we have only computed the vacuum energy in flat spacetime and it would be contradictory to apply our QFT-in-flat-spacetime solution to Einstein's equations with a non-trivial metric. Of course, Minkowski's spacetime do not necessarily have to be a global solution of Einstein's equations in vacuum and, in fact, the need of a non-null effective CC to explain the cosmological observations seems to indicate that certainly  it is not the case.}, without any link with a physical running. We cannot associate $\mu$ to an external physical energy scale that describes the state of the system. It is just a formal parameter. This is the simplest case. The scale $\mu$ can only ``induce'' a physical running of the CC when the metric describing the geometry of spacetime is non-trivial, so before going on with the interpretation of $\mu$ in the cosmological context, we are enforced to see which is the expression for $\beta_\Lambda$ obtained from QFT in curved spacetime. This is the main goal of the following section.

%%%%%%%%%%%%%%%%%%%%%%%%%%%%%%%%%%%%%%%%%%%%%%%%%%%%%%%%%
%%%%%%%%%%%%%%%%%%%%%%%%%%%%%%%%%%%%%%%%%%%%%%%%%%%%%%%%%
%%%%%%%%%%%%%%%%%%%%%%%%%%%%%%%%%%%%%%%%%%%%%%%%%%%%%%%%%

\subsection{Dynamical vacuum from QFT in curved spacetime}
\label{subsec:RVMintro}

In order to study how is the vacuum energy altered in the presence of a non-flat metric I will make use of some general results obtained from QFT in curved spacetime \cite{BirrellDavies,ParkerToms,MukhanovQGbook}. This is a semiclassical theory, in which the metric is treated classically (like a background field), whilst matter fields are quantized. This is a natural approach (previous to the eventual achievement of a successful theory of quantum gravity), which can be considered as a viable road to the study of quantum effects in gravity when the curvature is weak enough, i.e. when $R\ll M_p^2$. This is the case of the late-time Universe. In fact, this is fulfilled during all the post-inflationary eras, so it seems to constitute a good enough framework. 

In QFT in curved spacetime the total classical action reads,
\be              
S_{total}=S_{EH}+S_{HD}+S_{matter}\,,
\ee 
with
\be\label{eq:EHaction}
S_{EH}[g_{\mu\nu}]=-\int d^4x\,\sqrt{-g}\left(\frac{R+2\Lambda_b}{16\pi G_b}\right)\,,
\ee
\be\label{eq:HDaction}
S_{HD}[g_{\mu\nu}]=-\int d^4x\,\sqrt{-g} \left(\alpha^{(b)}_1C^2+\alpha^{(b)}_2R_{GB}^2+\alpha^{(b)}_3\Box R+\alpha^{(b)}_4R^2\right)\,.
\ee
$S_{EH}$ is the Einstein-Hilbert (EH) action with cosmological term, and $S_{HD}$ is the minimal set of higher derivative terms needed to ensure the renormalizability of the theory at one-loop level (see below). In the literature the sum $S_{EH}+S_{HD}$, which only contains geometric terms, is also called the vacuum action, i.e. $S_{vac}$, for obvious reasons. Notice that all the constants appearing in $S_{vac}$, i.e. $\Lambda_b$, $G_b$ and $\alpha^{(b)}_i$'s, must be understood as the bare ones. Later on we will see that they are in charge of the absorption of the infinities arising in the quantum theory. In \eqref{eq:HDaction}, 
\be                              
C^2\equiv R_{\mu\nu\rho\sigma}R^{\mu\nu\rho\sigma}-2R_{\mu\nu}R^{\mu\nu}+\frac{R^2}{3}
\ee 
is the square of the Weyl tensor, and $R_{GB}^2$ is the Gauss-Bonnet term defined before in \eqref{eq:GaussBonnet}. $S_{matter}$ is the classical action of the matter sector, which is a functional of the metric and also of the various matter fields. At this stage, though, I only consider the contribution of a scalar field without self-interaction in $S_{matter}$. This will allow us to compare in a more direct way the results of Sect. \ref{subsec:EffectivePot} with the ones that will be derived now. Moreover, in the one-loop approximation the interaction parameter does not affect the CC and, therefore, the inclusion of this self-interaction term does not add any extra information to the problem we want to analyze. Thus,
\be\label{eq:SmatterCurved}
S_{matter}[g_{\mu\nu},\phi]=\frac{1}{2}\int d^4x\,\sqrt{-g}\left(\partial_\mu\phi\partial^\mu\phi-m^2\phi^2-\xi R\phi^2\right)\,,
\ee
where the term $\xi R\phi^2$ is also needed to provide renormalizability. From the total action one can get the modified Einstein field equations, which obviously will include terms of second order in the curvature tensor, due to $S_{HD}$. These higher derivative corrections to the original equations do not play a significant role in the post-inflationary Universe. In the semiclassical approach, one cannot simply plug the total energy-momentum tensor in the {\it r.h.s.} of these equations because now this object has quantum nature, it is an operator which acts on states. One has to consider the expectation value $<T_{\mu\nu}>$, i.e. the quantum average in the path integral formalism, 
\be
R_{\mu\nu}-\frac{1}{2}Rg_{\mu\nu}-\Lambda_bg_{\mu\nu}+{\rm HD\,terms}=8\pi G_b <T_{\mu\nu}>   \,.
\ee
The main goal of QFT in curved spacetime is the obtainment of $<T_{\mu\nu}>$ or, more concretely, the effective action $W$ for the quantum matter fields that leads to this object. The starting point, as in flat spacetime, is the generating functional of the Green functions, \eqref{eq:GeneratingZ}, with $S_m[g_{\mu\nu},\phi]$ given by \eqref{eq:SmatterCurved}. It is easy to show that if we define $Z[0]\equiv e^{\frac{i}{\hbar}W}$, then 
\be
<T_{\mu\nu}>=\frac{<{\rm out},0|T_{\mu\nu}|0,{\rm in}>}{<{\rm out},0|0,{\rm in}>}=\frac{2}{\sqrt{-g}}\frac{\delta W}{\delta g^{\mu\nu}}\,.
\ee
The computation of the effective action in curved spacetime is a rather technical one. I will omit here these steps and refer the reader to the books \cite{BirrellDavies,ParkerToms,MukhanovQGbook} and the review \cite{MartinReview} for the details of this involved calculation. I just show here the result for the one-loop effective action in the case under study,
\be
\Gamma=S_{EH}+S_{HD}+W=S_{EH}+S_{HD}+S_{m}[\phi_{cl}]+\Gamma^{(1)}_{eff}\,,
\ee
where
\be\label{eq:EAcurved}
\Gamma^{(1)}_{eff}=\frac{\hbar}{2(4\pi)^2}\int d^4x \sqrt{-g}\left[\frac{1}{\epsilon}+\frac{3}{2}-\gamma+\ln\left(\frac{4\pi \mu^2}{m^2}\right)\right]\left(\frac{1}{2}m^4a_0(x)-m^2a_1(x)+a_2(x)\right)
\ee
is the one-loop quantum correction to the classical action, and $a_{0,1,2}$ are the so-called Schwinger-DeWitt coefficients (coming from the adiabatic expansion of the matter field propagator in the curved background, \cite{BirrellDavies,ParkerToms}). They read,
\be
a_0(x)=1\qquad a_1(x)=\left(\xi-\frac{1}{6}\right)R\,,
\ee
\be
a_2(x)=\frac{1}{2}\left(\frac{1}{6}-\xi\right)^2R^2+\frac{1}{6}\left(\frac{1}{5}-\xi\right)\Box R+\frac{1}{180}\left(R^{\mu\nu\sigma\delta}R_{\mu\nu\sigma\delta}-R^{\mu\nu}R_{\mu\nu}\right)\,.
\ee
A similar result can be obtained by using the canonical quantization formalism, instead of the path integral one (cf. Appendix \ref{ch:appZPEcurved} for the explicit calculations). Actually, it leads to the same functions $a_0(x)$ and $a_1(x)$. The difference arise in some of the HD terms of $a_2(x)$, which turn out not to be present in the result obtained with this alternative approach. Expression \eqref{eq:EAcurved} is obtained upon applying, again, the dimensional regularization technique. In order to renormalize the theory, we have to identify the following relations,
\be\label{eq:curved1}
\rho_{\Lambda,{\rm eff}}=\rho_{\Lambda,b}-\frac{\hbar m^4}{64\pi^2}\left[\frac{1}{\epsilon}+\frac{3}{2}-\gamma+\ln\left(\frac{4\pi \mu^2}{m^2}\right)\right]\,,
\ee
\be\label{eq:curved2}
\frac{1}{16\pi G_{\rm eff}}=\frac{1}{16\pi G_b}-\frac{\hbar m^2}{32\pi^2}\left(\frac{1}{6}-\xi\right)\left[\frac{1}{\epsilon}+\frac{3}{2}-\gamma+\ln\left(\frac{4\pi \mu^2}{m^2}\right)\right]\,.
\ee
The corresponding expressions are also found for the bare couplings appearing in the higher-derivative terms of \eqref{eq:EAcurved}, but we are not that interested in them because their impact in the low-energy Universe is negligible. The renormalization of the cosmological term and $G$ requires us to split the bare quantities as follows, 
\begin{figure}
\centering
\includegraphics[angle=0,width=0.6\linewidth]{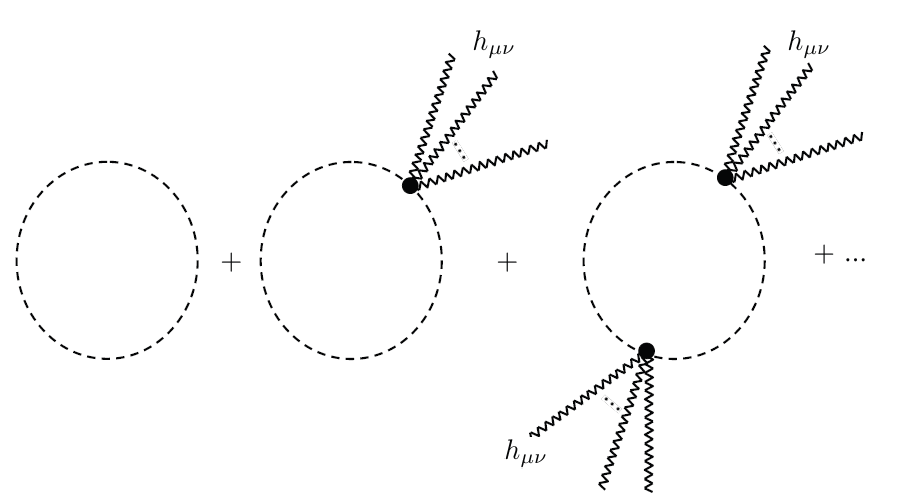}
\caption[One-loop vacuum-to-vacuum diagrams of the scalar matter field in the presence of an external gravitational field.]{\label{fig:VacuumDiag}%
\scriptsize{The one-loop vacuum-to-vacuum diagrams of the scalar matter field in the presence of an external gravitational field. From \cite{SolaReview2013}.}}
\end{figure}
\be
\rho_{\Lambda,b}=\rho_\Lambda+\delta\rho_\Lambda\qquad \frac{1}{G_b}=\frac{1}{G}+\delta\left(\frac{1}{G}\right)\,.
\ee
The counterterms $\delta\rho_\Lambda$ and $\delta\left(\frac{1}{G}\right)$ do the job of absorbing the infinities of \eqref{eq:curved1} and \eqref{eq:curved2}, respectively. But before renormalizing $\rho_{\Lambda,b}$ and $G_b$, it is important to detect which is the source of the divergences appearing in \eqref{eq:curved1} and \eqref{eq:curved2}. The pole of \eqref{eq:curved1} comes from the ``bald blob'' with no external tails of Fig. \ref{fig:VacuumDiag}, which is quartically divergent. The other contribution are the ``haired blobs'' with one or more external field insertions, where an arbitrary number of $h_{\mu\nu}$-tails of the background field are attached to one or more points. They appear from the expansion of the metric in the form $g_{\mu\nu}=\eta_{\mu\nu}+h_{\mu\nu}$, and the corresponding determinant in the action: $\sqrt{-g}=1+\frac{h}{2}+...$, with $h\equiv \eta_{\mu\nu}h^{\mu\nu}$. Notice that the expression \eqref{eq:curved1} coincides with the one that has been obtained in Minkowski, \eqref{eq:regZPEdim}. This is because the diagrams from which they are computed are exactly the same! In both cases, there is no communication between the blob and the background gravitational field, so it is totally normal to obtain the same result, since for the bald blob spacetime is Minkowskian even if it lives embedded in a curved one. Thus, \eqref{eq:curved1} cannot give us any hint about the possible physical running of $\rho_\Lambda$ because the gravitational field does not influence the calculation of the ZPE. After renormalization of \eqref{eq:curved1}, we obtain again,
\be\label{eq:renormCurved}
\rho_{\Lambda,{\rm eff}}=\rho_\Lambda(\mu)-\frac{\hbar m^4}{64\pi^2}\ln\left(\frac{4\pi\mu^2}{m^2}\right)\,.
\ee
A quite natural renormalization condition can be implemented by demanding $\rho_{\Lambda,{\rm eff}}$ to vanish \cite{SolaReview2013,SolaHonorableMention2015}. This is equivalent to force $\eta_{\mu\nu}$ to be a solution of Einstein's equations in vacuum. Of course, in the case under study we are only taking into account the ZPE of the scalar field. In case we wanted to include other contributions, the effective vacuum energy in curved spacetime would be obtained by subtracting the effective vacuum energy in Minkowski to the one obtained in the presence of a curved background. In this subtraction, the CC and the ZPE terms would cancel, and only the differences in other sectors would effectively contribute in the vacuum part of the {\it r.h.s.} of Einstein's equations. In principle, nothing prevents the resulting difference to run with the expansion (this would give rise to dynamical vacuum energy), although at this point we do not have any hint in favor of a particular $\mu$-dependence. It will be deduced later in a more indirect way, by demanding the fulfillment of the covariant conservation of the total energy of the system. Notice that the aforementioned renormalization condition also helps to alleviate the CC problem, in the sense that by using it we can get rid of the fine-tuning issue that affects the ZPE. If the ZPE's weren't exactly canceled, one would have to somehow explain why the (in general huge) contributions of \eqref{eq:renormCurved} proportional to $m_i^4$ are not observed\footnote{Actually, there is another way to alleviate the existing fine-tuning problems in the ZPE sector. This is done in the context of the renormalization group formalism in curved spacetime, by demanding the soft-decoupling of massive particles in the running of $\rho_\Lambda$. See the following pages for details and the corresponding discussion on this important point.}. In addition, the fact that the effective vacuum energy density results from $\rho_{\rm vac}^{\rm curved}-\rho_{\rm vac}^{\rm Mink}$ (with the ZPE's exactly canceled) is also very interesting, since this subtraction can suppress drastically other problematic contributions, as the one coming from the EW symmetry breaking. This cancellation mechanism resembles the Casimir effect \cite{Casimir1948}, in which vacuum contributions become observable thanks to a change in the boundary or geometrical conditions of the problem. In this case, though, vacuum effects gravitate due to a change in the background metric with respect to the flat one.

The study of the other two diagrams of Fig. \ref{fig:VacuumDiag} can also give us some insight in the problem we are facing. The first haired vacuum diagram with one insertion is quadratically divergent, as there is one propagator of the scalar field. The bunch of tails organize themselves in a covariant way to generate an action term of the form $\sqrt{-g}m^2R$, and hence renormalize the inverse gravitational Newton's coupling $1/G$ in the EH action \eqref{eq:EHaction}. The third type of diagram is the ``doubly haired blob'' and contains two insertion points. With two propagator lines, it is only logarithmically divergent. It renormalizes the coefficients of the HD-action \eqref{eq:HDaction}. The diagrams with three or more insertions of the external field are perfectly finite and thus do not alter the running of the coupling constants.

Let me renormalize now \eqref{eq:curved2} making use of the $\overline{MS}$ scheme. The result is,
\be\label{eq:Grenorm}
\frac{1}{16\pi G_{\rm eff}}=\frac{1}{16\pi G(\mu)}-\frac{\hbar m^2}{32\pi^2}\left(\frac{1}{6}-\xi\right)\ln\left(\frac{4\pi \mu^2}{m^2}\right)\,,
\ee
																																							from which one can easily compute the corresponding $\beta$-function for $1/G$,
\be\label{eq:betaG}
\beta^{(1)}_{G^{-1}}\equiv \frac{d}{d\ln\mu}\left(\frac{1}{16\pi G(\mu)}\right)=\frac{\hbar m^2}{16\pi^2}\left(\frac{1}{6}-\xi\right)\,.
\ee																							
This beta function can allow us to endow Newton's coupling with some physical running because, as it has been stated before, the renormalized expression \eqref{eq:Grenorm} is obtained from haired diagrams and, therefore, the gravitational background must leave a mark in this result. But we must take into account that we live in a low-energy Universe. The characteristic cosmological scale at present is given by $H_0\sim 10^{-33}$ eV, so $m_i\gg H_0$ for all the known particles of the SM. Thus, we have to address the following question: does this $\beta$-function describe in the IR regime the real running of $G$? We have seen before, in the QED example, that the $\overline{MS}$ renormalization scheme provides the correct running in the UV, but not in the IR, where the soft-decoupling of massive particles takes place. Despite this, we cannot ensure that the behavior of the gravitational coupling is the same as the electric charge. From the theoretical side, there does not exist any mass-dependent renormalization scheme like momentum-subtraction able to provide the low-energy running of the CC and $G$ \cite{ShapiroSola2008,ShapiroSola2008p}, although it does exist for the HD couplings\footnote{In \cite{GorbarShapiro2003} the authors obtained  a low-energy $\beta$-function for these parameters with a soft-decoupling behavior, following the Appelquist-Carazzone theorem.}. The absence of a theoretical prediction of the low-energy $\beta$-function for the CC and $G$ enforces us to face the problem from a phenomenological perspective at this stage. Let me analyze the casuistry of the problem, by considering the various possibilities that we have. At this point, we cannot postpone anymore the discussion about the interpretation of the physical energy scale $\mu_{phy}$\footnote{Here I use $\mu_{phy}$ instead of $\mu$. The former parametrizes the physical running, whereas the latter is just the parameter coming from the dimensional regularization scheme, with no direct physical meaning. The dependence of the various physical quantities on $\mu$ motivate the form in which they depend on $\mu_{phy}$, but it is important to remark the existing difference. In reality, the $\mu$-invariance of the effective action is an automatic property which holds for all kinds of renormalizable theories and, definitely, cannot be used to derive any direct conclusions on the running. Despite this, we have mentioned some examples (in QED and QCD) in which the physical running can be inferred in the UV by an identification $\mu_{phy}\leftrightarrow\mu$ in the corresponding $\beta$-function.} that is supposed to govern the running, since the aforementioned casuistry also depends on our choice of this energy scale, simply because it controls the decoupling ``border''. Obviously, in the cosmological context, in which one deals with the FLRW metric, $\mu_{phy}$ must be a dynamical quantity, which somehow must inform us about the gravitational background in which particles live in. Regarding to the association of $\mu_{phy}$ with some energy scale characteristic of the stage of evolution of the Universe, many different options have been explored in the literature. Some examples are: the critical energy density of the Universe, $\mu_{phy}\sim \rho_c^{1/4}$ \cite{BabicEtAl2002,ShapiroSola2000,GuberinaEtAl2003}; the temperature of the Universe, $\mu_{phy}\sim T$; or the inverse of the cosmic time, $\mu_{phy}\sim t^{-1}$ \cite{BonanoReuter2002a,BonanoReuter2002b}. Some of these alternatives are very similar or even equivalent at some stages of the Universe's evolution. For instance, during the radiation-dominated epoch $T\propto\rho_r^{1/4}\sim\rho_c^{1/4}$, so the first two approaches coincide in this case. A natural option would also consist in associating $\mu^2_{phy}$ with the curvature scalar $R$ or, alternatively, with the Hubble function, i.e. $\mu_{phy}\sim H$. $H$ can be thought of as the typical energy scale of the cosmological expansion described by the FLRW metric. In this thesis, I will make use of this identification and explore the phenomenological behavior of the corresponding models\footnote{A brief comment on the differences of decoupling border induced by alternative options of the scale setting. Let me compare the case in which $\mu_{phy}=H$ and the case in which $\mu_{phy}=\rho_c^{1/4}$. Notice that $H_0\sim 10^{-42}$ GeV, whereas $(\rho_c^{0})^{1/4}\sim 1$ meV. In the first case, one would expect a neutrino with a meV mass to be softly or completely decoupled in the current Universe, since $\mu_{phy}\ll m_\nu$, whereas in the second case this same neutrino would not be in principle decoupled due to the fact that $\mu_{phy}\approx m_\nu$.}. Upon the RG scale setting we also set the decoupling border. Basically, all the massive particles are much heavier than $H$ in the low-energy Universe. Depending on which type of decoupling we consider we will find a different behavior of the running at low energies. Let us see the different possibilities that we could study in our phenomenological analysis:

\begin{itemize}
\item {\it Ansatz 1: Sharp-cutoff for $G^{-1}$.} There is no massive particles contributing to the $\beta$-function, so $\beta_{G^{-1}}=0$ and, therefore, there is no running for $G$.

\item {\it Ansatz 2: Soft-decoupling for $G^{-1}$.} In this case the $\beta$-function reads,
\be
\beta_{G^{-1}}=\frac{\hbar m^2}{16\pi^2}\left(\frac{1}{6}-\xi\right)\left[c_1\left(\frac{H}{m}\right)^2+c_2\left(\frac{H}{m}\right)^4+...\right]\,,
\ee
Notice that we have only included even powers of $H$ in order to respect the general covariance of the theory. Thus, the leading term is given by
\be
\beta_{G^{-1}}=\frac{\hbar c_1}{16\pi^2}\left(\frac{1}{6}-\xi\right)H^2+\mathcal{O}(H^4/m^2)\,,
\ee
and, therefore, neglecting the higher order corrections we obtain, 
\be
G(H)=\frac{G_0}{1+\frac{\hbar c_1}{2\pi}\left(\frac{1}{6}-\xi\right)\left(\frac{H^2-H_0^2}{M_p^2}\right)}\,.
\ee
The running in this case is negligible and does not leave any trace at the phenomenological level in the post-inflationary Universe.

\item {\it Ansatz 3: No decoupling of massive particles for $G^{-1}$.} The dominant contribution to $\beta_{G^{-1}}$ is given by \eqref{eq:betaG}, $\beta_{G^{-1}}\sim m^2$. The integration of this expression yields,
\be\label{eq:G(H)fp}
G(H)=\frac{G_0}{1+\nu\ln\left(\frac{H^2}{H_0^2}\right)}\,,
\ee
where $G_0=G(H_0)$. Here $G$ should be understood as $G_{\rm eff}$, which is the real ``observable'' entity. From now on I will use $G$ instead of $G_{\rm eff}$, but bear in mind its real meaning! Obviously, the Universe is not only filled with one scalar field, so $\nu$ must encapsulate the running effects induced by all the fields, also those induced by fermionic and other bosonic fields,
\be
\nu=8\pi G_0\sum_{i=f,b}\beta_{G^{-1},i}=\sum_{i=f,b}\sigma_i\frac{m_i^2}{M_P^2}\,,
\ee
where the $m_i$'s denote the masses of the particles and $\sigma_i$ are characteristic constants, e.g. for a scalar particle we have seen that $\sigma_i=\frac{\hbar}{2\pi}\left(\frac{1}{6}-\xi_i\right)$. Of course, the prediction for $\nu$ depends on the particle content of the theory under consideration, but this coefficient is predicted to be naturally small since $m_i^2\ll M_P^2$, even for the heavy particles of a Grand Unified Theory (GUT) below the Planck scale. In the latter case, $\nu$ is typically found to be in the range $|\nu|=10^{-6}-10^{-3}$ \cite{BasiLimaSola2013}, and curiously enough, according to the most updated cosmological observations this is precisely the range in which $\nu$ lies, see the fitting tables presented in Chapters \ref{chap:Atype}-\ref{chap:H0tension}.

\end{itemize}

Now, we should think about the possible implications that these ansatzs have regarding to the vacuum energy. The various scenarios will ultimately depend on the specific type of decoupling considered for $\rho_\Lambda$.

\begin{enumerate}[1.]
        \item {\bf Ansatzs 1 and 2}
				
Due to the negligible running of $G$ found in ansatz 2, it can be considered for all practical purposes on an equal footing with ansatz 1. 
     \begin{enumerate}[a)]
		\item {\it Sharp-cutoff for the vacuum energy.} This is the trivial case, since $\beta_\Lambda=0$ and, therefore, we recover the standard $\Lambda$CDM model (there is no running for $\Lambda$ nor $G$). 
\item {\it Soft-decoupling for the vacuum energy.} The running is described in this case by the following $\beta$-function\footnote{See the recent reference \cite{Antipin2017}, where the authors deduce a $\beta$-function for the vacuum energy with this soft-decoupling behavior seemingly using a mass-dependent scheme.},
		\be\label{eq:betaLambda}
		\frac{d\rho_\Lambda}{d\ln H^2}=\frac{1}{16\pi^2}\sum_{i=f,b} \left[a_im_i^2H^2+b_iH^4+\mathcal{O}\left(\frac{H^6}{m_i^2}\right)\right]\,.
		\ee	
		where the dimensionless coefficients $a_i$, $b_i$, etc. receive contributions from loop corrections of boson and fermion matter fields with different masses $m_i$. The energy density $\rho_\Lambda$ should be understood as $\rho_{\Lambda,{\rm eff}}$, analogously to what I have mentioned before about $G_{\rm eff}$ and $G$. Again, no odd powers of $H$ are taken into account in order not to break the general covariance of the theory. In this particular case $G$ remains constant, so the variation of the vacuum energy must be accompanied by an anomalous conservation law for matter and/or radiation. These kind of models will be referred to from now on as type-A models \cite{Grande2011,JCAPnostre1,ApJnostre}. They will be studied in detail in the subsequent chapters, but let me pinpoint one relevant physical aspect that can be already grasped from \eqref{eq:betaLambda}. The leading correction of the cosmological term takes the form $\delta\rho_\Lambda\sim m_i^2H^2$. Thus, contrary to what one could naively think, the physics below the GUT scale is likely irrelevant for the potential scale dependence of the vacuum energy! 
		
		\item {\it Non-decoupling for the vacuum energy}. It does not lead us to any interesting case. This would lead to the $\beta_\Lambda\sim m^4$ \eqref{eq:betaFunc}, but I have explained before that this contribution could be removed by using the appropriate renormalization condition, consisting in subtracting the vacuum energy in Minkowski to the total vacuum energy in curved spacetime. Notice that even if we did not use this particular renormalization condition, we would have to exclude the contribution of these terms scaling like $\beta_\Lambda\sim m^4$, simply because this scenario is phenomenologically non-viable. It gives rise to a too large running for $\rho_\Lambda$.
		
\end{enumerate}

\item {\bf Ansatz 3}

\begin{enumerate}[a)]
		\item {\it If matter and radiation are self-conserved.} In this case, the running of $G$ described by the law \eqref{eq:G(H)fp} induces the dynamical evolution of the vacuum energy. This is natural, since the covariant conservation of the total energy must be preserved. Due to the Bianchi identity \eqref{eq:BianchiIndentity} we find,
\be
\nabla^\mu[G(H)T_{\mu\nu}]=0\,.
\ee
As matter and radiation are covariantly conserved, we find the following relation between the variation of $\rho_{\Lambda}$ and $G$,
\be
\dot{G}(\rho_m+\rho_r+\rho_{\Lambda})+G\dot{\rho}_{\Lambda}=0\,.
\ee
By using the Friedmann equation, together with \eqref{eq:G(H)fp} we finally find,
\be\label{eq:Lambda(H)fp}
\rho_{\Lambda}(H)=C_0+\frac{3\nu}{8\pi G_0}H^2\longleftrightarrow \beta_\Lambda=\frac{3\nu}{4\pi}H^2M_P^2\,.
\ee  
The law that governs the evolution of $\rho_\Lambda$ takes the same form as the leading term of \eqref{eq:betaLambda}, and the product $H_0^2/G_0=(H_0M_P)^2$ is not that far from the order of magnitude of the measured value of the CC, and thus the time variation of $\rho_\Lambda$ is in this framework something susceptible to be measured. These models will be referred to from now on as type-G models \cite{Grande2011,ApJLnostre,ApJnostre}. 
		\item {\it If matter and/or radiation are NOT self-conserved.} There are two possible scenarios: 
	\begin{enumerate}
	\item {\it Sharp-cutoff for the vacuum energy.} In this case the running of $G$ must induce an anomalous conservation law for matter and/or radiation. Now we do not have a running vacuum, but this model is also very rich from the phenomenological point of view. It can explain the potential link between the variation of Newton's coupling and the variation of the masses of particles and/or the running of other fundamental ``constants'' of Nature, as $\alpha_{\rm em}$ or $\Lambda_{\rm QCD}$ \cite{FritzschSola2012,FritzschSola2015,FritzschSolaNunes2017}. 
	\item {\it Soft-decoupling for the vacuum energy.} This scenario is quite general. One must make use of \eqref{eq:G(H)fp} in combination of \eqref{eq:betaLambda}. The anomalous conservation law for matter and/or radiation is obtained from the covariant energy conservation equation. 
	\end{enumerate}
\end{enumerate}
\end{enumerate}

Notice that the Hubble function gives us information about the rate at which the Universe expands. Our choice $\mu_{phy}=H$ seems quite reasonable, since when we do $H\to 0$ in the expression of $\beta_\Lambda$ it cancels, what means that the running of the CC is triggered by the non-trivial evolution of the background. On the other hand, the vacuum energy density can be obtained upon direct integration of $\beta_\Lambda$. As a result one generates an integration constant, giving rise to an expression like $\rho_\Lambda=C_0+C_1 H^2+\mathcal{O}(H^4)$. But is the presence of a non-zero $C_0$ consistent with the renormalization condition explained before? We have seen that the effective vacuum energy acting in a non-trivial spacetime could result from the subtraction of the vacuum energy computed in flat spacetime from the one obtained in that non-trivial spacetime. Thus, we would expect the effective vacuum energy density to cancel in the limit $H\to 0$, since the Hubble function parametrizes in the FLRW metric the deviation from Minkowski's one. Therefore, we would expect to find $C_0=0$, i.e. $\rho_\Lambda\propto H^2$. But our phenomenological studies show that the case $C_0=0$ is strongly excluded by observations (cf. Chapters \ref{chap:Atype} and \ref{chap:DynamicalDE} of this dissertation and references therein). This is pointing out that conceiving the effective vacuum energy as the subtraction $\rho_{\Lambda,{\rm eff}}=\rho_{\rm vac}^{\rm curved}-\rho_{\rm vac}^{Mink}$ is probably the result of an oversimplification of the problem. Nevertheless, we have seen that the fine-tuning required to match the measured value of $\rho_{\Lambda}$ with the theoretical prediction of QFT can also be alleviated by considering the soft decoupling of the massive particles in the running of the CC, which generates a dynamical correction to the cosmological constant that is not $\delta\rho_ \Lambda\propto m_i^4$, but $\delta\rho_\Lambda \propto m_i^2H^2$. The latter is a quantity that provides smoother dynamics to $\rho_{\Lambda}$ than the former. These facts soften the CC problem in the ZPE sector, but many other questions are still pending an answer. For instance, what does it happen with the induced QCD vacuum contributions or those coming from the Higgs potential? Is the CC problem also weakened in these sectors? Certainly, it should, but how do these contributions conspire to produce such a tiny constant? Is there indeed a definite answer to these questions? Maybe we must think of the CC as other renormalized quantities in QFT, e.g. $\alpha_{\rm em}$, which despite being fundamental can only be determined after measurement. It could be that only an eventual ``theory of everything'' could have the predictive power required to pin the values of all the fundamental constants down from first principles. Or maybe it is impossible even in such hopeful scenario, who knows? The lack of answers shows that the CC problem is still alive and kicking. 

In this thesis, I will not face the CC problem as such. The main aim is much more modest. Taking into account the current amount and quality of the cosmological data, I wish to put to the test the possibility that the $\Lambda$-term and its associated vacuum energy density, $\rho_\Lambda$, could actually be dynamical (``running'') quantities whose rhythms of variation might be linked to the Universe's expansion rate, $H$. The idea is to check if this possibility helps to improve the description of the overall cosmological data as compared to the rigid assumption $\rho_\Lambda={\rm const.}$ inherent to the concordance $\Lambda$CDM model.                                                                                                                                       
In the subsequent parts of this manuscript I will present the various phenomenological analysis carried out during my PhD research period. I have ordered them respecting as much as possible the chronological order in which they were written. In this way, the reader will be able to see the evolution of the these studies, not only appreciating the improvement in the statistical methods employed in the fitting procedure, but also in the data sets used to constrain the models. This continuous depuration of the analysis has led us to detect a $\sim 3.5-4\sigma$ evidence in favor of the dynamical nature of the dark energy. Remarkably, the reported level of evidence is, to the best of my knowledge, significantly higher than in any previous work in the literature.

\section{Main bibliography of the chapter}

Section \ref{chap:historyCC}, which deals with the history of the CC and the possible link between it and the vacuum energy, is based on the contents of the various historical references cited along the text, the exhaustive review \cite{SolaReview2013}, and the book \cite{KraghOverduinBook}. The main bibliography for the rest of the sections is also quite extensive. See the corresponding references therein.

\thispagestyle{empty}
\null
\newpage

\part[Firsts phenomenological studies. Dynamical vacuum models and the $\mathcal{D}$-class of dark energy models]{{\huge Firsts phenomenological studies. Dynamical vacuum models \\ and the \\ $\mathcal{D}$-class of dark energy models}}

\thispagestyle{empty}
\null
\newpage

\epigraph{<<By doubting indeed we come to the inquiry; and by inquiring, we perceive the truth.>>}{\itshape - Pierre Abailard, in ``Sic et Non'' (1125)}
\thispagestyle{empty}
\null
\newpage

\pagestyle{fancy}
\fancyhf{}
\fancyhead[CO]{\nouppercase{\leftmark}}
\fancyhead[CE]{\nouppercase{\rightmark}}
\cfoot{\thepage}

\chapter[A, B and C-type dynamical vacuum models]{A, B and C-type dynamical vacuum models. A dedicated study}
\label{chap:Atype}

In this chapter we focus on a large class of dynamical models of the vacuum energy inspired in QFT in curved spacetime, in which the vacuum energy has an explicit dependence on $H$ and its time derivative, i.e.  $\rL(H,\dot{H})$. We do not only solve their background cosmology but also perform a detailed analysis of the corresponding cosmic perturbations and their implications in the structure formation processes. For example, it is well-known that the so-called linear growth rate (the logarithmic derivative of the linear growth factor $\delta_m=\delta\rmr/\rmr$ with respect to $\ln a$, i.e. $f=d\ln \delta_m/d\ln a$), can be in some cases a
good indicator of clustering, together with the growth rate index $\gamma$ (used as the effective parametrization of the growth rate through a power of the density parameter\,\cite{Peebles1993}). In general, $\gamma$ is a function of the scale factor (or the redshift), and the relation with the growth rate reads: $f(z)\simeq\Om(z)^{\gamma(z)}$. The asymptotic value of the growth rate index parameter takes distinctive values for different gravity models\,\cite{Athina2014}. In the case of the concordance $\CC$CDM model, $\gamma(0)\simeq 0.55-0.60$. In this chapter we examine these linear indicators of structure formation for the new
models of the vacuum energy, but we find also very convenient to study  nonlinear effects, such as the theoretically predicted cluster-size halo redshift distributions. These ``number count'' observables can help to break degeneracies between the vacuum models and can be especially useful in the context of realistic and future X-ray and Sunyaev-Zeldovich cluster surveys such as eROSITA\,\cite{eROSITA} and SPT\,\cite{SPT1,SPT2}, as shown in \cite{BPS09,Grande2011}. For other implications of dynamical models in the astrophysical domain, see e.g. \cite{BPS09b}.

We shall discuss the virtues and troubles associated to some of these vacuum models according to the Hubble terms involved in the dynamical structure of the vacuum energy. Let us emphasize that not all of the $\rho_\CC(H,\dot{H})$ models are equally favored from the theoretical point of view.  Interestingly, those that are
theoretically more favored are in fact the ones that best fit the structure formation data in combination with the other cosmic observables such as type Ia supernovae, the Cosmic Microwave Background and the Baryonic Acoustic Oscillations. At the same time we find models that perform comparatively
not so good, and others that can be simply excluded by the current observations.

The plan of this chapter is as follows. In Sect. \ref{sec:TimeEvolvingVacuum} we discuss the general dynamical vacuum models (DVM's) that depend on powers of the Hubble function and its time derivative, and single out the class of the running vacuum models (RVM's), more closely related to QFT. The corresponding cosmological background solutions of these models is presented in Sect. \ref{sec:solving}. In the next section, Sect. \ref{sec:perturbations}, we formulate the linear cosmic perturbations for general dynamical vacuum models. In Sect. \ref{sec:radiation} we discuss the inclusion of radiation. The fitting of these models to the cosmic data is put forward in Sect. \ref{sec:fitting}, where we also briefly address the implications that our dynamical vacuum models could have on possible evidence recently found on dynamical dark energy. In Sect. \ref{sec:Number counts} we discuss how to distinguish the DVM's through the cluster-size halo redshift distributions. Finally, in Sect. \ref{sec:conclusions} we
provide the discussion and our conclusions. In Appendix \ref{ch:appCollapse} we summarize the calculation of the linear density threshold for collapse, $\delta_c$, for the models under consideration, a quantity that is crucially needed for the determination of the cluster-size halo redshift distributions.

%%%%%%%%%%%%%%%%%%%%%%%%%%%%%%%%%%%%%%%%%%%%%%%%%%%%%%%%%%%%%%%%%
%%%%%%%%%%%%%%%%%%%%%%%%%%%%%%%%%%%%%%%%%%%%%%%%%%%%%%%%%%%%%%%%%
%%%%%%%%%%%%%%%%%%%%%%%%%%%%%%%%%%%%%%%%%%%%%%%%%%%%%%%%%%%%%%%%%

\section[Time-evolving vacuum energy]{Time-evolving vacuum energy in an expanding Universe}
\label{sec:TimeEvolvingVacuum}

Let us consider the expanding Universe as a perfect fluid with matter-radiation density $\rho_{M}$ and vacuum energy density $\rL$. The latter is usually associated to the value of the cosmological term through $\CC=8\pi\,G\,\rL$, where $G$ is Newton's constant. While we assume in this chapter that $G$ remains strictly constant, we do not make the same assumption for $\rL$. The full energy-momentum tensor of the cosmic fluid can be written as  $\tilde{T}_{\mu\nu}\equiv T_{\mu\nu}^M+T_{\mu\nu}^{\CC} $, where $T_{\mu\nu}^M$ is the ordinary matter+radiation energy-momentum tensor and $T_{\mu\nu}^{\CC}$ describes the vacuum part. The EoS for the various components have been written in Sect. \ref{subsec:BasicsLCDM}

In the flat FLRW metric, on which we will hereafter exclusively concentrate,  the two independent gravitational field equations derived from Einstein's equations sourced by $\tilde{T}_{\mu\nu}$ are given by \eqref{eq:FriedmannLCDM} and \eqref{eq:PressureLCDM}. From these equations (and setting $k=0$) one can derive the rate of change of the
Hubble function,
\begin{equation}\label{rchangeH}
\dot{H}=-4\pi\,G\left[\rmr+\frac{4}{3}\rho_r\right]\,.
\end{equation}
A useful equation (actually a first integral) that follows from the original system (\ref{eq:FriedmannLCDM}-\ref{eq:PressureLCDM}) is the following:
\begin{equation}\label{Bianchi1}
\dot{\rho}_{m}+\dot{\rho}_{r}+3 H \rho_{m}+4H\rho_r=-\dot{\rho}_{\Lambda}\,. 
\end{equation}
In the frequent situation where there is a dominant matter component (e.g. cold matter, $\omega_M=\omega_m$, or relativistic matter, $\omega_M=\omega_r$), it is possible to obtain the evolution law for the Hubble function in terms of the vacuum term and that matter component:
\be
\label{eq:EHmonocomp}
\dot{H}+\frac{3}{2}\,(1+\omega_M)\,H^2=4\pi\,G\left(1+\omega_M\right)\,\rL=\frac12\,\left(1+\omega_M\right)\,\CC\,.
\ee
Formally the above equations are valid whether the vacuum term $\rL$ is a rigid quantity or is represented by a dynamical variable $\rL=\rL(t)$. However, here we will assume that $\rL(H,\dot{H})$ is a true dynamical vacuum term whose time evolution is exclusively associated to the quantum effects on $\CC$, see Sect. \ref{subsec:RVMintro} and the review \cite{SolaReview2013} for details and references. From this point of view, the corresponding EoS is still $\wCC=-1$. This will be henceforth taken for granted. By integrating the last equation we can obtain $H$ in the relevant epoch of the cosmic evolution where that matter component dominates. This procedure will be frequently used in our analysis.

Equation (\ref{Bianchi1}) constitutes the local conservation of the full energy-momentum tensor $\tilde{T}_{\mu\nu}$ in the presence of all contributions of matter and vacuum energy, namely it expresses explicitly the covariant conservation law $\nabla^{\mu}\tilde{T}_{\mu\nu}=0$ in the FLRW metric. Such law
holds for strictly constant $G$ since the left hand side of Einstein's equations must have zero covariant derivative by virtue of the Bianchi identity. Being the result of an identity, equation (\ref{Bianchi1}) is not independent of the fundamental system of Friedmann-Lema\^\i tre equations (\ref{eq:FriedmannLCDM}-\ref{eq:PressureLCDM}), but it is useful to provide a more physical interpretation of them.

In the present study the dynamical cosmological term is represented by a power series of the Hubble function and its time derivative:
\begin{equation}\label{powerH}
\CC(H)=c_0+\sum_{k=1}\alpha_k H^k+\sum_{k=1}\beta_k\dot{H}^k+...
\end{equation}
where the leading term $c_0$ describes in good approximation the current Universe and the other terms introduce a mild dynamical evolution. The expression (\ref{powerH}) generalizes the formula that is motivated within the class of running vacuum models, see Sect. \ref{subsec:RVMintro}\footnote{In Sect. \ref{subsec:RVMintro} I have made the correspondence $\mu_{phy}\sim H$, but actually it can be extended to a more general framework by associating $\mu^2_{phy}$ not only with  $H^2$, but also with $\dot{H}$. This is what we consider now.}. Notice that in this case we are considering odd powers of $H$ in the above expansion. They have been included in order to explore the phenomenological implications of this (more general) scenario. 

Lately these models have been successfully applied to describe the complete history of the Universe, as they involve the ingredients capable of yielding a smooth transition from an early de Sitter stage to a proper radiation and matter epochs\,\cite{BasiLimaSola2013}. In practice, since structure formation is a relatively recent
phenomenon, we limit ourselves to consider the lowest powers of $H$. Indeed, recall that the current vacuum energy density is of order $M_P^2\,H_0^2$, where $H_0\sim 10^{-42}$ GeV is the current value of the Hubble parameter and  $M_P=1/\sqrt{G}\sim 10^{19}$ GeV is the Planck mass (in natural units). It follows that the power terms of (\ref{powerH}) beyond $H^2$ and $\dot{H}$ are completely irrelevant for the present Universe. Thus, terms of the form $H^3$, $\dot{H} H$, $H^4$, $\dot{H}^2$, $H^2 \dot{H}$, $\ddot{H}$, etc. will be ignored for the present study, although they can be important for the early Universe\,\cite{BasiLimaSola2013,Essay2014}.

The nonvanishing {\it r.h.s.} of Eq. \ref{Bianchi1} signals a transfer of energy between vacuum and matter. Needless to say, this transfer is absent in the $\CC$CDM model for which $\rL=$const. The nonvanishing time derivative of $\rL$ in the above conservation law involves the relevant powers of the Hubble function in the equation (\ref{powerH}).

In this study we would like to check the effect of all terms that can be phenomenologically significant for the recent Universe. Therefore, we will consider the linear term in $H$ as well as the $H^2$ and $\dot{H}$ terms. The linear terms, however, have a different status. They are not expected to have a fundamental origin
within in QFT in curved spacetime, a fact that actually applies to all the odd powers of the Hubble function as they are, in principle, incompatible with the general covariance of the effective action. However, the linear terms appear in various dark energy models in the presence of phenomenological bulk viscosity, see e.g.\,\cite{RenMeng06a,RenMeng06b,Komatsu2013}. The linear term in $H$ was also theoretically motivated as a possible fundamental description of the cosmological vacuum energy in terms of QCD, see e.g.\,\cite{Schutzhold2002,KlinkhamerVolovik2009,ThomasUrbanZhitnitsky2009,Ohta2011}. Phenomenological analysis claiming its possible interest for the description of the current Universe were carried out in e.g. \cite{Borges2008,Carn08,Alcaniz2012,Carn14}.

Therefore, following our aim to explore the various existing possibilities from the phenomenological point of view, we will test here the following list of six types or classes of DVM's effectively leading to time-evolving $\CC$ scenarios:
\begin{eqnarray}\label{A1A2B1B2C}
A1: \phantom{XXX} \Lambda&=&a_0+a_2 H^2\nonumber\\
A2: \phantom{XXX}\Lambda&=&a_0+a_1 \dot{H}+a_2 H^2\nonumber\\
B1: \phantom{XXX}\Lambda&=&b_0+b_1 H\nonumber\\
B2:\phantom{XXX} \Lambda&=&b_0+b_1 H+b_2 H^2 \\
C1:\phantom{XXX} \Lambda&=&c_1 H+c_2 H^2\nonumber\\
C2:\phantom{XXX} \Lambda&=&c_1 \dot{H}+c_2 H^2\nonumber
\end{eqnarray}
All of them can be, in principle, relevant for the study of the current and recent
past cosmic history. Type A and B models, despite their differences,
share one important feature, to wit: they all have a well-defined
$\CC$CDM limit since they all tend to a constant value of $\Lambda$
when the coefficients of the powers of $H$ or $\dot{H}$ tend to
zero. In contrast, models C1 and C2 can never behave as a rigid
$\CC$ term. As we will see, this has important consequences for the  phenomenological consistency of these models, when faced against the expansion and structure formation data.

If we have a look to the list of vacuum models under study, Eq.\,(\ref{A1A2B1B2C}), we observe that the vacuum energy density in model A2 can be written as follows, 
\be\label{GRVE}
\rho_\Lambda(H,\dot{H})=\frac{3}{8\pi G}\left(C_0+C_H H^2+C_{\dot{H}}\dot{H}\right)\,.
\ee
Notice that model A1 is a particular case of A2. On the other hand, models B1 and B2 involve a linear term in $H$ which, as previously indicated, is not expected on fundamental grounds but could appear as an effective contribution.

Let us first turn our attention to model A1 in the list. It is the simplest model containing the expected ingredients. It is convenient to normalize the additive term such that it coincides with the value of the current CC density for $H=H_0$, and in addition we introduce a (dimensionless) parameter $\nu$, which plays the role of
coefficient of the $\beta$-function for the running of the vacuum energy. In this way the CC density for model A1 can be cast as follows:
\begin{equation}\label{RGlaw2}
 \rL(H)=\rLo+ \frac{3\nu}{8\pi}\,M_P^2\,(H^{2}-H_0^2)\,.
\end{equation}
As desired, $\rL(H=H_0)=\rLo$. For $\nu=0$ the vacuum energy remains strictly constant at all times, $\rL=\rLo$, whereas for non-vanishing $\nu$  there is an obvious evolution of the vacuum energy that departs as $H^2$ from a strictly constant value. This is a mild evolution provided $\nu$ is small enough. Obviously this model is a particular case of (\ref{GRVE}) with $C_H=\nu$ and $C_{\dot{H}}=0$. Substituting (\ref{RGlaw2}) in the general acceleration law for a FLRW-like Universe in the presence of a vacuum energy density $\rL$, we find 
\begin{equation}\label{vacuuma}
\frac{\ddot{a}}{a}=-\frac{4\pi\,G}{3}\,(\rho_M+3p_M-2\rL)=-\frac{4\pi\,G}{3}\,\,(1+3\omega_M)\rho_M+C_0+\nu\,H^2\,,
\end{equation}
where in this case
\begin{equation}\label{C0}
C_0=\frac{8\pi G}{3}\,\rLo-\nu\,H_0^2\,.
\end{equation}
Let us next consider model A2. It generalizes the previous one by introducing the $\dot{H}$  contribution. Recall that the homogeneous terms $H^2$ and $\dot{H}$ are in general independent variables. From the two Friedmann's equations \eqref{eq:FriedmannLCDM} and \eqref{eq:PressureLCDM} it is easy to show that 
\begin{equation}\label{eq:ratioH2dotH}
\frac{H^2}{\dot{H}}=-\frac{2}{3}\,\frac{1+r}{1+\omega_M}\,,
\end{equation}
where $r=\rL/\rho_M$ is the ratio between the vacuum and the dominant matter energy densities. Even for the $\CC$CDM model (where $\rL$ is strictly constant) $r$ is a dynamical variable. At present $r\sim {\cal O}(1)$ ($r\sim 7/3$), whereas in the past $r\to 0$. In the radiation epoch $H^2$ was just minus half the value of $\dot{H}$,
whereas at present  $H_0^2$ is roughly minus twice the value of $\dot{H}_0$.

The acceleration equation for the scale factor of the model class A2 reads,
\begin{equation}\label{vacuumgeneral}
\frac{\ddot{a}}{a}=-\frac{4\pi\,G}{3}\,(1+3\omega_M)\rho_M+C_0+\nu
H^2+\CHd\,\Hd\,.
\end{equation}
We still denote $\CH\equiv\nu$ since this parameter is closely related to the simplest running vacuum model (\ref{RGlaw2})-(\ref{vacuuma}). Equation (\ref{vacuumgeneral}) can be rewritten in terms of the deceleration parameter $q$ and the usual cosmological parameters $\Omega_M=\rho_M/\rc$ and $\OL=\rL/\rc$, where $\rc$ stands for the critical density $\rc=3H^2/8\pi G$. We find:
\begin{equation}\label{eq:decelparam}
q=-\frac{\ddot{a}}{aH^2}=-1-\frac{\dot{H}}{H^2}=\frac12\,(1+3\omega_M)\Omega_M-\Omega_{\CC}\,.
\end{equation}
In the current epoch (where radiation can be safely neglected) we obtain the following relation, $\CHd\,\dot{H}_0=-\CHd\,(q_0+1)\,H_0^2=-(3/2)\Omo\,\CHd\,H_0^2$, which
is now helpful to determine $C_0$ in (\ref{vacuumgeneral}) after we impose the boundary condition $\rL(H_0)=\rLo$ in (\ref{GRVE}):
\begin{equation}\label{eq:C0CHd}
C_0=H_0^{2}\left(\OLo-\nu+\frac32\,\,\Omo\CHd\right)\,.
\end{equation}
This relation clearly generalizes Eq.\,(\ref{C0}) for nonvanishing $\CHd$.

While models A (whether version A1 or the extended A2) and C2 are directly related to the RG approach, models B and C1 are more phenomenological by the reasons explained before. We will solve the cosmological equations for all these models in the next sections.

Before solving these models, let us stress that in the case of the running vacuum models deriving from the generalization of the RG equation \eqref{eq:betaLambda}, which includes also the terms with $\dot{H}$, the solution compatible with the general covariance of the effective action is of the form 
\begin{equation}\label{eq: GeneralRG}
\rL(t)=c_0+\sum_{k=1} \alpha_{k} H^{2k}(t)+\sum_{k=1}
\beta_{k}\dot{H}^{k}(t)\,,
\end{equation}
This is the particular form that Eq.\,(\ref{powerH}) takes for the running vacuum case\,\cite{SolaReview2013}. That is to say, one obtains in general
an ``affine'' function constructed out of  powers of $H^2=\left(\dot{a}/a\right)^2$ and $\dot{H}=\ddot{a}/a-H^2$, hence with an even number of time derivatives of the scale factor $a$. The higher order powers once more are irrelevant for the current Universe and for this reason the model types A, B, and C singled out above have been cut off at the lowest significant powers. As previously emphasized, the higher powers of $H$ can play a very significant role in the early Universe. This role has been studied in detail in Refs. \,\cite{BasiLimaSola2013,Essay2014}  where it is shown in particular that they can lead to an inflationary scenario with graceful exit of the vacuum phase (de Sitter regime) into the radiation phase. It means that in this kind of dynamical vacuum fraweworks one can formulate a unified model of the cosmological evolution covering both the early, the recent and the present Universe. From now on we focus on the last two stretches of the cosmic history.

%%%%%%%%%%%%%%%%%%%%%%%%%%%%%%%%%%%%%%%%%%%%%%%%%%%%%%%%%%%%
%%%%%%%%%%%%%%%%%%%%%%%%%%%%%%%%%%%%%%%%%%%%%%%%%%%%%%%%%%%%
%%%%%%%%%%%%%%%%%%%%%%%%%%%%%%%%%%%%%%%%%%%%%%%%%%%%%%%%%%%%
%%%%%%%%%%%%%%%%%%%%%%%%%%%%%%%%%%%%%%%%%%%%%%%%%%%%%%%%%%%%

\section[Background solutions]{Background solution of the cosmological equations}
\label{sec:solving}

Whenever possible we will solve the background equations for the
models (\ref{A1A2B1B2C}) by providing the matter and vacuum energy
densities, as well as the Hubble function, in terms both of the
scale factor $a$  and the cosmic time. In general the most useful
form is in terms of the scale factor since this is the variable
which can be more easily related with the cosmological redshift \eqref{eq:reshift}. However this will not always be possible and in some
cases the analytic solution can be given only in terms of the cosmic
time. In this section we provide the analytical solution of the
background cosmologies corresponding to these models, obtained by extending the
analysis of e.g. Refs.\,\cite{BPS09,BasSola14a}. The perturbation equations will be
analyzed in subsequent sections.

%%%%%%%%%%%%%%%%%%%%%%%
%%%%%%%%%%%%%%%%%%%%%%%

\subsection{Models A1 and A2} 
\label{subsec:solvingA1A2}

Let us consider the local covariant conservation law (\ref{Bianchi1})
in the matter dominated epoch and let us insert Eq.\,(\ref{GRVE}) on
its {\it r.h.s.} A straightforward calculation making use of
(\ref{rchangeH}) and its time derivative yields: 
\be 
{\dot \rho}_m
+3 H \frac{ 1 - C_H}{ 1 - \frac{3}{2} C_{\dot H} }~\rho_m=0\,.
\label{eq:generalizedConserv} 
\ee 
Trading the cosmic time variable
for the scale factor through $d/dt=aH d/da$ we can solve for the
energy densities as a function of the scale factor as follows:
\begin{equation}\label{eq:MatterdensityCCtCDM}
\rho_m(a) =  \rmo~a^{-3 \xi}
\end{equation}
and
\begin{equation}\label{eq:CCdensityCCtCDM}
\rL(a)=\rLo+{\rmo}\,\,(\xi^{-1} - 1) \left(a^{-3\xi} -1  \right)\,,
\end{equation}
with
\begin{equation}\label{eq:defxiM}
\xi= \frac{ 1 - \nu }{ 1 - \alpha }\,,
%\simeq 1+\alpha-\nu\,,
\end{equation}
where $\nu=C_H$ as before, and we have introduced  $\alpha=3\CHd/2$.
Obviously, for $\alpha=0$ model A2 becomes model A1 (for which
$\xi=1-\nu$).

The corresponding Hubble function can now be obtained from the above
energy densities using the Friedmann equation, resulting in the following
expression:
\begin{equation}\label{HubbleA}
H^2(a) =H_0^2\,\left[1+\frac{\Omo}{\xi}\,\left(~a^{-3
\xi}-1\right)\right] \,.
\end{equation}
It satisfies the correct normalization $H^2(a=1)=H_0^2$.

The solution can also be given in terms of the cosmic time by
solving the differential equation\,(\ref{eq:EHmonocomp}) in terms of
hyperbolic functions. The final result in the matter dominated epoch
can be brought to the form
\begin{equation}
\label{eq:Ht} H(t)=H_0\,\sqrt{\frac{\xi-\Omo}{\xi}} \;
\coth\left[\frac{3}{2}\,H_{0}\sqrt{\xi\,(\xi-\Omo)}\;t\right]\,.
\end{equation}
One more integration renders the scale factor in terms of the cosmic
time:
\begin{equation}\label{eq:at}
a(t)=\left(\frac{\Omo}{\xi-\Omo}\right)^{\frac{1}{3\,\xi}}\,
\sinh^{\frac{2}{3\,\xi}}
\left[\frac{3}{2}\,H_{0}\sqrt{\xi\,(\xi-\Omo)}\;t\right]\,.
\end{equation}
We can readily check that if we eliminate the cosmic time between
these two equations we recover Eq.\,(\ref{HubbleA}). Furthermore, if
$\xi=1$ (implying $\nu=\alpha=0$) the above formulae adopt the
$\CC$CDM form, as it should be.

Let us also compute the corresponding inflection point where
there is a transition of the Hubble expansion from the decelerating
to the accelerating regime. From the definition
(\ref{eq:decelparam}) of deceleration parameter it is easy to show
that it can be rewritten as follows:
\begin{equation}\label{eq:defqa}
q=-1-\frac{a}{2 H^2(a)}\,\frac{dH^2(a)}{da}\,.
\end{equation}
From this expression we can comfortably compute the point where
$q=0$ from Eq.\,(\ref{HubbleA}). Let us deliver the final result for
model A2 in terms of the redshift value at the transition point:
\begin{equation}\label{eq:zIA}
z_{tr}=\left[\frac{2(\xi-\Omo)}{(3\xi-2) \Omo}\right]^{1/3\xi}-1\,.
\end{equation}
The result for model A1 is just obtained by setting $\alpha=0$
(hence $\xi=1-\nu$). The standard $\CC$CDM result $z_{tr}^{\CC{\rm
CDM}}$ is, as always, recovered for $\xi=1$:
\begin{equation}\label{eq:zILCDM}
z_{tr}^{\CC{\rm CDM}}=\left[\frac{2\OLo}{\Omo}\right]^{1/3}-1\,.
\end{equation}
The numerical value is $z_{tr}^{\CC{\rm CDM}}\simeq 0.726 (0.645)$ for
$\OMo= 0.28 (0.31)$. As we can see the result is quite sensitive to the precision of $\Omo$ from
observations\,\footnote{$\Omo h^2 = 0.1426 \pm 0.0025$, with
$h=0.673\pm 0.012$
for the standard $\Lambda$CDM model from
Planck+WP\,\cite{Planck2013}.}.  Computing the departure of the
new transition point (\ref{eq:zIA}) from the standard result for
small $\nu$ and $\alpha$, we obtain:
\begin{eqnarray}\label{aIRG2}
z_{tr}-z_{tr}^{\CC{\rm
CDM}}=(\nu-\alpha)\left[\frac{2\OLo}{\Omo}\right]^{1/3}\,
\left[1+\frac13\left(\ln\frac{2\OLo}{\Omo}-\frac{1}{\OLo}\right)\right]\,.
\end{eqnarray}
The numerical difference will be small to the extend that $\nu$ and
$\alpha$ are small. The corresponding fit to these parameters will
be made in Sect. \ref{sec:fitting}.

%%%%%%%%%%%%%%%%%%%%%%%%%%%%%%%%%%%%
%%%%%%%%%%%%%%%%%%%%%%%%%%%%%%%%%%%%
%%%%%%%%%%%%%%%%%%%%%%%%%%%%%%%%%%%%

\subsection{Models B1 and B2} 
\label{subsec:solvingB1B2}

The vacuum energy density for model B2 can be parameterized
as follows:
\begin{equation}\label{B1B2}
\rho_{\Lambda}(H)=\frac{3}{8\pi G}(C_0+C_{1}H+C_{2}H^2)
\end{equation}
The solution of model B1 obviously
ensues by simply setting $C_2=0$ in the background solution of model B2. However, the technical
difficulty in solving these models resides already in the simplest
``affine'' model B1 since the linear term in $H$ is harder to handle
than the quadratic one. On the other hand, phenomenologically the
reason to single out the case B1 is because this model is able to
reasonable fit the data provided we maintain the additive term
$C_0\neq 0$. For $C_0=0$, in contrast, the pure lineal model $\rL\propto
H$ is unable to accommodate the data on structure
formation (cf. Refs. \cite{BPS09,BasSola14a}, and Sect. \ref{subsec:growthrate}). This feature has an important
impact on some theoretical and phenomenological proposals in the
literature\,\cite{Schutzhold2002,Ohta2011,Carn08,Alcaniz2012,Carn14}.

In order to solve model (\ref{B1B2}) analytically we proceed as
follows. First of all it is convenient to re-express $\rL(H)$ in
terms of dimensionless coefficients. We set $C_1\equiv\epsilon H_0$
(where $\epsilon$ is dimensionless) and we continue identifying
$C_2$ with $\nu$, that is $C_2\equiv\nu$ as the basic parameter of
the simplest viable model (\ref{RGlaw2}) compatible with the RG
formulation. With this notation the expression that relates $C_0$
with $\nu$ and $\epsilon$ is:
\begin{equation}\label{eq:C0TypeB}
C_0=\frac{8\pi
G}{3}\rLo-H_0^2(\epsilon+\nu)=H_0^2(\OLo-\epsilon-\nu)
\end{equation}
If we substitute (\ref{B1B2}) in the basic differential equation
(\ref{eq:EHmonocomp}) for $H(t)$ in the matter dominated epoch, we
find
\begin{equation}\label{eq:diffEq.TypeB}
\frac{2}{3}\dot{H}+\zeta\,H^2-\epsilon H_0 H=C_0
\end{equation}
where we have defined $\zeta=1-\nu$, not to be confused with
Eq.\,(\ref{eq:defxiM}).

Upon direct integration we determine the Hubble function for this
model explicitly in terms of the cosmic time:
\begin{equation} \label{eq:Hubblef}
H(t)=\frac{H_0}{2\,\zeta}\left[\mathcal{F}\,\coth\left(\frac{3}{4}H_0\mathcal{F}\,t\right)+\epsilon\right]\,,
\end{equation}
with 
\begin{equation}\label{defmathF}
\mathcal{F}(\OLo,\epsilon,\nu)\equiv\sqrt{\epsilon^2+4\,\zeta(\OLo-\epsilon-\nu)}\,.
\end{equation}
It is easy to check that for $\nu=\epsilon=0$
Eq.\,(\ref{eq:Hubblef}) boils down to the same result  as
(\ref{eq:Ht}) for $\nu=\alpha=0$ (and both equal to the $\CC$CDM
solution in this limit, as they should).

Notice the presence of an additive constant term  in
(\ref{eq:Hubblef}) (proportional to $\epsilon$) apart from the
hyperbolic one. This feature is precisely what makes the treatment
of the type-B models for $\rho_\Lambda(H)$ (the models with the
linear term in $H$) more complicated from the technical point of
view.

The time evolution of the  pressureless matter density can be
obtained from (\ref{rchangeH}) and the explicit form of the Hubble
function (\ref{eq:Hubblef}), with the result:
\begin{equation}\label{eq:rhom}
\rho_m(t)=-\frac{\dot{H}(t)}{4\pi G}=\frac{3H_0^2}{32\pi
G\,\zeta}\mathcal{F}^2\,{\rm
csch}^2\left(\frac{3}{4}H_0\mathcal{F}t\right)
\end{equation}
Similarly, from the previous equation and with the help of
(\ref{eq:Hubblef}) and \eqref{eq:FriedmannLCDM}, we infer the expression of
the vacuum energy density:
\begin{equation}\label{eq:rhoL}
\rho_\Lambda(t)=\frac{3H_0^2}{32\pi
G\zeta^2}\left[\mathcal{F}^2+\epsilon^2+2\epsilon\mathcal{F}\,\coth\left(\frac{3}{4}H_0\mathcal{F}t\right)+\nu
\mathcal{F}^2\,{\rm
csch}^2\left(\frac{3}{4}H_0\mathcal{F}t\right)\right]\,.
\end{equation}
Let us note that in the far past ($t\to 0$) we have
$\rL(t)/\rmr(t)\simeq \nu/(1-\nu)$ and therefore since $|\nu|\ll 1$
the vacuum energy is suppressed in this period, as expected. The
same conclusion applies in the radiation period, if we would include
the corresponding radiation term. As for the behavior in the remote future
($t\to\infty$) we have $\rmr(t)\to 0$ from (\ref{eq:rhom}). The
asymptotic form of the vacuum energy ensues from (\ref{eq:rhoL}):
\begin{equation}\label{eq:rhoLfuture}
\rho_\Lambda(t\to\infty)\to\frac{3H_0^2}{32\pi
G}\left(\frac{\mathcal{F}+\epsilon}{\zeta}\right)^2\,.
\end{equation}
For this model it is impossible to obtain analytically the
matter and vacuum energy densities in terms of the scale factor,
except for $C_0=0$ (which is, as previously warned,
phenomenologically problematic).  To see this let us obtain $a(t)$
by direct integration of the Hubble function (\ref{eq:Hubblef}).
After a straightforward calculation we find:
\begin{equation}\label{atBtype}
a(t)=\left(\frac{[(2\zeta-\epsilon)^2-\mathcal{F}^2]^{1+\frac{\epsilon}{\mathcal{F}}}}{\mathcal{F}^2(\mathcal{F}+2\zeta-\epsilon)^{\frac{2\epsilon}
{\mathcal{F}}}}\right)^{\frac{1}{3\zeta}}e^{\frac{\epsilon
H_0}{2\,\zeta}t}\,\sinh^{\frac{2}{3\zeta}}\left(\frac{3}{4}H_0\mathcal{F}t\right)\,,
\end{equation}
where the complicated normalization factor is fixed by using
$H(t_0)=H_0$ and $a(t_0)=1$. We can check that for $\epsilon=\nu=0$
it reduces to the standard $\CC$CDM one.

In view of the result (\ref{atBtype}) it is impossible to eliminate
the cosmic time variable in terms of the scale factor since the
isolated exponential term in the above expression cannot be combined
with the hyperbolic function.  Only in the particular case $C_0=0$
this combination is possible, as it will be shown in the next
section. Let us note that the troublesome exponential factor would
be there even if $\nu=0$, i.e. even if no $H^2$ term would be
present in the expression of the vacuum energy density (\ref{B1B2}).

The difficulty, therefore, stems entirely from the additive
$\epsilon$-term in the Hubble rate (\ref{eq:Hubblef}), i.e. from the
linear $H$ dependence in (\ref{B1B2}).  We cannot circumvent this
situation by trading the cosmic time by the scale factor in the
original Eq.\,(\ref{eq:diffEq.TypeB}) either, as the corresponding
solution provides $H(a)$ only as an implicit function. These
complications, which did not occur for the type-A models, completely
block off the possibility to analytically express the Hubble
function and the energy densities of the type-B ones in terms of the
scale factor. In practice this means that we have to find the
functions $H(a)$, $\rmr(a)$ and $\rL(a)$ numerically in this case.

We remark that the two basic parameters $(\epsilon,\nu)$,
or equivalently $(\epsilon,\zeta$), of type-B models are completely independent and cannot be mimicked by a single effective parameter in a given matter-dominated or radiation-dominated epoch. This is different from type-A
models, which can effectively be described by the unique parameter
$\xi$, defined in (\ref{eq:defxiM}), in the matter-dominated epoch.  However, as indicated above, the main analytical difficulty of type-B would remain even if the model would have the single parameter $\epsilon\neq 0$, with
$\nu=0$.

Finally, we note that type-B models have also an inflection point in
the recent past where deceleration changes into acceleration. Using
(\ref{eq:decelparam}) and (\ref{eq:Hubblef}) we can compute the
deceleration parameter:
\begin{equation}\label{eq:decparamTypeB}
q=-1+\frac32\,\frac{\zeta\,\mathcal{F}^2}{\left[\mathcal{F}\cosh\left(\frac{3}{4}H_0\mathcal{F}\,t\right)
+\epsilon\,\sinh\left(\frac{3}{4}H_0\mathcal{F}\,t\right)\right]^2}\,.
\end{equation}
In the pure matter epoch (sufficiently small $t$) it simplifies to $q_m\simeq-1+3\zeta/2\simeq
+0.5$ (since $\zeta=1-\nu\simeq 1$), whereas its approximate value
at present can be obtained by expanding
Eq.\,(\ref{eq:decparamTypeB}) linearly on the small $\epsilon$ and
$\nu$ parameters. By explicit calculation one can check that the
final result is the same as in the standard model, i.e.
$q_0=\Omo/2-\OLo\simeq -0.55$ (for $\Omo\simeq 0.3$). Because of the
difficulties mentioned above the exact determination of the
transition point in the recent past is harder in this case, but the existence of such point is as clear as for type-A ones. In both model types the transition point is
guaranteed thanks to the fact that these models have a smooth
$\CC$CDM limit for sufficiently small $\epsilon$ and $\nu$. In
particular, for type-B2 with $\epsilon\ll\nu$ the transition point
is essentially given by Eq.(\ref{eq:zIA}) with $\xi=1-\nu$. As we
will see in the next section, the situation with type-C models
concerning the transition point deserves some attention.

%%%%%%%%%%%%%%%%%%%%%%%%%%%%%%%%%%%%
%%%%%%%%%%%%%%%%%%%%%%%%%%%%%%%%%%%%
%%%%%%%%%%%%%%%%%%%%%%%%%%%%%%%%%%%%

\subsection{Models C1 and C2} 
\label{subsec:solvingC1C2}

Type C1 and C2 models are actually quite different from the previous
ones and at the same time different from each other. They have in
common the fact that do not have a well defined $\CC$CDM limit for
any value of the parameters, and therefore can never behave
sufficiently close to a pure $\CC$CDM model. This raises some doubts
about their possible viability, but they have nevertheless been
discussed in the literature for different theoretical and
phenomenological reasons.

For example, recently they have been discussed from the point of
view of their possible relation with the ``entropic-force''
scenario\,\cite{Verlinde10} and its implications for the dark
energy\,\cite{Easson10,Easson10b}\footnote{Some serious inconsistencies in Verlinde's proposal of emergent gravity have been provided in the very recent Ref. \cite{DaiStojkovic2017}. In this thesis, though, we want to explore the viability of these models from the phenomenological perspective. We will see that, in fact, they suffer from serious problems when we confront them with the cosmological data at our disposal.} . Sometimes type-C models are
presented in some drastically simplified forms as e.g. when only one
of the two terms is present (say, models of the form $\CC\propto H$
or $\CC\propto H^2$), see e.g.
\cite{Schutzhold2002,Ohta2011,Carn08,Alcaniz2012,Carn14} and
\cite{ArcuriWaga94}. Here we will summarize the situation for
completeness and also to highlight the important differences of type C with respect to type A and type B models.

Let us briefly comment on the model subclass C2 first. This is the canonical version of the mentioned class of models that acquired some
relevance recently in connection to the entropic-force scenario and
its possible cosmological implications\,\cite{Easson10,Easson10b}.
Many authors have analyzed recently this scenario for dark energy and
generalizations thereof, cf. e.g.
\,\cite{JapaneseHolog1a,JapaneseHolog1b,JapaneseHolog2,GrandaOliveros2008,KoivistoVisserA,KoivistoVisserB} and
\cite{BasPolarSola12,BasSola14a}.

The corresponding background cosmological solution for C2 can be
obtained as a particular case of our analysis of the A2 models in
the limit $C_0\to 0$ of Eq.\,(\ref{GRVE}), where $C_0$ is explicitly
given by Eq.\,(\ref{eq:C0CHd}). In this model $\OLo$ becomes
determined in terms of the other parameters:
\begin{equation}\label{eq:C0Omegam}
\OLo=\frac{\nu-\alpha}{1-\alpha}\,.
\end{equation}
Since $\Omo+\OLo=1$ it follows that $\Omo=\xi$, with $\xi$ as in
Eq.\,(\ref{eq:defxiM}). As a result it is impossible that the
parameters $|\nu|$ and $|\alpha|$ can be simultaneously small here.
The corresponding Hubble rate is easily found,
\begin{equation}\label{Hentropic}
H(a)=H_0\,a^{-3\xi/2}\,,
\end{equation}
together with the energy densities:
\begin{equation}\label{rhomrhoLentropic}
\rmr(a)=\rco\,\Omo\,a^{-3\xi}\,, \ \ \rL(a)=\rco\,\OLo\,a^{-3\xi}\,,
\end{equation}
where $\rco$ is the current critical density.  Worth noticing is the
deceleration parameter of this model. From (\ref{eq:defqa}) and
(\ref{Hentropic}) we find $q=-1+3\xi/2$, which remains constant,
i.e., it does not change with the expansion. Thus this kind of model
cannot have an inflection point from deceleration to acceleration:
if $\xi<2/3$ the Universe will always be accelerating since the
start of the matter-dominated epoch; if, instead, $\xi>2/3$ it will
always decelerate; and, finally, if $\xi=2/3$ the Universe expands
always at constant rate. Obviously none of these possibilities is
phenomenologically acceptable and hence model C2 must be
rejected\,\cite{BasPolarSola12,BasSola14a}.

In particular, models of the form $\rL\propto H^2$ or
$\rL\propto\dot{H}$ are definitely ruled out. Recently it has been
shown that even if the constant acceleration regime of model C2
would just be a partial description of a more complete model of
expansion where a transition point would exist, the corresponding
linear growth of cosmic perturbations is also strongly disfavored by
the current data\,\cite{BasSola14a}. This result inflicts a final
blow to this specific kind of entropic-force cosmologies. From now on we do not consider model C2 as a viable
possibility and hence we will not further analyze it.

Let us now move to model C1. It is a particular case of B2 in the
limit $C_0\to 0$ of Eq.\,(\ref{B1B2}), where in this case $C_0$ is
defined in Eq.\,(\ref{eq:C0TypeB}). Thus for this model we have
$\OLo=\epsilon+\nu$, which implies that $\epsilon$ and $\nu$ cannot
be both arbitrarily small. Upon imposing this constraint the
quantity $\mathcal{F}$ defined in (\ref{defmathF}) boils down to
$\mathcal{F}\to\epsilon$, and this allows to combine the exponential
factor and the hyperbolic function in the expression
(\ref{atBtype}), with the result: \be \label{eq:atC1}
a(t)=\left(\frac{\Omo}{\zeta-\Omo}\right)^{2/3\zeta}\left[
e^{3\,(\zeta-\Omo) H_0\,t/2}-1 \right]^{2/3\zeta}\,, \ee where we
recall that for type-B models $\zeta=1-\nu$. Obviously we must have
$\zeta>\Omo$ (and hence $\nu<\OLo$) for a realistic description of
the expansion. The cosmic time can now be given explicitly in terms
of the scale factor:
\begin{equation}\label{eq:taC1}
t=\frac{2}{3(\zeta-\Omo)\,H_0}\,\ln\left[1+\frac{\zeta-\Omo}{\Omo}\,a^{3\zeta/2}\right]\,,
\end{equation}
and from (\ref{eq:Hubblef}) the Hubble function can also be
expressed in that variable: \be\label{E2aC1} E(a)=1+
\frac{\Omo}{\zeta} \left(a^{-3\zeta/2}-1\right)\,, \ee where we have
defined the normalized Hubble rate $E(a)\equiv {H(a)}/{H_{0}}$. For
completeness we provide also the matter and vacuum energy densities
in terms of that rate:
\begin{equation}\label{eq:rhomaC1}
\rmr(a)=\rco\left[\zeta\,E^2(a)-(\zeta-\Omo) E(a)\right]
\end{equation}
and
\begin{equation}\label{eq:rhoLaC1}
\rL(a)=\rco\left[(1-\zeta)\,E^2(a)+(\zeta-\Omo) E(a)\right]\,.
\end{equation}
In the remote future ($a\to\infty$) we have $\rmr(a)\to 0$ and
$\rL(a)\to (1-\Omo/\zeta)^2\rco$. This is consistent with
Eq.\,(\ref{eq:rhoLfuture}) since $\mathcal{F}=\epsilon=\zeta-\Omo$
for model C1. In the past
$\rL/\rmr\propto(1-\zeta)/\zeta=\nu/(1-\nu)$, and as a result we
must require $|\nu|$ to be sufficiently small to avoid domination of
vacuum energy. Now, since $\nu$ and $\epsilon$ cannot be
simultaneously small for C1, we must have $|\nu|\ll\epsilon$ for
this specific model class.

We note that, in contraposition to model C2, model C1 has a
well-defined inflection point in the expansion regime, despite
$C_0=0$. It is found to be
\begin{eqnarray}
      \label{inflectionC1}
z_{tr}^{\rm
(C1)}=\left[\frac{2(\zeta-\Omo)}{(3\zeta-2)\Omo}\right]^{2/3\zeta}-1\,.
\end{eqnarray}
Two particular cases emerge. For $\epsilon=0$ we necessarily have $\nu=\OLo$ and the model becomes $\rL=\rLo\,H^2/H_0^2$, which was already ruled out since it has no inflection point. Another particular case is the pure linear model $\rL\propto H$, obtained
by setting $\nu=0$ (equivalently $\zeta=1$), which enforces
$\epsilon=\OLo$. Its vacuum energy density is given by
\begin{equation}\label{eq:rLproptoH}
\rL(H)=\frac{3}{8\pi G}\,\epsilon\,H_0\,H=\frac{3\OLo}{8\pi G}\,H_0\,H=\rLo\,\frac{H}{H_0}\,.
\end{equation}
At variance with the purely quadratic model $\rL\propto H^2$, the pure linear model does have a transition redshift. It is completely determined by the usual cosmological parameters:
\begin{eqnarray}
      \label{inflectionC1v2}
z_{tr}=\left[\frac{2\OLo}{\Omo}\right]^{2/3}-1\,.
\end{eqnarray}
We can evaluate it numerically assuming that the model
matches the correct values of these parameters. We find e.g.
$z_{tr}\simeq 1.979 (1.706)$ for $\OMo= 0.28 (0.31)$. These predictions
are substantially different from the $\CC$CDM ones, see
Eq.\,(\ref{eq:zILCDM}), and they are shared by essentially all C1
models owing to the aforementioned  $|\nu|\ll\epsilon$ restriction.
However this is not the only problem with the class C1. From
(\ref{eq:rhomaC1}) and (\ref{E2aC1}) we see that, in the past, the
behavior of the matter density for this model is abnormal:
$\rmr(a)\sim\rco \zeta\,E^2(a)\sim
\rco\,\left(\Omo\right)^2\,\zeta^{-1}\,a^{-3\zeta}$. Even for
$\nu=0$ (equivalently, $\zeta=1$, corresponding to the pure linear
model $\rL\propto H$) there is an extra factor of $\Omo$ as compared
to the $\CC$CDM. Such anomaly causes that when this model is
confronted with the high-redshift cosmological data the preferred value of $\Omo$
for this model is significantly larger than in the $\CC$CDM, see
\cite{BasSola14a} and \cite{BPS09}. By the same token one is forced to take anomalously large values of $\Omo$ in order to retrieve a transition redshift (\ref{inflectionC1v2}) that can be reasonably close to the $\CC$CDM one. Typically one has to take $\Omo\simeq 0.47-0.49$, for which $z_{tr}\simeq 0.63-0.72$. Not surprisingly model C1 (and
in particular the $\rL\propto H$ version) will not be for various
reasons among our favorite dynamical vacuum
models.

Needless to say, for a full assessment of the possibilities of the
various models we have to consider the analysis of cosmic
perturbations. We do this in the next section.

%%%%%%%%%%%%%%%%%%%%%%%
%%%%%%%%%%%%%%%%%%%%%%%

\section[Linear perturbations for the DVM's]{Linear perturbations for dynamical vacuum  models}
\label{sec:perturbations} 

After discussing the most relevant
aspects of the background cosmological solutions, the next essential
step in our study is to analyze the linear perturbation equations.
The structure formation properties obviously play an essential role
to discriminate between the three kinds A, B and C of dynamical
vacuum models that we are considering. In this section we discuss
the perturbations in the presence of a variable vacuum energy. While
we are not going to introduce perturbations for the vacuum energy
itself, only for matter, we incorporate the dynamical character of
$\rL$ in the matter perturbation equations. In other words,
$\rL=\rL(t)$ is time evolving, but homogeneous in first
approximation (and at small scales). This approach will suffice to clarify the fingerprint
differences between the A, B and C model types as far as structure
formation is concerned. The equation governing the linear matter density perturbations for the dynamical vacuum models with fixed $G$ and with a non-null interaction between the vacuum and matter sectors is explicitly derived in \ref{sec:LPmatter-vacuum}. More concretely, it is given by \eqref{eq:DensityContrastEq} in terms of the conformal time. We can rewrite the equation for the density contrast $\delta_m=\delta\rho_m/\rho_m$ in terms of the scale factor as independent variable as follows:
\begin{equation}\label{diffeqDa}
\delta_m^{\pp}(a)+\left[\frac{3}{a}+\frac{H'(a)}{H(a)}+\frac{\Psi(a)}{aH(a)}\right]\,{\delta^\p_m}(a)-\left[\frac{4\pi
G\rmr(a)}{H^2(a)}-\frac{2\Psi(a)}{H(a)}-a\frac{\Psi'(a)}{H(a)}\right]\,\frac{\delta_m(a)}{a^2}=0\,,
\end{equation}
where the primes here indicate $d/da$ differentiation, and we have traded the cosmic time variable for the scale factor through $d/dt=aH(a) d/da$. We have also defined 
\be\label{defQ}
\Psi\equiv-\frac{\dot{\rho}_\Lambda}{\rho_m}=3H+\frac{\dot{\rho}_m}{\rho_m}\,.
\ee
In particular, notice
that when the parameters $p_i=\nu,\alpha,\epsilon$ of the various models are set to $0$ we have $\Psi=0$ and (\ref{diffeqDa}) reduces to
the standard perturbation equation for the
$\CC$CDM\,\cite{KolbTurnerBook}:
\begin{equation}\label{diffeqDaLCDM}
\delta_m^{\pp}(a)+\left[\frac{3}{a}+\frac{H'(a)}{H(a)}\right]\,\delta_m^\p(a)-\frac32\,\Omo\,\frac{H_0^2}{H^2(a)}\,\frac{\delta_m(a)}{a^5}=0\,,
\end{equation}
where in the last term of the {\it l.h.s.} we have used the fact
that matter is covariantly conserved in the $\CC$CDM and hence
$\rmr=\rmo a^{-3}$. The decaying mode solution can be shown to be $\delta_m\propto H$, but this is not the one we want. The growing mode solution of
(\ref{diffeqDaLCDM}), which is the relevant one, is well-known and reads as follows:
\begin{equation}\label{eq:soldiffeqDaLCDM}
\delta_m(a)=\frac52\,\Omo\,E(a)\,\int_0^a\,\frac{da'}{\left(a'\,E(a')\right)^3}\,,
\end{equation}
where as before $E(a)\equiv H(a)/H_0$. Early on in the matter
dominated epoch, when $E(a)=\sqrt{\Omo}a^{-3/2}$,
Eq.(\ref{eq:soldiffeqDaLCDM}) yields the standard result for the
linear growth factor: $\delta_m(a)=a$. The effect of a
nonvanishing $\rL>0$ is to suppress this linear growing rate. To
better assess the physical meaning in a very particular situation,
suppose that $|\OL|\ll 1$ and that it remains approximately constant
during some period of the expansion.  One can show from
(\ref{diffeqDa}) -- with $\Psi=0$ -- that the power law solution is
then $\delta_m(a)\sim a^{1-6\OL/5}$ (cf. Ref. \cite{Opher07}), which clearly
displays the suppression behavior for $\OL>0$.  For a time variable
$\rL$ the evolution is, of course, more complicated and is described
by the full equation (\ref{diffeqDa}), in which $\Psi\neq 0$ is a
function (\ref{defQ}) evolving with expansion according to the given
dynamical vacuum model.

Let us remark that although the form (\ref{diffeqDa}) of the effective
perturbation equation can be useful to solve models A1, A2 and C1
(recall that model C2 was discarded in the previous section), it turns out that for models B1 and B2 we will have to use the equation in terms of the cosmic time, 
\be\label{diffeqD}
\ddot{\delta}_m+(2H+\Psi)\dot{\delta}_m-(4\pi G\rho_m-2H\Psi-\dot{\Psi})\delta_m=0\,, 
\ee
as in this case it is not possible to
write down the various cosmological functions in terms of the scale
factor, but only through the cosmic time (cf. previous section).
This makes the phenomenological discussion of type-B models a bit
more cumbersome since the natural variable to contact with
observations is the scale factor/redshift. We accomplish the
perturbation analysis in the next two sections.

%%%%%%%%%%%%%%%%%%%%%%%
%%%%%%%%%%%%%%%%%%%%%%%

\subsection{Perturbations for type-A models}
\label{sec:perturbationsTypeA}

In the following we solve the perturbation equation (\ref{diffeqDa})
for model A2 and then derive the solution of A1 as a particular
case. Recall that for these models the background solution can be
fully expressed in terms of the combined parameter $\xi$, which
depends on $\nu$ and $\alpha$ as indicated in (\ref{eq:defxiM}). It
is convenient to introduce the following change of independent
variable, which can be operated on the cosmic time or the scale
factor as follows:
\begin{equation}\label{changet}
x=\coth\left[\frac{3}{2}\,H_{0}\sqrt{\xi\,(\xi-\Omo)}\;t\right]
\end{equation}
and
\begin{equation}\label{changea}
x^2=\frac{\xi}{\xi-\Omo}E^2(a)=1+\frac{\Omo}{\xi-\Omo}a^{-3\xi}\,.
\end{equation}
These changes of variable are associated, respectively, to the
perturbation equations (\ref{diffeqD}) and (\ref{diffeqDa}).
Starting from any of these equations and applying the corresponding
change of variable (\ref{changet}) or (\ref{changea}) we arrive,
after some lengthy algebra, at the following final result:
\begin{eqnarray}\label{eq:eqdiff}
3\xi^2 (x^{2}-1)^{2}\,\frac{d^2\delta_m(x)}{dx^2}&+&2\,\xi\,(6\xi-5)\,
x(x^{2}-1)\,\frac{d\delta_m(x)}{dx}\\ &-& 2\,[(2-\xi)\,(3\xi-2)\,x^{2}-\xi
(4-3\xi)]\,\delta_m(x)=0\,. \nonumber\end{eqnarray}
Notice that for $\alpha=0$ (hence $\xi=\zeta=1-\nu$) the previous
equation reduces to the one for the type-A1 model, first analyzed in
Ref.\,\cite{BPS09}. Thus, we have extended the perturbations
analysis of that reference so as to include the more general class
of models A2, which had not been considered before. We find that the
basic parameter of models A1 and A2 are $\zeta$ and $\xi$
respectively, and this holds both at the background and perturbation
levels. Remarkably, the basic perturbation equation turns out to be
formally the same in each model after exchanging the parameters
$\zeta ({\rm A1})\leftrightarrow\xi ({\rm A2})$. Not only so, this
means that type A2 model effectively behaves as a single parameter
model $\xi$ for all purposes.

The solution of \eqref{eq:eqdiff} can be expressed as follows:
\begin{equation}\label{eq:solperturb}
\delta_m(x)=(x^2-1)^{\frac{5-3\xi}{6\xi}}Q_n^m(x),
\end{equation}
where $Q_n^m(x)$ is the associated Legendre's function of the second
kind, and
\begin{equation}
m=\frac{1}{3\xi}-1\qquad n=\frac{1}{3\xi}.
\end{equation}
Using standard properties of the Legendre
functions\,\cite{Gradshteyn} (see also Appendix B of
Ref.\,\cite{BPS09}) and restoring the scale factor variable through
Eq.\,(\ref{changea}), we can finally express the solution within the
natural parameter domain (\ref{naturalness}) in the following way:
\begin{equation}
\label{eq:solchangextoa} \delta_m(a)=A_1 a^{\frac{9\xi-4}{2}}E(a)\
F\left(\frac{1}{3\xi}+\frac{1}{2},\, \frac{3}{2};
\,\frac{1}{3\xi}+\frac{3}{2};\,-\frac{\xi-\Omo}{\Omo}\,\,a^{3\xi}
\right)\,,
\end{equation}
where $A_1$ is a constant to be adjusted by an initial condition,
and $F$ is the conventional hypergeometric series\,\,\cite{Gradshteyn}.

Although the above equations provide the exact perturbation solution
for arbitrary values of $\xi$, and hence of the original parameters
$\nu$ and $\alpha$, let us recall that from the theoretical point of
view the natural range for these parameters, for type-A models, is
\begin{equation}\label{naturalness}
|\nu|\ll 1\,,\ \ \ \ \ |\alpha|\ll 1\,.
\end{equation}
In this range the overall parameter $\xi$ can be expressed, in
linear approximation, as
\begin{equation}\label{eq:smalllimit}
\xi=\frac{1-\nu}{1-\alpha}\simeq 1-(\nu-\alpha)\,.
\end{equation}
We conclude that, in the relevant portion of the parameter space, we
can replace $\xi\to\zeff$ in the equation (\ref{eq:eqdiff}) and its solution (\ref{eq:solchangextoa}), where
\begin{equation}\label{nueff}
\zeff=1-(\nu-\alpha)\equiv1-\nueff\,.
\end{equation}
Let us note that, in the particular case $\nu=\alpha=0$ (hence
$\zeff=1$), the formula (\ref{eq:solchangextoa}) leads to the
$\CC$CDM solution (\ref{eq:soldiffeqDaLCDM}) after using a standard
integral representation of the hypergeometric
series\,\cite{Gradshteyn}. In the early epochs of matter domination,
i.e. for sufficiently large values of $a$ when the cosmological term
can be neglected and $E(a)\simeq \sqrt{\Omo}a^{-3\xi/2}$, the solution
(\ref{eq:solchangextoa}) takes on the simple form $\delta_m(a)\sim a^{3\xi-2}$. This
behavior can be used as initial condition to determine $A_1$.

Finally, let us mention that in Sect. \ref{sec:Number counts}, we will extend the structure formation analysis to the
nonlinear regime (specifically to the formation of collapsed
structures). It will be shown that the models are quite sensitive to
the number of formed clusters at a given redshift in a suitable
range of masses, and thanks to this feature they can be better discriminated.
Before addressing the nonlinear regime, in the next section we
complete the linear perturbations analysis of the remaining models.

%%%%%%%%%%%%%%%%%%%%%%%
%%%%%%%%%%%%%%%%%%%%%%%

\subsection{Perturbations for type-B and C models}
\label{sec:perturbationsTypeB} 

In Sect.\,\ref{subsec:solvingB1B2} we
have amply addressed the cosmological solution of type-B models at
the background level, and discussed the shortcomings in connection
to the possibility to present the solution in terms of the scale
factor. The corresponding perturbations analysis is no exception
and hence we have to attempt a solution once more in terms
of the cosmic time, i.e. we have to necessarily start from
Eq.\,(\ref{diffeqD}), not (\ref{diffeqDa}). It proves convenient the
following change of independent variable
\begin{equation}\label{eq:defx}
y(t)=\coth\left(\frac{3}{4}H_0\mathcal{F}t\right)\,,
\end{equation}
or, equivalently,
\begin{equation}\label{eq:defxv2}
y=\frac{2\zeta E-\epsilon}{\mathcal{F}}\,,
\end{equation}
where we have used  (\ref{eq:Hubblef}) with $E=H/H_0$. Let us employ
that variable to rewrite the matter density (\ref{eq:rhom}) in terms of it. The
result is:
\begin{equation}\label{eq:rhoy}
\nonumber \rho_m(y)=\frac{3H_0^2}{32\pi
G\zeta}\mathcal{F}^2(y^2-1)\,.
\end{equation}
Consider next the function $\Psi$ defined in (\ref{defQ}). With the
help of (\ref{eq:rhoL}) and after straightforward algebra we
can write it also in terms of $y$ and express the result in a rather
compact form:
\begin{equation}\label{Qdot2}
\Psi(y)=-\frac{\dot{\rho}_{\CC}(t)}{\rmr(t)}=\frac{3
H_0}{2\zeta}\,\left[\epsilon+(1-\zeta)\mathcal{F}\coth\left(\frac{3}{4}H_0\mathcal{F}t\right)\right]=
\frac{3H_0}{2\zeta}\,\left[y(1-\zeta)\,\mathcal{F}+\epsilon\right]\,.
\end{equation}
With the above formulae and performing the substitution
(\ref{eq:defxv2}) in the original Eq.\,(\ref{diffeqD}) we arrive at
the differential equation for the growth factor $\delta_m(y)$ in the
new variable:
\begin{eqnarray}\label{eq:perturbTypeB2}
&&(y^2-1)\frac{d^2\delta_m(y)}{dy^2}+\left[\frac{2y}{3\zeta}(6\zeta-5)-\frac{10\epsilon}{3\zeta\mathcal{F}}\right]\frac{d\delta_m(y)}{dy}+\phantom{XXXXXXXXXXXXXX}\\
&&\phantom{XXXXXXXx}+\frac{2}{3\zeta}\left[\frac{4}{\zeta(y^2-1)}\left(\frac{\epsilon^2}{\mathcal{F}^2}+(1-\zeta)y^2+\frac{\epsilon}{\mathcal{F}}(2-\zeta)y\right)+3\zeta-4\right]\delta_m(y)=0\,.\nonumber
\end{eqnarray}
This equation cannot be solved analytically, not even in terms of
standard special functions. In the particular case of the type-B1
model, for which $\nu=0$ - implying no quadratic term $H^2$ in the
expression of the vacuum energy density (\ref{B1B2}) - the above
equation simplifies slightly:
\begin{eqnarray}\label{eq:perturbTypeB1}
&&(y^2-1)\frac{d^2\delta_m(y)}{dy^2}+\left(\frac{2y}{3}-\frac{10\epsilon}{3\mathcal{F}_1}\right)\frac{d\delta_m(y)}{dy}+\phantom{XXXXXXXXXXXXXX}\nonumber\\
&&\phantom{XXXXXXXx}
+\frac{2}{3}\left[\frac{4\epsilon}{\mathcal{F}_1(y^2-1)}\left(\frac{\epsilon}{\mathcal{F}_1}+y\right)-1\right]\delta_m(y)=0\,
\end{eqnarray}
where  $\mathcal{F}_1=\mathcal{F}(\OLo,\epsilon,\nu=0)$,  see
(\ref{defmathF}). Unfortunately, this equation is still too
complicated for an analytic solution. As always the intrinsic
difficulty with type-B models resides already at the level of the linear term in $H$. Thus, for both types B1 and B2 we are forced to use
the numerical techniques, for instance the method of finite
differences. Namely, take a differential equation of the type
\begin{equation}\label{eq:eqsim}
a(y)\frac{d^2\delta_m(y)}{dy^2}+b(y)\frac{d\delta_m(y)}{dy}+c(y)\delta_m(y)=0
\end{equation}
whose basic coefficients $a(y)$, $b(y)$ and $c(y)$ can be readily
identified from (\ref{eq:perturbTypeB2}) and
(\ref{eq:perturbTypeB1}) in the case of type-B models. We must fix a
sufficiently small step $\Delta y=y_{n+1}-y_n$ so that we can take
Taylor's expansion up to second order in $\Delta y$ as a good
approximation of the linear growth factor in a neighborhood of $y_n$. If
we choose the step as indicated before we can relate the value
of the growth factor in $y_n$ and $y_{n+1}$ as follows:
\begin{equation}\label{eq:Taylor}
\delta_m(y_{n+1})=\delta_m(y_n)+\Delta y \delta_m^\p(y_n)+\frac{\Delta y^2}{2}\delta_m^{\pp}(y_n)\,.
\end{equation}
Isolating $\delta_m^{\pp}(y)$ in \eqref{eq:eqsim} and introducing it
into this Taylor expansion we find:
\begin{equation}\label{eq:recu1}
\delta_m(y_{n+1})=\left(1-\frac{c(y_n)}{a(y_n)}\frac{\Delta
y^2}{2}\right)\delta_m(y_n)+\Delta
y\left(1-\frac{b(y_n)}{a(y_n)}\frac{\Delta y}{2}\right)\delta_m^\p(y_n)\,.
\end{equation}
In order to compute the points of the curve $\delta_m(y)$ we also need to
obtain a recurrence formula which gives us $\delta_m^\p(y_n)$.
Differentiating \eqref{eq:Taylor} we get
\begin{equation}
\delta_m^\p(y_{n+1})=\delta_m^\p(y_n)+\Delta y \delta_m^{\pp}(y_n)\,,
\end{equation}
and, therefore, we obtain:
\begin{equation}\label{eq:recu2}
\delta_m^\p(y_{n+1})=\left(1-\frac{b(y_n)}{a(y_n)\Delta
y}\right)\delta_m^\p(y_n)+\Delta y \frac{c(y_n)}{a(y_n)}\delta_m(y_n)\,.
\end{equation}
With the use of \eqref{eq:recu1}, \eqref{eq:recu2} and two initial
conditions i.e. $\delta_m(y_i)$ and $\delta_m^\p(y_i)$, we can finally obtain
computationally not only the curve of the growth factor $\delta_m(y)$,
but also the one of the growth rate $f(y)$ (cf. Sect.
\ref{subsec:growthrate}) by just doing
\begin{equation}
f_{n}=\frac{\ln \delta_m(y_{n+1})-\ln \delta_m(y_n)}{\ln a(y_{n+1})-\ln
a(y_n)}\,.
\end{equation}
During the matter-dominated epoch $\delta_m(a)\approx a$, so we take as
initial conditions the value of the growth factor and its
derivative at very high redshifts $y_i=y(z_{tr}\gg z^*)$ (recall that
at redshifts of order $z^*={\cal O}(1)$ the vacuum energy density
begins to dominate the Universe's dynamics for the $\CC$CDM and the
other models considered here, see Sect. \ref{sec:solving}). The
scale factor can be expressed in terms of $y$ by substituting the relation extracted from \eqref{eq:defx},
\begin{equation}
e^{-\frac{3}{4}H_0t}=\sqrt{\frac{y-1}{y+1}}, 
\end{equation}
in (\ref{atBtype}):
\begin{equation}
a(y)=\left(\frac{[(2\zeta-\epsilon)^2-\mathcal{F}^2]^{1+\frac{\epsilon}{\mathcal{F}}}}{\mathcal{F}^2(\mathcal{F}+2\zeta-\epsilon)^{\frac{2\epsilon}
{\mathcal{F}}}}\right)^{\frac{1}{3\zeta}}(y^2-1)^{-\frac{1}{3\zeta}}\left(\frac{y+1}{y-1}\right)^{\frac{\epsilon}{3\zeta\mathcal{F}}}\,.
\end{equation}
We normalize the growth factor by taking $\delta_m(a)=a$ at very high redshifts. The initial
conditions at $y_i=700$ ($z=98.83$), are the
following. For the growth factor we have $\delta_m(y_i)=a(y_i)$, and for
its derivative with respect to the $y$-variable, we obtain
\begin{equation}
\delta_m^\p(y_i)=\left.\frac{d\delta_m(y)}{dy}\right|_{y_i}=\left.\frac{d\delta_m(a)}{da}\right|_{a_i}\left.\frac{da(y)}{dy}\right|_{y_i}=
\left.\frac{da(y)}{dy}\right|_{y_i}=\frac{-2a(y_i)}{3\zeta(y_i^2-1)}\left(y_i+\frac{\epsilon}{\mathcal{F}}\right)\,.
\end{equation}
For the analysis of perturbations we have to compute $\delta_m(z)$ and
$f(z)$ from $z=98.83$ up to $z=0$, and for this we need the value of
$y(z=0)$. Notice that it immediately follows from setting $E(t_0)=1$
in \eqref{eq:defxv2}, thus $y(t_0)=(2\zeta-\epsilon)/{\mathcal{F}}$.

For type-C1 models the perturbation equations can be readily
obtained by setting, once more,  $C_0\to 0$ in Eqs.\,(\ref{B1B2})
and (\ref{eq:C0TypeB}), which amounts to set
$\mathcal{F}\to\epsilon$ in the above perturbation equations for the
type-B2 model. In this limit, the variable (\ref{eq:defxv2}) reads
$y=-1+2\zeta E/\epsilon$. Introducing now
\begin{equation}\label{eq:defy1}
y_1=\frac{y+1}{y-1}\,,
\end{equation}
the differential equation (\ref{eq:perturbTypeB2}) can be brought to
the form
\begin{equation}\label{eq:perturbC1}
3\zeta^2\,y_1\,(y_1-1)^2\,\frac{d^2\delta_m}{dy_1^2}+2\,\zeta\,(y_1-1)\,(5y_1-3\zeta)\,\frac{d
\delta_m}{dy_1}-2\,(2-\zeta)\,(3\zeta-2y_1)\,\delta_m=0\,.
\end{equation}
This result is consistent with that of Ref.\,\cite{BasSola14a},
which appears here as a particular case of the general equation
(\ref{eq:perturbTypeB2}) for type-B models. A power-like solution of
Eq.\,(\ref{eq:perturbC1}) immediately ensues: $\delta_{m,{-}}(y_1)\sim
(y_1-1)^{(\zeta-2)/\zeta}$. While an explicit relation of the
variable $y$ with the scale factor is impossible for the general
type-B models, for C1 models the variable $y_1$ defined in
(\ref{eq:defy1}) permits such relation:
\begin{equation}
y_1=\frac{\zeta E}{\zeta
E-\epsilon}=1+\frac{\zeta-\Omo}{\Omo}\,a^{3\zeta/2} \,,
\end{equation}
with $E$ given by (\ref{E2aC1}). Thanks to this the previously found
solution can be rewritten as $\delta_{m,-}(a)\sim a^{3(\zeta-2)/2}$. The
latter is the decaying mode (since $\zeta\leq1$) and, therefore,
must be  rejected. However, from it we can obtain the growing mode
solution of the C1 model following the standard procedure: \be
\label{eqff2} \delta_{m,+}(a)=C_1\, a^{3(\zeta-2)/2}\int_{0}^{a} \frac{{\rm
d}a'}{a'^{3\zeta/2}E^{2}(a')}\,, \ee with $C_1$ a constant. The
behavior of Eq.\,(\ref{eqff2}) in the early epoch, namely when
$E(a)\sim (\Omo/\zeta) a^{-3/2}$, is $\delta_m(a)\sim a$. This is the same
limiting behavior as that of the $\CC$CDM model -- which we inferred
before from (\ref{eq:soldiffeqDaLCDM}) -- with the proviso mentioned
in Sect. \ref{subsec:solvingC1C2}  that this model has an extra factor
of $\Omo$ in the matter density. This anomaly has consequences: when
one tries to fit that model to the background data and to the linear
growth of matter perturbations one finds a poor quality fit, see
Sect. \ref{subsec:growthrate} and Ref. \cite{BasSola14a}.  The anomaly
disappears if the model is ``protected'' with an additive term in
the structure of the vacuum energy density, such as for general
type-A and  type-B models. Finally, concerning model C2 we stress that in Ref.\,\cite{BasSola14a} the perturbation analysis was considered and it was shown that it does not lead to a growing mode solution for any reasonable value of the cosmological parameters. We do not provide details here (we refer the reader to the aforementioned reference) and we recall that the model was already excluded at the background level.

%%%%%%%%%%%%%%%%%%%%%%%%%%%%%%%%%%%%%%%%
%%%%%%%%%%%%%%%%%%%%%%%%%%%%%%%%%%%%%%%%

\section{Including the effect of radiation} 
\label{sec:radiation}

Up to now we have not considered the effect of relativistic matter
in the solution of the background cosmological equations since we
have focused on the form of the solution near our time. For the
study of cosmological perturbations the effect of radiation was not
necessary either. However, in the next section we perform a fit of
the current models to the Baryonic Acoustic Oscillations (BAO)
and the Cosmic Microwave Background (CMB). To this end we do have to
include the effect of radiation since at the time when the CMB was
released (corresponding to $z\simeq 1100$) the amount of radiation
was not negligible ($\sim 24\%$). As we will see below, including
the effect of radiation in a consistent way is relatively simple for
type-A models, but is more cumbersome for type-B ones. Type C1 is a
particular case of them.

%%%%%%%%%%%%%%%%%%%%%%%%%%%%%%%%%%%%%%%%
%%%%%%%%%%%%%%%%%%%%%%%%%%%%%%%%%%%%%%%%

\subsection{Radiation component for type-A models}
\label{subsec:radiationA}

Let us write again Eq.\,(\ref{Bianchi1}),
\begin{equation}\label{eq:RadA}
{\dot \rho}_m + {\dot \rho}_r + {\dot \rho}_{\Lambda} + 3 H \rho_m +
4 H \rho_r=0\,.
\end{equation}
For type-A vacuum models, where $\rL$ is given by (\ref{GRVE}), we
obtain the following generalized form of
Eq.\,(\ref{eq:generalizedConserv}):
\be\label{eq:BianchitypeA2} {\dot \rho}_m +
3\,H\,\xi\,\rmr+\dot{\rho}_r+4\,H\,\xiR\rho_r=0\,, \ee 
where we used the
familiar parameter $\xi$, defined in (\ref{eq:defxiM}), and we have now
introduced a new one that is related with the radiation component:
\begin{equation}\label{eq:defxiR}
\xiR= \frac{ 1 - \nu }{ 1 - 4\alpha/3 }\,.
\end{equation}
Equation (\ref{eq:BianchitypeA2}) can be split as
\begin{eqnarray}
{\dot \rho}_m + 3 H \xi~\rho_m &=& Q  \label{eq:splitQA1}\\
{\dot \rho}_r + 4 H \xiR~\rho_r&=& - Q\,, \label{eq:splitQA2}
\end{eqnarray}
where the source function $Q(t)$ describes the exchange of energy
between relativistic and non-relativistic matter. We do not want to
describe in detail this exchange dynamics here (which was certainly relevant
in some epoch of the Universe), but only the situation when one of
the components dominates. Thus we will assume $Q\simeq 0$ and solve
separately each equation, (\ref{eq:splitQA1}) and (\ref{eq:splitQA2}). This is, of course, the
same kind of assumption as in the standard model case. In this way
we obtain:
\begin{eqnarray}\label{splitAsolution}
\rho_m &=& \rho_m^0 ~a^{-3 \xi} \label{splitAsolutionM} \\
\rho_r &=& \rho_r^0 ~a^{-4 \xiR}\,.\label{splitAsolutionR}
\end{eqnarray}
The presence of $\xi$ and $\xiR$ denotes the anomaly in the
corresponding conservation laws.  Of course such anomaly must be
small (i.e. $\xi\simeq \xiR\simeq 1$) in the natural physical region
(\ref{naturalness}). The small deviation from $1$, however, is
exactly what permits the vacuum energy to evolve with the expansion. It is remarkable that for type-A models the two parameters $\nu$ and $\alpha$ combine in such a way that in practice we have, in each relevant epoch (matter-dominated and radiation-dominated), only one effective parameter, $\xi$ and $\xiR$ respectively.

By repeating the integration procedure it is not difficult to show
that the vacuum energy density in the presence of radiation evolves
as
\begin{equation}\label{eq:rLArad}
\rL(a)=\rLo+{\rmo}\,\,(\xi^{-1} - 1) \left( a^{-3\xi} -1  \right) +
{\rRo}\,\,(\xiR^{-1} - 1) \left( a^{-4\xiR} -1\right)\,.
\end{equation}
We confirm that only for $\xi=\xiR=1$ the vacuum energy density
remains strictly constant. Similarly, the corresponding normalized
Hubble rate in the presence of radiation reads
\begin{equation}\label{eq:HArad}
 E^2(a) =1+\frac{\Omo}{\xi}\left(a^{-3\xi}-1\right)+\frac{\Oro}{\xiR}\left(a^{-4\xiR}-1\right)\,,
\end{equation}
where the various normalized cosmological parameters satisfy now the
constraint
\begin{equation}\label{eq:sumruleRad}
 \Omo + \Oro+\OLo = 1\,.\end{equation}
Obviously we have generalized the equations of
Sect.\,\ref{subsec:solvingA1A2} for when the radiation component is
taken into account.

%%%%%%%%%%%%%%%%%%%%%%%%%%%%%%%%%%%%%
%%%%%%%%%%%%%%%%%%%%%%%%%%%%%%%%%%%%%

\subsection{Radiation component for type-B models}
\label{subsec:radiationB}

Here the situation is a bit more complicated owing to the
impossibility to solve this type of models analytically in terms of
the scale factor variable. We start by inserting Eq.\,(\ref{B1B2})
in the generalized conservation law (\ref{eq:RadA}), and by making use of (\ref{rchangeH}) to express $\dot{H}$ in terms of $\rho_m$ and $\rho_r$. As in the
previous section we split the result into the matter and radiation
components with the help of an interaction source $Q(t)$. Using the standard
notations that we have introduced for type-B  models, we find:
\begin{eqnarray}\label{eq:splitQB}
\dot{\rho}_m+3\rho_m \left[\zeta H -\frac{\epsilon}{2} H_0\right]&=&Q(t)\nonumber\\
\dot{\rho}_r+4\rho_r\left[\zeta H -\frac{\epsilon}{2}
H_0\right]&=&-Q(t)\,,
\end{eqnarray}
with $\zeta=1-\nu$. Once more we focus on situations characterized by pure cold (i.e.
non-relativistic) matter or pure radiation-dominated epochs,
meaning that we search for a decoupled solution with $Q(t)\approx
0$. In this limit we can solve these equations and express the
energy densities as a function of the scale factor and the cosmic
time:
\begin{equation}\label{eq:rhomNR}
\rho_m(t,a)=\rmo\,e^{\frac{3}{2}\epsilon H_0(t-t_0)}\,a^{-3\zeta}
\end{equation}
\begin{equation}\label{eq:rhorNR}
\rho_r(t,a)=\rRo\,e^{2\epsilon H_0(t-t_0)}\,a^{-4\zeta},
\end{equation}
where $t_0$ is the value of the cosmic time at present. Clearly, only for $\epsilon=0$ we can get a solution fully expressed in terms of the scale factor variable. In addition, the previous equations also show the feature first indicated by the end of Sect.\,\ref{subsec:solvingB1B2}, namely that the two basic parameters $(\epsilon,\nu$) of type-B models are completely independent and cannot be mimicked by a single effective parameter in a matter-dominated or radiation-dominated epoch, in stark contrast to type-A models. Therefore, for models of class B there is no possible election of $\epsilon$ and $\nu$ by which e.g. matter or radiation behaves in a completely standard manner, despite the fact that for $\nu$ and $\epsilon$ sufficiently small these models behave, of course, arbitrarily close to the $\CC$CDM.  From the above
results we can also find the time-evolving vacuum energy density by
integrating the equation:
\begin{equation}\label{eq:rhoVB}
\dot{\rho}_\Lambda=-\left[(1-\zeta) \,H+\frac{\epsilon}{2}
H_0\right](3\rho_m+4\rho_r)\,.
\end{equation}
For type-B models the scale factor in terms of the cosmic time can
be computed by first solving Eq.\,(\ref{eq:EHmonocomp}), where $\omega_M$ is as in previous sections the EoS of a single dominant matter component (relativistic or non-relativistic). The result is:
\begin{equation}\label{eq:HtomegatyepB}
H(t)=\frac{H_0}{2\,\zeta}\left[\mathcal{F}\,\coth\left(\frac{3}{4}H_0\mathcal{F}\,(1+\omega_M)\,t\right)+\epsilon\right]\,,
\end{equation}
and from here we obtain:
\begin{equation}\label{eq:atomegatyepB}
a(t)=B^{-\frac{1}{3\zeta(1+\omega_M)}}\,e^{\frac{\epsilon
H_0t}{2\zeta}}\,\sinh^{\frac{2}{3\zeta(1+\omega_M)}}\left(\frac{3}{4}H_0\mathcal{F}(1+\omega_M)t\right)\,,
\end{equation}
with the normalization constant
\begin{equation}\label{eq:defCtypeBomega}
B=\left[\frac{[(2\zeta-\epsilon)^2-\mathcal{F}^2]^{1+\frac{\epsilon}{\mathcal{F}}}}{\mathcal{F}^2(\mathcal{F}+2\zeta-\epsilon)^{\frac{2\epsilon}{\mathcal{F}}}}\right]^{-1}\,.
\end{equation}
These results generalize those of Sect. \ref{subsec:solvingB1B2}, but as advertised they apply only for monocomponent situations where the Universe is
purely dominated either by radiation ($\omega_M=\omega_r=1/3$) or by matter
($\omega_M=\omega_m=0$). In this model we cannot invert the cosmic time in the
form $t=t(a)$ so as to deliver the energy-densities explicitly in
terms of the scale factor. Even if it was possible it would not help to
find the needed correction that the scale factor undergoes when both
components (non-relativistic matter and radiation) are
simultaneously present in nonnegligible amounts.

The problem, however, can be solved if the effects of radiation are relatively small and hence cannot modify dramatically the
matter-dominated solution. A relevant period of the cosmic evolution to which this situation applies is around the time when the CMB was released. To generate a consistent solution
within this period, we start from the expression of the scale factor (\ref{eq:atomegatyepB}) for  $\omega_M=0$, i.e. we initially assume that the scale factor evolves with time as in the non-relativistic epoch. Introducing the result in  \eqref{eq:rhomNR} and \eqref{eq:rhorNR} we obtain:
\begin{equation}\label{eq:rhomNR2}
\rho_m(t)=\rmo\,B\,e^{-\frac{3}{2}\epsilon H_0 t_0}\,{\rm
csch}^2\left(\frac{3}{4}H_0\mathcal{F}t\right)\,,
\end{equation}
\begin{equation}\label{eq:rhorNR2}
\rho_r(t)=\rro\,B^{4/3}\,e^{-2\epsilon H_0 t_0}\,{\rm
csch}^{8/3}\left(\frac{3}{4}H_0\mathcal{F}t\right)\,.
\end{equation}
Notice that we must demand the following normalization condition to
be satisfied so that the energy densities take the present values at
$t=t_0$:
\begin{equation}\label{eq:condition}
B\,e^{-\frac{3}{2}\epsilon H_0 t_0}\,{\rm
csch}^2\left(\frac{3}{4}H_0\mathcal{F}t_0\right)=1\,.
\end{equation}
One can compute the value of $t_0$ by solving numerically
this equation. However, we can get it more easily by noticing that
equations (\ref{eq:rhom}) and (\ref{eq:rhomNR2}) must be the same
and therefore the following relation ensues:
\begin{equation}\label{eq:toBF}
B\,e^{-\frac{3}{2}\epsilon H_0
t_0}=\frac{\mathcal{F}^2}{4\zeta\,\Omo}\,,
\end{equation}
which determines the exponential factor that appears in the energy densities in
terms of the parameters of the model and the current cosmological
parameters.

The Hubble function including the effects of radiation in the
matter-dominated epoch can now be computed from
\begin{equation}\label{eq:HtypeBRad}
H^2(t)=\frac{8\pi G}{3}\left[\rmr(t)+\rR(t)+\rL(t)\right]=\frac{8\pi
G}{3}\left[\rmr(t)+\rR(t)\right]+C_0+C_1\,H(t)+C_2\,H^2(t)\,,
\end{equation}
where we have used (\ref{B1B2}) and we understand that $\rmr(t)$ and
$\rR(t)$ are given by (\ref{eq:rhomNR2}) and (\ref{eq:rhorNR2})
respectively. It is easy to check that thanks to the condition
\eqref{eq:condition} the formula (\ref{eq:HtypeBRad}) for the Hubble
function for mixed matter and radiation satisfies the extended
cosmic sum rule (\ref{eq:sumruleRad}). This is of course a necessary
condition to insure a consistent treatment of the two components.

Expressing (\ref{eq:HtypeBRad}) in terms of the usual parameters
$\zeta$ and $\epsilon$ and rearranging terms, we find:
\begin{equation}\label{eq:solvingHalgebraic}
\zeta H^2-\epsilon H_0 H- H_0^2\,s(t)=0,
\end{equation}
where
\begin{eqnarray}
&&s(t)=\zeta-\epsilon-\Omo-\Oro+\frac{\mathcal{F}^2}{4\zeta}\,{\rm
csch}^2\left(\frac{3}{4}H_0\mathcal{F}t\right)
+\phantom{XXXXXXXXXXXXXX}\\
&&\phantom{XXXXXXXXXXXXXXx}+\Oro\,\left(\frac{\mathcal{F}^2}{4\zeta\,\Omo}\right)^{4/3}\,{\rm csch}^{8/3}\left(\frac{3}{4}H_0\mathcal{F}t\right)\,.\nonumber
\end{eqnarray}
Finally, the sought-for  Hubble function of the matter-dominated epoch in the presence of a relatively small amount of radiation can be explicitly furnished in terms of the cosmic time:
\begin{equation}\label{eq:hubbleimpro}
H(t)=\frac{H_0}{2\zeta}\,\left[\epsilon +\sqrt{\epsilon^2+4\zeta
s(t)}\right]=\frac{H_0}{2\zeta}\,\left[\epsilon
+\mathcal{F}\,\coth{\left(\frac{3}{4}H_0\mathcal{F}t\right)\,\sqrt{1+\Delta(t)}}\right]\,,
\end{equation}
where
\begin{equation}\label{eq:Delta}
\Delta(t)=\frac{\Oro}{\Omo}\,\left(\frac{\mathcal{F}^2}{4\zeta\Omo}\right)^{1/3}\,{\rm
csch}^{2/3}\left(\frac{3}{4}H_0\mathcal{F}t\right){\rm
sch}^2\left(\frac{3}{4}H_0\mathcal{F}t\right)\,.
\end{equation}
The second equality in Eq.\,(\ref{eq:hubbleimpro}) makes use of the relation
\begin{equation}\label{eq:identity1}
\zeta-\epsilon-\Omo-\Oro=\OLo-\nu-\epsilon=\frac{\mathcal{F}^2-\epsilon^2}{4\zeta}\,.
\end{equation}
The form of the result (\ref{eq:hubbleimpro}) enables us to better compare with the original form\,(\ref{eq:Hubblef}). The correction from the radiation is
clearly represented by the function ${\Delta}(t)$. Obviously this
effect is negligible at the present time since the prefactor in it reads $\Oro/\Omo=\left(1+0.227\,N_{\nu}\right)\,\Omega_{\gamma}^0/\Omo=4.15\times
10^{-5} h^{-2}/\Omo\simeq 3\times 10^{-4}$ (including photons and
$N_{\nu}=3$ neutrino species, and assuming $\Omo\,h^2\simeq 0.14$).
However, at decoupling (i.e. at the time $t_{*}$ of last scattering
of radiation with matter) the hyperbolic function in
(\ref{eq:Delta}) rockets into a numerical value of order $\sim 10^3$ and $\Delta(t_{*})$ can become quite
sizeable. In fact, the fraction of radiation at decoupling can be
around $\sim 23\%$ in the $\CC$CDM case. This is of course also the
case for the generalized type A and B vacuum models deviating mildly
with respect to the $\CC$CDM (i.e. for small values of the
parameters).

The integration of (\ref{eq:hubbleimpro}) provides the corresponding
improved version of the scale factor. Normalized to $a(t_0=1$), it
reads:
\begin{equation}\label{eq:a(t)Ind}
a(t)=e^{\int_{t_0}^{t}H(\hat{t})d\hat{t}}\,.
\end{equation}
From the above procedure it is clear that we can obtain the points
of the curve $H(a)$ numerically by  computing $H(t)$ by means of
\eqref{eq:hubbleimpro} and  $a(t)$ through \eqref{eq:a(t)Ind}.  Once
$H(a)$ is determined with this method -- and similarly with
$\rmr(a)$, $\rR(a)$ and $\rL(a)$ -- we can immediately confront the
model with experiment since the data inputs are given in terms of
the redshift, whose relation with the scale factor is simply
$z=(1-a)/a$.

%%%%%%%%%%%%%%%%%%%%%%%%%%%%%%%%%%%%%%%
%%%%%%%%%%%%%%%%%%%%%%%%%%%%%%%%%%%%%%%

\subsection{Radiation component for type-C1 models}
\label{subsec:radiationC1}

As we have seen, for type-C1 model (and in particular for type-C1 with $C_0=0$)
one can derive the energy densities (\ref{eq:rhomNR}) and
(\ref{eq:rhorNR}) in terms of the scale factor. The answer for
non-relativistic matter is given in Eq.\,(\ref{eq:rhomaC1}). For
radiation we can proceed in a similar way, and the result is:
\be\label{eq:rhorC1} \rho_r(a)=\rho_r^0\,f^{4/3}(a)\,a^{-4\zeta}\,,
\ee
where $f(a)$ is defined as $f(a)=a^{3\zeta/2}E(a)$. The modified Hubble rate
for type-C1 models can be expressed in terms of the scale factor by
solving (\ref{eq:HtypeBRad}) after setting $C_0=0$ and using
(\ref{eq:rhomaC1}) and (\ref{eq:rhorC1}):
\be\label{eq:EaIIrhomrhoR}
E(a)=\frac{\zeta-\Omo+\sqrt{(\zeta-\Omo)^2+4\zeta\left[\frac{\rho_m(a)+\rho_r(a)}{\rho_c^{0}}\right]}}{2\zeta}\,.
\ee
Substituting this expression in (\ref{GRVE}), with $C_0=0$, we find
the corresponding vacuum energy density $\rho_\CC(a)$ including the
effect of radiation, which is a cumbersome expression.

We can also estimate the equality time,  $t_{eq}$, between the
radiation and the non-relativistic matter energy densities  for
models of type-C1. Equating (\ref{eq:rhomaC1}) and
(\ref{eq:rhorC1}) and taking into account that $a_{eq}\ll 1$ we
obtain:
\be\label{eq:aeq}
a_{eq}=\left[\frac{\Omega^{0}_r}{\zeta^{1/3}\left(\Omo\right)^{2/3}}\right]^{1/\zeta}\,.
\ee
For the typical values that $\zeta$ and $\Omo$ shown in Tables \ref{tableFitBAOA} and \ref{tableExtra}, $a_{eq}$ deviates significantly from the $\CC$CDM
prediction value $a_{eq}=\Omega^0_r/\Omo$. In contrast, for models
of type-B (which have $C_0\ne 0$) one can use the
concordance value as a very good approximation. In these cases one
can show that
$a_{eq}=\Omega^0_r/\Omo[1+x\,\ln(\Omega^0_r/\Omo)+\mathcal{O}(x^2)]$,
where $x(\epsilon,\nu)\ll 1$ and the deviations from the $\CC$CDM
model value are only at the few percent level. Needless to say,
additional important differences of the $C_0=0$ models are expected
to appear in connection to the photon decoupling and baryon drag
epochs, and in the value of the comoving Hubble scale,
$k^{-1}_{eq}$, at the redshift of matter-radiation equality.

The  ratio between
the vacuum and radiation energy densities at $a\ll 1$ can be
estimated from the foregoing analysis, with the result:
\begin{equation}\label{eq:ratioDens}
\frac{\rho_\CC}{\rho_r}\approx
\left(\frac{1-\zeta}{\zeta}\right)\left(\frac{a}{a_{eq}}\right)^\zeta\,,
\end{equation}
where use has been made of (\ref{eq:aeq}). Taking into account the fitted values of the
parameters presented in Table \ref{tableFitBAOA}, we find
$\zeta=\epsilon+\Omo=1.189$. Therefore, at $a=a_{eq}$ the ratio
(\ref{eq:ratioDens}) yields $\rho_\CC\approx-0.16\rho_r$. We learn from this estimate that
type-C1 model predicts a negative value of $\rho_\CC$ in the past and, in addition, the proportion of vacuum energy is excessively large in comparison with the 
values encountered in the $\Lambda$CDM-like models. This fraction could of course be made smaller by
decreasing $\nu$ (i.e. approaching $\zeta\to 1$) but this would
worsen the quality of the fit. We will show later, that this fact has dramatic consequences for the predicted amount of structure formation in type-C1 models.

%%%%%%%%%%%%%%%%%%%%%%%%%%%%%%%%%%%%%%%
%%%%%%%%%%%%%%%%%%%%%%%%%%%%%%%%%%%%%%%

\section[Fitting the models to the data]{Fitting the models to the observational data}
\label{sec:fitting}

\begin{figure}
\centering
\includegraphics[angle=0,width=0.75\linewidth]{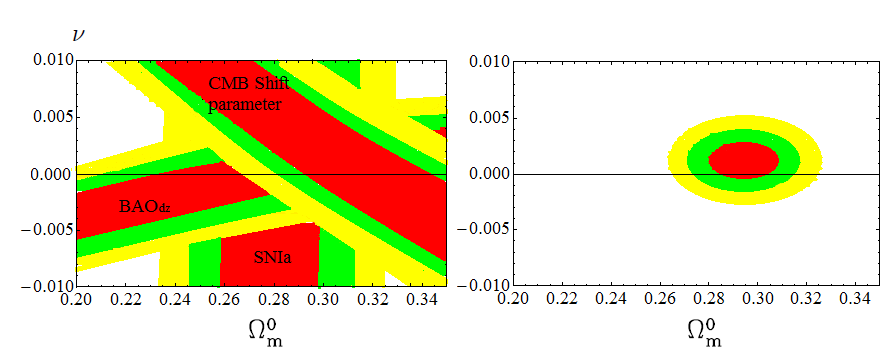}
\caption[Contour lines for model A1. Fitting analysis with BAO$_{dz}$ of Sect. \ref{chap:Atype}.]{\label{fig:contoursA1 BAOdz}%
\scriptsize {Likelihood contours (for $-2{\rm ln}{\cal L}/{\cal L}_{\rm max}$ equal to 2.30, 6.16 and 11.81, corresponding to 1$\sigma$, 2$\sigma$ and $3\sigma$ confidence levels) in the $(\Omega_{m}^{0},\nu)$ plane for the A1 vacuum model ($\alpha\equiv 0$). The left panel shows the contours based on the SNIa data (represented by approximate vertical bands), BAO$_{dz}$ (diagonal bands) and CMB shift parameter (antidiagonal bands). On the right panel we show the corresponding
contours based on the joint statistical analysis (SNIa+CMB+BAO$_{dz}$ data. From the inner to the outer regions (successively in red, green and yellow) we find the aforementioned 1$\sigma$, 2$\sigma$ and $3\sigma$ confidence levels, respectively.}}
\end{figure}

\begin{figure}
\centering
\includegraphics[angle=0,width=0.75\linewidth]{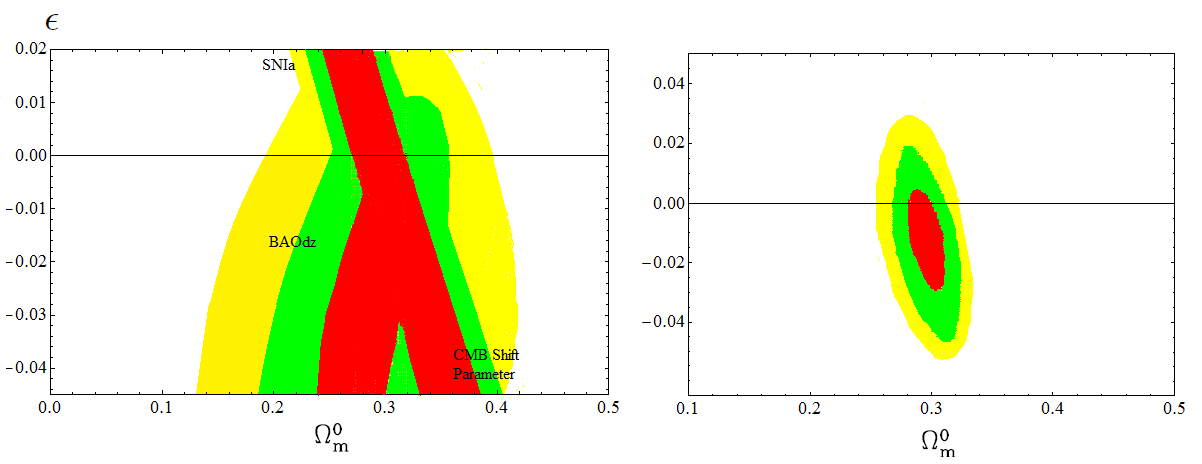}
\caption[Contour lines for model B1. Fitting analysis with BAO$_{dz}$ of Sect. \ref{chap:Atype}.]{\label{fig:contoursB1 BAOdz}%
\scriptsize {Likelihood contours for the B1 vacuum model, with BAO$_{dz}$ data. The left panel shows the contours based on the SNIa data, BAO$_{dz}$ and CMB shift parameter. The meaning of the shaded regions is as in Fig.\,\ref{fig:contoursA1 BAOdz}. The various bands follow a similar pattern as in the previous figure, but here the overlapping of the SNIa and BAO regions is larger. The right panel shows the joint contours of SNIa+CMB+BAO$_{dz}$.}}
\end{figure}

\begin{figure}
\centering
\includegraphics[angle=0,width=0.75\linewidth]{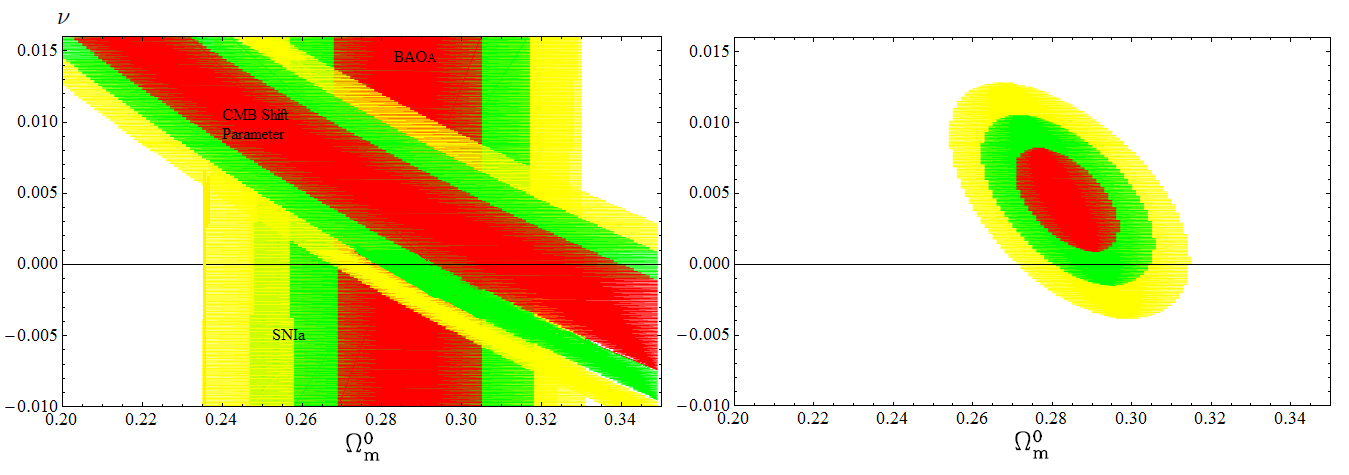}
\caption[Contour lines for model A1. Fitting analysis with BAO$_{A}$ of Sect. \ref{chap:Atype}.]{\label{fig:contoursA1 BAOA}%
\scriptsize {Likelihood contours for the A1
vacuum model, this time with BAO$_A$ data.
The left panel shows the contours based on the SNIa data, BAO$_{A}$ and CMB shift parameter indicated in a similar way as in Fig.\,\ref{fig:contoursA1 BAOdz}. The SNIa data and BAO$_{A}$ bands appear as almost vertical and with significant overlap. The right panel shows
the joint contours of SNIa+CMB+BAO$_{A}$. A well defined final region is projected with $\nu>0$ at $\sim 1.5\sigma$ level.}}
\end{figure}

\begin{figure}
\centering
\includegraphics[angle=0,width=0.75\linewidth]{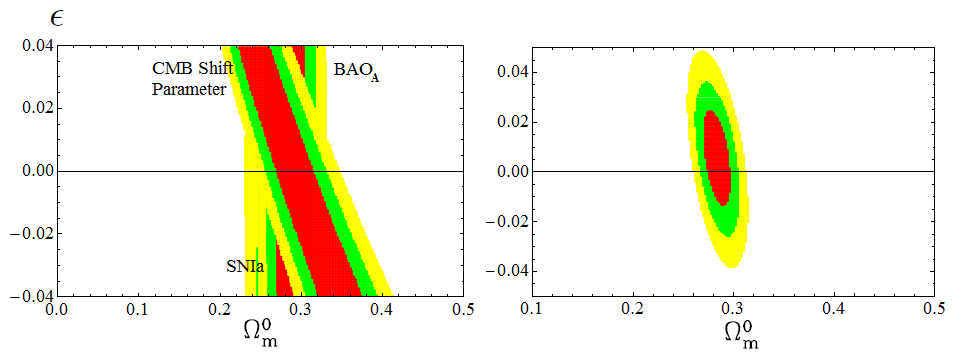}
\caption[Contour lines for model B1. Fitting analysis with BAO$_{A}$ of Sect. \ref{chap:Atype}.]{\label{fig:contoursB1 BAOA}%
\scriptsize {Likelihood contours for the B1
vacuum model, in this case with BAO$_A$ data.
The meaning of the shaded regions in the two panels is as in the previous figures.}}
\end{figure}

In the following, we describe the observational data samples and the statistical method that will be
adopted to constrain the parameters of the dynamical vacuum models
presented in the previous sections. We extract our fit from the
combined data on type Ia supernovae (SNIa), the shift parameter of
the Cosmic Microwave Background (CMB), and the data on the Baryon
Acoustic Oscillations (BAO's). The basic fitting results to the dynamical models under consideration are presented in a nutshell in Figs.\,\ref{fig:contoursA1 BAOdz}-\ref{fig:contoursB1 BAOA} and Tables \ref{tableFitBAOdz} and \ref{tableFitBAOA}. We devote the rest of this section to explain these results and also to analyze the implications for linear structure formation.

%%%%%%%%%%%%%%%%%%%%%%%%%%%%%%%%%%%%%%%%%%%%%%%%%%%%%%%%%%%%%%%%%%%%%%%%%%%%%%%

\begin{table}[t]
\centering
    \begin{scriptsize}
\begin{tabular}{| c | c | cc | ccc |}
\multicolumn{1}{c}{} & \multicolumn{1}{c}{6dF} & \multicolumn{2}{c}{SDSS}
 & \multicolumn{3}{c}{WiggleZ}  \\\hline
$z$ & 0.106 & 0.2 & 0.35 & 0.44 & 0.6 & 0.73  \\\hline
$d_z(\pm \sigma_{z})$ & $0.336(\pm 0.015)$ & $0.1905(\pm 0.0061)$ & $0.1097(\pm 0.0036)$ & $0.0916(\pm 0.0071)$ &
$0.0726(\pm 0.0034)$ & $0.0592(\pm 0.0032)$  \\\hline
$A(\pm \sigma_{A})$ & $0.526(\pm 0.028)$ & $0.488(\pm 0.016)$ & $0.484(\pm 0.016)$ & $0.472(\pm 0.034)$ &
$0.442(\pm 0.020)$ & $0.424(\pm 0.021)$  \\\hline
 \end{tabular}
 \end{scriptsize}
		\caption[BAO data used in the fitting analysis of Sect. \ref{chap:Atype}.]{{\scriptsize The BAO data points used in the current analysis, which correspond to those collected in Table 3 of \cite{Blake11}. Apart from the WiggleZ data points at $z_i=0.44,\,0.6$ and $0.73$ originally presented in \cite{Blake11}, we have also made use of the data at $z=0.106$ from the 6dGF galaxy survey \cite{Beutler2011} and the ones at $z_i=0.2$ and $0.35$ from the SDSS-DR7 one \cite{Percival2010}. The second and third rows provide, respectively, the BAO$_{dz}$ and BAO$_{A}$ measurements. See the text for details.}}
		\label{table1}
\end{table}

%%%%%%%%%%%%%%%%%%%%%%%%%%%%%%%%%%%%%%%%%%%%%%%%%%%%%%%%%%%%%%%%%%%%%%%%%%%%%%%

\subsection{The global fit to SNIa, CMB and BAO's}

First of all, we use the {\em Union 2.1} set of 580 type Ia supernovae
of Suzuki et al.~\cite{Suzuki:2011hu}.\footnote{Note that the data can be found in: \href{http://supernova.lbl.gov/Union/}{http://supernova.lbl.gov/Union/}.}.
The corresponding
$\chi^{2}_{\rm SNIa}$ function, to be minimized, is:
\begin{equation}\label{eq:xi2SNIa}
\chi^{2}_{\rm SNIa}({\bf p})=\sum_{i=1}^{580} \left[ \frac{ {\cal
\mu}_{\rm th} (z_{i},{\bf p})-{\cal \mu}_{\rm obs}(z_{i}) }
{\sigma_{i}} \right]^{2} \;,
\end{equation}
where $z_{i}$ is the observed redshift for each data point. The
fitted quantity ${\cal \mu}$ is the distance modulus, defined as
${\cal \mu}=m-M=5\log{d_{L}}+25$, in which $d_{L}(z,{\bf p})$ is the
luminosity distance:
\begin{equation}\label{eq:LumDist}
d_{L}(z,{\bf p})={c}{(1+z)} \int_{0}^{z} \frac{{\rm d}z'}{H(z')}\;,
%d_{L}(z,{\bf p})=\frac{c}(1+z) \int_{0}^{z} \frac{{\rm
%d}x}{H(x)} \;,
\end{equation}
with $c$ the speed of light (now included explicitly in some of these formulae for better clarity) and ${\bf p}$ a vector containing the
cosmological parameters of the models that we wish to fit for. In
equation (\ref{eq:xi2SNIa}), the theoretically calculated distance
modulus $\mu_{\rm th}$ for each point follows from using
(\ref{eq:LumDist}), in which the Hubble function is the one
corresponding to each model, see Sect.\,\ref{sec:solving}. Finally,
$\mu_{\rm obs}(z_{i})$ and $\sigma_i$ stand for the measured
distance modulus and the corresponding $1\sigma$ uncertainty for
each SNIa data point, respectively. The previous formula
(\ref{eq:LumDist}) for the luminosity distance applies only for
spatially flat Universes, which we are assuming throughout.

Furthermore, a very accurate and deep geometrical probe of dark
energy is the angular scale of the sound horizon at the last
scattering surface (i.e. at the time of decoupling of radiation from
matter). The probe is described by the  CMB ``shift
parameter''~\cite{BookAmendolaTsujikawa} and is encoded in the
location $l_1^{TT}$ of the first peak of the Cosmic Microwave
Background (CMB) temperature perturbation spectrum. It provides the reduced distance to the last scattering surface. For spatially
flat cosmologies it is given by
\begin{equation}\label{eq:shiftparameter}
R=\sqrt{\Omega_{m}^0}\int_{0}^{z_{*}} \frac{dz}{E(z)}\,.
\end{equation}
The measured shift parameter according to the Planck
data~\cite{Planck2013} is $R=1.7499\pm 0.0088$ at the redshift of decoupling (viz. at the last
scattering surface), $z_{*}$. Its precise value is given by the fitting formulas \ref{zlastscatt} and \ref{zlastscatt2}. In this case, the $\chi^{2}_{\rm CMB}$ function reads:
\begin{equation}
\chi^{2}_{\rm CMB}({\bf p})=\frac{[R({\bf
p})-1.7499]^{2}}{0.0088^{2}}\;.
\end{equation}
As emphasized in the previous section, when dealing with the CMB
shift parameter we have to include both the matter and radiation
terms in the total normalized matter density entering the $E(z)$
function in (\ref{eq:shiftparameter}) since the total radiation
contribution at the last scattering amounts to some $\sim 23\%$ of
the total energy density associated to matter and is therefore not entirely negligible. It means that when we
compute the CMB shift parameter we have to use the modified formulas
for the Hubble function that we  have found in Sections
\ref{subsec:radiationA} and \ref{subsec:radiationB}
for type A and B
models, respectively.

%%%%%%%%%%%%%%%%%%%%%%%%%%%%%%%%%%%%%%%%%%%%%%%%%%%%%%%%%%%%%%%%%

%

%%%%%%%%%%%%%%%%%%%%%%%%%%%%%%%%%%%%%%%%%%%%%%%%%%%%%%%%%%%%%%%%%%%%%%%%%%%%%

Finally, we also consider the BAO scale produced by the competition between the pressure of the
coupled baryon-photon fluid and gravity in the pre-recombination epoch. The resulting acoustic
waves leave (in the course of the evolution) an overdensity
signature at certain length scales of the matter distribution.
They appear as regular, periodic fluctuations of visible matter density in large-scale structure (LSS) resulting from sound waves propagating
in the early Universe.
Evidence of this excess has been found in the clustering properties
of luminous galaxies\footnote{See the paper reporting the first BAO detection in \cite{Eisenstein2005} by the SDSS galaxy survey.} and, in recent years, BAO has proven useful as a ``standard ruler'' that we can employ to constrain dark energy models.

%%%%%%%%%%%%%%%%%%%%%%%%%%%%%%%%%%%%%%%%%%%%%%%%%%%%%%%%%%%%%%%%%%%%%%%%%%%%%%%%

%%%%%%%%%%%%%%%%%%%%%%%%%%%%%%%%%%%%%%%%%%%%%%%%%%%%%%%%%%%%%%%%%%%%%%%%%%%%%

In this work we use the results collected in Table 3 of \cite{Blake11}
which are given in terms of the
parameter $d_z(z_{i})=r_{s}(z_{d})/D_{\rm V}(z_{i})$ (the sample contains 6
entries, see Table \ref{table1}), where
$D_{V}(z_{i})$ is the effective distance measure
\cite{Eisenstein2005} and $z_{i}$ is a reference redshift for
observations. Moreover $r_{s}(z_d)$ is the comoving sound horizon
size at the baryon drag epoch\,\cite{EisensteinHu98} (i.e. the epoch at which
baryons are released from the Compton drag of photons), and
$z_{d}\sim {\cal O}(10^{3})$ is the corresponding redshift of that
epoch, closely related to that of last scattering-- the precise
expression is given below, see Eq.\,(\ref{eq:zdrag}). Since $r_{s}(z_d)$ is the comoving distance that light can
travel prior to redshift $z_d$, it can be computed as follows:
\begin{equation}
\label{drag}
r_{s}(z_{d})=\int_{0}^{t(z_d)}\,\frac{c_s\,dt}{a}=\int_{0}^{a_{d}}
\frac{c_s(a)\,da}{a^{2} H(a)}=\int_{z_d}^{\infty}
\frac{c_s(z)\,dz}{H(z)}\;,
\end{equation}
where $a_{d}=(1+z_{d})^{-1}$, and
\begin{equation}\label{eq:cs2}
c_s(a)=c\,\left(\frac{\delta {p}_{\gamma}}{\delta
{\rho}_{\gamma}+\delta {\rho}_b}\right)^{1/2}=
\frac{c}{\sqrt{3\,\left(1+{\cal R}(a)\right)}}
\end{equation}
is the sound speed in the baryon-photon plasma. Here we assume
adiabatic perturbations and we have used $\delta {p}_b=0$ and
$\delta {p}_{\gamma}=(1/3)\,\delta{\rho}_{\gamma}$, and defined
${\cal R}(a)=\delta{\rho}_b/\delta{\rho}_{\gamma}$. If the scaling
laws for non-relativistic matter and radiation were those of the
standard model, we would have ${\cal
R}(a)=3\rho_b/4\,\rho_{\gamma}$, which can be finally cast as ${\cal
R}^{\CC
CDM}(a)=\left({3\Omega_{b}^{0}}/{4\Omega_{\gamma}^0}\right)\,a$,
where $\Omega_{b}^{0}h^{2}\simeq 0.02205$ and
$\Omega_{\gamma}^0\,h^2\simeq 2.46\times 10^{-5}$ are the current
values of the normalized baryon and photon
densities\footnote{We use
$\Omega_{r}^{0}=4.153\times 10^{-5}h^{-2}$,
$\Omega_{\gamma}^{0}=\frac{\Omega_{r}^{0}}{1+0.2271N_{\nu}}$
(with $N_{\nu}\simeq 3.04$ and $h=0.673$ \cite{Planck2013}).}.
However, when we consider cosmologies beyond the $\CC$CDM a modification of these
formula for ${\cal R}(a)$ has to be implemented.  We explain the
details in the next section.

%%%%%%%%%%%%%%%%%%%%%%%%%%%%%%%%%%%%%%%%%%%%%%%%%%%%%%%%%%%%%%%%%%%%%%%%%%%%%

%%%%%%%%%%%%%%%%%%%%%%%%%%%%%%%%%%%%%%%%%%%%%%%%%%%%%%%%%%%%%%%%%%%%%%%%%%%%%

The remaining ingredients of the BAO analysis are presented now. In particular the effective distance is (see \cite{Eisenstein2005}):
\begin{equation}
D_{\rm V}(z)\equiv \left[ (1+z)^{2} D_{A}^{2}(z)
\frac{cz}{H(z)}\right]^{1/3}\;,
\end{equation}
where $D_{A}(z)=(1+z)^{-2} d_{L}(z,{\bf p}) $ is the angular
diameter distance. It follows from the foregoing that the $d_z$ estimator for BAO analysis explicitly reads:
\begin{equation}\label{eq:fromula dz}
d_z(z_i)=\frac{r_s(z_d)}{\displaystyle{\left[\left(\int_0^{z_i}\frac{cd{z}}{H({z})}\right)^2\,\left(\frac{cz_i}{H(z_i)}\right)\right]^{1/3}}}\,.
\end{equation}
The fitted formula for the baryon drag redshift, $z_d$, is
given by\,\cite{EisensteinHu98}:
\begin{equation}\label{eq:zdrag}
z_d=1291\,\frac{(\Omo\,h^2)^{0.251}}{1+0.659(\Omo\,h^2)^{0.828}}[1+\beta_1(\Omega_b^{0}\,h^2)^{\beta_2}]\,,
\end{equation}
with
\begin{equation}
\beta_1=0.313(\Omo\,h^2)^{-0.419}[1+0.607(\Omo\,h^2)^{0.674}]\,,\ \
\qquad \beta_2=0.238(\Omo\,h^2)^{0.223}\,.
\end{equation}
As we see the drag epoch ends at a redshift which is somewhat more
strongly dependent on the parameters than the decoupling
redshift $z_{*}$. Numerically, $z_d$  is not very different from
$z_{*}$, both being of order $10^3$ (with $z_{*}>z_d$). Typically $z_{*}\simeq 1090$ and $z_d\simeq 1060$ for the Planck results\,\cite{Planck2013}.

At this point, we would like to stress that
we have also made use of
BAO measurements in terms of the acoustic parameter $A(z)$ \ref{table1},
first introduced by Eisenstein et al. \cite{Eisenstein2005}. Acoustic oscillations in the photon-baryon plasma prior to recombination give rise to a peak in the correlation function of galaxies, whose value is given by the mentioned $A(z)$-estimator for BAO analysis:
\begin{equation}\label{eq:defBAOA}
A({z_i,\bf p})=\frac{\sqrt{\Omo}}{[z_i^{2} E(a_{i})]^{1/3}}
\left[\int_{a_{i}}^{1} \frac{da}{a^{2}E(a)} \right]^{2/3}=\frac{\sqrt{\Omo}}{E^{1/3}(z_{i})}
\left[\frac{1}{z_i}\int_{0}^{z_i} \frac{dz}{E(z)} \right]^{2/3}\,,
\end{equation}
with $a_{i}=(1+z_{i})^{-1}$, and $z_i$ is the redshift at which the acoustic scale has been measured.

According to \cite{Blake11} the $A(z)$ measurements are
approximately uncorrelated with respect to $\Omega_{m}^{0}h^{2}$, while
this is not the case for the $d_{z}$ measurements.
Therefore, it is natural to use both BAO estimators, $d_{z}$ and $A(z)$,
in our statistical analysis in order to check the
range of validity of the free parameters included in the various
vacuum models. Specifically, here and henceforth
we consider the following
two notations: BAO$_{dz}$ for the $d_{z}$ measurements, and
BAO$_A$ for those based on the $A(z)$ estimator.
Therefore, the corresponding $\chi^{2}$-functions for BAO analysis are defined as:
\begin{equation}\label{eq:BAOd}
\chi^{2}_{\rm BAO_{dz}}({\bf p})=\sum_{i=1}^{6} \left[ \frac{ d_{z,\rm th} (z_{i},{\bf p})-d_{z,\rm obs}(z_{i})}
{\sigma_{z,i}} \right]^{2}
\end{equation}
and
\begin{equation}\label{eq:BAOA}
\chi^{2}_{\rm BAO_{A}}({\bf p})=\sum_{i=1}^{6} \left[ \frac{ A_{\rm th} (z_{i},{\bf p})-A_{\rm obs}(z_{i})}
{\sigma_{A,i}} \right]^{2} \;,
\end{equation}
where $z_{i}$, $d_{z,\rm obs}$, $\sigma_{z,i}$, $A_{\rm obs}$ and $\sigma_{A,i}$
can be found in Table \ref{table1}.

%%%%%%%%%%%%%%%%%%%%%%%%%%%%%%%%%%%%%%%%%%%%%%%%%%%%%%%%%%%%%%%%%%%%%%%%%%%%%%

%%%%%%%%%%%%%%%%%%%%%%%%%%%%%%%%%%%%%%%%%%%%%%%%%%%%%%%%%%%%%%%%%%%%%%
%%%%%%%%%%%%%%%%%%%%%%%%%%%%%%%%%%%%%%%%%%%%%%%%%%%%%%%%%%%%%%%%%%%%%%

\subsection{Adapting the BAO analysis for dynamical vacuum models}
\label{subsec:adaptBAO}

In this section we describe the necessary modifications to ${\cal
R}(a)$ for the BAO$_{dz}$  analysis when the cosmological vacuum is dynamical. The modifications are
necessary since the scaling laws for non-relativistic matter and
radiation are slightly different as compared to the $\CC$CDM. For
example, for type-A models the generalization is simple \cite{BasPolarSola12}. From the anomalous scaling laws
(\ref{splitAsolutionM}) and (\ref{splitAsolutionR}) we easily find
the following modification of the ${\cal R}^{\CC CDM}(a)$
function:
\begin{equation}\label{eq:Ra2}
{\cal R}^{({\rm type
A})}(a)=\frac34\,\frac{\xi}{\xiR}\,\frac{{\rho}_b(a)}{\,{\rho}_{\gamma}(a)}
=\frac34\,\frac{1-4\alpha/3}{1-\alpha}\frac{{\Omega}_{b}^{0}}{{\Omega}_{\gamma}^0}\,
a^{4\xiR-3\xi}\,.
\end{equation}
Of course for $\nu=0$ and $\alpha=0$ ($\xi=\xiR=1$) we recover ${\cal R}^{\CC CDM}(a)$ as given in the previous
section.

For type-B models the corresponding ${\cal R}(a)$ also
differs from the standard one, but is more complicated. Fortunately, from the discussion in
Sect.\,\ref{subsec:radiationB} the ground is prepared to derive the
result straightforwardly. For this model we have to use the cosmic
time rather than the scale factor, i.e.
$R(t)={\delta\rho_b(t)}/{\delta\rho_{\gamma}(t)}$. Thus, we differentiate
(\ref{eq:rhomNR2}) and (\ref{eq:rhorNR2}) with respect to $t$:
\begin{equation}
\frac{\delta\rho_b}{\delta
t}=-\frac{3}{2}\rho_b^{0}B\,e^{-\frac{3}{2}\epsilon
H_0t_0}H_0\mathcal{F}\frac{\cosh\left(\frac{3}{4}H_0\mathcal{F}t\right)}{\sinh^3\left(\frac{3}{4}H_0\mathcal{F}t\right)}
\end{equation}
and
\begin{equation}
\frac{\delta\rho_{\gamma}}{\delta t}=-2\rho_{\gamma}^{0}B^{4/3}e^{-2\epsilon
H_0t_0}H_0\mathcal{F}\frac{\cosh\left(\frac{3}{4}H_0\mathcal{F}t\right)}{\sinh^{11/3}\left(\frac{3}{4}H_0\mathcal{F}t\right)}\,.
\end{equation}
Using now the constraint (\ref{eq:condition}), the final result
simplifies considerably:
\begin{equation}
{\cal R}^{({\rm type
B})}(t)=\frac{3\Omega_b^{0}}{4\Omega_{\gamma}^{0}}\left[\frac{\sinh\left(\frac{3}{4}\,H_0\,\mathcal{F}\,t\right)}{\sinh\left(\frac{3}{4}\,H_0\,\mathcal{F}\,t_0\right)}\right]^{2/3}.
\end{equation}
One can easily check that for $\nu=\epsilon=0$, we have
$\mathcal{F}\to 2\sqrt{\OLo}$ and we recover the corresponding
$\CC$CDM result:
\begin{equation}
\left.{\cal R}^{({\rm type B})}(t)\right|_{\epsilon=\nu=0}
=\frac{3\Omega_b^{0}}{4\Omega_{\gamma}^{0}}\left[\frac{\sinh\left(\frac{3}{2}\,H_0\,\OLo\,t\right)}{\sinh\left(\frac{3}{4}\,H_0\,\OLo\,t_0\right)}\right]^{2/3}=\frac{3\Omega_b^{0}}{4\Omega_{\gamma}^{0}}\,a(t)={\cal
R}^{\CC CDM}(a)\,.
\end{equation}
Despite the simplification, the solution is given here in terms of
the cosmic time and therefore we are forced to integrate numerically
the cosmological equations to relate the cosmic time and the scale
factor, following the usual procedure for this kind of models. Let us finally mention the following specification for fitting the BAO$_{dz}$ observable (\ref{eq:BAOd}) for type-B models. In this case the computation of the parameter $r_{s}(z_{d})$ -- the comoving distance traveled by light to the drag epoch, Eq.\, (\ref{drag}) --
is performed in two steps: in the first step we integrate from $a=0$ up to the decoupling (or last scattering) point $a_{*}$ by neglecting the vacuum energy corrections since dark energy effects are negligible for $a<a_{*}$; in the second step we integrate from $a=a_{*}$ to the drag epoch $a=a_{d}$ using the correction from the radiation component discussed above. This was unnecessary for the type-A models as in this case the solution is given directly in terms of the scale factor and these features are taken automatically into account. But for type B scenarios, where the cosmic time variable is the starting point before mapping it to the scale factor, the numerical analysis is more complicated in the region $t\simeq 0$ and the previous prescription is convenient. For type-C1 models we do not make use of BAO$_{dz}$ nor CMB data due to the fact that the usual fitting formulas for the redshifts at decoupling and baryon drag epochs presented before are not good approximations for those models with no constant term $C_0$ in the expression of $\rho_\Lambda$. They are tailor-made for the $\Lambda$CDM model and, in general, for $\Lambda$CDM-like models. For this reason we have used only the BAO$_A$ data for them (based on the aforementioned acoustic parameter $A(z)$ whose computation does not
involve any integration in the very high redshift range), and of course the SNIa data.

%%%%%%%%%%%%%%%%%%%%%%%%%%%%%%%%%%%%%%%%%%%%%%%%%%
%%%%%%%%%%%%%%%%%%%%%%%%%%%%%%%%%%%%%%%%%%%%%%%%%%

\subsection{Combined likelihood function}
\label{subsec:combined likelihood}

In order to place tighter constraints on the corresponding parameter
space of our model, the probes described above must be combined
through a joint likelihood analysis\footnote{Likelihoods are
normalized to their maximum values. In the present analysis we
always report $1\sigma$ uncertainties on the fitted parameters. Note
also that if the total number of data points is
$N$ the associated degrees of freedom is:
$dof = N-n_p$, where $n_p$ is
the model-dependent number of fitted parameters.}, given by the
product of the individual likelihoods according to:
\begin{equation}\label{eq:overalllikelihood} {\cal L}_{\rm tot}({\bf p})=
{\cal L}_{\rm SNIa} \times {\cal L}_{\rm
CMB}\times {\cal L}_{\rm BAO_{X}}\,. \end{equation}
This translates into an addition of the joint
$\chi^2$ function:
\begin{equation}\label{eq:overalllikelihoo}
\chi^{2}_{\rm tot}({\bf p})=\chi^{2}_{\rm SNIa}+\chi^{2}_{\rm CMB}+\chi^{2}_{\rm
BAO_{X}}\;,
\end{equation}
where $X$ denotes the kind of BAO measurements used in the statistical
analysis, namely $X=d_z$ or $X=A(z)$.
In our $\chi^2$ gridded minimization procedure, for the vacuum models
(running and concordance $\Lambda$CDM) we use the following range
and steps of the fitted parameters: $\Omega_{m}^{0} \in [0.1,1]$ in
steps of 0.001 and $\nu \in [-0.02,0.02]$  in steps of
$10^{-4}$.

%%%%%%%%%%%%%%%%%%%%%%%%%%%%%%%%%%%%%%%%%%%%%%%%%%%%%%%%%%%%

\begin{table}
\begin{center}
\vspace{0.5cm}
%\begin{scriptsize}
\begin{tabular}{| c |  c |c | c | c | c |}
\multicolumn{1}{c}{Model} & \multicolumn{1}{c}{$\Omo$} & \multicolumn{1}{c}{$\nu$} & \multicolumn{1}{c}{$\epsilon$} & \multicolumn{1}{c}{$\chi^2/dof$} &\multicolumn{1}{c}{AIC} \\\hline
$\CC$CDM & $0.293\pm 0.013$ & - & - & $567.8/586$ & $569.8$\\\hline
$A1$ & $0.292\pm 0.014$ & $+0.0013\pm 0.0018 $ & - & $566.3/585$ & $570.3$ \\\hline
$A2$ & $0.290 \pm 0.014$ & $+0.0024\pm  0.0024 $ & - & $565.6/585$ & $569.6$ \\\hline
$B1$ & $ 0.297^{+0.015}_{-0.014}$ & -  & $-0.014^{+0.016}_{-0.013}$ & $587.2/585$ & $591.2$ \\\hline
$B2$ & $ 0.300^{+0.017}_{-0.003}$ &$ - 0.0039^{+0.0020}_{-0.0021}$  & $- 0.0039^{+0.0020}_{-0.0021}$ & $583.1/585$ & $587.1$ \\\hline
 \end{tabular}
% \end{scriptsize}
\caption[Fitting results for the various models of Sect. \ref{chap:Atype} using SNIa+CMB+BAO$_{dz}$ data.]{{\scriptsize The fit values for the various models using SNIa+CMB+BAO$_{dz}$ data, together with their statistical significance according to $\chi^2$ and AIC statistical tests. Notice that for type-A2 models the quoted value of $\nu$ stands actually for  $\nueff$ under the conditions explained in the text, and for B2 we have set $\nu=\epsilon$ (see the text as well).} \label{tableFitBAOdz}}
\end{center}
\end{table}

%%%%%%%%%%%%%%%%%%%%%%%%%%%%%%%%%%%%%%%%%%%%%%%%%%%%%%%%%%%%

Since the current vacuum models contain different number of free
parameters, as a further statistical test we use the
({\em corrected}) Akaike Information Criterion (AIC)
relevant to our case ($N/n_p>40$)\,\cite{Akaike1974,Sugiura1978}, which is defined,
for the case of Gaussian errors, as follows:
\begin{equation}\label{eq:AkaikeShort}
{\rm AIC}=\chi^2_{\rm tot}+2n_{p}\,,
\end{equation}
where $n_{fit}$ is the number of free parameters. It is well known that
a smaller value of AIC points to a better model-data fit. In this context,
we have to mention that small
differences in AIC are not necessarily significant and therefore, in order
to test the effectiveness of the models themselves, it is
important to calculate the model pair difference
$\Delta$AIC$ = {\rm AIC}_{y} - {\rm AIC}_{x}$. The larger the
value of $|\Delta{\rm AIC}|$, the higher the evidence against the
model with larger value of ${\rm AIC}$, with a
difference $|\Delta$AIC$| \ge 2$ indicating a positive
such evidence and $|\Delta$AIC$| \ge 6$
indicating a strong such evidence, while a value $\le 2$ indicates
consistency among the two comparison models.

Let us next present the basic results of the overall statistical analysis.

\begin{enumerate}
\item {\it We first consider SNIa+CMB+BAO$_{dz}$.}

\begin{itemize}

\item In the case of the concordance $\Lambda$CDM cosmology
we find $\Omega_{m}^{0}=0.293\pm 0.013$ with statistic significance $\chi_{\rm
tot}^{2}(\Omega_{m}^{0})/dof \simeq 567.8/586$
(AIC$_{\Lambda}$$\simeq 569.8$). For comparison, the determination by Planck+WP\,\cite{Planck2013} reads: $\Omo\,h^2=0.1426\pm0.0025$. With $h=0.673\pm 0.012$ (also from the same source) this yields $\Omo\simeq0.315\pm 0.016$, which agrees with our central value within slightly more than $1\sigma$. We shall comment further on the possible implications of the $\Omo$ determination in the next section.
Another important cosmological parameter that will enter later
our analysis is the rms mass fluctuation
on $R_{8}=8 h^{-1}$ Mpc scales at $z=0$. In this work we use as a prior for such parameter the value $\sigma_{8,\Lambda}=0.829$ in the $\CC$CDM, i.e. the Planck+WP result\,\cite{Planck2013}.
Then, with the aid of this value,
we can calculate the corresponding $\sigma_{8}$ values for the vacuum models
through Eq. (\ref{s88general}) below [see discussion in section \ref{sec:PSformalism} and Tables \ref{TableModelsBAOdz} and \ref{TableModelsBAOA}].

\item In Fig.\,\ref{fig:contoursA1 BAOdz} we present the results of our analysis for the type-A1 running
vacuum model, which is characterized by the $\nu$-parameter. We have sampled this parameter in the interval $[-0.02,0.02]$  in steps of
$10^{-4}$.  The left panel in that figure shows the fitted regions at 1$\sigma$,
2$\sigma$ and $3\sigma$ confidence levels in the
$(\Omega_{m}^{0},\nu)$ plane, from the  SNIa,  BAO$_{dz}$ and CMB shift parameter data. The right panel in the figure shows  the fit contours when the three types of data are intersected. Using the SNIa data alone it is evident (from the left panel)  that although the $\Omega_{m}^{0}$ parameter is tightly  constrained ($\sim 0.29$), the $\nu$ parameter remains completely unconstrained (in the shown interval). However, as it is manifest from the  right panel of that figure, the above degeneracy is broken when we use the joint likelihood analysis with all the
cosmological data. Indeed the overall likelihood function peaks at
$\Omega_{m}^{0}=0.292\pm 0.014$ and
$\nu=+0.0013\pm 0.0018$ with $\chi_{\rm
tot}^{2}(\Omega_{m}^{0},\nu) \simeq 566.3$ (AIC$_{A1}$$\simeq 570.3$)
for $585$ degrees
of freedom\footnote{Note that in \cite{Grande2011} the authors used the earlier
BAO results of Percival \cite{Percival2010} and the {\em Constitution} set
of 397 SNIa \cite{Hic09}. We would like to mention here that those
results are in agreement with the current results within $1\sigma$
uncertainties.}.
The $\CC$CDM value of AIC ($\simeq 569.8$) is smaller
with respect to that of the A1 model. This indicates that the
fit of the concordance model to the combined data is slightly better than that of the A1 vacuum model.
However, the differential AIC value $|\Delta {\rm AIC}|$=$|{\rm AIC}_{\Lambda}-{\rm AIC}_{A1}|=1.5$ is actually $\le 2$. This tells us that
the cosmological data are perfectly consistent with the A1 model in a way comparable to the $\CC$CDM.
Furthermore, we find that $\sigma_{8}=0.813$.

\item Let us now address the more general case of the type-A2 model,whose statistical vector ${\bf p}$ contains 3 free parameters, namely
${\bf p}=(\Omega_{m}^{0},\nu,\alpha)$.
Our minimization analysis provides strongly degenerate results between
$\nu$  and  $\alpha$, rendering impossible to put any
significant constraints on their values\footnote{This is similar to the situation with the CPL parameterization\,\cite{CPL1,CPL2} of the dark energy, where in the statistical vector ${\bf p}=(\Omega_{m}^{0},w_0, w_1)$ one actually has to fix a parameter to fit the other two in an efficient way.}.  Note, however that
the value of $\Omega_{m}^{0}$ is well constrained ($\simeq 0.29$).
Therefore we adopt -- as in Ref.\,\cite{BasPolarSola12}-- the
additional setting $\xiR \equiv 1$, which is tantamount to assume that there is no modification in the scaling law of radiation and hence radiation behaves in the strict standard way, in contraposition to dust. This occurs  when
$\alpha= \frac{3}{4} \nu$ (see equation \ref{eq:defxiR}), implying
from Eqs. (\ref{eq:smalllimit}) and (\ref{nueff})
that $\xi\simeq 1-\nueff$, in which $\nueff\simeq (\nu-\alpha)=\nu/4$.
The statistical vector
reduces in this case to ${\bf p}=(\Omega_{m}^{0},\nueff)$, and we sample $\nueff \in [-0.02,0.02]$  in steps of
$10^{-4}$.
The joint minimization provides now $\Omega_{m}^{0}=0.29\pm 0.014$,
$\nueff=0.0024 \pm 0.0024$ ($\nu\simeq 0.0096$ and $\alpha \simeq 0.0072$)
with $\chi_{\rm tot}^{2}(\Omega_{m}^{0},\nueff)/dof \simeq 565.6/585$
and AIC$_{A2}$$\simeq 569.6$. Notice that in the present case, opposite to the A1 model, the fit to the combined data from the A2 model is slightly better than in the $\CC$CDM model (cf. Table \ref{tableFitBAOdz}). However, utilizing the AIC information criterion, and because
AIC$_{\Lambda}\simeq$AIC$_{A2}$ we have that
the A2 vacuum model is statistically equivalent ($|\Delta {\rm AIC}|\leq 2$) with the $\Lambda$CDM model.
In this case we obtain $\sigma_{8}=0.797$.

%%%%%%%%%%%%%%%%%%%%%%%%%%%%%%%%%%%%%%%%%%%%%%%%%%%%%%%%%%%%

\begin{table}
\begin{center}
\vspace{0.5cm}
%\begin{scriptsize}
\begin{tabular}{| c |  c |c | c | c | c |}
\multicolumn{1}{c}{Model} & \multicolumn{1}{c}{$\Omo$} & \multicolumn{1}{c}{$\nu$} & \multicolumn{1}{c}{$\epsilon$} & \multicolumn{1}{c}{$\chi^2/dof$} &\multicolumn{1}{c}{AIC} \\\hline
$\CC$CDM & $0.292\pm 0.011$ & - & - & $567.5/586$ & $569.5$\\\hline
$A1$ & $0.282\pm 0.012$ & $+0.0048^{+0.0032}_{-0.0031} $ & - & $563.8./585$ & $567.8$ \\\hline
$A2$ & $0.283\pm 0.012$ & $+0.0048\pm 0.0031$ & - & $563.8/585$ & $567.8$ \\\hline
$B1$ & $ 0.283^{+0.012}_{-0.011}$ & -  & $+0.005^{+0.018}_{-0.015}$ & $563.7/585$ & $567.7$ \\\hline
$B2$ & $ 0.283^{+0.011}_{-0.012}$ & $+0.0015^{+0.0025}_{-0.0030}$ & $ +0.0015^{+0.0025}_{-0.0030}$ & $563.8/585$ & $567.8$ \\\hline
$C1$ & $ 0.296\pm 0.017$ & $-0.189\pm 0.008$ & - & $568.3/585$ & $570.3$ \\\hline
 \end{tabular}
% \end{scriptsize}
\caption[Fitting results for the various models of Sect. \ref{chap:Atype} using SNIa+CMB+BAO$_{A}$ data.]{{\scriptsize The fit values for the various models using the same data and statistical tests as before except for the BAO observable, which now is BAO$_{A}$, i.e., overall we use SNIa+CMB+BAO$_{A}$ data. In this case we include also the C1 model (see text). Same notation as the previous table.}\label{tableFitBAOA}}
\end{center}

\end{table}

%%%%%%%%%%%%%%%%%%%%%%%%%%%%%%%%%%%%%%%%%%%%%%%%%%%%%%%%%%%%

\item   We next face model B1 with BAO$_{dz}$ data. As in the A1 model, here we have only one characteristic parameter, in this case $\epsilon$  (apart from the generic one $\Omo$). We have dealt with the fitting procedure of $\epsilon$ in a similar way as $\nu$ for the A1 model. The fitting regions in the $(\Omo,\epsilon)$-plane are shown in Fig.\,\ref{fig:contoursB1 BAOdz}, left panel, whereas the  1$\sigma$, 2$\sigma$ and $3\sigma$ contour lines for the combined SNIa+CMB+BAO$_{dz}$ data are displayed on the right panel of that figure. As we can see, the determination of $\Omo$ is rather sharp around $0.3$, to be precise: $\Omo= 0.297^{+0.015}_{-0.014}$. However the parameter $\epsilon$ is not so well bounded by the data as it was the case of $\nu$, specifically we find $\epsilon=-0.014^{+0.016}_{-0.013}$. The central value and the errors are roughly one order of magnitude bigger than before, and the fit quality is significantly poorer. This is clear from the $\chi^2/dof$ and AIC statistical diagnostics in Table \ref{tableFitBAOdz}, which give substantially larger values than those in the $\CC$CDM model. If we attend strictly the AIC statistical criterion we should conclude that the type-B1 model does not fit at all the combined data in a way comparable to type-A1 or A2 models and the $\CC$CDM. Concerning the rms mass fluctuation for this model, we find $\sigma_{8}=0.859$.

\item As for the B2 model, we have once more a situation with three fit parameters $(\Omo,\nu,\epsilon)$. But to avoid similar difficulties as those mentioned with the A2 model, we fix a correlation between the two model parameters $\nu$ and $\epsilon$. Of course, both are small in absolute value, but let us note that for $\nu\ll\epsilon$ model B2 must reduce to B1, whereas for $\epsilon\ll\nu$ it essentially behaves as A1. Therefore, the parameter region which is left unexplored is when $\epsilon\simeq\nu$, and for definiteness we will fix $\epsilon=\nu$ and shall perform a two-parameter fit in $(\epsilon,\Omo)$. Under these conditions we find  $\epsilon\simeq -0.0039$ and  $\Omo\simeq 0.30$ (cf. Table \ref{tableFitBAOdz} for more precise values and errors). As with the B1 case, the quality of the fit is worse than for the $\CC$CDM or any of the type-A models.  Moreover, for the B2 model, we find $\sigma_{8}=0.896$.

\end{itemize}

\item {\it In the following we are based on SNIa+CMB+BAO$_{A}$ data.}

 We describe now the situation concerning the various models for when we use the alternative BAO option. The corresponding results are clearly displayed in Table\,\ref{tableFitBAOA}. 

\begin{itemize}

\item For the $\Lambda$CDM model
we obtain $\Omega_{m}^{0}=0.292\pm 0.011$
with $\chi_{\rm tot}^{2}(\Omega_{m}^{0})/dof \simeq 567.5/586$
(AIC$_{\Lambda}$$\simeq 569.5$). As it could be expected, we obtain almost the same
results with those of the previous fit with SNIa+CMB+BAO$_{dz}$ data.

\item A1 vacuum model: Here the overall likelihood function peaks at a lower value of the mass parameter
$\Omega_{m}^{0}=0.282\pm 0.012$ (see the concluding comments of this section) and higher running parameter
$\nu=0.0048^{+0.0032}_{-0.0031}$, with $\chi_{\rm
tot}^{2}(\Omega_{m}^{0},\nu) \simeq 563.8$ (AIC$_{A1}$$\simeq 567.8$)
for $585$ degrees of freedom.
The shape of the BAO$_A$-contours in Fig.\,\ref{fig:contoursA1 BAOA}
are quite different from those of BAO$_{dz}$ (Fig.\,\ref{fig:contoursA1 BAOdz}). This is somehow related
with the fact that unlike for the case of $A(z)$, the
$d_{z}$ BAO measurements are correlated with the
$\Omega_{m}^{0}h^{2}$ \cite{Blake11} as well as with the
necessary modifications to ${\cal R}(a)$
introduced in the BAO$_{dz}$ analysis (see Sect. \ref{subsec:adaptBAO}).
The rms mass fluctuation is found to be $\sigma_{8}=0.758$. Let us note the remarkable fact that using the BAO$_A$ observable, instead of BAO$_{dz}$, the value of $\nu$ is not compatible with zero at $1\sigma$, showing a slight tendency to favor nonvanishing (positive) values of $\nu$ rather than the $\CC$CDM result.

\item A2 vacuum model:
the overall minimization provides $\Omega_{m}^{0}=0.283\pm 0.012$,
$\nueff=+0.0048 \pm 0.0031$ ($\nu\simeq 0.019$ and $\alpha \simeq 0.014$)
with $\chi_{\rm tot}^{2}(\Omega_{m}^{0},\nueff)/dof \simeq 563.8/585$
and AIC$_{A2}$$\simeq 567.8$.
The rms mass fluctuation is $\sigma_{8}=0.757$.

At this point we would like to make some
comments for the A1-A2 models.
Generally, the $\Omega_{m}^{0}$ values are in agreement (with 1$\sigma$ errors)
with those of SNIa+CMB+ BAO$_{dz}$. However, as far as $\nueff$
(or $\nu$) is concerned we find differences among the parameters which
could reach up to a factor of $\sim 3.7$.
In this context, the SNIa+CMB+BAO$_{A}$ data
analysis highlights the fact that
the values of AIC$_{A1-A2}$($\simeq 567.8$) are actually smaller
with respect to those of the concordance $\CC$CDM cosmology. In other words, it turns out that the
type-A1 and A2 vacuum models appear now to
fit slightly better than the $\Lambda$CDM the
observational data.
Still, the $|\Delta {\rm AIC}|$=$|{\rm AIC}_{A1-A2}-{\rm AIC}_{\Lambda}|$
values (ie., $\le 2$) indicate that
the cosmological data are simultaneously consistent with the A1, A2 and the
$\Lambda$CDM models.

\item B1 model with BAO$_A$ data. We find  $\Omo=0.283^{+0.012}_{-0.011}$, so it remains similar to the type-A models, with $\epsilon=+0.005^{+0.018}_{-0.015}$. Interestingly, the fit quality is also in this case slightly better than in the $\CC$CDM model, but still with $|\Delta {\rm AIC}|\leq 2$, and hence statistically comparable.
The rms mass fluctuation is $\sigma_{8}=0.820$.

\item For the B2 model we proceed here with a similar strategy as with the BAO$_{dz}$ case, and we find $\Omo=0.283$ and $\epsilon\simeq \nu\simeq +0.0015$. The quality of the fit is once more comparable, but better, than for the concordance $\CC$CDM model.
The rms mass fluctuation is $\sigma_{8}=0.791$.

\item Finally, the observational viability of the C1 model has been
tested previously in Basilakos and Sol{\`a}
\cite{BasSola14a}. We use their
SNIa+CMB+BAO$_{A}$ analysis of the
$(\Omega_{m}^{0},\zeta)$ pair in Table \ref{tableFitBAOA} and Fig. \ref{sigma8TypeA}.  Specifically,
we remind the reader that  the following results were found\, \cite{BasSola14a}:
$\Omega_{m}^{0}=0.296\pm 0.017$, $\zeta=1.189
\pm 0.008$ with $\chi_{\rm tot}^{2}(\Omo,\zeta)/dof\simeq 568.3/585$. Because of the absence of the constant additive term for the C1 model (cf. Sect. \ref{subsec:solvingC1C2}), the value of $\nu$ is forced to be much larger than in the other models.
Notwithstanding, the corresponding AIC value is $570.3$, which is statistically comparable to that of the $\CC$CDM model, and therefore at least from the point of view of the Hubble expansion data and the shift parameter, the C1 model seems to present a respectable status. But it is only an ostensible good status. The situation for this model will undergo a radical change when we test the structure formation data at low redshifts, as we shall see in Sect. \ref{subsec:growthrate}. The trouble is related to the aforementioned absence of the additive term. A first hint of decline of this model appears when we compute the corresponding rms mass fluctuation, namely $\sigma_{8}=1.365$, which is clearly anomalously large.

\end{itemize}
\end{enumerate}

%%%%%%%%%%%%%%%%%%%%%%%%%%%%%%%%%%%%%%%%%%%%%%%%%%%%%%
%%%%%%%%%%%%%%%%%%%%%%%%%%%%%%%%%%%%%%%%%%%%%%%%%%%%%%
%%%%%%%%%%%%%%%%%%%%%%%%%%%%%%%%%%%%%%%%%%%%%%%%%%%%%%

\subsection{Discussion of the fitting results and implications for dynamical DE}

From the previous analysis one could tentatively say that type-A models are preferred to type-B ones from the point of view of the quality fits to the combined data. This is indeed suggested by the results involving BAO$_{dz}$. In contrast,  BAO$_{A}$ data does not seem to point so strongly to this conclusion. This is somehow understandable if we take into account that the BAO$_{A}$ data are exclusively based on the imprints of baryonic acoustic oscillations left at low redshifts during the early epochs of galaxy clusters formation, which means at relatively recent times, whereas the BAO$_{dz}$ data is also sensitive to the model behavior of these oscillations at earlier epochs in between the decoupling and baryon drag epochs. More observational work will be necessary to decide about the best vacuum models. 

We come now to a point mentioned in passing in the previous section concerning the fitting values of $\Omo$. For the $\CC$CDM we have found $\Omo\simeq 0.293$ (virtually independent of the type of BAO data used). This is smaller than the Planck+WP value $\Omo\simeq0.315$. Similarly, for the vacuum models A and B we have found $\Omo$ smaller than the Planck+WP value. This holds not only for BAO$_{dz}$ data (cf. Table \ref{tableFitBAOdz}) but even more pronounced when the fit is done using  BAO$_{A}$ data, where the value of $\Omo$ lessens significantly for all vacuum models at around $\Omo\simeq0.282-0.283$ (cf. Table \ref{tableFitBAOA}). At the same time one obtains, in the last case, a slightly improved fit quality with respect to the $\CC$CDM for all the dynamical vacuum models A and B. The difference with respect to the Planck+WP value of $\Omo$ is now larger and, as we will see in Sect.\,\ref{sec:Number counts}, it does matter as far as the possible implications on the predicted cluster number counts for the dynamical models. At this stage of precision cosmology it is difficult to make a final selection between the two types of BAO  data, and therefore we have decided to present the results separately for each BAO set. 

%%%%%%%%%%%%%%%%%%%%%%%%%%%%%%%%%%%%%%%%%%%%%%%%%%%%%%%%%%%%%%%%%%%%%%%%%%%%%
\begin{figure}[t]
\begin{center}
\includegraphics[scale=0.46]{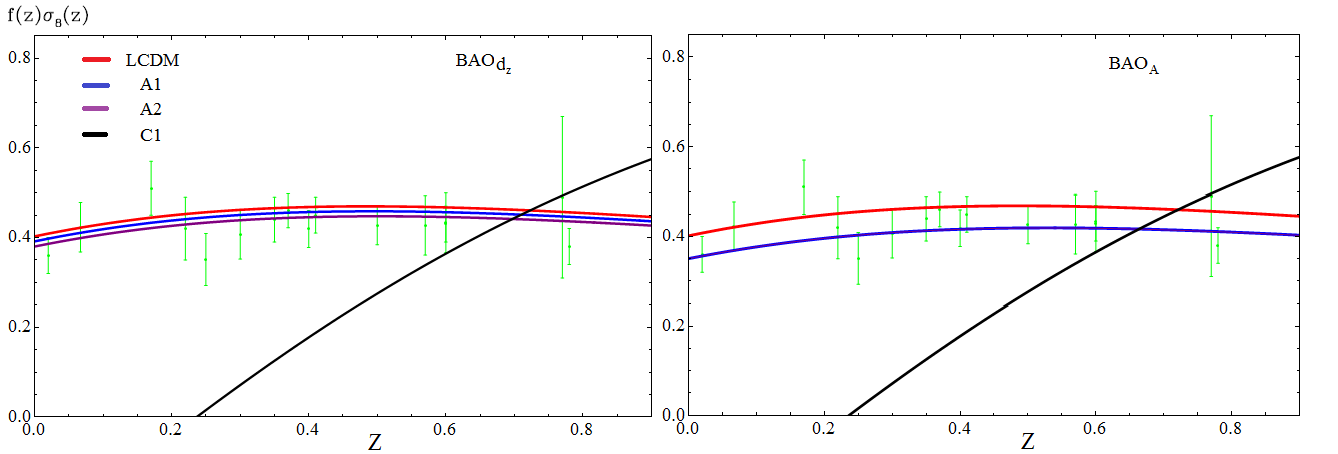}
\caption[$f(z)\sigma_8(z)$ for type-A and C1 models of Sect. \ref{chap:Atype}.]{\scriptsize{
Comparison of the observed (solid points with vertical error bars) and
theoretical evolution of the weighted growth
rate $f(z)\sigma_{8}(z)$ for the models A1, A2 and C1. The uppermost (red) line corresponds to the $\CC$CDM model, used as a reference. The subsequent ones (from top to bottom) correspond to the vacuum models A1 (blue line) and A2 (purple line). On the right panel, the lines for A1 and A2 are one essentially overlapping. The curve that deviates significantly from the others in the two panels and loses power quickly near our time corresponds to model
C1 (black line). The curves have been obtained for the best
fit values of the cosmological parameters as discussed in Sect.\,\ref{subsec:combined likelihood} (for a summary, cf. Tables \ref{tableFitBAOdz} and \ref{tableFitBAOA}).
The left panel shows the results based on the
SNIa+CMB+BAO$_{dz}$ fitting while the right panel those of the
SNIa+CMB+BAO$_{A}$ analysis. The C1 curve is obtained only for SNIa+CMB+BAO$_{A}$ data (and therefore is the same in both panels).
 \label{sigma8TypeA}}
}
\end{center}
\end{figure}
%%%%%%%%%%%%%%%%%%%%%%%%%%%%%%%%%%%%%%%%%%%%%%%%%%%%%%%%%%%%%%%%%%%%%%%%%%%%

The importance of the BAO's measurements cannot be underemphasized. They are sensitive to the physics of large scales and hinge primarily on the well-known principles of the linear regime of gravitational instability. Recall the recent hints of dynamical dark energy based on BAO's data mentioned in Sect. \ref{subsec:OPLCDM} of the Introduction -- cf. Ref.\,\cite{SahniShafielooStarobinsky}. These authors utilize the ``$Om(z)$-diagnostic''\,\cite{SahniShaStaro2008} and its improved version \eqref{eq:Omh2Diagnostic} \cite{SahniShafielooStarobinsky} and conclude that there exists a tension between the BAO observations and the CMB measurements that cannot be solved assuming the concordance $\CC$CDM model. According to the authors, their test provides model-independent evidence in favor of dynamical DE. Although at this stage it is probably too early to draw definite conclusions from these results before we get more statistics on $H(z)$ at high redshifts (cf. e.g. \cite{HuMiaoZhang14}), we can at least say that this kind of scenario is roughly consistent with the results of the current analysis. We have indeed found that our vacuum dynamical framework, when confronted with the presently available SNIa+CMB+BAO data, tends to emphasize significantly smaller values of $\Omo$. Therefore, in case that the  claims of dynamical DE would be confirmed at some point, the vacuum models presented here could provide an explanation.

We can understand analytically the possible origin of these results in our theoretical framework. Let us take e.g. a general type-A model. From the formulae of Sect. \ref{subsec:solvingA1A2} we can easily compute the corresponding $Om(z)$-diagnostic. The result is:
\begin{equation}\label{OmzA}
Om(z)=\frac{\Omo}{\xi}\,\frac{\left(1+z\right)^{3\xi}-1}{(1+z)^3-1}\,,
\end{equation}
with $\xi$ given in Eq.\,(\ref{eq:defxiM}). It is pretty obvious that for $\xi=1$ we recover the $\CC$CDM result, which remains pegged to $Om(z)=\Omo\ (\forall z)$. However, as soon we allow a small dynamical running of vacuum (meaning $\nu$ and/or $\alpha$ different from zero) we obtain a small departure of $\xi$ from $1$ and therefore the DE diagnostic $Om(z)$ deviates from $\Omo$. Actually, in this case (\ref{OmzA}) evolves with time (or redshift). According to the $Om(z)$ diagnostic this implies that the vacuum energy is dynamical. By the same token, the two-point diagnostic for type-A models can be computed:
\begin{equation}\label{Omh2zA}
Omh^2(z_2,z_1^2)=\frac{\Omo h^2}{\xi}\,\frac{\left(1+z_2\right)^{3\xi}-\left(1+z_1\right)^{3\xi}}{(1+z_2)^3-1+z_1)^3}\,.
\end{equation}
Clearly, the result depends on $z_i$ for any $\xi\neq 1$. Only for the $\CC$CDM case ($\xi=1$) it remains anchored to $\Omo\,h^2$ for any $z_i$.
One can perform similar considerations for type-B models. The upshot is that the detailed numerical analysis of both model types, in the light of the available observations, confirms that such vacuum dynamics leads to smaller $\Omo$ -- cf. Sect. \ref{subsec:combined likelihood}.

%%%%%%%%%%%%%%%%%%%%%%%%%%%%%%%%%%%%%%%%%%%%%%%%%%%%%%%%%%%%%%%%%%%%%%%%%%%%

\begin{figure}[t]
\begin{center}
\includegraphics[scale=0.50]{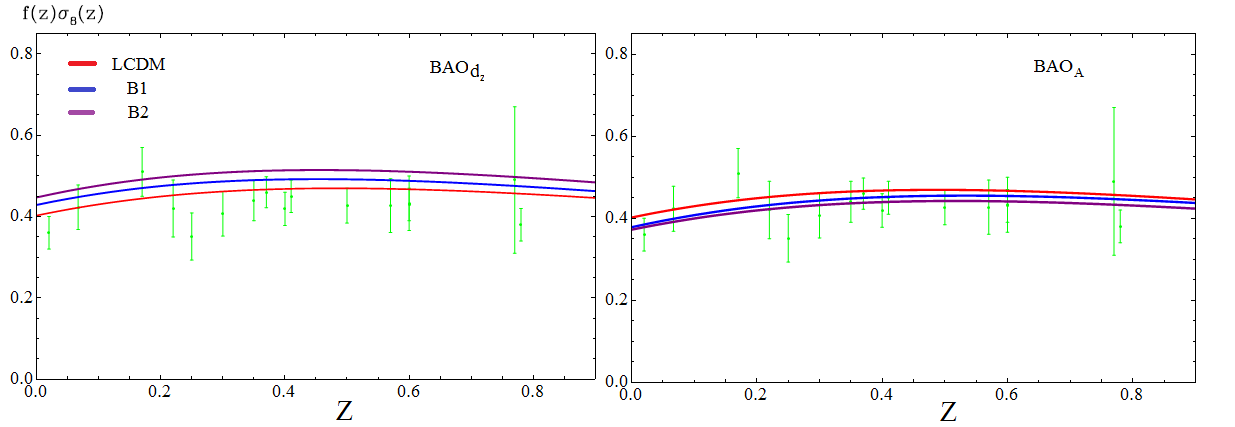}
\caption[$f(z)\sigma_8(z)$ for type-B models of Sect. \ref{chap:Atype}.]{\scriptsize{
Comparison of the observed and
theoretical evolution of the weighted growth
rate $f(z)\sigma_{8}(z)$ for the models B1 and B2. As in Fig. \ref{sigma8TypeA}, the left and right panels show the results based on the SNIa+CMB+BAO$_{dz}$ and
SNIa+CMB+BAO$_{A}$ best fit values, respectively.  On the left panel, the lowermost (red) line corresponds to the $\CC$CDM model, used as a reference. The closest one on top of it at $z=0$ corresponds to model B1 (blue line) and the next to closest at this point is model B2 (purple line). On the right panel, the B1 and B2 lines essentially overlap below the $\CC$CDM one, the B1 being in the middle.
\label{sigma8TypeB}}
}
\end{center}
\end{figure}

%%%%%%%%%%%%%%%%%%%%%%%%%%%%%%%%%%%%%%%%%%%%%%%%%%%%%%%%%%%%%%%%%%%%%
%%%%%%%%%%%%%%%%%%%%%%%%%%%%%%%%%%%%%%%%%%%%%%%%%%%%%%%%%%%%%%%%%%%%%
%%%%%%%%%%%%%%%%%%%%%%%%%%%%%%%%%%%%%%%%%%%%%%%%%%%%%%%%%%%%%%%%%%%%%

\subsection{The linear growth rate of clustering and the $\gamma$ index}
\label{subsec:growthrate}

In this section we analyze the linear perturbations growth
regime for the various models.
Although one could do it by means of the
power spectrum, we follow the approach of\,\cite{BPS09} and will
test herein the implications of the various models on structure
formation through the study of the linear growth rate of clustering\,\cite{Peebles1993}.
This important (dimensionless)
indicator is defined as the logarithmic derivative of the linear growth factor $\delta_m(a)$ with respect to the variable $\ln a$. Therefore,
\begin{equation}\label{eq:growingfactor}
f(a)\equiv \frac{1}{\delta_m}\frac{d\delta_m}{d\ln a}=\frac{d{\rm ln}\delta_m}{d{\rm ln}a}=-(1+z) \frac{d{\rm
ln}\delta_m}{dz}\,,
\end{equation}
where $\delta_m(a)$ is obtained from solving
the differential equation (\ref{diffeqDa}) for each model. The physical significance of $f(a)$ is that it determines the amplitude of redshift distortions, and also of the peculiar velocity flows. The latter can be seen by witting $f(a)=(\dot{\delta}_m/\delta_m)/(\dot{a}/a)=\dot{\delta}_m/(\delta_m\,H)$, which is the ratio of the peculiar flow rate to the Hubble rate.

In order to investigate the performance of our vacuum models, we
compare the theoretical growth prediction with the latest growth data
(as collected e.g. by \cite{BasilakosNes2013} and references therein),
which are based on the combined observable $f(z)\sigma_{8}(z)$, viz. the ordinary growth rate weighted by the rms mass fluctuation field.
It has been found that this estimator is almost a model-independent
way of expressing the observed growth history
of the Universe, in particular it is found to be independent of the galaxy density bias (see \cite{Song09}).
The theoretical functional form
of $\sigma_{8}(z)$ will be studied in Sect. \ref{sec:PSformalism} -- see Eq. (\ref{ss88}.

In Figs.\,\ref{sigma8TypeA} and \ref{sigma8TypeB} we display the predicted $f(z)\sigma_{8}(z)$
together with the observed linear growth data, for the various vacuum models A, B and C1. No information is provided on C2, which we already discarded.
Notice, that the theoretical curves on the left and the right panels correspond
to fitted values of the cosmological parameters
derived from SNIa+CMB+BAO$_{dz}$ and SNIa+CMB+BAO$_{A}$ respectively.
Obviously, despite the fact that the C1 model fits well the expansion data (cf. Sect. \ref{subsec:combined likelihood}),
it is finally ruled out by the growth data.
The lack of structure formation near our time
is exceedingly evident (confer Fig.\,\ref{sigma8TypeA}) as compared to the other vacuum models.
For these reasons we come to the conclusion that
the entire C1 class of models (in particular the linear sort $\CC\propto H$) is strongly unfavored. Further analysis can be found in Ref. \cite{BasSola14a} and Sect. \ref{sec:ExtraTypeB}.

Overall, we can see that the A1-A2 vacuum models
with the single parameter $\nu$ (or $\nueff$) match quite well the
growth data, that is to say in a way which is
comparable to the $\CC$CDM model (red line). We confirm this fact via
a $\chi^{2}_{\rm growth}$ minimization statistical test.
%and a Kolmogorov-Smirnov (KS)
%statistical test respectively.
In particular, for the vacuum
models (including the $\Lambda$CDM) we find that
$\chi^{2}_{\rm growth}/16$ lies in the interval $[0.52-1.25]$\footnote{The growth
sample contains 16 entries \cite{BasilakosNes2013}.}.

%%%%%%%%%%%%%%%%%%%%%%%%%%%%%%%%%%%%%%%%%%%%%%%%%%%%%%%%%%%%%%%%%%%%%%%%%%%%%%%%%%%%%

\begin{figure}[t!]
\begin{center}
\includegraphics[scale=0.45]{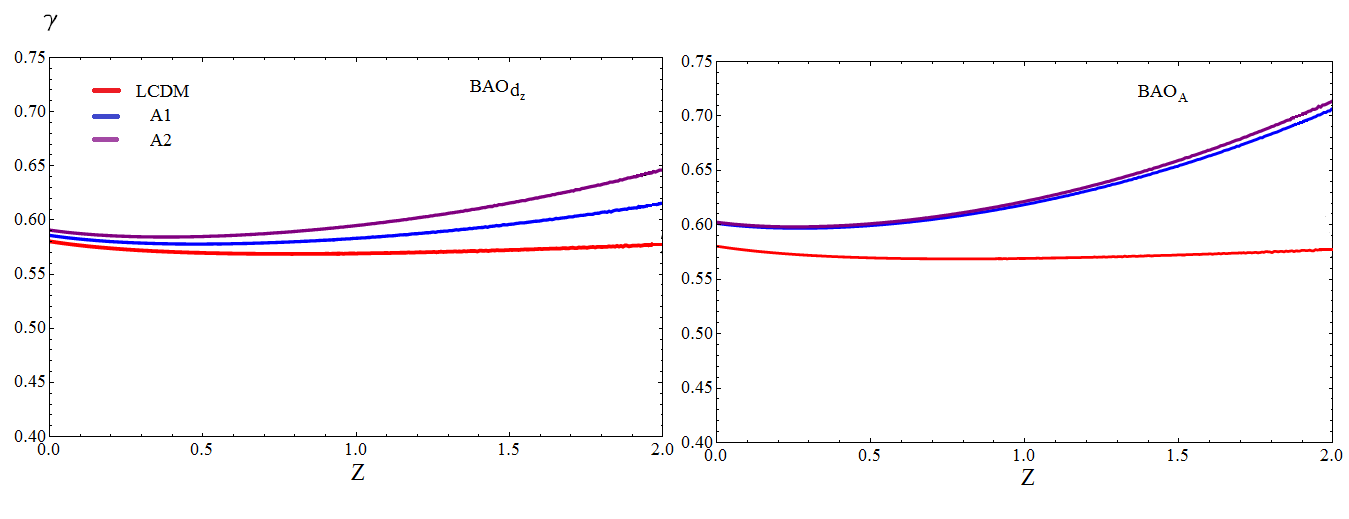}
\caption[Growth index $\gamma(z)$ for type-A models of Sect. \ref{chap:Atype}.]{\scriptsize{The evolution of the growth rate index, Eq.\,(\ref{eq:growingfactor2}).
The lines correspond to the
A1 and A2 vacuum models for the best
fit cosmological values discussed in Sect.\,\ref{subsec:combined likelihood}. From top to bottom (in both panels): A2 (purple line), A1 (blue line) and $\CC$CDM (red line), used as reference.
The left panel shows the results based on the
SNIa+CMB+BAO$_{dz}$ fitting while the right panel those of the
SNIa+CMB+BAO$_{A}$ analysis.
\label{GrowthIndexTypeA}}
}
\end{center}
\end{figure}

%%%%%%%%%%%%%%%%%%%%%%%%%%%%%%%%%%%%%%%%%%%%%%%%%%%%%%%%%%%%%%%%%%%%%%%%%%%%%%%%%%%%

On inspecting once more Figs.\,\ref{sigma8TypeA} and \ref{sigma8TypeB}, the data clearly shows that the
growth of structure is hindered near our time, which is evidence of
a positive cosmological constant exerting a negative pressure
against the process of matter collapse. This is well described by
the $\CC$CDM. But it is also comparably well described by the
running vacuum models carrying 
an additive constant term in their functional
form [see Eq. (\ref{A1A2B1B2C})]
and a relatively small value of $\nu$ or $\epsilon$ of order $\sim
10^{-3}$. All these features can be seen very clearly in that
figure.

Finally, let us finish with a short discussion concerning the
growth rate index $\gamma$. As we have already mentioned in the introduction
we can express the linear growth rate of clustering
in terms of $\Omega_{m}(z)$ as follows: $f(z)\simeq \Omega_{m}(z)^{\gamma(z)}$, where
$\gamma$ is the linear growth rate index. For the usual $\Lambda$CDM model, such index is approximated by $\gamma_{\CC} \simeq 6/11\simeq 0.545$. This result is a particular case (for $\omega_D=-1$)  of the theoretical formula $\gamma\simeq 3(\omega_D-1)/(6\omega_D-5)$ corresponding to DE models with a slowly varying  equation of state $\omega_D$\,\cite{WangSteinhardt98}.

To obtain the linear growth index for the dynamical vacuum models studied here we have to use the corresponding linear growth factor, $\delta_m(z)$, and
from Eq. (\ref{eq:growingfactor}) we easily
obtain:
\begin{equation}\label{eq:growingfactor2}
\gamma(z)\simeq \frac{\ln\left[-(1+z) \frac{d\ln \delta_m}{dz}\right]}
{\ln\Omega_{m}(z)} \;,
\end{equation}
where $\delta_m(z)$ for the different vacuum models is given in sections
\ref{sec:perturbationsTypeA} and \ref{sec:perturbationsTypeB}.

\begin{figure}[t!]
\begin{center}
\includegraphics[scale=0.48]{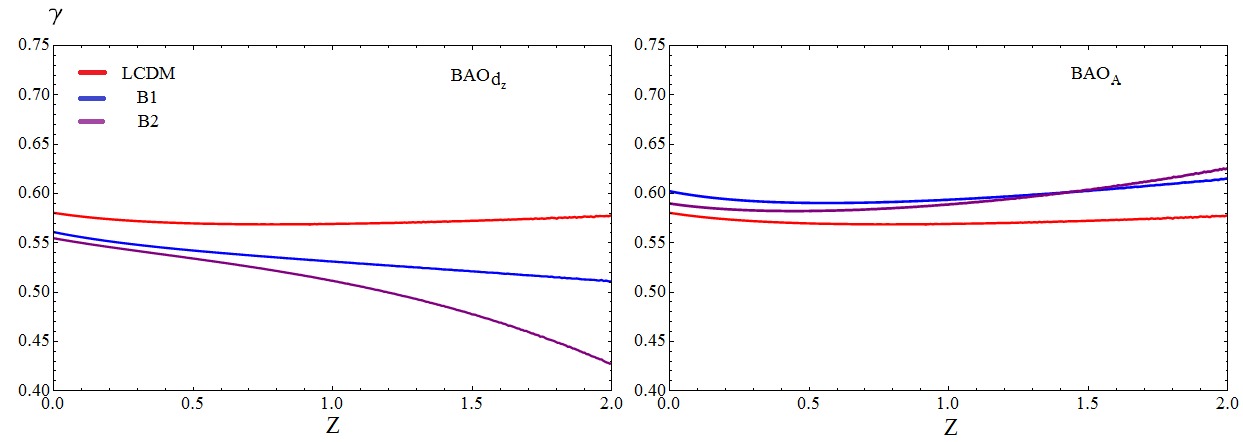}
\caption[Growth index $\gamma(z)$ for type-B models of Sect. \ref{chap:Atype}.]{\scriptsize{The evolution of the growth rate index.
The lines correspond to the B1 and B2 vacuum models for the best
fit cosmological values discussed in Sect.\,\ref{subsec:combined likelihood}.
The left panel shows the results based on the
SNIa+CMB+BAO$_{dz}$ fitting while the right panel those of the
SNIa+CMB+BAO$_{A}$ analysis. The uppermost curve (in red) on the left panel corresponds to the $\CC$CDM, the middle one is for B1 (blue line) and the lowest one (in purple) is for B2. On the right panel the $\CC$CDM curve is the lowest one. Near $z=0$ the highest one (in blue) is for B1 and the middle curve (in purple) is for B2.
\label{GrowthIndexTypeB}}
}
\end{center}
\end{figure}

In Figs.\,\ref{GrowthIndexTypeA} and \ref{GrowthIndexTypeB} we present the evolution of the linear growth index
for the A and B type of vacuum models, respectively
(on the left with SNIa+CMB+BAO$_{dz}$ data, and on the right with SNIa+CMB+BAO$_{A}$ data). In the same figures we can also see our determination of $\gamma_{\CC}(z)$ as a function of the redshift, and in particular we find $\gamma_{\CC}(0)\simeq 0.58$.

The comparison shown in the mentioned figures indicates that the growth index of the type-A vacuum models with SNIa+CMB+BAO$_{dz}$ data is well approximated
by the $\Lambda$CDM constant value
for $z \le 1$, while at large redshifts there are deviations. When SNIa+CMB+BAO$_{A}$ data is used, instead, there is a visible deviation from above in all the range, which becomes smaller (at the level $5\%$) for $z \le 1$
%We can appreciate positive deviations of the type-A1 and A2 models with respect to the $\CC$CDM value
%$\gamma(0)_{\CC CDM}\simeq 6/11\simeq 0.545$
%ranging $5-10\%$ at $z=0$
(see Fig. \ref{GrowthIndexTypeA}, right panel).  Let us note that other vacuum models, such as e.g. B1 and B2, depart also from the $\CC$CDM result (in this case from below) when using  SNIa+CMB+BAO$_{dz}$ data (cf. Fig.\,\ref{GrowthIndexTypeB}, left panel). We find that for $z \le 1$ the departure can be of order $5-10\%$. The deviation, on the other hand, is not so pronounced (and of opposite sign) when  SNIa+CMB+BAO$_{A}$ data are used (right panel of the same figure).

%%%%%%%%%%%%%%%%%%%%%%%%%%%%%%%%%%%%%%%%%%%%%%%%%%%%%%%%%%%%%%%%%%%%%%%%%%%%%%%%%%%%

%%%%%%%%%%%%%%%%%%%%%%%%%%%%%%%%%%%%%%%%%%%%%%%%%%%%%%%%%%%%%%%%%%%%%%%%%%%%%%%%%%%%%%

It is worth mentioning that the differences we have found with respect to the $\CC$CDM are near the edge of the present experimental limits. For example, in a recent analysis of the clustering properties of Luminous Red Galaxies and the growth rate data provided by the various galaxy surveys it is found that $\gamma=0.56\pm 0.05$ and $\Omo=0.29\pm0.01$\,\cite{Athina2014}. The prediction of $\gamma$ for all our vacuum models lies within $1\sigma$ of that range.

Since the experimental error on
the $\gamma$-index is of order $10\%$ and some of the vacuum models are bordering these limits, it opens the possibility that the deviations presented by these models might be resolved in the future when more accurate data will be available. This is quite evident from the results presented in Figs.\,\ref{GrowthIndexTypeA} and \ref{GrowthIndexTypeB} of our analysis. Combining the analysis of the growth rate with the study of cluster number counts (see the next section), it should be possible to further pin down the nature of these dynamical vacuum models. 

%%%%%%%%%%%%%%%%%%%%%%%%%%%%%%%%%%%%%%%%%%%%%%%%%%%%%%%
%%%%%%%%%%%%%%%%%%%%%%%%%%%%%%%%%%%%%%%%%%%%%%%%%%%%%%%
%%%%%%%%%%%%%%%%%%%%%%%%%%%%%%%%%%%%%%%%%%%%%%%%%%%%%%%

\section[DVM's and the number counts method]{Testing the dynamics of vacuum through the cluster number counts method}
\label{sec:Number counts}

In the foregoing part of our analysis we have shown that the A and B types of vacuum models can successfully
fit the background cosmological data and the growth of linear
perturbations in a way which in some cases is perfectly comparable to the
$\CC$CDM. This is not so with type-C models, which fail seriously in regard to the expansion data or the structure formation data or both. We have also shown that the A and B vacuum classes have different predictions concerning the linear growth rate index $\gamma$, which in the future may be resolved. In that case we could distinguish between these two sort of vacuum models and also with respect to the $\CC$CDM.

%%%%%%%%%%%%%%%%%%%%%%%%%%%%%%%%%%%%%%%%%%%%%%%%%%%%%%%%%%%%%%%%%%%%%%%%%%%%

 \begin{table}[t!]
\tabcolsep 5pt \vspace {0.2cm}
\hspace{3cm}\begin{tabular}{|c|c|ccc|c|c|} \hline
Model    & $\Omega_m^0$& $\nu$ &\phantom{X}& $\epsilon$ &
$\sigma_{8}$
&$\delta_c$ \\
 \hline
$\Lambda$CDM     & 0.292 & 0 & \phantom{X} & 0 & 0.829 & 1.675 \\
\hline
A1 & 0.292 & +0.0013 & \phantom{X} & 0  & 0.813 & 1.666\\
A2  & 0.290 & +0.0024 & \phantom{X} & 0 & 0.797 &1.659
\\ \hline
B1  & 0.297 & 0 & \phantom{X} & -0.014 & 0.859 &1.696\\
B2 & 0.300 & -0.0039 & \phantom{X}
 & -0.0039 & 0.896 & 1.705 \\ \hline
\end{tabular}
\caption[Numerical results for type-A and B models from fitting SNIa+CMB+BAO$_{dz}$ data (in correspondence with Table \ref{tableFitBAOdz})]{{\scriptsize Numerical results from fitting SNIa+CMB+BAO$_{dz}$ data (in correspondence with Table \ref{tableFitBAOdz}). The $1^{st}$ column indicates the vacuum
energy model. The $2^{nd}$ shows the central fit value of  $\Omo$. The $3^{rd}$ and $4^{th}$ display the best fit values of the parameters $\nu$ and $\epsilon$, with the understanding that $\nu$ is to be taken $\nueff$ for model A2. Finally, the
$5^{th}$ and $6^{th}$ columns list the computed values of $\sigma_8$ and $\delta_c\equiv\delta_c(z=0)$, respectively. The procedure to compute the
collapse density threshold $\delta_c(z)$ for each model is explained in Appendix \ref{ch:appCollapse}.}} \label{TableModelsBAOdz}
\end{table}

%%%%%%%%%%%%%%%%%%%%%%%%%%%%%%%%%%%%%%%%%%%%%%%%%%%%%%%%%%%%%%%%%%%%%%%%%%%%

In the meanwhile and in an attempt to define further observational criteria capable of distinguishing the realistic model variants A
and B from the concordance $\Lambda$CDM cosmology, we analyze now their theoretically predicted cluster-size halo redshift
distributions, i.e. the expected cluster number counts of each model as a function of the redshift. As it turns, this is an efficient method to separate vacuum
models which perform outstanding at the linear perturbations level but differ very little in the values of the parameters. 

In previous works some of us have described and tested this methodology
for simpler versions of the dynamical vacuum models, see
Refs. \cite{Grande2011} and \cite{BPS09}. The method has also been used to place bounds on cosmological parameters and on different types of dark energy models, see e.g.\,\cite{BasilakosPlionisLima2010,Campanelli2011,ChandrachaniDevi1,ChandrachaniDevi2}.  The basic tool is the Press-Schechter formalism and its generalization. In the following
we briefly summarize the basics of this method and refer the reader
to the aforesaid references for more details. A crucial
ingredient of the cluster number counts method is the linearly
extrapolated density threshold above which structures collapse,
$\delta_c$. The computation of this model-dependent parameter is a
rather demanding task as it requires to solve the perturbations
equations beyond the linear approximation. In the Appendix \ref{ch:appCollapse} we
compute $\delta_c$ for the models under consideration.

%%%%%%%%%%%%%%%%%%%%%%%%%%%%%%%%%%%%%%%%%%%%%%%%%%%%%%%%%%%%%%%%%%%%%%%%%%%%%%%%%%

 \begin{table}[t!]
\tabcolsep 5pt \vspace {0.2cm}
\hspace{3cm} \begin{tabular}{|c|c|ccc|c|c|} \hline
Model   & $\Omega_m^0$& $\nu$ &\phantom{X} & $\epsilon$ &
$\sigma_{8}$
&$\delta_{c}$ \\
 \hline
$\Lambda$CDM    & 0.292 & 0  &\phantom{X} & 0 & 0.829 & 1.675 \\
\hline
A1 & 0.282 & +0.0048 & \phantom{X} & 0  &0.758 & 1.644\\
A2  & 0.283 & +0.0048 &\phantom{X}  & 0 & 0.757 &1.642
\\ \hline

B1  & 0.283 & 0 &\phantom{X}  & +0.005 & 0.820 &1.667\\
B2 & 0.283 & +0.0015 &\phantom{X}
 & +0.0015 & 0.791  & 1.662\\ \hline
\end{tabular}
\caption[Numerical results for type-A and B models from fitting SNIa+CMB+BAO$_{A}$ data (in correspondence with Table \ref{tableFitBAOA})]{\scriptsize{As in Table \ref{TableModelsBAOdz}, but using the fitting results from SNIa+CMB+BAO$_{A}$ data (in correspondence with Table \ref{tableFitBAOA}).}} \label{TableModelsBAOA}
\end{table}

%%%%%%%%%%%%%%%%%%%%%%%%%%%%%%%%%%%%%%%%%%%%%%%%%%%%%%%%%%%%%%%%%%%%%%%%%%%%

%%%%%%%%%%%%%%%%%%%%%%%%%%%%%%%%%%%%%%%%%%%%%%%%%%%%%%%%%%
%%%%%%%%%%%%%%%%%%%%%%%%%%%%%%%%%%%%%%%%%%%%%%%%%%%%%%%%%%
%%%%%%%%%%%%%%%%%%%%%%%%%%%%%%%%%%%%%%%%%%%%%%%%%%%%%%%%%%

\subsection{Generalized Press-Schechter formalism}
\label{sec:PSformalism}

The Press and Schechter  (hereafter PSc) formalism to compute the
fraction of matter in the Universe that has formed bounded
structures and its redshift distribution was developed in a pioneering work of these authors more than $40$ years ago\,\cite{press} and has been generalized and improved since then.
One introduces the so-called halo mass function, $F(M,z)$, representing
the fraction of the Universe that has collapsed by the redshift $z$
in halos above some mass $M$, where the primordial density
fluctuation for a given mass $M$ of the dark matter fluid is
described by a random Gaussian field. With this  function and
assuming a mean (comoving) background mass density $\bar{\rho}$ one may
estimate the (comoving) number density of virialized halos, $n(M,z)$, with
masses within the range $(M, M+\delta M)$: 
\be n(M,z) dM=
\frac{\partial F(M,z)}{\partial M} \frac{{\bar \rho}}{M} dM\,. \ee
This expression can be rewritten as follows:
\begin{eqnarray}\label{MF}
n(M,z) dM &=& \frac{\bar{\rho}}{M} \frac{d{\rm \ln}\sigma^{-1}}{dM}
f_{\rm PSc}(\sigma) dM,
\end{eqnarray}
where $f_{\rm PSc}(\sigma)=\sqrt{2/\pi} (\delta_c/\sigma)
\exp(-\delta_c^2/2\sigma^2)$. Note that in this approach all the
mass is locked inside halos, according to the normalization
constraint: 
\be\label{eq:PSnormalization} \int_{-\infty}^{+\infty} f_{\rm PSc}(\sigma) d{\rm
\ln}\sigma^{-1} = 1\;. \ee
%

%%%%%%%%%%%%%%%%%%%%%%%%%%%%%%%%%%%%%%%%%%%%%%%%%%%%%%%%%%%%%%%%%%%%%%%%%%%%
\begin{figure}[t!]
\begin{center}
\includegraphics[scale=0.55]{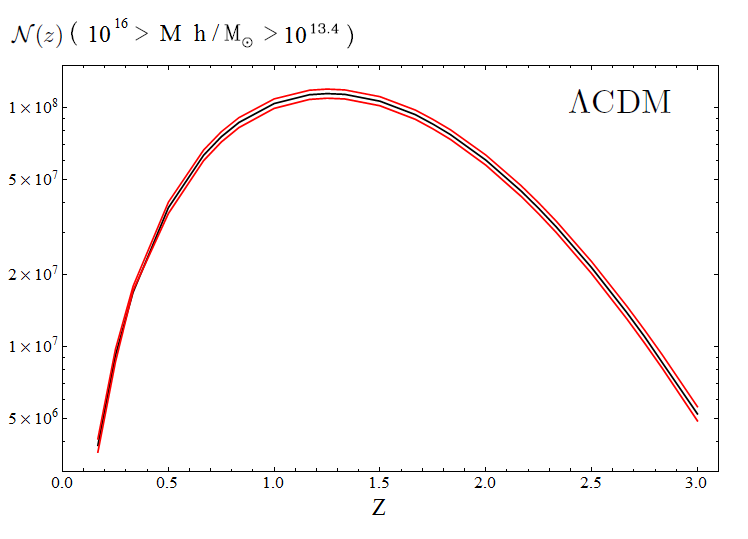}
\caption[Predicted redshift distribution of the total number counts, ${\cal N}(z)$, in the $\Lambda$CDM with masses in the range  $10^{13.4}\,h^{-1}\lesssim M/M_{\odot}\lesssim 10^{16}\,h^{-1}$]{\scriptsize{The theoretically predicted redshift distribution of the total number of cluster counts, ${\cal N}(z)$, with masses in the range  $10^{13.4}\,h^{-1}\lesssim M/M_{\odot}\lesssim 10^{16}\,h^{-1}$, corresponding to the concordance $\CC$CDM model and using the generalized Press-Schechter function (\ref{PS-Reed}). The curves correspond to the best fit value $\Omo=0.293\pm 0.013$  within $1\sigma$ (cf. Table \ref{tableFitBAOdz}). }}\label{fig:NLCDM}
\end{center}
\end{figure}

%%%%%%%%%%%%%%%%%%%%%%%%%%%%%%%%%%%%%%%%%%%%%%%%%%%%%%%%%%%%%%%%%%%%%%%%

\noindent The parameter $\delta_{c}$ is the collapse density threshold, i.e. the linearly extrapolated density
threshold above which structures collapse \cite{eke} (see the Appendix \ref{ch:appCollapse}
for more details), while $\sigma^2(M,z)$ is the mass variance of the
smoothed linear density field, which depends on the redshift $z$ at
which the halos are identified. It is given in Fourier space by: 
\be
\label{sig88} \sigma^2(M,z)=\frac{\delta_m^2(z)}{2\pi^2} \int_0^\infty k^2
P(k) W^2(kR) dk \,. 
\ee
In this expression, $\delta_m(z)$ is the linear growth
factor of perturbations, which we have computed before for our
models, $P(k)$ is the power-spectrum of the linear density field,
and finally we have the smoothing function $W(kR)=3({\rm
  sin}kR-kR{\rm cos}kR)/(kR)^{3}$, which is the Fourier image of the following geometric top-hat function with spherical symmetry:  $f_{\rm TH}(r)=3/(4\pi R^3)\,\theta(1-r/R)$.
It contains on average a mass $M$ within a comoving radius $R=(3M/ 4\pi
\bar{\rho}_m)^{1/3}$. The current mean mass density at redshift $z$ reads $\bar{\rho}_m^{0}=\Omo\,\rco=2.78 \times
10^{11}\Omega_{m}^{0}h^{2}M_{\odot}$Mpc$^{-3}$. We use the CDM power
spectrum:
\begin{equation}\label{eq:Pk}
P(k)=P_{0} k^{n_s} T^{2}(\Omega_{m}^{0},k)\,,
\end{equation}
where $P_0$ is a normalization constant (see below), $n_s=0.9603\pm 0.0073$ is the value of the spectral index
measured by Planck+WP\,\cite{Planck2013}; and $T(\Omega_{m}^{0},k)$
the BBKS transfer function \cite{Bard86,LiddleLyth,BookGorbunovRubakov}. Introducing the dimensionless variable $x=k/k_{eq}$, in which
$k_{eq}=a_{eq}H(a_{eq})$ is the value of the wavenumber at the
equality scale of matter and radiation, we can write the transfer
function as follows:
\be\label{jtf}T(x) =  \frac{\ln (1+0.171 x)}{0.171\,x}\Big[1+0.284 x + (1.18 x)^2 + \, (0.399 x)^3+(0.490
x)^4\Big]^{-1/4}\,.
\ee
It is important to emphasize that $k_{eq}$ is a model
dependent quantity. For type-A and type-B models we have used the
same formula that is obtained in the $\CC$CDM, due to the fact that
the deviations are small in these cases,
\be\label{eq:keqLCDM}
k^\Lambda_{eq}=H_0\,\Omo\,\sqrt{\frac{2}{\Omega^0_r}}\,e^{-\Omega_b^{0}-\sqrt{2h}\frac{\Omega_b^{0}}{\Omo}}\,.
\ee
It is traditional to parameterize the mass variance in terms of
$\sigma_8$, the rms mass fluctuation amplitude on scales of $R_{8}=8
\; h^{-1}$ Mpc at redshift $z=0$ [$\sigma_{8} \equiv \sigma_8(0)$]. This allows us to normalize the power spectrum, i.e. to determine $P_0$. Indeed, using equations (\ref{sig88}) and (\ref{eq:Pk})
we have: 
\be \label{s888} \sigma^2(M,z)=\sigma^2_8(z)
\frac{\int_{0}^{\infty} k^{n_s+2} T^{2}(\Omega_{m}^{0}, k) W^2(kR)
dk} {\int_{0}^{\infty} k^{n_s+2} T^{2}(\Omega_{m}^{0}, k)
W^2(kR_{8}) dk}\,, \ee where \be \label{ss88}
\sigma_8(z)=\sigma_8\frac{\delta_m(z)}{\delta_m(0)} \;. 
\ee
Equivalently,
\begin{equation}\label{eq:P0}
P_0=2\pi^2\,\frac{\sigma_8^2}{\delta_m^2(0)}\,\left[{\int_{0}^{\infty} k^{n_s+2} T^{2}(\Omega_{m}^{0}, k)
W^2(kR_{8}) dk}\right]^{-1}\,.
\end{equation}
The Planck+WP value of
$\sigma_8$, which we use for our analysis, is $\sigma_{8,\Lambda}=0.829\pm
0.012$\,\cite{Planck2013}. The $\sigma_8$ value for the different dynamical vacuum models can
be estimated by scaling the present time $\CC$CDM value\footnote{In
the following discussion, the quantities referred to the $\CC$CDM
model are distinguished by the subscript `$\CC$' ($\sigma_{8,\CC}$;
$D_\CC$; $\Omega_{m,\Lambda}^{0}$) whereas the corresponding
quantities in the dynamical vacuum models carry no subscript.}
($\sigma_{8, \Lambda}$) using once more equations (\ref{sig88}) and (\ref{eq:Pk}):
\be
\sigma_{\rm 8}=\sigma_{8, \Lambda} \frac{\delta_m(0)}{\delta_{m,\Lambda}(0)}
\left[\frac{\int_{0}^{\infty} k^{n_s+2}
T^{2}(\Omega_{m}^{0},k) W^2(kR_{8}) dk}
{\int_{0}^{\infty} k^{n_s+2} T^{2}(\Omega_{m,
\Lambda}^{0},k) W^2(kR_{8}) dk} \right]^{1/2}\,.\label{s88general}
\ee
%
%%%%%%%%%%%%%%%%%%%%%%%% FIGURE  2 %%%%%%%%%%%%%%%%%%%%%%%%%%%%%%%%%%%%%
%
\begin{figure}[t!]
\begin{center}
{\includegraphics[scale=0.50]{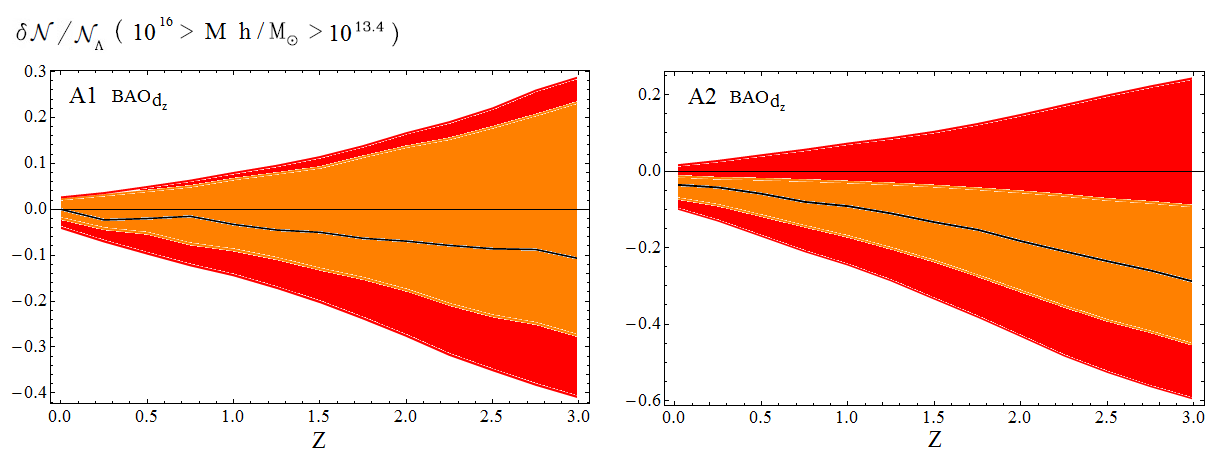}}
\end{center}
\caption[$\delta {\cal N}/{\cal N}$ for the
type-A model using the SNIa+CMB+BAO$_{dz}$ data]{\scriptsize{\textbf{Left}: Fractional difference $\delta {\cal N}/{\cal N}$ in the number of counts of clusters between the
vacuum model A1 and the concordance $\CC$CDM model (cf. Fig.\,\ref{fig:NLCDM}) using SNIa+CMB+BAO$_{dz}$ data from Table \ref{tableFitBAOdz}. The continuous solid line represents $\delta {\cal N}/{\cal N}$ for the best fit value from that table,  whereas the innermost (resp. outermost) band comprises the $\delta {\cal N}/{\cal N}$ prediction for the points within the $\pm 1\sigma$ (resp. $\pm 3\sigma$) values around it. \textbf{Right}: As before, but for model A2.}} \label{NC A1A2 BAO_dz}
\end{figure}
%
%%%%%%%%%%%%%%%%%%%%%%%%%%%%%%%%%%%%%%%%%%%%%%%%%%%%%%%%%%%%%%%%%%%%%%%%
%

\noindent Overall it follows from the foregoing formulae that the mass variance of the linear density field is determined from
\begin{equation}\label{eq:variance}
\sigma^2(M,z)=\sigma^2_{8,\Lambda}\,\frac{\delta_m^2(z)}{\delta^2_{m,\Lambda}(0)}
\frac{\int_{0}^{\infty} k^{n_s+2} T^{2}(\Omega_{m}^{0}, k) W^2(kR)
dk} {\int_{0}^{\infty} k^{n_s+2} T^{2}(\Omega_{m,\Lambda}^{0}, k)
W^2(kR_{8}) dk}\,,
\end{equation}
Furthermore, the numerical value of $\sigma_8$ for the $\CC$CDM and the various vacuum models under consideration has been collected in the last but one column of Tables \ref{TableModelsBAOdz} and \ref{TableModelsBAOA} together with the best fitting values of the parameters according to  each BAO type that we have used (cf. Tables \ref{tableFitBAOdz} and \ref{tableFitBAOA}).

%%%%%%%%%%%%%%%%%%%%%%% FIGURE  2 %%%%%%%%%%%%%%%%%%%%%%%%%%%%%%%%%%%%%

\begin{figure}[t!]
\begin{center}
{\includegraphics[scale=0.50]{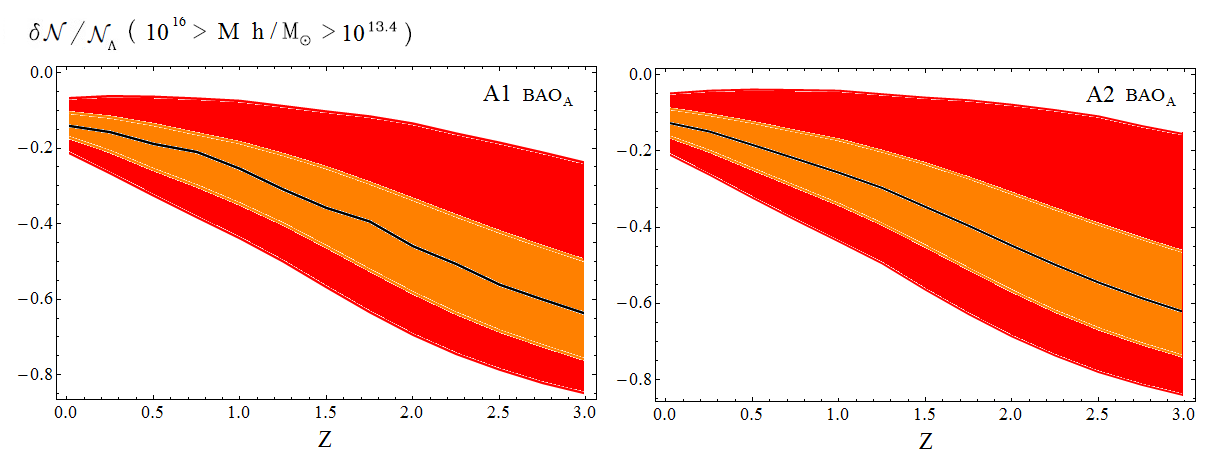}}
\end{center}
\caption[$\delta {\cal N}/{\cal N}$ for the
type-A model using the SNIa+CMB+BAO$_{A}$ data]{\scriptsize{Fractional difference $\delta {\cal N}/{\cal N}$ in the number of counts of clusters between the
vacuum models A1 (left) and A2 (right) and the concordance $\CC$CDM model, but using SNIa+CMB+BAO$_{A}$ data from Table \ref{tableFitBAOA}. Same notation as in Fig.\,\ref{NC A1A2 BAO_dz}.}} \label{NC:A1A2 BAO_A}
\end{figure}

%%%%%%%%%%%%%%%%%%%%%%%%%%%%%%%%%%%%%%%%%%%%%%%%%%%%%%%%%%%%%%%%%%%%%%%%
%

The original Press-Schechter function $f_{\rm PSc}$ was shown to
provide a relatively good first approximation to the halo mass function
obtained by numerical simulations. In Sect. \ref{subsec:understandingNumberCount} we use $f_{\rm
PSc}$ to assess in detail why the number count method is an
efficient one to separate models that may be difficult to
distinguish at the linear perturbation regime. The method, however,
is not tied to the particularly simple form of the original
Press-Schechter function $f_{\rm PSc}$.  More recently  a large
number of works have provided better fitting functions for
$f(\sigma)$. In practice, in our analysis for the various dynamical
vacuum models under consideration we will adopt the generalized one
proposed by  Reed et al. \cite{Reed2007}:
\begin{eqnarray}\label{PS-Reed}
f_R(\sigma, n_{\rm eff}) = A\sqrt{\frac{2b}{\pi}}
\left[1+\left(\frac{\sigma^2}{b\delta_c^2}\right)^p + 0.6G_1 + 0.4G_2\right]\,\left(\frac{\delta_c}{\sigma}\right)\nonumber\\
\times\exp\left[-\frac{cb\delta_c^2}{2\sigma^2} -\frac{0.03}{(n_{\rm
eff}+3)^2} \left(\frac{\delta_c}{\sigma}\right)^{0.6} \right]\,,
\end{eqnarray}
where $A = 0.3222,\, p = 0.3,\, b = 0.707,\, c=1.08$, while
$G_1,\,G_2$ and $n_{\rm eff}$, {the slope of the non-linear
power-spectrum at the halo scale}, are given by:
\begin{equation}
G_1 = \exp\left[-\frac{(\ln\sigma^{-1}-0.4)^2}{2(0.6)^2}\right],\; G_2 = \exp\left[-\frac{(\ln\sigma^{-1}-0.75)^2}{2(0.2)^2}\right],\;
n_{\rm eff} = 6 \frac{{\rm d}\ln\sigma^{-1}}{{\rm d}\ln M\phantom{+}}-3.
\end{equation}
The previous generalized PSc function is a refined variant of an
older function that was used to improve the original PSc-formalism by
Sheth and Tormen \,\cite{STfunction1999}:
\begin{equation}\label{PS-ST}
f_{ST}(\sigma) = A'\sqrt{\frac{2b}{\pi}} \left[1+\left(\frac{\sigma^2}{
b\delta_c^2}\right)^p \right]\,\left(\frac{\delta_c}{\sigma}\right)
\,\exp\left[-\frac{b\delta_c^2}{2\sigma^2} \right]\,,
\end{equation}
where the parameters $b$ and $p$ are the same as in
(\ref{PS-Reed}). Once more $A'$ must be fixed from the normalization condition (\ref{eq:PSnormalization}). Let us, however, note that the value of the normalization constant cancels in the ratio $\delta\mathcal{N}/\mathcal{N}_{\CC{\rm CDM}}$, where $\delta\mathcal{N}=\mathcal{N}-\mathcal{N}_{\CC{\rm CDM}}$ represents the deviations of the number counts of the given vacuum model with respect to the $\CC$CDM. In fact, the fractional difference $\delta\mathcal{N}/\mathcal{N}_{\CC{\rm CDM}}$ will be the main observable in our test analysis of the number counts for dynamical vacuum models.
%and we find  $A'=0.1611$.
While we have also made use of the parameterization (\ref{PS-ST}) to test the sensibility of our results to the generalized Press-Schechter functions, we will for definiteness only present the final results in terms of the more complete function (\ref{PS-Reed}).

To use that function we need to know the value of the collapse density threshold parameter $\delta_c$. In Appendix \ref{ch:appCollapse} we compute $\delta_c$ by solving the corresponding nonlinear perturbation equations for each vacuum model.  The resulting values are listed in the last column of Tables \ref{TableModelsBAOdz} and \ref{TableModelsBAOA}, where we have separated them according to the type of BAO observable used in the best fitting to the SNIa+CMB+BAO cosmological data.

%%%%%%%%%%%%%%%%%%%%%%%%%%%%%%%%%%%%%%%%%%%%%%%%
%%%%%%%%%%%%%%%%%%%%%%%%%%%%%%%%%%%%%%%%%%%%%%%%
%%%%%%%%%%%%%%%%%%%%%%%%%%%%%%%%%%%%%%%%%%%%%%%%

\subsection{Numerical results: number counts of the dynamical vacuum
models}
\label{subsec:HaloMasFunction}

From the halo mass function (\ref{MF}) we can derive for each vacuum
model the redshift distribution of clusters, ${\cal N}(z)$, within
some determined mass range, say $M_1\le M\le M_2$. This can be
estimated by integrating the expected differential halo mass
function, $n(M,z)$, with respect to mass, namely
\be\label{eq:Nz} 
{\cal N}(z)=\frac{dV}{dz}\;\int_{M_{1}}^{M_{2}}
n(M,z)dM, 
\ee 
where $dV/dz$ is the comoving volume element, which in
a flat Universe takes the form:
\be \frac{dV}{dz} =4\pi r^{2}(z)\frac{dr(z)}{dz}, \ee with $r(z)$
denoting the comoving radial distance out to redshift $z$: \be
r(z)=\frac{c}{H_{0}} \int_{0}^{z} \frac{dz'}{E(z')}.
\ee
It follows that
\begin{equation}\label{eq:Nzv2}
\mathcal{N}(z)=4\pi r^2(z)\frac{dr}{dz}\int_{M_1}^{M_2}n(M,z)dM=-\frac{4\pi r^2\,\bar{\rho}(z)}{H_0 E(z)}\,\int_{M_1}^{M_2}\frac{1}{M}\left(\frac{1}{\sigma}\frac{d\sigma}{dM}\right)f(\sigma)dM\,.
\end{equation}
In practice, as we have said, we will use the function (\ref{PS-Reed}) for $f(\sigma)$ in the above expression.

%%%%%%%%%%%%%%%%%%%%%%%%%%%%%%%%%%%%%%%%%%%%%%%%%%%%%%%%%%%%%

\begin{figure}[t!]
\begin{center}
{\includegraphics[scale=0.50]{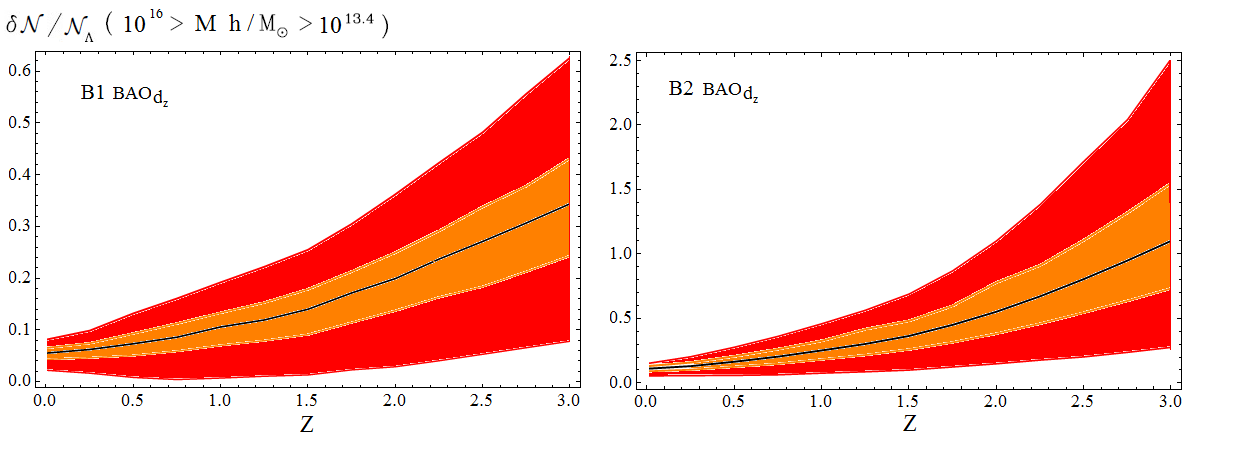}}
\end{center}
\caption[$\delta {\cal N}/{\cal N}$ for the
type-B model using the SNIa+CMB+BAO$_{dz}$ data]{\scriptsize{Fractional difference $\delta {\cal N}/{\cal N}$ in the number of counts of clusters between the
vacuum models B1 (left) and B2 (right) and the concordance $\CC$CDM model, using SNIa+CMB+BAO$_{dz}$ data from Table \ref{tableFitBAOdz}. Same notation as in Fig.\,\ref{NC A1A2 BAO_dz}.}} \label{NC:B1B2 BAO_dz}
\end{figure}

%%%%%%%%%%%%%%%%%%%%%%%%%%%%%%%%%%%%%%%%%%%%%%%%%%%%%%%%%%%%%%%%%%%%%%%%

In Fig.~\ref{fig:NLCDM}, we show the theoretically predicted redshift distribution of the total number of cluster counts, ${\cal N}(z)$, with masses in the range  $10^{13.4}\,h^{-1}\lesssim M/M_{\odot}\lesssim 10^{16}\,h^{-1}$, corresponding to the concordance $\CC$CDM model. Notice that there is no significant difference in the best fitted $\CC$CDM value of $\Omo$ when we employ SNIa+CMB+BAO$_{dz}$ or SNIa+CMB+BAO$_{A}$  data, as can be seen in Tables \ref{TableModelsBAOdz} and \ref{TableModelsBAOA}, and therefore the number of counts in Fig.~\ref{fig:NLCDM} does not depend on the BAO data used in the fit. We can see that the total number of counts increases with the redshift up to a maximum point and then decreases steadily, meaning that from that point onwards the larger is the redshift the smaller is the number of counts
of virialized halos with a mass $M$ in the indicated range. The decrease of number counts near our time is clearly caused by the repulsive effect of the DE, which becomes very important at redshifts $z\lesssim\mathcal{O}(1)$. The curves shown in that figure (which include the $1\sigma$ error in the fitted value of $\Omo$) define the fiducial $\CC$CDM prediction. We will use it to compare with the corresponding outcome from the dynamical vacuum models under study. Recall that we denote the deviations of the number counts of a given vacuum model with respect to the $\CC$CDM as $\delta\mathcal{N}=\mathcal{N}-\mathcal{N}_{\CC{\rm CDM}}$.

%%%%%%%%%%%%%%%%%%%%%%%%%%%%%%%%%%%%%%%%%%%%%%%%%%%%%%%%%%%%%

\begin{figure}[t!]
\begin{center}
{\includegraphics[scale=0.50]{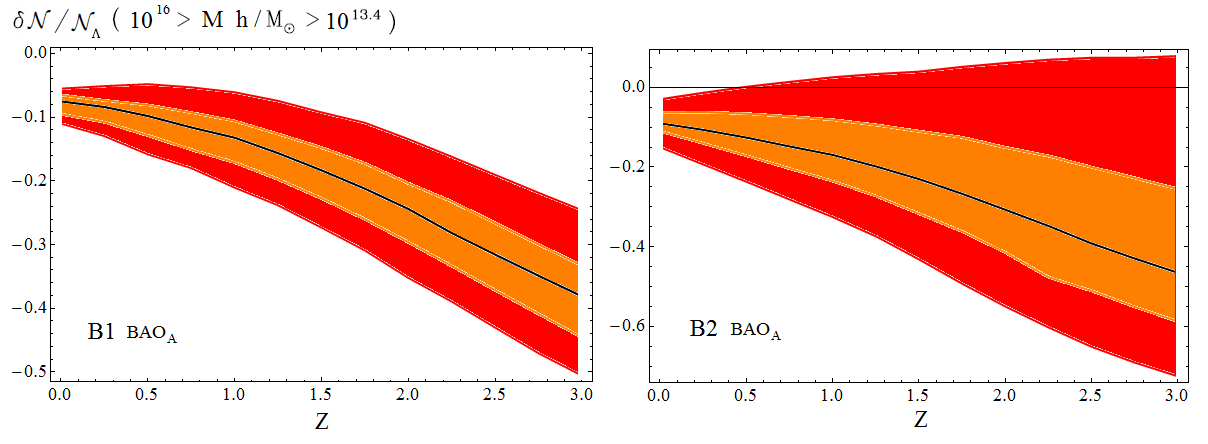}}
\end{center}
\caption[$\delta {\cal N}/{\cal N}$ for the
type-B model using the SNIa+CMB+BAO$_{A}$ data]{\scriptsize{Fractional difference $\delta {\cal N}/{\cal N}$ in the number of counts of clusters between the
vacuum models B1 (left) and B2 (right) and the concordance $\CC$CDM model, using SNIa+CMB+BAO$_{A}$ data from Table \ref{tableFitBAOA}. Same notation as in Fig.\,\ref{NC A1A2 BAO_dz}.}} \label{NC:B1B2 BAO_A}
\end{figure}

%%%%%%%%%%%%%%%%%%%%%%%%%%%%%%%%%%%%%%%%%%%%%%%%%%%%%%%%%%%%%%%%%%%%%%%%

We start our comparison by considering Fig.\,\ref{NC A1A2 BAO_dz}, where we display (on the left plot of it) the fractional difference $\delta {\cal N}/{\cal N}$ in the number of counts of clusters between the
vacuum model A1 and the concordance $\CC$CDM model (cf. Fig.\,\ref{fig:NLCDM}) using SNIa+CMB+BAO$_{dz}$ fitting data from Table \ref{TableModelsBAOdz}.
The continuous solid line in the figure represents the predicted deviation $\delta {\cal N}/{\cal N}$ for the best fit value from that table,  whereas the inner and outer bands comprise the $\delta {\cal N}/{\cal N}$ prediction for the points within $\pm 1\sigma$ and  $\pm 3\sigma$ values around it, respectively. The plot on the right of Fig.\,\ref{NC A1A2 BAO_dz} is similar, but for model A2. The corresponding results for models A1 and A2 when SNIa+CMB+BAO$_{A}$ fitting data from Table \ref{TableModelsBAOA} are used can be seen in the two plots of Fig. \ref{NC:A1A2 BAO_A}.

We can summarize the analysis presented in Figs.\,\ref{NC A1A2 BAO_dz} and \ref{NC:A1A2 BAO_A} by saying that $\delta {\cal N}/{\cal N}$  can have both signs in the case of using BAO$_{dz}$ data, provided we consider the points in the $\pm 3\sigma$ band. The narrower $\pm 1\sigma$ band is nevertheless more predominantly bent into the negative sign. As for the  BAO$_{A}$ data, the prediction for  $\delta {\cal N}/{\cal N}$ is negative for all points, even for those in the $\pm 1\sigma$ band. It means that, all in all, models A1 and A2 tend to predict a smaller number of counts as compared to the $\CC$CDM. The fractional decrease can be as significant as $30-60\%$.

The corresponding deviations in the number of counts for models B1 and B2 are depicted in Figs.\,\ref{NC:B1B2 BAO_dz} and \ref{NC:B1B2 BAO_A}. Here we find a feature that was not present for type-A models, namely we observe from these figures that the deviations with respect to the $\CC$CDM are all positive when the BAO$_{dz}$ data are used (cf. Fig.\,\ref{NC:B1B2 BAO_dz}), whilst they are negative when the BAO$_{A}$ data are utilized (cf. Fig. \,\ref{NC:B1B2 BAO_A}). This may seem surprising, but is related to the sensitivity of the number counts to the best fit value of $\Omo$ employed in the analysis, which is different for each type of BAO. As we have seen from Table \ref{tableFitBAOdz}, the BAO$_{dz}$ fitting data projects a value of $\Omo$ that is closer to the $\CC$CDM value than in the case of BAO$_A$ (cf. Table \ref{tableFitBAOA}). In the latter, $\Omo$ is significantly smaller than in the $\CC$CDM model. The sign of $\delta{\cal N}/{\cal N}$ is tied to this fact. As it is shown in Sect. \ref{subsec:understandingNumberCount}, if a given vacuum model has the same $\Omo$ value (or very similar), the sign of $\delta{\cal N}/{\cal N}$ is opposite to the sign of the vacuum parameter $\nu$ or $\epsilon$ that dominates the model. For models B1 and B2 with BAO$_{dz}$ data, the best fit value of $\Omo$ is indeed very close to the fitted value for the $\CC$CDM.  Thus, since for these models  $\epsilon<0$ we find $\delta{\cal N}/{\cal N}>0$ and moreover this fraction is growing quite fast, up to $50-100\%$ and more (Fig.\,\ref{NC:B1B2 BAO_dz}). At variance with this situation, with BAO$_{A}$ data these models predict a substantial depletion in the number counts as compared to the $\CC$CDM, as shown in Fig.\,\ref{NC:B1B2 BAO_A}, the reason being the smaller preferred value of $\Omo$ as compared to the concordance model.

\begin{figure}[t!]
\begin{center}
{\includegraphics[scale=0.50]{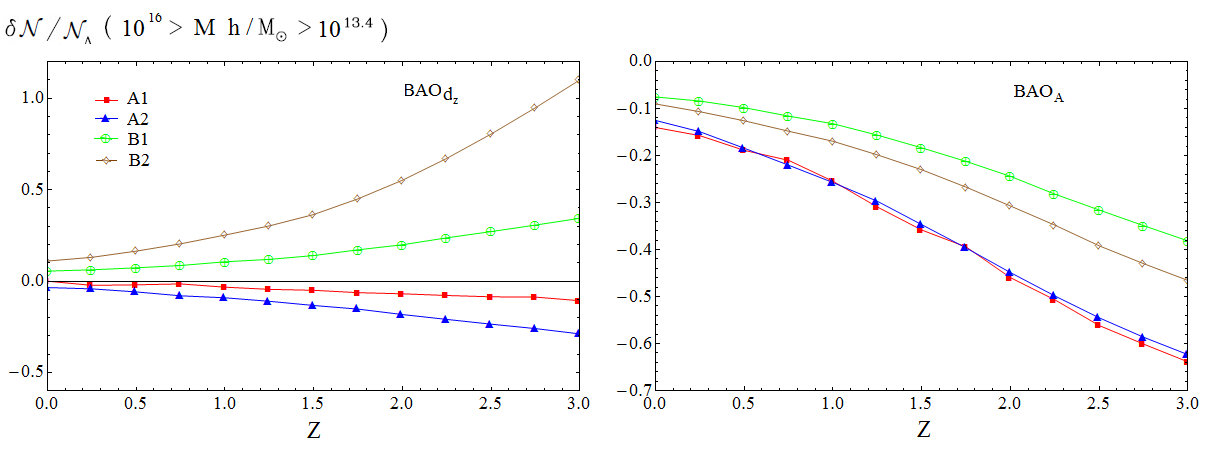}}
\end{center}
\caption[$\delta {\cal N}/{\cal N}$ for all the vacuum models]{\scriptsize{{\bf Left}: Comparison of the fractional difference $\delta {\cal N}/{\cal N}$ in the redshift distribution of cluster number counts of all the
vacuum models with respect to the concordance $\CC$CDM model using the central fit values of the SNIa+CMB+BAO$_{dz}$ data (cf. Table \ref{tableFitBAOdz}); {\bf Right}: As before, but for the SNIa+CMB+BAO$_{A}$ data (cf. Table \ref{tableFitBAOA}).}} \label{NC:CentralFits}
\end{figure}

In Fig.\,\ref{NC:CentralFits} we have put in a nutshell the essential results of our number counts analysis. Namely, we have displayed the fractional differences $\delta{\cal N}/{\cal N}$ with respect to the $\CC$CDM by using only the best fit values of all the vacuum models in the two BAO modalities. Obviously we need an improvement of the two sorts of BAO measurements to see it they can eventually provide a more coincident best fit value of $\Omo$, as this is essential to decide on the sign of $\delta{\cal N}/{\cal N}$. From our point of view perhaps the least model-dependent BAO results are those from BAO$_{A}$, as they are based on low-z data only and therefore are not so tied to the specific behavior of the models around the drag epoch.

%%%%%%%%%%%%%%%%%%%%%%% FIGURE  2 %%%%%%%%%%%%%%%%%%%%%%%%%%%%%%%%%%%%%

%%%%%%%%%%%%%%%%%%%%%%%%%%%%%%%%%%%%%%%%%%%%%%%%%%%%%%%%%%%%%%%%%%%%%%%%

Finally, as a particular case of our general treatment of type-A and type-B models, we briefly mention the situation with the class of C1 models (where e.g.  the linear model $\rL\propto H$ is also included). We have emphasized in Sect. \ref{subsec:growthrate} that the C1 models perform a rather bad fit to the linear growth of density perturbations. Recently, however, the number count analysis of the linear model $\rL\propto H$ has been considered in Ref.\,\cite{Carn14}, in which a significant excess in the number of counts is reported as compared to the $\CC$CDM model. As previously indicated in Sect. \ref{subsec:solvingC1C2}, the dynamics of this model just follows from that of the general type-B model by setting $C_0=0$ and $C_2=0$ in Eq.\,(\ref{B1B2}). The linear model has no free parameters apart from $\Omo$ or $\OLo$, and its vacuum energy density is completely determined by Eq.\,(\ref{eq:rLproptoH}). Although it is not part of our main purpose, we have computed in passing the corresponding number of counts for this model. Unfortunately, we do not concur with the results of\,\cite{Carn14}. Namely, we do not find an excess in the number of counts as compared to the $\CC$CDM. On the contrary, we find a large deficit. This shortage in the number of collapsed structures could actually be expected from the apparent lack of power of these models with respect to the $\CC$CDM, as can be seen in Fig. \ref{sigma8TypeA}. We can reconfirm this situation by computing the growing rate index of this model and verify the discrepancy with respect to the $\CC$CDM model or any of the type-A or B models. We have checked that the deviation is huge and hence unacceptable. On general physical grounds one may argue that the rapid increase of the (positive) vacuum energy density in the past for the linear model $\rL\propto H$ should severely prevent the formation of structures. Since the main task of this section is to report on the number counts for the dynamical models that pass all the basic tests on background expansion and linear structure formation, we do not extent further our considerations for models behaving otherwise.

To conclude, in view of the results found in our analysis of the cluster
halo redshift distribution presented in Figures \ref{NC A1A2 BAO_dz}-\ref{NC:CentralFits}, we can assert that it is an efficient method to distinguish the various sorts of dynamical vacuum models with respect to the $\CC$CDM and also among themselves, especially when the different sources of BAO data will become more precise. The sensitivity of the method to the parameters $(\nu,\alpha,\epsilon,...)$ of the vacuum models is large if we take into account that they are relatively small. We have found that the fractional differences $\delta{\cal N}/{\cal N}$ with respect to the $\CC$CDM can typically be as large as $\pm 50\%$ despite the fact that the (absolute) values of those parameters are typically of order $10^{-3}$.

%%%%%%%%%%%%%%%%%%%%%%%%%%%%%%%%%%%%%%%%%%%%%%%%%%%%%
%%%%%%%%%%%%%%%%%%%%%%%%%%%%%%%%%%%%%%%%%%%%%%%%%%%%%
%%%%%%%%%%%%%%%%%%%%%%%%%%%%%%%%%%%%%%%%%%%%%%%%%%%%%

\subsection{Understanding how the cluster number counts method works}
\label{subsec:understandingNumberCount}

We have shown that type A and B models of the vacuum energy
successfully fit all known cosmological data, including linear
structure formation, in a way comparable to the $\CC$CDM. However we
would like to find a way to lift their alike performance and be able
to distinguish them in a practical way. We have shown that the
number counts method is a good method to accomplish this aim. To
understand semianalytically why the method works, it will suffice to
consider the original Press-Schechter function defined in
Sect.\,\ref{sec:PSformalism}. For convenience let us define the
ratios
\begin{equation}
{\cal
T}(M)\equiv\frac{\int_{0}^{\infty}{k^{n+2}T^2(\Omega_m^{(0)},k)W^2(kR)dk}}{\int_{0}^{\infty}{k^{n+2}T^2(\Omega_m^{(0)},k)W^2(kR_8)dk}}\quad
{\rm and}\quad D_N(z)\equiv\frac{D(z)}{D(0)}\,.
\end{equation}
In this way from (\ref{s888}) we have
$\sigma^2(z)=\sigma^2_8\,D_N^2(z)\,{\cal T}(z)$, and we can rewrite
(\ref{eq:Nzv2}) as follows:
\begin{equation}\label{eq:N(z)}
\mathcal{N}(z)=-\frac{4\pi r^2\,\bar{\rho}}{H_0 E(z)}\,\int_{M_1}^{M_2}\frac{1}{M}\left(\frac{1}{\sigma}\frac{d\sigma}{dM}\right) f_{PSc}(\sigma)dM
=-4\sqrt{2\pi}\bar{\rho}\frac{c}{H_0}\left(\frac{\delta_cr^2(z)}{E(z)\sigma_8
D_N(z)}\right)\,I^{(\nu)}\,.
\end{equation}
In the last step we have used explicitly the original form of the
Press-Schechter function $ f_{PSc}(\sigma)$, and we have defined the
integral
\begin{equation}\label{eq:defIntegral}
I^{(\nu)}\equiv \int_{M_1}^{M_2}\frac{dM}{M}\frac{1}{{\cal
T}}\frac{d\sqrt{{\cal T}}}{dM}e^{-\frac{\delta_c^2}{2\sigma_8^2D_N^2
{\cal T}}}\,.
\end{equation}
Using the generalized forms (\ref{PS-Reed}) or (\ref{PS-ST}) does
not alter the explanation why the method works, and for this reason
we restrict ourselves to the canonical one.

The variations with respect to the $\CC$CDM model  should come from
the variations in the terms in the big parenthesis on the
\textit{r.h.s.} of Eq.\,(\ref{eq:N(z)}), as well as from the
integral (\ref{eq:defIntegral}). The other ingredients of
$\mathcal{N}(z)$ should not depend on the model details in a
significant way. Let us take model A1 and assume that there is only one parameter in
this dynamical vacuum model, say $\nu$. Expanding around $\nu=0$,
i.e. around the $\CC$CDM case, we can get the departure terms:
\begin{equation}
\left(\frac{r^2(z)}{E(z)}\right)^{(\nu)}=\left(\frac{r^2(z)}{E(z)}\right)^{(0)}
+\delta a_1\,;\ \ \ (\sigma_8 D_N(z))^{(\nu)}=(\sigma_8
D_N(z))^{(0)}+\delta a_2\,;\ \ \
\delta_c^{(\nu)}=\delta_c^{(0)}+\delta a_3 \,.
\end{equation}
Notice that all the $\delta a_i$ in the previous expression are
proportional to $\nu$, and therefore very small compared to the
leading terms.  Let us warn the reader that it would be
inappropriate to expand the exponential in the integrand of
(\ref{eq:N(z)}) in the same way, as the linear approximation would
be insufficient for the typical values of $\nu$ found in our
analysis.  The number counts formula (\ref{eq:N(z)}) therefore
yields
\begin{eqnarray}
\mathcal{N}(z)&=&-4\sqrt{2\pi}\bar{\rho}\frac{c}{H_0}\left(\frac{\delta_cr^2(z)}{E(z)\sigma_8 D_N(z)}\right)^{(0)}\nonumber\\
&&\times\left[1+\delta
a_1\left(\frac{E(z)}{r^2(z)}\right)^{(0)}-\frac{\delta
a_2}{(\sigma_8 D_N(z))^{(0)}}+\frac{\delta
a_3}{\delta_c^{(0)}}+\mathcal{O}(\nu^2)\right]\, I^{(\nu)}\,.
\end{eqnarray}
In this way we can compute the  variation in the number of clusters
(at a given redshift) with respect to the $\CC$CDM, and assuming that the value of $\Omega_m^{0}$ is the same in the two models, i.e.
$\delta\mathcal{N}=\mathcal{N}(\nu)-\mathcal{N}(\nu=0)$. The
corresponding relative variation can be cast as
\begin{equation}\label{eq:N(z)2}
\frac{\delta\mathcal{N}}{\mathcal{N}}=\underbrace{\frac{I^{(\nu)}-I^{(0)}}{I^{(0)}}}_{T0}+\underbrace{\delta
a_1\left(\frac{E}{r^2}\right)^{(0)}\frac{I^{(\nu)}}{I^{(0)}}}_{T1}-\underbrace{\frac{\delta
a_2}{(\sigma_8D_N)^{(0)}}\frac{I^{(\nu)}}{I^{(0)}}}_{T2}+\underbrace{\frac{\delta
a_3}{\delta_c^{(0)}}\frac{I^{(\nu)}}{I^{(0)}}}_{T3}\,.
\end{equation}
\begin{table}[t]
\centering
\begin{tabular}{|c|c|c|c|c|c|c|c|c|}
\hline
$\nu$ & $\sigma_8$ & $D_N(z=2)$ & $\delta_c(z=2)$ & $T0$ & $T1$ & $T2$ & $T3$ & $\frac{\delta\mathcal{N}}{\mathcal{N}}$ \\
\hline
-0.0017 & 0.829 & 0.4315 & 1.695 & 0.096 & -0.0034 & -0.015 & 0.0065 & 0.084\\
\hline
0.0017 & 0.794 & 0.4362 & 1.675 & -0.139 & 0.0027 & 0.016 & -0.0051 & -0.125\\
\hline
-0.004 & 0.854 & 0.4282 & 1.710 & 0.276 & -0.0092 & -0.047 & 0.0189 & 0.239\\
\hline
0.004 & 0.770 & 0.4394 & 1.660 & -0.278 & 0.0052 & 0.030 & -0.0107 & -0.254 \\
\hline
\end{tabular}
\caption[Understanding how the cluster number counts method works.]{{\scriptsize Numerical evaluation of
${\delta\mathcal{N}}/{\mathcal{N}}$, i.e. the relative variation in
the number of counts as compared to the $\CC$CDM, see Eq.
(\ref{eq:N(z)2}). We consider different values of the $\nu$
parameter at fixed  $z=2$ and provide also the breakdown of the
result in the individual contributions $T0-T3$. For the numerical
computation we have used $\Omega_b(z=0)=0.022242h^{-2}$,
$\Omega_m(z=0)=0.284$, $\delta_c^{(0)}=1.675$,
$\sigma_8^{(0)}=0.811$ and $D_N^{(0)}(z=2)=0.435088$.}}
\label{Appe1}
\end{table}

\noindent The numerical evaluation of the various terms of this expression is
displayed in Table \ref{Appe1}. It clearly shows that the dominant
term is $T0$ in (\ref{eq:N(z)2}). The terms $T1-T3$ are all of them
proportional to $\delta a_i$ and hence to $\nu$. Since $\nu={\cal
O}(10^{-3})$ all the terms proportional to it are of the same order
of magnitude. The $T0$-term is not, and it becomes the leading one.
Here is where the main contribution comes from, which is typically
two orders of magnitude larger than $\nu$ and hence it can reach the
order $10\%$ rather than $1$ per mil. This feature is at the root of
the main difference of this method with respect to the linear
perturbations analysis. In the latter  the deviations of the
dynamical vacuum models with respect to the linear growth rate
of the $\CC$CDM are proportional to $\nu$ and therefore cannot be
distinguished. Here, instead, the relative differences become magnified thanks to the nonperturbative effects associated to (\ref{eq:defIntegral}). In addition, we note from Table \ref{Appe1} that
there are significant differences for different values of $\nu$
within the same order of magnitude, which are also sensitive to sign
changes of the parameter. In the present case the sign of $\delta{\cal N}$ is opposite
to the sign of $\nu$, but this is because the value of $\Omo$ for the dynamical vacuum model that we have analyzed is the same as in the $\CC$CDM, but in general there is no such sign correlation. What is important is that using the number count method we expect visible effects that would remain almost invisible in the linear approach owing to the small values of the model parameters. This
is corroborated in the numerical analysis presented in Sect.\,\ref{subsec:HaloMasFunction}.

%%%%%%%%%%%%%%%%%%%%%%%%%%%%%%%%%%%%%%%%%%%%
%%%%%%%%%%%%%%%%%%%%%%%%%%%%%%%%%%%%%%%%%%%%
%%%%%%%%%%%%%%%%%%%%%%%%%%%%%%%%%%%%%%%%%%%%

\section{More about type-B (and C1) models} 
\label{sec:ExtraTypeB}

In this section we analyze in deeper detail type-B and C1 models. We go a step further in their characterization. The output of this study allows us to reinforce even more the results obtained in previous sections. We can definitely discard type-C1 models, since they are incapable of explaining the large scale structure formation in the Universe at both, the linear and non-linear regimes. In particular, we focus our attention on type-C1 models with $\nu=0$ (and $\epsilon\ne 0$). From now on this model will be referred to as C1B, whereas C1A will denote the model with $\nu\ne 0$ and $\epsilon\ne 0$,
\be
{\rm C1A}:\quad\epsilon+\nu=\Omega^0_\Lambda \quad \ \ \ \
{\rm C1B}:\quad\epsilon=\Omega^0_\Lambda\quad (\nu=0)
\,.\label{eq:constraintsIandII}
\ee
As mentioned before, the linear type-C1 model has been amply discussed in the literature by different authors from the theoretical and phenomenological perspective. We hope that the present analysis will serve to definitely rule out not only C1B, but the larger class of type-C1 models with no constant additive term in the formula of $\rho_\Lambda$.

%%%%%%%%%%%%%%%%%%%%%%%%%%%%%%%%%%%%%%%%%%%%%%%%%%%%%
\begin{table*}
\centering
\begin{tabular}{| c |  c |c | c | c | }
\multicolumn{1}{c}{Model} & \multicolumn{1}{c}{$\Omo$} &
\multicolumn{1}{c}{$\nu=1-\zeta$} & \multicolumn{1}{c}{$\epsilon$} &
\multicolumn{1}{c}{$\chi^2/dof$}  \\\hline $\CC$CDM & $0.293\pm
0.013$ & - & - & $567.8/586$ \\\hline C1B & $0.302^{+0.015}_{-0.014} $
& -  & $1-\Omo$ & $575.5/585$ \\\hline C1A & $0.295^{+0.016}_{-0.011}
$ &  $1-\Omo-\epsilon$  & $0.93^{+0.01}_{-0.02}$ & $567.7/584$
\\\hline B1 & $ 0.297^{+0.015}_{-0.014}$ & -  &
$-0.014^{+0.016}_{-0.013}$ & $587.2/585$  \\\hline ${\rm B2a}$ & $
0.300^{+0.017}_{-0.003}$ &$ - 0.004\pm 0.002 $ & $- 0.004\pm 0.002$
& $583.1/585$ \\ \hline ${\rm B2b}$ & $ 0.297^{+0.005}_{-0.015}$ &$
- 0.002\pm 0.002 $ & $-0.001\pm0.001$ & $579.5/585$ \\ \hline
 \end{tabular}
\caption[Fitting results for type-B and C1 models.]{{\scriptsize The fit values for the various models, together with their
statistical significance according to a $\chi^2$-test. We have
performed a joint statistical analysis of the SNIa+CMB+BAO$_{dz}$
data for the $\CC$CDM and type-B models. For type-C1 models, instead, we have used SNIa+BAO$_{A}$ data for the
reasons explained in the text. To break parameter degeneracies we
present the fitting results for two different cases: the one
indicated as B2a (resp. B2b) corresponds to $\nu=\epsilon$ (resp.
$\nu=2\epsilon$). Recall that because of the constraints
(\ref{eq:constraintsIandII}) model C1B has $\Omo$ as the sole free
parameter, whereas for model C1A one can adopt $\Omo$ and $\epsilon$.
\label{tableExtra}}}
\end{table*}
%%%%%%%%%%%%%%%%%%%%%%%%%%%%%%%%%%%%%%%%%%%%%%%%%%%%%%%%%%%%%%%%%%%

We have seen in section \ref{subsec:solvingC1C2} that type-C1 models suffer from severe problems already at the background level. Due to the anomalous scaling law for the matter energy density at high redshifts, i.e. $\rho_m\propto (\Omo)^2$, these models demand a value of $\Omo$ significantly larger ($\sim 70\%$) than the standard one. This is not only a requirement to be fulfilled so as to fit properly the CMB data, but also in order to obtain a value of the transition redshift $z_{tr}$ close to the one encountered in the $\Lambda$CDM-like models. In Table \ref{tableExtra} we show the fitting results for all the models under study. For type-B models we have used a
joint statistical analysis involving the SNIa data, BAO$_{dz}$ and the
CMB shift parameter (see Sect. \ref{sec:fitting} for details). We have proceeded in a different manner with type-C1 models due to the fact that the usual fitting formulas for computing the redshifts at decoupling and the baryon drag epochs
provided in \cite{HuSugiyama} and \cite{EisensteinHu98}, respectively, work only as good approximations for those models with small departures from the $\Lambda$CDM. While this is the
case for type-B models, this is not so for type-C1, for which the additive term is  $C_0= 0$. For this reason, for the type-C1 models we have
implemented the fitting procedure by just concentrating on the low
and intermediate redshifts, that is to say, we have used the type Ia
supernovae data but avoided using CMB data. At the same time for
these models we have used Eisenstein's BAO parameter $A(z)$ (cf. again Sect. \ref{sec:fitting} for details). So notice that for type-C1 models we are not forced to consider the radiation effects that become important at large redshifts. We just have to take into account the small corrections introduced in a matter-dominated Universe and do not worry about the dominance of relativistic matter over the non-relativistic one in the RD epoch. 

Proceeding in this way we can see from Table \ref{tableExtra} that the fitting
values of $\Omo$ associated to type-C1 models are not very
different from those of type-B models, and all of them are
reasonably close to the $\CC$CDM model (which is also included in
that table and fitted from the same data). This is because $\rho_m\propto\Omo$ at low redshifts in type-C1 models, and therefore, the aforementioned anomaly is invisible if only low and intermediate redshift observables are used in the fitting analysis. From this point of view
(and attending also to the $\chi^2$ values per d.o.f.) we can say
that all the models perform an acceptable fit to the cosmological
data. For type-C1 models, however, we can attest this fact only for
the low and intermediate redshift data. If we include the CMB shift
parameter and the BAO$_{dz}$ data, type-C1 models then peak at
around $\Omo\sim 0.5$ (and with a bad fit quality).
Such poor performance is caused by the 
$\rmr\propto\left(\Omo\right)^2$ anomalous behavior of these models
at large redshift.

\begin{figure}
\centering
\includegraphics[scale=0.7]{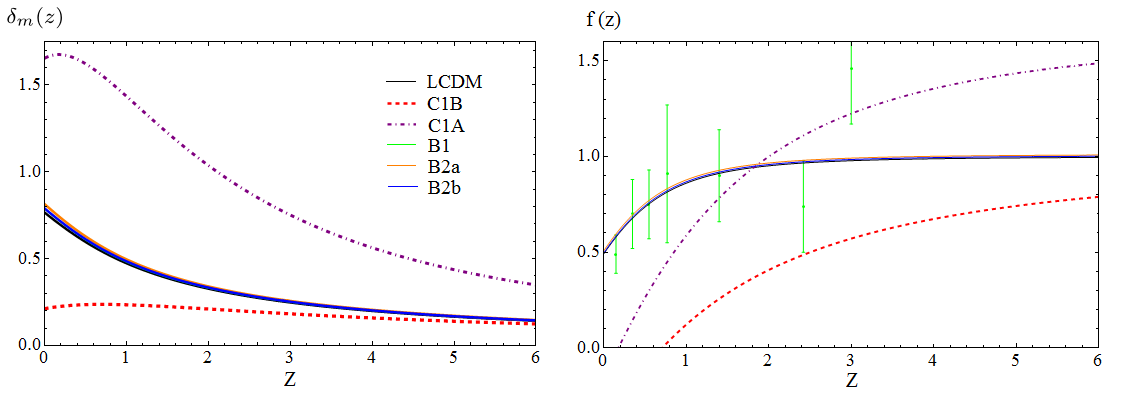}
\caption[$\delta_m(z)$ and $f(z)$ for type-B a C1 models.]{\scriptsize{{\bf Left:} The density contrast
$\delta_m(z)$ predicted by the various models under study using the fit values collected in Table \ref{tableExtra};
{\bf Right:} Comparison of the observational data (see text) --
with error bars depicted in green --  and the theoretical evolution
of the linear growth rate of clustering $f(z)$, confer
Eq.\,(\ref{eq:growingfactor}), for each vacuum model. Type-B models are
almost indistinguishable from the $\CC$CDM one. \label{fig:f(z)MNRAS}}}
\end{figure}

Even if we restrain to the low and intermediate redshift data for type-C1
models, which as we have seen lead to an acceptable value
of $\Omo\simeq 0.3$ (cf. Table \ref{tableExtra}), they nevertheless clash violently
with a serious difficulty, namely they are bluntly unable to account
for the linear structure formation data, as it is plain at a glance
on Fig.\,\ref{fig:f(z)MNRAS}, where both $\delta_m(z)$ and $f(z)$ have been plotted for the models under study together with the
$\CC$CDM. The observational data points in the right plot have been taken from Table 1 of \cite{JesusEtAl2011}, see references therein.

The obvious departure of type-C1 models from the linear growth data
is an important drawback for these models. It implies that the
initial success in fitting the Hubble expansion data cannot be
generalized to all low redshift data. Such situation is in contrast
to type-B models, which are able to successfully fit the
linear growth data at a similar quality level as the $\CC$CDM, as
can also be appreciated in Fig.\,\ref{fig:f(z)MNRAS}. In fact, the three
curves corresponding to type-B and $\CC$CDM models (for the
best-fit values of the parameters in Table \ref{tableExtra}) lie almost on top of
each other in that figure, whereas the curves for type-C1 models depart very openly from the group of $\CC$CDM-like models.  For the type-$C1B$
model there is an evident defect of structure formation with
respect to the $\CC$CDM, whilst for the C1A one there is a notable
excess.

\begin{figure}
\centering
\includegraphics[scale=0.6]{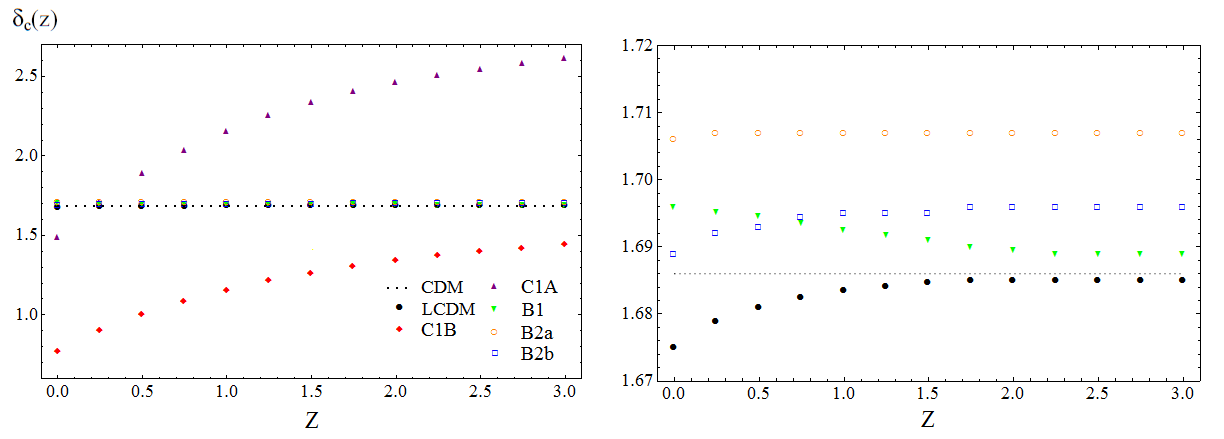}
\caption[$\delta_c(z)$ for type-B and C1 models obtained from Table \ref{tableExtra}.]{\scriptsize{Computation of the collapse density threshold $\delta_c(z)$
using the best fit values shown in Table \ref{tableExtra}, see Appendix \ref{ch:appCollapse} for details. In both plots we include the fiducial constant CDM
value $\delta_c=\frac{3}{20}(12\pi)^{2/3}\approx 1.686$ (horizontal
dotted line) and the $\CC$CDM curve (solid points, in black). The
models with $C_0\ne 0$ (i.e. type-B models) provide $\delta_c(z)$ very
close to the $\CC$CDM model and the corresponding curves are
cluttered in the plot on the left. In the right plot we zoom in the
relevant region $\delta_c^{CDM}$$^{+0.024}_{-0.016}$ in order to
clearly appreciate the differences between them. In the plot on the
left these differences cannot be seen owing to the large deviations
shown by type-C1 models ($C_0=0$) which required to use a large
span for the vertical axis.
\label{fig:deltacExtra}}}
\end{figure}

The large differences can be explained as follows. As we have seen
before the ratio $\rho_\CC/\rho_r$ for the type-C1A model is far from 0,
and negative, in the far past. Now, let us take the acceleration equation during the matter-dominated epoch the
acceleration of the expansion is given by $\ddot{a}/{a}=(4\pi
G/3)(2\rL-\rmr)$. Thus, a negative value of the vacuum energy
density, $\rL<0$, helps to slow down the expansion (it actually
cooperates with gravitation and enhances the aggregation of matter
into clusters). In point of fact, the vacuum energy of model C1A did not
become positive until $H(\tilde{z})=-\epsilon
H_0/\nu\approx 4.13 H_0$, what corresponds to a redshift
$\tilde{z}=3.204$. This is why it yields larger values of the density
contrast in comparison with the models that take $C_0\ne 0$ (cf.
Fig.\,\ref{fig:f(z)MNRAS}). Later on the universe
started to speed up, and the transition value from deceleration to acceleration is given by \ref{inflectionC1}. From the values of the fitted parameters in Table \ref{tableExtra}, we find $z_{tr}=1.057$. Numerically, it is significantly larger than in the $\CC$CDM ($z_{tr}\simeq 0.69$, for the central fit value of $\Omo$ quoted in Table \ref{tableExtra}). From this point onwards the type-C1A vacuum has been accelerating the universe and restraining the gravitational
collapse, but it has left behind a busy history of structure
formation triggered by the large growth rate $\delta_m(a)\sim a^{3\zeta-2}= a^{1.675}$ (cf. the fit value $\zeta=1.225$ from Table \ref{tableExtra}). Such history is difficult to reconcile with the (much more
moderate) one indicated by observations.

In the other extreme we have type-C1B model, showing a serious lack of
structure formation as compared to the $\CC$CDM  (cf.
Fig.\,\ref{fig:f(z)MNRAS}), despite for both models $\delta_m(a)\sim a$.  We can also understand the reason as follows. Let us assume a common value of the density parameter $\Omo$ (which is a good
approximation under the fitting strategy we have followed in Table
\ref{tableExtra}). In that case the ratio of their vacuum energy
densities is: $\rho^I_\CC/\rho_\CC^{\CC CDM}=1+\Omo(a^{-3/2}-1)$. Thus, during the past cosmic history the vacuum energy density for the type-C1B model  is positive and always larger
than in the concordance model, so we should expect a reduced growth rate as compared to the $\CC$CDM. This is confirmed in Fig.\,\ref{fig:f(z)MNRAS}.

\begin{figure}
\centering
\includegraphics[scale=0.55]{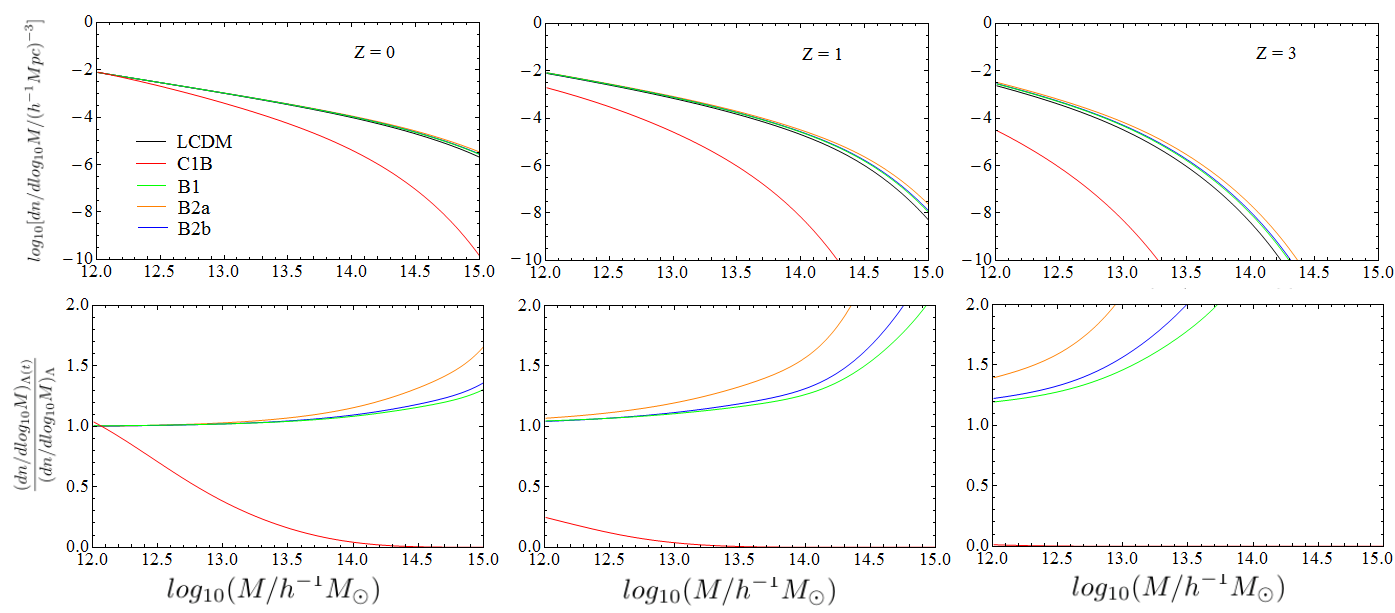}
\caption[Differential comoving number density as a function of the halo mass for the type-B, C1 and $\Lambda$CDM models]{{\scriptsize {\bf Upper plots:} The differential comoving number density
as a function of the halo mass for the type-B, C1 and $\Lambda$CDM models at redshifts $z=0$,
$z=1$ and $z=3$, respectively. {\bf Lower plots:} Corresponding
differences in the comoving number density with respect to the
$\CC$CDM model.\label{fig:dndlogM}}}
\end{figure}

\begin{figure}
\centering
\includegraphics[scale=0.5]{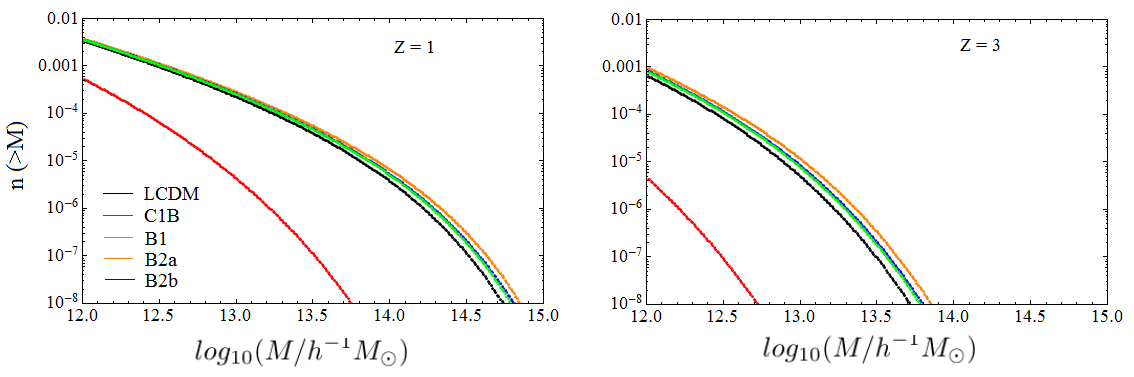}
\caption[Comoving number density for
the $\Lambda$CDM, B and C1B-type models]{{\scriptsize The comoving number density at two different redshifts for
the different models.\label{fig:n}}}
\end{figure}

\begin{figure}
\centering
\includegraphics[scale=0.68]{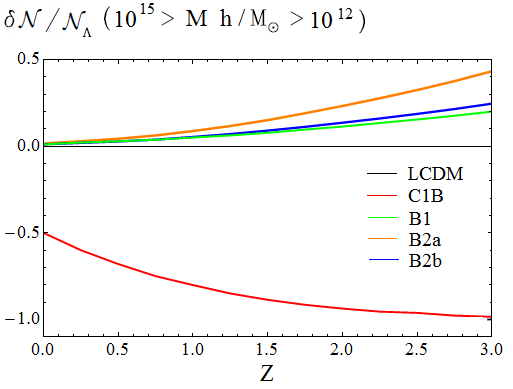}
\caption[$\delta\mathcal{N}/\mathcal{N}$ for the B and C1B type models]{{\scriptsize The fractional difference
$\delta\mathcal{N}/\mathcal{N}_{\CC{\rm CDM}}$ with respect to the
$\CC$CDM model (where we have defined
$\delta\mathcal{N}\equiv\mathcal{N}-\mathcal{N}_{\CC{\rm CDM}}$).
The curve for the type-C1A model is not plotted because it is out of
range, i.e. $\delta\mathcal{N}/\mathcal{N}> 1$ .\label{fig:DeltaN}}}
\end{figure}

Now, we analyze the nonlinear perturbations effects
at small scales and consider the different capability of the vacuum
models under study to produce cluster-size halo structures in the
Universe. This study will give strength to the results obtained at
the linear level. It is important to emphasize that $k_{eq}$ appearing in \ref{jtf} is a model
dependent quantity. For type-B models one can use the
same formula that is obtained in the $\CC$CDM \eqref{eq:keqLCDM} as a good approximation, due to the fact that
the deviations are small in these cases. On
the contrary, with type-C1 models we are not allowed to
do that. We must derive the corresponding expression for $k_{eq}$ (recall that $k_{eq}=a_{eq}H(a_{eq})$). The final
results read as follows:
\be\label{eq:keqC1Bmodel}
({\rm C1B})\qquad
k_{eq}=H_0\,\sqrt{\frac{2}{\Omega_r^{0}}}\,\left(\Omo\right)^{4/3}\,,\ee
\be\label{eq:keqC1Amodel}
 ({\rm C1A})\qquad
k_{eq}=\frac{H_0\sqrt{2}}{\zeta^{\frac{1}{3\zeta}+\frac{1}{2}}}\left(\Omo\right)^{
2-\frac{2}{3\zeta}}
\left(\Omega_r^0\right)^{\frac{1}{\zeta}-\frac{3}{2}}\,.
\ee
Using the artillery of Sect. \ref{sec:Number counts}, along with the best-fit values of Table
\ref{tableExtra} and the numerically determined collapse density
$\delta_c(z)$ (cf. Fig.\,\ref{fig:deltacExtra}), we have
computed the fractional difference $\delta\mathcal{N}/\mathcal{N}$ for the number counts of clusters between the dynamical
vacuum models and the concordance $\Lambda$CDM one. The differential
comoving number density of predicted cluster-size structures at
particular values of the redshift ($z=0,z=1$ and $z=3$), as well as
the normalized results with respect to the corresponding $\CC$CDM
prediction, are presented in Fig.\, \ref{fig:dndlogM}, whereas in
Fig.\,\ref{fig:n} we show the differences in the halo mass function
through the comoving number density for the various models at two
fixed redshifts ($z=1$ and $z=3$).  Finally, in
Fig.\,\ref{fig:DeltaN} we plot the redshift distribution of the
total number of counts.

These figures encapsulate all the main information on the number
counts analysis. They display the number of counts for each
model per mass range at fixed redshift, and the total number of
structures at each redshift within the selected mass range. The
upshot from our analysis is that the models with $C_0\ne 0$ predict
either a very small (type-C1B) or a very large (type-C1A) number of
clusters as compared to the $\CC$CDM. This is not surprising if we
inspect the power for structure formation of these models in the
linear perturbations regime (see Fig.\,\ref{fig:f(z)MNRAS}). As a result
we deem (again) unrealistic the situation for both the type-C1 models. 
When we translate this situation to the corresponding
prediction for the number counts we find that, for model C1B,  ${\cal
N}^{\rm C1B}/{\cal N}_{\CC{\rm CDM}}\ll 1$,   whereas for model C1A
${\cal N}^{\rm C1A}/{\cal N}_{\CC{\rm CDM}}\gg 1$ in the whole range.
As a result, the former yields $\delta{\cal N}/{\cal N}_{\CC{\rm
CDM}}\to -1$ at increasing redshifts (as can be appreciated in
Fig.\,\ref{fig:DeltaN}), whereas the latter is out of the
window under study.


In contrast, the situation with the $\CC$CDM-like models B
is quite encouraging. These models represent viable alternatives, at
least from the phenomenological point of view, to the strictly rigid
situation of the $\CC$CDM  (in which $\rL=$const. for the entire
cosmic history).  While these models depart only mildly from the
$\CC$CDM predictions near our time, the differences become sizeable
deep in the past, but still within bound. Concerning the number
counts differences with respect to the concordance model we
recognize from Fig.\,\ref{fig:DeltaN}  significant ($\sim 20-30\%$)
positive departures at moderate redshift ranges, where the total
number of counts is still sizeable. Therefore the predicted
deviations can be measured, in principle, and could be used as an
efficient method to separate type-B1 and B2 models.

%%%%%%%%%%%%%%%%%%%%%%%%%%%%%%%%%%%%%%%%%%%%
%%%%%%%%%%%%%%%%%%%%%%%%%%%%%%%%%%%%%%%%%%%%
%%%%%%%%%%%%%%%%%%%%%%%%%%%%%%%%%%%%%%%%%%%%

\section{Discussion and conclusions} 
\label{sec:conclusions}

In this chapter, we have analyzed in great detail several classes of dynamical vacuum
models in which the vacuum energy density can be expressed as a power series of
the Hubble function and its cosmic time derivative.
We have singled out model types which are particularly
attractive from the theoretical point of view, namely vacuum models for which the number of time derivatives of the scale
factor is even:
$\rL(t)=c_0+\sum_{k=1} \alpha_{k} H^{2k}(t)+\sum_{k=1}
\beta_{k}\dot{H}^{k}(t)$. These can be well motivated within the
context of quantum field theory (QFT) in curved spacetime since their structure is manifestly compatible with the general covariance
of the effective action and can be linked to the notion of
renormalization group. For the study of the current Universe the series naturally terminates at the level of the $H^2$ and $\dot{H}$ terms, but the higher order ones can be very important for a proper description of the early Universe and the inflationary phase.

We have stressed the need for the nonvanishing
additive (constant) term, $c_0\neq 0$, in the above class of models. It guarantees a smooth limit converging to the standard $\CC$CDM model when the
coefficients of the dynamical terms go to zero. For instance,  we have considered vacuum models of the form
$\CC=a_0+a_1\dot{H}+a_2H^2$, with $c_0\neq 0$ (the class of models that we have called
type-A). They are well-behaved and if the (dimensionless) coefficients $a_1$ and $a_2$ are
sufficiently small, the cosmological term develops just a mild dynamical behavior around
the $\CC$CDM model. Such framework could compete as a good candidate for a consistent description of the Universe in terms of dynamical vacuum energy, an option that should be considered
natural in QFT in curved spacetime. 

In contrasdistinction, we have verified
that models with  $c_0=0$ are generally in conflict with
observations, whether with the background data, or with the structure
formation data, or both. In the context of QFT in curved spacetime, the bad phenomenological behavior 
of models with $c_0=0$ has important implications, since it automatically forbids the use of the encouraging renormalization condition 
$\rho_{\Lambda,{\rm eff}}^{\rm Mink}=0$. We have seen in Sect. \ref{subsec:RVMintro} that this renormalization condition 
would strongly alleviate the old CC problem. Unfortunately, the possibility of having a zero $c_0$ is strongly excluded by observations. This fact implies that Minkowski spacetime cannot be an exact (global) solution of Einstein's equations, even in the ideal vacuum case. This means that vacuum automatically forces the spacetime to be dynamical. Put in other words, the non-zero value of $c_0$ endows the spacetime itself with an irremovable curvature not connected with matter.

In our analysis we have also admitted the possibility that some
terms in the effective structure of $\CC(H,\dot{H})$ could mildly violate
the covariance requirement on phenomenological grounds.
Notwithstanding, we considered this possibility viable only when the expected terms are also present. We do not deem
theoretically sound those vacuum models exclusively constructed from
noncanonical terms (i.e. unexpected terms not satisfying the above mentioned conditions), such as e.g. the model $\CC\propto H$. A model of this
sort has the double inconvenience that $c_0=0$ and that the number
of time derivatives of the scale factor is odd (one derivative in this case). Not surprisingly when
such model is confronted with observations fails on
several accounts. When we add up to it the power $H^2$, we reach
the model $\CC=c_1H+c_2H^2$ (referred to in this work as the type-C1 model). In this extended form the situation of the new model improves at the background level, but is still troublesome at the perturbations level since the model fails to describe the linear growth of structure formation. Similarly, the pure quadratic model
$\CC\propto H^2$ is problematic, but for a different reason. While this
model contains an even power of the Hubble rate, it is actually not sufficient to comply with the phenomenological requirement in the absence of a constant
additive term. The reason is that it does not admit an
inflection point from deceleration to acceleration, and moreover it
does not have a growing mode for structure formation (for reasonable
values of the cosmological parameters). The pure linear and quadratic models, $\CC\propto H$ and $\CC\propto H^2$, are therefore strongly excluded; and their combination, $\CC=c_1H+c_2H^2$, provides a model still crippled to account for the structure formation data. Similar (though not identical) criticisms can be applied to vacuum models of the sort $\CC=c_1\dot{H}+c_2H^2$ (type-C2). Hence, all type-C models are  unfavored, strongly disfavored or simply ruled out.

We have already mentioned that type-A dynamical vacuum models are in very good shape inasmuch as they are perfectly comparable to the concordance $\CC$CDM model when the dynamical components are subdominant in the current Universe. Interestingly, another viable variant that we have considered are the type-B models. These are obtained by including an additive term to the type-C1 models, i.e. they have the structure $\CC=b_0+b_1H+b_2H^2$ with $b_0\neq 0$. This class is the most difficult one to deal with, as the simultaneous presence of $b_0\neq 0$ and the linear term $\sim H$ complicates considerably the analytic treatment. We have nevertheless presented a full-fledged analysis of these models as well, and we have found that they are also, in principle, phenomenologically admissible.
For them the presence of the linear component is not so determinant as in the case of the type-C1, because it can be interpreted as a correction (e.g. a bulk viscosity effect) to the main structure. Most important, the type-B models have a smooth $\CC$CDM limit as in the type-A case, and this fact is again crucial to protect them from departing exceedingly from the concordance $\CC$CDM model near our time.

In the present chapter we have solved the background and perturbations
cosmology for all these vacuum models and confronted them with
observations. In the light of the recent observational data on type Ia supernovae,
the cosmic microwave background and the baryon acoustic
oscillations, we have obtained a fit to their basic parameters $(\nu, \alpha, \epsilon)$.  From the fitted values we have computed the linear growth factor of structure formation for each model and compared with the observed
linear growth rate of clustering measured by several galaxy surveys.
Subsequently we have moved to the nonlinear regime and considered
the predicted redshift distribution of cluster-size collapsed
structures as a powerful method to distinguish the models. We have computed the corresponding fractional deviation $\delta {\cal N}/{\cal N}$ in the number of counts of clusters with respect to the $\CC$CDM prediction.

The general conclusion we have reached is that the studied dynamical vacuum models (type-A and
type-B with nonvanishing additive constant term) are able to pass (with some differences) the combined observational tests, including the structure formation data, with a statistical significance that in some cases is comparable or even better than that of the
concordance $\CC$CDM model. The current Universe appears in all these models as FLRW-like, except that the vacuum energy is not a rigid quantity but a mildly evolving one. In fact, the typical values we have obtained for the coefficients $\nu$, $\alpha$ and $\epsilon$ responsible for the
time evolution of $\rL$ in these models lie in the ballpark of $\sim
10^{-3}$. This order of magnitude value is roughly consistent with the theoretical expectations, some of them interpreted in QFT as one-loop $\beta$-function coefficients of the running cosmological constant.

Despite the two types of viable dynamical vacuum models remain close to the $\CC$CDM model, the overall fit from type-B models is not so good as the type-A ones. We have pointed out that this may be due to the fact that the presence of the linear term $\sim \epsilon H$ (characteristic of type-B models, especially the type-B1 ones) is unexpected in the general structure of the effective action in QFT in curved spacetime. This is in contradistinction to the vacuum structure of type-A models, where all included terms are expected. Overall, this feature might be indicative that the A-class of models are both theoretically and phenomenologically preferred to the B-ones. However, it is too early for a final verdict, and more observational work may be necessary to decide. In the meanwhile we have shown that the two types of models could be distinguished from the point of view of the measured redshift distribution of cluster-sized collapsed structures in the Universe. We have found that they can show significant deviations (of order $\pm 50\%$) from the predicted redshift distribution in the concordance $\CC$CDM model. Our expectation is that when the upcoming and present X-ray and Sunyaev-Zeldovich surveys (such as eROSITA and SPT) will have collected enough statistics, it should be possible to decide about the best type of dynamical vacuum model from the phenomenological point of view.

In the course of our analysis we have also briefly pointed out the fact that generally the dynamical models under consideration in this chapter, and especially when fitted using the BAO$_A$ observable (which depends on low-redshift data on the acoustic $A(z)$-parameter), tend to provide a value of $\Omo$ significantly smaller than in the $\CC$CDM model. This would seem to be consistent with the possible dynamical character of the dark energy recently claimed in the literature on the basis of model-independent DE diagnostics\,\cite{SahniShafielooStarobinsky}.

To summarize, the dynamical vacuum models of the cosmic evolution may
offer an appealing and phenomenologically consistent perspective
for describing dynamical dark energy without introducing
extraneous dark energy fields. In that framework, dark energy is reinforced as
being nothing more, but nothing less, than dynamical $\CC$. This
could help to better understand the origin of the $\CC$-term and the vacuum energy density in the fundamental context of QFT in curved spacetime, and ultimately, it could shed light on the existing problems around the cosmological constant discussed in Chapter \ref{chap:Introduction} .

\section{Main bibliography of the chapter}

This chapter, together with Appendix \ref{ch:appCollapse}, are based on the contents of the papers \cite{JCAPnostre1} and \cite{MNRASnostre}:
\vskip 0.5cm
\noindent
{\it Dynamical vacuum energy in the expanding Universe confronted with observations: a dedicated study.}\newline
A. G\'omez-Valent, J. Sol\`a, and S. Basilakos\newline
JCAP {\bf 1501}, 004 (2015) ; arXiv:1409.7048

\newpage
\noindent
{\it  Vacuum models with a linear and a quadratic term in H: structure formation and number counts analysis.}\newline
A. G\'omez-Valent and J. Sol\`a\newline
Mon. Not. R. Astron. Soc. {\bf 448}, 2810 (2015) ; arXiv:1412.3785
\newpage

\chapter[${\cal D}$-class of dynamical DE models ]{Background history and cosmic perturbations for a general system of self-conserved dynamical dark energy and matter}
\label{chap:DynamicalDE}

In this chapter we determine the Hubble expansion and the general cosmic perturbation equations for a general system consisting of self-conserved matter, $\rmr$, and self-conserved dark energy (DE), $\rho_D$. While at the background level the two components are non-interacting, they do interact at the perturbations level. We show that the coupled system of matter and  DE perturbations can be transformed into a single, third order,  matter perturbation equation, which reduces to the (derivative of the) standard one in the case that the DE is just a cosmological constant. As a nontrivial application we analyze a class of dynamical models whose DE density
$\rho_D(H)$ consists of a constant term, $C_0$, and a series of powers of the Hubble rate. These models were analyzed in the previous chapter from the point of view of dynamical vacuum models, but here we treat them as self-conserved DE models with a dynamical equation of state parameter. We fit them to the wealth of expansion history and linear structure formation data and compare their fit quality with that of the concordance $\CC$CDM model. Those with $C_0=0$ include the so-called ``entropic-force'' and ``QCD-ghost'' DE models, as well as the pure linear model $\rD\sim H$, all of which appear strongly disfavored. The models with $C_0\neq 0$, in contrast, emerge as promising dynamical DE candidates whose phenomenological performance is highly competitive with the rigid $\CC$-term inherent to the $\CC$CDM.

The DE models under consideration in this chapter are closely related to those analyzed in the previous one, although with an important difference: now we consider that they describe a self-conserved dark energy, $\rD(H)$, with a dynamical equation of state evolving itself with the expansion history: $\wD=\wD(H)$. It means that $\rD(H)$ does not exchange energy with matter (which therefore remains also covariantly conserved). These assumptions imply a completely different new class of DE scenarios which requires an independent cosmological analysis. We call them the ``${\cal D}$-class'' of dynamical DE models, to distinguish them from the vacuum class (for which $\wD=-1$ at any time of the cosmic expansion). We undertake the task of examining in detail the ${\cal D}$-class here. Of particular interest is the analysis of the effective EoS parameter of the models in this class near our time.

In this chapter we provide detailed considerations not only on the background cosmology of the new DE models but also on the perturbation equations and their implications on the structure formation. The new perturbation equations are indeed formally different from the equations when the models are treated as dynamical vacuum models in interaction with matter. Of special significance is the formal proof that we provide according to which the coupled system of matter and DE perturbation equations for general models with self-conserved DE and matter components can be described by a single, third order, perturbation equation for the matter component. As a nontrivial application we subsequently solve (numerically) that equation for the  ${\cal D}$-class models and compare the results with the situation when the background DE density is present but the corresponding DE perturbations are neglected. The main, and very practical, outcome of our work is that some of these dynamical DE models can provide a highly competitive fit to the overall cosmological data as compared to the performance of the concordance  $\CC$CDM model -- based on a rigid $\CC$-term.

The layout of the chapter is as follows. We present our dynamical DE models in Sect. \ref{sec:DEmodels} and address their background solution and equation of state analysis in Section \ref{sec:background}. The matter and dark energy perturbations are considered in Section \ref{sec:LSF}.
The confrontation with the background history and linear growth rate data is performed in Sect. \ref{sec:FittingDE}. Finally, in Sect. \ref{sec:conclusionsDyn} we present our conclusions.

%%%%%%%%%%%%%%%%%%%%%%%%%%%%%%%%%%%%%%%%
%%%%%%%%%%%%%%%%%%%%%%%%%%%%%%%%%%%%%%%%
%%%%%%%%%%%%%%%%%%%%%%%%%%%%%%%%%%%%%%%%

\section{Dynamical DE models: the \texorpdfstring{$\mathcal{D}$}--class versus the vacuum class}
\label{sec:DEmodels}

Let us consider a (spatially) flat FLRW universe. Einstein's field equations read $G_{\mu\nu}=8\pi G\tilde{T}_{\mu\nu}$, with $G_{\mu\nu}$ the Einstein tensor and $\tilde{T}_{\mu\mu}=T^m_{\mu\nu}+T^D_{\mu\nu}$ the total energy-momentum tensor involving matter and dark energy densities. We take both the matter and DE parts of the energy-momentum tensor in the form of a perfect fluid characterized by isotropic pressures and proper energy densities $(p_i,\rho_i)$. One can then derive Friedmann's equation and the pressure equation by taking
the $00$ and the $ii$ components of Einstein's equations, respectively:
\be\label{eq:FriedmannEqMPLA} 3H^2=8\pi\,G\,(\rho_D+\rho_m+\rho_r)\,, \ee
\be\label{eq:PressureEqMPLA} 3H^2+2\dot{H}=-8\pi\,G\,(p_D+p_m+p_r)\,. \ee
\noindent As fluid components we consider cold matter, $p_m=0$, radiation, $p_r=\rho_r/3$, and a general dark energy (DE) fluid with dynamical equation of state (EoS): $p_D=\omega_D\rho_D$  ($\dot{\omega}_D\ne 0$). If the matter and DE densities are separately conserved (in the local covariant form which we will indicate explicitly) we shall speak of self-conserved densities. Assuming also that there is no transfer of energy between non-relativistic matter and radiation (which is certainly the case in the epoch under study), the equation of local covariant conservation of the total energy density following from (\ref{eq:FriedmannEqMPLA}) and (\ref{eq:PressureEqMPLA}) can be split into three equations -- reflecting the Bianchi identities of the Einstein tensor --  as follows:
\be\label{eq:MatterConsEq}
\dot{\rho}_m+3H\rho_m=0\,, \ee
\be\label{eq:RadConsEq}
\dot{\rho}_r+4H\rho_r=0\,,\ee
and
\be\label{eq:DEConsEq}
\dot{\rho}_D+3H\rho_D(1+\omega_D)=0\,. \ee
Similarly, we can e.g. check that from these local conservation laws and the pressure equation (\ref{eq:PressureEqMPLA}) we can reconstruct Friedmann's equation \eqref{eq:FriedmannEqMPLA}. It follows that as independent set of equations we can take either the pair (\ref{eq:FriedmannEqMPLA}) and (\ref{eq:PressureEqMPLA}), or Friedman's equation together with the local conservation laws. In the last case the pressure equation (\ref{eq:PressureEqMPLA}) is not independent. This will be our strategy in practice.

Trading now the cosmic time derivative for the derivative with respect to the scale factor, through the simple relation $d/dt=aH\,d/da$, the above conservation laws can be easily solved:
\be\label{eq:MatterConsSol}
\rho_m(a)=\rho_m^{(0)}a^{-3}\,, \ee
\be\label{eq:RadConsSol}
\rho_r(a)=\rho_r^{(0)}a^{-4}\,, \ee
\be\label{eq:DEConsEq2} \rho_D(a)=\rho_D^{(0)}\,\exp{\left\{-3\int_{1}^{a}\frac{da^\prime}{a^\prime}[1+\omega_D(a^\prime)]\right\}}\,, \ee
where the densities $\rho_i^{(0)}$ denote, again, the respective current values. In the last case it is assumed that $\wD(a)$ has been computed. However, our method will be actually the opposite, we will compute the DE density $\rD(a)$ first and then use (\ref{eq:DEConsEq}) to compute the function $\wD(a)$ -- see Eq.\,(\ref{eq:EoS}) below.

As it is transparent from (\ref{eq:MatterConsSol}) and (\ref{eq:RadConsSol}), the pressureless (non-relativistic) and the relativistic matter energy densities are described by the standard $\Lambda$CDM laws, but the Universe's evolution depends on the specific dynamical
nature of the dark energy density $\rho_D$. The following basic DE models
will be considered\,:
\begin{eqnarray}\label{eq:ModelsA}
\mathcal{D}A1:\phantom{XX} \rho_D(H)&=&\frac{3}{8\pi
G}\left(C_0+\nu H^2\right)\nonumber \\
\mathcal{D}A2:\phantom{XX} \rho_D(H)&=&\frac{3}{8\pi
G}\left(C_0+\nu H^2+\frac{2}{3}\alpha \dot{H}\right)\\
\mathcal{D}A3:\phantom{XX} \rho_D(H)&=&\frac{3}{8\pi
G}\left(C_0+\frac{2}{3}\alpha \dot{H}\right)\nonumber\\
\mathcal{D}C1: \phantom{XX}\rho_D(H)&=&\frac{3}{8\pi G}(\epsilon H_0H+\nu H^2)\nonumber\\
\mathcal{D}C2: \phantom{XX}\rho_D(H)&=&\frac{3}{8\pi G}(\nu H^2+\frac{2}{3}\alpha \dot{H})\label{eq:ModelsC}\,.
\end{eqnarray}
Notice that the constant, additive, parameter $C_0$ has
dimension $2$ (i.e. mass squared) in natural units. As in the previous chapter, we have
introduced the dimensionful constant $H_0$ as a part of the linear term in $H$ (for \DCU) as in this way the free parameter $\epsilon$ in front of it can be dimensionless.
Similarly $\nu$ and $\alpha$ are dimensionless parameters since they are the coefficients of
$H^2$ and $\dot{H}$, both of dimension $2$. In the case of $\alpha$ we have extracted again an explicit factor of $2/3$ for convenience. These free parameters will be fitted to the observational data. For all models we consider two free parameters  at most.

Models \DAU\ and \DAT\ in the list above are, of course, particular cases of \DAD\ corresponding to $\alpha=0$ and $\nu=0$, respectively, but we have given them different labels for convenience and for further reference in the subsequent parts of this chapter.  Similarly, model \DC2\ is, too, a particular case of \DA2\ (for $C_0=0$), but as we shall see it has some particular features that advice a separate study. Model \DC1\,, in contrast, is \textit{not} a particular case of any of the others. A particular realization of the \DC1-model is the case $\nu=0$ (i.e. the purely linear DE model in $H$), which will be denoted
\be
\label{eq:linH}
\mathcal{D}H: \phantom{XX}\rho_D(H)=\frac{3\epsilon H_0}{8\pi G}H\,.
\ee
This ostensibly simple model of the DE has been proposed in the literature on different accounts. It will be analyzed here in detail, along with the rest of the models, to ascertain its phenomenological viability (cf. Sect. \ref{sec:FittingDE}).

We remark that some of the above models are very similar to the ones we formerly called A1, A2 and C1, C2 in the comprehensive study of Chapter \ref{chap:Atype}. Formally the expressions for the DE densities are the same, the ``only'' difference being that in the previous chapter they were treated as vacuum models (therefore with constant EoS parameter, $\wD=-1$) in interaction with matter, whereas here the effective EoS parameter is a function of the cosmological redshift, $\wD=\wD(z)$, with the additional feature that matter and DE are both conserved.
The ``$\mathcal{D}$'' in front of their names reminds us of the fact that these DE densities will be treated here in the fashion of dynamical DE models with a nontrivial EoS, the latter being determined by the equation of local covariant conservation of the DE density, namely Eq. (\ref{eq:DEConsEq}). We will refer the cosmological DE models (\ref{eq:ModelsA}-\ref{eq:linH}) solved under these specific conditions as the ``${\cal D}$-models'', or the models in the ``${\cal D}$-class'', whereas we reserve the denomination of ``dynamical vacuum models'' when the same DE expressions are solved under the assumption that the EoS is $\wD=-1$ at all times\,\footnote{In this case, in order to fulfill the Bianchi identity, one has to assume that there is an interaction with matter\, \cite{JCAPnostre1} and/or that there is an additional dynamical component interacting with the DE (as e.g. in the $\CC$XCDM model\,\cite{LXCDM}), and/or that the gravitational coupling $G$ is running\,\cite{Grande2011,ApJLnostre}.}. Let us also mention that the case of the pure linear DE model (\ref{eq:linH}) was analyzed in detail in the previous chapter from the point of view of a dynamical vacuum model, but here we will reassess its situation as a ${\cal D}$-model\,\footnote{In analogy with Chapter \ref{chap:Atype} we could additionally have introduced the ${\cal D}$B$_i$ models, namely the ${\cal D}$-class analogous of the vacuum counterparts B$_i$ defined there. The former are the model types obtained by replacing the $\dot{H}$ term of (\ref{eq:ModelsA}) with the linear term in $H$ when $C_0\neq0$. We will not address here the solution of the general models containing the linear term in $H$ since it is not necessary at this point (see Chapter \ref{chap:Atype} for details in the vacuum case). It will suffice for our purposes to study \DC1\ and the pure linear model \DHlin.}.

The cosmological solution of the above ${\cal D}$-models both at the background and perturbations levels turns out to be very different from their vacuum counterparts, as we shall show in this study. We refer the reader to Chapter \ref{chap:Atype} for the details on the vacuum models.

A few additional comments are in order before presenting the calculational details. On inspecting the various forms for the ${\cal D}$-models indicated in (\ref{eq:ModelsA}-\ref{eq:linH}), it may be questioned if all these possibilities are theoretically admissible. For
example, the presence of a linear term $\propto H$ in some particular form of $\rD(H)$ -- models \DC1\ and, of course, \DHlin -- deserves some attention. As commented in Chapter \ref{chap:Atype}, such term does not respect
the general covariance of the effective action of QFT in curved
spacetime\,\cite{SolaReview2013}. The reason is that it involves
only one time derivative with respect to the scale factor. In
contrast, the terms $H^2$ and $\dot{H}$ involve two derivatives ($\dot{a}^2$ or $\ddot{a}$)
and hence they can be consistent with covariance. From this point of
view one expects that the terms $\propto H^2$ and $\propto\dot{H}$ are primary structures
in a dynamical $\rD$ model, whereas $\propto H$ is not. Still, we
cannot exclude {\it a priori} the presence of the linear term since it
can be of phenomenological interest. Let us also mention again the connection of the \DHlin\ and \DC-type models with the attempts to understand the DE from the point of view of QCD and the so-called ``QCD ghost dark energy'' models and related ones, see references in Chapter \ref{chap:Atype}. In the previous chapter the type-C models (with $C_0=0$) were shown to be phenomenologically problematic. In fact, models \DC1, \DC2\ and \DHlin\ present some phenomenological difficulties which we will elucidate here in the specific context of the ${\cal D}$-class. In this respect let us emphasize that in order to identify the nature of these difficulties it is not enough to judge from the structure of the DE density, e.g. equations (\ref{eq:ModelsC}-\ref{eq:linH}), as the potential problems may reside also in the assumed behavior of matter (e.g. whether matter is conserved or in interaction with the DE). That is why the list of pros and cons of the troublesome models examined in the previous references have to be carefully reassessed in the light of the new assumptions. We will accomplish this task here. As we will see, some of the old problems persist, while others become cured or softened, but new problems also appear. At the same time we will show that the only trouble-free models are indeed those in the large \DA-subclass, i.e. the models possessing a well-defined $\CC$CDM limit. This was shown to be the case as  dynamical vacuum models and it will be shown to be so, too, here as ${\cal D}$-models.

%%%%%%%%%%%%%%%%%%%%%%%%%%%%%%%%%%%%%%%%
%%%%%%%%%%%%%%%%%%%%%%%%%%%%%%%%%%%%%%%%
%%%%%%%%%%%%%%%%%%%%%%%%%%%%%%%%%%%%%%%%

\section{Cosmological background solutions}
\label{sec:background}

The first task for us to undertake in order to analyze the above cosmological scenarios is to determine the background cosmological history. From the above equations it is possible, for all the models (\ref{eq:ModelsA}-\ref{eq:linH}), to obtain a closed analytical form for the Hubble function and the energy densities in terms of the scale factor $a$ or, equivalently, in terms of the redshift $z=(1/a-1)$, i.e. $H(z)$. We use them to derive also the EoS and the deceleration parameters, which are very useful to investigate consistency with observational data. Thanks to the self-conservation of dark energy, the EoS parameter can be extracted from the derivative of the DE density with respect to the cosmic redshift,
\be\label{eq:EoS}
\omega_D(z)=-1+\frac{1+z}{3\rho_D(z)}\frac{d\rho_D(z)}{dz}\,.
\ee
Similarly, the deceleration parameter emerges from the corresponding derivative of the Hubble rate, $H$:
\be\label{eq:qz}
q(z)=-1+\frac{1+z}{2H^2(z)}\frac{dH^2(z)}{dz}=-1+\frac{1+z}{2E^2(z)}\frac{dE^2(z)}{dz}\,.
\ee
In the last expression we have used the normalized Hubble rate with respect to the current value, $E(z)=H(z)/H_0$, a dimensionless quantity that will be useful throughout our analysis.
The transition redshift point $z=z_{tr}$ from cosmic deceleration to acceleration can then be computed by solving the algebraic equation
\be\label{eq:zt}
q(z_{tr})=0\,.
\ee
It is important to compute this transition point for the various models, as in some cases there may be significant deviations from the $\CC$CDM prediction.

In the subsequent sections we systematically solve the background cosmologies for the \DA- and \DC-type models. The more complicated details concerning the solution at the perturbations level is presented right next.

%%%%%%%%%%%%%%%%%%%%
%%%%%%%%%%%%%%%%%%%%
%%%%%%%%%%%%%%%%%%%%

\subsection{\texorpdfstring{\DA}--models: background cosmology and equation of state analysis}
\label{sec:BackgroundCosmology}

For the general \DAD-model we have to solve the following differential equation:

\be\label{eq:WHEq}
3(1-\nu)H^2=3H_0^2(\Omega_r^{(0)}a^{-4}+\Omega_m^{(0)}a^{-3})+3C_0+\alpha a\,\frac{dH^2}{da}\,,
\ee
which follows from inserting the corresponding expression (\ref{eq:ModelsA}) of the DE density into Friedmann's equation and trading the cosmic time variable for the scale factor. We use also the fact that for the models under consideration the matter is locally self-conserved. The differential equation (\ref{eq:WHEq}) is very different from the one obtained when the model is treated as a vacuum model in interaction with matter (cf. Chapter \ref{chap:Atype}) and therefore we expect that the \DA-models should have a cosmic history different from the A-type ones.
The constant $C_0$ in (\ref{eq:WHEq}) is fixed by imposing once more the current value of the DE density to be $\rho_D^{(0)}$. This yields
\be
C_0=H_0^2\left[\Omega_D^{(0)}-\nu+\alpha\left(1+\omega_D^{(0)}\Omega_D^{(0)}+\frac{\Omega_r^{(0)}}{3}\right)\right]\,,
\ee
where we have used the relations $\dot{H}_0=-(q_0+1)\,H_0^2$ and
\begin{equation}
q_0=\frac{\Omega_m^{(0)}}{2}+\Omega_r^{(0)}+(1+3\omega_D^{(0)})\,\frac{\Omega_D^{(0)}}{2}\,,
\end{equation}
with $\wDo$ the EoS value of the DE today. Upon integration of \eqref{eq:WHEq} we obtain the normalized Hubble rate $E(a)=H(a)/H_0$, whose square in this case reads:
\be\label{eq:E2DA2}
E^2(a)=a^{3\beta}+\frac{C_0}{H_0^2(1-\nu)}(1-a^{3\beta})+\frac{\Omega_m^{(0)}}{1-\nu+\alpha}(a^{-3}-a^{3\beta})+\frac{\Omega_r^{(0)}}{1-\nu+4\alpha/3}(a^{-4}-a^{3\beta})\,,
\ee
\begin{figure}
\begin{center}
\includegraphics[scale=0.5]{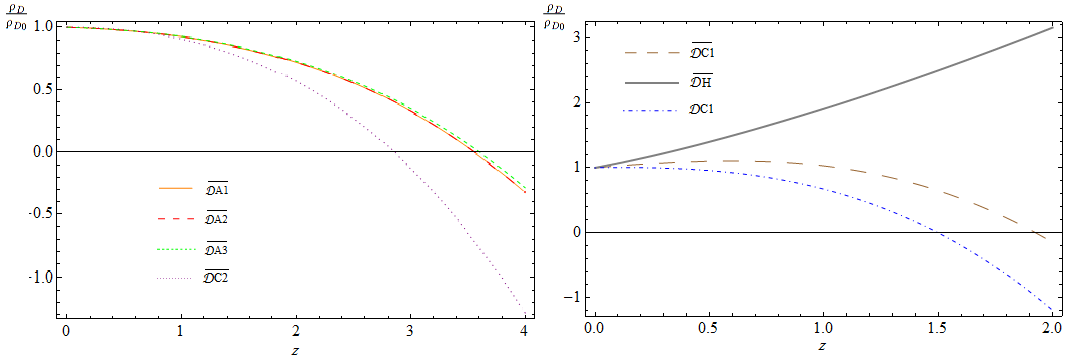}
\caption[DE densities for the various DE models of Chapter \ref{chap:DynamicalDE}]{\footnotesize{
DE densities normalized to their current value for the various dynamical DE models \DA\ and \DC\ under consideration. The behavior of \DC1\ is different and has been plotted apart, together with the
pure linear model \DHlin. We note that in all plots (unless stated otherwise) we use the best fit values of Table \ref{tableFit2} corresponding to the barred quantities, i.e. those obtained when the structure formation data have also been taken into account in the fit.
\label{DEdensity}}
}
\end{center}
\end{figure}

\noindent where $\beta\equiv(1-\nu)/\alpha$. We note the correct normalization $E(1)=1$. It is important to emphasize that we must have $\alpha\geq0$ for these models (in contrast to the situation with the A2-vacuum type considered in Chapter \ref{chap:Atype}) since otherwise the term proportional to the derivative $dH^2/da$ on the \textit{r.h.s.} of Eq. (\ref{eq:WHEq}) could become arbitrarily large and negative in the past, which would violate the non-negativity of $H^2$ in the corresponding \textit{l.h.s.} of that expression. As a matter of fact, for the models with $C_0\neq0$ (for which $|\nu,\alpha|\ll 1$) we have $\beta\gg1$  since $\alpha$ cannot be negative. Thus, for the term $a^{3\beta}$ appearing in \eqref{eq:E2DA2} we obtain the null effective behavior $a^{3\beta}=(1+z)^{-(1-\nu)/\alpha}\simeq 0$  for most of the cosmic history, namely unless $z$ is extremely close to zero. This observation does not apply for the models \DC2, Eq.\,(\ref{eq:ModelsC}), since $C_0=0$ for them and hence the parameters $\nu,\alpha$ can no longer be simultaneously small in absolute value. 

For \DA1-type models we take the lateral limit $\alpha\to 0^+$ (i.e. we approach $0$ from the right) in Eq.\,\eqref{eq:E2DA2} (implying $\beta\to+\infty$), from which we can verify that all of the $a^{3\beta}$ terms vanish since in this limit $a^{3\beta}\to 0$ for $a<1$. For $a=1$ (the current time) these terms also cancel because the overall coefficient of $a^{3\beta}$ is $0$ in that limit. The Hubble function becomes, in this case, pretty much simpler:
\be\label{eq:E2DC1}
E^2(a)=1+\frac{\Omega_m^{(0)}}{1-\nu}(a^{-3}-1)+\frac{\Omega_r^{(0)}}{1-\nu}(a^{-4}-1)\,.
\ee
This result for \DA1\ can, of course,  be obtained also from  Eq.\,(\ref{eq:WHEq}), which for $\alpha=0$ just becomes a simple algebraic equation. This shows the internal consistency of the obtained results. For $\nu=0$ we recover of course the $\CC$CDM result.

For illustrative purposes let us write down explicitly the evolution of the DE density in the last case (i.e. for the \DAU-model). Using the redshift variable we find:
\begin{equation}\label{eq:DEa for DA1}
\rD(z)=\frac{\rho_D^{(0)}-\nu\rho_c^{(0)}}{1-\nu}+\frac{\nu}{1-\nu}\left[\rho_m^{(0)}\,(1+z)^3+\rho_r^{(0)}\,(1+z)^4\right]
\end{equation}
It is easily checked that it satisfies $\rD(0)=\rho_D^{(0)}$, as it should, thanks to the cosmic sum rule involving radiation: $\Omega_m^{(0)}+\Omega_r^{(0)}+\Omega_D^{(0)}=1$. And of course it also boils down identically to $\rho_D^{(0)}$ for $\nu=0$.

\begin{figure}
\begin{center}
\includegraphics[scale=0.42]{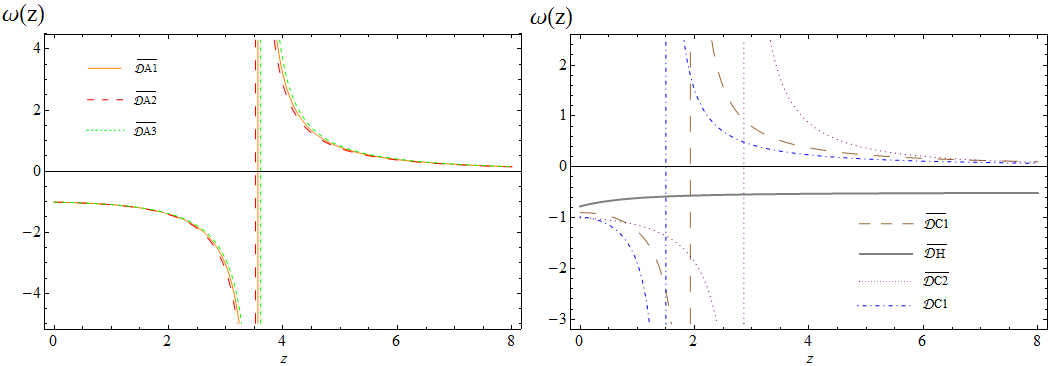}
\caption[EoS $\omega_{D}(z)$ for the DE models \DA, \DC1\ and \DC2]{\footnotesize{
EoS function $\omega_{D}(z)$ for the DE models \DA\  (left) and for  \DC1\ and \DC2\ (right).  The vertical asymptotes are located at the points where the DE density vanishes (compare with Fig. \ref{DEdensity}). The linear \DHlin\ model is seen not to present any asymptote (cf. right plot).
\label{w(z)z1}}
}
\end{center}
\end{figure}

\begin{figure}
\begin{center}
\includegraphics[scale=0.55]{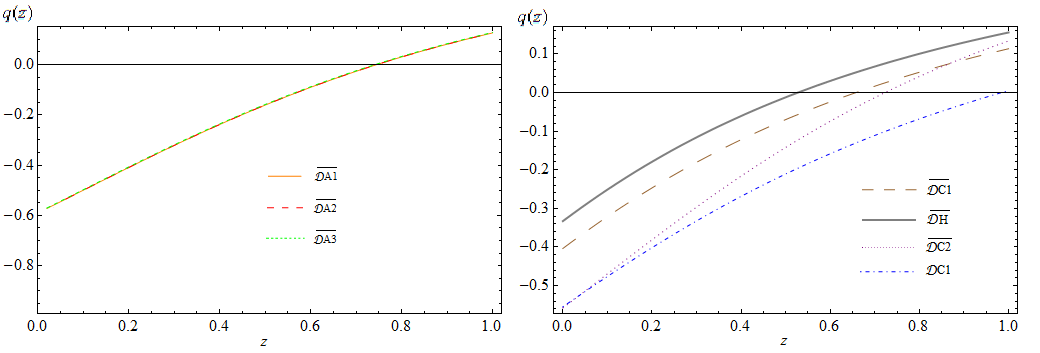}
\caption[Deceleration parameter, $q(z)$, for the various \DA\ and \DC\ models]{\footnotesize{
Deceleration parameter, $q(z)$, and transition point from deceleration to acceleration for the various \DA\ and \DC\ models. Notice that the transition point $q(z_{tr})=0$ from deceleration to acceleration changes significantly for \DC1-models depending on whether we use barred or unbarred fitting parameters in Table 1. In contrast, \DA\ and \DC2-models have a more similar transition redshift $z_{tr}$ which is not far away from that of the $\CC$CDM.
\label{q(z)}}
}
\end{center}
\end{figure}

Let us come back to the more general model \DAD. We neglect radiation at this point, as we want to focus now on features of the current Universe, such as the equation of state of the DE near our time. The normalized Hubble function (\ref{eq:E2DA2}) can then be cast in terms of the redshift as follows:
\begin{equation}
\label{eq:hw}
E^2(z)=\frac{C_{0}}{H_0^2(1-\nu)}+\frac{\Omega_{m}^{(0)}}{1-\nu+\alpha}(1+z)^{3}-\eta\left(1+z\right)^{-3\beta}.
\end{equation}
where
\begin{equation}
\label{eq:eta}
\eta=\frac{C_{0}}{H^2_0(1-\nu)}+\frac{\Omega_{m}^{(0)}}{1-\nu+\alpha}-1.
\end{equation}
The evolution of the DE density for the \DA2-models in the matter-dominated and current epoch can be computed with the help of the previous result, yielding
\begin{equation}\label{eq:rDaDA2}
\rD(z)=\frac{\rho_c^{(0)}\,C_0}{H_0^2(1-\nu)}+\rho_c^{(0)}\,\Omega_m^{(0)}\frac{\nu-\alpha}{1-\nu+\alpha}\,(1+z)^{3}-\rho_c^{(0)}\,{\eta}\,(1+z)^{-3\beta}\,.
\end{equation}
As can be checked, it satisfies $\rD(0)=\rho_c^{(0)}\,(1-\Omega_m^{(0)})=\rho_D^{(0)}$ and it identically reduces to $\rho_D^{(0)}$ for $\nu=\alpha=0$, i.e. we recover in this limit the $\CC$CDM result. In the last part we use the fact that $(1+z)^{-3\beta}\to 0$ in the limit  $\alpha\to 0^+$ for any $z>0$. One can verify that the \DA1-type solutions \eqref{eq:E2DC1} and \eqref{eq:DEa for DA1} are particular cases of the last results in the matter-dominated and current epochs, as they should. Similarly, the \DA3 -type solution is the particular case obtained from the above formulas for $\nu\to 0$. The numerical evolution of the DE energy densities for these models (normalized to the current value $\rho_D^{(0)}$) is shown in Fig.\,\ref{DEdensity} for the best-fit values of Table \ref{tableFit2} (see Sect. \ref{sec:FittingDE} for the details of the fitting procedure leading to the results of that table). In Figures \ref{w(z)z1}-\ref{q(z)} we provide the corresponding behavior of the dark energy EoS and deceleration parameters for these models, which will be commented below in turn.

\begin{table}
\begin{center}
\resizebox{1\textwidth}{!}{
\begin{tabular}{| c  |c | c | c | c | c | c |c | c | c  | c |c |}
\multicolumn{1}{c}{Model}  & \multicolumn{1}{c}{$\Omega_m^{(0)}$} & \multicolumn{1}{c}{$\overline{\Omega}_m^{(0)}$}  &  \multicolumn{1}{c}{{\small$\nueff=\nu-\alpha$} }  & \multicolumn{1}{c}{{\small$\bar{\nu}_{\rm eff}$}}    & \multicolumn{1}{c}{$\sigma_{8}$} & \multicolumn{1}{c}{$\overline{\sigma}_{8}$} &
\multicolumn{1}{c}{$\chi^2_{r}/dof$}  &
\multicolumn{1}{c}{$\chi^2/dof$} &
\multicolumn{1}{c}{$\overline{\chi}^2/dof$} &
\multicolumn{1}{c}{AIC} &
\multicolumn{1}{c}{$\overline{\rm AIC}$}
\\\hline {\small $\CC$CDM}  & {\small$0.291^{+0.008}_{-0.007}$} & {\small$0.286\pm 0.007$} & - & -  & {\small$0.815$} & {\small$0.815$} & {\small$569.21/592$} & {\small$584.91/608$} & {\small$584.38/608$} & {\small$586.91$} & {\small$586.38$}
\\\hline
{\small $\mathcal{D}$A1}  & {\small$0.286^{+0.012}_{-0.011}$} & {\small$0.281\pm 0.005$} & {\small$-0.024\pm 0.018$} & {\small $-0.028\pm 0.016$} & {\small$0.773$} & {\small$0.770$} & {\small$565.50/591$} & {\small$573.02/607$} & {\small$573.31/607$} & {\small$577.02$} & {\small$577.31$}
\\\hline
{\small $\mathcal{D}$A2}  & {\small$0.286^\pm 0.011$} & {\small$0.281\pm 0.005$} & {\small$-0.024\pm 0.018$} & {\small $-0.028\pm 0.016$} & {\small$0.772$}  & {\small$0.769$} & {\small$565.57/591$}& {\small$573.03/607$} & {\small$573.40/607$} & {\small$577.03$} & {\small$577.40$}
\\\hline
{\small $\mathcal{D}$A3}  & {\small$0.287^\pm 0.011$} & {\small$0.282\pm 0.005$} & {\small$-0.023^{+0.017}_{-0.018}$} & {\small $-0.027\pm 0.015$} & {\small$0.777$} & {\small$0.773$} & {\small$565.63/591$} & {\small$573.44/607$} & {\small$573.47/607$} & {\small$577.44$} & {\small$577.47$}
\\\hline
{\small $\mathcal{D}$C1}  & {\small$0.286\pm 0.014$} & {\small$0.335\pm 0.007$} & {\small$-0.64\pm 0.13$} & {\small$-0.35\pm 0.05$} & {\small$0.440$}   & {\small$0.735$} & {\small$563.86/584$} & {\small$880.74/600$} &{\small$635.23/600$} & {\small$884.74$} & {\small$639.23$}
\\\hline
{\small \DHlin}  & {\small$0.242\pm 0.008$} & {\small$0.286\pm 0.005$} & - & - & {\small$0.513$} & {\small$0.729$} & {\small$639.85/585$} & {\small$809.61/601$} & {\small$677.11/601$} & {\small$811.61$} & {\small$679.11$}
\\\hline
{\small $\mathcal{D}$C2}  & {\small$0.285\pm 0.013$} & {\small$0.295\pm 0.006$} & {\small$1.03^{+0.09}_{-0.06}$} & {\small$1.02\pm 0.01$} & {\small$0.666$}   & {\small$0.752$} & {\small$563.53/584$} & {\small$594.13/600$} & {\small$572.17/600$} & {\small$598.13$} & {\small$576.17$}
\\\hline
 \end{tabular}}
\end{center}
\caption[Best-fit values for the various DE models of \ref{chap:DynamicalDE}]{\scriptsize The best-fitting values for the various models and their
statistical  significance ($\chi^2$-test and Akaike information criterion AIC \cite{Akaike1974}, see Sect. \ref{sect:Discussion}).
All quantities with a bar involve a fit to the total input data, i.e. the expansion history (BAO$_A$+BAO$_{d_z}$+SNIa) and CMB shift parameter data, as well as the linear growth data. Those without bar correspond to a fit in which we use all data but exclude the growth data points from the fitting procedure. The value $\chi^2_{r}$ is the reduced $\chi^2$, which does not include the linear growth $\chi^2_{f\sigma_8}$ contribution.  For models \DA1 (resp. \DA3) $\nueff=\nu$ (resp. $-\alpha$); for \DA2\ we have fixed $\alpha=-\nu$ to break degeneracies (see text). For \DC\ and \DHlin\ models we have not used the BAO$_{d_z}$ and CMB data  for the reasons explained in the text. In addition, for the \DC\ models the given value of $\nueff$ must be understood as the value of $\nu$ since $\alpha$ is not defined for \DC1\ and becomes determined for \DC2  (see text). The quoted number of degrees of freedom ($dof$) is equal to the number of data points minus the number of independent fitting parameters. Details of the fitting observables are given in Sect. \ref{sec:FittingDE}. \label{tableFit2}}
\end{table}

From \eqref{eq:rDaDA2} and  (\ref{eq:EoS}) we may derive
the EoS parameter for the general subclass of \DA2\ models. It can be expressed in the following compact form:
\begin{equation}
\label{eq:wdw}
\omega_{D}(z)=-\frac{\rho_c^{(0)}\,C_0}{H_0^2(1-\nu)\rD(z)}+\rho_c^{(0)}\,\frac{\eta(1+\beta)}{\rD(z)}(1+z)^{-3\beta}\,,
\end{equation}
with $\rD(z)$ given by (\ref{eq:rDaDA2}). For $C_0\neq0$ the parameters $\nu$ and $\alpha$ are small (this is confirmed by the fitting values in Table \ref{tableFit2}) and therefore we can apply a similar argument to the limiting situation just explained above to prove that the second term on the \textit{r.h.s.} of \eqref{eq:wdw} does not contribute significantly except in the origin. In practice, for any $z>0$ (even for points very close to the origin) the effective EoS parameter is actually given by the first term on the \textit{r.h.s.} of \eqref{eq:wdw}. This is indeed the result that is connected by continuity with the EoS of the \DA1-models in the limit $\alpha\to 0^+$, as we shall show in a moment below. Therefore, for all practical situations related to redshift points around our current epoch we establish as effective EoS for the \DA2-models the following expression:
\begin{equation}
\label{eq:EoSDA2}
\omega_{D}(z)=-\frac{1}{1+\frac{H_0^2(1-\nu)}{C_0}\Omega_m^{(0)}\frac{\nu-\alpha}{1-\nu+\alpha}\,(1+z)^3}\,,
\end{equation}
where by the same token we have disregarded the last term of Eq.\,(\ref{eq:rDaDA2}). There is however one proviso related to the fact that $\rD(z)$ in the denominator of the two terms in (\ref{eq:wdw}) could vanish, and in fact does vanish in our case for some particular redshift value (see below). This causes the presence of a vertical asymptote at a finite $z$ point.  In these cases, one would think of using Eq. (\ref{eq:wdw}) to better describe the behavior around the asymptote.  In actual fact, not even this possibility affects in any significant way the practical use of Eq.\, \eqref{eq:EoSDA2}, as we have checked. Being the parameters $\nu$ and $\alpha$  small in absolute value  we can expand the expression (\ref{eq:EoSDA2}) linearly in them. The result can be cast in the following suggestive form for redshift points near our time (typically valid in the more accessible region  $0<z\lesssim2$):
\begin{equation}\label{eq:EoS approx}
\omega_{D}(z)\simeq-1+\frac{H_0^2(1-\nu)}{C_0}\,\Omo\,\nueff\,(1+z)^3\simeq -1+\frac{\Omo}{1-\Omo}\,\nueff\,(1+z)^3\,.
\end{equation}
In this expression, the dimensionless quantity
\begin{equation}\label{eq:nueff}
\nueff=\nu-\alpha
\end{equation}
is the basic fitting parameter. Using its value and that of $\bar{\Omega}_m^{(0)}$ from Table \ref{tableFit2} we can evaluate the current EoS parameter of the \DA2\ model, Eq.\,\eqref{eq:EoS approx}:
\begin{equation}\label{eq:EoScurrentDA2}
\omega_{D}^{(0)}\equiv \omega_D(0)\simeq -1.011\,,
\end{equation}
which turns out to be remarkably close to the $\CC$CDM behavior.
Let us also stress at this point that $\nueff$ is indeed the effective fitting parameter for the entire set of \DA-models at the background level in the matter-dominated and current epochs. It is a small parameter since $|\nu,\alpha|\ll 1$ (owing to $C_0\neq 0$). In particular, let us note that because $\omega_{D}^{(0)}$ is very close to $-1$, the coefficient carrying $C_0$ in the Hubble function (\ref{eq:E2DA2}) can be written in first order as $\ODo-\Omo\nueff$ in the limit $\omega_{D}^{(0)}\to -1$. On the other hand the remaining $\nu$ and $\alpha$-dependence in $\beta$ virtually disappears for the reasons discussed above.

In the radiation epoch, however, the dependence on $\nu$ and $\alpha$ is different than (\ref{eq:nueff}), as it is clear from the radiation term in \eqref{eq:E2DA2}. Also, in Sect. \ref{sec:FittingDE} we will see that this produces corrections to the transfer function which depend separately on $\nu$ and $\alpha$, and in these cases we must unavoidably fix some relation between these parameters. It has been indicated in the caption of Table \ref{tableFit2}, and more details are given in Sect. \ref{sec:FittingDE}.

There is one more, worth noticing, feature to stand out in connection to the numerical value (\ref{eq:EoScurrentDA2}) and the effective EoS function \eqref{eq:EoS approx}. They suggest that the dynamical DE models under study can provide, in principle, a reason for the quintessence and phantom-like character of the DE without necessarily using fundamental scalar fields. Indeed, we find that for $\nu>\alpha$ (i.e. $\nueff>0$) the \DA-models behave effectively as quintessence whereas for $\nu<\alpha$ (i.e. $\nueff<0$) they behave phantom-like near our time. In practice, after fitting the overall cosmological data, we have seen that the latter is the observed situation and also the predicted theoretical result. Recall that the current observational evidence on the dark energy EoS, for example from Planck results, leads to $\omega_{D}^{(0)}=-1.006\pm 0.045$\,\cite{Planck2015}. This is perfectly compatible with the estimate (\ref{eq:EoScurrentDA2}) both quantitatively and qualitatively. This said, we are not suggesting that the current data on the EoS of the dark energy implies that the DE is phantom-like, as the accuracy around the central value is still insufficient\footnote{Actually, in Chapters \ref{chap:AandGRevisited}-\ref{chap:H0tension} we will see that the use of a more complete data set (than the one used by the Planck 2015 or WMAP teams) in the fitting analysis of, e.g. the XCDM parametrization \cite{XCDM}, leads to quintessence values of the corresponding EoS parameter. But these investigations were carried out almost a year later of the publication of the work presented in this chapter.}. However, for years the central value of the $\wD^{(0)}$ measurement has shown some tilt into the region below $-1$, even during the long period of WMAP observations\,\cite{WMAP9}. Here we are merely saying that if such kind of measurement would be reinforced in the future, the general class of ${\cal D}$-models encodes the ability for providing an explanation of the phantom character of the observed DE without need of invoking real phantom fields at all.  Being $\nueff={\cal O}(10^{-2})$ rather small (and negative) the departure of $-1$ from below is very small, which is precisely the kind of result compatible with observations. To produce the plots of Fig. \ref{w(z)z1} (left) we have used both the exact expression (\ref{eq:wdw}) for the effective EoS of the \DA2-model and the effective one (\ref{eq:EoSDA2}), and have found no appreciable numerical differences.

Let us now say some words on the presence of vertical asymptotes in Fig. \ref{w(z)z1}. All models under study display these asymptotes, except \DHlin\ (cf. next section). Such pole-like singularity is observed also in other contexts, for example in non-parametric reconstructions of the EoS function $\wD(z)$\,\cite{Shafieloo2006}, in certain brane-model cosmologies\,\cite{Sahni2003} and in other situations, such as e.g. in mimicking quintessence and phantom DE through a variable $\CC$ \cite{BasSola2014b,SS0506a,SS0506b}. In our case the pole-like feature is related to the fact that the denominator of (\ref{eq:EoSDA2}) vanishes for some finite value of $z$, which is to be expected since $\nueff<0$ (cf. Table \ref{tableFit2}). Physically this means that the DE density $\rD(z)$ vanishes at these values of $z$ (around $3.5$ for models \DA\,, and near $1.5$ or $2$ for \DC1\ depending on the fit options indicated in Table \ref{tableFit2}), as it is confirmed from the behavior of $\rD(z)$ in Fig. \ref{DEdensity}. As a consequence of this fact the EoS function (\ref{eq:EoSDA2}) develops a singularity at this point. Notice, however, that the late-time expansion
of the Universe in the wide span $0<z<2$ (possibly comprising all relevant supernovae data) is free from these exotic behaviors, if using the most optimized fit values that include the structure formation data. Obviously the latter are not associated with inherent pathologies of the model since the values of the fundamental physical quantities, such as the energy densities, stay finite, e.g. $\rD(z)$ simply vanishes at these ``singular'' points. Interestingly, the very existence of these points might carry valuable information, for if the DE would be described by the ${\cal D}$-models and we could eventually explore the EoS behavior at high redshifts ($z>2$) we should be able to pin down these features, which would manifest through an apparent flip from phantom-like behavior into quintessence-like one when observing points before and after, respectively, the one where the DE density vanishes (cf. Fig. \ref{DEdensity}). Observing such phenomenon could be a spectacular signature of these models\,\footnote{It should be stressed that the present situation is different from the EoS studies previously entertained in Refs.\,\cite{BasSola2014b,SS0506a,SS0506b}, in the sense that in the latter the effective EoS was only a representation (the so-called ``DE picture'') of the original vacuum model (which in turn was called the ``CC-picture''\,\cite{SS0506a,SS0506b}), whereas here the EoS under study stands for the ``physical'' one associated with the original  ${\cal D}$-model. Therefore, if these models are to provide a correct description of the DE we should be able to observe the exotic EoS patterns shown in Fig.\,\ref{w(z)z1}, much in the same way as in the alternative frameworks proposed in Refs.\, \cite{Shafieloo2006,Sahni2003}.}.

The EoS for the simpler subclass of \DA1-models can now be obtained from the limit $\alpha\to 0^+$ of Eq.\,(\ref{eq:wdw}). In this limit the second term on the \textit{r.h.s} exactly cancels for any $z>0$ and we are left only with the first term. The result is simply Eq.\,(\ref{eq:EoSDA2}) for $\alpha=0$. In particular, we can see that for $z\to 0$ we obtain a prediction for the current EoS value for this model:
\be\label{eq:omegaDo for DA1}
\omega_D^{(0)}=-\left[\frac{1-\Omega_m^{(0)}-\nu}{(1-\nu)(1-\Omega_m^{(0)})}\right]\simeq -1+\nu\,\frac{\Omo}{1-\Omo}\,,
\ee
where the first expression is exact and the second is valid for $|\nu|\ll1$. The latter is seen to be consistent with the limit $z\to 0$ of Eq.\,(\ref{eq:EoS approx}). Since in this case the fitting values of $\nu$ and $\nueff$ in Table \ref{tableFit2} are the same, we retrieve the numerical result (\ref{eq:EoScurrentDA2}) also for \DA1\,. Interestingly, we can explicitly verify that the result (\ref{eq:omegaDo for DA1}), which we have first obtained by taking  the limit $\alpha\to 0^+$ in Eq.\,(\ref{eq:wdw}), can also be worked out from (\ref{eq:EoS}) using the specific Eq. (\ref{eq:DEa for DA1}) of the \DA1-model.

The deceleration parameter for \DA-type models reads:
\begin{equation}
\label{eq:qw}
q(z)=-1+\frac{1}{2E^{2}(z)}\left(\frac{3\Omega_{m}^{(0)}}{1-\nu+\alpha}\left(1+z\right)^{3}+3\beta \eta\left(1+z\right)^{-3\beta}\right)\,.
\end{equation}
Solving Eq. (\ref{eq:zt}) in this case we may find the transition redshift for a general \DA2-type model. Once more we neglect the last term of Eq. \eqref{eq:qw} since it gives a negligible correction, and we arrive at

\be\label{eq:ztrDA2}
z_{tr}=\left[\frac{2(1+\alpha-\nu)[1-\Omega_m^{(0)}-\nu+\alpha(1+\omega_D^{(0)}\Omega_D^{(0)})]}{\Omega_m^{(0)}(1-\nu)}\right]^{1/3}-1\,.
\ee
The corresponding result for \DA1\ is obtained by setting $\alpha=0$ in the above expression. Numerically, the deceleration and  EoS parameters at the current time for $\mathcal{D}$A1  read respectively as follows: $q^{(0)}=-0.590$ and $z_{tr}=0.745$. For $\mathcal{D}$A3 models, the results are $q^{(0)}=-0.589$ and $z_{tr}=0.742$. For $\nu=\alpha=0$ the formula (\ref{eq:ztrDA2}) naturally returns the $\CC$CDM result:
\begin{equation}\label{eq:ztrLCDM}
z_{tr}=\left(\frac{2\Omega_D^{(0)}}{\Omega_m^{(0)}}\right)^{1/3}-1\,,
\end{equation}
whose numerical value for the fitting parameters in Table \ref{tableFit2} is $z_{tr}\simeq 0.709$.  The plot of $q(z)$ for the \DA\ models is depicted in Fig. \ref{q(z)} (left), where we can also read the transition point $z_{tr}$.

%%%%%%%%%%%%%%%%%%%%%%%%%%%%%%%%%%%%%%%%%%%%
%%%%%%%%%%%%%%%%%%%%%%%%%%%%%%%%%%%%%%%%%%%%
%%%%%%%%%%%%%%%%%%%%%%%%%%%%%%%%%%%%%%%%%%%%

\subsection{Models \texorpdfstring{$\mathcal{D}$}{DC1}C1, \texorpdfstring{$\mathcal{D}$}{DH}H and \texorpdfstring{$\mathcal{D}$}{DC2}C2}
\label{sect:DCModels}

Let us start with \DCU. The normalized Hubble rate $E(z)= H(z)/H_0$ can easily be found starting from Friedman's equation upon inserting in it the corresponding DE density from Eq.\,\eqref{eq:ModelsC}:
\be\label{eq:HubbleDC1}
E(z)=\frac{\epsilon+\Sigma(z)}{2(1-\nu)}\,,\ee
with
\be
\Sigma(z)=\sqrt{\epsilon^2+4(1-\nu)[\Omega_r^{(0)}\,(1+z)^{4}+\Omega_m^{(0)}(1+z)^{3}]}\,
\ee
The parameters in the above relation are constrained to satisfy $1-\nu=\epsilon+\Omega_m^{(0)}+\Omega_r^{(0)}$ so as to insure that $E(0)=1$. Equivalently, $\epsilon=\ODo-\nu$. Thus we can take e.g. $\nu$ as the single free model-parameter, given the values of the ordinary parameters $\Omega_m^{(0)}$ and $\Omega_r^{(0)}$. For this model indeed, $\nueff$ in Table \ref{tableFit2} is meant to be $\nu$. Furthermore, the same constraint shows that $\epsilon$ and $\nu$ cannot be both small parameters (in absolute value) in the case of the \DCU\ models, i.e. they do \emph{not} satisfy $|\epsilon,\nu|\ll 1$, in contrast to ${\cal D}$A models, the reason being that the \DCU\ models do not have a well-defined $\CC$CDM limit for any value of $\epsilon$ and $\nu$. As a result one of the two parameters can be of order $\mathcal{O}(1)$. For example, the fit in Table \ref{tableFit2} for the barred quantities indicates that $\bar{\nu}\simeq -0.35$ and $\bar{\Omega}_m^{(0)}\simeq 0.335$, and therefore since the radiation component is negligible at present we obtain $\epsilon\simeq 1.02$. For the case when the structure formation data are not used in the fit (unbarred parameters) we have ${\nu}\simeq -0.64$, which is much larger in absolute value, and then $\epsilon$ is also larger: $\epsilon\simeq 1.35$.

The corresponding expression for the DE density of this model reads
\be\label{eq:rDforDC1}
\rD(z)=\rho_c^{(0)}\,\left[\epsilon\,E(z)+\nu\,E^2(z)\right]\,,
\ee
where $E(z)$ is given by (\ref{eq:HubbleDC1}). At $z=0$ the above function correctly renders the DE density now: $\rD(0)=\rho_c^{(0)}\,(\epsilon+\nu)=\rho_c^{(0)}\,(1-\Omega_m^{(0)}-\Omega_r^{(0)})=\rho_c^{(0)}\,\Omega_D^{(0)}=\rho_D^{(0)}$ upon using the mentioned constraint between the parameters of the model.
For considerations in the matter-dominated and current epochs we can ignore of course the radiation component. A plot of the function (\ref{eq:rDforDC1}), normalized to its current value $\rho_D^{(0)}$, is shown in Fig. \ref{DEdensity} (right).  In the same figure we plot the case of \DC1\ when $\nu=0$, i.e. the linear model \DHlin, Eq.\,(\ref{eq:linH}). This model has no free parameter, apart from $\Omega_m^{(0)}$, since the above mentioned constraint enforces the relation  $\epsilon=1-\Omega_m^{(0)}=\Omega_D^{(0)}$ in the matter-dominated epoch. Therefore,
\begin{equation}\label{eq:EzDH}
E(z)=\frac12\,\left(\Omega_D^{(0)}+\left.\Sigma(z)\right|_{\nu=0}\right)=\frac12\,\left(\Omega_D^{(0)}+\sqrt{\Omega_D^{(0)2}+4\Omega_m^{(0)}\,(1+z)^3}\right)\,.
\end{equation}
The corresponding DE density is:
\begin{equation}\label{eq:rDzDH}
\rD(z)=\rho_c^{(0)}\,\Omega_D^{(0)}\,E(z)=\frac12\,\rho_D^{(0)}\left(\Omega_D^{(0)}+\sqrt{\Omega_D^{(0)2}+4\Omega_m^{(0)}\,(1+z)^3}\right)\,.
\end{equation}
Notice the correct normalization $\rD(0)=\rho_D^{(0)}$ since at $z=0$ the argument in the square root becomes $(2-\ODo)^2$ after using the sum rule $\Omega_m^{(0)}+\Omega_D^{(0)}=1$.
It is interesting to remark at this point that the behavior of the \DC1\ and \DHlin\ models is rather different from their vacuum counterparts in interaction with matter. Let us e.g. focus on the linear vacuum model, i.e. the model (\ref{eq:linH}) when the EoS is $\wD=-1$ and interacting with matter. As it has been shown in Chapter \ref{chap:Atype}, the corresponding DE and matter densities are
\begin{eqnarray}\label{eq:rhomaLinear1}
\rho_{\CC}(z)&=&\rho_\Lambda^{(0)}\,\left\{1+\Omega_m^{(0)}\left[(1+z)^{3/2}-1\right]\right\}\\
\rho_m(z)&=&\rho_m^{0}\left[\Omega_m^{(0)}+(1-\Omega_m^{(0)})(1+z)^{-3/2}\right]\,(1+z)^{3}\label{eq:rhomaLinear2}\,.
\label{eq:rhoLaLinear2}
\end{eqnarray}
In the matter density formula we note an extra factor of $\Omega_m^{(0)}$ in front of the term $(1+z)^3$ for large enough $z$. Thus, the deviation with respect to the $\CC$CDM behavior becomes increasingly large when we explore deeper our past. This is in contrast to its \DHlin\  counterpart since matter is conserved for this model and hence it stays with the standard law $\rmr(z)=\rho_m^{(0)} (1+z)^{3}$. On the other hand, if we compare the dynamical DE densities \eqref{eq:rDzDH} and \eqref{eq:rhomaLinear1}, both deviate with respect to the $\CC$CDM  linearly with $z$ near our time (as can be seen by expanding these expressions for low $z$), but in in the \DHlin\ case the relative deviation is a factor $1-2/(1+\Omo)>50\%$ larger. This fact is actually determinant to explain the inability of the \DHlin-model to correctly describe the low $z$ data, and it reflects in a bad quality fit in Table \ref{tableFit2}. We conclude that the linear model (\ref{eq:linH}) is essentially inefficient for a correct description of the cosmological data, both as a vacuum model and as a $\mathcal{D}$-model.
A similar situation occurs for the  \DC1\ and C1 models (cf. Table \ref{tableFit2} and Chapter \ref{chap:Atype}). This is a clear sign that the ${\cal D}$-models generally depart significantly from their vacuum analogs and both of them also with respect to the $\CC$CDM. In Sect. \ref{sec:FittingDE} we will retake this important issue in more detail, as these models have been repeatedly called for in the literature from different points of view and it may be appropriate to further discuss the reason for their  delicate phenomenological status, which is made quite evident from the statistical analysis presented in the last four columns of Table \ref{tableFit2}. As we will see, it is not only caused by their background cosmological behavior but also by their troublesome prediction concerning structure formation.

Let us now address the EoS analysis of these models. Using \eqref{eq:EoS} and \eqref{eq:rDforDC1} we are lead to the following expression for the dynamical EoS of the \DC1\ and \DHlin-models:
\be
\label{eq:wEoS}
\omega_D(z)=-1+\frac{\Omega_m^{(0)}(1+z)^3[\epsilon+2\nu E(z)]}{E(z)\,[\epsilon+\nu E(z)]\,[2(1-\nu)E(z)-\epsilon]}\,.
\ee
As we are interested in the behavior of $\omega_D(z)$ and $q(z)$ in the low-redshift period (i.e. near our time), we have neglected the radiation contribution when we compute them. The corresponding EoS plot for \DC1\ and \DHlin\ are shown in Fig.\,\ref{w(z)z1} (right). In the case of \DC1\ ($\nu\neq0$), it displays an asymptote at large $z$ near $2$. The asymptote appears because the fitted value of $\nu$ is negative in Table \ref{tableFit2}, so the last term of the expression (\ref{eq:rDforDC1}) takes over at sufficiently large $z$ and $\rD$ vanishes near $z=2$ (cf. Fig. \ref{DEdensity}, right). There is no asymptote for \DHlin\ because the function (\ref{eq:rDzDH}) is monotonically increasing with $z$. Moreover, one can show analytically that for large $z$ the EoS for this model tends to $-1/2$, as can be confirmed graphically by looking at Fig. \ref{w(z)z1}.

Of particular importance is the present-day value of the EoS parameter for \DC1\,, i.e. $\omega_D^{(0)}$. Working out the result from the previous equations one finds:
\be
\label{eq:wEoSz0}
\omega_D^{(0)}=-\frac{\epsilon}{(1-\Omega_m^{(0)})(\epsilon+2\Omega_m^{(0)})}\,.
\ee
Notice that this is an exact result and we recall that $\epsilon$ and $\nu$ are \emph{not} small parameters for DC1 models.
Substituting the best-fit values  of the parameters in the above formula, according to the results shown in Table \ref{tableFit2}, the current value of the EoS parameter achieved for $\mathcal{D}$C1  reads $\omega_D^{(0)}=-0.906$\,\footnote{Here and hereafter the numerical estimates use always the barred quantities in Table \ref{tableFit2}, which are the most optimized ones since they are obtained from fitting all the expansion and structure formation data.}. This value can be spotted in Fig.\,\ref{w(z)z1} (right) and deviates significantly from the expected, narrow, range around $-1$\,\,\cite{Planck2015}.
As for the pure linear model \DHlin\, (corresponding to $\nu=0$), the constraint given above enforces $\epsilon=\Omega_D^{(0)}$ (when we neglect the radiation component) and therefore Eq.\,(\ref{eq:wEoSz0}) yields
\be
\label{eq:wEoSz0H2}
\omega_D^{(0)}=-\frac{1}{1+\Omega_m^{(0)}}\simeq -0.778
\ee
for the best-fit value of $\Omega_m^{(0)}$  collected in Table \ref{tableFit2}. Such EoS result is out of the typical range of current observations (even more than for \DC1) and puts the \DHlin\ model once more against the wall.
Let us also note that another particular case of \DC1 models, namely the pure quadratic $\rD\sim H^2$ model (which is obtained for $\epsilon=0$), would have $\wD^{(0)}=0$ according to Eq.\,\eqref{eq:wEoSz0}, and therefore it is completely excluded. The pure quadratic model was used in the past motivated by holographic ideas. But unfortunately the simplest ideas of this sort are actually ruled out\,\cite{MiaoLi04} and some other generalizations (which include \DC1-type models) too\,\cite{Komatsu2013,JapaneseHolog1a,JapaneseHolog1b,JapaneseHolog2,BasPolarSola12,BasSola2014b}.

The deceleration parameter for the \DCU-model can be computed from \eqref{eq:qz} and \eqref{eq:HubbleDC1}, and we find
\begin{equation}
\label{qz1}
q=-1+\frac{3\Omega_{m}^{(0)}H_{0}^{2}\left(1+z\right)^{3}}{\left[2\left(1-\nu\right) H^{2}-\epsilon H_{0}H\right]}\,.
\end{equation}
The best-fit values indicated in Table \ref{tableFit2} lead to the following estimates for the current acceleration parameters for the general $\mathcal{D}$C1 and \DHlin, respectively: $q^{(0)}=-0.404$, $q^{(0)}=-0.333$. These models are significantly less accelerated now than the $\CC$CDM (for which $q^{(0)}=-0.571$). The behavior of $q(z)$ is plotted in Fig. \ref{q(z)} (right). As it is seen in this figure, for the \DCU-model the Universe has a transit from a decelerated phase ($q > 0$) to an accelerated one ($q < 0$), which occurs precisely at the following transition redshift:
\be
z_{tr}=\left[\frac{2\epsilon^2}{\Omega_m^{(0)}(\epsilon+\Omega_m^{(0)})}\right]^{1/3}-1\,.
\ee
Once more we borrow the fitting results from Table \ref{tableFit2} and for $\mathcal{D}$C1 models we obtain $z_{tr}=0.658$, whilst for \DHlin\ we have $z_{tr}=0.528$. These values are clearly smaller than for the $\CC$CDM ($z_{tr}\simeq 0.691$), meaning that the transition is accomplished much more recently than in the standard case.

Let us now finally deal with \DCD-models. This model is sometimes also related with the entropic-force formulations, see \cite{Easson10}. However it is not enough to give the structure of the DE to know if the model is phenomenologically allowed or excluded.  Treated as a vacuum C2-model ($\wD=-1$) it was shown in Ref. \cite{BasSola14a} (see also Chapter \ref{chap:Atype}) to be excluded, but as a ${\cal D}$-model it has some more chances to survive. Let us describe here the main traits of its background evolution and leave the issues of structure formation for the next section. The Hubble function for \DC2\ follows from (\ref{eq:E2DA2}) for $C_0=0$. In the matter-dominated and current epochs reads
\be\label{eq:E2DC2}
E^2(a)=a^{3\beta}+\frac{\Omega_m^{(0)}}{1-\nu+\alpha}(a^{-3}-a^{3\beta})\,.
\ee
As before $\beta\equiv(1-\nu)/\alpha$, but we should recall that in this case $\alpha$ and $\nu$ cannot be both small in absolute value. This is confirmed by  explicitly displaying the constraint $C_0=0$ that they satisfy in the matter-dominated epoch, namely $1-\Omega_m^{(0)}-\nu+\alpha=-\alpha\wDo\Omega_D^{(0)}$, in which $\wDo\simeq -1$. The evolution of the DE density for these models simply follows from (\ref{eq:rDaDA2}) by setting $C_0=0$, and we find:
\begin{equation}\label{eq:rDaDC2}
\rD(z)=\rho_c^{(0)}\frac{1-\nu+\alpha-\Omega_m^{(0)}}{1-\nu+\alpha}\,(1+z)^{-3\beta}+\rho_c^{(0)}\,\Omega_m^{(0)}\frac{\nu-\alpha}{1-\nu+\alpha}\,(1+z)^{3}\,.
\end{equation}
As can be checked it satisfies $\rD(0)=\rho_c^{(0)} (1-\Omega_m^{(0)})=\rho_D^{(0)}$ and it identically reduces to $\rho_D^{(0)}$ for $\nu=\alpha=1$, which is the $\CC$CDM result. Another way to see it is that, in this same limit, we have $\beta\to 0$ and the normalized Hubble function (\ref{eq:E2DC2}) becomes exactly the $\CC$CDM one. Ultimately the reason for this is the following: for a DE density of the form \DC2\ in \eqref{eq:ModelsC}, the second Friedmann Eq.\,(\ref{eq:PressureEqMPLA}) can be satisfied identically if we set $\omega_D=-1$ and $\alpha=\beta=1$, and at the same time neglect the radiation component. Needless to say, the self-conserved DE equation (\ref{eq:DEConsEq}) is also automatically satisfied, because it is not independent of the two Friedmann's equations. None of the other models could fulfill these conditions and consequently a nontrivial EoS different from $\wD=-1$ is required for them. It is thus not surprising that the fitting procedure singles out a parameter region for \DC2\ very close to the $\CC$CDM (cf. Table \ref{tableFit2}). In actual fact the fitting values of the parameters do carry some small deviations from unity, as this helps to slightly improve the quality of the fit as compared to the strict $\CC$CDM. This in turn induces a nontrivial evolution of the EoS, which eventually departs from $\wD=-1$ when we move to higher redshifts, so in practice \DC2\ mimics the $\CC$CDM only near our time and quickly deviates from it for $z\gtrsim1$ (cf. Fig. \ref{w(z)z1}, right).

In contrast to its vacuum counterpart -- viz. the C2 model studied in \cite{BasSola2014b} and Chapter \ref{chap:Atype} -- the \DC2-model does have a transition point from deceleration to acceleration.  A straightforward calculation yields
\be
z_{tr}=\left[-\omega_D^{(0)}\frac{\Omega_D^{(0)}}{\Omega_m^{(0)}}(3-3\nu+2\alpha)\right]^{\frac{1}{3(1+\beta)}}-1
=\left[\frac{(1-\nu+\alpha-\Omo)(3-3\nu+2\alpha)}{\alpha\Omega_m^{(0)}}\right]^{\frac{\alpha}{3(1-\nu+\alpha)}}-1\,.
\ee
As expected, for $\nu=\alpha=1$ (which implies $\wDo=-1$ from the aforementioned constraint) we recover exactly the $\CC$CDM result (\ref{eq:ztrLCDM}). As for the effective EoS of this model, $\wD(z)$, it can be directly computed with the help of \eqref{eq:EoS} and (\ref{eq:rDaDC2}). In particular one can check that for $z=0$ we obtain $\wD(0)=-(1-\Omega_m^{(0)}-\nu+\alpha)/(\alpha\Omega_D^{(0)})=\wDo$, which is consistent with the constraint obeyed by the parameters of this model.  For the numerical analysis presented in Table \ref{tableFit2} we have fixed $\wDo=-1$ for this model and we have fitted $\Omega_m^{(0)}$ and $\nu$, and in this way $\alpha$ becomes determined:
$\alpha=(\nu+\Omega_m^{(0)}-1)/\Omega_m^{(0)}$. For this reason we can use $\nu$ as the only vacuum free parameter for \DC2\,, and indeed this is the value that is meant for $\nueff$ in Table \ref{tableFit2}.
The plots for the densities, the EoS behavior, as well as for the deceleration parameter and the transition point from deceleration to acceleration for \DC2\ are included in Figures \ref{DEdensity}-\ref{q(z)}. The model works well at low $z$ but we should warn the reader that it may present serious difficulties in describing the radiation epoch, and in fact this is the reason why we have used only the low $z$ observables in its fit. This fact puts us on guard concerning the eventual viability of \DC2\,. We will come back to this important point in Sect. \ref{sect:Discussion}.

%%%%%%%%%%%%%%%%%%%%%%%%%%%%%%%%%%%%%%%%%%%%%%%%%%%%%%%%%%%%%%%%%
%%%%%%%%%%%%%%%%%%%%%%%%%%%%%%%%%%%%%%%%%%%%%%%%%%%%%%%%%%%%%%%%%
%%%%%%%%%%%%%%%%%%%%%%%%%%%%%%%%%%%%%%%%%%%%%%%%%%%%%%%%%%%%%%%%%
\section{Linear structure formation  with self-conserved DE and matter}
\label{sec:LSF}

In this section we deal with the cosmic perturbations for linear structure formation, and we commence with a study of the set of perturbation equations for a general system of self-conserved dynamical dark energy and matter. Subsequently we specialize our results for the concrete  ${\cal D}$-models whose background behavior has been elucidated in the previous sections, all of them falling in that category.

\subsection{Linear perturbations for matter and dark energy}

Let us consider a general system in which the various components (matter and DE) are covariantly self-conserved and the gravitational coupling G remains constant throughout the cosmic history. In the study of linear structure for this system we take into account both the matter perturbations, $\delta\rho_m$, and the perturbations in the DE component, $\rho_D$. Although the DE is not coupled to the matter sector at the background level owing to the assumption of separate covariant conservation, the two sectors develop some kind of interaction in the linear perturbations regime, which we are going to compute. We start by perturbing the $(0,0)$ component of Einstein's equations, together with the $(0,i)$ components of the covariant energy conservation equation, i.e. $\nabla_\mu G^{\mu\nu}=0$, using the synchronous gauge, i.e. $ds^2=dt^2+(-a^2\delta_{ij}+h_{ij})d\vec{x}^2$, and taking into account that all the components of the cosmic fluid are covariantly self-conserved. We obtain the following coupled system of diferential equations\,\footnote{See e.g. Refs.\,\cite{GrandePelinsonSola08,GSFS10} for a formulation of the perturbation equations in a system of several components, including the DE. Here we have introduced explicitly the perturbations in the dynamical EoS variable of the DE.}:

\be\label{eq:a}
\dot{\hat{h}}+2H\hat{h}=8\pi G\left[\prm+\prD(1+3\omega_D)+3\rho_D\poD\right]
\ee

\be\label{eq:b}
\rho_m\left(\theta_m-\frac{\hat{h}}{2}\right)+3H\prm+\dot{\prm}=0
\ee

\be\label{eq:c}
\rho_D(1+\omega_D)\left(\theta_D-\frac{\hat{h}}{2}\right)+3H[(1+\omega_D)\prD+\rho_D\poD]+\dot{\prD}=0
\ee

\be\label{eq:d}
\rho_m(\dot{\theta_m}+5H\theta_m)+\theta_m\dot{\rho}_m=0
\ee
\be\label{eq:e}
\left\{\rho_D(1+\omega_D)(\dot{\theta}_D+5H\theta_D)+\theta_D\left[\dot{\rho}_D(1+\omega_D)+\rho_D\dot{\omega}_D\right]\right\}\frac{a^2}{k^2} -(\omega_D\prD+\rho_D\poD)=0\,,
\ee
where $\theta_N\equiv\nabla_\mu\delta U_N^\mu$ is the divergence of the perturbed velocity for each component (matter and DE), $\hat{h}\equiv \frac{d}{dt}\left(\frac{h_{ii}}{a^2}\right)$ and $k$ stands for the wavenumber (recall that these equations are written in Fourier space).

%%%%%%%%%%%%%%%%%%%%%%%%%
%%%%%%%%%%%%%%%%%%%%%%%%%
%%%%%%%%%%%%%%%%%%%%%%%%%

\subsection{Reduction to a single third-order differential equation for matter perturbations}
\label{sect:ThirdOrder}

Notice that we have six unknowns and five equations, so at this stage is not possible to solve the system in order to find each of the perturbed functions $\theta_m$, $\theta_D$, $\hat{h}$, $\poD$, $\prD$ and $\prm$.  However, we will now show that under reasonable assumptions and working at sub-horizon scales we can eliminate the perturbations in the DE, i.e. $\delta\rD$, in favor of a single higher order equation for the matter part, $\delta\rmr$. This is characteristic of the coupled systems of matter and DE perturbations for cosmologies with self-conserved dynamical DE and matter. For the particular case $\CC=$const. the obtained third-order equation boils down to the (derivative of the) second order one of the $\CC$CDM, as we will show. Let us proceed step by step. Firstly, we make use of \eqref{eq:MatterConsEq} and \eqref{eq:d} so as to obtain $\theta_m(a)$:
\be\label{eq:theta}
\dot{\theta}_m(a)+2H(a)\theta_m(a)=0\rightarrow \theta_m(a)=\theta_{m}^{(0)} a^{-2}\,.
\ee
It follows that the matter velocity gradient decreases fast with the expansion. We shall adopt the conventional initial condition that $\theta_{m}^{(0)}$ is negligible or zero at present and hence $\theta_m\simeq0$ throughout the evolution. Since the scales relevant to the matter power spectrum remain always well below the horizon, i.e. $k\gg H$, we expect small DE perturbations at any time, i.e. the DE should naturally be smoother than matter, wherefore the velocity gradient of the DE perturbations is also naturally set to $\theta_D=0$. In this way we can confine our study to the perturbation $\delta_D$ in the DE sector and $\delta_m$ in the matter sector. This simplification looks reasonable and it will allow us to solve the coupled system of matter and DE perturbations without introducing further assumptions and/or additional parameters.
Under this setup it is clear that the terms that are not proportional to $k^2$ in \eqref{eq:e} can be neglected and we find $\omega_D\prD+\rho_D\poD=0$, which is tantamount to say that we are keeping the perturbations in the DE density but neglecting the perturbations in its pressure, similar as for matter.
From \eqref{eq:b} we can isolate $\hat{h}$,

\be\label{eq:h}
\hat{h}=\frac{2}{\rho_m}\left(3H\prm+\dot{\prm}\right)\,,
\ee
and differentiating this expression we arrive at:
\be\label{eq:hdot}
\dot{\hat{h}}=\frac{2}{\rho_m}\left(3\dot{H}\prm+3H\dot{\prm}+\ddot{\prm}\right)-\frac{2\dot{\rho}_m}{\rho^2_m}\left(3H\prm+\dot{\prm}\right)\,.
\ee
Now we insert the last two equations in \eqref{eq:a} and find:
\be\label{eq:Bfunc}
B(\prm,\dot{\prm},\ddot{\prm},H,\dot{H})=8\pi G\left[\prm+\prD(1+3\omega_D)+3\rho_D\poD\right]\,,
\ee
where we have defined

\begin{multline}\label{eq:defB}
B(\prm,\dot{\prm},\ddot{\prm},H,\dot{H})\equiv\frac{6\dot{H}\prm+6H\dot{\prm}+2\ddot{\prm}}{\rho_m}-\\
-\frac{\dot{\rho}_m(6H\prm+2\dot{\prm})}{\rho^2_m}+\frac{4H(3H\prm+\dot{\prm})}{\rho_m}\,. 
\end{multline}
On the other hand from \eqref{eq:h} and \eqref{eq:c} we gather
\be\label{eq:ct}
-\frac{\rho_D}{\rho_m}(1+\omega_D)(3H\prm+\dot{\prm})+3H[(1+\omega_D)\prD+\rho_D\poD]+\dot{\prD}=0\,.
\ee
We have three independent equations for the three unknowns: $\poD$, $\prD$ and $\prm$. We first take $\poD=-\omega_D\delta\rho_D/\rho_D$ from the above mentioned equation and substitute it in \eqref{eq:Bfunc} and \eqref{eq:ct}. We find:
\be\label{eq:att}
B(\prm,\dot{\prm},\ddot{\prm},H,\dot{H})=8\pi G\left(\prD+\prm\right)
\ee
and
\be\label{eq:ctt}
-\frac{\rho_D}{\rho_m}(1+\omega_D)(3H\prm+\dot{\prm})+3H\,\prD+\dot{\prD}=0\,.
\ee
At this point we can use the last two equations to get rid of the explicit $\prD$ dependence on the DE perturbations, and we obtain:
\be\label{eq:cttt}
-\frac{\rho_D}{\rho_m}(1+\omega_D)(3H\prm+\dot{\prm})+3H\left(\frac{B}{8\pi G}-\prm\right)+\frac{\dot{B}}{8\pi G}-\dot{\prm}=0\,.
\ee
This is the preliminary expression of the final differential equation for the linear density perturbations of the matter field, but in order to make it operative we must still do some algebra. Let us first rewrite the expression $B$, Eq.\,(\ref{eq:defB}), as follows:
\be\label{eq:B}
B=\frac{\prm}{\rho_m}(6\dot{H}+30H^2)+\frac{16H}{\rho_m}\dot{\prm}+2\frac{\ddot{\prm}}{\rho_m}\,,
\ee
and compute its time derivative,
\be\label{eq:Bdot}
\dot{B}=\frac{\prm}{\rho_m}(6\ddot{H}+78H\dot{H}+90H^3)+
\frac{\dot{\prm}}{\rho_m}(78H^2+22\dot{H})+22H\frac{\ddot{\prm}}{\rho_m}+\frac{2}{\rho_m}\dddot{\prm}\,,
\ee
where in the previous formulas we have made repeated use of the matter conservation \eqref{eq:MatterConsEq}.
Introducing the last two expressions in \eqref{eq:cttt}, we obtain the first explicit form of the sought-for third order equation for $\delta\rho_m$:
$$
\frac{\prm}{\rho_m}\left[\frac{6\ddot{H}+96\dot{H}H+180H^3}{8\pi G}-3H\left[\rho_m+\rho_D(1+\omega_D)\right])\right]+\frac{2\dddot{\prm}}{8\pi G\rho_m}+\frac{28H\ddot{\prm}}{8\pi G\rho_m}+
$$
\be\label{eq:inter}
+\frac{\dot{\prm}}{\rho_m}\left[\frac{126H^2+22\dot{H}}{8\pi G}-\left[(\rho_m+\rho_D (1+\omega_D)\right]\right]=0\,.
\ee
A final simplification of Eq.\,\eqref{eq:inter} can still be performed. From the pair of Friedmann's equations \eqref{eq:FriedmannEqMPLA} and \eqref{eq:PressureEqMPLA} we find the relation
\be
\dot{H}=-4\pi G[\rho_m+\rho_D(1+\omega_D)]\,.
\ee
Using it in \eqref{eq:inter} we finally arrive at the desired form of the third order differential equation for the perturbations of the matter field:
\be\label{eq:inter1}
 \dddot{\prm}+14H\ddot{\prm}+3\dot{\prm}(4\dot{H}+21H^2)+3\prm(\ddot{H}+30H^3+17\dot{H}H)=0\,.\ee
It is convenient to reexpress it in terms the density contrast $\delta_m\equiv \prm/\rho_m$. After some algebra we find the following rather compact form:

\be\label{eq:DifEqCosmicTime}
%\boxed{\dddot{\delta}_m+5H\ddot{\delta}_m+3\dot{\delta}_m(\dot{H}+2H^2)=0}
\dddot{\delta}_m+5H\ddot{\delta}_m+3\dot{\delta}_m(\dot{H}+2H^2)=0\,.
\ee
This is actually the final form in terms of the cosmic time.
However, to make contact with the observations it is highly advisable to rewrite the previous equation in terms of the scale factor, as it has a simple relation with the cosmic redshift. To this end we have to trade the time derivatives for the scale factor derivatives (indicating the latter by a prime), starting from $\dot{\delta}_m=aH\delta_m^\prime$ and its subsequent second and third order derivatives. After a simple calculation we obtain:

\be\label{eq:DifEqScaleFactor}
\delta_m^{\prime\prime\prime}+\delta_m^{\prime\prime}\left(\frac{8}{a}+\frac{3H^\prime}{H}\right)+\delta_m^\prime\left(\frac{12}{a^2}+\frac{12H^{\prime}}{aH}+\frac{H^{\prime 2}}{H^2}+\frac{H^{\prime\prime}}{H}\right)=0\,.
\ee
In general, this differential equation cannot be solved analytically, but we can proceed numerically. In any case we need to set up the initial conditions at a high redshift, when dust dominates over the dark energy i.e. typically at $z={\cal O}(100)$. The initial conditions are, of course, model-dependent, but it can be shown that for  \DA-models they reduce to the situation that comprises the well-known standard $\Lambda$CDM ones for the function and the first derivative, viz. $\delta_m(a_i)=a_i$ and $\delta_m^\prime(a_i)=1$, to which we have to add $\delta_m^{\prime\prime}(a_i)=0$.

Let us recall that the obtained third order differential equation for the matter perturbations does automatically incorporate the leading effect of the DE perturbations and is valid under the assumption that the values of the perturbed matter and DE velocity gradients are negligibly small. This is the simplest assumption we can make in order to solve the initial system of differential equations without introducing additional parameters; and it is certainly well justified if we are interested in the physics at scales deep inside the Hubble radius (sub-horizon scales), i.e $k\gg H$. From Eq.\,(\ref{eq:e}) we can see that deviations from this framework should scale roughly as $\sim H^2/k^2$. We will estimate them in the next section \ref{sect:DEorNot}.

%%%%%%%%%%%%%%%%%%%%%%%%%%%%%%%%%%%%%%%%
%%%%%%%%%%%%%%%%%%%%%%%%%%%%%%%%%%%%%%%%

\subsection{Recovering the \texorpdfstring{$\CC$}{Lambda}CDM perturbations as a particular case}
\label{subsec:RecoveringLCDM}

Let us consider the situation of the $\CC$CDM model deep in the matter-dominated epoch when we can neglect the $\CC$-term. In this case $H^2\propto a^{-3}$ and hence $H^{\prime}(a)/H(a)=-3/(2a)$ and $H^{\prime\prime}(a)/H(a)=15/(4a^2)$. It follows that Eq.\,\eqref{eq:DifEqScaleFactor} boils down in this case to the very simple form
\begin{equation}\label{eq:ReducedForm}
{\delta}_m^{\prime\prime\prime}+\frac72\,\frac{{\delta}_m^{\prime\prime}}{a}=0\,,
\end{equation}
whose elementary solution is $\delta_m=c_1\,a+c_2+c_3\,a^{-3/2}$. Neglecting the decaying mode and setting $c_2=0$ we are led to the standard growing mode solution of the $\CC$CDM, i.e. $\delta\propto a$. This is a clear indication that the $\CC$CDM result is contained in the generalized perturbation equation \eqref{eq:DifEqScaleFactor}.

Alternatively we can prove in a more formal way, in this case using the cosmic time variable, that the well-known $\Lambda$CDM linear perturbation equation\,\cite{Peebles1993,LiddleLyth},
\be\label{eq:LCDMperturbations}
\ddot{\delta}_m+2H\dot{\delta}_m-4\pi G\rho_m\delta_m=0\,,
\ee
or, written only in terms of the Hubble function,
\be\label{eq:DifLCDM0}
\ddot{\delta}_m+2H\dot{\delta}_m+\dot{H}\delta_m=0\,,
\ee
is a particular case of \eqref{eq:DifEqCosmicTime} if we treat the dark energy component as a rigid cosmological term with the vacuum constant EoS ($\omega_D=-1$). First, we perform the time derivative of the last equation in order to have a third order one:
\be\label{eq:DifLCDM}
\dddot{\delta}_m+2H\ddot{\delta}_m+3\dot{H}\dot{\delta}_m+\ddot{H}\delta_m=0\,.
\ee
To prove that \eqref{eq:DifEqCosmicTime} and \eqref{eq:DifLCDM} are the same when the DE is the traditional $\CC$-term, let us compute its difference. We find:
\be\label{eq:defDelta}
3H\ddot{\delta}_m+6H^2\dot{\delta}_m-\ddot{H}\delta_m=3H\left(\ddot{\delta}_m+2H\dot{\delta}_m-\frac{\ddot{H}}{3H}\delta_m\right)\,.
\ee
Upon differentiating the pressure equation (\ref{eq:PressureEqMPLA}) with respect to the cosmic time under the $\CC$CDM conditions (viz. $\dot{\rho}_{D}=0$) we find $\ddot{H}+3H\dot{H}=0$, so the difference \eqref{eq:defDelta} actually reads
\be 3H\left(\ddot{\delta}_m+2H\dot{\delta}_m+\dot{H}\delta_m\right)=0\,,\ee
where in the last step \eqref{eq:DifLCDM0}  has been used. This obviously proves our contention. At the same time it becomes clear from the proof that if the DE was dynamical (with $\dot{\omega}_D\neq 0$) we could not have obtained the previous result, which means that the third-order differential equation \eqref{eq:DifEqCosmicTime} is indeed more general than the standard one \eqref{eq:LCDMperturbations} and covers the entire class of dynamical DE models with separate (local and covariant) conservation of matter and dark energy densities, at fixed gravitational coupling $G$.

%%%%%%%%%%%%%%%%%%%%%%%%%%%%%%%%%%%%%%%%
%%%%%%%%%%%%%%%%%%%%%%%%%%%%%%%%%%%%%%%%

\subsection{Running  \texorpdfstring{$\CC$}{Lambda} and running \texorpdfstring{$G$}{G} with matter conservation}\label{subsec:ThirdOrderDifEq}

There is another interesting situation that we would like to mention in passing here in which a third order perturbation equation of the same sort is also needed, but under different conditions. It was first encountered in Ref.\,\cite{GSFS10} and is characterized by dynamical vacuum energy  (hence $\wD=-1$) and conserved matter. This case is, in principle, different from the class of models covered in the previous section because the EoS is of the vacuum type. However, it differs also in a second aspect, namely in the fact that  the gravitational coupling $G$ is running or time-evolving along with the vacuum energy density. In this way the Bianchi identity can be satisfied since there is a dynamical interplay between the vacuum and $G$ that permits the conservation of matter -- see \,\cite{Grande2011,GSFS10} for details. We will show now that the cosmic perturbations of the matter field for this system are also governed  by Eq.\,\eqref{eq:DifEqCosmicTime}, or equivalently by \eqref{eq:DifEqScaleFactor}. In Ref.\,\,\cite{GSFS10} (see also Appendix \ref{sec:AppenGpert}) it was shown that the matter perturbations follow the third order equation:
\be\label{eq:GPertEq}
{\delta}_m^{\prime\prime\prime}+\frac{{\delta}_m^{\prime\prime}}{2a}(16-9\Omega_m)+\frac{3\delta^\prime_m}{2a^2}(8-a\Omega^\prime_m+3\Omega_m^2-11\Omega_m)=0\,,
\ee
in which  $\Omega_m(a)={8\pi\,G(a)}\rho_m(a)/{3H^2(a)}$.
Equation \eqref{eq:GPertEq} also includes the effects of the perturbations in $\rho_\Lambda$, which have been eliminated in favor of the matter perturbations following a similar procedure as described in the previous section, except that in this case the dynamics of $G$ enters explicitly too. Taking into account that $\dot{H}=aH(a)H^\prime(a)=-4\pi G(a)\rho_m(a)$  it follows that $\Omega_m(a)=(-2/3)a\, H^{\prime}(a)/H(a)$; and using now this relation in \eqref{eq:GPertEq} we immediately recover \eqref{eq:DifEqScaleFactor}. This confirms that \eqref{eq:DifEqScaleFactor} is, in fact, also valid for models with self-conserved matter and a dynamical vacuum triggered by the time variation of G. We conjecture that \eqref{eq:DifEqScaleFactor} may also be appropriate for more general models with self-conserved matter and different possibilities for
the evolutions of $G$ and the DE, but we will not further pursue the framework of $G$-variable models here. See Chapters \ref{chap:Gtype} and \ref{chap:AandGRevisited} for recent analyses of the G-type models.

%%%%%%%%%%%%%%%%%%%%%%%%%%%%%%%%%%%%%%%%
%%%%%%%%%%%%%%%%%%%%%%%%%%%%%%%%%%%%%%%%
%%%%%%%%%%%%%%%%%%%%%%%%%%%%%%%%%%%%%%%%

\section{Fitting the  \texorpdfstring{$\mathcal{D}$}{DA}A,  \texorpdfstring{$\mathcal{D}$}{DC}C and  \texorpdfstring{$\mathcal{D}$}{DH}H models to the observational data}
\label{sec:FittingDE}

Up to this point we have carried out a thorough study of both the background equations as well as of the linear perturbation equations that are obeyed by the class of models under consideration. Therefore we are in position to understand the fitting results collected from the beginning in Table \ref{tableFit2} and that were used throughout to evaluate a number of observables associated with the background cosmology in Figs.\,\ref{DEdensity}-\ref{q(z)}. At the same time we need to further elaborate on the implications of these dynamical DE models on the structure formation. But let us first summarize the methodology employed in our statistical analysis of the cosmological data.

\subsection{Global fit observables}

To carry out the fit we have used the expansion history  (SNIa+BAO$_A$+BAO$_{d_z}$) and the CMB shift-parameter, as well as the linear growth data, through the following joint likelihood analysis:
\be
\mathcal{L}_{tot}(\vec{p})=\mathcal{L}_{SNIa}\times\mathcal{L}_{BAO}\times\mathcal{L}_{CMB}\times\mathcal{L}_{f\sigma_8},
\ee
where $\vec{p}=(\Omega^{(0)}_m,\nu)$. By maximizing the total likelihood or, equivalently, by minimizing the joint $\chi^2_{tot}$ function with respect to the elements of $\vec{p}$:
\be
\mathcal{\chi}^2_{tot}(\vec{p})=\mathcal{\chi}^2_{SNIa}\times\mathcal{\chi}^2_{BAO}\times\mathcal{\chi}^2_{CMB}\times\mathcal{\chi}^2_{f\sigma_8},
\ee
we have obtained the best-fit values for the various parameters. In the current analysis we have made use of the same data points described in Sect. \ref{sec:fitting} for the SNIa, BAO$_A$, BAO$_{d_z}$ and CMB observables. In the expression of the luminosity distance \eqref{eq:LumDist} entering the SNIa sector we have plugged the value $H_0=70$ km/s/Mpc, as it is used in the Union 2.1 sample. In all the theoretical expressions that are not the SNIa distance modulus ones, we have used the current value of the Hubble function that has been found by the Planck Collaboration\,\cite{Planck2015}, i.e. $H_0=67.8$ km/s/Mpc. 

Finally, we have also taken into account the data on the linear growth rate of clustering provided from different sources, see specifically \cite{Percival2004,Tegmark2006,Guzzo2008,Song09,Blake2011LSS,Hudson2012,Samushia2012,Beutler2012,Tojeiro2012, Reid2012}. More concretely, we have used $f(z)\sigma_8(z)$ and the corresponding $\chi^2$-function:
\be
\chi_{f\sigma_{8}}^2(\vec{p})=\sum_{i=1}^{16}\left(\frac{f\sigma_{8}(\vec{p},z_i)-f\sigma_{8,obs}(z_i)}{\sigma_{f\sigma_{8},i}}\right)^2\,.
\ee
In the rest of Sect.\,\ref{sec:FittingDE} we focus on the observables used to fit the structure formation data, as we feel that they play a very important role in our analysis after we have derived in the previous section the general perturbation equation for the class of self-conserved matter and DE models; and of course we report also on the impact on our models. In particular, we wish to compare the results of our fitting analysis in the situation when the DE perturbations are included (as analyzed in the previous section) and when the DE perturbations are not included and the effects of the DE act only on the background history.

%%%%%%%%%%%%%%%%%%%%%%%%%%%%%%%%%%%%%%%%
%%%%%%%%%%%%%%%%%%%%%%%%%%%%%%%%%%%%%%%%
%%%%%%%%%%%%%%%%%%%%%%%%%%%%%%%%%%%%%%%%

\subsection{Structure formation with and without DE perturbations}
\label{sect:DEorNot}

\begin{figure}[!t]
\begin{center}
\includegraphics[scale=0.8]{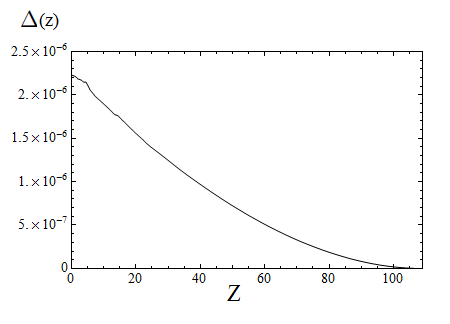}
\caption[Effect of the DE perturbations at subhorizon scales]{\footnotesize{
Relative differences $\Delta(z)$, Eq.\, (\ref{eq:DefDelta}), between the matter density contrasts $\delta_m^{\rm DE}(z)$ computed with the third order differential equation \eqref{eq:DifEqScaleFactor}, which includes DE perturbations, and the result $\delta_m(z)$ from the second order one \eqref{eq:LambdaDifeq}, where the DE density enters only at the background level. The curve shows the essentially degenerate result valid for all the $\mathcal{D}$A models.
\label{DifEqComparison}}
}
\end{center}
\end{figure}

As we have explained in Sect. \ref{sect:ThirdOrder},
we compute the linear matter perturbations $\delta_m$ of the models under consideration through the third-order equation \eqref{eq:DifEqScaleFactor}, which already encodes the effect of the DE perturbations. When the DE perturbations are not included we use the $\CC$CDM form (\ref{eq:LCDMperturbations}), which in terms of the scale factor variable can equivalently be written as
\be\label{eq:LambdaDifeq}
\delta^{\prime\prime}_m+\left(\frac{3}{a}+\frac{H^\prime}{H}\right)\delta^\prime_m+\frac{H^\prime}{aH}\delta_m=0\,,
\ee
with the understanding that the Hubble function in this expression involves the DE density of the corresponding ${\cal D}$-model in Sect. \ref{sec:DEmodels}. Thus, when using this approximation, the DE effects enter only at the background level. It is surely interesting to compare the results obtained in this approximate way with those generated from the more accurate treatment when the DE perturbations are compute from Eq.\,\eqref{eq:DifEqScaleFactor}. We do not foresee, however, large differences between the two treatments since the DE is expected to be much smoother than matter, at least at the scales where linear structure formation takes place. Still, the fact that we have been able to provide a combined treatment of the DE and matter perturbations in these models gives us a good opportunity to test these expectations.

In Fig. \ref{DifEqComparison} we provide a quantitative test. We have plotted the relative difference
\be\label{eq:DefDelta}
\Delta (z)=\frac{\delta_m^{\rm DE}(z)-\delta_m(z)}{\delta_m(z)}
\ee
as a function of the redshift. The density contrast $\delta_m^{\rm DE}(z)$ corresponds to the solution of the third order equation \eqref{eq:DifEqScaleFactor} and therefore involves the effect of the DE perturbation, whereas $\delta_m(z)$ is the solution to the more conventional second order equation (\ref{eq:LambdaDifeq}), expressed both in terms of the cosmic redshift $z=(1-a)/a$. As we can see the differences well inside the horizon are very small, of order $\Delta\sim 10^{-6}$. Let us recall that the observational data concerning the linear regime of the matter power spectrum lie in the approximate wave number range $0.01 h {\rm Mpc}^{-1} \lesssim k \lesssim 0.2  h {\rm Mpc}^{-1}$, whereas  the current horizon is given by $H_0^{-1}\simeq 3000\,h^{-1} {\rm Mpc}$ and hence $H_0\simeq 3.33\times 10^{-4} h {\rm Mpc}^{-1}$. From our discussion in Sect. \ref{sect:ThirdOrder} we expect that when we take larger and larger scales (i.e. smaller values of $k$ as compared to $H$) potentially important contributions scaling as $\sim H^2/k^2$ can develop. Clearly the ratio squared of $H_0$ to the minimum and maximum values of the wave number in the linear regime satisfies $H_0^2/k^2\simeq 10^{-6}-10^{-3}\ll 1$  and hence it is natural to expect that the DE perturbations are highly suppressed. Our calculation confirms explicitly this fact. Only if we would approach scales of the order of the horizon such suppression would no longer hold and we could expect sizeable effects from them.

With this nontrivial test we can say that the DE perturbations for the  dynamical ${\cal D}$-models under consideration became fully under control; and in fact we find that at subhorizon scales they do not trigger large deviations from the zeroth order (smooth) DE effects that each model already imprints on the background cosmology. Only at scales comparable to the horizon and for non-negligible velocity gradients the DE perturbations can play a role.

%%%%%%%%%%%%%%%%%%%%%%%%%%%%%%%%%%%%%%%%
%%%%%%%%%%%%%%%%%%%%%%%%%%%%%%%%%%%%%%%%
%%%%%%%%%%%%%%%%%%%%%%%%%%%%%%%%%%%%%%%%

\subsection{Linear growth and growth index}
\label{sec:LinearGrowth}

In practice it is convenient to investigate the linear structure formation of our models through the so-called linear growth rate \eqref{eq:growingfactor}. A related quantity is the $\gamma$-index of matter perturbations, $f(z)\simeq \Om(z)^{\gamma(z)}$. For the $\CC$CDM we have $\gamma\simeq 6/11\simeq 0.545$, and one typically expects $\gamma(0)=0.56\pm 0.05$ for $\CC$CDM-like models\,\cite{Athina2014}. The growth index can obviously be reexpressed as in \eqref{eq:growingfactor2},
\be\label{eq:gamma}
\gamma(z)\cong\frac{\ln\,f(z)}{\ln\,\Omega_m(z)}\,.
\ee
By definition $\Om(z)=\rmr(z)/\rc(z)$, and hence for the class of self-conserved models under study it reads
\begin{equation}\label{eq:OmzDef}
\Om(z)=\frac{\rho_m^{(0)}\,(1+z)^3}{\rho_m^{(0)}\,(1+z)^3+\rD(z)}\,,
\end{equation}
 in which $\rD(z)$ is the DE density of the corresponding ${\cal D}$-model. Obviously $\Om(0)=\rho_m^{(0)}/\rho_c^{(0)}=\Omega_m^{(0)}$, with $\rho_c^{(0)}=\rho_m^{(0)}+\rho_D^{(0)}=3H_0^2/(8\pi G)$ the current critical density. The explicit form for $\rD(z)$ for the models under consideration has been determined in Sect. \ref{sec:background}.

\begin{figure}
\begin{center}
\includegraphics[scale=0.55]{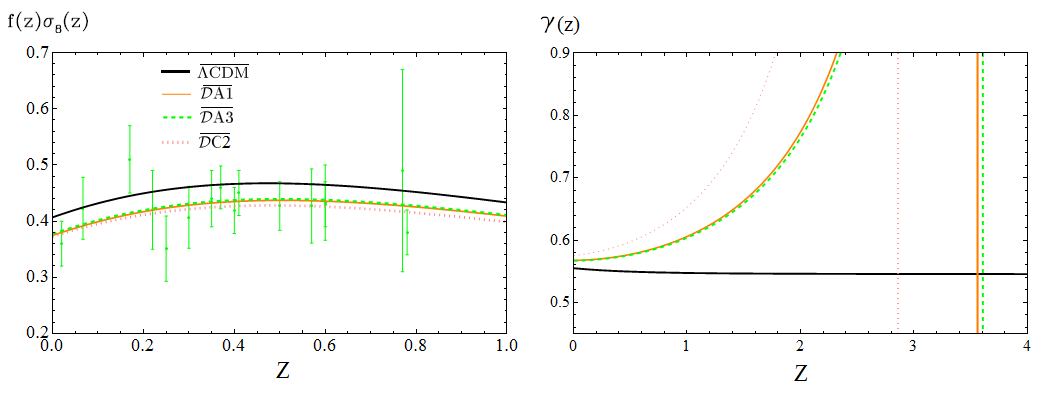}
\caption[$f(z)\sigma_{8}(z)$ and $\gamma(z)$ for the \DA\ and DC2\ models]{\footnotesize{
{\bf Left}: Comparison of the observed and
theoretical evolution of the weighted growth
rate $f(z)\sigma_{8}(z)$ for the $\Lambda$CDM versus the \DA\ and DC2\ models; {\bf Right}: The evolution of the growth index \eqref{eq:gamma} for the same models in a wide span of $z$ so as to display the vertical asymptotes at the points where the DE density  $\rD(z)$ vanishes in each case and the $\gamma$-index (\ref{eq:gamma}) becomes singular. The bar above the label for the various models corresponds to using the barred fitting parameters of Table \ref{tableFit2}, whose meaning is explained there.
\label{sigma8gamma}}
}
\end{center}
\end{figure}

In recent times, however, a more useful quantity is found to be the weighted growth rate of clustering, i.e. $f(z)\sigma_8(z)$. This quantity has the advantage of being independent of the galaxy density bias\,\cite{Song09}, see Sect. \ref{subsec:growthrate} for details. Now, we use the Planck 2015 value $\sigma_{8,\Lambda}=0.815\pm 0.009$ (cf. \cite{Planck2015}) in the normalization of the power spectrum, instead of the Planck 2013 one used in Chapter \ref{chap:Atype}. 

Notice, again, that $k_{eq}$ is a model-dependent quantity that must be derived for each of the models we are studying in this chapter. For the $\CC$CDM we have the standard result \eqref{eq:keqLCDM}. However, for the \DA\ and \DC1 we have to introduce a correction factor, which is analytically calculable. We find:
\begin{eqnarray}
(\mathcal{D}A):\qquad\qquad
k_{eq}&=&k^{\Lambda}_{eq}\left[\frac{3/2}{4\alpha+3(1-\nu)}+\frac{1/2}{1+\alpha-\nu}\right]^{1/2}\label{eq:keqDA}\\
(\mathcal{D}C1):\qquad\qquad k_{eq}&=&\frac{k^{\Lambda}_{eq}}{\sqrt{1-\nu}}\label{eq:keqDC1}\,.
\end{eqnarray}
As we can see, the correction factor in (\ref{eq:keqDA}) depends separately from $\nu$ and $\alpha$, not just on the difference.
The corresponding formulas for $DA1$ and $DA3$ models is found by setting $\alpha=0$ and $\nu=0$, respectively, in the first expression. Notice that the first formula above reduces to the second for $\alpha=0$, as expected from the fact that at high $z$ the model \DC1 behaves as \DA1. This also explains why the effect of the $\epsilon$-term for \DC1 is negligible in this regime and does not appear in (\ref{eq:keqDC1}). As for the correction factor for model \DC2\ it is derived also from (\ref{eq:keqDA}). Recall that once $\nu$ is fixed from the fit value, the parameter $\alpha$ becomes determined for this model under the conditions discussed at the end of Sect.\, \ref{sect:DCModels}.

The model-dependent corrections to $k_{eq}$ do modify the shape of the transfer function and can be of importance in some cases.  It can be instructive for the reader to compare the above correction formulas with the ones determined for the corresponding dynamical vacuum models in Sect. \ref{sec:ExtraTypeB} -- see Eqs. \eqref{eq:keqC1Bmodel} and \eqref{eq:keqC1Amodel}. One may be a bit surprised at first that the formula for the correction factor found there for C1A is considerably more complicated than the one found here for \DC1. The reason is that the matter density for the former model has, at variance with the latter, an anomalous behavior at high redshift  -- cf. our discussion in  Sect.\, \ref{sect:DCModels}, particularly Eq.\,(\ref{eq:rhomaLinear2}). Even so the correction indicated in (\ref{eq:keqDC1}) for \DC1\ can be significant because for this model $\nu$ is not {\it a priori} a negligible parameter (cf. Table \ref{tableFit2}). As a matter of fact, even for models of the \DA-subclass the correction to $k_{eq}$ is of some significance for the overall fit, e.g. it diminishes slightly the $\chi^2$ value with respect to the situation without correction.

%%%%%%%%%%%%%%%%%%%%%%%%%%%%%%%%%%%%%%%%
%%%%%%%%%%%%%%%%%%%%%%%%%%%%%%%%%%%%%%%%
%%%%%%%%%%%%%%%%%%%%%%%%%%%%%%%%%%%%%%%%

\subsection{Results}

\begin{figure}[!t]
\begin{center}
\includegraphics[scale=0.75]{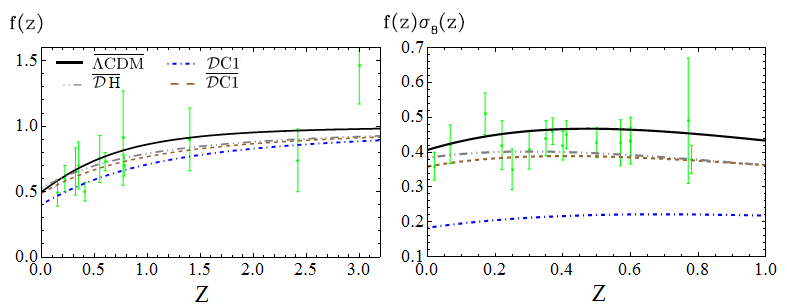}
\caption[Linear and weighted growth rate for the various $\mathcal{D}$-models]{\footnotesize{
{\bf Left}: The evolution of the linear growth rate of clustering, Eq. \eqref{eq:growingfactor}, specifically for the general \DC1-model and the \DHlin\ one (i.e. the linear model $\rD\sim H$) both for the barred and unbarred fitting parameters indicated in Table \ref{tableFit2}. For reference, the $\Lambda$CDM model is also included; {\bf Right}: The corresponding curves for the weighted growth rate $f(z)\sigma_{8}(z)$. The dash-dotted line (in blue) in both plots corresponds to the best-fit values of the parameters  for the situation when the structure formation data are not used. This plot makes apparent the delicate situation of the \DC1-model when the unbarred fit parameters are used, particularly in terms of $f(z)\sigma_{8}(z)$ rather than just in terms of $f(z)$ -- see also the text.
\label{sigma8DC1}}
}
\end{center}
\end{figure}

\begin{figure}[!t]
\begin{center}
\includegraphics[scale=0.7]{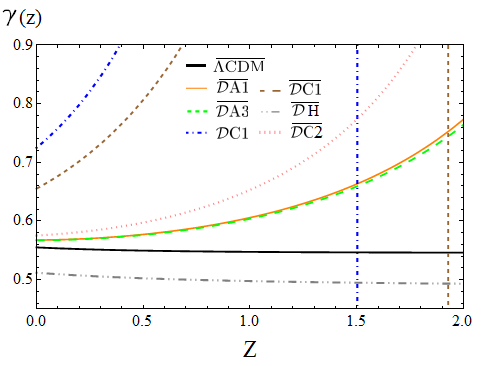}
\caption[$\gamma(z)$ in the range $0<z<2$ for the various $\mathcal{D}$ models]{\footnotesize{Magnified view of the growth index plot (cf. Fig. \ref{sigma8gamma}, right) for the local range $0<z<2$. We have superimposed also the growth index for the general
$\mathcal{D}$C1-model (dash-dotted blue line for the unbarred case and dashed brown line for the barred one) and the linear model $\mathcal{D}$H (dash-dotted gray line). The curves have been obtained under the same conditions of the two previous figures.
\label{sigma}}
}
\end{center}
\end{figure}

\begin{figure}[!t]
\begin{center}
\includegraphics[scale=0.65]{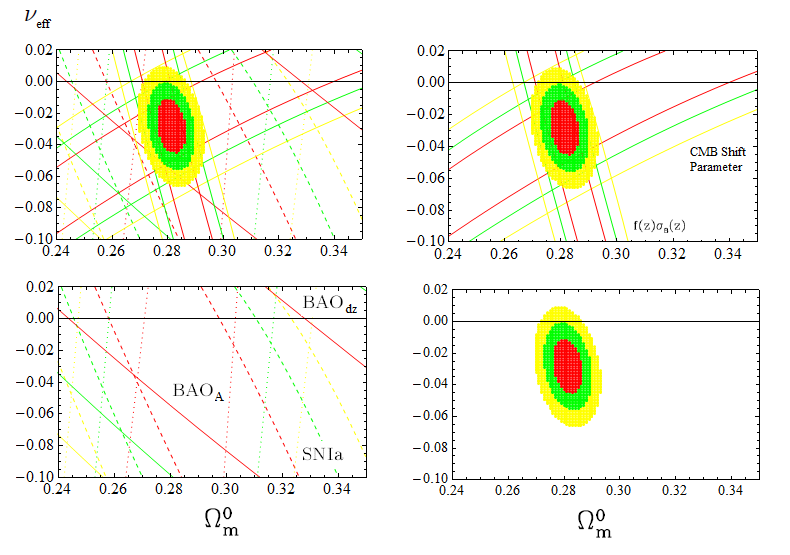}
\caption[Contour lines in the $(\Omega_m^{(0)},\nueff)$ plane for model \DA2\ ]{\footnotesize{
Likelihood contours in the $(\Omega_m^{(0)},\nueff)$ plane (for the values $-2\ln\mathcal{L}/\mathcal{L}_{max}=2.30$, $6.16, 11.81$, corresponding to 1$\sigma$, 2$\sigma$ and $3\sigma$ confidence levels) for the \DA1-model using the full expansion history and CMB shift parameter data (BAO+SNIa+CMB). Models \DA2\ and \DA3\ present very similar plots, as can be expected from the statistical analysis of Table \ref{tableFit2}, and for this reason we have denoted $\nueff$ the vertical axis of the plot. In all cases the $\nueff=0$ region ($\CC$CDM) is disfavored at $\sim 2\sigma$ level.
\label{CLW1}}
}
\end{center}
\end{figure}

Let us now further discuss the specific results obtained for the ${\cal D}$-models under study. Table \ref{tableFit2}, repeatedly referred to in the previous sections, collects in a nutshell the main numerical output describing the outcome of our analysis. The basic traits of the background history of these models have been presented in Sect.\,\ref{sec:background}. Here we mainly focus on the structure formation results. They are represented in Figs. \ref{sigma8gamma}-\ref{sigma}. Furthermore, the contour plots combining all the observational data (including of course the structure formation data extracted from the linear growth rate and associated quantities) is indicated in Fig. \ref{CLW1}.  Next we discuss these figures in turn.

In Fig. \ref{sigma8gamma} (left) we consider models \DA\ and \DC2, together with the concordance model $\CC$CDM. We plot their theoretical prediction of $f(z)\sigma_8(z)$ versus the data points (provided in the above mentioned references\,\cite{Percival2004,Tegmark2006,Guzzo2008,Song09,Blake2011LSS,Hudson2012,Samushia2012,Beutler2012,Tojeiro2012, Reid2012}) as a function of the redshift and for the best-fit values of the parameters in Table \ref{tableFit2}. We use the barred parameters of the table, as in this case we take into account the structure formation data in the fit. Notice that since model \DA2\ interpolates between \DA1\ and \DA3\ we have plotted the last two only in order to show the range that is covered. The fitting results for \DA2\ recorded in Table \ref{tableFit2} correspond to fixing $\alpha=-\nu>0$. In fact, since $\alpha$ must be positive (cf. Sect. \ref{sec:BackgroundCosmology}) the previous setting is illustrative of how to break degeneracies among the parameters in a way that is consistent with the sign of $\alpha$. Notice that, at this point, this is necessary since the $\nu$ and $\alpha$ dependence in e.g. equations (\ref{eq:keqDA}) and (\ref{eq:keqDC1}) is no longer of the form $\nu-\alpha$. Other choices compatible with $\alpha>0$ give very similar results.

The reason why we have included \DC2\ also in Fig. \ref{sigma8gamma}  stems from our discussion in Sect.\,\ref{sect:DCModels}, where we showed that this model can mimic the $\CC$CDM near our time, so it is natural to joint it in the same group of models approaching the concordance cosmology now. We can see that the three curves thrive quite well and stay together slightly below the $\CC$CDM line. They actually provide a better fit than the $\CC$CDM (cf. Table \ref{tableFit2}). We will further qualify the magnitude of this improvement in the next section. On the other hand, in Fig. \ref{sigma8gamma} (right) we depict the growth index, $\gamma(z)$, defined above. Here the three models are seen to approach together only when $z\to 0$, i.e. around the current time. While \DA1\ and \DA3 still remain quasi-glued at all points, \DC2\ departs significantly from them at larger redshifts. So the mimicking of the $\CC$CDM by \DC2\ is actually not better than that of the \DA-models. Let us also note the presence of vertical asymptotes, which are connected with the vanishing of the denominator in Eq.\,(\ref{eq:gamma}). These are the points where the DE density $\rD(z)$ becomes zero (confer Fig. \ref{DEdensity}), and hence $\Omega_m\to 1$ from  Eq.\,(\ref{eq:OmzDef}). Recall also from Fig.\, \ref{w(z)z1} that at these same points the EoS parameter of the DE becomes singular and develops a corresponding asymptote.

In Fig. \ref{sigma8DC1} we compare the situation of the \DC1\ and \DHlin\ models (together with the $\CC$CDM as a reference). We have separated these ${\cal D}$-models from the rest because we know that they behave differently. Let us first focus on the plot on the left, where we display the linear growth function $f(z)$ for them. In the case of \DC1\ we include both the curve using the barred parameters and the unbarred ones (cf. Table \ref{tableFit2}). We can see that in the latter case the \DC1\ curve is displaced only slightly below the others.  However, when we inspect the figure on the right, corresponding to the weighted function $f(z)\sigma_8(z)$, i.e. the growth function appropriately rescaled with the rms mass fluctuation amplitude at each redshift, the same \DC1\ curve now strays downwards quite significantly from the rest. As it turns, the weighted linear growth $f(z)\sigma_8(z)$ appears specially sensitive to that. It demonstrates that when we attempt to fit \DC1\ without using the structure formation data the model immediately gets in tension with them, quite in contrast with the situation with the \DA\ models.

The anomalous behavior of \DC1\  (and, {\it a fortiori}, of \DHlin) is particularly clear from the numbers in Table \ref{tableFit2}, where the difference between the unbarred and barred values of the fitting parameter (viz. $\nu\simeq -0.64$ and $\bar{\nu}\simeq-0.35$) is abnormally large, in contrast to the other models. Such behavior is also reflected in the large increase of $\chi^2$, recorded in Table \ref{tableFit2}, for this model when we compare $\chi_r^2$, that is to say, the ``reduced $\chi^2$ value'' (the value which is computed without including
the linear growth $\chi^2_{f\sigma_8}$ contribution)  with the unreduced one -- when the growth data are counted in the $\chi^2$ computation. The latter can be computed either without optimizing the fit to the growth data  (yielding $\chi^2/dof=880.74/600$) or after optimizing it (rendering $\bar{\chi}^2/dof=635.23/600$). In both cases the result is much larger than $\chi_r^2/dof=563.86/584$. Obviously the second result is better than the first, but both are extremely poor since they have to be compared with the respective  $\CC$CDM values,  $\chi^2/dof=584.91/608$ and  $\bar{\chi}^2/dof=584.38/608$.  This is  in contrast to the situation with the \DA\ models, where we can check that the difference between the reduced $\chi_r^2$ value and the corresponding $\chi^2$ values computed with or without optimizing the fit to the growth data is by no means as pronounced as in the \DC1\ and \DHlin\ cases.

We can actually retrace the same features described above if we look at Fig. \ref{sigma}. This is a magnified view of the growth index plot in Fig. \ref{sigma8gamma} (right) for the local range $0<z<2$, where we have superimposed also the growth index for the general
\DC1-model and the linear model \DHlin\,. These lines are in correspondence with the ones in Fig. \ref{sigma8DC1} and highlight once more the anomalous behavior of the \DC1-model and its exceeding sensitivity to the structure formation data. The current growth index value $\bar{\gamma}(0)\gtrsim 0.66$ (resp. $\gamma(0)\gtrsim 0.72$) that one can read off in Fig. \ref{sigma} for \DC1\ when the structure formation data are used (resp. not used) in the fit, is clearly too large to be acceptable. The vertical asymptote at $z\simeq 1.5$ (where $\rD(z)$ vanishes) is much closer here because the best-fit value for the unbarred $\nu$ is much bigger in absolute value than  $\bar{\nu}$. As for $\mathcal{D}$H it has no asymptote because the corresponding $\rD(z)$ never vanishes (cf. Fig. \ref{DEdensity}). This model also separates from below the $\CC$CDM line. Models \DA\,, instead, perform quite well and meet sufficiently close the $\CC$CDM horizontal line at $\gamma\simeq 6/11\simeq 0.545$  for $z\to 0$.

In short, the unfavorable verdict for \DC1\ seems to have no cure: when we enforce the adjustment of the growth points, then it is the expansion history data that becomes maladjusted; and conversely, if we strive to adjust the expansion history data, then it is the structure formation data that get strayed. In both cases the $\chi^2$ value of \DC1\ becomes too large in comparison to the other models. The exceeding sensitivity to the growth rate and associated quantities is symptomatic of its delicate status as a candidate model to describe the overall observations. And, of course, everything we have said for \DC1\ applies to \DHlin\,, which is the particular case $\nu=0$ of the latter. The model \DHlin\ has no free parameter in the vacuum sector and therefore there is nothing we can do to improve its delicate situation, which in the light of Table \ref{tableFit2} appears to be the less favorable one among the models under study.

From the plots in Figs. \ref{sigma8gamma}-\ref{sigma} and the statistical results of Table \ref{tableFit2} we can say that the \DA\ models are capable of describing the overall set of data in a way which is perfectly competitive with the $\CC$CDM. Model \DA1\,, in particular, represents the simplest realization with one parameter, $\nu$. The same can be said of model \DA3\ with the parameter $\alpha$. These models are similar because the time evolutions induced by $H^2$ and $\dot{H}$ are comparable. As for model \DA2\ it is the most general of the \DA-subclass and interpolates between the two previous cases. All of them offer a better fit quality than the $\CC$CDM. In the next section we further qualify this assertion.

%%%%%%%%%%%%%%%%%%%%%%%%%%%%%%%%%%%%%%%%%%%%%%%%%%%%%%%%%%%%%%%%%
%%%%%%%%%%%%%%%%%%%%%%%%%%%%%%%%%%%%%%%%%%%%%%%%%%%%%%%%%%%%%%%%%
%%%%%%%%%%%%%%%%%%%%%%%%%%%%%%%%%%%%%%%%%%%%%%%%%%%%%%%%%%%%%%%%%

\subsection{Discussion}\label{sect:Discussion}

Let us now focus on Fig. \ref{CLW1}, where we display a detailed representation of the fitting procedure we have followed in order to pin down the most favorable region of the various DE models under study. In that figure we specifically consider the case of model \DA1\,, where we can carefully appraise the way we have combined all the data on expansion history and structure formation mentioned in Table \ref{tableFit2}. Thanks to the juxtaposition of the permitted regions defined by each observable we find that a well delimited domain emerges. This bounded, elliptically shaped, area is what we may call the physical region for each model. The upper plot on the left of that figure displays all the lines involved in the projection of the final physical region; right next we show a plot with the boundary lines corresponding to the CMB shift parameter and the weighted growth rate $f(z)\sigma_8(z)$; then the lower plot on the left highlights the edges associated with the two types of BAO data involved, together with the borders of the area allowed by the supernovae (SNIa) observations; finally, the net intersection of all of them is represented by the lower plot on the right, in which we depict the $1\sigma$, $2\sigma$ and $3\sigma$ domains within the physical region.
The corresponding plots for models \DA2\ and \DA3\ are very similar and are not shown.

A most remarkable feat emerging from our analysis is the fact that the \DA-subclass of dynamical DE models is capable of improving the fit quality of the $\CC$CDM. This observation is specially significant if we bare in mind that the physical region in Fig. \ref{CLW1} is mostly located outside the domain of the concordance model, namely it lies roughly $\sim 2\sigma$ away from the $\nueff=0$ line representing the $\CC$CDM; specifically, we can estimate with accuracy that $84.27\%$ of the area of the  $3\sigma$ domain is below the $\nueff=0$ line.
Models \DA\,, therefore, represent a challenge to the $\CC$CDM since they are phenomenologically quite competitive. Somehow they ``break the ice'' inherent to the rigid character of the cosmological $\CC$-term, showing that the possibility of a DE is not only more natural from the theoretical point of view but also more instrumental for adjusting the cosmological observations.

In order to assess the statistical quality of our fits, and the competition score between the models, we have displayed their global $\chi^2$ value per degree of freedom in Table \ref{tableFit2}. However, in the same table we provide also (in the last columns) the Akaike Information Criterion (AIC)\,\cite{Akaike1974}, which is particularly useful for comparing competing models describing the same data and is amply used in many areas of science --- see e.g. the book\,\cite{Burnham}. We wish to use the AIC here to compare the ${\cal D}$-models with the $\CC$CDM. Such criterion is formulated directly in terms of the maximum of the likelihood function, ${\cal L}$, and is defined as follows. Let $n_p$ be the number of independent estimable parameters in a given model and $N$ the sample size of data points entering the fit. If ${\cal L_{\rm max}}$ is the maximum value of the likelihood function, the corresponding AIC value is defined as follows:
\begin{equation}\label{eq:IACDef}
{\rm AIC}=-2\ln{\cal L_{\rm max}}+2n_{p}+\,\frac{2n_p(n_p+1)}{N-n_p-1}\,.
\end{equation}
This is the generalization of formula \eqref{eq:AkaikeShort}, which is also valid for $N/n_p<40$. As mentioned in Sect. \ref{subsec:combined likelihood}, the comparison criterion based on this statistics is the following: given two competing models describing the same data, the model that does better is the one with smaller AIC value. Notice that there is a kind of tradeoff between the first term on the {\it r.h.s.} of \eqref{eq:IACDef} and the other two. The first term (the one carrying the maximum likelihood value) tends to decrease as more parameters are added to the approximating model (since ${\cal L_{\rm max}}$ becomes larger), while the second and third terms increase with the number of parameters and therefore represent the penalties applied to our modeling when we add more and more parameters. In particular, the second term ($2n_p$) represents an universal penalty related only to the increase in the total number of independent parameters, whereas the third term is an extra penalty applied when the sample is not sufficiently large as compared to the number of independent parameters.
For large samples $N\gg n_p$ (typically for $N/n_p > 40$) the third term becomes negligible and the above formula simplifies. This is actually the situation that holds good in our case. Indeed, for the $\CC$CDM we have one independent parameter ($n_p=1$) represented by $\Omega_m^{(0)}$, whereas for \DA1\,, for instance, we have $n_p=2$, namely $(\Omega_m^{(0)},\nu)$. In the case of \DA2\ we have $(\Omega_m^{(0)},\nu,\alpha)$ and hence $n_p=3$, but since we fix a relation between $\nu$ and $\alpha$ we have again $n_p=2$. On the other hand we can see from Table \ref{tableFit2} that $N\simeq 590$ in all cases, and therefore the condition $N/n_p > 40$ is amply satisfied and the use of the more simplified formula is fully justified in this case. For Gaussian errors, the maximum of the likelihood function can be expressed in terms of the minimum of $\chi^2$ and therefore Eq. \eqref{eq:IACDef} can be reexpressed as in \eqref{eq:AkaikeShort}.

As in Chapter \ref{chap:Atype}, to test the effectiveness of models $M_i$ and $M_j$, one considers the pairwise difference (AIC increment)
$(\Delta$AIC$)_{ij} = ({\rm AIC})_{i} - ({\rm AIC})_{j}$. The larger the
value of $\Delta_{ij}\equiv|\Delta({\rm AIC})_{ij}|$, the higher the evidence against the
model with larger value of ${\rm AIC}$, with $\Delta_{ij} \ge 2$ indicating a positive such evidence and $\Delta_{ij}\ge 6$
denoting significant such evidence. Finally the evidence ratio (ER) against one model whose AIC value is larger than another is judged by the relative likelihood of model pairs, ${\cal L}_i/{\cal L}_j$, or equivalently by the ratio of Akaike weights $w_i=e^{({\rm AIC})_{i}/2}$. Thus, given two models $M_i$ and $M_j$ whose AIC increment is $\Delta_{ij}$, the evidence ratio (ER) against the model whose AIC value is larger is computed from ${\rm ER}=w_i/w_j=e^{{\Delta}_{ij}/2}$.

From Table \ref{tableFit2} we can derive the AIC increments $(\Delta$AIC$)_{ij}$ and evidence ratios when we compare the fit quality of models $i=$\DA, \DC, \DHlin\ with that of  $j=\CC$ ($\CC$CDM).  We find that for all \DA\ models $(\Delta$AIC$)_{i\CC}\gtrsim 9$  against the $\CC$CDM. It follows that when we compare any \DA\ model with the $\CC$CDM the typical evidence ratio against the latter is typically of order ER$\simeq 90$.  Worth noticing is also the result of the fit when we exclude the growth data from the fitting procedure but still add their contribution to the total $\chi^2$ -- it corresponds to the unbarred parameters in Table \ref{tableFit2}. This fit is of course less optimized, but allows us to risk a prediction for the linear growth and hence to test the level of agreement with these data points. It is noteworthy that in the case of the \DA\ models the corresponding AIC pairwise differences with the $\CC$CDM remain similar to the outcome when the growth data enters the fit. Therefore, the $\CC$CDM appears significantly disfavored versus the \DA-models in both cases (i.e. with barred and unbarred parameters in Table \ref{tableFit2}).

Quite another story is the situation with the \DC1-model, which does not seem capable of adjusting simultaneously the expansion history observables and the structure formation data. A glance at Table \ref{tableFit2} shows that when we compare it with the $\CC$CDM, with barred parameters, we find  $(\Delta$AIC$)_{1\CC}\gtrsim 53$  against \DC1\,, whereas with unbarred ones we get the even worse result $(\Delta$AIC$)_{1\CC}\gtrsim 296$ . The respective evidence ratios against \DC1\ versus the $\CC$CDM under these two circumstances are of course extremely large, the smallest one being ER$\sim 10^{11}$. Thus, quite obviously, from the point of view of the Akaike Information Criterion the situation of the \DC1\ model is  desperate, and that of \DHlin\ is even more dramatic. Not so for the \DC2\ model, but we will assess its viability more carefully at the end.

The upshot after making use of the AIC is that the competitive position of the \DA\ models became even more prominent from the point of view of phenomenology, whilst the status of \DC1\ and \DHlin\ against observations appear essentially as terminal. This fact may have nontrivial implications for cosmological model building.  For instance, the \DC1\ model has been studied in some formulations relating QCD with cosmology and it goes sometimes under the name of  ``QCD ghost dark energy model'', see references in Chapter \ref{chap:Atype}.  Although some formulations of these models are certainly of theoretical interest and could possibly be modified to become in better harmony with observations, the strict phenomenological analysis of models of DE containing a linear power of the Hubble function and a quadratic power of it, with no additive constant term ($C_0=0$) -- i.e. the strict \DC1-form indicated in our model list (\ref{eq:ModelsC}) --- leads to the inescapable conclusion that they are unable to account for the observations. The particular case in which the DE contains only the linear term in $H$, i.e. $\rD\sim H$, suffers of an even more hapless fate. Such linear form was first proposed in Ref.\,\cite{Schutzhold2002} and subsequently  was adopted and adapted to different purposes by different authors, until it eventually gave rise to the extended \DC1\ form. Incidentally, the two \DC\ models (\ref{eq:ModelsC}) and the linear (\ref{eq:linH}) were previously shown to be phenomenologically problematic as well when treated as vacuum models, see the previous chapter. Here we have shown that the phenomenological conflict also persists when they are treated as ${\cal D}$-models with self-conserved dynamical DE. If we compare Fig. \ref{sigma8TypeA} with the Fig. \ref{sigma8DC1} we can see that the \DC1-model actually does better than their vacuum counterpart, the C1-model (whose lack of power for structure formation near our time is quite dramatic, as shown in Fig. \ref{sigma8TypeA}. Still, the overall performance of the \DC1-model remains helpless in front of the $\CC$CDM and its \DA\ companions.

We would like to point out here that our assessment about the \DC1\ and \DHlin\ models is definitely much more pessimistic as compared to the analysis presented e.g. in\,\cite{QCDghostCai2011,QCDghostCai2012}. Take e.g. \DC1\,; the basic difference that we have been able to trace with respect to these authors is that we do not find that such model can pass the structure formation test when \DC1\ is adjusted to the background data, and vice versa. We suspect that in the aforementioned references the linear growth data points have directly been fitted with the \DC1\ model without perhaps verifying if the density perturbations of the model were able to meet the $\CC$CDM behavior at large redshift, namely $\delta\propto a$. In our case we fitted the data after imposing such boundary condition, and we found that the model failed (by far) to reproduce the observations, as we have reported in detail throughout our work. The results exhibited in Figs. \ref{sigma8DC1}-\ref{sigma}, combined with the comparatively poor statistical output of the \DC1-model recorded in Table \ref{tableFit2}, altogether make a quite eloquent case against the delicate health of that model in front of observations. The situation with \DHlin\,, which is a particular case of \DC1\,, is even worse since it has one less degree of freedom to maneuver. In short, from the comprehensive study carried out in this chapter we conclude that the \DC1\ and \DHlin\ models are ruled out. This was already the firmly inferred conclusion when these models were treated within the vacuum class in Chapter \ref{chap:Atype}, and here we can confirm that it is also the same unfavorable situation in the context of the ${\cal D}$-class.

Let us finally comment on the viability of the \DC2-subclass, which was shown to fit the data reasonably well (cf. Table \ref{tableFit2}). It is sometimes associated with the so-called entropic-force models\,\cite{Easson10}. We have seen in Sect. \ref{sect:DCModels} that it can provide a reasonable fit to the low redshift data (including linear growth) and therefore performs better than the \DC1\ and \DHlin\ models. However, this conclusion cannot be placed out of context, meaning that the structure of the \DC2\ model -- as in fact of any other one in the list of DE models given in (\ref{eq:ModelsA}-\ref{eq:linH}) -- is not sufficient to conclude whether the model is viable or excluded; for it also depends on whether there is exchange of energy with matter or not. Thus, despite the \DC2\ model (as a self-conserved DE model) provides a reasonable fit to the low-$z$ data in Table \ref{tableFit2}, its vacuum counterpart -- i.e. the model C2 with $\wD=-1$ and matter non-conservation studied in detail in the previous chapter -- was shown to be ruled out, already at the background level, even though it has exactly the same structure as a function of $H$.

Unfortunately \DC2\ has ultimately, too, a very serious drawback as self-conserved DE model, which we have preliminary announced at the end of Sect.\,\ref{sect:DCModels}. While \DC2\ tends to mimic the $\CC$CDM for a while, which is of course a positive attribute, such feature is limited to a period around the current epoch and cannot be extended to the radiation-dominated era. The reason can be seen on inspecting the radiation term in Eq.\,(\ref{eq:E2DA2}), namely the one proportional to $\Oro/(1-\nu+4\alpha/3)$. This term does apply to \DC2\ as well (with $C_0=0$), and the problem appears because the low-$z$ data naturally choose $\nu\simeq\alpha\simeq 1$ (cf. Table \ref{tableFit2}) so as to mimic as much as possible the $\CC$CDM. However, this is only possible at the price of ``renormalizing'' the size of the radiation coefficient from $\Oro$ to  $\sim 3\Oro/4$. Obviously this is a rather significant modification, in which $25\%$ of the normal content of the radiation is missing. Thus, deep in the radiation-dominated epoch, the model becomes anomalous and one can foresee a significant departure from the $\CC$CDM. For this reason and despite its ostensible success at low energies we do not place this model in the list of our favorite ${\cal D}$-models. A more detailed analysis (which we do not deem necessary here) would require to appropriately modify the standard formulas existing in the literature e.g. for computing the decoupling and baryon drag redshifts, as well as the corresponding modification of the BAO$_{dz}$ observable (cf. the discussion of these formulas in Chapter \ref{chap:Atype} and references therein).

We close this section by mentioning the analysis of Ref.\,\cite{MathewDC2} on a similar model as our \DC2\,. The authors do not seem to have detected the above mentioned important problem, as in fact they did not compute the radiation contribution. In addition, these authors incorrectly compare their results with those of \cite{BasSola2014b} without realizing, or at least clarifying, that the two models are very different; namely, in one case it is a C1-type vacuum model in interaction with matter, whilst in the other it is a \DC1-model with self-conserved matter and DE. Be as it may, in the light of the results presented here it becomes clear that all the  models with $C_0=0$, whether in the vacuum class or in the ${\cal D}$-class, are actually excluded and cannot be used to support an alternative formulation of holographic DE.

%%%%%%%%%%%%%%%%%%%%%%%%%%%%%%%%%%%%%%%%%%%%%%%%%%%%%%%%%%%%%%%%%
%%%%%%%%%%%%%%%%%%%%%%%%%%%%%%%%%%%%%%%%%%%%%%%%%%%%%%%%%%%%%%%%%
%%%%%%%%%%%%%%%%%%%%%%%%%%%%%%%%%%%%%%%%%%%%%%%%%%%%%%%%%%%%%%%%%

\section{Conclusions}
\label{sec:conclusionsDyn}

In this work we have thoroughly analyzed the ${\cal D}$-class of self-conserved dynamical dark energy models. These models are based on a dynamical DE density  $\rD$ which takes the form of a series of powers of the Hubble rate, $H$, and its derivatives. To be consistent with the general covariance of the effective action of quantum field theory in curved spacetime, we can expect that the series of powers must involve an even number of cosmic time derivatives of the scale factor only. Thus, if we should apply strictly this recipe for the current Universe, only the first powers up to $\dot{H}$ and $H^2$ would be involved in the structure of $\rD$, including or not a constant additive term $C_0$. However, in order to cover a wider variety of possibilities that have also been addressed in the literature from different theoretical perspectives, we have admitted also a subclass of models in which the linear term in $H$ is included for $C_0=0$. This greater flexibility in the general form of $\rD$ has allowed us to compare the performance of this restricted set of models with respect to those that are theoretically most preferred, namely the $C_0\neq 0$ ones.

An additional characteristic of the ${\cal D}$-class is that $\rD$ is locally and covariantly self-conserved. At fixed value of the gravitational constant, this entails the simultaneous local covariant conservation of matter as a necessary condition to satisfy the Bianchi identity. The ${\cal D}$-class models are thus enforced to have a dynamical equation of state with a nontrivial evolution with the cosmic expansion, $\wD=\wD(H)$, and therefore it is different from the previously studied dynamical vacuum class in Chapter \ref{chap:Atype}, in which $\wD=-1$ and the vacuum exchanges energy with matter. The two dynamical DE types share the same formal functional dependence of $\rD$ on $H$, but the behaviors are substantially different, already at the background level, as attested by the analytical and numerical solutions in each case.

We have also studied for the first time (to the best of our knowledge) the generic set of linear matter and DE perturbations for a system of self-conserved cosmic components, and studied the conditions by which the system can be transformed into an equivalent third order differential equation for the matter perturbations. We have demonstrated that the situation of a rigid $\CC$-term, i.e. the $\CC$CDM concordance model, appears as a particular case of that general framework. We have subsequently applied it to the entire ${\cal D}$-class of DE models and upon solving the general perturbation equations we have compared the solution with the corresponding results obtained when the DE enters only at the background level. In this way we have been able to have good control on the reach of the DE perturbations and explicitly confirmed that they play a small role when the considered scales are well under the horizon.

After a detailed study of the background history and the cosmic perturbations of the various models in the ${\cal D}$-class, we have been able to identify the subclass of the \DA-models as a most promising one insofar as it provides an excellent fit to the overall data on Hubble expansion and structure formation. These are the theoretically favored ${\cal D}$-models for which $C_0\neq0$ -- cf. Eq.\,(\ref{eq:ModelsA}). The fit quality rendered by them has been shown to be significantly better than that of the $\CC$CDM. Most conspicuously, using the Akaike Information Criterion\,\cite{Akaike1974} as a method to compare competing models describing the same data, we find that the evidence ratio in favor of the \DA-subclass (namely the relative likelihood of these models as compared to the concordance model) is of order of a hundred. We also find that the physical region of the parameter space for the \DA-models lies $\sim 2\sigma$ away from the $\CC$CDM region, and therefore the two models can be clearly distinguished.

Using the same testing tools we have reached the firm conclusion that all of the ${\cal D}$-models with $C_0=0$ are strongly disfavored, in particular the linear model $\rD\sim H$. Furthermore, among the theoretical models existing in the literature that become automatically excluded by our analysis we have the so-called entropic-force models and the QCD-ghost formulations of the DE (cf. Sect. \ref{sect:Discussion} for details and references).

At the end of the day the most distinguished dynamical  ${\cal D}$-models, both theoretically and phenomenologically, are those in the \DA-subclass -- viz. the set of models which endow the DE with a mild dynamical character and at the same time have a well-defined $\CC$CDM limit. These models improve significantly the fit quality of the $\CC$CDM, showing that a moderate dynamical DE behavior is better than having a rigid  $\CC$-term for the entire cosmic history.

We have found that the favored \DA-subclass has also the ability to mimic the quintessence behavior and it could even provide a possible explanation for the phantom character of the DE at present, as suggested by the persistent region projected below the $\wD=-1$ line (the so-called phantom divide) from the fits to the recent and past cosmological data. There is of course no significant evidence of this phantom character, as the physical region includes the $\wD=-1$ line. However, if in the future more accurate observations would insist on singling out the domain below the phantom divide, then the dynamical DE models in our \DA-subclass could provide a simple explanation without need of invoking true phantom fields, which are of course abhorred in QFT.

Let us conclude by emphasizing our main message.  The dynamical DE models treat the DE density as a cosmic variable on equal footing to the matter density. In a context of an expanding universe this option may be seen as more reasonable than just postulating an everlasting and rigid cosmological term for the full cosmic history. Furthermore, the subclass of \DA-models favored by our analysis furnishes a theoretically consistent and phenomenologically competitive perspective for describing the dark energy of our Universe as a dynamical quantity evolving with the cosmic expansion. The structure of the DE density in these models as a series of powers of the Hubble rate and its time derivatives is suggestive of a close connection with the QFT formulation in curved spacetime. We hope that they will help to better understand the origin of the cosmological term from a more fundamental perspective, and hopefully they might eventually shed some light as to the nature of the cosmological vacuum energy and its relation with the quantum vacuum of modern gauge field theories.

%%%%%%%%%%%%%%%%%%%%%%%%%%%%%%%%%%%%%%%%%%%%%%%%%%%%%%%%%%%%%%%%%
%%%%%%%%%%%%%%%%%%%%%%%%%%%%%%%%%%%%%%%%%%%%%%%%%%%%%%%%%%%%%%%%%
%%%%%%%%%%%%%%%%%%%%%%%%%%%%%%%%%%%%%%%%%%%%%%%%%%%%%%%%%%%%%%%%%

\section{Main bibliography of the chapter}

This chapter is mainly based on the contents of the paper \cite{JCAPnostre2}:
\vskip 0.6cm
\noindent
{\it  Background history and cosmic perturbations for a general system of self-conserved dynamical dark energy and matter.}\newline
A. G\'omez-Valent, E. Karimkhani, and J. Sol\`a\newline
JCAP {\bf 1512}, 048 (2015) ; arXiv:1509.03298

\newpage

\chapter[G-type dynamical vacuum models]{G-type dynamical vacuum models}
\label{chap:Gtype}

In this chapter we consider another interesting phenomenological scenario, which has also been motivated in Sect. \ref{subsec:RVMintro} in the context of QFT in curved spacetime. In the G-type models, the (dynamical) cosmological term also takes the general structure $\Lambda(H)=c_0+c_H H^2+c_{\dot{H}} \dot{H}$, but now matter and radiation are conserved, and the variation of $\Lambda$ is now possible thanks to the dynamical evolution of the Newtonian coupling $G$, a possibility which is perfectly consistent with the Bianchi identity. Again, we will differentiate between subtype-1 and subtype-2 models,
\begin{eqnarray}
G1: \phantom{XXX} \Lambda(H)&=&3(C_0+\nu
H^2)\label{eq:G1}\\
G2: \phantom{Xx}\Lambda(H,\dot{H})&=&3\left(C_0+\nu H^2+\frac{2}{3}\alpha\,\dH \right)\,.\label{eq:G2}
\end{eqnarray}
Notice that here we have redefined $c_0=3C_0$ $c_H= 3\nu$ and $c_{\dot{H}}=2\alpha$ for convenience, in analogy with what has been done in previous chapters for other models. Model G1 is of course a particular case of model G2, but it will be useful to distinguish between them. 

%%%%%%%%%%%%%%%%%%%%%%%%%%%%%%%%%%%%%%%%%%%%%%%%
%%%%%%%%%%%%%%%%%%%%%%%%%%%%%%%%%%%%%%%%%%%%%%%%
%%%%%%%%%%%%%%%%%%%%%%%%%%%%%%%%%%%%%%%%%%%%%%%%

\section{Background cosmological solutions}
\label{sect:BackgroundGtype}
The field equations for the dynamical vacuum energy density in the FLRW metric in flat space are
derived in the standard way and are formally similar to the ones
with strictly constant $G$ and $\CC$ terms:
\begin{eqnarray}\label{eq:FriedmannEqApJL}
&&3H^2=8\pi\,G(H)\,(\rho_m+\rR+\rho_\Lambda(H))\\
&&3H^2+2\dot{H}=-8\pi\,G(H)\,(p_\Lambda+p_r)\,, \label{eq:PressureEqApJL}\,
\end{eqnarray}
where  $\rL(H)={\Lambda(H)}/(8\pi G(H))$ is the dynamical vacuum energy density, $p_\Lambda(H)=-\rho_\Lambda(H)$, and
$G(H)$ is the dynamical gravitational coupling. We can
combine \eqref{eq:FriedmannEqApJL} and \eqref{eq:PressureEqApJL} to obtain
the equation of local covariant conservation of the energy, i.e.
$\nabla^\mu(GT_{\mu 0})=0$. Explicitly, since we assume matter
conservation (meaning $\dot\rho_m+3H\rmr=0$ and $\dot\rho_r+4H\rR=0$), it leads to a
dynamical interplay between the vacuum and the Newtonian coupling:
\be\label{eq:Bianchi}\dG(\rma+\rr+\rL)+G\drL=0. \ee
Trading the cosmic time for the scale factor $a$,
the previous equations amount to determine $G$ as a function of $a$. Using the matter conservation equations, we arrive at
\be\label{eq:Star} G(a)=- G_0\,\left[\frac{a\,\left(E^2(a)\right)^{\prime}}{3 \Omega_m^{(0)}\,a^{-3}+4\Omega_r^{(0)}\,a^{-4}}\right]\,, \ee
where $G_0\equiv G(a=1)$ is the present value of $G$, and the prime stands for $d/da$. Inserting (\ref{eq:G2}) and the above result for $G(a)$ in Eq.\,(\ref{eq:FriedmannEqApJL}) and integrating, we obtain:

\begin{table}
\begin{center}
\resizebox{1\textwidth}{!}{
\begin{tabular}{| c  |c | c | c | c | c |c | c | c | c | c |c |}
\multicolumn{1}{c}{Model} & \multicolumn{1}{c}{$|\frac{\Delta G}{G_0}|$ (BBN,CMB), Omh$^2$}  & \multicolumn{1}{c}{$\Omega_m$} & \multicolumn{1}{c}{$\overline{\Omega}_m$} ({\scriptsize all data}) &  \multicolumn{1}{c}{{\small$\nu$}} &  \multicolumn{1}{c}{{\small$\bar{\nu}$}}  & \multicolumn{1}{c}{$\sigma_8$}  & \multicolumn{1}{c}{$\overline{\sigma}_{8}$}  &
\multicolumn{1}{c}{$\chi^2/dof$} &
\multicolumn{1}{c}{$\overline{\chi}^2/dof$} &
\multicolumn{1}{c}{AIC} &
\multicolumn{1}{c}{$\overline{\rm AIC}$}
\\\hline {\small $\CC$CDM} & -, Yes & {\small$0.278^{+0.005}_{-0.004}$} & {\small$0.276\pm 0.004$} & - & - & {\small$0.815$} & {\small$0.815$} & {\small$828.84/1010$} & {\small$828.69/1010$} & {\small$830.84$} & {\small$830.69$}
\\\hline
{\small G1} & (10\%,5\%), Yes  & {\small$0.278\pm 0.006$} & {\small$0.275\pm 0.004$} & {\small$0.0015^{+0.0017}_{-0.0015}$} & {\small$0.0021^{+0.0014}_{-0.0016}$}  & {\small$0.797$} & {\small$0.784$}  & {\small$822.82/1009$} & {\small$821.97/1009$} & {\small$826.82$} & {\small$825.97$}

\\\hline
{\small $\CC$CDM} & -, No & {\small$0.292\pm 0.008$} & {\small$0.286\pm 0.007$} & - & - & {\small$0.815$} & {\small$0.815$} & {\small$583.38/604$} & {\small$582.74/604$} & {\small$585.38$} & {\small$584.74$}
\\\hline
{\small G1} & (10\%,5\%), No  & {\small$0.290\pm 0.011$} & {\small$0.281\pm 0.005$} & {\small$0.0008^{+0.0016}_{-0.0015}$} & {\small$0.0015\pm 0.0014$}  & {\small$0.795$} & {\small$0.771$}  & {\small$577.62/603$} & {\small$575.70/603$} & {\small$581.62$} & {\small$579.70$}
\\\hline {\small $\CC$CDM*} & -, Yes$^*$ & {\small$0.297\pm 0.006$} & {\small$0.293\pm 0.006$} & - & - & {\small$0.815$} & {\small$0.815$} & {\small$806.68/982$} & {\small$806.17/982$} & {\small$808.68$} & {\small$808.17$}
\\\hline
{\small G1*} & (10\%,5\%), Yes$^*$  & {\small$0.296\pm 0.009$} & {\small$0.287\pm 0.004$} & {\small$0.0006\pm 0.0015$} & {\small$0.0012^{+0.0014}_{-0.0013}$}  & {\small$0.803$} & {\small$0.770$}  & {\small$802.66/981$} & {\small$799.15/981$} & {\small$806.66$} & {\small$803.15$}
\\\hline
 \end{tabular}}
\end{center}
\caption[Best-fit values for the type-G1 model, from the analysis of Chapter \ref{chap:Gtype}]{{\scriptsize The best-fitting values for the G1-type models and their
statistical  significance ($\chi^2$-test and Akaike information criterion AIC, see the text).
All quantities with a bar involve a fit to the total input data, i.e. the expansion history  (Omh$^2$+BAO+SNIa), CMB shift parameter, the indicated constraints on the value of $\Delta G/G_0$ at BBN and at recombination, as well as the linear growth data. Those without bar correspond to a fit in which we use all data  but exclude the growth data points from the fitting procedure. ``Yes'' or ``No'' indicates if Omh$^2$ enters or not the fit. The starred scenarios correspond to removing the high redshift point $z=2.34$ from Omh$^2$ (see text). The quoted number of degrees of freedom ($dof$) is equal to the number of data points minus the number of independent fitting parameters. The fitting parameter $\nu$ includes all data.}\label{tableFit41}}
\end{table}

\be\label{eq:DifEqHApJL} E^2(a)=1+\frac{\Omega_m^{(0)}}{\xi}\left[-1+a^{-4\xi'}\left(a+\frac{\xi\,\Omega_r^{(0)}}{\xi'\Omega_m^{(0)}}\right)^{\frac{\xi'}{1-\alpha}}\right]\,, \ee
where again $\xi$ and $\xi'$ have been defined as in \eqref{eq:defxiM} and \eqref{eq:defxiR}, respectively. For small $|\nu,\alpha|\ll1$ (the expected situation), we can use the approximations $\nueff\simeq \nu-\alpha$ and $\nueffp\simeq \nu-(4/3)\alpha$. Note that, in order to simplify the presentation, we have removed from Eq.\,(\ref{eq:DifEqHApJL}) terms proportional to $\Omega_r^{(0)}\ll\Omega_m^{(0)}$ that are not relevant here\footnote{The full expression is presented in Eq. \eqref{eq:DifEqH}, in Chapter \ref{chap:AandGRevisited}. \label{footnote1}}. We can check e.g. that in the radiation-dominated epoch the leading term in the expression (\ref{eq:DifEqHApJL}) is $\sim \Omega_r^{(0)}\,a^{-4\xi'}$, whilst in the matter-dominated epoch is $\sim \Omega_m^{(0)}\,a^{-3\xi}$. Furthermore, we find that the (full) expression for $E^2(a)$  reduces to the $\CC$CDM form, $1+\Omega_m^{(0)}\,(a^{-3}-1)+\Omega_r^{(0)}(a^{-4}-1)$, in the limit $\nu,\alpha\to 0$ (i.e. $\xi,\xi'\to 1$).
Notice also the constraint among the parameters,
\be
C_0=H_0^2\left[\Omega_\Lambda^{(0)}-\nu+\alpha\left(\Omega_m^{(0)}+\frac43\,\Omega_r^{(0)}\right)\right]\,,
\ee
which follows
from matching the vacuum energy density $\rL(H,\dot{H})$ to its present
value $\rho_\Lambda^{(0)}$ for $H=H_0$ and using $\Omega_m^{(0)}+\Omega_r^{(0)}+\Omega_\Lambda^{(0)}=1$.  The explicit scale factor dependence of the Newtonian coupling
ensues upon inserting  \eqref{eq:DifEqHApJL} in \eqref{eq:Star} and computing the derivative, 
\begin{multline}\label{eq:GApJL}
G(a)=G_0a^{-4\xi^\p}\left(\Omega_r^{(0)}+\frac{\xi^\p}{\xi}\Omega_m^{(0)}\right)\left(4-\frac{\xi^\p}{1-\alpha}\frac{a}{a\xi^\p+\xi\frac{\Omega_r^{(0)}}{\Omega_m^{(0)}}}\right)\times\\\times\left(\frac{a\xi^\p+\xi\frac{\Omega_r^{(0)}}{\Omega_m^{(0)}}}{\xi^\p+\xi\frac{\Omega_r^{(0)}}{\Omega_m^{(0)}}}\right)^{\xi^\p/(1-\alpha)}\frac{1}{4\Omega_r^{(0)}a^{-4}+3\Omega_m^{(0)}a^{-3}}\,.
\end{multline}
One can check that in the limit $a\to 0$ (relevant for the Big Bang Nucleosynthesis epoch) it behaves as

\begin{table}
\begin{center}
\resizebox{1\textwidth}{!}{
\begin{tabular}{| c  |c | c | c | c | c |c | c | c | c | c |c |}
\multicolumn{1}{c}{Model} & \multicolumn{1}{c}{$|\frac{\Delta G}{G_0}|$ (CMB), Omh$^2$}  & \multicolumn{1}{c}{$\Omega_m$} & \multicolumn{1}{c}{$\overline{\Omega}_m$} ({\scriptsize all data}) &  \multicolumn{1}{c}{{\small$\nueff$}} &  \multicolumn{1}{c}{{\small$\bar{\nu}_{\rm eff}$}}  & \multicolumn{1}{c}{$\sigma_8$}  & \multicolumn{1}{c}{$\overline{\sigma}_{8}$} &
\multicolumn{1}{c}{$\chi^2/dof$} &
\multicolumn{1}{c}{$\overline{\chi}^2/dof$} &
\multicolumn{1}{c}{AIC} &
\multicolumn{1}{c}{$\overline{\rm AIC}$}
\\\hline {\small $\CC$CDM} & -, Yes & {\small$0.278^{+0.005}_{-0.004}$} & {\small$0.276\pm 0.004$} & - & - & {\small$0.815$} & {\small$0.815$} & {\small$828.84/1009$} & {\small$828.69/1009$} & {\small$830.84$} & {\small$830.69$}
\\\hline
{\small G2} & 5\%, Yes & {\small$0.278\pm 0.006$} & {\small$0.277\pm 0.004$} & {\small$0.0038^{+0.0025}_{-0.0023}$} & {\small$0.0043^{+0.0018}_{-0.0020}$}  & {\small$0.774$} & {\small$0.773$} & {\small$817.17/1008$} & {\small$817.26/1008$} & {\small$821.17$} & {\small$821.26$}
\\\hline
{\small $\CC$CDM} & -, No & {\small$0.292\pm 0.008$} & {\small$0.286\pm 0.007$} & - & - & {\small$0.815$} & {\small$0.815$} & {\small$583.38/603$} & {\small$582.74/603$} & {\small$585.38$} & {\small$584.74$}
\\\hline
{\small G2} & 5\%, No  & {\small$0.287\pm 0.011$} & {\small$0.283\pm 0.005$} & {\small$0.0025^{+0.0026}_{-0.0025}$} & {\small$0.0030^{+0.0021}_{-0.0018}$} & {\small$0.763$} & {\small$0.767$} & {\small$572.68/602$} & {\small$572.99/602$} & {\small$576.68$} & {\small$576.99$}
\\\hline {\small $\CC$CDM*} & -, Yes$^*$ & {\small$0.297\pm 0.006$} & {\small$0.293\pm 0.006$} & - & - & {\small$0.815$} & {\small$0.815$} & {\small$806.68/981$} & {\small$806.17/981$} & {\small$808.68$} & {\small$808.17$}
\\\hline
{\small G2*} & 5\%, Yes$^*$  & {\small$0.295\pm 0.009$} & {\small$0.289\pm 0.005$} & {\small$0.0015^{+0.0026}_{-0.0025}$} & {\small$0.0028^{+0.0018}_{-0.0021}$} & {\small$0.789$} & {\small$0.765$} & {\small$798.85/980$} & {\small$797.05/980$} & {\small$802.85$} & {\small$801.05$}
\\\hline
 \end{tabular}}
\end{center}
\caption[Best-fit values for the type-G2 model, from the analysis of Chapter \ref{chap:Gtype}]{\scriptsize{As in Table \ref{tableFit41}, but for G2 models with $\xi'=1$ so as to maximally preserve the BBN bound (see text). The effective G2-model fitting parameter in this case is  $\nueff=\nu/4$. The constraint on $|\Delta G/G_0|$ from CMB anisotropies at recombination is explicitly indicated.}\label{tableFit42}}
\end{table}

\be\label{eq:GNR} G(a)=G_0\,a^{4(1-\xi')}\simeq G_{0}(1+4\nueffp\,\ln\,a)\,. \ee
Thus, the gravitational coupling
evolves logarithmically with the scale factor and hence changes very slowly. This
logarithmic law was motivated previously in \cite{Fossil07,SolaReview2013} within the context of the renormalization group of QFT in curved spacetime.

For $\nu=\alpha=0$  we obtain
$G=G_0$ identically, i.e. the current value of the gravitational coupling. However the situation $G=G_0$ is also attained in the radiation-dominated epoch for $\nu=(4/3)\alpha$ (i.e. $\xi'=1)$; and indeed we shall adopt this setting hereafter in order to maximally preserve the BBN constraint for the G2 model. The effective fitting parameter will be $\nueff=\nu/4$.  Obviously this setting is impossible for G1, so in this case we will adopt the average  BBN restriction $|\Delta G/G|<10\%$ in the literature\,\cite{Chiba2011,NagataChibaSugiyama2004,Uzan2011}. At the same time we require  $|\Delta G/G|<5\%$ at recombination ($z\simeq 1100$) for both G1 and G2 from the CMB anisotropy spectrum\,\cite{Chiba2011}.

\begin{figure}
\centering
\includegraphics[angle=0,width=0.9\linewidth]{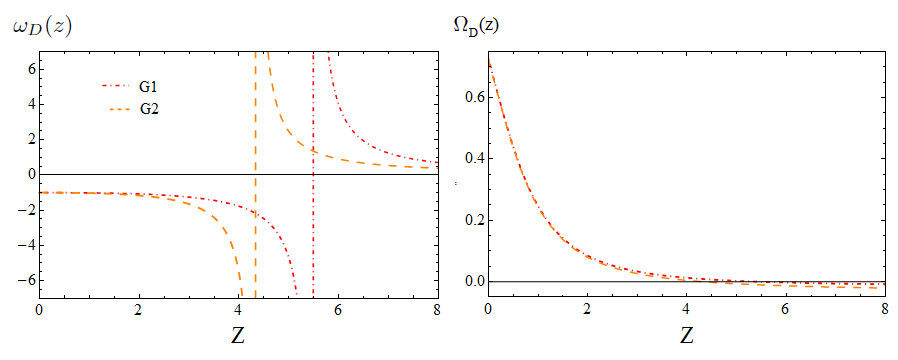}
\caption[Effective EoS parameter $\omega_D(z)$ and effective DE density associated to the type-G models of Chapter \ref{chap:Gtype}]{\label{fig:DE}\scriptsize{{\bf Left:} Evolution of the effective DE EoS parameter $\omega_D(z)$, Eq.\,(\ref{eq:effEoS}), for the models under consideration. {\bf Right:} The corresponding evolution of the effective DE density $\OD(z)$ normalized to the critical density (see text).}}
\end{figure}

The expression for the dynamical vacuum energy density can be
obtained from Friedmann's equation (\ref{eq:FriedmannEqApJL}), in combination with the explicit form of $G(a)$. We quote here only the simplified expression valid for the matter-dominated epoch:
\be\label{eq:RhoLNRApJL} \rL(a)=\rho_{c}^{(0)}\,
a^{-3}\left[a^{3\xi}+\frac{\Omega_m^{(0)}}{\xi}(1-\xi-a^{3\xi})\right]\,.
\ee
For $\xi\to 1$ we have $\rho_\Lambda\to\rho_{c}^{(0)}(1-\Omega_m^{(0)})=\rho_c^{(0)}\Omega_\Lambda^{(0)}$  and
we retrieve the $\CC$CDM case with strictly constant
$\rL$. The form (\ref{eq:RhoLNRApJL}) is sufficient to obtain an effective DE density $\rD(z)$ and effective equation of state (EoS) for the (non-interacting) DE at fixed $G=G_0$, as conventionally used in different places of the literature -- see e.g. \,\cite{Shafieloo2006,BasSola2014b,SS0506a,SS0506b}. We find:
\begin{equation}\label{eq:effEoS}
\omega_D(a)=-\frac{1}{1+\frac{\rho_m(a)}{\rho_\Lambda(a)}\frac{G(a)-G_0}{G(a)}}\,.
\end{equation}
In Fig. \ref{fig:DE} (left) we plot $\omega_D$ as a function of the cosmic redshift $z=-1+1/a$ for models G1 and G2.
Near our time, $\omega_D$ stays very close to $-1$ (compatible with the $\CC$CDM), but at high $z$ it departs. In the same Fig. \ref{fig:DE} (right) we plot $\OD(z)=\rD(z)/\rc(z)$, i.e. the normalized DE density with respect to the critical density  at constant $G_0$.  The asymptotes of $\omega_D$ for each model at  $z>4$ are due to the vanishing of $\OD(z)$ at the corresponding point (as clearly seen in the figure)-- confer the aforementioned references and the previous chapter for similar features.

%%%%%%%%%%%%%%%%%%%%%%%%%%%%%%%%%%%%%%%%%%%%%%%%%%%%%%%%%%%%%%%%%
%%%%%%%%%%%%%%%%%%%%%%%%%%%%%%%%%%%%%%%%%%%%%%%%%%%%%%%%%%%%%%%%%
%%%%%%%%%%%%%%%%%%%%%%%%%%%%%%%%%%%%%%%%%%%%%%%%%%%%%%%%%%%%%%%%%

%

\section{Fitting the models to the observational data}\label{sect:FitGtype}

Let us now test these models versus observation. First of all, we
use the available measurements of the Hubble function as collected in \cite{Ding2015}. These are essentially the data points of \cite{FarooqRatra2013} in the redshift range $0\leqslant z\leqslant 1.75$ and the BAO
measurement at the largest redshift $H(z = 2.34)$ taken after
\cite{Delubac2015} on the basis of BAO's in the Ly$\alpha$ forest of BOSS DR11 quasars. We define the following $\chi^2$ function, to be
minimized:
\begin{equation}\label{xi2Omh2}
\chi^2_{Omh^2}=\sum_{i=1}^{N-1}\sum_{j=i+1}^{N}\left[\frac{Omh^2_{th}(H_i,H_j)-Omh^2_{obs}(H_i,H_j)}{\sigma_{Omh^2\,i,j}}\right]^2\,,
\end{equation}
where $N$ is the number of points $H(z)$ contained in the data set,
$H_i\equiv H(z_i)$, and the two-point diagnostic
$Omh^2(z_2,z_1)$ given in \eqref{eq:Omh2Diagnostic} was defined in\,\cite{SahniShafielooStarobinsky}, and $\sigma_{Omh^2\,i,j}$ is the
uncertainty associated to the observed value $Omh^2_{obs}(H_i,H_j)$ for
a given pair of points $ij$, viz.\\
\begin{equation}
\sigma^2_{Omh^2\,i,j}=\frac{4\left[h^2(z_i)\sigma^2_{h(z_i)}+h^2(z_j)\sigma^2_{h(z_j)}\right]}{\left[(1+z_i)^3-(1+z_j)^3\right]^2}\,.
\end{equation}
For the  $\CC$CDM the two-point diagnostic boils down to $Omh^2(z_2,z_1)=\Om\,h^2$, which is constant for any pair
$z_1$, $z_2$. Using this testing tool and the
known observational information on $H(z)$ at the three redshift
values $z=0,0.57,2.34$  the aforementioned authors observed
that the average result is: $Omh^2=0.122\pm0.010$, with very little
variation from any pair of points taken. The obtained result is
significantly smaller than the  corresponding  Planck value of the
two-point diagnostic, which is constant and given by
$Omh^2=\Omo\,h^2=0.1415\pm0.0019$ \cite{Planck2015}.

A departure of $Omh^2$ from the Planck result should, according to
\cite{SahniShafielooStarobinsky}, signal that the DE cannot be described by a rigid
cosmological constant.
For the $\CC$CDM we obtain $Omh^2=0.1250\pm0.0039$, and
$Omh^2=0.1402\pm0.0059$, by taking all data points, and excluding the
high redshift one, respectively.  Since there is {\it a priori} no reason to exclude the high-redshift point\,\cite{Delubac2015}, whose
uncertainty is one of the lowest in the full data sample, relaxing the tension with data may
require the dynamical nature of the DE. The vacuum models G1 and G2 considered here, Eqs.\,(\ref{eq:G1},\ref{eq:G2}), aim at cooperating in this task.

\begin{figure}[!t]
\centering
\includegraphics[scale=0.35]{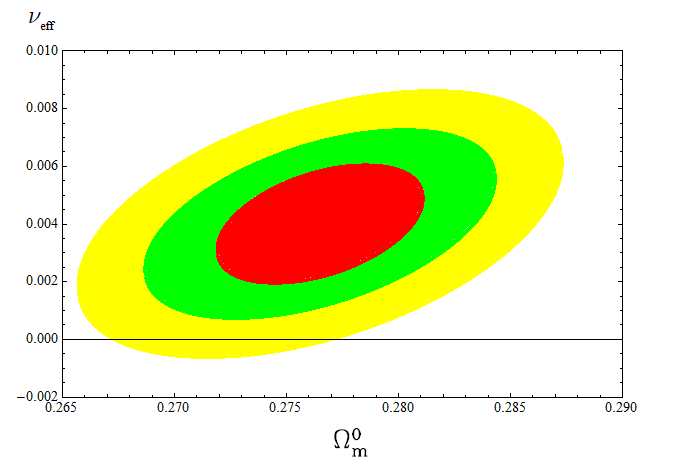}
\caption[Contour lines for the G2 model in Chapter \ref{chap:Gtype}]{\scriptsize Likelihood contours in the $(\Omega_m,\nueff)$ plane (for the values $-2\ln\mathcal{L}/\mathcal{L}_{max}=2.30$, $6.16, 11.81$, corresponding to 1$\sigma$, 2$\sigma$ and $3\sigma$ confidence levels for the G2 model using the full data analysis indicated in Table \ref{tableFit42}. The $\nueff=0$ region ($\CC$CDM) is disfavored at $\sim 2.5\sigma$.\label{fig:CL}}
\end{figure}

For these models  the theoretical value  $Omh_{th}^2$ of the two-point
diagnostic entering (\ref{xi2Omh2}) can be computed, in the matter-dominated epoch (relevant for such observable), as follows:
\begin{equation}\label{eq:Omh2zA}
Omh_G^2(z_i,z_j)=\frac{\Omo\,h^2}{\xi}\,\frac{\left(1+z_i\right)^{3\xi}-\left(1+z_j\right)^{3\xi}}{(1+z_i)^3-(1+z_j)^3}\,.
\end{equation}
It is evident that for $\xi=1$ we recover the $\CC$CDM
result, which remains anchored at $Omh^2(z_i,z_j)=\Omo h^2\ (\forall
z_i,\forall z_j)$. However, when we allow some small vacuum dynamics
(meaning $\nu$ and/or $\alpha$ different from zero) we
obtain a small departure of $\xi$ from $1$ and therefore the DE
diagnostic $Omh^2$ deviates from $\Omo h^2$. In this case $Omh^2$
evolves with cosmic time (or redshift).

To the above Hubble parameter data we add the recent supernovae type Ia data, the CMB shift parameter, the baryon acoustic oscillations, the growth rate for structure formation (see next section) and the BBN and CMB anisotropy bounds. Contour lines for $\nueff=1-\xi$ are shown in Fig. \ref{fig:CL} for model G2 at fixed $\xi'=1$. The $\chi^2$ functions associated to SNIa distance modulus
$\mu(z)$, the BAO $A$-parameter and the CMB shift parameter can be
found in Sect. \ref{sec:fitting}. Therein, one can also find the
corresponding references of the data sets that we have used in the
present analysis.

\begin{figure}
\centering
\includegraphics[angle=0,width=0.9\linewidth]{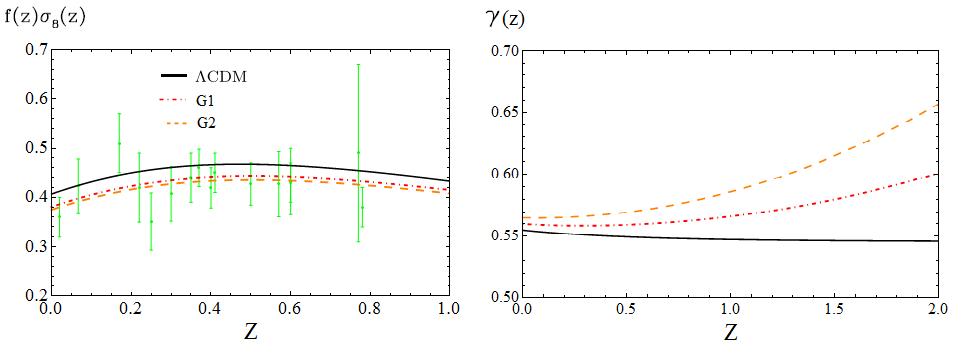}
\caption[$f(z)\sigma_8(z)$ and $\gamma(z)$ for the G-type models in Chapter \ref{chap:Gtype}]{\label{fig:LinearStructure}%
\scriptsize {\bf Left:} Comparison of the observed data with error bars (in green) and the theoretical evolution of the weighted growth rate
of clustering $f(z)\sigma_8(z)$ for each dynamical vacuum model and the $\CC$CDM. {\bf Right:} The corresponding evolution of the linear growth index $\gamma(z)$.
\vspace{0.3cm}
}
\end{figure}

%%%%%%%%%%%%%%%%%%%%%%%%%%%%%%%%%%%%%%%%%%%%%%%%%%%%%%%%%%%%%%%%%
%%%%%%%%%%%%%%%%%%%%%%%%%%%%%%%%%%%%%%%%%%%%%%%%%%%%%%%%%%%%%%%%%
%%%%%%%%%%%%%%%%%%%%%%%%%%%%%%%%%%%%%%%%%%%%%%%%%%%%%%%%%%%%%%%%%

\section{Linear structure formation}\label{sect:LinStructure}

Finally, we take into consideration the data on the linear structure formation. For the G1 and G2 models the calculation of $\delta_m=\delta\rho_m/\rho_m$ is significantly  more
complicated than in the $\CC$CDM case and follows from
applying linear perturbation theory to Einstein's field equations
and the Bianchi identity \eqref{eq:Bianchi}, see \cite{GSFS10} and Appendix \ref{sec:AppenGpert}. The final result reads\footnote{The third-order feature of this equation is characteristic of the coupled systems of matter and DE perturbations for cosmologies with matter conservation, after eliminating the perturbations in the DE in favor of a single higher order equation for the matter part -- cf. Sect. \ref{subsec:ThirdOrderDifEq} for details. For $\CC=$const. Eq.\,\eqref{eq:thirdorder} boils down to the (derivative of the) second order one of the $\CC$CDM, see Sect. \ref{subsec:RecoveringLCDM}.}:
\begin{equation}\label{eq:thirdorder}
\delta_m^{\prime\prime\prime}+\frac{\delta_m^{\prime\prime}}
{2a}(16-9\Omega_m)+\frac{3\delta_m^\prime}{2a^2}(8-11\Omega_m+3\Omega_m^2-a\Omega_m^\prime)=0\,,\end{equation}
with $\Omega_m(a)={8\pi\,G(a)}\rho_m(a)/{3H^2(a)}$. Notice that the $(\nu,\alpha)$  model-dependence is encoded in $H(a)$ -- cf. Eq. (\ref{eq:DifEqHApJL}). To solve the above equation (numerically) we have to fix the initial
conditions for $\delta_m$, $\delta_m^\prime$ and
$\delta_m^{\prime\prime}$.
We  take due account of the fact that for these models at small $a$ (when non-relativistic matter
dominates over the vacuum) we have $\delta_m(a)=a^s$, where $s=3\xi-2=1-3\nueff$. If $\xi=1$ ($\nueff=0$), then $\delta_m(a)\sim a$
and we recover the $\CC$CDM behavior.
Thus, the initial conditions set at a high redshift
$z_i=(1-a_i)/a_i$, say $z_i=100$ (or at any higher value), are the
following. For the growth factor we have $\delta_m(a_i)=a_i^s$, and
for its derivatives: $\delta_m^\prime(a_i)=sa_i^{s-1}$,
$\delta_m^{\prime\prime}(a_i)=s(s-1)a_i^{s-2}$. In practice we investigate the agreement with the structure formation data by comparing the theoretical weighted growth rate $f(z)\sigma_8(z)$, as in Sect. \ref{sec:FittingDE}. The values of $\sigma_8\equiv\sigma_8(z=0)$ for the various models are collected in Tables \ref{tableFit41} and \ref{tableFit42}, and in Fig. \ref{fig:LinearStructure} we plot $f(z)\sigma_8(z)$ and $\gamma(z)$ for them.

The joint likelihood analysis is performed on the set of Omh$^2$+BAO+SNIa+CMB, BBN and linear growth data, involving one ($\Omega_m$) or two $(\Omega_m,\nueff)$ independently adjusted parameters depending on the model. For the $\CC$CDM we have one parameter ($n_p=1$) and for G1 and G2 we have $n_p=2$. Recall that for G2 we have fixed $\xi'=1$.

%%%%%%%%%%%%%%%%%%%%%%%%%%%%%%%%%
%%%%%%%%%%%%%%%%%%%%%%%%%%%%%%%%%
%%%%%%%%%%%%%%%%%%%%%%%%%%%%%%%%%

\section{Discussion}
The main results of this chapter are synthesized in Tables \ref{tableFit41}-\ref{tableFit42} and Figures \ref{fig:DE}-\ref{fig:LinearStructure}. In particular, from Fig. \ref{fig:CL} we see that the model parameter $\nueff$ for G2 is clearly projected onto the positive region, which encompasses most of the $3\sigma$ range. Remarkably, the $\chi^2$-value of the overall fit is smaller than that of $\CC$CDM for both G1 and G2, cf. Tables \ref{tableFit41}-\ref{tableFit42}.
To better assess the distinctive quality of the fits we apply again the well-known Akaike Information Criterion (AIC)\cite{Akaike1974}, \eqref{eq:AkaikeShort}. To test the effectiveness of models $M_i$ and $M_j$, one considers the pairwise difference
$(\Delta$AIC$)_{ij} = ({\rm AIC})_{i} - ({\rm AIC})_{j}$. The larger the
value of $\Delta_{ij}\equiv|\Delta({\rm AIC})_{ij}|$, the higher the evidence against the
model with larger value of ${\rm AIC}$, with $\Delta_{ij} \ge 2$ indicating a positive such evidence and $\Delta_{ij}\ge 6$
denoting significant such evidence.

From Tables \ref{tableFit41}-\ref{tableFit42} we see that when we compare the fit quality of models $i=$G1, G2 with that of  $j=\CC$CDM, in a situation when we take all the data for the fit optimization, we find $\overline{\Delta}_{ij}\equiv\Delta\overline{\rm AIC}\gtrsim 9.43$ for G2 and $4.72$ for G1, suggesting significant evidence in favor of these models (especially G2) against the $\CC$CDM -- the evidence ratio\,\cite{Akaike1974} being ${\rm ER}=e^{\overline{\Delta}_{ij}/2}\gtrsim111.6$ for G2 and $10.6$ for G1. Worth noticing is also the result of the fit when we exclude the growth data from the fitting procedure but still add their contribution to the total $\chi^2$. This fit is of course less optimized, but allows us to risk a prediction for the linear growth and hence to test the level of agreement with these data points (cf. Fig. \ref{fig:LinearStructure}). It turns out that the corresponding AIC pairwise difference with the $\CC$CDM are similar as before. Therefore, the $\CC$CDM appears significantly disfavored versus the dynamical vacuum models, especially in front of G2, according to the Akaike Information Criterion. Let us mention that if we remove all of the $H(z)$ data points from our analysis the fit quality weakens, but it still gives a better fit than the $\CC$CDM (cf. the third and fourth row of Tables \ref{tableFit41} and \ref{tableFit42}). If, however, we keep these data points but remove \textit{only} the high redshift point $z=2.34$\,\cite{Delubac2015}, the outcome is not dramatically different from the previous situation (confer the starred scenarios in Tables \ref{tableFit41} and \ref{tableFit42}), as in both cases the significance of $\nueff\neq0$ is still close to $\sim 2\sigma$ with $\Delta_{ij}>7$ for G2 (hence still strongly favored, with ${\rm ER}> 33$). In this sense the high $z$ point may not be so crucial for claiming hints in favor of dynamical vacuum, as the hints themselves seem to emerge more as an overall effect of the data. While we are awaiting for new measurements of the Hubble parameter at high redshift to better assess their real impact, we have checked that if we add to our analysis the points $z=2.30$\,\cite{Busca2013} and  $z=2.36$\,\cite{FontRibera2014}, not included in either \cite{SahniShafielooStarobinsky} or \cite{Ding2015}, our conclusions remain unchanged. Ditto if using the three high $z$ points only.

%%%%%%%%%%%%%%%%%%%%%%%%%%%%%%%%%%%%%%%%%%%%%%%%%%%%%%%%%%%%%%%%%
%%%%%%%%%%%%%%%%%%%%%%%%%%%%%%%%%%%%%%%%%%%%%%%%%%%%%%%%%%%%%%%%%
%%%%%%%%%%%%%%%%%%%%%%%%%%%%%%%%%%%%%%%%%%%%%%%%%%%%%%%%%%%%%%%%%

\section{Main bibliography of the chapter}

This chapter is based on the contents of the letter \cite{ApJLnostre}:
\vskip 0.5cm
\noindent
{\it Hints of dynamical vacuum energy in the expanding Universe.}\newline
J. Sol\`a, A. G\'omez-Valent, and J. de Cruz P\'erez\newline
Astrophys. J. Lett. {\bf 811}, L14 (2015) ; arXiv:1506.05793

\thispagestyle{empty}
\null
\newpage
\thispagestyle{empty}
\null
\newpage

\pagestyle{fancy}
\fancyhf{}
%\fancyhead[LO,RE]{\thepage}
\fancyhead[CO]{Summary of Part I}
\fancyhead[CE]{Summary of Part I}
\cfoot{\thepage}

\chapter*{Brief note on the obtained results and the data used in Part I (Summary)}
\label{chap:BriefNote}

\addcontentsline{toc}{chapter}{Brief note on the obtained results and the data used in Part I (Summary)}

At this point, I consider it is worth making a short summary with the most relevant results obtained in the first part of this dissertation. We have analyzed in  quite detail several dynamical vacuum models, and some DE models inside the $\mathcal{D}$-class. They have been motivated from QFT in curved spacetime. These studies have been carried out not only at the background level, but also at the linear perturbations one. In some cases, even at the nonlinear regime, making use of the time-honored Press-Schechter formalism with an improved version for the halo mass function. More concretely, we have focused our attention on the following models:

\begin{itemize}
\item DVM's in which the variation of the vacuum energy density is directly linked to an anomalous matter and radiation conservation laws, in Chapter \ref{chap:Atype}. This is the case of type-A and B models. The firsts receive corrections $\delta\rho_\Lambda(H,\dot{H})=a_1H^2+a_2\dot{H}$ to the rigid CC term of the standard $\Lambda$CDM model. These corrections respect the general covariant form that is expected if they do arise from the effective action of QFT in curved spacetime. The second ones, type-B models, in which $\delta\rho_\Lambda(H)=b_1H+b_2 H^2$, also incorporate a linear term in $H$ that is not as theoretically favorable as those encountered in the A-type ones. The linear term in $H$ can be thought of as one of pure phenomenological nature, which can account for e.g. bulk-viscosity effects. As particular cases of this kind of models we find the type-C models, with no constant term in the expression of $\rho_\Lambda$. We have seen in Chapter \ref{chap:Atype} that type-A and B models are phenomenologically viable, whilst type-C ones can be ruled out for several reasons. In the case of type-C2 models, which are obtained from type-A ones by doing $C_0=0$, they can be discarded simply because they are unable to generate the needed transition from a decelerated to a positively accelerated Universe. On the other hand, type-C1 models, obtained in the same limit from the type-B ones, are excluded for different reasons. They cannot explain the observed LSS of the Universe, either giving rise to an excess (in C1A model) or a defect (in the pure linear C1B case) of structure formation when the best-fit values for the various parameters entering the models are inferred from the fitting analysis of the low and intermediate redshift observables, as SNIa and BAO$_{A}$.

\item Dynamical dark energy models inside the $\mathcal{D}$-class, in Chapter \ref{chap:DynamicalDE}. In this case, we have considered a self-conserved DE component, ruled by an energy density of the form $\rho_D(H,\dot{H})=C_0+C_H H^2+C_{\dot{H}}\dot{H}$ in the general $\mathcal{D}$A case. We have shown that these models are also viable from the phenomenological perspective. When LSS data are considered in the fitting analysis, we find that the $\Lambda$CDM model is disfavored at a $\sim 2\sigma$ c.l. in front of the $\mathcal{D}$A models. Moreover, we are in position to discard $\mathcal{D}$C-type models, in which $C_0=0$. The $\mathcal{D}$C1-type models are absolutely unable to fit the LSS data, whereas the $\mathcal{D}$C2 ones, although are able to fit the low and intermediate redshift data, suffer from a large deficit of radiation that completely spoils the desired behavior of the Hubble rate during e.g. the BBN epoch. In Chapter \ref{chap:DynamicalDE} we have also derived the differential equation that governs the evolution of matter density perturbations at subhorizon scales, by taking also into account the DE perturbations when the DE is self-conserved. This has allowed us to quantify the effect of DE perturbations and conclude that their effect is almost insignificant at subhorizon scales.  

\item DVM's in which the variation of the vacuum energy density is due to an accompanying  variation of the Newtonian coupling $G$, the so-called type-G models. This study has been carried out in Chapter \ref{chap:Gtype}. In this case we have found some hints sitting on top of the data in favor of the dynamical nature of the vacuum energy density. We have seen that in the case of the type-G2 model there is a $\sim 2.5\sigma$ c.l. evidence in favor of a varying cosmological term. It is important to remark, though, that contrary to the analysis performed for type-A and B models in Chapter \ref{chap:Atype}, now we have incorporated to our database the LSS $f(z)\sigma_8(z)$ data points. Thus, the fitting conditions have been quite different, since now we have included information which does not only requires the use of the background formulas, but also demands the use of the linear perturbations ones. Moreover, we have also taken into account the Hubble function data through the $Omh^2(z_i,z_j)$  diagnostic. We have also tested the impact of some of the high redshift Hubble data points (of $z>2.3$) on our results. These differences on the data sets used in the fitting analyses might be the reason that explains why we have not detected such hints in favor of the vacuum dynamics in type-A and B models. 

\end{itemize} 

Now, let me list the main outcomes of the first part of this thesis:

\begin{itemize}
\item The rigid cosmological term is something one cannot avoid in the context of the DVM's and the $\mathcal{D}$-class of DE models under study. It seems to be a crucial ingredient that forces the spacetime to be intrinsically dynamical, even in absence of matter and radiation. 

\item All the analyzed models with a well-defined $\Lambda$CDM-like limit have a good phenomenological behavior. This is the case of type-A, B, G and $\mathcal{D}$A models.

\item From the point of view of QFT in curved spacetime, the fact that $C_0\ne 0$ implies the impossibility of using the simple renormalization condition $\rho_{\Lambda,{\rm eff}}^{\rm Mink}=0$. This is pointing out that considering the overall vacuum energy in curved spacetime arising from a sort of Casimir gravitational effect is, probably, insufficient. If it was really the case, we would only be able to measure the dynamical terms appearing in the expression of the vacuum energy density, and we have seen that this is not what happens. We also ``observe'' a non-null additive (and rigid) constant term. The latter cannot arise from the aforesaid gravitational Casimir effect. There must be something more. See the discussion on these issues in Sect. \ref{subsec:RVMintro}.  

\item Another question should be addressed now. Is the vacuum energy or the DE ersatz an evolving-in-time entity? This is a more modest issue than the one concerning the solution of the old CC problem, and it is probably within easier reach in view of the big advances that observational Cosmology is undergoing (and will undergo in the future). We have detected some small hints of evolving DE when LSS data is used in combination with BAO, CMB, SNIa, and $H(z)$ data. But is this signal real, or is it just a spurious statistical fluctuation? At this point, we are not in position of pinning this issue down. In fact, a $2-2.5\sigma$ evidence is not significant enough from the statistical point of view to claim the existence of DE dynamics. Thus, more efforts must be made so as to resolve this question.

\end{itemize}

A couple of crucial points that should me improved with respect to the previous studies:

\begin{itemize}

\item First of all, it is convenient to understand in a deeper way the data set used in the fitting of the various models under study. This means, for instance, to pinpoint possible correlations among the data. This has not been exhaustively done in the analysis carried out in the previous chapters of the thesis. In practice, this implies to avoid the use of correlated data when the corresponding correlations are strong and we do not know the numerical values of the correlation coefficients, and to include their effects through non-diagonal covariance matrices when they are known.

\item Up to now, we have applied a gridded minimization procedure of the total $\chi^2$ functions so as to search for the best-fit values of the parameters. This is a very slow method, very expensive in terms of computational time. This has forced us to restrict the dimensionality of the fitting vectors to be of dimension lower or equal than 2. We have typically given freedom to $\Omega_m^{(0)}$ and the $\nu$ (or more generally $\nu_{\rm eff}$) parameter in the dynamical vacuum and $\mathcal{D}$-class models. We have fixed the other parameters to the best-fit values provided e.g. by the Planck collaboration. This procedure might be considered good enough to grasp the general features of the models we are considering, such as whether they are able or not to roughly explain the available data. This method, though, is clearly insufficient if one wants to go a step farther and see whether the $2-2.5\sigma$ c.l. signal observed in favor of such dynamical models is something really trustworthy. Thus, we are forced to improve the minimization procedure by using a more efficient one. This will allow us to enlarge the dimensionality of our fitting vector and, by doing this, we will be in disposition of determining whether the statistical signal is kept untouched under these changes. This improvement in our methodology is a sort of minimal requirement that we need to fulfill in order to make the fitting procedure more robust.  

\end{itemize}

The second part of this thesis is focused on these issues. We will refine the computational programs used to fit the data together with the statistical methods used to analyze the results. Let us now see in Part II where do all these improvements lead us.

\thispagestyle{empty}
\null
\newpage

\part[Refinement of the fitting analysis. Firsts significant evidences in favor of dynamical DE]{{\huge Refinement of the fitting analysis. Firsts significant evidences in favor of dynamical DE}}

\thispagestyle{empty}
\null
\newpage

\thispagestyle{empty}
\epigraph{<<If you would be a real seeker after truth, it is necessary that at least once in your life you doubt, as far as possible, all things.>>}{\itshape - Ren\'e Descartes, in his {\it Principia philosophiae} (1644)}

\thispagestyle{empty}
\null
\newpage

\pagestyle{fancy}
\fancyhf{}
\fancyhead[CO]{\nouppercase{\leftmark}}
\fancyhead[CE]{\nouppercase{\rightmark}}
\cfoot{\thepage}
\chapter[First Evidence of Running Cosmic Vacuum]{First Evidence of Running Cosmic Vacuum: Challenging the Concordance Model}
\label{chap:AandGRevisited}

In this chapter we dedicate our efforts to extend the phenomenological analysis carried out in the first part of this dissertation. We consider the string of cosmological data on SNIa+BAO+$H(z)$+LSS+BBN+CMB, and put to the test the possibility that the vacuum energy density could be a ``running'' quantity varying with the Universe's expansion rate, $H$. Our goal is to check if this possibility improves the description of the cosmological data offered by the the concordance $\CC$CDM model. For the class of models being considered we do not make any direct association of the $\CC$ and $G$ running with the dynamical evolution of scalar fields. We re-analyze the type-A and G models studied in Chapters \ref{chap:Atype} and \ref{chap:Gtype}, respectively, but this time focusing even more on the fitting procedure and trying to improve each of the points previously mentioned in the summary of Part I. Although a simple Lagrangian description of these models at the level of standard scalar fields is not available, attempts have been made in the literature\,\cite{Fossil07,SolaReview2013} and in any case this is of course something that one would eventually hope to find. There is, however, no guarantee that such description is possible in terms of a simple local action\,\cite{Fossil07}.

Our main aim here is phenomenological. We will argue upon carefully confronting theory and observations that the idea of running vacuum models (RVM's) can be highly competitive, if not superior, to the traditional $\CC$CDM framework.
The first indications of dynamical vacuum energy (at the $\sim 2.5\sigma$ c.l.) were reported in \cite{ApJLnostre}, see Chapter \ref{chap:Gtype}. Earlier comprehensive studies hinted also at this possibility but remained at a lower level of significance, see e.g. \cite{BPS09,Grande2011,JCAPnostre1}\footnote{Recent claims that the $\Lambda$CDM may not be the best description of our Universe can also be found in e.g. \cite{SahniShafielooStarobinsky}, \cite{Ding2015}, \cite{Zheng2016} and \cite{Ferreira2017}; see, however, Section \ref{sect:FitApJ}, point S4).}. Remarkably, in the present work the reported level of evidence is significantly higher than in any previous work in the literature (to the best of our knowledge).
While Occam's razor says that ``Among equally competing models describing the same observations, choose the simplest one'', the point we wish to stress here is that the RVM's are able to describe the current observations better than the $\CC$CDM, not just alike.  For this reason we wish to make a case for the RVM's, in the hope that they could shed also some new light on the CC problem, e.g. by motivating further theoretical studies on these models or related ones.

The plan of this chapter is as follows. In section \ref{sect:RVMs} we make a brief description (at the background level) of the different types of running vacuum models (RVM's) that will be considered in this study. In section \ref{sect:FitApJ} we fit these models to a large set of cosmological data on distant type Ia supernovae (SNIa), baryonic acoustic oscillations (BAO's), the known values of the Hubble parameter at different redshift points, the large scale structure (LSS) formation data, the BBN bound on the Hubble rate, and, finally, the CMB distance priors from WMAP and Planck. We include also a fit of the data with the standard XCDM parametrization, which serves as a baseline for comparison. In section \ref{sec:discussionApJ} we present a detailed discussion of our results. Finally, in section \ref{sec:conclusionsApJ}, we deliver our conclusions.

%%%%%%%%%%%%%%%%%%%%%%%%%%%%%%%%%%%%%%%%%%%%%%%%%%%%%%%%%%%%%%%%%
%%%%%%%%%%%%%%%%%%%%%%%%%%%%%%%%%%%%%%%%%%%%%%%%%%%%%%%%%%%%%%%%%
%%%%%%%%%%%%%%%%%%%%%%%%%%%%%%%%%%%%%%%%%%%%%%%%%%%%%%%%%%%%%%%%%

\begin{table*}
\begin{center}
\resizebox{1\textwidth}{!}{
\begin{tabular}{| c | c |c | c | c | c | c | c | c | c |}
\multicolumn{1}{c}{Model} &  \multicolumn{1}{c}{$h$} &  \multicolumn{1}{c}{$\omega_b= \Omega_b h^2$} & \multicolumn{1}{c}{{\small$n_s$}}  &  \multicolumn{1}{c}{$\Omega_m$}&  \multicolumn{1}{c}{{\small$\nu_{eff}$}}  & \multicolumn{1}{c}{$w$}  &
\multicolumn{1}{c}{$\chi^2_{\rm min}/dof$} & \multicolumn{1}{c}{$\Delta{\rm AIC}$} & \multicolumn{1}{c}{$\Delta{\rm BIC}$}\vspace{0.5mm}
\\\hline
$\Lambda$CDM  & $0.693\pm 0.003$ & $0.02255\pm 0.00013$ &$0.976\pm 0.003$& $0.294\pm 0.004$ & - & $-1$  & 90.44/85 & - & - \\
\hline
XCDM  & $0.670\pm 0.007$& $0.02264\pm0.00014 $&$0.977\pm0.004$& $0.312\pm0.007$ & - &$-0.916\pm0.021$  & 74.91/84 & 13.23 & 11.03 \\
\hline
A1  & $0.670\pm 0.006$& $0.02237\pm0.00014 $&$0.967\pm0.004$& $0.302\pm0.005$ &$0.00110\pm 0.00026 $ &  $-1$ & 71.22/84 & 16.92 & 14.72 \\
\hline
A2   & $0.674\pm 0.005$& $0.02232\pm0.00014 $&$0.965\pm0.004$& $0.303\pm0.005$ &$0.00150\pm 0.00035 $& $-1$  & 70.27/84 & 17.87 & 15.67\\
\hline
G1 & $0.670\pm 0.006$& $0.02236\pm0.00014 $&$0.967\pm0.004$& $0.302\pm0.005$ &$0.00114\pm 0.00027 $& $-1$  &  71.19/84 & 16.95 & 14.75\\
\hline
G2  & $0.670\pm 0.006$& $0.02234\pm0.00014 $&$0.966\pm0.004$& $0.303\pm0.005$ &$0.00136\pm 0.00032 $& $-1$  &  70.68/84 & 17.46 & 15.26\\
\hline
\end{tabular}}
\end{center}
\caption[Best-fit values for $\CC$CDM, XCDM and the various running vacuum models (RVM's) using the full data set S1-S7 of Chapter \ref{chap:AandGRevisited}]{\scriptsize{The best-fit values for the $\CC$CDM, XCDM and the RVM's, including their statistical  significance ($\chi^2$-test and Akaike and Bayesian information criteria, AIC and BIC, see the text). The large and positive values of $\Delta$AIC and $\Delta$BIC strongly favor the dynamical DE options (RVM's and XCDM) against the $\CC$CDM (see text). We use $90$ data points in our fit, to wit: $31$ points from the JLA sample of SNIa, $11$ from BAO, $30$  from $H(z)$, $13$ from linear growth, $1$ from BBN, and $4$ from CMB (see S1-S7 in the text for references). In the XCDM model the EoS parameter $w$ is left free, whereas for the RVM's and $\CC$CDM is fixed at $-1$.  The specific RVM fitting parameter is $\nueff$, see Eq.\,(\ref{eq:xixip}) and the text. For G1 and A1 models, $\nueff=\nu$. The remaining parameters are the standard ones ($h,\omega_b,n_s,\Omega_m$).  The quoted number of degrees of freedom ($dof$) is equal to the number of data points minus the number of independent fitting parameters ($5$ for the $\CC$CDM, $6$ for the RVM's and the XCDM. The normalization parameter M introduced in the SNIa sector of the analysis is also left free in the fit, cf. \cite{BetouleJLA}, but it is not listed in the table). For the CMB data we have used the marginalized mean values and standard deviation for the parameters of the compressed likelihood for Planck 2015 TT,TE,EE + lowP data from \cite{Huang}, which provide tighter constraints to the CMB distance priors than those presented in \cite{PlanckDE2015}.}\label{tableFitApJ}}
\end{table*}

%%%%%%%%%%%%%%%%%%%%%%%%%%%%%%%%%%%%%%%%%%%%%%%%%%%%%%%%%%%%%%%%%
%%%%%%%%%%%%%%%%%%%%%%%%%%%%%%%%%%%%%%%%%%%%%%%%%%%%%%%%%%%%%%%%%
%%%%%%%%%%%%%%%%%%%%%%%%%%%%%%%%%%%%%%%%%%%%%%%%%%%%%%%%%%%%%%%%%

\section{Two basic types of RVM's}\label{sect:RVMs}
Let us make a quick review of type-A and G models. If the reader has dived carefully through chapters \ref{chap:Atype} and \ref{chap:Gtype}, then he/she can skip this section without no information loss. For those readers that have previously skipped chapters \ref{chap:Atype} and \ref{chap:Gtype}, this quick review will allow them to follow all the subsequent sections of this chapter without any problem, since it is self-contained.  

In an expanding Universe we may expect that the vacuum energy density and the gravitational coupling are functions of the cosmic time through the Hubble rate, thence $\rL=\rL(H(t))$ and $G=G(H(t))$. Adopting the canonical equation of state $p_\Lambda=-\rho_\Lambda(H)$ also for the dynamical vacuum,
the corresponding field equations in the
Friedmann-Lema\^\i tre-Robertson-Walker (FLRW) metric in flat space become formally identical to those
with strictly constant $G$ and $\CC$:
\begin{eqnarray}\label{eq:FriedmannEqApJ}
&&3H^2=8\pi\,G(H)\,(\rho_m+\rR+\rho_\Lambda(H))\\
&&3H^2+2\dot{H}=-8\pi\,G(H)\,(p_r-\rho_\Lambda(H))\,. \label{eq:PressureEqApJ}\,
\end{eqnarray}
The equations of state for the densities of relativistic ($\rho_r$) and dust matter ($\rho_m$) read, as usually, $p_r=(1/3)\rho_r$ and $p_m=0$, respectively. We consider now the characteristic RVM structure
of the dynamical vacuum energy:
\begin{eqnarray}\label{eq:rhoLApJ}
\rL(H;\nu,\alpha)=\frac{3}{8\pi G}\left(C_0+\nu H^2+\frac{2}{3}\alpha\,\dH\right)+{\cal O}(H^4)\,,\label{eq:rL}
\end{eqnarray}

\begin{table*}
\begin{center}
\resizebox{1\textwidth}{!}{
\begin{tabular}{| c | c |c | c | c | c | c| c | c | c |}
\multicolumn{1}{c}{Model} &  \multicolumn{1}{c}{$h$} &  \multicolumn{1}{c}{$\omega_b= \Omega_b h^2$} & \multicolumn{1}{c}{{\small$n_s$}}  &  \multicolumn{1}{c}{$\Omega_m$}&  \multicolumn{1}{c}{{\small$\nu_{eff}$}}  & \multicolumn{1}{c}{$w$}  &
\multicolumn{1}{c}{$\chi^2_{\rm min}/dof$} & \multicolumn{1}{c}{$\Delta{\rm AIC}$} & \multicolumn{1}{c}{$\Delta{\rm BIC}$}\vspace{0.5mm}
\\\hline
{\small $\Lambda$CDM} & $0.692\pm 0.004$ & $0.02254\pm 0.00013$ &$0.975\pm 0.004$& $0.295\pm 0.004$ & - & $-1$   & 86.11/78 & - & -\\
\hline
XCDM  &  $0.671\pm 0.007$& $0.02263\pm 0.00014 $&$0.976\pm 0.004$& $0.312\pm 0.007$& - & $-0.920\pm0.022$  & 73.01/77 & 10.78 & 8.67 \\
\hline
A1  & $0.670\pm 0.007$& $0.02238\pm0.00014 $&$0.967\pm0.004$& $0.302\pm0.005$ &$0.00110\pm 0.00028 $ & $-1$  & 69.40/77 & 14.39 & 12.27 \\
\hline
A2   & $0.674\pm 0.005$& $0.02233\pm0.00014 $&$0.966\pm0.004$& $0.302\pm0.005$ &$0.00152\pm 0.00037 $& $-1$  & 68.38/77 & 15.41 & 13.29\\
\hline
G1 & $0.671\pm 0.006$& $0.02237\pm0.00014 $&$0.967\pm0.004$& $0.302\pm0.005$ &$0.00115\pm 0.00029 $& $-1$  &  69.37/77 & 14.42 & 12.30\\
\hline
G2  & $0.670\pm 0.006$& $0.02235\pm0.00014 $&$0.966\pm0.004$& $0.302\pm0.005$ &$0.00138\pm 0.00034 $& $-1$  &  68.82/77 & 14.97 & 12.85\\
\hline
\end{tabular}}
\end{center}
\caption[Best-fit values for the various vacuum models and the XCDM using the Planck 2015 results and removing the BAO and LSS data from WiggleZ]{\scriptsize{Same as in Table \ref{tableFitApJ}, but excluding from our analysis the BAO and LSS data from WiggleZ, see point S5) in the text.}\label{tableFitApJ2}}
\end{table*}

\noindent where $G$ can be constant or a function $G=G(H;\nu,\alpha)$ depending on the particular model. The above expression is the form that has been suggested in the literature from the quantum corrections of QFT in curved spacetime (cf. \cite{SolaReview2013,SolGom2015} and references therein). The terms  with higher powers of the Hubble rate have recently been used to describe inflation,
see e.g. \cite{BasiLimaSola2013,Essay2014} and \cite{SolaHonorableMention2015}, but these terms play no role at present and will be hereafter omitted. The coefficients $\nu$ and $\alpha$ have been defined dimensionless. They are responsible for the running of $\rL(H)$ and $G(H)$, and so for $\nu=\alpha=0$ we recover the $\CC$CDM, with $\rL$ and $G$ constants. The values of $\nu$ and $\alpha$ are naturally small in this context since they can be related to the $\beta$-functions of the running. An estimate in QFT indicates that they are of order $10^{-3}$ at most \cite{Fossil07}, but here we will treat them as free parameters of the RVM and hence we shall determine them phenomenologically by fitting the model to observations. As previously indicated, a simple Lagrangian language for these models that is comparable to the scalar field DE description may not be possible, as suggested by attempts involving the anomaly-induced action\,\cite{Fossil07,SolaReview2013}.

Two types of RVM will be considered here: i) type-G models,  when matter is conserved and the running of $\rL(H)$ is compatible with the Bianchi identity at the expense of a (calculable) running of $G$; ii) type-$A$ models, in contrast, denote those with $G=$const. in which the running of $\rL$ must be accompanied with a (calculable) anomalous conservation law of matter. Both situations are described by the generalized local conservation equation $\nabla^{\mu}\left(G\,\tilde{T}_{\mu\nu}\right)=0$, where $\tilde{T}_{\mu\nu}=T_{\mu\nu}+\rL\,g_{\mu\nu}$ is the total energy-momentum tensor involving both matter and vacuum energy. In the FLRW metric, and summing over all energy components, we find
\begin{equation}\label{BianchiGeneralApJ}
\frac{d}{dt}\,\left[G(\rho_m+\rho_r+\rL)\right]+3\,G\,H\,\sum_{i=m,r}(\rho_i+p_i)=0\,.
\end{equation}
If $G$ and $\rL$ are both constants, we recover the canonical  conservation law $\dot{\rho}_m+\dot{\rho}_r+3H\rho_m+4H\rho_r=0$ for the combined system of matter and radiation.
%This law actually follows from the generalized field equations (\ref{eq:FriedmannEq}) and (\ref{eq:PressureEq}).
For type-G models Eq.\,(\ref{BianchiGeneralApJ}) boils down to $\dG(\rma+\rr+\rL)+G\drL=0$
since $\dot\rho_m+3H\rmr=0$ and $\dot\rho_r+4H\rR=0$ for separated conservation of matter and radiation, as usually assumed.  Mixed type of RVM scenarios are possible, but will not be considered in this thesis.

We can solve analytically the type-G and type-A models by inserting equation \,(\ref{eq:rL}) into (\ref{eq:FriedmannEqApJ}) and (\ref{eq:PressureEqApJ}), or using one of the latter two and the corresponding conservation law (\ref{BianchiGeneralApJ}). It is convenient to perform the  integration using the scale factor $a(t)$  rather than the cosmic time. For type-G models the full expression for the  Hubble function normalized to its current value, $E(a)=H(a)/H_0$, can be found to be
\begin{equation}\label{eq:DifEqH} 
\left.E^2(a)\right\vert_{\rm type-G}=1+\left(\frac{\Omega_m}{\xi}+\frac{\Omega_r}{\xi^\prime}\right)\left[-1+a^{-4\xi^\prime}\left(\frac{a\xi^\prime+\xi\Omega_r/\Omega_m}{\xi^\prime+\xi\Omega_r/\Omega_m}\right)^{\frac{\xi^\prime}{1-\alpha}}\right]\,, 
\end{equation}
where $\Omega_{i}=\rho_{i0}/\rho_{c0}$ are the current cosmological parameters for matter and radiation, and we have defined, as in previous chapters,
\be\label{eq:xixip}
\xi=\frac{1-\nu}{1-\alpha}\equiv 1-\nueff\,,\ \ \  \xi^\prime=\frac{1-\nu}{1-\frac{4}{3}\alpha}\equiv 1-\nueffp\,.
\ee
\begin{figure}
\centering
\includegraphics[angle=0,width=1.0\linewidth]{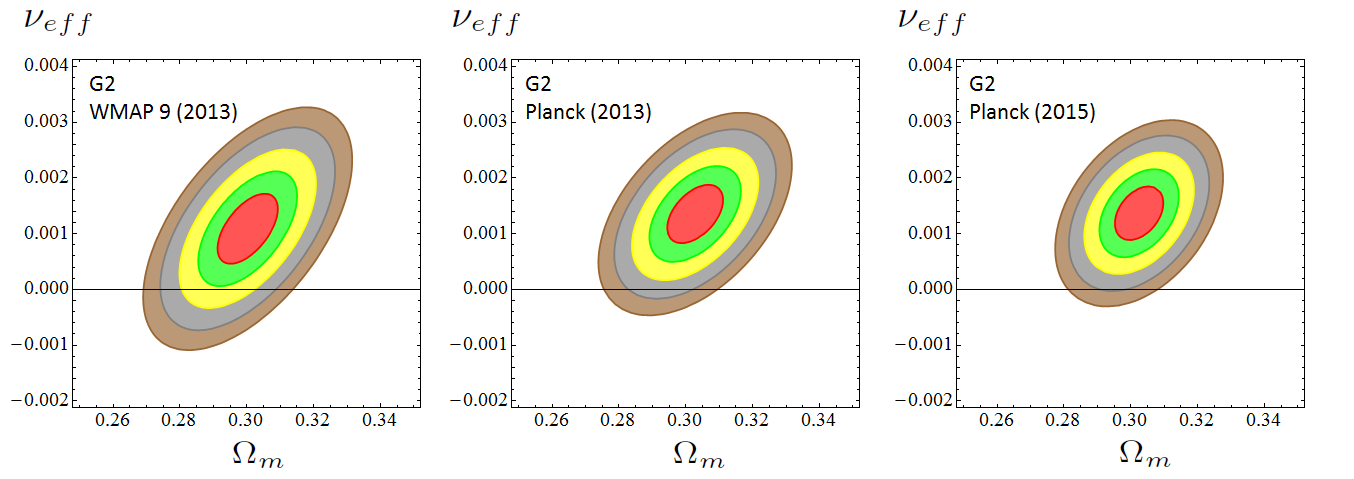}
\caption[Contour lines for model G2 using WMAP9, Planck 2013 and Planck 2015 CMB data. Analysis of chapter \ref{chap:AandGRevisited}]{{\scriptsize Likelihood contours in the $(\Omega_m,\nueff)$ plane for the values $-2\ln\mathcal{L}/\mathcal{L}_{max}=2.30$, $6.18, 11.81$, $19.33$, $27.65$ (corresponding to 1$\sigma$, 2$\sigma$, 3$\sigma$, 4$\sigma$ and 5$\sigma$ c.l.) after marginalizing over the rest of the fitting parameters indicated in Table \ref{tableFitApJ}. We display the progression of the contour plots obtained for model G2 using the 90 data points on SNIa+BAO+$H(z)$+LSS+BBN+CMB, as we evolve from the high precision CMB data from WMAP9, Planck 2013 and Planck 2015 -- see text, point S7). In the sequence, the prediction of the concordance  model ($\nueff=0$) appears increasingly more disfavored, at an exclusion c.l. that ranges from  $\sim 2\sigma$ (for WMAP9), $\sim 3.5\sigma$ (for Planck 2013) and up to  $4\sigma$ (for Planck 2015). Subsequent marginalization over $\Omega_m$ increases slightly the c.l. and renders the fitting values indicated in Table \ref{tableFitApJ}, which reach a statistical significance of $4.2\sigma$ for all the RVM's. Using numerical integration we can estimate that $\sim99.81\%$ of the area of the $4\sigma$ contour for Planck 2015 satisfies $\nueff>0$. We also estimate that $\sim95.47\%$ of the $5\sigma$ region also satisfies $\nueff>0$. The corresponding AIC and BIC criteria (cf. Table \ref{tableFitApJ}) consistently imply a very strong support to the RVM's against the $\CC$CDM.}\label{fig:G2Evolution}}
\end{figure}

\noindent Note that $E(1)=1$, as it should. Moreover, for  $\xi,\xi'\to 1$ (i.e. $|\nu,\alpha|\ll 1$)  $\nueff\simeq \nu-\alpha$ and $\nueffp\simeq \nu-(4/3)\alpha$.
In the radiation-dominated epoch, the leading behavior of Eq.\,(\ref{eq:DifEqH}) is $\sim \Omega_r\,a^{-4\xi'}$, while in the matter-dominated epoch is $\sim \Omega_m\,a^{-3\xi}$. Furthermore, for $\nu,\alpha\to 0$,  $E^2(a)\to 1+\Omega_m\,(a^{-3}-1)+\Omega_r(a^{-4}-1)$. This is the $\CC$CDM form, as expected in that limit. Note
that the following constraint applies among the parameters:
$C_0=H_0^2\left[\Omega_\Lambda-\nu+\alpha\left(\Omega_m+\frac43\,\Omega_r\right)\right]$,
as the vacuum energy density $\rL(H)$ must reproduce the current
value $\rLo$ for $H=H_0$, using $\Omega_m+\Omega_r+\Omega_\Lambda=1$.  The explicit scale factor dependence of the vacuum energy density, i.e. $\rL=\rL(a)$,
ensues upon inserting  \eqref{eq:DifEqH} into \eqref{eq:rL}. In addition, since the matter is conserved for type-G models, we can use the obtained expression for $\rL(a)$ to also infer the explicit form for $G=G(a)$ from (\ref{eq:FriedmannEqApJ}).
We refrain from writing here these cumbersome expressions and we limit ourselves to quote some simplified forms\footnote{See Eq. \eqref{eq:GApJL} for the complete formula of $G(a)$.}. For instance, the expression for $\rL(a)$ when we can neglect the radiation contribution is simple enough:
\be\label{eq:RhoLNR} \rL(a)=\rho_{c0}\,
a^{-3}\left[a^{3\xi}+\frac{\Omega_m}{\xi}(1-\xi-a^{3\xi})\right]\,,
\ee
where  $\rho_{c0}=3H_0^2/8\pi\,G_0$ is the current critical density and $G_0\equiv G(a=1)$ is the current value of the gravitational coupling. Quite obviously for $\xi=1$ we recover the $\CC$CDM form: $\rL=\rho_{c0}(1-\Omega_m)=\rho_{c0}\Omega_\Lambda=$const. As for the gravitational coupling, it
evolves logarithmically with the scale factor and hence changes very slowly\footnote{This is a welcome feature already expected in particular  realizations of type-G models in QFT in curved spacetime\,\cite{Fossil07,SolaReview2013}. See also \cite{Grande2011}.}. It suffices to say that it behaves as
\begin{equation}\label{Gafunction}
G(a)=G_0\,a^{4(1-\xi')}\,f(a)\simeq G_{0}(1+4\nueffp\,\ln\,a)\,f(a)\,,
\end{equation}
where $f(a)=f(a;\Omega_m,\Omega_r; \nu,\alpha)$ is a smooth function of the scale factor. We can dispense with the full expression here, but let us mention that $f(a)$ tends to one at present irrespective of the values of the various parameters $\Omega_m,\Omega_r,\nu,\alpha$ involved in it; and $f(a)\to1$ in the remote past ($a\to 0$) for $\nu,\alpha\to 0$ (i.e. $\xi,\xi'\to 1$). As expected, $G(a)\to G_0$ for $a\to 1$, and $G(a)$ has a logarithmic evolution for $\nueffp\neq 0$.
Notice that the limit $a\to 0$ is  relevant for the BBN (Big Bang Nucleosynthesis) epoch and therefore $G(a)$ should not depart too much from $G_0$ according to the usual bounds on BBN. We shall carefully incorporate this restriction in our analysis of the RVM models, see later on.

Next we quote the solution for type-A models. As indicated, in this case we have an anomalous matter conservation law. Integrating (\ref{BianchiGeneralApJ}) for $G=$const. and using \eqref{eq:rL} in it one finds $\rho_t(a)\equiv\rho_m(a)+\rho_r(a)=\rho_{m0}a^{-3\xi}+\rho_{r0}a^{-4\xi'}$. We have assumed, as usual, that there is no exchange of energy between the relativistic and non-relativistic components. The standard expressions for matter and radiation energy densities are  recovered for $\xi,\xi'\to 1$. The normalized Hubble function for type-A models is simpler than for type-G ones. The full expression including both matter and radiation reads:
\begin{equation}\label{eq:HubbleA}
\left.E^2(a)\right|_{\rm type-A}=1+\frac{\Omega_m}{\xi}\left(a^{-3\xi}-1\right)+\frac{\Omega_r}{\xi^\prime}\left(a^{-4\xi^\prime}-1\right)\,.
\end{equation}
From it and the found expression for $\rho_t(a)$ we can immediately derive the corresponding $\rL(a)$:
\begin{equation}\label{rLaTypaA}
\rho_\CC(a)=\rho_{\Lambda 0}+\rho_{m0}(\xi^{-1}-1)(a^{-3\xi}-1)+\rho_{r0}(\xi^{\prime-1}-1)(a^{-4\xi^\prime}-1)\,.
\end{equation}
Once more for $\nu,\alpha\to 0$ (i.e. $\xi,\xi'\to 1$) we recover the $\CC$CDM case, as easily checked. In particular one finds $\rL\to\rho_{\Lambda 0}=$const. in this limit.

\begin{figure}
\centering
\includegraphics[angle=0,width=1.0\linewidth]{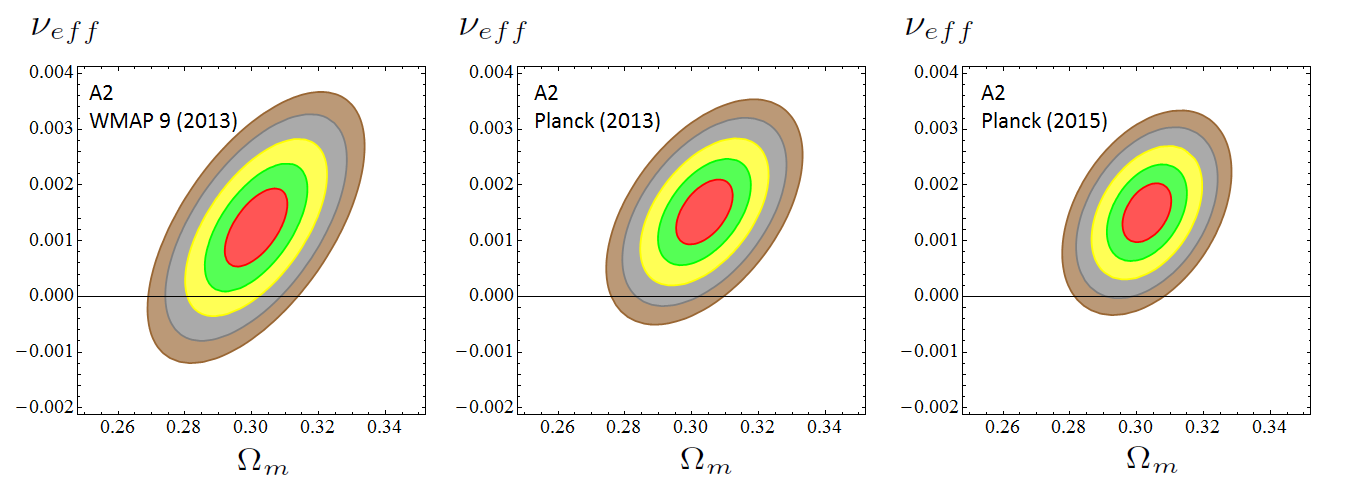}
\caption[Contour lines for model A2 using WMAP9, Planck 2013 and Planck 2015 CMB data. Analysis of chapter \ref{chap:AandGRevisited}]{{\scriptsize As in Fig. \ref{fig:G2Evolution}, but for model A2. Again we see that the contours tend to migrate to the $\nueff>0$ half plane as we evolve from WMAP9 to Planck 2013 and Planck 2015 data. Using the same method as in Fig. \ref{fig:G2Evolution}, we find that $\sim99.82\%$ of the area of the $4\sigma$ contour for Planck 2015 (and  $\sim95.49\%$ of the corresponding $5\sigma$ region) satisfies $\nueff>0$.  The $\CC$CDM  becomes once more excluded at $\sim 4\sigma$ c.l.  ({cf. Table \ref{tableFitApJ} for Planck 2015}).}\label{fig:A2Evolution}}
\end{figure}

\section{Fitting the vacuum models to the data}\label{sect:FitApJ}

\begin{figure}
\centering
\includegraphics[angle=0,width=0.4\linewidth]{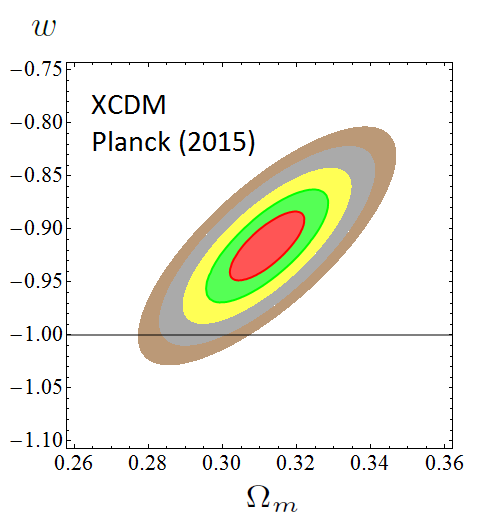}
\caption[Contour lines for the XCDM. Analysis of Chapter \ref{chap:AandGRevisited}]{{\scriptsize As in Fig.\,\ref{fig:G2Evolution} and \ref{fig:A2Evolution}, but for model XCDM and using Planck 2015 data. The $\CC$CDM is excluded at $\sim 4\sigma$ c.l.  ({cf. Table \ref{tableFitApJ}}).
}\label{fig:XCDMEvolution}}
\end{figure}

\begin{figure}
\centering
\includegraphics[angle=0,width=0.6\linewidth]{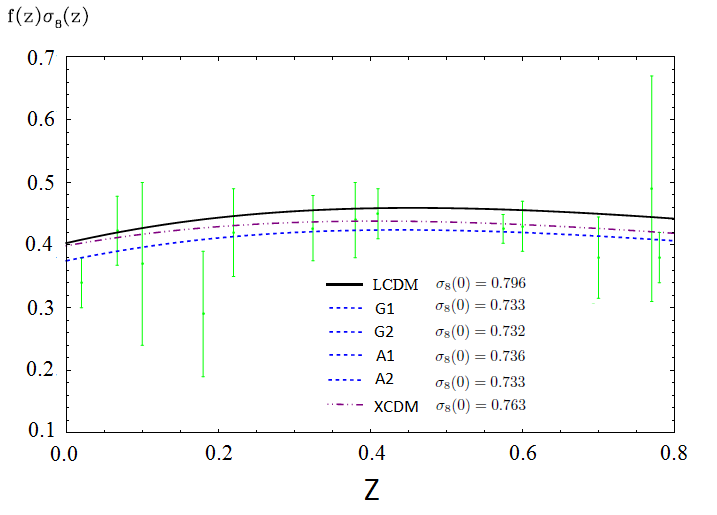}
\caption[$f(z)\sigma_8(z)$ curves for the RVM's, XCDM and the $\CC$CDM, together with the observational data points. Analysis of Chapter \ref{chap:AandGRevisited}]{{\scriptsize {The $f(z)\sigma_8(z)$ data (Table \ref{compilationLSS}) and the predicted curves by the RVM's, XCDM and the $\CC$CDM, using the best-fit values in Table \ref{tableFitApJ}. Shown are also the values of $\sigma_8(0)$ that we obtain for all the models. The theoretical prediction of all the RVM's are visually indistinguishable and they have been plotted using the same (blue) dashed curve.}\label{fig:fsigma8}}
}
\end{figure}

In order to better handle the possibilities offered by the type-G and type-A models as to their dependence on the two specific vacuum parameters $\nu,\alpha$, we shall refer to model G1 (resp. A1) when we address type-G (resp. type-A) models with $\alpha=0$ in Eq.\,(\ref{eq:rL}), as in previous chapters. In these cases $\nueff=\nu$.  When, instead, $\alpha\neq 0$ we shall indicate them by G2 and A2, respectively. This classification scheme is used in Tables \ref{tableFitApJ}-\ref{tableFitApJ2} and \ref{tableFitRla}-\ref{tableFitPlanck}, and in Figs. \ref{fig:G2Evolution}-\ref{fig:XCDMreconstruction}. In the tables we are including also the XCDM (cf. Sect. \ref{sec:discussionApJ}) and the $\CC$CDM.

To this end, we fit the various models to the wealth of cosmological data compiled from distant type Ia supernovae (SNIa), baryonic acoustic oscillations (BAO's), the known values of the Hubble parameter at different redshift points, $H(z_i)$, the large scale structure (LSS) formation data encoded in $f(z_i)\sigma_8(z_i)$, the BBN bound on the Hubble rate, and, finally, the CMB distance priors from WMAP and Planck, with the corresponding correlation matrices in all the indicated cases. Specifically, we have used $90$ data points (in some cases involving compressed data) from $7$ different sources S1-S7, to wit:

\begin{table}
\begin{center}
\begin{tabular}{| c | c | c |}
\multicolumn{1}{c}{$z$} &  \multicolumn{1}{c}{$H(z)$} & \multicolumn{1}{c}{{\small References}}
\\\hline
$0.07$ & $69.0\pm 19.6$ & \cite{Zhang}
\\\hline
$0.09$ & $69.0\pm 12.0$ & \cite{Jimenez}
\\\hline
$0.12$ & $68.6\pm 26.2$ & \cite{Zhang}
\\\hline
$0.17$ & $83.0\pm 8.0$ & \cite{Simon}
\\\hline
$0.1791$ & $75.0\pm 4.0$ & \cite{Moresco2012}
\\\hline
$0.1993$ & $75.0\pm 5.0$ & \cite{Moresco2012}
\\\hline
$0.2$ & $72.9\pm 29.6$ & \cite{Zhang}
\\\hline
$0.27$ & $77.0\pm 14.0$ & \cite{Simon}
\\\hline
$0.28$ & $88.8\pm 36.6$ & \cite{Zhang}
\\\hline
$0.3519$ & $83.0\pm 14.0$ & \cite{Moresco2012}
\\\hline
$0.3802$ & $83.0\pm 13.5$ & \cite{Moresco2016}
\\\hline
$0.4$ & $95.0\pm 17.0$ & \cite{Simon}
\\\hline
$0.4004$ & $77.0\pm 10.2$ & \cite{Moresco2016}
\\\hline
$0.4247$ & $87.1\pm 11.2$ & \cite{Moresco2016}
\\\hline
$0.4497$ & $92.8\pm 12.9$ & \cite{Moresco2016}
\\\hline
$0.4783$ & $80.9\pm 9.0$ & \cite{Moresco2016}
\\\hline
$0.48$ & $97.0\pm 62.0$ & \cite{Stern}
\\\hline
$0.5929$ & $104.0\pm 13.0$ & \cite{Moresco2012}
\\\hline
$0.6797$ & $92.0\pm 8.0$ & \cite{Moresco2012}
\\\hline
$0.7812$ & $105.0\pm 12.0$ & \cite{Moresco2012}
\\\hline
$0.8754$ & $125.0\pm 17.0$ & \cite{Moresco2012}
\\\hline
$0.88$ & $90.0\pm 40.0$ & \cite{Stern}
\\\hline
$0.9$ & $117.0\pm 23.0$ & \cite{Simon}
\\\hline
$1.037$ & $154.0\pm 20.0$ & \cite{Moresco2012}
\\\hline
$1.3$ & $168.0\pm 17.0$ & \cite{Simon}
\\\hline
$1.363$ & $160.0\pm 33.6$ & \cite{Moresco2015}
\\\hline
$1.43$ & $177.0\pm 18.0$ & \cite{Simon}
\\\hline
$1.53$ & $140.0\pm 14.0$ & \cite{Simon}
\\\hline
$1.75$ & $202.0\pm 40.0$ & \cite{Simon}
\\\hline
$1.965$ & $186.5\pm 50.4$ & \cite{Moresco2015}
\\\hline
\end{tabular}
\end{center}
\caption[Compilation of $H(z)$ data points used in all the analysis of Part II]{{\scriptsize Current published values of $H(z)$ in units [km/s/Mpc] obtained using the differential-age technique (see the quoted references and point S4 in the text).}\label{compilationH}}

\end{table}

\begin{itemize}

\item[] S1) The SNIa data points from the SDSS-II/SNLS3 Joint Light-curve Analysis (JLA) \cite{BetouleJLA}. We have used the $31$ binned distance modulus fitted to the JLA sample and the compressed form of the likelihood with the corresponding covariance matrix.

\item[] S2) 5 points on the isotropic BAO estimator $r_s(z_d)/D_V(z_i)$: $z=0.106$ \cite{Beutler2011}, $z=0.15$\, \cite{Ross}, $z_i=0.44, 0.6, 0.73$ \cite{Kazin2014}, with the correlations between the last 3 points.

\item[] S3) 6 data points on anisotropic BAO estimators: 4 of them on  $D_A(z_i)/r_s(z_d)$ and $H(z_i)r_s(z_d)$ at $z_i=0.32, 0.57$, for the LOWZ and CMASS samples, respectively. These data are taken from \cite{GilMarin2OLD}, based on the Redshift-Space Distortions (RSD) measurements of the power spectrum combined with the bispectrum, and the BAO post-reconstruction analysis of the power spectrum (cf. Table 5 of that reference), including the correlations among these data encoded in the provided covariance matrices. We also use  2 data points  based on  $D_A(z_i)/r_s(z_d)$ and $D_H(z_i)/r_s(z_d)$ at $z=2.34$, from the combined LyaF analysis \cite{Delubac2015}. The correlation coefficient among these 2 points are taken from \cite{Aubourg2015} (cf. Table II of that reference). We also take into account the correlations among the  BAO data and the corresponding $f\sigma_8$ data of \cite{GilMarin2OLD} -- see S5) below and Table \ref{compilationLSS}.

\end{itemize}

\begin{savenotes}
\begin{table}
\centering
\begin{center}
\begin{tabular}{| c | c |c | c |}
\multicolumn{1}{c}{Survey} &  \multicolumn{1}{c}{$z$} &  \multicolumn{1}{c}{$f(z)\sigma_8(z)$} & \multicolumn{1}{c}{{\small References}}
\\\hline
6dFGS & $0.067$ & $0.423\pm 0.055$ & \cite{Beutler2012}
\\\hline
SDSS-DR7 & $0.10$ & $0.37\pm 0.13$ & \cite{Feix}
\\\hline
GAMA & $0.18$ & $0.29\pm 0.10$ & \cite{Simpson}
\\ \cline{2-4}& $0.38$ & $0.44\pm0.06$ & \cite{Blake2013}
\\\hline
DR12 BOSS & $0.32$ & $0.427\pm 0.052$  & \cite{GilMarin2OLD} \footnote{The analysis of this chapter was carried out before the publication of \cite{GilMarin2OLD}. In the published version of this paper, \cite{GilMarin2}, there is an increase of the $\sim 8\%$ ($\sim 26\%$) in the $f(z=0.32)\sigma_8(z=0.32)$ uncertainty (resp. in the $f(z=0.57)\sigma_8(z=0.57)$ one). In Chapter \ref{chap:PRDbased} we analyze the changes induced by this fact in the fitting results. Basically, there is a loss in the statistical confidence level at which the RVM is preferred over the $\Lambda$CDM, $4.1\sigma\to 3.8\sigma$, but it is kept close to the $\sim 4\sigma$ c.l. See the aforementioned chapter for more details.}\\ \cline{2-3}
 & $0.57$ & $0.426\pm 0.023$ & \\\hline
 WiggleZ & $0.22$ & $0.42\pm 0.07$ & \cite{Blake2011LSS} \tabularnewline
\cline{2-3} & $0.41$ & $0.45\pm0.04$ & \tabularnewline
\cline{2-3} & $0.60$ & $0.43\pm0.04$ & \tabularnewline
\cline{2-3} & $0.78$ & $0.38\pm0.04$ &
\\\hline
2MTF & $0.02$ & $0.34\pm 0.04$ & \cite{Springob}
\\\hline
VIPERS & $0.7$ & $0.380\pm0.065$ & \cite{Granett}
\\\hline
VVDS & $0.77$ & $0.49\pm0.18$ & \cite{Guzzo2008}\tabularnewline
 & & &\cite{Song09}
\\\hline
 \end{tabular}
\end{center}
\caption[Compilation of $f(z)\sigma_8(z)$ data points used in Chapter \ref{chap:AandGRevisited}]{{\scriptsize Current published values of $f(z)\sigma_8(z)$. See the text, S5).}\label{compilationLSS}}
\end{table}
\end{savenotes}

\begin{itemize}
\item[] S4) $30$ data points on $H(z_i)$ at different redshifts, listed in Table \ref{compilationH}. We use only $H(z_i)$ values obtained by the so-called differential-age techniques applied to passively evolving galaxies. These values are  uncorrelated with the BAO data points. See also \cite{FarooqRatra2013}, \cite{SahniShafielooStarobinsky}, \cite{Ding2015}, \cite{Zheng2016} and \cite{ChenKumarRatra2016}, where the authors make only use of Hubble parameter data in their analyses. We find, however, indispensable to take into account the remaining data sets to derive our conclusions on dynamical vacuum, specially the BAO, LSS and CMB observations. This fact can also be verified quite evidently in {Figures \ref{fig:A2reconstruction}-\ref{fig:XCDMreconstruction}}, to which we shall turn our attention in Section \ref{sec:discussionApJ}.

\item[] S5) $f(z)\sigma_8(z)$: 13 points. These are referred to in the text as LSS (large scale structure formation). {The actual fitting results shown in Table \ref{tableFitApJ} make use of the LSS data listed in Table \ref{compilationLSS}, in which we have carefully avoided possible correlations among them (see below). Let us mention that although we are aware of the existence of other LSS data points in the literature concerning some of the used redshift values in our Table \ref{compilationLSS} -- cf. e.g. \cite{Percival2004}; \cite{Turnbull2012,Hudson2012}; \cite{Johnson2014} -- we have explicitly checked that their inclusion or not in our numerical fits has no significant impact on the main results of this chapter, that is to say, it does not affect the attained $\gtrsim4\sigma$ level of evidence in favor of the RVM's. This result is definitely secured in both cases, but we have naturally presented our final results sticking to the most updated data.}

The following observation is also in order. We have included both the WiggleZ and the CMASS data sets in our analysis. We are aware that there exists some overlap region between the CMASS and WiggleZ galaxy samples. But the two surveys have
been produced independently and the studies on the existing correlations among these observational results \cite{Beutler2016,Marin2016} show that the correlation is small. The overlap region of the CMASS and WiggleZ galaxy samples is actually not among the galaxies that the two surveys pick up, but between the region of the sky they explore. Moreover, despite almost all the WiggleZ region (5/6 parts of it) is inside the CMASS one, it only takes a very small fraction of the whole sky region covered by CMASS, since the latter is much larger than the WiggleZ one (see, e.g. Figure 1 in \cite{Beutler2016}). In this paper, the authors are able to quantify the correlation degree among the BAO constraints in CMASS and WiggleZ, and they conclude that it is less than 4\%. Therefore, we find it justified to include the WiggleZ data in the main table of results of our analysis (Table \ref{tableFitApJ}), but we provide also the fitting results that are obtained when we remove the WiggleZ data points from the BAO and $f(z)\sigma_8(z)$ data sets (see Table \ref{tableFitApJ2}). The difference is small and the central values of the fitting parameters and their uncertainties remain intact. Thus the statistical significance of Tables \ref{tableFitApJ} and \ref{tableFitApJ2} is the same.

\item[] S6) BBN:  we have imposed the average bound on the possible variation of the BBN speed-up factor, defined as the ratio of the expansion rate predicted in a given model versus that of the  $\CC$CDM model at the BBN epoch ($z\sim 10^9$). This amounts to the limit $|\Delta H^2/H_\Lambda^2|<10\%$ \,\cite{Uzan2011}.

\item[] S7) CMB distance priors:  $R$ (shift parameter) and $\ell_a$ (acoustic length) and their correlations  with $(\omega_b,n_s)$. For WMAP9 and Planck 2013 data we used the covariance matrix from the analysis of \cite{WangWang}, while for Planck 2015 data those of \cite{Huang}. Our fitting results for the last case are recorded in all our tables (except in Table \ref{tableFitRla} where we test our fit in the absence of CMB distance priors $R$ and $\ell_a$). We display the final contour plots for all the cases, see Figs. \ref{fig:G2Evolution}-\ref{fig:A2Evolution}. Let us point out that in the case of the Planck 2015 data we have checked that very similar results ensue for all models if we use the alternative CMB covariance matrix from \cite{PlanckDE2015}. We have, however, chosen to explicitly present the case based on  \cite{Huang} since it uses the more complete compressed likelihood analysis for Planck 2015 TT,TE,EE + lowP data whereas \cite{PlanckDE2015} uses Planck 2015 TT+lowP data only.

\end{itemize}

Notice that G1 and A1 have one single vacuum parameter ($\nu$) whereas G2 and A2 have two ($\nu,\alpha$). There is nonetheless a natural alignment between $\nu$ and $\alpha$ for general type-G and A models, namely $\alpha=3\nu/4$, as this entails $\xi'=1$ (i.e. $\nueffp=0$) in Eq.\,(\ref{eq:xixip}). Recall that for G2 models we have $G(a)\sim G_0\,a^{4(1-\xi')}$ deep in the radiation epoch, cf. Eq.\,(\ref{Gafunction}), and therefore the condition $\xi'=1$ warrants $G$ to take the same value as the current one, $G=G_0$,  at BBN. For model G1 this is not possible (for $\nu\neq 0$) and we adopt the aforementioned $|\Delta H^2/H_{\CC}^2|<10\%$ bound. We apply the same BBN restrictions to the A1 and A2 models, which have constant $G$. With this setting all the vacuum models contribute only with one single additional parameter as compared to the $\CC$CDM: $\nu$, for G1 and A1; and $\nueff=\nu-\alpha=\nu/4$, for G2 and A2.

For the statistical analysis, we define the joint likelihood function as the product of the likelihoods for all the data sets. Correspondingly, for Gaussian errors the total $\chi^2$ to be minimized reads:
\be
\chi^2_{tot}=\chi^2_{SNIa}+\chi^2_{BAO}+\chi^2_{H}+\chi^2_{f\sigma_8}+\chi^2_{BBN}+\chi^2_{CMB}\,.
\ee
Each one of these terms is defined in the standard way, for some more details see e.g. Sect. \ref{sec:fitting}, although we should emphasize that here the correlation matrices have been included. The BAO part was split as indicated in S2) and S3) above. Also, in contrast to the previous analysis of Chapter \ref{chap:Gtype}, we did not use here the correlated $Omh^2(z_i,z_j)$ diagnostic for $H(z_i)$ data. Instead, we use
\begin{equation}\label{chi2HApJ}
\chi^{2}_{\rm H}({\bf p})=\sum_{i=1}^{30} \left[ \frac{ H(z_{i},{\bf p})-H_{\rm obs}(z_{i})}
{\sigma_{H,i}} \right]^{2}\,.
\end{equation}
%
%%%%%%%%%%%%%%%%%%%%%%%%%%%%%%%%%%%%%%%%%%%%%%%%%%%%%%%%%%%%%%%%%%%%%%%%%%%%%%%%%%%%%%%%%%%%%%
%%%%%%%%%%%%%%%%%%%%%%%%%%%%%%%%%%%%%%%%%%%%%%%%%%%%%%%%%%%%%%%%%%%%%%%%%%%%%%%%%%%%%%%%%%%%%%%
\noindent As for the linear structure formation data we have computed the density contrast $\delta_m=\delta\rho_m/\rho_m$ for each vacuum model by adapting the cosmic perturbations formalism for type-G and type-A vacuum models.
The matter perturbation, $\delta_m$, obeys a generalized equation which depends on the RVM type. For type-A models it reads (as a differential equation with respect to the cosmic time)
\begin{equation}\label{diffeqDApJ}
\ddot{\delta}_m+\left(2H+\Psi\right)\,\dot{\delta}_m-\left(4\pi
G\rmr-2H\Psi-\dot{\Psi}\right)\,\delta_m=0\,,
\end{equation}
where $\Psi\equiv-\frac{\dot{\rho}_{\CC}}{\rmr}$. For $\rL=$const. we have $\Psi=0$ and Eq.\,(\ref{diffeqDApJ}) reduces to the $\CC$CDM form. For type-G models the matter perturbation equation is explicitly given in \ref{sect:LinStructure} \footnote{For details on these equations, confer the corresponding sections in the first part of this dissertation, an also Appendix \ref{ch:appPert}.}. From here we can derive the weighted linear growth $f(z)\sigma_8(z)$, where $f(z)=d\ln{\delta_m}/d\ln{a}$ is the growth factor and $\sigma_8(z)$ is the rms mass fluctuation amplitude on scales of $R_8=8\,h^{-1}$ Mpc at redshift $z$. {It is computed from}
\begin{equation}
\begin{small}\sigma_8^2(z)=\frac{\delta_m^2(z)}{2\pi^2}\int_{0}^{\infty}k^2\,P(k,\vec{p})\,W^2(kR_8)\,dk\,,\label{s88generalNNApJ}
\end{small}\end{equation}
with $W$ a top-hat smoothing function (see Sect. \ref{sec:PSformalism} for
details) \footnote{We revisit here the calculation of the rms mass fluctuation amplitude and the normalization of the power spectrum, since we have improved now the calculational procedure with respect to the one used in previous chapters.}. The linear matter power spectrum reads $P(k,\vec{p})=P_0k^{n_s}T^2(k,\vec{p})$, where $\vec{p}=(h,\omega_b,n_s,\Omega_m,\nueff)$ is the fitting vector for the vacuum models we are analyzing (including the $\Lambda$CDM, for which $\nueff=0$ of course), and $T(\vec{p},k)$ is the transfer function, which we take from\,\cite{Bard86}, {upon introducing the baryon density effects through the modified shape parameter $\Gamma$\,\cite{PeacockDodds,Sugiyama}. We have also explicitly checked that the use of the effective shape of the transfer function provided in \cite{EisensteinHu98} does not produce any change in our results.}

{The expression (\ref{s88generalNNApJ}) at $z=0$ allows us to write $\sigma_8(0)$ in terms of the power spectrum normalization factor $P_0$ and the primary parameters that enter our fit for each model (cf. Table \ref{tableFitApJ}). We fix $P_0$ from}
\begin{equation}
\begin{small}P_0=2\pi^2\frac{\sigma_{8,\Lambda}^2}{\delta^2_{m,\Lambda}(0)}\left[\int_0^\infty k^{2+n_{s,\Lambda}}T^2(k,\vec{p}_\Lambda)W^2(kR_{8,\Lambda})dk\right]^{-1}\,,\label{P0ApJ}
\end{small}\end{equation}
{in which we have introduced the vector of fiducial parameters $\vec{p}_\CC=(h_{\CC},\omega_{b,\CC},n_{s,\CC},\Omega_{m,\CC},0)$. This vector is defined in analogy with the fitting vector introduced before, but all its parameters are fixed and taken to be equal to those from the Planck 2015 TT,TE,EE+lowP+ lensing analysis\,\cite{Planck2015} with $\nueff=0$. The fiducial parameter $\sigma_{8,\Lambda}$ is also taken from the aforementioned Planck 2015 data.  However, $\delta_{m,\Lambda}(0)$ in (\ref{P0ApJ}) is computable: it is the value of $\delta_m(z=0)$ obtained from solving the perturbation equation of the $\CC$CDM using the mentioned fiducial values of the other parameters. Finally, from $\sigma_8(z) = \sigma_8(0)\delta_m(z)/\delta_m(0)$ and plugging \eqref{P0ApJ} in \eqref{s88generalNNApJ} one finds:}
\begin{equation}
\begin{small}\sigma_{\rm 8}(z)=\sigma_{8, \Lambda}
\frac{\delta_m(z)}{\delta_{m,\CC}(0)}
\left[\frac{\int_{0}^{\infty} k^{2+n_s} T^{2}(k,\vec{p})
W^2(kR_{8}) dk} {\int_{0}^{\infty} k^{2+n_{s,\CC}} T^{2}(k,\vec{p}_\Lambda) W^2(kR_{8,\Lambda}) dk}
\right]^{1/2}\,.
\end{small}\end{equation}
{Computing next the weighted linear growth rate  $f(z)\sigma_8(z)$ for each model under consideration, including the $\CC$CDM, all models become normalized to the same fiducial model defined above. The results for $f(z)\sigma_8(z)$ in the various cases are displayed in Fig. \ref{fig:fsigma8} together with the LSS data measurements (cf. Table \ref{compilationLSS}). We will further comment on these results in the next section.}

\section{Discussion}
\label{sec:discussionApJ}

Table \ref{tableFitApJ} and Figures \ref{fig:G2Evolution}-\ref{fig:A2Evolution} present in a nutshell our main results. We observe that the effective vacuum parameter, $\nueff$, is neatly projected non null and positive for all the RVM's. The presence of this effect can be traced already in the old WMAP9 data (at $\sim 2\sigma$), but as we can see it becomes strengthened at $\sim 3.5\sigma$ c.l. with the  Planck 2013 data  and at $\sim4\sigma$ c.l.  with the Planck 2015 data -- see Figs. \ref{fig:G2Evolution} and \ref{fig:A2Evolution}. For Planck 2015 data it attains up to $\gtrsim4.2\sigma$ c.l. for all the RVM's after marginalizing over the other fitting parameters.

%%%%%%%%%%%%%%%%%%%%%%%%%%%%%%%%%%%%%%%%%%%%%%%%%%%%%%%%%%%%%%%%%%%%%%%%%%%%%%%%%%%%%%%%%%%%%%
%%%%%%%%%%%%%%%%%%%%%%%%%%%%%%%%%%%%%%%%%%%%%%%%%%%%%%%%%%%%%%%%%%%%%%%%%%%%%%%%%%%%%%%%%%%%%%%
\begin{table}
\begin{center}
\resizebox{1\textwidth}{!}{
\begin{tabular}{| c | c |c | c | c | c | c| c | c | c |}
\multicolumn{1}{c}{Model} &  \multicolumn{1}{c}{$h$} &  \multicolumn{1}{c}{$\omega_b= \Omega_b h^2$} & \multicolumn{1}{c}{{\small$n_s$}}  &  \multicolumn{1}{c}{$\Omega_m$}&  \multicolumn{1}{c}{{\small$\nu_{eff}$}}  & \multicolumn{1}{c}{$w$} &
\multicolumn{1}{c}{$\chi^2_{\rm min}/dof$} & \multicolumn{1}{c}{$\Delta{\rm AIC}$} & \multicolumn{1}{c}{$\Delta{\rm BIC}$}\vspace{0.5mm}
\\\hline
$\Lambda$CDM  & $0.679\pm 0.005$ & $0.02241\pm 0.00017$ &$0.968\pm 0.005$& $0.291\pm 0.005$ & - & $-1$ & 68.42/83 & - & - \\
\hline
$X$CDM  & $0.673\pm 0.007$& $0.02241\pm 0.00017 $&$0.968\pm 0.005$& $0.299\pm 0.009$& - & $-0.958\pm0.038$  & 67.21/82  & -1.10 & -3.26 \\
\hline
A1  & $0.679\pm 0.010$& $0.02241\pm0.00017 $&$0.968\pm0.005$& $0.291\pm0.010$ &$-0.00001\pm 0.00079 $ & $-1$  & 68.42/82 & -2.31 & -4.47 \\
\hline
A2   & $0.676\pm 0.009$& $0.02241\pm0.00017 $&$0.968\pm0.005$& $0.295\pm0.014$ &$0.00047\pm 0.00139 $& $-1$  & 68.31/82 & -2.20 & -4.36 \\
\hline
G1 & $0.679\pm 0.009$& $0.02241\pm0.00017 $&$0.968\pm0.005$& $0.291\pm0.010$ &$0.00002\pm 0.00080 $& $-1$  &  68.42/82 & -2.31  &  -4.47\\
\hline
G2  & $0.678\pm 0.012$& $0.02241\pm0.00017 $&$0.968\pm0.005$& $0.291\pm0.013$ &$0.00006\pm 0.00123 $& $-1$  &  68.42/82 & -2.31  &  -4.47\\
\hline
\end{tabular}}
\end{center}
\caption[Best-fit values for the DVM's and the XCDM using the Planck 2015 results and removing the $R$-shift parameter and the acoustic length $l_a$]{{\scriptsize Same as in Table \ref{tableFitApJ}, but removing both the $R$-shift parameter and the acoustic length $l_a$ from our fitting analysis.}\label{tableFitRla}}
\end{table}
\begin{table}
\begin{center}
\resizebox{1\textwidth}{!}{
\begin{tabular}{| c | c |c | c | c | c | c| c | c | c |}
\multicolumn{1}{c}{Model} &  \multicolumn{1}{c}{$h$} &  \multicolumn{1}{c}{$\omega_b= \Omega_b h^2$} & \multicolumn{1}{c}{{\small$n_s$}}  &  \multicolumn{1}{c}{$\Omega_m$}&  \multicolumn{1}{c}{{\small$\nu_{eff}$}}  & \multicolumn{1}{c}{$w$}  &
\multicolumn{1}{c}{$\chi^2_{\rm min}/dof$} & \multicolumn{1}{c}{$\Delta{\rm AIC}$} & \multicolumn{1}{c}{$\Delta{\rm BIC}$}\vspace{0.5mm}
\\\hline
$\Lambda$CDM  & $0.685\pm 0.004$ & $0.02243\pm 0.00014$ &$0.969\pm 0.004$& $0.304\pm 0.005$ & - & $-1$  & 61.70/72 & - & - \\
\hline
XCDM  & $0.683\pm 0.009$ & $0.02245\pm 0.00015$ &$0.969\pm 0.004$& $0.306\pm 0.008$ & - & $-0.991\pm0.040$  & 61.65/71 & -2.30 & -4.29 \\
\hline
A1  & $0.685\pm 0.010$& $0.02243\pm0.00014 $&$0.969\pm0.004$& $0.304\pm0.005$ &$0.00003\pm 0.00062 $ & $-1$  & 61.70/71 & -2.36 & -4.34 \\
\hline
A2   & $0.684\pm 0.009$& $0.02242\pm0.00016 $&$0.969\pm0.005$& $0.304\pm0.005$ &$0.00010\pm 0.00095 $ & $-1$  & 61.69/71 & -2.35 & -4.33 \\
\hline
G1 & $0.685\pm 0.010$& $0.02243\pm0.00014 $&$0.969\pm0.004$& $0.304\pm0.005$ &$0.00003\pm 0.00065 $ & $-1$  & 61.70/71 & -2.36 & -4.34 \\
\hline
G2  & $0.685\pm 0.010$& $0.02242\pm0.00015 $&$0.969\pm0.004$& $0.304\pm0.005$ &$0.00006\pm 0.00082 $ & $-1$  & 61.70/71 & -2.36 & -4.34 \\
\hline
\end{tabular}}
\end{center}
\caption[Best-fit values for the DVM's and the XCDM using the Planck 2015 results and removing the LSS data set]{{\scriptsize Same as in Table \ref{tableFitApJ}, but removing the points from the LSS data set from our analysis,  i.e. all the 13 points on $f\sigma_8$.}\label{tableFitLSS}}
\end{table}
%%%%%%%%%%%%%%%%%%%%%%%%%%%%%%%%%%%%%%%%%%%%%%%%%%%%%%%%%%%%%%%%%%%%%%%%%%%%%%%%%%%%%%%%%%%%%%%
%

It is also interesting to gauge the dynamical character of the DE by performing a fit to the overall data in terms of the well-known XCDM parametrization, in  which the DE is mimicked through the density $\rho_X(a)=\rho_{X0}\,a^{-3(1+w)}$ associated to some generic entity X, which acts as an ersatz for the $\CC$ term; $\rho_{X0}$ being the current energy density value of X and therefore equivalent to $\rho_{\CC 0}$, and  $w$ is the (constant) equation of state (EoS) parameter for X. The XCDM trivially boils down to the rigid $\CC$-term for $w=-1$, but by leaving $w$ free it proves a useful approach to roughly mimic a (non-interactive) DE scalar field with constant EoS. The corresponding fitting results are included in all our tables along with those for the RVM's and the $\CC$CDM. {In Table \ref{tableFitApJ} (our main table) and in Fig. \ref{fig:XCDMEvolution}, we can see that the best fit value for $w$ in the XCDM is $w=-0.916\pm0.021$. Remarkably, it departs from $-1$ by precisely $4\sigma$.}

{Obviously, given the significance of the above result it is highly convenient to compare it with previous analyses of the XCDM reported by the Planck and BOSS collaborations. The Planck 2015 value for the EoS parameter of the XCDM reads $w = -1.019^{+0.075}_{-0.080}$  \cite{Planck2015} and the BOSS one is $w = -0.97\pm 0.05$\,\cite{Aubourg2015}. These results are perfectly compatible with our own result for $w$ shown in Table \ref{tableFitApJ} for the XCDM, but in stark contrast to our result their errors are big enough as to be also fully compatible with the $\CC$CDM value $w=-1$. This is, however, not too surprising if we take into account that none of these analyses included LSS data in their fits, as explicitly indicated in their papers\,\footnote{Furthermore, at the time these analyses appeared they could not have used the important LSS and BAO results from \cite{GilMarin2OLD}, i.e. those that we have incorporated as part of our current data set, not even the previous ones from\,\cite{GilMarin1}. The latter also carry a significant part of the dynamical DE signature we have found here, as we have checked.}. In the absence of LSS data we would find a similar situation. In fact, as our Table \ref{tableFitLSS} clearly shows, the removal of the LSS data set in our fit induces a significant increase in the magnitude of the central value of the EoS parameter, as well as the corresponding error. This happens because the higher is $|w|$ the higher is the structure formation power predicted by the XCDM, and therefore the closer is such prediction with that of the $\CC$CDM (which is seen to predict too much power as compared to the data, see Fig. \ref{fig:fsigma8}). In these conditions our analysis renders $w = -0.991\pm 0.040$, which is definitely closer to (and therefore more compatible with) the central values obtained by Planck and BOSS teams. In addition, this result is now fully compatible with the $\CC$CDM, as in the Planck 2015 and BOSS cases, and all of them are unfavored by the LSS observations. This is consistent with the fact that both information criteria, $\Delta$AIC and $\Delta$BIC, become now slightly negative in Table \ref{tableFitLSS}, which reconfirms that if the LSS data are not used the $\CC$CDM performance is  comparable or even better than the other models. So in order to fit the  observed values of $f\sigma_8$, which are generally lower than the predicted ones by the $\CC$CDM, $|w|$ should decrease. This is exactly what happens for the XCDM, as well as for the RVM's, when the LSS data are included in our analysis (in combination with the other data, particularly with BAO and CMB data). It is apparent from Fig. \ref{fig:fsigma8} that the curves for these models are  then shifted below and hence adapt significantly better to the data points. Correspondingly, the quality of the fits increases dramatically, and this is also borne out by the large and positive values of  $\Delta$AIC and $\Delta$BIC, both above $10$ (cf. Table \ref{tableFitApJ}).}

{The above discussion explains why our analysis of the observations through the XCDM is sensitive to the dynamical feature of the DE, whereas the previous results in the literature  are not. It also shows that the size of the effect found with such a parametrization of the DE essentially concurs with the strength of the dynamical vacuum signature found for the RVM's using exactly the same data. This is remarkable, and it was not obvious {\it a priori}} since for some of our RVM's (specifically for A1 and A2) there is an interaction between vacuum and matter that triggers an anomalous conservation law, whereas for others (G1 and G2) we do not have such interaction (meaning that matter is conserved in them, thereby following the standard decay laws for relativistic and non-relativistic components). The interaction, when occurs, is however proportional to $\nueff$ and thus is small because the fitted value of $\nueff$ is small. This probably explains why the XCDM can succeed in nailing down the dynamical nature of the DE with a comparable performance. However not all dynamical vacuum models describe the data with the same efficiency, see e.g \,\cite{Salvatelli2014}, \cite{Murgia2016}, \cite{Li2016}. A detailed comparison is made among models similar (but different) from those addressed here in Refs. \cite{PRLnostre,PRDnostre,PLBnostre,IJMPA2nostre}, and in Chapters \ref{chap:PRDbased} and \ref{chap:H0tension}. In the XCDM case the departure from the $\CC$CDM takes the fashion of  ``effective quintessence'', whereas for the RVM's it appears as genuine vacuum dynamics. In all cases, however, we find unmistakable signs of DE physics beyond the $\CC$CDM (cf. Table \ref{tableFitApJ}), and this is a most important result of this chapter (and also of this thesis!).

As we have discussed in Section \ref{sect:RVMs}, for models A1 and A2 there is an interaction between vacuum and matter. Such interaction is, of course, small because the fitted values of $\nueff$  are small, see Table \ref{tableFitApJ}. The obtained values are in the ballpark of $\nueff\sim {\cal O}(10^{-3})$ and therefore this is also the order of magnitude associated to the anomalous conservation law of matter. For example, for the non-relativistic component we have
\begin{equation}
\rho_m(a)=\rho_{m0}a^{-3\xi}=\rho_{m0}a^{-3(1-\nueff)}\,.
\end{equation}
This behavior has been used in the works by \cite{FritzschSola2012,FritzschSola2015} as a possible explanation for the hints on the time variation of the fundamental constants, such as coupling constants and particle masses, frequently considered in the literature. The current observational values for such time variation are actually compatible with the fitted values we have found here. This is an intriguing subject that is currently of high interest in the field, see e.g.\,\cite{Uzan2011} and \cite{Sola2015editor}.
For models G1 and G2, instead, the role played by $\nueff$ and $\nueffp$ is different. It does not produce any anomaly in the traditional matter conservation law (since matter and radiation are conserved for type-G models), but now it impinges a small (logarithmic) time evolution on $G$ in the fashion sketched in Eq.\,(\ref{Gafunction}). Thus we find, once more, a possible description for the potential variation of the fundamental constants, in this case $G$, along the lines of the above cited works, see also \cite{FritzschSolaNunes2017}. There are, therefore, different phenomenological possibilities to test the RVM's considered here from various points of view.

\begin{table}
\begin{center}
\resizebox{1\textwidth}{!}{
\begin{tabular}{| c | c |c | c | c | c | c| c | c | c |}
\multicolumn{1}{c}{Model} &  \multicolumn{1}{c}{$h$} &  \multicolumn{1}{c}{$\omega_b= \Omega_b h^2$} & \multicolumn{1}{c}{{\small$n_s$}}  &  \multicolumn{1}{c}{$\Omega_m$}&  \multicolumn{1}{c}{{\small$\nu_{eff}$}}  & \multicolumn{1}{c}{$w$}  &
\multicolumn{1}{c}{$\chi^2_{\rm min}/dof$} & \multicolumn{1}{c}{$\Delta{\rm AIC}$} & \multicolumn{1}{c}{$\Delta{\rm BIC}$}\vspace{0.5mm}
\\\hline
$\Lambda$CDM  & $0.693\pm 0.006$ & $0.02265\pm 0.00022$ &$0.976\pm 0.004$& $0.293\pm 0.007$ & - & $-1$  & 39.35/38 & - & - \\
\hline
XCDM  & $0.684\pm 0.010$ & $0.02272\pm 0.00023$ &$0.977\pm 0.005$& $0.300\pm 0.009$ & - & $-0.960\pm0.033$  & 37.89/37 & -1.25 & -2.30 \\
\hline
A1  & $0.681\pm 0.011$& $0.02254\pm0.00023 $&$0.972\pm0.005$& $0.297\pm0.008$ &$0.00057\pm 0.00043 $ & $-1$  & 37.54/37 & -0.90 & -1.95 \\
\hline
A2   & $0.684\pm 0.009$& $0.02252\pm0.00024 $&$0.971\pm0.005$& $0.297\pm0.008$ &$0.00074\pm 0.00057 $ & $-1$  & 37.59/37 & -0.95 & -2.00 \\
\hline
G1 & $0.681\pm 0.011$& $0.02254\pm0.00023 $&$0.972\pm0.005$& $0.297\pm0.008$ &$0.00059\pm 0.00045 $ & $-1$  & 37.54/37 & -0.90 & -1.95 \\
\hline
G2  & $0.682\pm 0.010$& $0.02253\pm0.00024 $&$0.971\pm0.005$& $0.297\pm0.008$ &$0.00067\pm 0.00052 $ & $-1$  & 37.61/37 & -0.97 & -2.02 \\
\hline
\end{tabular}}
\end{center}
\caption[Best-fit values for the DVM's and the XCDM using the Planck 2015 data set]{{\scriptsize Fitting results using the same data as in \cite{PlanckDE2015}.}\label{tableFitPlanck}}
\end{table}

%%%%%%%%%%%%%%%%%%%%%%%%%%%%%%%%%%%%%%%%%%%%%%%%%%%%%%%%%%%%%%%%%%%%%%%%%%%%%%%%%%%%%%%%%%%%%%

We may reassess the quality fits obtained in this chapter from a different point of view. While the $\chi^2_{\rm min}$ value of the overall fit for any RVM and the XCDM is seen to be definitely smaller than the $\CC$CDM one, it proves very useful to reconfirm our conclusions with the help of the time-honored Akaike and Bayesian information criteria, AIC and BIC, see\,\cite{Akaike1974,Sugiura1978,Schwarz,Burnham}.
They read as follows:
\begin{equation}\label{eq:AICandBIC}
{\rm AIC}=\chi^2_{\rm min}+\frac{2nN}{N-n-1}\,,\ \ \ \ \ {\rm BIC}=\chi^2_{\rm min}+n\,\ln N\,.
\end{equation}
In both cases, $n$ is the number of independent fitting parameters and $N$ the number of data points used in the analysis.
To test the effectiveness of a dynamical DE model (versus the $\CC$CDM) for describing the overall data, we evaluate the pairwise differences $\Delta$AIC ($\Delta$BIC) with respect to the model that carries smaller value of AIC (BIC) -- in this case, the RVM's or the XCDM. The larger these differences the higher is the evidence against the
model with larger value of  AIC (BIC) -- the $\CC$CDM, in this case.
For $\Delta$AIC and/or $\Delta$BIC in the range $6-10$ one may claim ``strong evidence'' against such model; and, above 10, one speaks of ``very strong evidence''\,\cite{Akaike1974,Burnham}. The evidence ratio associated to rejection of the unfavored model is given by the ratio of Akaike weights, $e^{\Delta{\rm AIC}/2}$. Similarly, $e^{\Delta{\rm BIC}/2}$ estimates the so-called Bayes factor, which gives the ratio of marginal likelihoods between the two models\,\cite{BookAmendolaTsujikawa}.

Table \ref{tableFitApJ} reveals conspicuously that the $\CC$CDM appears very strongly disfavored (according to the above statistical standards) as compared to the running vacuum models. Specifically, $\Delta$AIC is in the range $17-18$ and $\Delta$BIC around $15$ for all the RVM's. These results are fully consistent and since both  $\Delta$AIC and  $\Delta$BIC are well above $10$ the verdict of the information criteria is conclusive. But there is another remarkable feature to single out at this point, namely the fact that the simple XCDM parametrization is now left behind as compared to the RVM's. While the corresponding XCDM values of $\Delta$AIC and $\Delta$BIC are also above $10$ (reconfirming the ability of the XCDM to improve the $\CC$CDM fit) they stay roughly $4$ points below the corresponding values for the RVM's. This is considered a significant difference from the point of view of the information criteria. Therefore, we conclude that the RVM's are significantly better than the XCDM in their ability to fit the data. In other words, the vacuum dynamics inherent to the RVM's seems to describe better the overall cosmological data than the effective quintessence behavior suggested by the XCDM parametrization.

Being the ratio of Akaike weights and Bayes factor  much bigger for the RVM's than for the $\CC$CDM, the former appear definitely much more successful than the latter. The current analysis undoubtedly reinforces the conclusions of the study of Chapter \ref{chap:Gtype}, with the advantage that the determination of the vacuum parameters is here much more precise and therefore at a higher significance level. Let us stand out some of the most important differences with respect to the study carried out in the previous chapter: 

\begin{enumerate}
\item To start with, we have used now a larger and fully updated set of cosmological data.
\item The selected data set is uncorrelated and has been obtained from independent analysis in the literature, see points S1-S7) above and references therein.
\item We have taken into account all the known covariance matrices among the data.
\item In this (and the subsequent) chapter(s), $h$, $\omega_b$ and $n_s$ are not fixed {\it a priori} (as we did in Chapter \ref{chap:Gtype}), we have now allowed them to vary in the fitting process. This is, of course, not only a more standard procedure, but also a most advisable one in order to obtain unbiased results. The lack of consensus on the experimental value of $h$ is the main reason why we have preferred to use an uninformative flat prior  -- in the technical sense -- for this parameter. This should be more objective in these circumstances, rather than being subjectively elicited -- once more in the technical sense -- by any of these more or less fashionable camps for $h$ that one finds in the literature, {\cite{RiessH02011}; \cite{ChenRatra2011}; \cite{Freedman2012}; \cite{WMAP9}, \cite{ACTSievers}; \cite{Aubourg2015}; \cite{Planck2015}; \cite{RiessH0}}, whose ultimate fate is unknown at present (compare, e.g. the value from \cite{Planck2015} with the one from \cite{RiessH0}, which is more than $3\sigma$ larger than the former). See Chapter \ref{chap:H0tension} for a detailed discussion on the $H_0$ tension in the light of vacuum dynamics.
\item But the most salient feature perhaps, as compared to our previous study, is that we have introduced here a much more precise treatment of the CMB, in which we used not only the shift parameter, $R$, (which was the only CMB ingredient in our previous studies) but the full data set indicated in S7) above, namely $R$ together with $\ell_a$ (acoustic length) and their correlations  with $(\omega_b,n_s)$.
\end{enumerate}
\begin{figure}
\centering
\includegraphics[angle=0,width=1.0\linewidth]{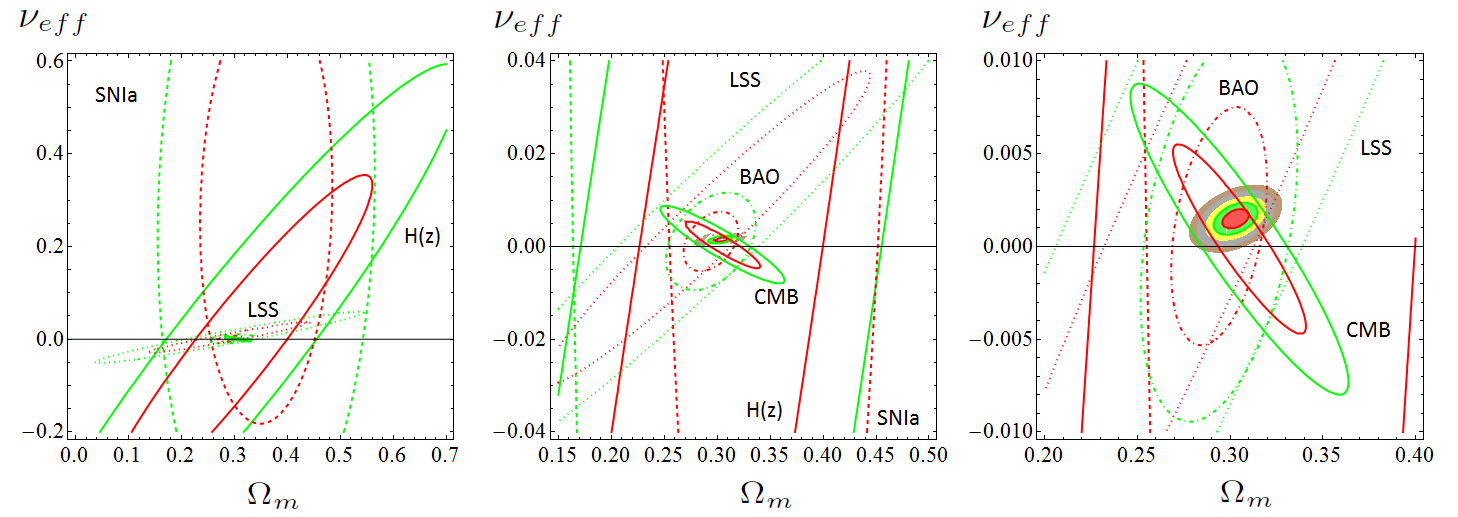}
\caption[Reconstruction of the contour lines for model A2]{{\scriptsize Reconstruction of the contour lines for model A2, under Planck 2015 CMB data (rightmost plot in Fig. \ref{fig:A2Evolution}) from the partial contour plots of the different SNIa+BAO+$H(z)$+LSS+BBN+CMB  data sources. The $1\sigma$ and $2\sigma$ contours are shown in all cases. For the reconstructed final contour lines we also plot the $3\sigma$, $4\sigma$ and $5\sigma$ regions.
}\label{fig:A2reconstruction}
}
\end{figure}

Altogether, this explains the substantially improved accuracy obtained in the current fitted values of the $\nueff$ parameter as compared to Chapter \ref{chap:Gtype}. In particular, in what concerns points 1-3) above  we should stress that for the present analysis we are using a much more complete and restrictive BAO data set. Thus, while in our previous work we only used 6  BAO data points based on the $A(z)$ estimator (cf. Table 3 of \cite{Blake11}), here we are using a total of 11 BAO points (none of them based on $A(z)$, see S2-S3). These include the recent results from\,\cite{GilMarin2OLD}, which narrow down the allowed parameter space in a more efficient way, not only because the BAO data set is larger but also owing to the fact that each of the data points is individually more precise and the known correlation matrices have been taken into account. Altogether, we are able to significantly reduce the error bars with respect to the ones we had obtained in our previous work. We have actually performed a practical test to verify what would be the impact on the fitting quality of our analysis if we would remove the acoustic length $l_a$ from the CMB part of our data and replace the current BAO data points by those used in the previous chapter. Notice that the CMB part is now left essentially with the $R$-shift parameter only, which was indeed the old situation. The result is that we recover the error bars' size shown in the previous chapter, which are $\sim 4-5$ times larger than the current ones, i.e. of order $\mathcal{O}(10^{-3})$ (cf. Tables \ref{tableFit41} and \ref{tableFit42}). We have also checked what would be the effect on our fit if we would remove both the data on the shift parameter and on the acoustic length; or if we would remove only the data points on LSS. The results are presented in Tables \ref{tableFitRla} and \ref{tableFitLSS}, respectively. We observe that the  $\Delta$AIC and $\Delta$BIC values become $2-4$ points negative. This means that the full CMB and LSS data are individually very important for the quality of the fit and that without any of them the evidence of dynamical DE would be lost. If we would restore part of the CMB effect on the fit in Table \ref{tableFitRla} by including the $R$-shift parameter in the fitting procedure we can recover, approximately, the situation of our previous analysis, but not quite since the remaining data sources used now are more powerful.

It is also interesting to explore what would have been the result of our fits if we would not have used our rather complete SNIa+BAO+$H(z)$+LSS+BBN+CMB data set and had restricted ourselves to the much more limited one used by the Planck 2015 collaboration in the paper \cite{PlanckDE2015}. {The outcome is  presented in Table \ref{tableFitPlanck}. In contrast to \cite{Planck2015}, where no LSS (RSD) data were used, the former reference uses some BAO and LSS data, but their fit is rather limited in scope since they use only 4 BAO data points, 1 AP (Alcock-Paczynski parameter) data point, and one single LSS point, namely $f\sigma_8$ at $z=0.57$}, see details in that paper. In contradistinction to them, in our case we used 11 BAO and 13 LSS data points, some of them very recent and of high precision\,\cite{GilMarin2OLD}.  From Table \ref{tableFitPlanck} it is seen that with only the data used in \cite{PlanckDE2015}  the fitting results for the RVM's are poor enough and cannot still detect clear traces of the vacuum dynamics. In particular, the $\Delta$AIC and $\Delta$BIC values in that table are moderately negative, showing that the $\Lambda$CDM does better with only these data. As stated before, not even the XCDM parametrization is able to detect any trace of dynamical DE with that limited data set, as the effective EoS is compatible with $w=-1$ at roughly $1\sigma$ ($w=-0.960\pm 0.033$). This should explain why the features that we are reporting here have been missed till now.

\begin{figure}[t!]
\centering
\includegraphics[angle=0,width=0.75\linewidth]{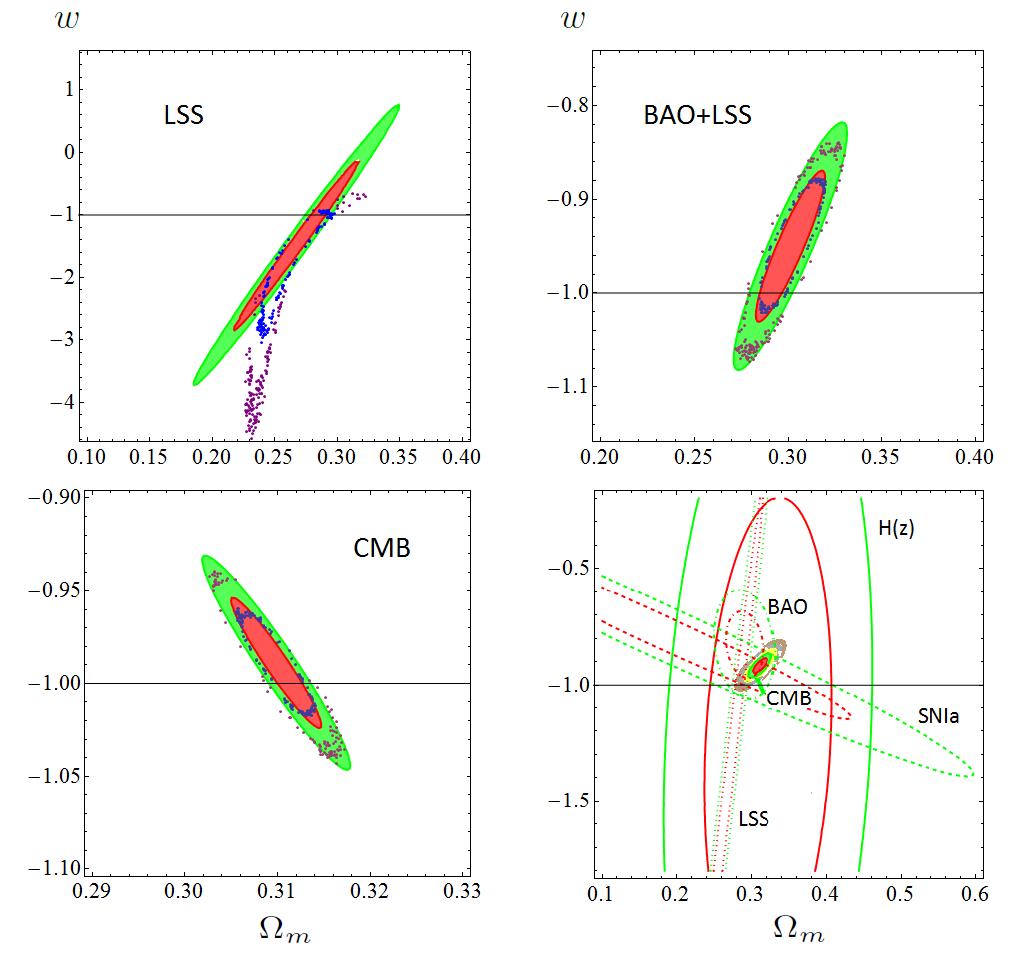}
\caption[Reconstruction of the contour lines for the XCDM parametrization]{{\scriptsize {\it Upper-left plot:} two-dimensional $\Omega_m-w$ contours at $1\sigma$ and $2\sigma$ c.l. obtained with only the LSS data set. The dotted contours in blue and purple are the exact ones, whilst the red and green ellipses have been obtained using the Fisher's approximation. {\it Upper-right plot:} Same, but for the combination BAO+LSS. {\it Lower-left plot:} As in the upper plots, but for the CMB data. {\it Lower-right plot:} The Fisher's generated contours at $1\sigma$ and $2\sigma$ c.l. for all the data sets: SNIa (dotted lines), H(z) (solid lines), BAO (dot-dashed lines), LSS (dotted, very thin lines) and CMB (solid lines, tightly packed in a very small, segment-shaped, region at such scale of the plot). The exact, final, combined contours (from $1\sigma$ up to $5\sigma$) can be glimpsed in the small colored area around the center. See the text for further explanations and Fig. \ref{fig:XCDMEvolution} for a detailed view.}\label{fig:XCDMreconstruction}
}
\end{figure}

We complete our analysis by displaying in a graphical way the contributions from the different data sets to our final contour plots in {Figs. \ref{fig:G2Evolution}-\ref{fig:XCDMEvolution}. We start analyzing the RVM's case.} For definiteness we concentrate on the rightmost plot for model A2 in Fig. \ref{fig:A2Evolution}, but we could do similarly for any other one in Figs. \ref{fig:G2Evolution}-\ref{fig:A2Evolution}. The result for model A2 is depicted in Fig. \ref{fig:A2reconstruction}, where we can assess the detailed reconstruction of the final contours in  terms of the partial contours from the different SNIa+BAO+$H(z)$+LSS+BBN+CMB data sources. This reconstruction is presented through a series of three plots made at different magnifications. In the third plot of the sequence we can easily appraise that the BAO+LSS+CMB data subset plays a fundamental role in narrowing down the final physical region of the $(\Omega_m,\nueff)$ parameter space, in which all the remaining parameters have been marginalized over. This reconstruction also explains in very obvious visual terms why the conclusions that we are presenting here hinge to a large extent on considering the most sensitive components of the data. While CMB obviously is a high precision component in the fit, we demonstrate in our study (both numerically and graphically) that the maximum power of the fit is achieved when it is combined with the wealth of BAO and LSS data points currently available.

In Fig. \ref{fig:XCDMreconstruction} we show the corresponding decomposition of the data contours for the XCDM model as well. In the upper-left plot we display the two-dimensional contours at $1\sigma$ and $2\sigma$ c.l. in the $(\Omega_m,w)$ plane, found using only the LSS data set. The elliptical shapes are obtained upon applying the Fisher matrix formalism\,\cite{BookAmendolaTsujikawa}, i.e. assuming that the two-dimensional distribution is normal (Gaussian) not only in the closer neighborhood of the best-fit values, but in all the parameter space. In order to obtain the dotted contours we have sampled the exact distribution making use of the Metropolis-Hastings Markov chain Monte Carlo algorithm \cite{Metropolis,Hastings}. We find a significant deviation from the ideal perfectly Gaussian case. In the upper-right plot we do the same for the combination BAO+LSS. The continuous and dotted contours are both elliptical, which remarkably demonstrates the Gaussian behavior of the combined BAO+LSS distribution. Needless to say the correlations among BAO and LSS data (whose covariance matrices are known) are responsible for that, i.e. they explain why the product of the non-normal distribution obtained from the LSS data and the Gaussian BAO one produces perfectly elliptical dotted contours for the exact BAO+LSS combination. Similarly, in the lower-left plot we compare the exact (dotted) and Fisher's generated (continuous) lines for the CMB data. Again, it is apparent that the distribution inferred from the CMB data in the $(\Omega_m,w)$ plane is a multivariate normal. Finally, in the lower-right plot we produce the contours at $1\sigma$ and $2\sigma$ c.l. for all the data sets in order to study the impact of each one of them. They have all been found using the Fisher approximation, just to sketch the basic properties of the various data sets, despite we know that the exact result deviates from this approximation and therefore their intersection is not the final answer. The final contours (up to $5\sigma$) obtained from the exact distributions can be seen in the small colored area around the center of the lower-right plot. The reason to plot it small at that scale is to give sufficient perspective to appreciate the contour lines of all the participating data. The final plot coincides, of course, with the one in Fig. \ref{fig:XCDMEvolution}, where it can be appraised in full detail.

{As it is clear from Fig. \ref{fig:XCDMreconstruction}, the data on the H(z) and SNIa observables are not crucial for distilling the final dynamical DE effect, as they have a very low constraining power. This was also so for the RVM case. Once more the final contours are basically the result of the combination of the crucial triplet of BAO+LSS+CMB data. The main conclusion is essentially the same as for the corresponding RVM analysis of combined contours in Fig. \ref{fig:A2reconstruction}, except that in the latter there are no significant deviations from the normal distribution behavior, as we have checked, and therefore all the contours in Fig. \ref{fig:A2reconstruction} can be accurately computed using the Fisher's matrix method. }

The net outcome is that using either the XCDM or the RVM's the  signal in favor of the DE dynamics is clearly pinned down and in both cases it is the result of the combination of all the data sets used in our detailed analysis, although to a large extent it is generated from the crucial BAO+LSS+CMB combination of data sets. In the absence of any of them the signal would get weakened, but when the three data sets are taken together they have enough power to capture the signal of dynamical DE at the remarkable level of $\sim 4\sigma$.

\section{Conclusions}
\label{sec:conclusionsApJ}

To conclude, the running vacuum models emerge as serious alternative candidates for the description of the current state of the Universe in accelerated expansion. These models have a close connection with the possible quantum effects on the effective action of QFT in curved spacetime, cf. \cite{SolaReview2013} and references therein. There were previous phenomenological studies that hinted in different degrees at the possibility that the RVM's could fit the data similarly as the $\CC$CDM, see e.g. the earlier works by \cite{BPS09}, \cite{Grande2011}, \cite{BasPolarSola12}, \cite{BasSola14a}, as well as the more recent ones presented in Chapter \ref{chap:Atype}, including of course the study of Chapter \ref{chap:Gtype}.  However, to our knowledge there is no devoted work comparable in scope to the one presented here for the running vacuum models under consideration. The significantly enhanced level of dynamical DE evidence attained with them is unprecedented, to the best of our knowledge, all the more if we take into account the diversified amount of data used. Our study employed for the first time the largest updated SNIa+BAO+$H(z)$+LSS+BBN+CMB data set of cosmological observations available in the literature. Some of these data (specially the BAO+LSS+CMB part) play a crucial role in the overall fit and are substantially responsible for the main effects reported here. Furthermore, recently the BAO+LSS components have been enriched by more accurate contributions, which have helped to further enhance the signs of the vacuum dynamics. At the end of the day it has been possible to improve the significance of the dynamical hints from a confidence level of roughly $2.5\sigma$, as reported in Chapter \ref{chap:Gtype}, up to the $4.2\sigma$ achieved here. Overall, the signature of dynamical vacuum energy density seems to be rather firmly supported by the current cosmological observations. Already in terms of the generic XCDM parametrization we are able to exclude, for the first time, the absence of vacuum dynamics ($\CC$CDM) at  $4\sigma$ c.l.

{It may be quite appropriate to mention at this point our analysis of Chapter \ref{chap:MPLAbased}, in which we have considered the well-known Peebles \& Ratra scalar field model with an inverse power law  potential $V(\phi)\propto \phi^{-{\alpha}}$ \cite{PeeblesRatra88a,PeeblesRatra88b}, where the power ${\alpha}$ here should, of course, not be confused with a previous use of $\alpha$ for model A2, cf. Sect. \ref{sect:RVMs}. In that study we consider the response of the Peebles \& Ratra model when fitted with the same data sets as those used in the current work. Even though there are other recent tests of that model, see e.g. the works \cite{Samushia2009,FarooqRatra2013,FarooqManiaRatra2013,Farooq2009thesis,Ratra2014,Samushia2014}, none of them used a comparably rich data set as the one we used here. This explains why the analysis of Chapter \ref{chap:MPLAbased} is able to show that a non-trivial scalar field model, such as the Peebles \& Ratra model, is able to fit the observations at a level comparable to the models studied here. In fact, the central value of the ${\alpha}$ parameter of the potential is found to be nonzero at $\sim 4\sigma$ c.l., and the corresponding equation of state parameter $w$ deviates consistently from $-1$ also at the $4\sigma$ level. These remarkable features are only at reach when the crucial triplet of BAO+LSS+CMB data are at work in the fitting analysis of the various cosmological models. The net outcome of these investigations is that several models and parametrizations of the DE do resonate with the conclusion that there is a significant ($\sim 4\sigma$) effect sitting in the current wealth of cosmological data. The effect looks robust enough and can be unveiled using a variety of independent frameworks. Needless to say, compelling statistical evidence conventionally starts at $5\sigma$ c.l. and so we will have to wait for updated observations to see if such level of significance can eventually be attained. In the meanwhile the possible dynamical character of the cosmic vacuum, as suggested by the present study, is pretty high and gives hope for an eventual solution of the old cosmological constant problem, perhaps the toughest problem of fundamental physics.

%%%%%%%%%%%%%%%%%%%%%%%%%%%%%%%%%%%%%%%%%%%%%%%%%%%%%%%%%%%%%%%%%
%%%%%%%%%%%%%%%%%%%%%%%%%%%%%%%%%%%%%%%%%%%%%%%%%%%%%%%%%%%%%%%%%
%%%%%%%%%%%%%%%%%%%%%%%%%%%%%%%%%%%%%%%%%%%%%%%%%%%%%%%%%%%%%%%%%

\section{Main bibliography of the chapter}

This chapter is based on the contents of the paper \cite{ApJnostre}:
\vskip 0.5cm
\noindent
{\it First evidence of running cosmic vacuum: challenging the concordance model.}\newline
J. Sol\`a, A. G\'omez-Valent, and J. de Cruz P\'erez\newline
Astrophys. J. {\bf 836}, 43 (2017) ; arXiv:1602.02103

\thispagestyle{empty}
\null
\newpage

\chapter[Dynamical dark energy: scalar fields and running vacuum]{Dynamical dark energy: scalar fields and running vacuum}
\label{chap:MPLAbased}

The RVM's have been carefully confronted against observations with significant success. For example, the analysis of Chapter \ref{chap:Gtype} reveals that they fit better the cosmological data than the $\CC$CDM at a confidence level of around $2.5\sigma$. This significance has been promoted in the last chapter to $4\sigma$. The next natural question that can be formulated is whether the traditional class of $\phi$CDM models, which do have a local Lagrangian description, and in which the DE is described in terms of a scalar field $\phi$ with some standard form for its potential $V(\phi)$, are also capable of capturing clear signs of dynamical DE using the same set of cosmological observations used for fitting the RVM.

We devote this chapter to show that, indeed, it is so. We compare these two kind of different models and also with the results obtained using the well-known XCDM\,\cite{XCDM} and CPL \cite{CPL1,CPL2} parametrizations of the DE. The upshot is that we are able to collect further evidence on the time evolution of the DE from different types of models at a confidence level of $\sim 4\sigma$. This result is very encouraging and suggests that the imprint of dynamical DE in the modern data is fairly robust and can be clearly decoded using independent formulations.

%%%%%%%%%%%%%%%%%%%%%%%%%%%%%%%%%%%%%%%%%%%%%%%%%%%%%%%%%%%%%%%%%
%%%%%%%%%%%%%%%%%%%%%%%%%%%%%%%%%%%%%%%%%%%%%%%%%%%%%%%%%%%%%%%%%
%%%%%%%%%%%%%%%%%%%%%%%%%%%%%%%%%%%%%%%%%%%%%%%%%%%%%%%%%%%%%%%%%

\section{$\phi$CDM with Peebles \& Ratra potential}\label{sect:phiCDM}
Suppose that the dark energy is described in terms of some scalar field $\phi$ with a standard form for its potential $V(\phi)$, see below. We wish to compare its ability to describe the data with that of the $\CC$CDM, and also with other models of DE existing in the literature. The data used in our analysis will be the same one used in the previous chapter, SNIa+BAO+$H(z)$+LSS+CMB.  Precise information on these data and corresponding observational references are given therein and are also summarized in Table \ref{tableFit1MPLA}. The main results of our analysis are displayed in Tables \ref{tableFit1MPLA}-\ref{tableFit3MPLA} and Figures \ref{fig:Parella2}-\ref{fig:Parella3b}, which we will account in detail throughout our exposition.

We start by explaining our theoretical treatment of the $\phi$CDM model in order to optimally confront it with observations. The scalar field $\phi$ is taken to be dimensionless, being its energy density and pressure given by
\begin{equation}\label{eq:rhophi}
\rho_\phi=\frac{M^2_{P}}{16\pi}\left[\frac{\dot{\phi}^2}{2}+V(\phi)\right]\,,\ p_\phi=\frac{M^2_{P}}{16\pi}\left[\frac{\dot{\phi}^2}{2}-V(\phi)\right]\,.
\end{equation}
Here $M_{P}=1/\sqrt{G}=1.22\times 10^{19}$ GeV is the Planck mass, in natural units.
As a representative potential we adopt the original Peebles \& Ratra (PR) form\,\cite{PeeblesRatra88b,PeeblesRatra88a}:
\begin{equation}\label{eq:PRpotential}
V(\phi)=\frac{1}{2}\kappa M_{P}^2\phi^{-\alpha}\,,
\end{equation}
in which $\kappa$ and $\alpha$ are dimensionless parameters. These are to be determined in our fit to the overall cosmological data. The motivation for such potential is well described in the original paper \cite{PeeblesRatra88b}. In a nutshell: such potential stands for the power-law tail of a more complete effective potential in which inflation is also comprised. We expect $\alpha$ to be positive and sufficiently small such that $V(\phi)$ can mimic an approximate CC that is decreasing slowly with time, in fact more slowly than the matter density. Furthermore, we must have $0<\kappa\ll 1$  such that $V(\phi)$ can be positive and of the order of the measured value $\rLo\sim 10^{-47}$ GeV$^4$. In the late Universe the tail of the mildly declining potential finally surfaces over the matter density (not far away in our past, at $z\sim\mathcal{O}(1)$) and appears as an approximate CC which dominates since then.  Recent studies have considered the PR-potential in the light of the cosmological data, see e.g. \cite{Samushia2009,FarooqRatra2013,FarooqManiaRatra2013,Farooq2009thesis,Ratra2014,Samushia2014}. Here we show that the asset of current observations indicates strong signs of dynamical DE which can be parametrized with such potential. In this way, we corroborate the unambiguous signs obtained with independent DE models shown in Chapter \ref{chap:Gtype} and more conspicuously in Chapter \ref{chap:AandGRevisited}, and with a similar level of confidence.

%%%%%%%%%%%%%%%%%%%%%%%%%%%%%%%%%%%%%%%%%%%%%%%%%%%%%%%%%%%%%%%%%%%%%%%%%%%%%%%%%%%%%%%%%%%%%
\begin{table}
\begin{center}
\resizebox{1\textwidth}{!}{
\begin{tabular}{| c | c |c | c | c | c |}
\multicolumn{1}{c}{Model} &  \multicolumn{1}{c}{$\Omega_m$} &  \multicolumn{1}{c}{$\omega_b= \Omega_b h^2$} & \multicolumn{1}{c}{{\small$n_s$}}  &  \multicolumn{1}{c}{$h$} &
\multicolumn{1}{c}{$\chi^2_{\rm min}/dof$}
\\\hline
{$\Lambda$CDM} & $0.294\pm 0.004$ & $0.02255\pm 0.00013$ &$0.976\pm 0.003$& $0.693\pm 0.004$ & 90.44/85    \\
\hline
 \end{tabular}
 }
\caption[Best-fit values for the $\CC$CDM parameters. Analysis Chapters \ref{chap:AandGRevisited} and \ref{chap:MPLAbased}]{{\scriptsize The best-fit values for the $\CC$CDM parameters $(\Omega_m,\omega_b,n_s,h)$. These values coincide with those presented in Table \ref{tableFitApJ}. We use a total of $89$ data points from SNIa+BAO+$H(z)$+LSS+CMB observables in our fit: namely $31$ points from the JLA sample of SNIa\,\cite{BetouleJLA}, $11$ from BAO\,\cite{Beutler2011,Ross,Kazin2014,GilMarin2OLD,Delubac2015,Aubourg2015}, $30$  from $H(z)$\,\cite{Zhang,Jimenez,Simon,Moresco2012,Moresco2016,Stern,Moresco2015}, $13$ from linear growth \cite{GilMarin2OLD,Beutler2012,Feix,Simpson,Blake2013,Blake2011LSS,Springob,Granett,Guzzo2008,Song09}, and $4$ from CMB\,\cite{Huang}. For a summarized description of these data, see Chapter \ref{chap:AandGRevisited}. The quoted number of degrees of freedom ($dof$) is equal to the number of data points minus the number of independent fitting parameters ($4$ for the $\CC$CDM). For the CMB data we have used the marginalized mean values and standard deviation for the parameters of the compressed likelihood for Planck 2015 TT,TE,EE + lowP data from \cite{Huang}. The parameter M in the SNIa sector\,\cite{BetouleJLA} was dealt with as a nuisance parameter and has been marginalized over analytically. The best-fit values and the associated uncertainties for each parameter in the table have been obtained by numerically marginalizing over the remaining parameters\,\cite{BookAmendolaTsujikawa}.}\label{tableFit1MPLA}}
\end{center}
\end{table}
The scalar field of the $\phi$CDM models satisfies the Klein-Gordon equation in the context of the Friedmann-Lema\^\i tre-Robertson-Walker (FLRW) metric: $\ddot{\phi}+3H\dot{\phi}+{dV}/{d\phi}=0$, where $H=\dot{a}/a$ is the Hubble function.
In some cases the corresponding solutions possess the property of having an attractor-like behavior, in which a large family of solutions are drawn towards a common trajectory\,\cite{PeeblesRatra88b,ZlatevWangSteinhardt99a,ZlatevWangSteinhardt99b}. If there is a long period of convergence of all the family members to that common trajectory, the latter is called a ``tracker solution''\,\cite{ZlatevWangSteinhardt99a,ZlatevWangSteinhardt99b}. When the tracking mechanism is at work, it funnels a large range of initial conditions into a common final state for a long time (or forever, if the convergence is strict). Not all potentials $V$ admit tracking solutions, only those fulfilling the ``tracker condition'' $\Gamma\equiv V\,V''/(V')^2>1$\,\cite{ZlatevWangSteinhardt99a,ZlatevWangSteinhardt99b}, where $V'=\partial V/\partial\phi$.  For the Peebles \& Ratra potential (\ref{eq:PRpotential}), one easily finds $\Gamma=1+1/\alpha$, so it satisfies such condition precisely for $\alpha>0$.

It is frequently possible to seek power-law solutions, i.e. $\phi(t)=A\,t^p$,
for the periods when the energy density of the Universe is dominated by some conserved matter component $\rho(a)=\rho_1\left({a_1}/{a}\right)^n$ (we may call these periods ``nth-epochs''). For instance,
$n=3$ for the matter-dominated epoch (MDE) and $n=4$ for the radiation-dominated epoch (RDE), with
$a_1$ the scale factor at some cosmic time $t_1$ when the corresponding component dominates. We define  $a=1$ as the current value. Solving Friedmann's equation in flat space,
$3H^2(a)=8\pi G\rho(a)$, we find  $H(t)={2}/(nt)$ as a function of the cosmic time in the nth-epoch. Substituting these relations in the  Klein-Gordon equation with the Peebles \& Ratra potential (\ref{eq:PRpotential}) leads to
\begin{equation}
p=\frac{2}{\alpha+2}\,,\qquad A^{\alpha+2}=\frac{\alpha(\alpha+2)^2M_{pl}^2\kappa n}{4(6\alpha+12-n\alpha)}\,.
\end{equation}
From the power-law form we find the evolution of the scalar field with the cosmic time:
\begin{equation}\label{eq:initialphi}
\phi(t)=\left[\frac{\alpha(\alpha+2)^2M_{pl}^2\kappa n}{4(6\alpha+12-n\alpha)}\right]^{1/(\alpha+2)}t^{2/(\alpha+2)}\,.
\end{equation}
In any of the nth-epochs the equation of state (EoS) of the scalar field remains stationary. A straightforward calculation from (\ref{eq:rhophi}), (\ref{eq:PRpotential}) and (\ref{eq:initialphi}) leads to a very compact form for the EoS:
\begin{equation}\label{eq:EoSphi}
w_\phi=\frac{p_{\phi}}{\rho_{\phi}}=-1+\frac{\alpha n}{3(2+\alpha)}\,.
\end{equation}
Since the matter EoS in the nth-epoch is given by $\omega_n=-1+n/3$, it is clear that (\ref{eq:EoSphi}) can be rewritten also as $w_\phi=(\alpha\omega_n-2)/(\alpha+2)$. This is precisely the form predicted by the tracker solutions\,\cite{ZlatevWangSteinhardt99a,ZlatevWangSteinhardt99b}, in which the condition $w_\phi<\omega_n$ is also secured since $|\alpha|$ is expected small. In addition, $w_\phi$ remains constant in the RDE and MDE, but its value does \textit{not} depend on $\kappa$, only on $n$ (or $\omega_n$) and $\alpha$. The fitting analysis presented in Table \ref{tableFit2MPLA} shows that $\alpha={\cal O}(0.1)>0$ and therefore $w_\phi\gtrsim-1$. It means that the scalar field behaves as quintessence in the pure RD and MD epochs (cf. the plateaus at constant values  $w_\phi\gtrsim-1$ in Fig.\,\ref{fig:Parella2}). Notice that the behavior of $w_\phi$ in the interpolating epochs, including the period near our time, is \textit{not} constant (in contrast to the XCDM, see next section) and requires numerical solution of the field equations. See also\, \cite{PodariuRatra} for related studies.

We can trade the cosmic time in (\ref{eq:initialphi}) for the scale factor. This is possible using $t^2=3/(2\pi G n^2\rho)$ (which follows from Friedmann's equation in the nth-epoch) and  $\rho(a)=\rho_{0}a^{-n}=\rho_{c0}\,\Omega\,a^{-n}$, where $\Omega=\Omega_m, \Omega_r$ are the present values of the cosmological density parameters for matter ($n=3$) or radiation ($n=4$) respectively, with $\rho_{c 0}=3H_0^2/(8\pi\,G)$  the current critical energy density. Notice that $\Omega_m=\Omega_{dm}+\Omega_b$ involves both dark matter and baryons. In this way we can determine $\phi$ as a function of the scale factor in the nth-epoch. For example, in the MDE we obtain
\begin{equation}\label{eq:Phia}
\phi(a)=\left[\frac{\alpha(\alpha+2)^2\bar{\kappa}}{9\times 10^4\omega_m(\alpha+4)}\right]^{1/(\alpha+2)}a^{3/(\alpha+2)}\,.
\end{equation}
Here we have conventionally defined the reduced matter density parameter $\omega_m\equiv\Omega_m\,h^2$, in which the reduced Hubble constant $h$ is defined as usual from $H_0\equiv 100h\,\varsigma$, with $\varsigma\equiv 1 Km/s/Mpc=2.133\times10^{-44} GeV$ (in natural units). Finally, for convenience we have introduced in (\ref{eq:Phia}) the dimensionless parameter $\bar{\kappa}$ through $\kappa\,M_P^2\equiv \bar{\kappa}\,\varsigma^2$.

%%%%%%%%%%%%%%%%%%%%%%%%%%%%%%%%%%%%%%%%%%%%%%%%%%%%%%%%%%%%%%%%%%%%%%%%%%%%%%%%%%%%%%%%%%%%

\begin{table}
\begin{center}
\resizebox{1\textwidth}{!}{
\begin{tabular}{| c | c |c | c | c | c | c | c | c | c|c|}
\multicolumn{1}{c}{Model} &  \multicolumn{1}{c}{$\omega_m=\Omega_m h^2$} &  \multicolumn{1}{c}{$\omega_b=\Omega_b h^2$} & \multicolumn{1}{c}{{\small$n_s$}}  &  \multicolumn{1}{c}{$\alpha$} &  \multicolumn{1}{c}{$\bar{\kappa}$}&  \multicolumn{1}{c}{$\chi^2_{\rm min}/dof$} & \multicolumn{1}{c}{$\Delta{\rm AIC}$} & \multicolumn{1}{c}{$\Delta{\rm BIC}$}\vspace{0.5mm}
\\\hline
$\phi$CDM  &  $0.1403\pm 0.0008$& $0.02264\pm 0.00014 $&$0.977\pm 0.004$& $0.219\pm 0.057$  & $(32.5\pm1.1)\times 10^{3}$ &  74.85/84 & 13.34 & 11.10 \\
\hline
 \end{tabular}
 }
\caption[Best-fit values for the Peebles \& Ratra $\phi$CDM parameters. Analysis Chapter \ref{chap:MPLAbased}]{{\scriptsize The best-fit values for the parameter fitting vector (\ref{eq:vfittingPhiCDMMPLA}) of the $\phi$CDM model with Peebles \& Ratra potential (\ref{eq:PRpotential}), including their statistical significance ($\chi^2$-test and Akaike and Bayesian information criteria, AIC and BIC, see the text). We use the same cosmological data set as in Table \ref{tableFit1MPLA}. The large and positive values of $\Delta$AIC and $\Delta$BIC strongly favor the $\phi$CDM model against the $\CC$CDM.  The specific $\phi$CDM fitting parameters are $\bar{\kappa}$ and $\alpha$. The remaining parameters  $(\omega_m,\omega_b,n_s)$ are standard (see text).  The number of independent fitting parameters is $5$, see Eq.\,(\ref{eq:vfittingPhiCDMMPLA})-- one more than in the $\CC$CDM. Using the best-fit values and the overall covariance matrix derived from our fit, we obtain: $h=0.671\pm 0.006$ and $\Omega_m=0.311\pm 0.006$, which allows direct comparison with Table \ref{tableFit1MPLA}. We find $\sim 4\sigma$ evidence in favor of $\alpha>0$. Correspondingly the EoS of $\phi$ at present appears quintessence-like at $4\sigma$ confidence level: $w_\phi= -0.931\pm 0.017$.}\label{tableFit2MPLA}}
\end{center}
\end{table}

Equation (\ref{eq:Phia}) is convenient since it is expressed in terms of the independent parameters that enter our fit, see below.  Let us note  that $\phi(a)$, together with its derivative $\phi^{\prime}(a)=d\phi(a)/da$, allow us to fix the initial conditions in the MDE (a similar expression can be obtained for the RDE). Once these conditions are settled analytically we have to solve numerically the Klein-Gordon equation, coupled to the cosmological equations, to obtain the exact solution. Such solution must, of course, be in accordance with (\ref{eq:Phia}) in the pure MDE. The exact EoS is also a function $w_\phi=w_\phi(a)$, which coincides with the constant value (\ref{eq:EoSphi}) in the corresponding nth-epoch, but interpolates nontrivially between them. At the same time it also interpolates between the MDE and the DE-dominated epoch in our recent past, in which the scalar field energy density surfaces above the nonrelativistic matter density, i.e. $\rho_{\phi}(a)\gtrsim \rho_m(a)$, at a value of $a$ near the current one $a=1$. %(which we normalize to $1$).
The plots for the deceleration parameter, $q=-\ddot{a}/aH^2$, and the scalar field EoS, $w_\phi(a)$, for the best fit parameters of Table \ref{tableFit2MPLA} are shown in Fig. \ref{fig:Parella2}.  The transition point from deceleration to acceleration ($q=0$) is at $z_{tr}=0.628$, which is in good agreement with the values obtained in other works \cite{FarooqRatra2013,Farooqztr2016}, and is also reasonably near the $\CC$CDM one ($z_{tr}^{\CC{\rm CDM}}=0.687$) for the best fit values in Tables \ref{tableFit1MPLA} and \ref{tableFit2MPLA}. The plots for $\phi(a)$ and the energy densities are displayed in Fig. \ref{fig:Parella3}. 
From equations \eqref{eq:PRpotential} and \eqref{eq:Phia} we can see that in the early MDE the potential of the scalar field decays as $V\sim a^{-3\alpha/(2+\alpha)}\sim  a^{-3\alpha/2} $, where in the last step we used the fact that  $\alpha$ is small. Clearly the decaying behavior of $V$ with the expansion is much softer than that of the matter density,  $\rho_m\sim a^{-3}$, and for this reason the DE density associated to the scalar field does not play any role until we approach the current time. This fact is apparent in Fig. \ref{fig:Parella3} (right), where we numerically plot the dimensionless density parameters $\Omega_i(a)=\rho_i(a)/\rho_c(a)$ as a function of the scale factor, where $\rho_c(a)=3H^2(a)/(8\pi G)$ is the evolving critical density.

{As indicated above, the current value of the EoS} can only be known after numerically solving the equations for the best fit parameters in Table \ref{tableFit2MPLA}, with the result $w_\phi(z=0)= -0.931\pm 0.017$ (cf. Fig. \ref{fig:Parella2}). Such result lies clearly in the quintessence regime and with a significance of $4\sigma$. It is essentially consistent with the dynamical character of the DE derived from the non-vanishing value of $\alpha$ in Table \ref{tableFit2MPLA}.

In regard to the value of $h$, there is a significant tension between non-local measurements of $h$, e.g. \cite{Planck2015,Aubourg2015,ChenRatra2011,ACTSievers,HuillierShafieloo,VerdeRiessH0,LukovicAgostinoVittorio}, and local ones, e.g. \cite{RiessH0}. Some of these values can differ by $3\sigma$ or more. For the $\CC$CDM model we find $h=0.693\pm 0.004$ (cf. Table \ref{tableFit1MPLA}), which is in between the ones of \cite{Planck2015} and \cite{RiessH0} and is compatible with the value presented in \cite{WMAP9}. For the $\phi$CDM, our best-fit value is $h=0.671\pm 0.006$ (cf. caption of Table \ref{tableFit2MPLA}), which differs by more than $3\sigma$ with respect to the $\CC$CDM one in our Table \ref{tableFit1MPLA}. Still, both remain perfectly consistent with the recent estimates of $h$ from Hubble parameter measurements at intermediate redshifts\,\cite{ChenKumarRatra2016}. At the moment it is not possible to distinguish models on the sole basis of $H(z)$ measurements. Fortunately, the combined use of the different sorts of SNIa+BAO+$H(z)$+LSS+CMB data offers nowadays a real possibility to elucidate which models are phenomenologically preferred.

%%%%%%%%%%%%%%%%%%%%%%%%%%%%%%%%%%%%%%%%%%%%%%%%%%%%%%%%%%%%%%%%%%%%%%%%%%%%%%%%%%%%%%%%%%%%%%%%

\begin{table}
\begin{center}
\resizebox{1\textwidth}{!}{
\begin{tabular}{| c | c |c | c | c | c | c | c | c | c|c|}
\multicolumn{1}{c}{Model} &  \multicolumn{1}{c}{$\Omega_m$} &  \multicolumn{1}{c}{$\omega_b= \Omega_b h^2$} & \multicolumn{1}{c}{{\small$n_s$}}  &  \multicolumn{1}{c}{$h$} &  \multicolumn{1}{c}{$\nu$}&  \multicolumn{1}{c}{$w_0$} &  \multicolumn{1}{c}{$w_1$} &
\multicolumn{1}{c}{$\chi^2_{\rm min}/dof$} & \multicolumn{1}{c}{$\Delta{\rm AIC}$} & \multicolumn{1}{c}{$\Delta{\rm BIC}$}\vspace{0.5mm}
\\\hline
{\small XCDM} & $0.312\pm 0.007$ & $0.02264\pm 0.00014$ &$0.977\pm 0.004$& $0.670\pm 0.007$ & - & $-0.916\pm 0.021$ & - & 74.91/84 & 13.28 & 11.04\\
\hline
{\small CPL} & $0.311\pm 0.009$ & $0.02265\pm 0.00014$ &$0.977\pm 0.004$& $0.672\pm 0.009$ & - & $-0.937\pm 0.085$ & $0.064\pm 0.247$  & 74.85/83 & 11.04 & 6.61\\
\hline
{\small RVM} & $0.303\pm 0.005$ & $0.02231\pm 0.00015$ &$0.965\pm 0.004$& $0.676\pm 0.005$ & $0.00165\pm 0.00038$ & -1 & - & 70.32/84 & 17.87 & 15.63\\
\hline
 \end{tabular}
 }
\caption[Best-fit values for the RVM, XCDM and CPL. Analysis Chapter \ref{chap:MPLAbased}]{{\scriptsize The best-fit values for the running vacuum model (RVM), together with the XCDM and CPL parametrizations, including also their statistical significance ($\chi^2$-test and Akaike and Bayesian information criteria, AIC and BIC) as compared to the $\CC$CDM (cf. Table \ref{tableFit1MPLA}). The values for the XCDM coincide with those presented in Table \ref{tableFitApJ} of the previous chapter. We use the same string of cosmological SNIa+BAO+$H(z)$+LSS+CMB data as in Tables \ref{tableFit1MPLA} and \ref{tableFit2MPLA}. The specific fitting parameters for these models are $\nu,w_0,$ and $(w_0,w_1)$ for RVM, XCDM and CPL, respectively.  The remaining parameters  are standard. For the models RVM and XCDM the number of independent fitting parameters is $5$, exactly as in the $\phi$CDM. For the CPL parametrization there is one additional parameter ($w_1$). The large and positive values of $\Delta$AIC and $\Delta$BIC strongly favor the RVM and XCDM against the $\CC$CDM. The CPL is only moderately favored as compared to the $\CC$CDM and much less favored than the $\phi$CDM, RVM and XCDM.}\label{tableFit3MPLA}}
\end{center}
\end{table}
%%%%%%%%%%%%%%%%%%%%%%%%%%%%%%%%%%%%%%%%%%%%%%%%%%%%%%%%%%%%%%%%%%%%%%%%%%%%%%%%%%%%%%%%%%%%%%%%%%

Let us now describe the computational procedure that we have followed for the $\phi$CDM model. The initial conditions must be expressed in terms of the parameters that enter our fit. These are defined by means of the following $5$-dimensional fitting vector:
\begin{equation}\label{eq:vfittingPhiCDMMPLA}
\vec{p}_{\phi{\rm CDM}}=(\omega_m,\omega_b,n_s,\alpha,\bar{\kappa})
\end{equation}
where $\omega_b\equiv\Omega_b\,h^2$ is the baryonic component and $n_s$ is the spectral index. These two parameters are specifically involved in the fitting of the CMB and LSS data ($\omega_b$ enters the fitting of the BAO data too), whereas the other three also enter the background analysis, see the previous chapter for more details in the methodology. For the $\phi$CDM we have just one more fitting parameter than in the $\CC$CDM, i.e. $5$ instead of $4$ parameters (cf. Tables \ref{tableFit1MPLA} and \ref{tableFit2MPLA}).
However, in contrast to the $\CC$CDM, for the $\phi$CDM we are fitting the combined parameter $\omega_m=\Omega_m h^2$ rather than $\Omega_m$ and $h$ separately.
The reason is that $h$ (and hence $H_0$) is not a direct fitting parameter in this case since the Hubble function values are determined from Friedmann's equation $3H^2=8\pi\,G(\rho_{\phi}+\rho_m)$, where $\rho_{\phi}$ is given in Eq.\,(\ref{eq:rhophi}) and $\rho_m=\rho_{c 0}\Omega_m a^{-3}=(3\times 10^4/8\pi G)\varsigma^2\,\omega_m\,a^{-3}$ is the conserved matter component. This is tantamount to saying that $h$ is eventually determined from the parameters of the potential and the reduced matter density $\omega_m$. For instance, in the MDE it is not difficult to show that
\begin{equation}\label{eq:barH2MPLA}
\bar{H}^2(a)=\frac{\bar{\kappa}\,\phi^{-\alpha}(a)+1.2\times 10^5\,\omega_m\,a^{-3}}{12-a^2\phi^{\prime 2}(a)}\,,
\end{equation}
where we have defined the dimensionless $\bar{H}=H/\varsigma$, and used $\dot{\phi}=a\,H\,\phi^{\prime}(a)$. As we can see from (\ref{eq:barH2MPLA}), the value of $h\equiv\bar{H}(a=1)/100$ is determined once the three parameters $(\omega_m,\alpha,\bar{\kappa})$ of the fitting vector (\ref{eq:vfittingPhiCDMMPLA}) are given, and then $\Omega_m=\omega_m/h^2$ becomes also determined.  Recall that $\phi(a)$ is obtained by solving numerically the Klein-Gordon equation under appropriate initial conditions (see below) which also depend on the above fitting parameters. As a differential equation in the scale factor, the Klein-Gordon equation reads
\begin{equation}\label{eq:KGaMPLA}
\phi^{\prime\prime}+\phi^\prime\left(\frac{\bar{H}^\prime}{\bar{H}}+\frac{4}{a}\right)-\frac{\alpha}{2}\frac{\bar{\kappa}\phi^{-(\alpha+1)}}{(a\bar{H})^2}=0\,.
\end{equation}
It can be solved after inserting (\ref{eq:barH2MPLA}) in it, together with
\begin{equation}
\bar{H}^\prime=-\frac{3}{2a\bar{H}}\left(\frac{a^2\bar{H}^2\phi^{\prime 2}}{6}+10^4\,\omega_m a^{-3}\right)\,.
\end{equation}
The last formula is just a convenient rephrasing of the expression $\dot{H}=-4\pi\,G(\rho_m+p_m$ $+p_{\phi})$ upon writing it in the above set of variables. According to (\ref{eq:rhophi}), the sum of density and pressure for $\phi$ reads $\rho_{\phi}+p_{\phi}=\dot{\phi}^2/(16\pi{G})=a^2 \bar{H}^2{\phi'}^2 \varsigma^2/(16\pi{G})$, and  of course $p_m=0$ for the matter pressure after the RDE.

The initial conditions for solving (\ref{eq:KGaMPLA}) are fixed in the mentioned nth-epochs of the cosmic evolution. They are determined from the values of the fitting parameters in (\ref{eq:vfittingPhiCDMMPLA}). For example if we set these conditions in the MDE they are defined from the expression of $\phi(a)$ in Eq.\,(\ref{eq:Phia}), and its derivative $\phi^{\prime}(a)$, both taken at some point deep in the MDE, say at a redshift $z>100$, i.e. $a<1/100$.  The result does not depend on the particular choice in this redshift range provided we do not approach too much the decoupling epoch  ($z\simeq 1100$), where the radiation component starts to be appreciable. We have also iterated our calculation when we take the initial conditions deep in the RDE ($n=4$), in which the radiation component $\rho_r$ dominates. In this case $\omega_m=\Omega_m h^2$ is replaced by $\omega_r=\Omega_r h^2$, which is a function of the radiation temperature and the effective number of neutrino species, $N_{eff}$. We find the same results as with the initial conditions settled in the MDE. In both cases the fitting values do agree and are those indicated in Table \ref{tableFit2MPLA}. Let us also mention that when we start from the RDE we find that $\rho_{\phi}(a)\ll\rho_r(a)$ at (and around) the time of BBN (Big Bang Nucleosynthesis), where $a\sim 10^{-9}$, thus insuring that the primordial synthesis of the light elements remains unscathed.

Consistency with BBN is indeed a very important point that motivates the Peebles \& Ratra's inverse power potential $\phi$CDM, Eq.\,(\ref{eq:PRpotential}), together with the existence of the attractor solution. Compared, say to the exponential potential, $V(\phi)=V_0\,e^{-\lambda\,\phi/M_P}$, the latter is inconsistent with BBN (if $\lambda$ is too small) or cannot be important enough to cause accelerated expansion at the current time (if $\lambda$ is too large) \,\cite{PeeblesRatra88b,CopelandLiddleWands98}. This can be cured with a sum of two exponentials with different values of $\lambda$\,\cite{DoubleExpBarreiro}, but of course it is less motivated since involves more parameters. Thus, the PR-potential seems to have the minimal number of ingredients to successfully accomplish the job. In point of fact, it is what we have now verified at a rather significant confidence level in the light of the modern cosmological data.

Finally, let us mention that we have tested the robustness of our computational program by setting the initial conditions out of the tracker path and recovering the asymptotic attractor trajectory. This is of course a numerical check, which is nicely consistent with the fact that the Peebles \& Ratra potential satisfies the aforementioned tracker condition $\Gamma>1$. More details will be reported elsewhere.

\begin{figure}
\begin{center}
\includegraphics[width=5.0in, height=2.32in]{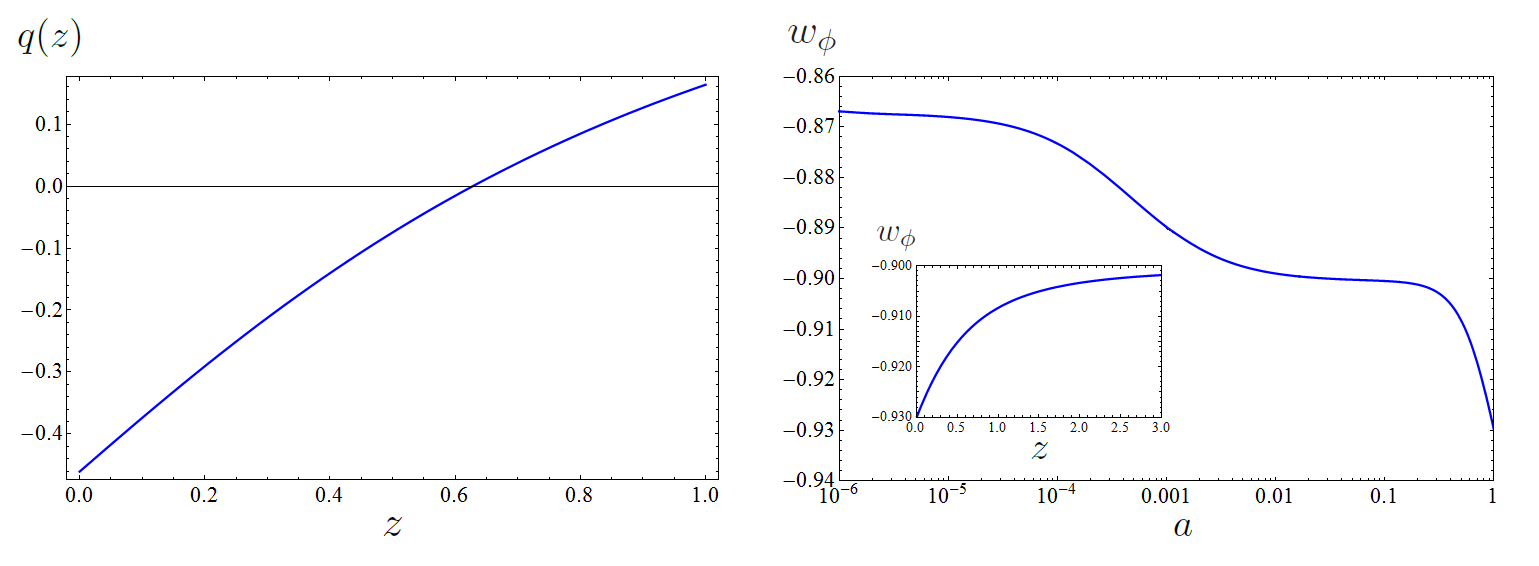}
\caption[$q(z)$ and EoS parameter, $w_{\phi}(a)$, for the $\phi$CDM]{\scriptsize{
{\bf Left:} The deceleration parameter $q(z)$ for the recent Universe. The transition point where $q(z_{tr})=0$ is at $z_{tr}=0.628$, for the best fit values of Table \ref{tableFit2MPLA}. {\bf Right:} The scalar field EoS parameter, $w_{\phi}(a)$, for the entire cosmic history after numerically solving the cosmological equations for the $\phi$CDM model with Peebles \& Ratra potential using the best-fit values of Table \ref{tableFit2MPLA}. The two plateaus from left to right correspond to the epochs of radiation and matter domination, respectively. The sloped stretch at the end, which is magnified in the inner plot in terms of  the redshift variable, corresponds to the recent epoch, in which the scalar field density (playing the role of DE) dominates. We find  $w_\phi(z=0)= -0.931\pm 0.017$.}\label{fig:Parella2}}
\end{center}
\end{figure}

%%%%%%%%%%%%%%%%%%%%%%%%%%%%%%%%%%%%%%%%%%%%%%%%%%%%%%%%%%%%%%%%%%%%%%%%%%%%%%%%%%%%%%%%%%%%%%%%%%%%%%%%
\section{XCDM and CPL parametrizations}
\label{sect:XCDMand CPL}
The XCDM parametrization was first introduced in\,\cite{XCDM} as the simplest way to track a possible dynamics for the DE. Here one replaces the $\CC$-term with an unspecified dynamical entity $X$, whose energy density at present coincides with the current value of the vacuum energy density, i.e. $\rho_X^0=\rLo$. Its EoS reads $p_X=w_0\,\rho_X$, with $w_0=$const. The XCDM mimics the behavior of a scalar field, whether quintessence ($w_0\gtrsim-1$) or phantom ($w_0\lesssim-1$), under the assumption that such field has an essentially constant EoS parameter around $-1$.  Since both matter and DE are self-conserved in the XCDM (i.e. they are not interacting), the energy densities as a function of the scale factor are given by $\rho_m(a)=\rho_m^0\,a^{-3}=\rho_{c 0}\Omega_m\,a^{-3}$ and $\rho_X(a)=\rho_X^0\,a^{-3(1+w_0)}=\rho_{c 0}(1-\Omega_m)\,a^{-3(1+w_0)}$.
Thus, the Hubble function in terms of the scale factor is given by
\begin{equation}\label{eq:HXCDM}
H^2(a)=%\frac{8\pi G}{3}\left[\rho_m^0\,a^{-3}+\rho_X^0\,a^{-3(1+\omega_0)}\right]=
H_0^2\left[\Omega_m\,a^{-3}+(1-\Omega_m)\,a^{-3(1+w_0)}\right]\,.
\end{equation}
A step further in the parametrization of the DE is the CPL prametrization\,\cite{CPL1,CPL2}, whose EoS for the DE is defined as follows:
\begin{equation}\label{eq:CPL}
w=w_0+w_1\,(1-a)=w_0+w_1\,\frac{z}{1+z}\,,
\end{equation}
where $z$ is the cosmological redshift.
In contrast to the XCDM, the EoS of the CPL is not constant and is designed to have a well-defined asymptotic limit in the early Universe. The XCDM serves as a simple baseline to compare other models for the dynamical DE. The CPL further shapes the XCDM parametrization at the cost of an additional parameter ($w_1)$ that enables some cosmic evolution of the EoS. The Hubble function for the CPL in the MDE is readily found:
\begin{eqnarray}
\label{Hzzzquint} H^2(z)&=&
H_0^2\,\left[\Omega_m\,(1+z)^3+(1-\Omega_m)
(1+z)^{3(1+w_0+w_1)}\,e^{-3\,w_1\,\frac{z}{1+z}}\right]
 \,.
\end{eqnarray}

%%%%%%%%%%%%%%%%%%%%%%%%%%%%%%%%%%%%%%%%%%%%%%%%%%%%%%%%%%%%%%%%%%%%%%%%%%%%%%%%%%
\begin{figure}
\begin{center}
\includegraphics[width=5.0in, height=2.32in]{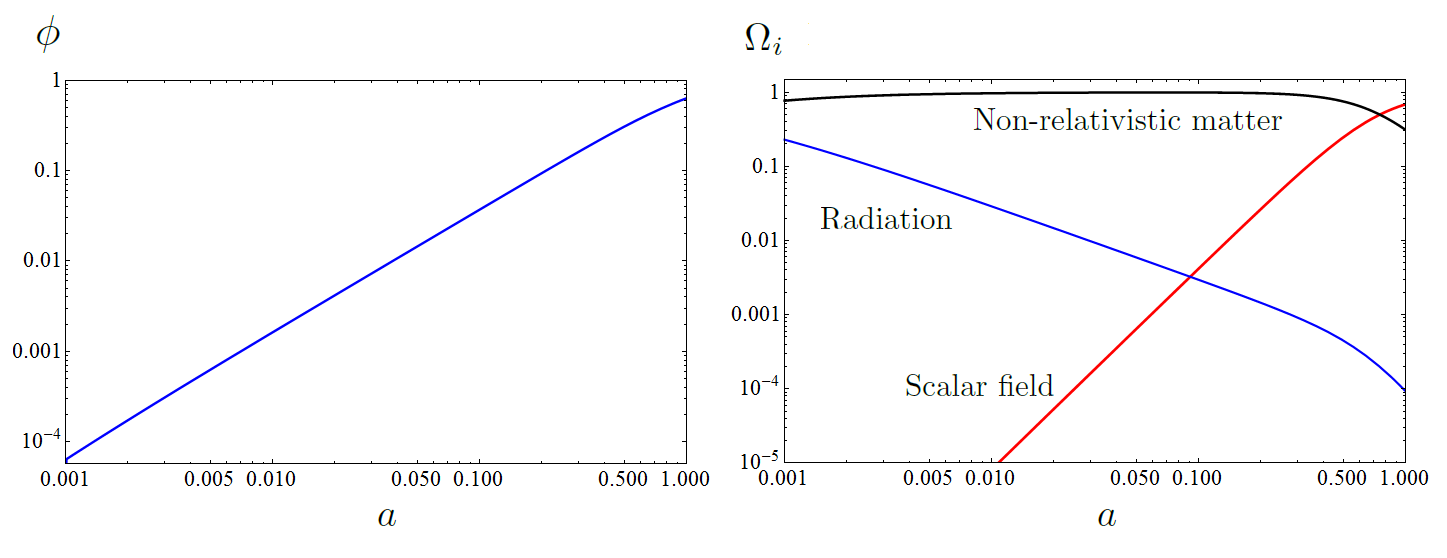}
\caption[Evolution of the scalar field $\phi(a)$ and the various density parameters $\Omega_i(a)$ in the PR-model]{{\scriptsize Same as Fig. \ref{fig:Parella2}, but for $\phi(a)$ and the density parameters $\Omega_i(a)$.  The crossing (or coincidence) point between the scalar field density and the non-relativistic matter density lies very close to our time (viz. $a_{coinc}=0.751$, equivalently $z_{coinc}=0.332$), as it should. This is the point where the tail of the PR potential becomes visible and appears in the form of DE. The point $z_{coinc}$ lies nearer our time than the transition redshift from deceleration to acceleration, $z_{tr}=0.628$ (cf. $q(z)$ in Fig. \ref{fig:Parella2}), similar to the $\CC$CDM.}\label{fig:Parella3}}
\end{center}
\end{figure}
%%%%%%%%%%%%%%%%%%%%%%%%%%%%%%%%%%%%%%%%%%%%%%%%%%%%%%%%%%%%%%%%%%%%%%%%%%%%%%%
\noindent It boils down to (\ref{eq:HXCDM}) for $w_1=0$, as expected. It is understood that for  the RDE  the term $\Omega_r (1+z)^4$ has to be added in the Hubble function. Such radiation term is already relevant for the analysis of the CMB data, and it is included in our analysis.
The fitting results for the XCDM and CPL parametrizations have been collected in the first two rows of Table \ref{tableFit3MPLA}.
Comparing with the $\phi$CDM model (cf. Table \ref{tableFit2MPLA}), we see that the XCDM parametrization also projects the effective quintessence option  at $4\sigma$ level, specifically $w_0=-0.916\pm0.021$.  The CPL parametrization, having one more parameter, does not reflect the same level of significance, but the corresponding AIC and BIC parameters (see below) remain relatively high as compared to the $\CC$CDM, therefore pointing also at clear signs of dynamical DE as the other models. Although the results obtained by the XCDM parametrization and the PR-model are fairly close (see Tables \ref{tableFit2MPLA} and \ref{tableFit3MPLA}) and both EoS values  lie in the quintessence region, the fact that the EoS of the XCDM model is constant throughout the cosmic history makes it difficult to foresee if the XCDM can be used as a faithful representation of a given nontrivial $\phi$CDM model, such as the one we are considering here. The same happens for the extended CPL parametrization, even if in this case the EoS has some prescribed mild cosmic evolution. In actual fact, both the XCDM and CPL parametrizations are to a large extent arbitrary and incomplete representations of the dynamical DE.

\section{RVM: running vacuum}
\label{sect:RVM}
The last model whose fitting  results are reported in Table \ref{tableFit3MPLA} is the running vacuum model (RVM). The RVM is a dynamical vacuum model, meaning that the corresponding EoS parameter is still $w=-1$ but the corresponding vacuum energy density is a ``running'' one, i.e. it departs  (mildly) from the rigid assumption $\rL=$const. of the $\CC$CDM. Specifically, the form of $\rL$ reads as follows:
\begin{equation}\label{eq:RVMvacuumdadensity}
\rho_\CC(H) = \frac{3}{8\pi{G}}\left(C_{0} + \nu{H^2}\right)\,.
\end{equation}
Here $C_0=H_0^2\left(1-\Omega_m-\nu\right)$ is fixed by the boundary condition $\rL(H_0)=\rLo=\rho_{c0}\,(1-\Omega_m)$. The dimensionless coefficient $\nu$ is expected to be very small, $|\nu|\ll1$, since the model must remain sufficiently close to the $\CC$CDM. The moderate dynamical evolution of $\rL(H)$ is possible at the expense of the slow decay rate of vacuum into dark matter. Here we assume that baryons and radiation are conserved, and this is something which has not been studied in the previous chapters. This vacuum-DM interaction in the context of the RVM's will be also studied in Chapters \ref{chap:PRDbased} and \ref{chap:H0tension}.

In practice, the confrontation of the RVM with the data is performed by means of the following $5$-dimensional fitting vector:
\begin{equation}\label{eq:vfitting}
\vec{p}_{\rm RVM}=(\Omega_m,\omega_b,n_s, h,
\nu)\,.
\end{equation}
The first four parameters are the standard ones as in the $\CC$CDM, while $\nu$ is the mentioned vacuum parameter for the RVM. Although it can be treated in a mere phenomenological fashion,
formally $\nu$ can be given a QFT meaning by linking it to the $\beta$-function of the running $\rL$ \cite{SolaReview2013}. We have mentioned in previous chapters that the theoretical estimates place its value in the ballpark of $\nu\sim 10^{-3}$ at most\,\cite{Fossil07}, and this is precisely the order of magnitude pinned down for it in Table \ref{tableFit3MPLA} from our overall fit to the data. The order of magnitude coincidence is reassuring.
The corresponding Hubble function in the MDE reads:
\begin{equation}\label{eq:H2RVM}
H^2(z)=H_0^2\,\left[1+\frac{\Omega_m}{1-\nu}\left((1+z)^{3(1-\nu)}-1\right)\right]\,.
\end{equation}
It depends on the basic fitting parameters $(\Omega_m, h,\nu)$, which are the counterpart of $(\omega_m,\alpha,\bar{\kappa})$ for the $\phi$CDM. The remaining two parameters are common and hence both for the RVM and the $\phi$CDM the total number of fitting parameter is five, see (\ref{eq:vfittingPhiCDMMPLA}) and (\ref{eq:vfitting}).
Note that for $\nu=0$ we recover the $\CC$CDM case, as it should be expected.

%%%%%%%%%%%%%%%%%%%%%%%%%%%%%%%%%%%%%%%%%%%%%%%%%%%%%%%%%%%%%%%%%%%%%%%%%%%%%%%

\begin{figure}
\begin{center}
\includegraphics[width=3.5in,height=2.4in]{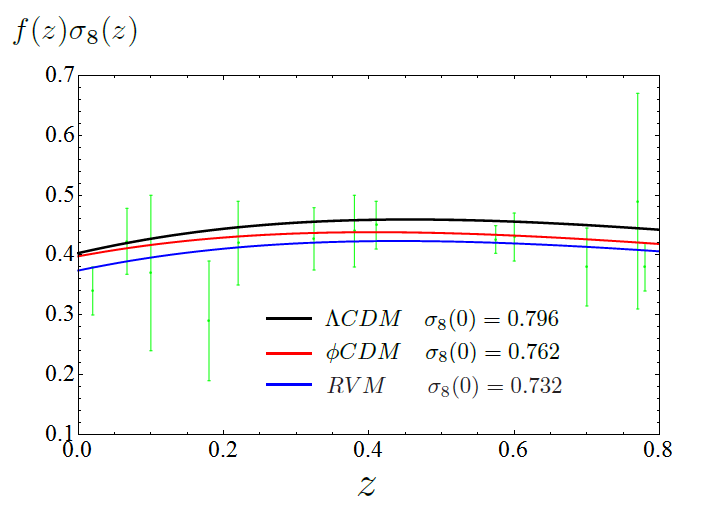}
\caption[$f(z)\sigma_8(z)$ for the $\Lambda$CDM, RVM and $\phi$CDM. Analysis Chapter \ref{chap:MPLAbased}]{{\scriptsize The LSS data on the weighted linear growth rate, $f(z)\sigma_8(z)$, and the predicted curves by the various models, using the best-fit values in Tables \ref{tableFit1MPLA}-\ref{tableFit3MPLA}. The XCDM and CPL lines are not shown since they are almost on top of the $\phi$CDM one. The values of $\sigma_8(0)$ that we obtain for the different models are also indicated.}\label{fig:Parella1}}
\end{center}
\end{figure}

%%%%%%%%%%%%%%%%%%%%%%%%%%%%%%%%%%%%%%%%%%%%%%%%%%%%%%%%%%%%%%%%%%%%%%%%%%%%%%%%%%%%

\section{Structure formation}
A few observations on the analysis of structure formation are in order, as it plays a significant role in the fitting results. On scales well below the horizon the scalar field perturbations are relativistic and hence can be ignored\,\cite{PeeblesRatra88b,PeeblesRatra88a}. As a result in the presence of non-interacting scalar fields the usual matter perturbation equation remains valid\,\cite{BookAmendolaTsujikawa}. Thus, for the $\phi$CDM, XCDM and CPL models we compute the perturbations through the standard equation\,\cite{Peebles1993}
\begin{equation}\label{diffeqLCDM}
\ddot{\delta}_m+2H\,\dot{\delta}_m-4\pi
G\rmr\,\delta_m=0\,,
\end{equation}
with, however, the Hubble function corresponding to each one of these models -- see the formulae in the previous sections.

For the RVM the situation is nevertheless different. In the presence of dynamical vacuum, the perturbation equation not only involves the modified Hubble function (\ref{eq:H2RVM}) but the equation itself becomes modified. One can make use of equation \eqref{diffeqD}, together with the appropriate initial conditions for the density contrast and its first derivative.

Let us also note that, in all cases, we can neglect the DE perturbations at subhorizon scales. We have already mentioned above that this is justified for the $\phi$CDM. For the RVM it can be shown to be also the case, see e.g. Sect. \ref{sect:DEorNot} and Appendix \ref{ch:appPert}. The situation with the XCDM and CPL is not different, and once more the DE perturbations are negligible at scales below the horizon. A detailed study of this issue can be found e.g. in Refs.\,\cite{LXCDM,GrandePelinsonSola08}, in which the so-called $\CC$XCDM model is considered in detail at the perturbations level. In the absence of the (running) component $\CC$ of the DE, the $\CC$XCDM model reduces exactly to the XCDM as a particular case. One can see in that quantitative study that at subhorizon scales the DE perturbations become negligible no matter what is the  assumed value for the sound velocity of the DE perturbations (whether adiabatic or non-adiabatic).

The analysis of the linear LSS regime is conveniently implemented with the help of the weighted linear growth $f(z)\sigma_8(z)$. In Fig. \ref{fig:Parella1} we display  $f(z)\sigma_8(z)$ for the various models using the fitted values of Tables \ref{tableFit1MPLA}-\ref{tableFit3MPLA}. We have applied the procedure discussed in Sect. \ref{sect:FitApJ}.

%%%%%%%%%%%%%%%%%%%%%%%%%%%%%%%%%%%%%%%%%%%%%%%%%%%%%%%%%%%%%%%%%%%%%%%%%%%%%%%
\begin{figure}
\begin{center}
\includegraphics[width=4.5in, height=2.4in]{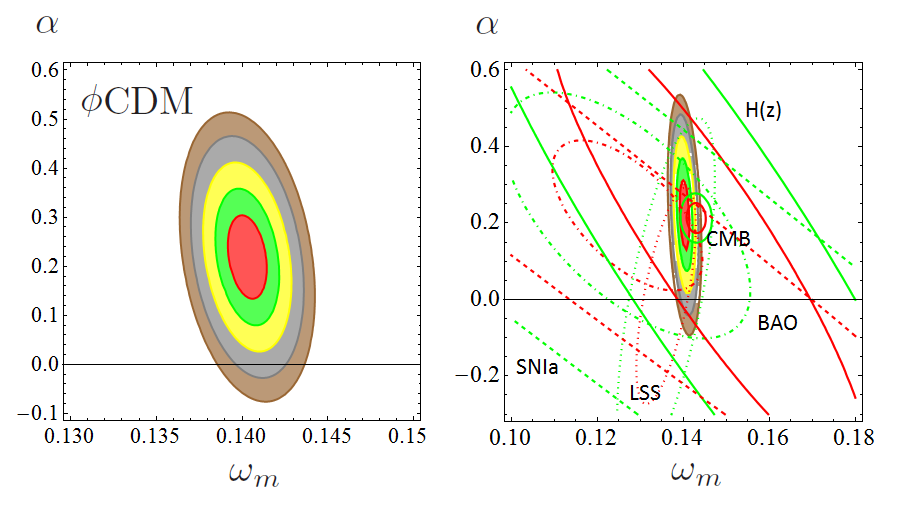}
\caption[Contour lines for the $\phi$CDM in the ($\omega_m$,$\alpha$)-plane, and its reconstruction]{{\scriptsize {\bf Left}: Likelihood contours for the $\phi$CDM model in the ($\omega_m$,$\alpha$)-plane after marginalizing over the remaining parameters (cf. Table \ref{tableFit2MPLA}). The various contours correspond to 1$\sigma$, 2$\sigma$, 3$\sigma$, 4$\sigma$ and 5$\sigma$  c.l. The line $\alpha=0$ corresponds to the concordance $\CC$CDM model. The tracker consistency region $\alpha>0$ (see the text) is clearly preferred, and we see that it definitely points to dynamical DE at $\sim4\sigma$ confidence level. {\bf Right}: Reconstruction of the aforementioned contour lines from the partial contour plots of the different SNIa+BAO+$H(z)$+LSS+CMB data sources using Fisher's approach\,\cite{BookAmendolaTsujikawa}. The $1\sigma$ and $2\sigma$ contours are shown in all cases, but for the reconstructed final contour lines we include the $3\sigma$, $4\sigma$ and $5\sigma$ regions as well. For the reconstruction plot we display a larger $\omega_m$-range to better appraise the impact of the various data sources.}\label{fig:Parella2b}}
\end{center}
\end{figure}

\begin{figure}
\begin{center}
\includegraphics[width=3.8in, height=2.25in]{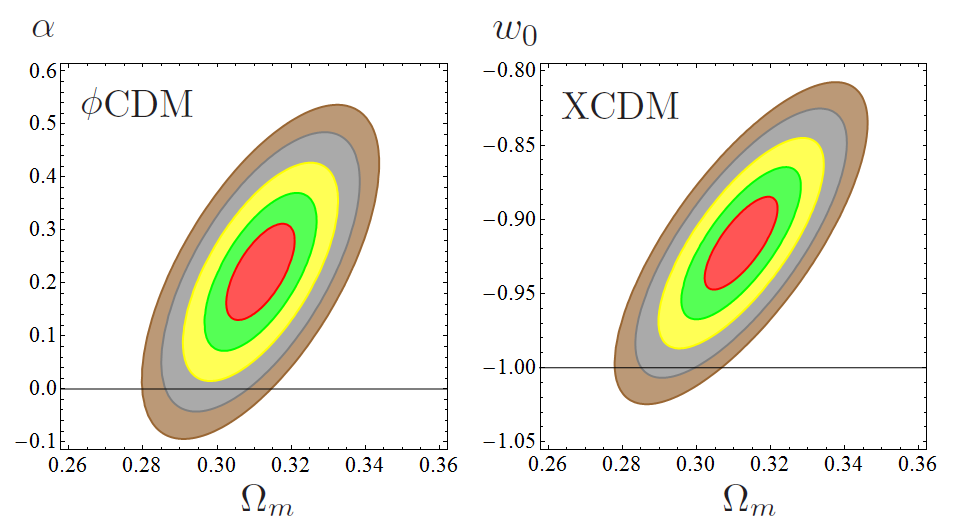}
\caption[Contour lines for the $\phi$CDM model and the XCDM in the ($\Omega_m$,$\alpha$)-plane]{{\scriptsize Likelihood contours for the $\phi$CDM model (left) and the XCDM parametrization (right) in the relevant planes after marginalizing over the remaining parameters in each case (cf. Tables \ref{tableFit2MPLA} and \ref{tableFit3MPLA}). The various contours correspond to 1$\sigma$, 2$\sigma$, 3$\sigma$, 4$\sigma$ and 5$\sigma$ c.l. The central values in both cases are $\sim4\sigma$ away from the $\CC$CDM, i.e. $\alpha=0$ and $w_0=-1$, respectively.}\label{fig:Parella3b}}
\end{center}
\end{figure}

%%%%%%%%%%%%%%%%%%%%%%%%%%%%%%%%%%%%%%%%%%%%%%%%%%%%%%%%%%%%%%%%%%%%%%%%%%%%%%%%

\section{Discussion and conclusions}
The statistical analysis of the various models considered in this study is performed in terms of a joint likelihood function, which is the product of the likelihoods for each data source
%\begin{equation}\label{eq:chi2}
%\chi^2_{tot}=\chi^2_{SNIa}+\chi^2_{BAO}+\chi^2_{H}+\chi^2_{f\sigma_8}+\chi^2_{CMB}\,.
%\end{equation}
and including the corresponding covariance matrices, following the standard procedure\,\cite{BookAmendolaTsujikawa}. The contour plots for the $\phi$CDM and XCDM models are shown in Figures \ref{fig:Parella2b} and \ref{fig:Parella3b}, where the the dynamical character of the DE is clearly demonstrated at $\sim 4\sigma$ c.l. More specifically, in the left plot of  Fig. \ref{fig:Parella2b} we display the final contour plots for $\phi$CDM in the plane $(\omega_m,\alpha)$ -- defined by two of the original parameters of our calculation, cf. Eq.\,(\ref{eq:vfittingPhiCDMMPLA}) --  together with the isolated contours of the different data sources (plot on the right). It can be seen that the joint triad of observables BAO+LSS+CMB conspire to significantly reduce the final allowed region of the ($\omega_m,\alpha$)-plane, while the constraints imposed by SNIa and $H(z)$ are much weaker. This is something that has also been pinpointed in Chapter \ref{chap:AandGRevisited} (cf. Sect. \ref{sec:discussionApJ} therein). Finally, for the sake of convenience, in Fig. \ref{fig:Parella3b} we put forward the final $\phi$CDM and the XCDM contours in the more conventional $(\Omega_m,\alpha)$-plane. As for the RVM, see the contours in the following chapter, where a dynamical DE effect $\sim 4\sigma$ is recorded.

As noted previously, the three models $\phi$CDM, XCDM and RVM have the same number of parameters, namely 5, one more than the $\CC$CDM. The CPL, however, has 6 parameters. Cosmological models having a larger number of parameters have more freedom to accommodate observations. Thus, for a fairer comparison of the various nonstandard models with the concordance $\CC$CDM we have to invoke a suitable statistical procedure that penalizes the presence of extra parameters. As in previous chapters, we make use of the Akaike information criterion (AIC) and the Bayesian information criterion (BIC), which are extremely valuable tools for a fair statistical analysis of this kind. These criteria are defined in \eqref{eq:AICandBIC} \cite{Akaike1974,Schwarz,Burnham}. The larger are the differences $\Delta$AIC ($\Delta$BIC) with respect to the model that carries smaller value of AIC (BIC) the higher is the evidence against the model with larger value of  AIC (BIC) -- the $\CC$CDM in all the cases considered in Tables \ref{tableFit1MPLA}-\ref{tableFit3MPLA}.
The rule applied to our case is the following\,\cite{Akaike1974,Schwarz,Burnham}: for $\Delta$AIC and $\Delta$BIC in the range $6-10$ we can speak of ``strong evidence'' against the $\CC$CDM, and hence in favor of the given nonstandard model. Above 10, one speaks of ``very strong evidence''. Notice that the Bayes factor is $e^{\Delta {\rm BIC}/2}$, and hence near 150 in such case.

A glance at Tables \ref{tableFit2MPLA} and \ref{tableFit3MPLA} tells us that for the models $\phi$CDM, XCDM and RVM, the values of $\Delta$AIC and $\Delta$BIC are both above 10. The CPL parametrization has only one of the two increments above 10, but the lowest one is above 6, therefore it is still fairly (but not so strongly) favored as the others. We conclude from the AIC and BIC criteria that the models $\phi$CDM, XCDM and RVM  are definitely selected over the $\CC$CDM as to their ability to optimally fit the large set of cosmological SNIa+BAO+$H(z)$+LSS+CMB data used in our analysis.  Although the most conspicuous model of those analyzed here appears to be the RVM {(cf. Tables \ref{tableFit2MPLA} and \ref{tableFit3MPLA})}, the scalar field model $\phi$CDM with Peebles \& Ratra potential also receives a strong favorable verdict from the AIC and BIC criteria. Furthermore, the fact that the generic XCDM and CPL parametrizations are also capable of detecting significant signs of evolving DE suggests that such dynamical signature is sitting in the data and is not privative of a particular model, although the level of sensitivity does indeed vary from one model to the other.

To summarize, the current cosmological data disfavors the hypothesis $\CC=$const. in a rather significant way. The presence of DE dynamics is confirmed by all four parametrizations considered here and with a strength that ranges between strong and very strong evidence, according to the Akaike and Bayesian information criteria. Furthermore, three of these parametrizations are able to attest such evidence at $\sim4\sigma$ c.l., and two of them ($\phi$CDM and RVM) are actually more than parametrizations since they are associated to specific theoretical frameworks.  The four approaches resonate in harmony with the conclusion that the DE is decreasing with the expansion, and therefore that it behaves effectively as quintessence.

%%%%%%%%%%%%%%%%%%%%%%%%%%%%%%%%%%%%%%%%%%%%%%%%%%%%%%%%%%%%%%%%%
%%%%%%%%%%%%%%%%%%%%%%%%%%%%%%%%%%%%%%%%%%%%%%%%%%%%%%%%%%%%%%%%%
%%%%%%%%%%%%%%%%%%%%%%%%%%%%%%%%%%%%%%%%%%%%%%%%%%%%%%%%%%%%%%%%%

\section{Main bibliography of the chapter}

This chapter is based on the contents of the paper \cite{MPLAnostre}:
\vskip 0.5cm
\noindent
{\it Dynamical dark energy: scalar fields and running vacuum.}\newline
J. Sol\`a, A. G\'omez-Valent, and J. de Cruz P\'erez\newline
Mod. Phys. Lett. A{\bf32}, 1750054 (2017) ; arXiv:1610.08965

\thispagestyle{empty}
\null
\newpage

\chapter[More compelling signs of vacuum (and DE) dynamics]{More compelling signs of vacuum (and DE) dynamics}
\label{chap:PRDbased}

Here we continue our study on the dynamical vacuum models, this time also including two of pure phenomenological nature. Apart from the latter, we analyze the same RVM studied in the previous chapter. It is similar (but not equal to) the A1-type model, since the vacuum energy density varies due to an interaction with dark matter only. Here baryons and radiation are conserved and, therefore, their energy densities are ruled by the standard $\Lambda$CDM laws. Again, we deem convenient to compare the capability of fitting the data of these DVM's with the one offered by the XCDM and CPL dark energy parametrizations, and the more elaborated $\phi$CDM Peebles \& Ratra model.   

The guidelines of the chapter are as follows. In Sect.\,\ref{sect:DVMs} we describe the different  dynamical vacuum models (DVM's) under consideration. In Sect.\,\ref{sect:Fit} we report on the main fitting results and the set of cosmological data used, on distant type Ia supernovae (SNIa), baryonic acoustic oscillations (BAO's), the Hubble parameter values at different redshifts, the LSS data, and the cosmic microwave background (CMB) distance priors from Planck 2015. In Sect.\,\ref{sect:perturbationsDVM} we discuss aspects of structure formation with dynamical vacuum. The numerical analysis of the DVM's and a comparison with the standard XCDM and CPL parametrizations is the object of Sect.\,\ref{sect:numerical results}. An ample discussion of the results along with a reanalysis under different conditions, including a comparison with the traditional $\phi$CDM model, is developed in Sect.\,\ref{sect:discussion}. Finally, in Sect.\,\ref{sect:conclusions} we present our conclusions. Appendix \ref{chap:App5}, on the cosmological observables and statistical analysis, complete the study carried out in this extensive chapter.

\section{Dynamical vacuum models}\label{sect:DVMs}
The gravitational field equations with cosmological term $\CC$ are
\begin{equation}\label{eq:EEs1}
G_{\mu\nu}-\CC\,g_{\mu\nu}=8\pi G\ {T}_{\mu\nu}\,,
\end{equation}
where $G_{\mu\nu}=R_{\mu \nu }-\frac{1}{2}g_{\mu \nu }R$ is the Einstein tensor.
Defining the vacuum energy density (in natural units) as $\rL=\CC/(8\pi G)$, the full energy-momentum tensor involving the effect
of both matter and the vacuum energy density, reads
$\tilde{T}_{\mu\nu}\equiv T_{\mu\nu}+g_{\mu\nu}\,\rL $.
The original field equations can then be brought to the same form as in the case without $\CC$-term:
%\begin{equation}\label{eq:EEs2}
$G_{\mu\nu}=8\pi G\ \tilde{T}_{\mu\nu}$\,.
%\end{equation}
Notice from the structure of $\tilde{T}_{\mu\nu}$ that the vacuum is dealt with as a perfect fluid, with  EoS  $p_{\CC}=-\rho_{\CC}$. When the matter can also be treated as an ideal fluid and is distributed homogeneously and isotropically, as postulated by the Cosmological Principle, we can write \begin{equation}
\tilde{T}_{\mu\nu}= (\rho_{\Lambda}-p_{m})\,g_{\mu\nu}+(\rho_{m}+p_{m})U_{\mu}U_{\nu}\,,
\label{TmunuFullideal}
\end{equation}
where $U_{\mu}$ is the $4$-velocity of the cosmic fluid, $\rho_m$ is the proper energy density of matter and $p_m$ its isotropic pressure.
We assume the standard cosmological framework grounded on the FLRW metric with flat three-dimensional slices: $ds^2=dt^2-a^2(t)\,d{\bf x}^2$, where $t$ is the cosmic time and $a(t)$ the scale factor. However, we admit that matter can be in interaction with vacuum, which is tantamount to saying that $\rL=\rL(\zeta)$ is a function of some cosmic variable evolving with time, $\zeta=\zeta(t)$. While this, of course, implies that $\dot{\rho}_{\CC}\equiv d\rL/dt\neq 0$ we assume that $\dot{G}=0$ in our study. Such dynamics of vacuum is compatible with the Bianchi identity (see below) provided there is some energy exchange between vacuum and matter. In other words, matter cannot be strictly conserved in these circumstances. The Friedmann and acceleration equations for the present Universe are nonetheless formally identical to the standard $\CC$CDM case:
\begin{eqnarray}
&&3H^2=8\pi\,G\,(\rho_m+\rho_r+\rho_\Lambda(\zeta))\label{eq:FriedmannEq}\\
&&3H^2+2\dot{H}=-8\pi\,G\,(p_r-\rho_\Lambda(\zeta))\label{eq:PressureEq}\,.
\end{eqnarray}
Here $H=\dot{a}/a$ is the usual Hubble rate,  $\rho_m=\rho_b+\rho_{dm}$ involves the pressureless contributions from baryons and cold dark matter (DM) in the current epoch, and $\rho_r$ is  the radiation density (with the usual EoS $p_r=\rho_r/3$). We emphasize once more that in the above equations we stick to the EoS, $p_{\CC}=-\rho_{\CC}$, although the vacuum is dynamical, $\rL(t)=\rL(\zeta(t))$, and its evolution is tied to the cosmic expansion. In all of the dynamical vacuum models (DVM's) being considered here, the cosmic variable $\zeta$ is either the scale factor or can be expressed analytically in terms of it, $\zeta=\zeta(a)$, or equivalently in terms of the cosmological redshift, $z=a^{-1}-1$, where we adopt the normalization $a=1$ at present.

From the basic pair of Friedmann and acceleration equations \eqref{eq:FriedmannEq}-\eqref{eq:PressureEq}, a first integral of the system can be derived, namely
\begin{equation}\label{BianchiGeneral}
\sum_{N=dm,b,r,\CC} \left[\dot{\rho}_N+3\,H(\rho_N+p_N)\right]=0\,.
\end{equation}
Of course, the last term being summed over (for the $\CC$ component) is zero owing to the vacuum EoS.
Such first integral ensues also from the divergenceless property of the full energy-momentum tensor $\tilde{T}_{\mu\nu}$, see Eq.\eqref{TmunuFullideal}, in the FLRW metric, i.e. $\nabla^{\mu}\tilde{T}_{\mu\nu}=0$. The last property is a consequence of the Bianchi identity satisfied by the Einstein tensor, $\nabla^{\mu} G_{\mu\nu}=0$, and the assumed constancy of the Newtonian coupling $G$. It reflects the local conservation law of the compound system formed by matter and vacuum, and the consequent nonconservation of each of these components when taken separately. For our purposes it will be more convenient to reexpress the combined conservation law in a way that reflects more explicitly the interaction between the particular components of the cosmic fluid. Thus, by introducing the vacuum-matter interaction source, $Q$, and using the vacuum EoS we can conveniently split (\ref{BianchiGeneral}) into two coupled equations:
\begin{eqnarray}
&&\dot{\rho}_{dm} + 3H\rho_{dm}+\dot{\rho}_b + 3H\rho_b+\dot{\rho}_r + 4H\rho_r = Q\,,\label{eq:GeneralCL1}\\
&&\dot{\rho}_\CC = -Q\,.\label{eq:GeneralCL2}
\end{eqnarray}
Rephrased in this form, it will be easier to sort out the types of DVM's we will be dealing with in terms of the different proposed forms for $Q$.
For example, in the $\CC$CDM case the vacuum energy density is $\rL=$const. and therefore $Q=0$. The concordance model assumes also that matter and radiation are self-conserved after equality. It also assumes that baryons and CDM are separately conserved. Therefore their respective energy densities satisfy $\dot{\rho}_b + 3H\rho_b=0$, $\dot{\rho}_r + 4H\rho_r=0$ and $\dot{\rho}_{dm} + 3H\rho_{dm}=0$. Nevertheless, it is obvious that in the presence of vacuum dynamics at least one of these equations cannot hold. Following our definite purpose to remain as close as possible to the $\CC$CDM, we would like to leave the most sensitive and accessible components of the cosmic fluid completely unaltered by the vacuum dynamics. Thus, in the considered DVM's, we will assume that the first two of the above conservation equations still hold good but that the last does not, and hence that the vacuum exchanges energy only with DM. The dilution laws for baryons and radiation as a function of the scale factor therefore take on the conventional $\CC$CDM forms:
\begin{equation}\label{eq:BaryonsRadiation}
\rho_b(a) = \rho_{b0}\,a^{-3}, \ \ \ \ \ \ \ \rho_r(a)=\rho_{r0}\,a^{-4}\,,
\end{equation}
where $\rho_{b0}$ and $\rho_{r0}$ are the corresponding current values. In contrast, the density of DM is tied to the dynamics of the vacuum through Eqs.\,(\ref{eq:GeneralCL1})-(\ref{eq:GeneralCL2}), which now become simplified:
\begin{equation}\label{eq:Qequations}
\dot{\rho}_{dm}+3H\rho_{dm}=Q\,,\ \ \ \ \ \ \ \dot\rho_{\CC}=-{Q}\,,
\end{equation}
 after the conservation laws (\ref{eq:BaryonsRadiation}) have been taken into account. The solution of these equations will depend on the particular form assumed for $Q$, which determines the leakage rate of vacuum energy into dark matter or vice versa. Such leakage must certainly be much smaller than the standard dilution rate of nonrelativistic matter associated to the cosmic expansion (i.e. much smaller than $\sim a^{-3}$), as otherwise these anomalous effects would be too sharp at the present time. Therefore, we must have   $0<|Q|\ll\dot{\rho}_m$. The various DVM's will be characterized by different functions $Q_i$  ($i=1,2,..$) that are proportional to a small dimensionless coupling, $|\nu_i|\ll1$. In some cases the interaction source $Q_i$ will be merely phenomenological, but in one of the cases (the running vacuum model) we  have a more theoretical motivation.

\subsection{The running vacuum model}
\label{sect:RVMPRD}

One of the DVM's under study is the so-called running vacuum model, which can be motivated in the context of QFT in curved spacetime (cf. \,\cite{SolaReview2013,SolGom2015} and references therein). Details on these model have been provided in the previous chapter, in Sect. \ref{sect:RVM}. In the RVM case, the source function $Q$ in \eqref{eq:Qequations} is not just put by hand (as in the case of the \textit{ad hoc} DVM's we will introduce in a moment). It is a calculable expression from \eqref{eq:RVMvacuumdadensity}, together with Friedmann's equation (\ref{eq:FriedmannEq}) and the generalized local conservation law \eqref{BianchiGeneral}, realized in the form (\ref{eq:GeneralCL1})-(\ref{eq:Qequations}). We find:
\begin{equation}\label{eq:QRVM}
{\rm RVM:}\qquad Q=-\dot{\rho}_{\Lambda}=\nu\,H(3\rho_{m}+4\rho_r)\,,
\end{equation}
where we recall that $\rho_{m}=\rho_{dm}+\rho_{b}$, and that $\rho_b$ and $\rho_r$ are known functions of the scale factor -- see Eq.\,(\ref{eq:BaryonsRadiation}). The remaining densities, $\rho_{dm}$ and $\rL$, must be determined upon further solving the model explicitly, see the next subsection. If baryons and radiation would also possess a small interaction with vacuum, their densities in Eq.\,(\ref{eq:QRVM}) would not follow the standard conservation laws and the cosmological solutions would be those encountered in type-A1 model (see e.g. Chapter \ref{chap:Atype}).

\subsection{Other dynamical vacuum models}\label{sect:otherDVM}

Next we include in our study two \textit{ad hoc} DVM's in which the source function $Q$ is introduced by hand, i.e. without any special theoretical motivation. Two possible phenomenological ansatzs are the following:
\begin{eqnarray}\label{eq:PhenModelQdm}
{\rm Model\ \ }Q_{dm}: \phantom{XX}Q_{dm}&=&3\nu_{dm}H\rho_{dm}\\
{\rm Model\ \ }Q_{\CC}:\phantom{XXx}Q_{\CC}&=&3\nu_{\CC}H\rho_{\CC}\,.\label{eq:PhenModelQL}
\end{eqnarray}
Model $Q_{\CC}$  was previously addressed in \cite{Salvatelli2014}, but as we shall see we do not agree with their analysis, in accordance also with\,\cite{Murgia2016}. Model $Q_{dm}$ was recently studied in\,\cite{Li2016}; it is closer to the RVM than $Q_{\CC}$, but certainly not identical, confer equations\,(\ref{eq:QRVM}) and (\ref{eq:PhenModelQdm}). There are many more choices for $Q$, see e.g. \cite{Bolotin2015} and \cite{Costa2017}, but it will suffice to focus on the RVM and on the two additional variants (\ref{eq:PhenModelQdm})-(\ref{eq:PhenModelQL}) to assess the possible impact of the DVM's in the light of the modern observational data.

The dimensionless parameters $\nu_{i}=(\nu,\nu_{dm},\nu_\CC)$ for each model  (RVM, $Q_{dm}$, $Q_\CC$) determine the strength of the dark-sector interaction in the sources $Q_i$ and enable the evolution of the vacuum energy density. In all cases we have a similar structure $Q_i\propto\nu_i H\,{\cal Q}_i$, but the density function ${\cal Q}_i$ varies from one model to the other, as it is plain from the above formulae. For $\nu_{i}>0$ the vacuum decays into DM (which is thermodynamically favorable) whereas for $\nu_{i}<0$ is the other way around. This will be an important argument (see Sect.\,\ref{subsect:SLT}) to judge the viability of these models, as only the first situation is compatible with the second law of thermodynamics (SLT).

\subsection{Solving explicitly the dynamical vacuum models}\label{sect:solvingDVM}

The matter and vacuum energy densities of the DVM's can be computed straightforwardly upon solving the coupled system of differential equations (\ref{eq:Qequations}), given the explicit forms (\ref{eq:QRVM})-(\ref{eq:PhenModelQL}) for the interacting source in each case and keeping in mind that, in the current framework, the baryon ($\rho_b$) and radiation ($\rho_r$) parts are separately conserved. After some calculations the equations for the DM energy densities $\rho_{dm}$ for each model (RVM, $Q_{dm}$,$Q_{\CC}$) can be explicitly solved in terms of the scale factor. Below we quote the final results for each case:
\begin{eqnarray}
{\rm\mathbf{RVM}}:\label{eq:rhoRVM}\\
\rho_{dm}(a) &=& \rho_{dm0}\,a^{-3(1-\nu)} + \rho_{b0}\left(a^{-3(1-\nu)} - a^{-3}\right)+\frac{4\nu}{1 + 3\nu}\,{\rho_{r0}}\,\left(a^{-3(1-\nu)} - a^{-4}\right)\nonumber 
\\
{\rm\mathbf{Q_{dm}}}:\label{eq:rhoQdm} \\
\rho_{dm}(a) &=& \rho_{dm0}\,a^{-3(1-\nu_{dm})}
 \nonumber
\\
{\rm\mathbf{Q_{\CC}}}:\label{eq:rhoQL}\\
\rho_{dm}(a) &=&\rho_{dm0}\,a^{-3} + \frac{\nu_\CC}{1-\nu_\CC}\rho_{\Lambda 0}\left(a^{-3\nu_\Lambda}-a^{-3}\right)\,. \nonumber
\end{eqnarray}
In solving the differential equations (\ref{eq:Qequations}) we have traded the cosmic time variable for the scale factor using the chain rule $d/dt=aH d/da$. The corresponding vacuum energy densities can also be solved in the same variable, and yield:

\begin{table}
\begin{center}
\begin{scriptsize}
\resizebox{1\textwidth}{!}{
\begin{tabular}{| c | c |c | c | c | c | c | c | c | c | c|}
\multicolumn{1}{c}{Model} &  \multicolumn{1}{c}{$h$} &  \multicolumn{1}{c}{$\omega_b= \Omega_b h^2$} & \multicolumn{1}{c}{{\small$n_s$}}  &  \multicolumn{1}{c}{$\Omega_m$}&  \multicolumn{1}{c}{{\small$\nu_i$}}  & \multicolumn{1}{c}{$w_0$} & \multicolumn{1}{c}{$w_1$} &
\multicolumn{1}{c}{$\chi^2_{\rm min}/dof$} & \multicolumn{1}{c}{$\Delta{\rm AIC}$} & \multicolumn{1}{c}{$\Delta{\rm BIC}$}\vspace{0.5mm}
\\\hline
{\small $\CC$CDM} & $0.692\pm 0.004$ & $0.02253\pm 0.00013$ &$0.976\pm 0.004$& $0.296\pm 0.004$ & - & -1 & - & 84.88/85 & - & -\\
\hline
XCDM  &  $0.672\pm 0.007$& $0.02262\pm 0.00014 $&$0.976\pm0.004$& $0.311\pm 0.007$& - & $-0.923\pm0.023$ & - & 74.08/84 & 8.55 & 6.31 \\
\hline
CPL  &  $0.673\pm 0.009$& $0.02263\pm 0.00014 $&$0.976\pm0.004$& $0.310\pm 0.009$& - & $-0.944\pm0.089$ & $0.063\pm0.259$ & 74.03/83 & 6.30 & 1.87 \\
\hline
RVM  & $0.677\pm 0.005$& $0.02231\pm 0.00014$&$0.965\pm 0.004$& $0.303\pm 0.005$ & $0.00158\pm 0.00042$ & -1 & - & 69.72/84 & 12.91 & 10.67 \\
\hline
$Q_{dm}$ &  $0.678\pm 0.005$& $0.02230\pm 0.00015 $&$0.965\pm0.004$& $0.302\pm 0.005 $ & $0.00216\pm 0.00060 $ & -1 & - &  70.50/84  & 12.13 & 9.89 \\
\hline
$Q_\CC$  &  $0.691\pm 0.004$& $0.02230\pm 0.00016 $&$0.966\pm0.005$& $0.298\pm 0.005$ & $0.00601\pm 0.00253$ & -1 & - &  79.22/84 & 3.41 & 1.17 \\
\hline \end{tabular}
}
\end{scriptsize}
\end{center}
\caption[Best-fit values for the various models analyzed in Chapter \ref{chap:PRDbased}. Main table]{\scriptsize Best-fit values for the $\CC$CDM, XCDM, CPL and the three dynamical vacuum models (DVM's), including their statistical  significance ($\chi^2$-test and Akaike and Bayesian information criteria, AIC and BIC). The  $\Delta$AIC and $\Delta$BIC increments clearly favor the dynamical DE options. In particular, the RVM and $Q_{dm}$ are strongly favored ($\sim 4\sigma$ c.l.). Our fit is grounded on a rich and fully updated SNIa+BAO+$H(z)$+LSS+CMB data set, see data sources DS1)-DS6) in the text for details and references. The specific fitting parameters for each DVM are $\nu_{i}=\nu $ (RVM), $\nu_{dm}$($Q_{dm}$) and $\nu_{\CC}$($Q_{\CC}$), whilst for the XCDM and CPL are the EoS parameters $w_0$ and the pair ($w_0$,$w_1$), respectively. For the vacuum models, including the $\CC$CDM, we have $w_0=-1$ and $w_1=0$. The remaining parameters are standard ($h,\omega_b,n_s,\Omega_m$).  The number of degrees of freedom ($dof$) is equal to the number of data points minus the number of fitting parameters ($4$ for the $\CC$CDM, $5$ for the DVM's and the XCDM, and $6$ for the CPL parametrization). The parameter M in the SNIa sector\,\cite{BetouleJLA} was dealt with as a nuisance parameter and has been marginalized over analytically (see Appendix \ref{chap:App5}).}
\label{tableFit1PRD}
\end{table}
\begin{eqnarray}
{\rm \textbf{RVM}}:\label{eq:rhoVRVM}\\
\rho_\CC(a) &=& \rho_{\Lambda 0} + \frac{\nu\,\rho_{m0}}{1-\nu}\left(a^{-3(1-\nu)}-1\right)+ \frac{\nu{\rho_{r0}}}{1-\nu}\left(\frac{1-\nu}{1+3\nu}a^{-4} + \frac{4\nu}{1+3\nu}a^{-3(1-\nu)} -1\right)  \nonumber\\ \nonumber\\
{\rm \ \mathbf{Q_{dm}}}:\label{eq:rhoVQdm}\\
\rho_\CC(a)&=& \rho_{\Lambda 0} + \frac{\nu_{dm}\,\rho_{dm0}}{1-\nu_{dm}}\,\left(a^{-3(1-\nu_{dm})}-1\right)\nonumber  \\
{\rm \ \mathbf{Q_{\CC}}}:\label{eq:rhoVQL}\\
\rho_\CC(a) &=&\rho_{\Lambda 0}\,{a^{-3\nu_\CC}}\,.\nonumber
\end{eqnarray}
One can easily check that for $a=1$ (i.e. at the present epoch) all of the energy densities (\ref{eq:rhoRVM})-(\ref{eq:rhoVQL}) recover their respective current values $\rho_{N0}$ ($N=dm,b,r,\CC$). In addition,
for $\nu_{i}\to 0$ we retrieve for the three DM densities the usual $\CC$CDM expression $\rho_{dm}(a)=\rho_{dm 0}a^{-3}$, and the corresponding vacuum energy densities $\rL(a)$ boil down to the constant value $\rho_{\Lambda 0}$ in that limit. The normalized Hubble rate $E\equiv H/H_0$ ($H_0$ being the current value) for each model can be easily obtained by plugging the above formulas, along with the radiation and baryon energy densities, into the Friedmann's equation (\ref{eq:FriedmannEq}). We find:
\begin{eqnarray}
{\rm \textbf{RVM}}:\label{HRVM}\\
E^2(a) &=& 1 + \frac{\Omega_m}{1-\nu}\left(a^{-3(1-\nu)}-1\right)+ \frac{\Omega_r}{1-\nu}\left(\frac{1-\nu}{1+3\nu}a^{-4} + \frac{4\nu}{1+3\nu}a^{-3(1-\nu)} -1\right) \nonumber\\ \nonumber\\
{\rm \ \mathbf{Q_{dm}}}:\label{HQdm}\\
E^2(a)&=& 1 + \Omega_b\left(a^{-3}-1\right) + \frac{\Omega_{dm}}{1-\nu_{dm}}\left(a^{-3(1-\nu_{dm})}-1\right)+ \Omega_r\left(a^{-4}-1\right)  \nonumber\\ \nonumber\\
{\rm \ \mathbf{Q_{\CC}}}:\label{HQL}\\
E^2(a) &=& \frac{a^{-3\nu_\CC}-\nu_\CC{a^{-3}}}{1-\nu_\CC} + \frac{\Omega_m}{1-\nu_\CC}\left(a^{-3}-a^{-3\nu_\CC}\right)+ \Omega_r\left(a^{-4} + \frac{\nu_\CC}{1-\nu_\CC}a^{-3} - \frac{a^{-3\nu_\CC}}{1-\nu_\CC}\right)\nonumber\,.
\end{eqnarray}
In the above expressions, we have used the cosmological parameters $\Omega_N=\rho_{N0}/\rco$ for each fluid component ($N=dm,b,r,\CC$), and defined $\Omega_m=\Omega_{dm}+\Omega_b$. Altogether, they satisfy the sum rule $\sum_N\Omega_N=1$. The normalization condition $E(1)=1$  in these formulas is apparent, meaning that the Hubble function for each model reduces to $H_0$ at present, as they should; and, of course, for $\nu_i\to 0$ we recover the $\CC$CDM form for $H$, as should be expected.

%%%%%%%%%%%%%%%%%%%%%%%%%%%%%%%%%%%%%%%%%%%%%%%%%%%%%%%%%%%%%%%%%
%%%%%%%%%%%%%%%%%%%%%%%%%%%%%%%%%%%%%%%%%%%%%%%%%%%%%%%%%%%%%%%%%
%%%%%%%%%%%%%%%%%%%%%%%%%%%%%%%%%%%%%%%%%%%%%%%%%%%%%%%%%%%%%%%%%

\begin{table}
\begin{center}
\begin{scriptsize}
\resizebox{1\textwidth}{!}{
\begin{tabular}{| c | c |c | c | c | c | c | c | c | c | c|}
\multicolumn{1}{c}{Model} &  \multicolumn{1}{c}{$h$} &  \multicolumn{1}{c}{$\omega_b= \Omega_b h^2$} & \multicolumn{1}{c}{{\small$n_s$}}  &  \multicolumn{1}{c}{$\Omega_m$}&  \multicolumn{1}{c}{{\small$\nu_i$}}  & \multicolumn{1}{c}{$w_0$} & \multicolumn{1}{c}{$w_1$} &
\multicolumn{1}{c}{$\chi^2_{\rm min}/dof$} & \multicolumn{1}{c}{$\Delta{\rm AIC}$} & \multicolumn{1}{c}{$\Delta{\rm BIC}$}\vspace{0.5mm}
\\\hline
{\small $\CC$CDM} & $0.691\pm 0.004$ & $0.02251\pm 0.00013$ &$0.973\pm 0.004$& $0.297\pm 0.005$ & - & -1 & - & 80.19/78 & - & -\\
\hline
XCDM  &  $0.673\pm 0.007$& $0.02261\pm 0.00014 $&$0.976\pm0.004$& $0.311\pm 0.007$& - & $-0.929\pm0.023$ & - &  72.00/77 & 5.92 & 3.78 \\
\hline
CPL  &  $0.674\pm 0.009$& $0.02261\pm 0.00015 $&$0.976\pm0.004$& $0.310\pm 0.009$& - & $-0.938\pm0.087$ & $0.026\pm0.258$ & 71.99/76 & 3.60 & -0.61 \\
\hline
RVM  & $0.677\pm 0.005$& $0.02232\pm 0.00015$&$0.965\pm 0.004$& $0.302\pm 0.005$ & $0.00159\pm 0.00048$ & -1 & - &  67.85/77 & 10.07 & 7.93 \\
\hline
$Q_{dm}$ &  $0.678\pm 0.005$& $0.02230\pm 0.00015 $&$0.965\pm0.004$& $0.302\pm 0.005 $ & $0.00215\pm 0.00066 $ & -1 & - &  68.59/77  & 9.33 & 7.19 \\
\hline
$Q_\CC$  &  $0.690\pm 0.004$& $0.02230\pm 0.00017 $&$0.966\pm0.005$& $0.299\pm 0.005$ & $0.00568\pm 0.00259$ & -1 & - &  75.40/77 & 2.52 & 0.38 \\
\hline
 \end{tabular}
 }
\end{scriptsize}
\end{center}
\caption[Same as in Table \ref{tableFit1PRD}, but excluding from the analysis the BAO and LSS data from WiggleZ]{\scriptsize Same as in Table \ref{tableFit1PRD}, but excluding from our analysis the BAO and LSS data from WiggleZ, see point DS5) in the text.}
\label{tableFit2PRD}
\end{table}
From the structure of equations (\ref{eq:rhoVRVM}) and (\ref{HRVM}) we can readily see that the vacuum energy-density for the RVM can be fully written as a function of a cosmic variable $\zeta$, which can be chosen to be not only the scale factor but the full Hubble function $\zeta=H$. The result is, of course, Eq.\,(\ref{eq:RVMvacuumdadensity}).  In contrast, for the $Q_{dm}$ and $Q_{\CC}$ models this is not possible, as it is clear on comparing equations (\ref{eq:rhoVQdm})-(\ref{eq:rhoVQL}) and the corresponding ones (\ref{HQdm})-(\ref{HQL}). For these models $\rL$ can only be written as a function of the scale factor. In fact, the RVM happens to have the greatest level of symmetry since its origin is an RG equation in $H$ whose solution is precisely (\ref{eq:RVMvacuumdadensity}).

%%%%%%%%%%%%%%%%%%%%%%%%%%%%%%%%%%%%%%%%%%%%%%%%%%%%%%%%%%%%%%%%%
%%%%%%%%%%%%%%%%%%%%%%%%%%%%%%%%%%%%%%%%%%%%%%%%%%%%%%%%%%%%%%%%%
%%%%%%%%%%%%%%%%%%%%%%%%%%%%%%%%%%%%%%%%%%%%%%%%%%%%%%%%%%%%%%%%%

\subsection{XCDM and CPL parametrizations}\label{sect:XCDMandCPL}

It will be convenient to fit also the data through the simple XCDM parametrization of the dynamical DE, first introduced by Turner and White\,\cite{XCDM} shortly after the discovery of the cosmic acceleration with type Ia distant supernovae. The basic background formulas are shown in Sect. \ref{sect:XCDMand CPL}, together with those of the CPL parametrization \cite{CPL1,CPL2}. Both parametrizations can be conceived as a kind of baseline frameworks to be referred to in the study of dynamical DE. They can be used as fiducial models to which we can compare other, more sophisticated, models for the dynamical DE, such as the DVM's under study. The mimicking of the vacuum dynamics through these parametrizations need not be perfect through these parametrizations, though, since the DVM's interact with matter whereas the XCDM and CPL do not involve any interaction. Still, since the vacuum dynamics inherent in the DVM's is mild enough, we expect that a significant part of the effects departing from the $\CC$CDM should be captured by these parametrizations, either in the form of effective quintessence behavior ($w\gtrsim -1$) or effective phantom behavior ($w\lesssim-1$).
The fact that the CPL comprises two parameters $(w_0,w_1)$ rather than just $w_0$ will, however, be in detriment of the effectiveness of the fit versus that of the XCDM.
The XCDM is in fact more appropriate for a fairer comparison with the DVM's, which also have one single vacuum parameter, $\nu_i$. We present the main fitting results with these parametrizations, along with the other models, in Table \ref{tableFit1PRD}. A more detailed discussion will be given later on in Sections \ref{sect:numerical results} and \ref{sect:discussion}.

\section{Fitting the vacuum models to the data}\label{sect:Fit}

In this chapter, we fit the $\CC$CDM, XCDM, CPL and the three DVM's to the cosmological data from type Ia supernovae, BAO's, the values of the Hubble parameter at various redshifts, $H(z_i)$, the LSS formation data embodied in the quantity $f(z_i)\sigma_8(z_i)$ and, finally, the CMB distance priors from Planck 2015. We denote this cosmological data set by SNIa+BAO+$H(z)$+LSS+CMB.
%In all cases we include the known correlation matrices.
A guide to the presentation of our results is the following. The various fitting analyses and contour plots under different conditions (to be discussed in detail in the next sections) are displayed in twelve fitting tables, Tables \ref{tableFit1PRD}-\ref{tableFit2PRD}, \ref{tableFit3PRD}-\ref{tableFitPhiCDM} and \ref{tableFitXIII}, and in ten figures, Figs.\,\ref{fig:PRD1}-\ref{fig:PRD10}. Table \ref{compilationLSSprd} contains the LSS data points used in the current study. The main results of our analysis are those recorded in Table \ref{tableFit1PRD} and Figs.\,\ref{fig:PRD1}-\ref{fig:fsigma8PRD}. The elliptical shapes in the contour plots have been obtained according to the standard statistical technique relying on the Fisher matrix formalism\,\cite{BookAmendolaTsujikawa}. This is perfectly legitimate in this case, since the deviation of the posterior distribution from an ideal Gaussian one is found to be negligible, as we have checked -- similarly as we did in Chapter \ref{chap:AandGRevisited} for other models. Details of the $\chi^2$ function to be minimized, as well as of  the covariance matrices for the most significant data sets are specified in the Appendix \ref{chap:App5}. For additional details the reader is referred to specific references of our bibliography, which we deem is quite generous.
A guide to the presentation of our results is the following. The various fitting analyses and contour plots under different conditions (to be discussed in detail in the next sections) are displayed in twelve fitting tables, Tables \ref{tableFit1PRD}-\ref{tableFit2PRD}, \ref{tableFit3PRD}-\ref{tableFitPhiCDM} and \ref{tableFitXIII}, and in ten figures, Figs.\,\ref{fig:PRD1}-\ref{fig:PRD10}. Table \ref{compilationLSSprd} contains the LSS data points used in the current study. The main results of our analysis are those recorded in Table \ref{tableFit1PRD} and Figs.\,\ref{fig:PRD1}-\ref{fig:fsigma8PRD}. The elliptical shapes in the contour plots have been obtained according to the standard statistical technique relying on the Fisher matrix formalism\,\cite{BookAmendolaTsujikawa}. This is perfectly legitimate in this case, since the deviation of the posterior distribution from an ideal Gaussian one is found to be negligible, as we have checked -- similarly as we did in Chapter \ref{chap:AandGRevisited} for other models. Details of the $\chi^2$ function to be minimized, as well as of  the covariance matrices for the most significant data sets are specified in the Appendix \ref{chap:App5}. For additional details the reader is referred to specific references of our bibliography, which we deem is quite generous.

The remaining tables and figures contain complementary information, which can be very helpful to gather a more detailed picture of our rather comprehensive study. In particular, Figs.\,\ref{fig:PRD5}-\ref{fig:PRD10} are quite revealing of different aspects of the dynamical dark energy signal, together with the complementary tables, which are in support of the figures.
A summarized road map to the content of the fitting tables is the following: Table \ref{tableFit2PRD} explores a possible effect of correlations in data sets; Tables \ref{tableFit3PRD}-\ref{tableFitX1} examine the influence of refitting the data by excluding a single one of the three basic observables CMB, LSS or BAO, respectively; Table \ref{tableFit6PRD} shows the results when we use only the fitting data employed by the Planck 2015 collaboration, where no dynamical DE signal was detected; Table \ref{tableFitX5} explores the difference between including or not including the data on the three-point correlation function, i.e. the bispectrum, in addition to the usual spectrum; Table \ref{tableFit8PRD} tests the effect on our results when we replace some of the BAO data in Table \ref{tableFit1PRD} with different  BAO data existing in the recent literature; Table \ref{tableFit9PRD} allows to compare the effect of using a recent (but earlier) release of the BOSS data on BAO and LSS (including the bispectrum) carrying different errors as compared to the most recent BOSS data that we used in Table \ref{tableFit1PRD}; Table \ref{tableFitPhiCDM} reports on the fitting results to the same cosmological data when we use the original Peebles \& Ratra model as a prototype $\phi$CDM model to explore the dynamical DE; and, finally, Tables \ref{tableSound} and \ref{tableFitXIII} check the impact on our results upon using alternative fitting formulas existing in the literature for the sound horizon at the redshift of the drag epoch. We will discuss all these results in detail in the corresponding sections and in Appendix \ref{chap:App5}. Below we start with a detailed description of the data used.

%
%%%%%%%%%%%%%%%%%%%%%%%%%%%%%%%%%%%%%%%%%%%%%%%%%%%%%%%%%%%%%%%%%%%%%%%%%%%%%%
%
\begin{table}[t]
\begin{center}
%\begin{scriptsize}
\begin{tabular}{| c | c |c | c |}
\multicolumn{1}{c}{Survey} &  \multicolumn{1}{c}{$z$} &  \multicolumn{1}{c}{$f(z)\sigma_8(z)$} & \multicolumn{1}{c}{{\small References}}
\\\hline
6dFGS & $0.067$ & $0.423\pm 0.055$ & \cite{Beutler2012}
\\\hline
SDSS-DR7 & $0.10$ & $0.37\pm 0.13$ & \cite{Feix}
\\\hline
GAMA & $0.18$ & $0.29\pm 0.10$ & \cite{Simpson}
\\ \cline{2-4}& $0.38$ & $0.44\pm0.06$ & \cite{Blake2013}
\\\hline
DR12 BOSS & $0.32$ & $0.427\pm 0.056$  & \cite{GilMarin2}\\ \cline{2-3}
 & $0.57$ & $0.426\pm 0.029$ & \\\hline
 WiggleZ & $0.22$ & $0.42\pm 0.07$ & \cite{Blake2011LSS} \tabularnewline
\cline{2-3} & $0.41$ & $0.45\pm0.04$ & \tabularnewline
\cline{2-3} & $0.60$ & $0.43\pm0.04$ & \tabularnewline
\cline{2-3} & $0.78$ & $0.38\pm0.04$ &
\\\hline
2MTF & $0.02$ & $0.34\pm 0.04$ & \cite{Springob}
\\\hline
VIPERS & $0.7$ & $0.38^{+0.06}_{-0.07}$ & \cite{Granett}
\\\hline
VVDS & $0.77$ & $0.49\pm0.18$ & \cite{Guzzo2008},\cite{Song09}
\\\hline
 \end{tabular}
% \end{scriptsize}
\caption[Published values of $f(z)\sigma_8(z)$]{{\scriptsize Published values of $f(z)\sigma_8(z)$, referred to in the text as the LSS data, see the quoted references and text in point DS5). The only difference with respect to Table \ref{compilationLSS} is in the uncertainties of the the DR12 BOSS data points. Now we use the updated (and finally published) values of Gil-Mar\'in et al. \cite{GilMarin2}, instead of the ones found in the previous version \cite{GilMarin2OLD}.}\label{compilationLSSprd}}
\end{center}
\end{table}
%%%%%%%%%%%%%%%%%%%%%%%%%%%%%%%%%%%%%%%%%%%%%%%%%%%%%%%%%%%%%%%%%%%%%%%%%%%%%%
%

\subsection{Data sets}\label{sect:DataSets}

The $89$ data points used in our fit stem from six data sources DS1-DS6 that are associated to the chain SNIa+BAO+$H(z)$+LSS+CMB of cosmological observations,  keeping nonetheless in mind that two of the data sources described below stem from BAO's of different kinds (isotropic and anisotropic). Notice that the selected BAO+LSS points actually encode the information of hundreds of thousands of galaxies corresponding to certain effective redshifts.
Specifically, the data sets used in our analysis are almost the same as the ones used in Chapter \ref{chap:AandGRevisited}, but we consider useful for the reader to make a quick review in the following lines:\\

DS1) Data on distant type Ia supernovae. We take the SNIa data points from the SDSS-II/SNLS3 Joint Light-curve Analysis (JLA) \cite{BetouleJLA}. We have used the $31$ binned distance modulus fitted to the JLA sample and the compressed form of the likelihood with the corresponding $31\times31$ covariance matrix.\\

DS2) Data on isotropic baryon acoustic oscillations.  We use 5 data points on the isotropic BAO at the following redshifts:  $z=0.106$\,\cite{Beutler2011}, obtained from the $D_V(z)$ estimator; and $z=0.15$\, \cite{Ross} and $z_i=0.44, 0.6, 0.73$ \cite{Kazin2014} from the alternative estimator $D_V(z)/r_s(z_d)$  (with the correlations between the last 3 points). See Appendix \ref{chap:App5} for more details on these estimators.\\

DS3) Data on anisotropic baryon acoustic oscillations. We use 6 data points on anisotropic BAO estimators: 4 of them on  $D_A(z_i)/r_s(z_d)$ and $H(z_i)r_s(z_d)$ at $z_i=0.32, 0.57$, for the LOWZ and CMASS samples, respectively. These data are taken from \cite{GilMarin2}, based on the Redshift-Space Distortions (RSD) measurements of the power spectrum combined with the bispectrum, and the BAO post-reconstruction analysis of the power spectrum (cf. Table 5 of that reference), including the correlations among these data encoded in the provided covariance matrices. We also use  2 data points involving the observables  $D_A(z_i)/r_s(z_d)$ and $D_H(z_i)/r_s(z_d)$ at $z=2.34$, from the combined LyaF analysis \cite{Delubac2015}. The correlation coefficient among these 2 points is taken from \cite{Aubourg2015} (cf. Table II of that reference). We have also taken due account of the correlations among the  BAO data and the corresponding $f\sigma_8$ data of \cite{GilMarin2} -- see DS5) below and Table \ref{compilationLSSprd} with the associated references. We refer the reader to Appendix \ref{chap:App5} and references therein for further technical details on the (isotropic and anisotropic) BAO observables.\\

DS4) Data on the Hubble parameter at different redshift points. We use $30$ data points on $H(z_i)$ at different redshifts, listed in Table \ref{compilationH}. We use only $H(z_i)$ values obtained by the so-called differential-age (or cosmic chronometer) techniques applied to passively evolving galaxies. That is to say, we use data where one estimates $H(z)$ from  $(1+z)H(z)=-dz/dt$. Here $dz/dt\simeq\Delta z/\Delta t$ is extracted from a sample of passive galaxies (i.e. with essentially no active star formation) whose age and redshift differences, $\Delta t$ and  $\Delta z$, are known. See the references also in Table \ref{compilationH}. The important point to remark here is that these $H(z)$ values, although relying on the theory of spectral evolution of galaxies, are  uncorrelated with the BAO data points. See also \cite{FarooqRatra2013,SahniShafielooStarobinsky,Ding2015,Zheng2016,ChenKumarRatra2016}, where the authors make only use of Hubble parameter data in their analyses. We find, however, indispensable in our analysis to take into account the remaining data sets to derive our conclusions on dynamical vacuum, specially the BAO, LSS and CMB observations. This fact can also be verified quite evidently in Figs. \ref{fig:PRD7}-\ref{RecRVMtriad}, to which we shall turn our attention in Sect. \ref{subsect:deconstruction}. See, however,\,\cite{Maurice2017} for other tests of dynamical DE using $H(z)$ data.\\

DS5) Data on large scale structure formation (LSS). In this paper, the LSS data specifically refers to data points on the product of the ordinary growth rate $f(z_i)$ with $\sigma_{8}(z_i)$ (the root mean square matter fluctuation on $R_8 = 8{h^{-1}}$ Mpc spheres) at different (effective) redshifts $z_i$, see Sect.\,\ref{sect:fsigma8} for more details.  We use 13 data points on $f(z_i)\sigma_8(z_i)$.  Not all are from RSD. The actual fitting results shown in Table \ref{tableFit1PRD} make use of the LSS data listed in Table \ref{compilationLSSprd}, in which we have carefully avoided possible correlations among them (see below). As an example, we do not include the 2dFGRS point at $z=0.17$ \cite{Song09,Percival2004} in our analysis because, as it is stated in \cite{AlamHoSilvestri}, it is strongly correlated with the 6dFGS one (cf. Table \ref{compilationLSSprd}). Let us mention that although we are aware of the existence of other LSS data points in the literature concerning some of the used redshift values in our Table \ref{compilationLSSprd} -- cf.\, \cite{Guzzo2008,Percival2004}; \cite{Turnbull2012,Hudson2012}; \cite{Johnson2014} -- we have explicitly checked that their inclusion in our numerical fits has no significant impact on the main results of our paper, that is to say, it does not alter in any significant way the attained level of evidence in favor of the main DVM's under study. Our results are definitely secured in both cases, but we have naturally presented our final results sticking to the most updated data.  See also Sect.\,\ref{sect:OtherDataSets} for more details on the data selection procedure.

%
%%%%%%%%%%%%%%%%%%%%%%%%%%%%%%%%%%%%%%%%%%%%%%%%%%%%%%%%%%%%%%%%%%%%%%%%%%%%%%%%%%%
%%%%%%%%%%%%%%%%%%%%%%%%%%%%%%%%%%%%%%%%%%%%%%%%%%%%%%%%%%%%%%%%%%%%%%%%%%%%%%%%%%%%
\begin{figure}
\centering
\includegraphics[angle=0,width=1.03\linewidth]{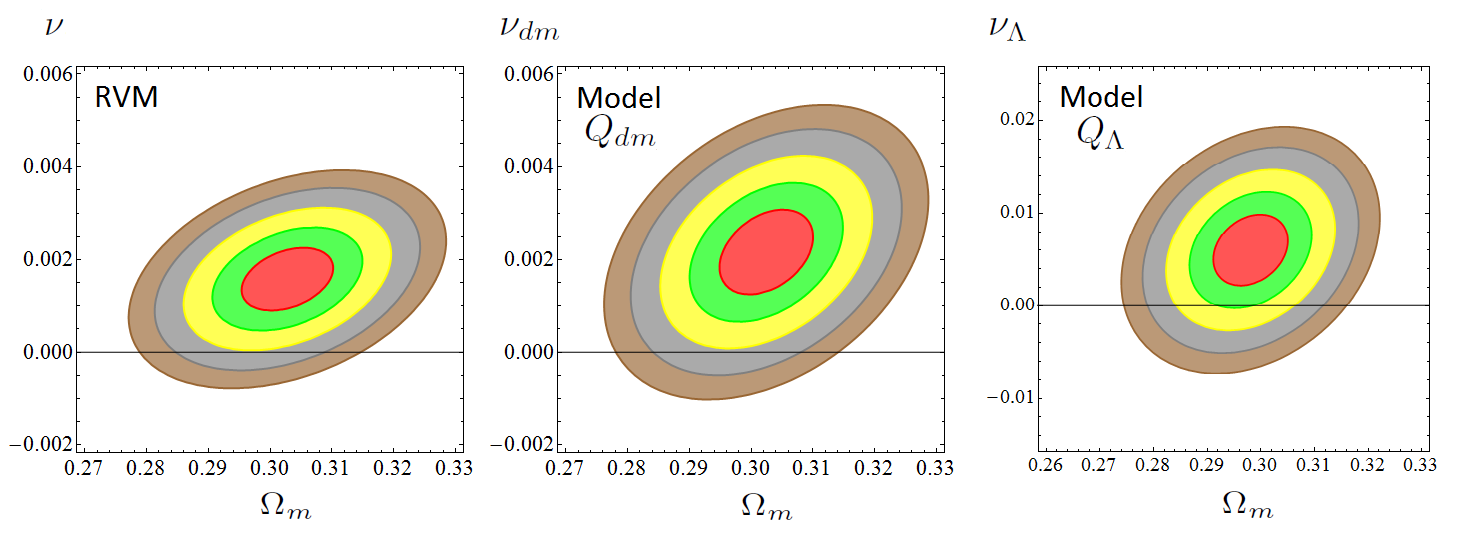}
\caption[Contour lines for the DVM's in the $(\Omega_m,\nu_i)$ plane]{\label{fig:PRD1}%
{\scriptsize Likelihood contours for the DVM's in the $(\Omega_m,\nu_i)$ plane for the values $-2\ln\mathcal{L}/\mathcal{L}_{max}=2.30$, $6.18, 11.81$, $19.33$, $27.65$ (corresponding to 1$\sigma$, 2$\sigma$, 3$\sigma$, 4$\sigma$ and 5$\sigma$ c.l.) after marginalizing over the rest of the fitting parameters indicated in Table \ref{tableFit1PRD}. The elliptical shapes have been obtained applying the standard Fisher matrix approach. We estimate that for the RVM, $94.80\%$ (resp. $89.16\%$) of the 4$\sigma$ (resp. 5$\sigma$) area is in the $\nu>0$ region. For the Q$_{dm}$ we find that $95.24\%$ (resp. $89.62\%$) of the 4$\sigma$ (resp. 5$\sigma$) area is in the $\nu_{dm}>0$ region. Finally, for the Q$_{\CC}$ we estimate that $99.45\%$ (resp. $90.22\%$) of the $2\sigma$ (resp. $3\sigma$) area is in the $\nu_{\CC}>0$ region. Subsequent marginalization over $\Omega_m$ increases slightly the c.l. and renders the fitting values collected in Table \ref{tableFit1PRD}. The $\CC$CDM ($\nu_i=0$) appears disfavored at $\sim 4\sigma$ c.l. in the RVM and $Q_{dm}$, and at $\sim 2.5\sigma$ c.l. for $Q_\CC$.}}
\end{figure}

Let us now reiterate the following observation concerning the possible correlations between data sets. In our analysis we have included both the WiggleZ and the CMASS data sets. We are aware that there exists some overlap region between these two galaxy samples. But the two surveys have been produced independently and the studies on the existing correlations among these observational results, see \cite{Beutler2016,Marin2016}, show that the correlation is small. The overlap region of the CMASS and WiggleZ galaxy samples is actually not among the galaxies that the two surveys pick up, but between the region of the sky they explore. Furthermore, in spite of the fact that almost all of the WiggleZ region (specifically, 5/6 parts of it) lies inside the CMASS one, it only takes a very small fraction of the whole sky region covered by CMASS, since the latter is much larger than the WiggleZ one (see, e.g. Fig.\,1 in \cite{Beutler2016}). In this paper, the authors are able to estimate the correlation degree among the BAO constraints in CMASS and WiggleZ, and they conclude that it is less than 4\%. For this reason we find it perfectly justified to keep the WiggleZ data as part of our BAO and LSS data sets. As an additional check, we have computed the fitting results that are obtained when we remove the WiggleZ data points from the BAO and $f(z)\sigma_8(z)$ data sets (see Table \ref{tableFit2PRD}). We can see that the differences in the uncertainties and in the central values of the parameters with respect to Table \ref{tableFit1PRD} are small. Therefore, no significant change in the statistical significance of the results is found. \\

DS6) Data on the CMB distance priors. These comprise:  $R$ (shift parameter), $\ell_a$ (acoustic length) and their correlations  with the reduced baryon mass parameter and the spectral index  $(\omega_b,n_s)$. It has been shown in \cite{Elgaroy2007,WangMukherjee2007,Mukherjee2008} that although this compression loses information compared to the full CMB likelihood, such information loss becomes negligible when more data is added, as in the current analysis. We have used the CMB covariance matrix from the analysis of the Planck 2015 data of \cite{Huang}. Our fitting results with such covariance matrix are compiled in all our tables (except in Table \ref{tableFit3PRD} where we test the behavior of our fit in the absence of the CMB distance priors $R$ and $\ell_a$). We display the final contour plots for the DVM's and the XCDM in  Figs.\,\ref{fig:PRD1}-\ref{fig:PRD3}. Let us point out that we have checked that very similar results ensue for all models if we use the alternative CMB covariance matrix from \cite{PlanckDE2015}. We have definitely chosen that of \cite{Huang} since it uses the more complete compressed likelihood analysis for Planck 2015 TT,TE,EE + lowP data whereas \cite{PlanckDE2015} uses Planck 2015 TT+lowP data only.\\

Finally, let us examine the Big Bang Nucleosynthesis (BBN) bound (see e.g. \cite{Chiba2011,Uzan2011}), namely the limitation in the value of the BBN speed-up factor, defined as the ratio of the expansion rate predicted in a given model versus that of the $\CC$CDM model at the BBN epoch. For models of constant $G$, it is tantamount to enforcing the condition that the vacuum energy density at the BBN epoch is sufficiently small in comparison to the radiation density, typically less than $5-10\%$. In the current study, the BBN bound need not be enforced since it is automatically fulfilled. The reason stems from our assumption that the radiation is conserved for the DVM's analyzed in this paper. As a result, the behavior of the ratio of the vacuum energy density to the radiation density in these models at BBN time ($a\to 0$) is, in each case, as follows:
\begin{eqnarray}
&&\left.\frac{\rL(a)}{\rho_r(a)}\right|_{\rm RVM}\longrightarrow \frac{\nu}{1+3\nu}\simeq\nu={\cal O}\left(10^{-3}\right)\ll1 \label{eq:BBN-VRVM}\\
&&\left.\frac{\rL(a)}{\rho_r(a)}\right|_{\rm Q_{dm}}\, \longrightarrow \frac{\nu_{dm}}{1-\nu_{dm}}\,\frac{\rho_{dm 0}}{\rho_{r 0}}\,a^{(1+3\nu_{dm})}\to 0\ \ \ \ \
\label{eq:BBN-Qdm} \\
&&\left.\frac{\rL(a)}{\rho_r(a)}\right|_{\rm Q_{\CC}}\ \ \longrightarrow \frac{\rho_{\CC 0}}{\rho_{r 0}}\,a^{(4-3\nu_{\CC})}\to 0\ \ \ \ \  \label{eq:BBN-QL}\,,
\end{eqnarray}
as it can be easily checked from the energy densities explicitly computed in Sect.\,\ref{sect:solvingDVM}. We use, of course, the fact that $0<\nu_i\ll 1$. In all cases the ratio is very small and therefore the BBN proceeds standard, with no significant difference versus the $\CC$CDM.
Note that this situation is in contradistinction to the A and G-type models,  addressed in previous chapters, where strict BBN bounds had to be applied in the analysis. In particular, when $G$ can be variable it generally produces an effective change in the BBN speed-up factor, which was duly restrained in these analysis. For more details, see e.g. Chapter \ref{chap:AandGRevisited}. Recently other models with variable $G$ have been considered aiming to cure a certain level of tension between the Planck 2015 measurements and some independent observations
in intermediate cosmological scales \cite{Perivolaropoulos2017}.  

%%%%%%%%%%%%%%%%%%%%%%%%%%%%%%%%%%%%%%%%%%%%%%%%%%%%%%%%%%%%%%%%%%%%%%%%%%%%%%%%%%%
%%%%%%%%%%%%%%%%%%%%%%%%%%%%%%%%%%%%%%%%%%%%%%%%%%%%%%%%%%%%%%%%%%%%%%%%%%%%%%%%%%%%
\begin{figure}[t!]
\centering
\includegraphics[width=15.0cm]{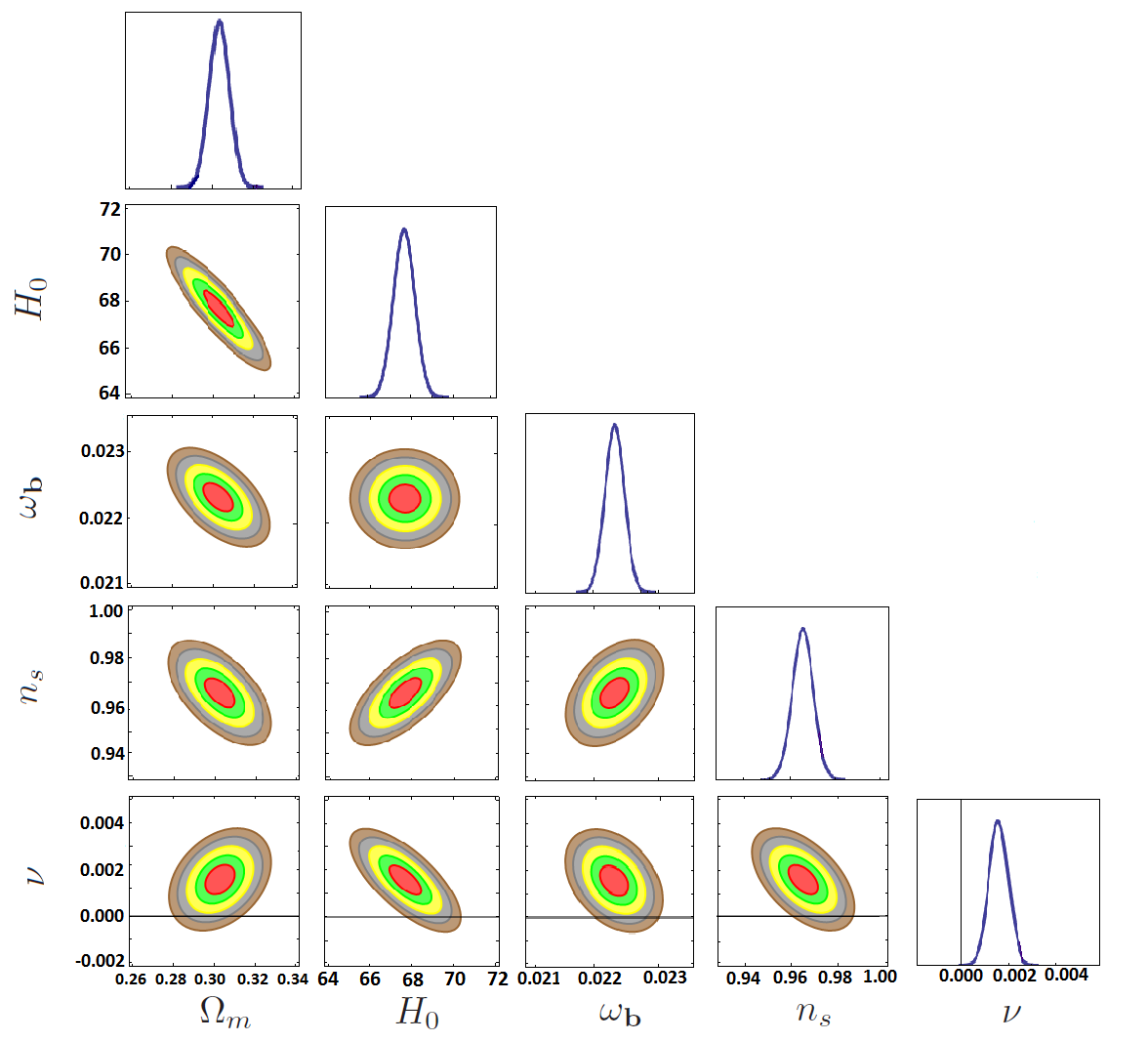}
\caption[CL's of the RVM onto the different planes of the involved parameters]{\label{fig:MatrixPlot}%
{\scriptsize As in Fig.\,\ref{tableFit1PRD}, but projecting the fitting results of the RVM onto the different planes of the involved parameters ($H_0$ is expressed in Km/s/Mpc). The horizontal line $\nu=0$ in the plots of the last row corresponds to the $\CC$CDM. It is apparent that it is significantly excluded at $\sim 4\sigma$ c.l. in all cases. The peak in the rightmost plot corresponds to the central value $\nu=0.00158$ indicated in Table \ref{tableFit1PRD}.
}}
\end{figure}
%
%%%%%%%%%%%%%%%%%%%%%%%%%%%%%%%%%%%%%%%%%%%%%%%%%%%%%%%%%%%%%%%%%%%%%%%%%%%%%%%%%%%

%%%%%%%%%%%%%%%%%%%%%%%%%%%%%%%%%%%%%%%%%%%%%%%%%%%%%%%%%%%%%%%%%%%%%%%%%%%%%%%%%%%%%%%%%%%%%%
%%%%%%%%%%%%%%%%%%%%%%%%%%%%%%%%%%%%%%%%%%%%%%%%%%%%%%%%%%%%%%%%%%%%%%%%%%%%%%%%%%%%%%%%%%%%%%%

\subsection{Background observables}\label{sect:Background}

With the above data sets the fitting results for the main background cosmological observables $(h,\Omega_m)$ are afforded in Table \ref{tableFit1PRD}, together with the reduced baryon mass parameter $\omega_b=\Omega_b\,h^2$, the spectral index $n_s$ and the vacuum parameter $\nu_i$, or alternatively the XCDM or CPL parameters. We will discuss in detail these parameters throughout the paper. Here we want to comment only on the basic two $(h,\Omega_m)$ we have just mentioned. The reduced Hubble constant $h$ is defined as usual from $H_0\equiv 100h\,\varsigma$, with $\varsigma\equiv 1\, {\rm Km/s/Mpc}=2.1332\times10^{-44} GeV$ (in natural units). In particular, the fitting values for $h$ in our tables correspond to using an uninformative flat prior (in the technical sense). Owing to some tension in the values of $h$ in the current literature (see Sect.\,\ref{subsect:Htension}), this procedure should be the fairest option rather than adopting some particular prior, see Sect. \ref{chap:H0tension} for a detailed study on the $H_0$ tension in light of vacuum dynamics in the Universe.

Concerning the radiation density parameter $\Omega_r$, it is taken to be a function of the fitted value of $h$ in our analysis through $\Omega_r=\omega_rh^{-2}$.
The value of $\omega_r$ can be computed from
the current temperature of the CMB photons, $T_{\gamma 0}=2.7255$ K \cite{Fixsen2009}, and the effective number of neutrino species, $N_{\textrm{eff}}=3.046$ \cite{Mangano2005}. Recall that the total radiation density at present is given by
\begin{equation}\label{eq:rhorTot}
\rho_{r 0}=\rho_{\gamma 0}\left[1+N_{\textrm{eff}}\,\frac78\,\left(\frac{4}{11}\right)^{4/3}\right]\,,
\end{equation}
where $T_{\nu 0}/T_{\gamma 0}=\left(4/11\right)^{1/3}$ is the ratio of the current neutrino and photon temperatures\,\cite{KolbTurnerBook}. Finally, the density of photons now is $\rho_{\gamma 0}=(\pi^2/15)\,T_{\gamma 0}^4$.
Using the aforementioned numerical values, we obtain
\begin{equation}\label{eq:omegar}
\omega_r= \frac{\rho_{r 0}}{\rho_{c 0}}\,h^2=\frac{8\pi\,G}{3\times 10^4\,\varsigma^2}\,\rho_{r 0}=4.18343\times 10^{-5}\,.
\end{equation}
This value will enter explicitly the computation of the equality point between radiation and nonrelativistic matter and will be used in the evaluation of the transfer function in Sect.\,\ref{sect:fsigma8}.

Finally, let us stress that the DVM's have a background cosmology similar to that of the $\CC$CDM. The main differences appear at the cosmic perturbations level, which will be discussed in the next section. But at the moment let us mention that all of them have a transition point from braking (deceleration) to acceleration. This is clear since they all describe some dynamical behavior around a dominant (nonvanishing) constant  term. The deceleration parameter $q=-\ddot{a}a/\dot{a}^2$ can be comfortably computed from
\begin{equation}\label{eq:qformula}
q(a)=-1-\frac{a}{2\,E^2(a)}\frac{d E^2(a)}{da}\,,
\end{equation}
where $E=H/H_0$ is again the normalized Hubble function. This function is available for the three DVM's under study, see Eqs.\,(\ref{HRVM})-(\ref{HQL}). Obviously we can neglect the radiation terms in these expressions for this kind of calculation. The transition point is the scale factor value $a_{tr}$ (or redshift value $z_{tr}=1/a_{tr}-1$) where $q(a_{tr})=0$. For instance, for the RVM one finds
\begin{equation}\label{eq:qformula2}
{\rm \textbf{RVM}}:\phantom{XX} a_{tr}=\left[\frac{(1-3\nu)\,\Om}{2(\OL-\nu)}\right]^{1/(3(1-\nu))}\,.
\end{equation}
From this expression we see that if the constant term  in Eq. (\ref{eq:RVMvacuumdadensity}) would be $C_0=0$, then $\Omega_\CC=\nu$ and $a_{tr}=\infty$, i.e. in the absence of $C_0$ there would be no transition.
One can reason similarly in the case of models $Q_{dm}$ and $Q_{\CC}$. However, for model $Q_{dm}$ it turns out that an explicit formula  for $a_{tr}$ is not possible since different powers of the scale factor appear mixed up with an additive term. As a result the equation  $q(a_{tr})=0$ cannot be solved analytically. For model $Q_{\CC}$, however, an explicit result can be obtained:
\begin{equation}\label{eq:qformula3}
{\rm \ \mathbf{Q_{\CC}}}:\phantom{XX} a_{tr}=\left[\frac{\Om-\nu_\CC}{\OL(2-3\nu_\CC)}\right]^{1/(3(1-\nu_\CC))}\,.
\end{equation}
%
%%%%%%%%%%%%%%%%%%%%%%%%%%%%%%%%%%%%%%%%%%%%%%%%%%%%%%%%%%%%%%%%%%%%%%%%%%%%%%%%%%%%
%
\begin{figure}[t!]
\centering
\includegraphics[angle=0,width=0.5\linewidth]{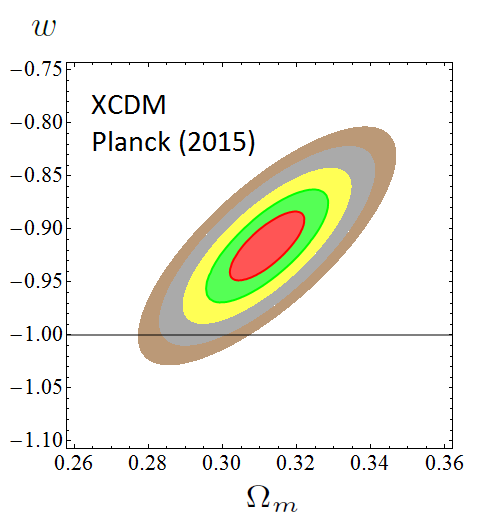}
\caption[Contour lines for the XCDM in the ($\Omega_m$,$\omega$) plane]{
{\scriptsize As in Fig.\,\ref{fig:PRD1}, but for model XCDM. The $\CC$CDM is excluded in this case at $\sim 3\sigma$ c.l. Marginalization over $\Omega_m$ increases the c.l. up to $3.35\sigma$ (cf. Table \ref{tableFit1PRD}).
}\label{fig:PRD3}}
\end{figure}
%
%%%%%%%%%%%%%%%%%%%%%%%%%%%%%%%%%%%%%%%%%%%%%%%%%%%%%%%%%%%%%%%%%%%%%%%%%%%%%%%%%%%
%%%%%%%%%%%%%%%%%%%%%%%%%%%%%%%%%%%%%%%%%%%%%%%%%%%%%%%%%%%%%%%%%%%%%%%%%%%%%%%%%%%%

\noindent For $\nu_i=0$ the above formulas boil down to the $\CC$CDM expression, as they should.  Let us quote the values that we obtain for the transition redshift using the fitted values in Table \ref{tableFit1PRD}. For the $\CC$CDM we obtain $z_{tr}^{\small \CC{\rm CDM}}=0.681\pm 0.012$, whereas from (\ref{eq:qformula2}) and (\ref{eq:qformula3}) we find:
\begin{equation}\label{eq:zts1}
z_{tr}^{RVM}= 0.667 \pm 0.013\,, \ \ \ z_{tr}^{{Q_{\CC}}}= 0.687 \pm 0.012\,.
\end{equation}
The value, obtained numerically, for $Q_{dm}$ is:
\begin{equation}\label{eq:zts2}
z_{tr}^{{Q_{dm}}} = 0.668\pm 0.013\,.
\end{equation}
Clearly, for both the RVM and $Q_{dm}$ the transition occurs essentially at the same point, and such point is closer to our time than in the  $\CC$CDM case. In contrast, for model $Q_{\CC}$ the transition occurs slightly earlier. Unfortunately, owing to the errors these differences are too small and cannot be disentangled; in fact they are compatible with the $\CC$CDM at $1\sigma$. We conclude that the transition redshift is not, at the moment, sufficiently accurate to be used as a tool for distinguishing among these models. The measurable differences will appear, instead, when we move from the background cosmology to the perturbed one. We prepare the ground for it in the next section.

\section{Structure formation with dynamical vacuum}\label{sect:perturbationsDVM}

The analysis of the linear structure formation data deserves some remarks. It hinges on the theory of cosmological perturbations, which has been discussed at length in several specialized textbooks, see e.g.\,\cite{Peebles1993} and \,\cite{LiddleLyth,LiddleLyth2,Dodelson}. Notwithstanding, the analysis of new types of models may introduce novel features that must be carefully taken into account, as it has been explained in previous chapters. At deep subhorizon scales and in the presence of dynamical vacuum energy one can show that the matter density contrast $\delta_m=\delta\rho_m/\rho_m$ obeys the following differential equation (see Appendix \ref{ch:appPert}):
\begin{equation}\label{diffeqDPRD}
\ddot{\delta}_m+\left(2H+\Psi\right)\,\dot{\delta}_m-\left(4\pi
G\rmr-2H\Psi-\dot{\Psi}\right)\,\delta_m=0\,,
\end{equation}
where $\Psi\equiv -\dot{\rho}_{\Lambda}/{\rmr}= Q/{\rmr}$, and the (vacuum-matter) interaction source $Q$ for each DVM is given by Eqs.\,(\ref{eq:QRVM})-(\ref{eq:PhenModelQL}). For $\rL=$const. and for the XCDM and CPL there is no such interaction, therefore $Q=0$, and Eq.\,(\ref{diffeqDPRD}) reduces to the $\CC$CDM form \,\cite{Peebles1993}:
\begin{equation}\label{diffeqDLCDM}
\ddot{\delta}_m+2H\,\dot{\delta}_m-4\pi
G\rmr\,\delta_m=0\,.
\end{equation}
We note that at the scales under consideration we are neglecting the perturbations of the vacuum energy density in front of the perturbations of the matter field (see Appendix \ref{ch:appPert} for further details).

Let us briefly justify by alternative methods the modified form (\ref{diffeqDPRD}), in which the variation of $\rho_\CC$ enters only through the background quantity $\Psi$ and not through any perturbed quantity. We shall conveniently argue in the context of two well-known gauges, the synchronous gauge and the Newtonian conformal gauge, thus providing a twofold justification. In the synchronous gauge, vacuum perturbations $\delta\rL$ modify the momentum conservation equation for the matter particles in a way that we can easily get e.g. from the general formulae of\,\cite{GrandePelinsonSola08}, with the result
\begin{equation}\label{eq:ModifiedMomentum}
\dot{v}_m+ \, H v_m=\frac{1}{a}\frac{\delta\rho_\Lambda}{\rho_m}-\Psi v_m\,,
\end{equation}
where  $\vec{v}=\vec{\nabla}v_m$ is the associated peculiar velocity, with potential $v_m$ (notice that this quantity has dimension of inverse energy in natural units). By setting $\delta\rho_\Lambda= a\,Q\,{v_m}=a\,\rho_m\,\Psi\,v_m$ the two terms on the {\it r.h.s.} of (\ref{eq:ModifiedMomentum}) cancel each other and we recover the corresponding equation of the $\CC$CDM.
On the other hand, in the Newtonian or conformal gauge\,\cite{Mukhanov92,MaBertschinger1995} we find a similar situation. The analog of the previous equation is the modified Euler's equation in the presence of dynamical vacuum energy. It reads as follows:
\begin{equation}\label{eq:Euler}
\frac{d}{d\eta}\left(\rho_mv_m\right)+4\mathcal{H}\rho_mv_m+\rho_m\phi-\delta\rho_\Lambda=0\,,
\end{equation}
where $\phi$ is the gravitational potential that appears explicitly in the Newtonian conformal gauge, and $\eta$ is the conformal time. Let an overhead circle denote a derivative with respect to the conformal time, i.e. $\mathring{f}=df/d\eta$ for any $f$. We define the quantities $\mathcal{H}=\mathring{a}/a=aH$ and $\bar{\Psi}=-\mathring{\rho}_{\Lambda}/\rho_m=a\Psi$, which are the analogues of $H$ and $\Psi$ in conformal time. Using the background local conservation equation (\ref{BianchiGeneral}) for the current epoch (neglecting therefore radiation) and rephrasing it in conformal time, i.e. $\mathring{\rho}_\Lambda+\mathring{\rho}_m+3\mathcal{H}\rho_m=0$, we can bring (\ref{eq:Euler}) to the simpler form
\begin{equation}\label{eq:ModifiedEuler}
\mathring{v}_m+ \, \mathcal{H} v_m+\phi=\frac{\delta\rho_\Lambda}{\rho_m}-\bar{\Psi} v_m\,.
\end{equation}
Once more the usual fluid equation (in this case Euler's equation) is retrieved if we arrange that $\delta\rho_\Lambda=\rho_m\,\bar{\Psi}\,v_m=a\,\rho_m\,\Psi\,v_m$, as then the two terms on the {\it r.h.s.} of (\ref{eq:ModifiedEuler}) cancel each other. Alternatively, one can use the covariant form $\nabla^{\mu} T_{\mu\nu}=Q_\nu$ for the local conservation law, with  the source 4-vector $Q_\nu= Q U_\nu$, where $U_\nu=(a,\vec{0})$ is the background matter 4-velocity in conformal time\,\cite{BookAmendolaTsujikawa}. By perturbing the covariant conservation equation one finds

\be
\delta\left(\nabla^{\mu} T_{\mu\nu}\right)=\delta Q_\nu=\delta Q\,U_\nu+Q\delta U_\nu\,,
\ee
where $\delta Q$ and $\delta U_\nu=a(\phi,-\vec{v})$ are the perturbations of the source function and the 4-velocity, respectively. Thus, we obtain

\be
\delta\left(\nabla^{\mu} T_{\mu\nu}\right)=a(\delta Q+Q\phi,-Q\vec{v})\,.
\ee
From the $\nu=j$ component of the above equation, we derive anew the usual Euler equation $\mathring{v}_m+ \, \mathcal{H} v_m+\phi=0$, which means that the relation $\delta\rho_\Lambda= a Q\,{v_m}=a\,\rho_m\,\Psi\,v_m$ is automatically fulfilled. So the analyses in the two gauges converge to the same final result for $\delta\rL$.

After we have found the condition that $\delta\rho_\CC$ must satisfy in each gauge so as to prevent that the vacuum modifies basic conservation laws of the matter fluid, one can readily show that any of the above equations (\ref{eq:ModifiedMomentum}) or (\ref{eq:ModifiedEuler}) for each gauge (now with their {\it r.h.s.} set to zero), in combination with the corresponding perturbed continuity equation and the perturbed $00$-component of Einstein's equations (giving Poisson's equation in the Newtonian approximation), leads to the desired matter perturbations equation\,(\ref{diffeqDPRD}). While our discussion holds good in a rigorous relativistic context, a derivation was first possible in the Newtonian formalism\,\cite{ArcuriWaga94}. Altogether, the above considerations formulated in the context of different gauges allow us to consistently neglect the DE perturbations at scales down the horizon. This justifies the use of Eq.\,(\ref{diffeqDPRD}) for the effective matter perturbation equation in our study of linear structure formation in the framework of the DVM's.  We refer the reader to chapter \ref{sect:DEorNot} and Ref. \cite{GrandePelinsonSola08} for additional discussions on the suppression of the DE perturbations at subhorizon scales.

It will be convenient for further use to rewrite Eq.\,(\ref{diffeqDPRD}) in terms of the scale factor variable rather than the cosmic time. A straightforward calculation leads to the following expression
\begin{equation}\label{diffeqDaPRD}
\delta''_m + \frac{A(a)}{a}\delta'_m + \frac{B(a)}{a^2}\delta_m = 0\,,
\end{equation}
where primes denote differentiation $d/da$, and the functions $A$ and $B$ of the scale factor are given by
\begin{align}
& A(a) = 3 + a\frac{H'(a)}{H(a)} + \frac{\Psi(a)}{H(a)}\,,\label{deffA}\\
& B(a) = - \frac{4\pi{G}\rho_m(a)}{H^2(a)} + \frac{2\Psi(a)}{H(a)} + a\frac{\Psi'(a)}{H(a)}\label{deffB}\,.
\end{align}

\subsection{Initial conditions}

To solve \eqref{diffeqDaPRD} we have to fix the initial conditions for $\delta_m(a)$ and $\delta'_m(a)$ for each model at high redshift, say at $z_i\sim100$ ($a_i\sim10^{-2}$), when nonrelativistic matter dominates both over the vacuum and the radiation contributions. We have to calculate the limit of the functions \eqref{deffA}-\eqref{deffB} for small values of the scale factor. In this limit, the leading form of the normalized Hubble rate (squared) for each model, see equations \eqref{HRVM}-\eqref{HQL}, can be written to within very good approximation as follows:
\begin{eqnarray}
{\rm RVM}:\phantom{X}E^2(a) &=& \frac{\Omega_m}{1-\nu}\,a^{-3(1-\nu)} \label{HubbleRatesmalla1} \\
{\rm \ Q_{dm}}:\phantom{X} E^2(a)&=& \frac{\Omega_m}{1-\nu_{dm}}\,a^{-3(1-\nu_{dm})} \label{HubbleRatesmalla2} \\
{\rm \ Q_{\CC}}:\phantom{X} E^2(a) &=& \frac{\Omega_m - \nu_{\CC}}{1-\nu_{\CC}}\,a^{-3}\,. \label{HubbleRatesmalla3}
\end{eqnarray}
These equalities are to be understood in an approximate sense, that is to say, as the leading expressions at the point where we take the initial conditions.
As it is clear from them, each model remains close to the $\CC$CDM behavior, given by $E^2(a)\simeq \Omega_{m}a^{-3} $ in the same limit, and perfectly reduces to it for $\nu_i\to 0$, as expected. The small departure, however, must be duly taken into account when fixing the initial conditions. Below we explain how this is done.

A key ingredient in the structure formation equation for the DVM's is $\Psi(a)$, since it encodes the dynamical character of the vacuum energy at the background level. It can be easily computed for each model and yields:
\begin{eqnarray}
{\rm RVM}:\phantom{X}\Psi(a) &=& 3\nu{H(a)} \label{PsiFunction1}\\
{\rm \ Q_{dm}}:\phantom{X}\Psi(a)&=& 3\nu_{dm}\frac{\Omega_{dm}}{\Omega_m}H(a) + \mathcal{O}(\nu^2_{dm}) \label{PsiFunction2} \\
{\rm \ Q_{\CC}}:\phantom{X}\Psi(a) &=& 3\nu_{\CC}\frac{\rho_{\CC}(a)}{\rho_m(a)}H(a) \label{PsiFunction3}\,,
\end{eqnarray}
where the first and third expressions are exact in $\nu_i$ since we disregard the radiation terms proportional to $\,\Omega_r$, whereas the second expression involves subleading contributions of ${\cal O}(\nu_{dm}^2)$, which will be neglected. With the above estimates we can assess the value of the correction terms $\Psi/H$ and $a\Psi'/H$ in \eqref{deffA}-\eqref{deffB}. They represent the extra terms in the matter perturbations equation with respect to the $\CC$CDM.
We immediately see that both for the RVM and $Q_{dm}$ we have $\Psi/H$= const. and such a constant is proportional to $\nu_i$, whereby  $\Psi/H$ is small but not necessary negligible. In contrast, for model $Q_{\CC}$ the ratio $\Psi/H \sim \nu_\CC\,a^{3(1-\nu_{\CC})}$ becomes smaller and smaller in the past, and therefore its value at $a\sim 10^{-2}$ is strongly suppressed as compared to the other DVM's. Thus, in leading order, there are no corrections to the matter perturbations equation for model $Q_{\CC}$ other than the background effect associated to the modified Hubble function. Similar considerations apply to the term $a\Psi'/H$ for each model.

From the above formulas we can straightforwardly obtain the leading form of the functions \eqref{deffA}-\eqref{deffB} for the different DVM's:
\begin{eqnarray}
{\rm RVM}:\phantom{X} A &=& \frac{3}{2}(1+3\nu) \label{AFunction1}\\
{\rm \ Q_{dm}}:\phantom{X} A &=& \frac{3}{2}\left(1 + \nu_{dm}\right) + 3\frac{\Omega_{dm}}{\Omega_m}\nu_{dm}+\mathcal{O}(\nu^2_{dm})  \label{AFunction2}\\
{\rm \ Q_{\CC}}:\phantom{X} A &=& \frac{3}{2}\,,
\label{AFunction3}
\end{eqnarray}
and
\begin{eqnarray}
{\rm RVM}:\phantom{X} B &=& -\frac{3}{2} +3\nu + \frac{9}{2}\nu^2  \label{BFunction1}\\
{\rm \ Q_{dm}}:\phantom{X} B &=& -\frac{3}{2}\left(1-\nu_{dm} - \frac{\Omega_{dm}}{\Omega_m}\nu_{dm}\right)+\mathcal{O}(\nu^2_{dm})  \label{BFunction2}\\
{\rm \ Q_{\CC}}: \phantom{X} B &=& -\frac{3}{2}\,,
\label{BFunction3}
\end{eqnarray}
For $\nu_i\rightarrow 0$, we recover the $\CC$CDM behavior $A \rightarrow \frac{3}{2}$ and $B \rightarrow -\frac{3}{2}$, as it should. This is already true for the $Q_{\CC}$ without imposing $\nu_{\CC}\rightarrow 0$, therefore its initial conditions are precisely the same as for the concordance model. Once the functions \eqref{deffA}-\eqref{deffB} take constant values (as it is the case here at the high redshifts where we fix the initial conditions), the differential equation \eqref{diffeqDaPRD} admits power-like solutions of the form $\delta_m(a_i)=a_i^{s}$. Of the two solutions, we are interested only in the growing mode solution, as this is the only one relevant for structure formation. For example, using (\ref{AFunction1}) and (\ref{BFunction1}) for the case of the RVM, the perturbations equation (\ref{diffeqDaPRD}) becomes
\begin{equation}\label{diffeqDaRVM}
\delta''_m + \frac{3}{2a}(1+3\nu)\delta'_m - \left(\frac{3}{2}-3\nu-\frac{9}{2}\,\nu^2\right)\frac{\delta_m}{a^2} = 0\,.
\end{equation}
The power-law solution for the growing mode gives the result $\delta_m=a^{1-3\nu}$, which is exact even keeping the ${\cal O}(\nu^2)$ term. Nevertheless, as warned previously, in practice we can neglect all ${\cal O}(\nu_i^2)$ contributions despite we indicate their presence.  Repeating the same procedure for the other models, the final outcome for the growing mode solution $\delta_m\sim a^s$ is the following:
\begin{eqnarray}
{\rm RVM}:\phantom{X} s &=& 1-3\nu \label{svalues1}\\
{\rm \ Q_{dm}}:\phantom{X} s &=& 1-\nu_{dm}\left(\frac{6\Omega_m + 9\Omega_{dm}}{5\Omega_m}\right) + \mathcal{O}(\nu^2_{dm}) \label{svalues2}\\
{\rm \ Q_{\CC}}:\phantom{X} s &=& 1\,.
\label{svalues3}
\end{eqnarray}
As expected the deviations with respect to $\CC$CDM are small (and for $Q_{\CC}$ there is no deviation at all). For the RVM and $Q_{dm}$ the departures from the $\CC$CDM are proportional to $\nu_i$, and therefore the dynamical vacuum energy induces small changes in the initial conditions. Imposing the above analytical results to fix the initial conditions, we are then able to solve numerically the full differential equation \eqref{diffeqDaPRD} from a high redshift $z_i\sim100$ ($a_i\sim10^{-2}$) up to our days. The result does not significantly depend on the precise value of $z_i$, provided it is large enough but still well below the decoupling time ($z\sim 10^3$), where the radiation component starts to be nonnegligible.

\subsection{Linear growth and growth index}

The linear growth rate of clustering is an important (dimensionless)
indicator of structure formation\,\cite{Peebles1993}, as it has been shown in previous chapters. It is defined as the logarithmic derivative of the linear growth factor $\delta_m(a)$ with respect to the log of the scale factor, $\ln a$. Therefore,
\begin{equation}\label{growingfactor}
f(a)\equiv \frac{a}{\delta_m}\frac{d\delta_m}{d a}=\frac{d{\rm ln}\delta_m}{d{\rm ln}a}\,,
\end{equation}
where $\delta_m(a)$ is obtained from solving
the differential equation (\ref{diffeqDaPRD}) for each model. Since $f(a)$ is equal to $(\dot{\delta}_m/\delta_m)/(\dot{a}/a)$ (the ratio of the peculiar flow rate to the Hubble rate), the physical significance of $f(a)$ is that it determines the peculiar velocity flows\,\cite{Peebles1993}. In terms of the redshift variable, we have $f(z)=-(1+z)\,d{\ln\delta_m}/{dz}$, and thus the linear growth can also be used to determine the amplitude of the redshift distortions. This quantity has been analytically computed for the RVM in \cite{BasiSola2015}. Here we shall take it into account for the study of the LSS data in our overall fit to the cosmological observations.

One usually expresses the linear growth rate of clustering
in terms of $\Omega_m(z)=\rho_m(z)/\rho_c(z)$, where $\rho_c(z)=3H^2(z)/(8\pi G)$ is the evolving critical density, as follows\,\cite{Peebles1993}:
\begin{equation}\label{eq:gammaIndex}
f(z)\simeq \left[\Omega_{m}(z)\right]^{\gamma(z)}\,,
\end{equation}
where $\gamma$ is the so-called linear growth rate index. For the usual $\Lambda$CDM model, such index is approximately given by $\gamma_{\CC} \simeq 6/11\simeq 0.545$. For models with a slowly varying  equation of state $w_D$ (i.e. approximately behaving as the XCDM, with $w_D\simeq w_0$) one finds the approximate formula $\gamma_D\simeq 3(w_D-1)/(6w_D-5)$ \cite{WangSteinhardt98} for the asymptotic value when $\Omega_m\to 1$. Setting $w_D=-1+\epsilon$, it can be rewritten
\begin{equation}\label{eq:GammaIndex}
\gamma_{D}\simeq \frac{6-3\epsilon}{11-6\epsilon}\simeq \frac{6}{11}\left( 1+\frac{1}{22}\,\epsilon\right)\,.
\end{equation}
Obviously, for $\epsilon\to 0$ (i.e. $\omega_D\to -1$) one retrieves the $\CC$CDM case.
Since the current experimental error on the $\gamma$-index is of order $10\%$, it opens the possibility to discriminate cosmological models using such index, see e.g. \cite{BasiNesseris2016,Athina2014}. In the case of the RVM and related models, the function $\gamma(z)$ has been computed numerically in Chapters \ref{chap:Atype}-\ref{chap:Gtype}. Under certain approximations, an analytical result can also be obtained for the asymptotic value\,\cite{BasiSola2015}:
\begin{equation}\label{eq:GammaIndexRVM}
\gamma_{\rm RVM}\simeq \frac{6+3\nu}{11-12\nu}\simeq \frac{6}{11}\left(1+\frac{35}{22}\,\nu\right)\,.
\end{equation}
This expression for the RVM is similar to (\ref{eq:GammaIndex}) for an approximate XCDM parametrization, and it reduces to the $\CC$CDM value for $\nu=0$, as it should. It is interesting to note that for $\epsilon>0$ and $\nu>0$ (which correspond to quintessence-like behavior in both pictures, XCDM and RVM) one consistently finds $\gamma$ slightly bigger than the $\CC$CDM value.

One hopes that models might be resolvable by this method in the future when more accurate data will be available\,\cite{BasiNesseris2016}. However, in the present analysis we will not concentrate on $f(z)$ nor on $\gamma(z)$, but on a related quantity on which we focus in the next section, $f(z)\sigma_{8}(z)$. This important quantity defines the LSS data used in our analysis.

\subsection{Weighted linear growth and power spectrum}\label{sect:fsigma8}

A most suitable observable to assess the performance of our vacuum models in regard to structure formation is the combined quantity $f(z)\sigma_{8}(z)$, viz. the ordinary growth rate weighted with $\sigma_{8}$, the rms total matter fluctuation
(baryons + CDM) on $R_8 = 8{h^{-1}}$ Mpc spheres at the given redshift $z$, computed in linear theory. It has long been recognized that this estimator is almost a model-independent way of expressing the observed growth history
of the Universe, in particular it is found to be independent of the galaxy density bias\,\cite{Guzzo2008}.

%%%%%%%%%%%%%%%%%%%%%%%%%%%%%%%%%%%%%%%%%%%%%%%%%%%%%%%%%%%%%%%%%%%%%%%%%
\begin{figure}[t!]
\centering
\includegraphics[angle=0,width=0.7\linewidth]{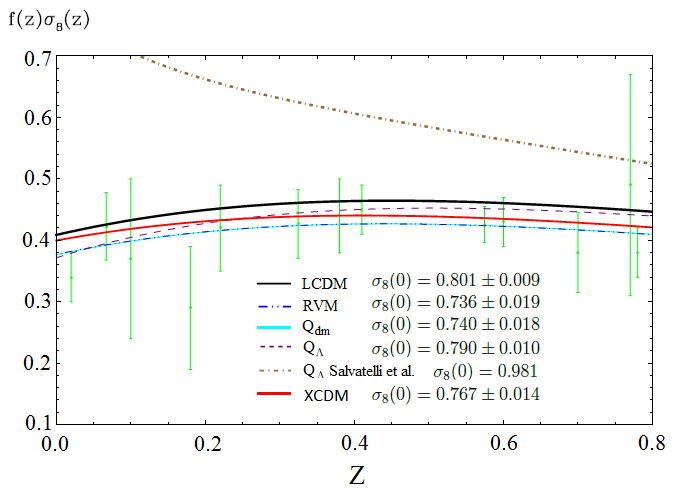}
\caption[$f(z)\sigma_8(z)$ data and predicted curves for the DVM's, XCDM and the $\CC$CDM, using the best-fit values in Table \ref{tableFit1PRD}]{{\scriptsize The $f(z)\sigma_8(z)$ data (cf. Table \ref{compilationLSSprd}) and the predicted curves by the DVM's, XCDM and the $\CC$CDM, for the best-fit values in Table \ref{tableFit1PRD}. The highest curve (in brown) corresponds to model $Q_\CC$ for the same best-fit values as \cite{Salvatelli2014} and assuming (as they do) that the dark sector interaction begins at $z=0.9$. Such strayed curve is unable to explain the LSS data (see the text for discussion) and corresponds to $Q_\CC<0$. As explained in the text, this sign of the interaction source violates the SLT.
This is in stark contrast to the well-behaved results of all the DVM's and XCDM when we use the fitting results of our own analysis, all of them plotted in the figure.
Shown are also the central values and uncertainties of $\sigma_8(0)$ that we have obtained for all the models.}\label{fig:fsigma8PRD}}
\end{figure}
%
%%%%%%%%%%%%%%%%%%%%%%%%%%%%%%%%%%%%%%%%%%%%%%%%%%%%%%%%%%%%%%%%%%%%%%%%%

Equipped with the above generalized matter perturbations equation (\ref{diffeqDaPRD}) and the appropriate initial conditions, the analysis of the linear LSS regime is implemented with the help of the weighted linear growth $f(z)\sigma_8(z)$, in which the variance of the smoothed linear density field on $R_8 = 8{h^{-1}}$ Mpc spheres at redshift $z$ is computed from
%
%\begin{equation}
%\begin{small}\sigma_8^2(z)=\frac{\delta_m^2(z)}{2\pi^2}\int_{0}^{\infty}k^2\,P(k,\vec{p})\,W^2(kR_8)\,dk\,,\label{s88generalNN}
%\end{small}\end{equation}
%
\begin{equation}
\sigma_8^2(z)=\delta_m^2(z)\int\frac{d^3k}{(2\pi)^3}\, P(k,\vec{p})\,\,W^2(kR_8)\,.
\label{s88generalNN}
\end{equation}
Here ${P}(k,\vec{p})={P}_0\,k^{n_s}T^2(k)$ is the ordinary linear matter power spectrum (viz. the coefficient of the two-point correlator of the linear perturbations), with $P_0$ a normalization factor, $n_s$ the spectral index and $T(k)$ the transfer function. Furthermore,
$W(kR_8)$ in the above formula is a top-hat smoothing function, which can be expressed in terms of the spherical Bessel function of order $1$, as follows:
\begin{equation}\label{eq:WBessel}
W(kR_8)=3\,\frac{j_1(kR_8)}{kR_8}=\frac{3}{k^2R_8^2}\left(\frac{\sin{\left(kR_8\right)}}{kR_8}-\cos{\left(kR_8\right)}\right)\,.
\end{equation}
Moreover,
\begin{equation}\label{eq:fittingvector}
\vec{p}=(h,\omega_b, n_s,\Omega_m,\nu_i)
\end{equation}
is the $5$-dimensional fitting vector of free parameters for the vacuum models we are analyzing. The concordance $\Lambda$CDM model can be viewed as the particular case  $\nu_i=0$. It is understood that for the XCDM and CPL parametrizations the fitting parameter $\nu_i$ is replaced by $w_0$ and $(w_0,w_1)$, respectively.

The power spectrum depends on all the components of the fitting vector (\ref{eq:fittingvector}). The dependence on the spectral index $n_s$ is quite obviously power-like, whereas the transfer function $T(k,\vec{q})$ depends in a a more complicated way on the rest of the fitting parameters (see below), and thus for convenience we collect them in the reduced $4$-dimensional vector
\begin{equation}\label{eq:fittingvector2}
\vec{q}=(h,\omega_b,\Omega_m,\nu_i)\,.
\end{equation}
In the expression (\ref{s88generalNN}) both $P(k)$ and $W(k)$ are functions of $k\equiv |\vec{k}|$. Therefore, it is possible to integrate  $d^3k\,P(k)/(2\pi)^3$ (i.e. the elementary bin power centered at $k$) over all orientations of $\vec{k}$. Writing
\begin{equation}
\int\frac{d^3k}{(2\pi)^3}\,P(k,\vec{p})=\int_{0}^{\infty}\frac{dk}{k}\,{\cal P}(k,\vec{p})\,,\label{binpower}
\end{equation}
where we have introduced the dimensionless linear matter power spectrum, ${\cal P}(k,\vec{p})=\left(k^3/2\pi^2\right)\,P(k,\vec{p})$, the variance (\ref{s88generalNN}) can finally be written as
\begin{equation}
\sigma_8^2(z)=\delta_m^2(z)\int_{0}^{\infty}\frac{dk}{k}\, {\cal P}(k,\vec{p})\,\,W^2(kR_8)\,,
\label{s88generalNN2}
\end{equation}
with
\begin{equation}\label{eq:PowerSpectrum}
{\cal P}(k,\vec{p})={\cal P}_0k^{n_s+3}T^2(k,\vec{q})\,.
\end{equation}
The normalization factor ${\cal P}_0=P_0/2\pi^2$  will be determined in the next section in view of defining a fiducial model.

For the transfer function, we have adopted the usual BBKS form\,\cite{Bard86} (see, however, below):
\begin{equation}\label{BBKS}
T(x) = \frac{\ln (1+0.171 x)}{0.171\,x}\Big[1+0.284 x + (1.18 x)^2+ \, (0.399 x)^3+(0.490x)^4\Big]^{-1/4}\,.
\end{equation}
Originally, $x=k/k_{eq}$, with
\be\label{keqDef}
k_{eq}=a_{eq}H(a_{eq})\,,
\ee
being the value of the comoving wavenumber at the equality scale $a_{eq}$ between matter and radiation densities: $\rho_r(a_{eq})=\rho_m(a_{eq})$. It is well-known that \eqref{BBKS} does not incorporate the effects produced by the tightly coupled photo-baryon plasma before the decoupling time. The fight between pressure and gravity in this coupled system generates the baryon acoustic oscillations in the matter power spectrum at ``small'' scales, i.e. for $k>k_{eq}$. The baryon density effects can be introduced in \eqref{BBKS} through the modified shape parameter $\tilde{\Gamma}$\,\cite{PeacockDodds,Sugiyama} in $x=k/(k_{eq}\tilde{\Gamma})$, with
\be
\tilde{\Gamma}=e^{-\Omega_b-\sqrt{2h}\frac{\Omega_b}{\Omega_m}}\,.
\ee
Here it is understood that $\Omega_b=\omega_b/h^2$ since we have to  express the remaining parameters in terms of the components of the basic free set in (\ref{eq:fittingvector}).

We remark that $k_{eq}$ is a model-dependent quantity, which departs from the $\CC$CDM expression in those models in which matter and/or radiation are governed by an anomalous continuity equation, as e.g. in the DVM's. In point of fact $k_{eq}$ depends on all the parameters of the reduced fitting vector (\ref{eq:fittingvector2}) since for a given model beyond the $\CC$CDM $a_{eq}$ and/or $H(a_{eq})$  depend on $\vec{q}$, as it will be clear below.

For the concordance model, $k_{eq}$ has the simplest expression, which depends on $h$ and $\Omega_m$ only:
\begin{equation}\label{keqCCprev}
k^\CC_{eq} = H_0\,\Omega_m\sqrt{\frac{2}{\Omega_r}}\,,
\end{equation}
where we recall that $\Omega_r=\omega_r/h^2$, with $\omega_r$ fixed from Eq.\,(\ref{eq:omegar}). The above expression for the  wavenumber at equality can also be written as follows:
\begin{equation}\label{keqCC}
k^\CC_{eq} = \frac{\Omega_mh^2}{3000}\sqrt{\frac{2}{\omega_r}}\,\textrm{Mpc}^{-1}\,,
\end{equation}
where we have used the accurate relation $H_0^{-1}=3000 h^{-1}$ Mpc.
For the DVM's  is not possible to find a formula as compact as \eqref{keqCC}, since either the corresponding expression for $a_{eq}$ is quite involved, as in the RVM case:
\begin{equation}\label{eq:aeqRVM}
\textrm{RVM}:\quad a_{eq}=\left[\frac{\Omega_r(1+7\nu)}{\Omega_m(1+3\nu)+4\nu\Omega_r}\right]^{\frac{1}{1+3\nu}}\,,
\end{equation}
or because $a_{eq}$ must be computed numerically, as for the models Q$_{dm}$ and Q$_\CC$. In all cases, for $\nu_i=0$ we retrieve the value of $a_{eq}$ in the $\CC$CDM. This is pretty obvious for the RVM from (\ref{eq:aeqRVM}), where $a_{eq}\to\Omega_r/\Omega_m$ for $\nu\to 0$.
After obtaining $a_{eq}$ for the various models, it is straightforward to calculate the corresponding value of $k_{eq}$ from  \eqref{keqDef}. Needless to say, for $\nu_i=0$ the $\CC$CDM value of $k_{eq}$ is recovered too, since $H(a)$ in Eq.\,(\ref{keqDef}) also adopts the $\CC$CDM form for $\nu_i=0$, as we know from Sect.\,\ref{sect:solvingDVM}. Similar comments apply for the XCDM and CPL parametrizations. For these two parametrizations, $a_{eq}$ is exactly the same as in the $\CC$CDM, and $k_{eq}$ presents only a negligible difference with respect to the $\CC$CDM. This is because the effect of the DE on the value of $H(a_{eq})$ is very small. The main correction on $k_{eq}$ indeed comes from the situations in which the modification impinges on the value of $a_{eq}$ itself -- see  \eqref{keqDef} -- such as in the case of the DVM's.

Let us stress the importance of correctly computing the characteristic comoving wavenumber $k_{eq}$ for all models, as we have done above, since it determines the borderline between small and large modes that entered the horizon in the radiation-matter equality time. Finally, let us mention that for the sake of completeness we have checked that the use of the alternative %effective shape parameter of the
matter transfer function furnished by Eisenstein and Hu \cite{EisensteinHu98} %(instead of the alternative transfer function\,\cite{Bardeen}, with the modified shape parameter of \cite{PeacockDodds,Sugiyama})
does not produce any significant change in our results.

\subsection{Fiducial model}\label{sect:FiducialModel}

Inserting  the dimensionless power spectrum \,(\ref{eq:PowerSpectrum}) into the variance (\ref{s88generalNN2}) at $z=0$ allows us to write $\sigma_8(0)$ in terms of the power spectrum normalization factor ${\cal P}_0$ and the primary parameters that enter our fit for each model. This is tantamount to saying that ${\cal P}_0$ can be fixed as follows:
%
%\begin{equation}\label{P0}
%\begin{small}
%P_0=2\pi^2\frac{\sigma_{8,\Lambda}^2}{\delta^2_{m,\Lambda}(0)}\left[\int_{0}^{\infty} %k^{2+n_{s,\Lambda}}T^2(k,\vec{p}_{\Lambda})W^2(kR_{8,\Lambda})dk\right]^{-1}\,,
%\end{small}
%\end{equation}
\begin{equation}\label{P0}
\begin{small}
{\cal P}_0=\frac{\sigma_{8,\Lambda}^2}{\delta^2_{m,\Lambda}}\left[\int_{0}^{\infty} k^{n_{s,\Lambda+3}}T^2(k,\vec{q}_{\Lambda})W^2(kR_{8,\Lambda})(dk/k)\right]^{-1}\,,
\end{small}
\end{equation}
where the chosen values of the parameters in this expression define our fiducial model. Specifically, we have set $\sigma_{8,\Lambda}\equiv\sigma_{8,\Lambda}(0)$ and $\delta_{m,\Lambda}\equiv\delta_{m,\Lambda}(0)$, and at the same time we have introduced the vector of fiducial parameters
\begin{equation}\label{eq:vectorfiducial}
\vec{p}_\CC=(h_{\CC},\omega_{b,\CC},n_{s,\CC},\Omega_{m,\CC},0)
\end{equation}
and the reduced one
\begin{equation}\label{eq:vectorfiducial2}
\vec{q}_\CC=(h_{\CC},\omega_{b,\CC},\Omega_{m,\CC},0)\,.
\end{equation}
These vectors are defined in analogy with the fitting vectors introduced before, see equations \,(\ref{eq:fittingvector})-(\ref{eq:fittingvector2}), but all their parameters are taken to be equal to those from the Planck 2015 TT,TE,EE+lowP+lensing analysis\,\cite{Planck2015} with $\nu_i=0$. The subindex $\CC$ in all these parameters denotes such setting.  In particular,  $\sigma_{8,\Lambda}$ in (\ref{P0}) is also taken from the aforementioned Planck 2015 data.  However, $\delta_{m,\Lambda}$ in the same formula is computable: it is the value of $\delta_m(z=0)$ obtained from solving the perturbation equation of the $\CC$CDM, Eq.\,(\ref{diffeqDLCDM}), using the mentioned fiducial values of the other parameters. Finally, plugging the normalization factor \eqref{P0} in \eqref{s88generalNN2} one finds:
%
%\begin{equation}
%\begin{small}
%\sigma_{\rm 8}(z)=\sigma_{8, \Lambda}
%\frac{\delta_m(z)}{\delta_{m,\CC}(0)}
%\sqrt{\frac{\int_0^\infty k^{2+n_s} T^{2}(k,\vec{p})W^2(kR_{8}) dk}{\int_0^\infty k^{2+n_{s,\CC}} T^{2}(k,\vec{p}_\Lambda) W^2(kR_{8,\Lambda}) dk}}\,.
%\end{small}\end{equation}
%
\begin{equation}
\begin{small}\label{eq:sigma8normalized}
\sigma_{8}(z)=\sigma_{8, \Lambda}
\frac{\delta_m(z)}{\delta_{m,\CC}}
\sqrt{\frac{\int_0^\infty k^{n_s+2} T^{2}(k,\vec{q})W^2(kR_{8})\, dk}{\int_0^\infty k^{n_{s,\CC}+2} T^{2}(k,\vec{q}_\Lambda) W^2(kR_{8,\Lambda})\, dk}}\,.
\end{small}\end{equation}
For the fiducial $\CC$CDM, this expression just gives the scaling of $\sigma_{8,\CC}(z)$ with the redshift in the linear theory, that is to say, $\sigma_{8,\CC}(z)/\sigma_{8,\CC}=\delta_{m,\CC}(z)/\delta_{m,\CC}$. But for an arbitrary model, Eq.\,(\ref{eq:sigma8normalized}) normalizes the corresponding $\sigma_{8}(z)$ with respect to the fiducial value, $\sigma_{8, \Lambda}$. This includes, of course, our fitted $\CC$CDM, which is not the same as the fiducial $\CC$CDM. So all fitted models are compared to the same fiducial model, and this should be the fairest procedure. If, in contrast, we would let the normalization (\ref{P0}) free for each model, we could not secure the Planck 2015 results and the corresponding normalization of the power spectrum. Notice that one cannot adjust the power spectrum and the $f\sigma_8$ values independently. Therefore, we first normalize with Planck 2015 results, as above described, and from here we fit the models to the data, in which the LSS  component takes an essential part. Similarly, upon computing with this method the weighted linear growth rate  $f(z)\sigma_8(z)$ for each model under consideration, (including the $\CC$CDM) the functions $f(z)\sigma_8(z)$ for all models become normalized to the same fiducial model.

In Fig. \ref{fig:fsigma8PRD} we exhibit the results for $f(z)\sigma_8(z)$ for the various models, together with the LSS data measurements (cf. Table \ref{compilationLSSprd}), using the fitted values of Table \ref{tableFit1PRD}. We indicate also in the figure the values that we find for $\sigma_8(0)$ for each model, with the corresponding uncertainties. In addition, we include the curve for model $Q_{\CC}$ under the conditions of Ref.\,\cite{Salvatelli2014}. We disagree both in magnitude and sign concerning their fitted parameter $q_V$ ($\equiv3\nu_{\CC}$ in our notation). We find $q_V>0$ whereas they find $q_V<0$, and our $q_V$ is roughly one order of magnitude smaller in absolute value. The curve with $q_V<0$ clearly has a wrong behavior in the face of the data points, and therefore we exclude such possibility.

\subsection{DVM's and the second law of thermodynamics}\label{subsect:SLT}

For fixed values of the present-day energy densities, if $q_V>0$ (equivalently $\nu_\CC>0$) implies that the vacuum transfers energy to the matter sector and hence there will be more vacuum energy (and less dark matter) in the past than what is predicted by the $\Lambda$CDM. If, on the contrary, $q_V<0$  (equivalently $\nu_\CC<0$), there will be less vacuum energy (and more dark matter) in the past than predicted by the $\Lambda$CDM. This last situation generally entails a violation of the second law of thermodynamics (SLT) since it implies the decay of matter into vacuum energy, and hence the disappearance of particles. The SLT can only be preserved when it is the vacuum that decays into matter rather than the other way around, as particle creation from vacuum introduces the only genuine source of irreversibility here. DM particles are created out of the vacuum at a rate given by $\Gamma_{dm}\sim\Psi_i\propto \nu_i\,H$, see (\ref{PsiFunction1})-(\ref{PsiFunction3}), the rate being positive only if $\nu_i>0$.  Thus, in ordinary conditions, namely in the absence of a significant asymmetry between the number of particles and antiparticles in the Universe (and hence with the chemical potential of the particles being vanishing or very small), the decaying of vacuum into matter is the only thermodynamically safe option\,\cite{SalimWaga93,Lima96,Pavon91}. In the specific case of model $Q_{\CC}$ the situation is particularly odd when  $\nu_\CC<0$, as the corresponding DM mass density  eventually becomes more and more negative:
\begin{equation}\label{eq:mnegative}
\rho_{dm}(a)\longrightarrow\,\frac{\nu_{\CC}}{1-\nu_\CC}\rho_{\CC 0}\,a^{3|\nu_\CC|}<0\,,
\end{equation}
as can be easily checked from Eq.\,(\ref{eq:rhoQL}). This is, of course, rather ugly as it leads to a highly unstable model of the Universe, whose matter content would cascade down into increasingly negative energy states. Both matter and vacuum energy densities would increase without end as $\rho_{m,\CC}\sim a^{3|\nu_\CC|}$ (in absolute value), but in the asymptotic limit (i.e. for $a\to\infty$) the ratio between the two stays constant and can be computed exactly: $\rL/\rho_m=(1-\nu_\CC)/\nu_\CC$. This is a large, but bounded, ratio since $\nu_\CC$ has to be small in absolute value. The ratio is positive for $\nu_\CC>0$ and negative for $\nu_\CC<0$. In the last case the Universe would accumulate an arbitrarily large amount of vacuum energy at the expense of acquiring an arbitrarily large negative mass density!

The asymptotic state of such Universe can be compared to that of the ``big rip'' in the phantom scalar field case\,\cite{CaldwellPDE2003}. In the latter case, $\rho_m\to 0$, whereas the DE density $\rho_D\to\infty$, so that $\rho_D/\rho_m\to \infty$. It is this unbounded ratio that is responsible for the final big rip. In the $Q_\CC$ case with $\nu_\CC<0$, the ratio $\rL/\rho_m$ stays bounded, but the Universe is left with an evermore increasing amount of vacuum energy and negative mass density. A planet bounded in an orbit of radius $R$ (or, for that matter, any bounded state of any sort) has a net negative energy density associated to the binding energy. However, in the $Q_\CC$ model with negative $\nu_\CC$, the Universe is found to acquire an unlimited amount of extra energy density of cosmological origin, given by
\begin{eqnarray}\label{eq:bigripQL}
&&-\frac{4\pi}{3}\sum_{i=m,\CC}\left(\rho_i+3p_i\right)\,R^3=-\frac{4\pi}{3}\left(\rho_m-2\rL\right)\,R^3\to\nonumber\\ &&+\frac{4\pi}{3}\frac{2-3\nu_\CC}{1-\nu_\CC}\rho_{\CC 0}\,a^{-3\nu_\CC}\,R^3\to +\infty\ \ \ \,(\nu_\CC<0)\,,
\end{eqnarray}
and therefore any bounded system eventually becomes dissociated. Such a Universe is in fact driven into a crazy new form of big rip final state, akin to that of usual phantom fields, but with negative mass. A rather weird Universe too!  For $\nu_\CC>0$ the problem disappears altogether since the extra energy that such Universe acquires tends to zero with the expansion, as it is clear from (\ref{eq:bigripQL}).

The $\nu_\CC<0$ version of the $Q_\CC$ Universe is not only at odds with the SLT and leads to a catastrophic end point in its evolution (which, in itself, is of course not forbidden {\it a priori}); its truly inadmissible behavior lies, however, on the fact that it does not fit the structure formation data at all. The highest curve in our Fig.\,\ref{fig:fsigma8PRD}  neatly shows that the instance $\nu_\CC<0$ is an outlier and is definitely ruled out. That curve is computed from the fit value of $q_V\equiv 3\nu_\CC<0$ given in Ref.\,\cite{Salvatelli2014}, and in contrast to these authors we find that such value is unacceptable. Moreover, using our own fit value $\nu_\CC$ (quoted in Table \ref{tableFit1PRD}), which is positive, the $Q_\CC$ model does no longer exhibit such an anomalous departure from the LSS data, see Fig.\,\ref{fig:fsigma8PRD}. For the other dynamical vacuum models under consideration (RVM and $Q_{dm}$) we definitely favor again the option $\nu_i>0$, which perfectly fits in with the expectations of the SLT. For these two models, the asymptotic values of the DM densities never become negative, irrespective of the sign of $\nu_i$, see Eqs.\,(\ref{eq:rhoRVM})-(\ref{eq:rhoQdm}). Thus, while the SLT would still be violated if the sign of the corresponding vacuum parameters were negative,  there is no analogous doomsday for these models.

Even if one would put in doubt the SLT on the grounds that it is applied here to an unusual global system  (the Universe as a whole) the bare fact is that the sign occurrence in $\nu_i$ that violates the SLT  does bluntly contradict the observational data, as we have shown.  For models $Q_{\CC}$ and $Q_{dm}$, our fitting results on $\nu_i>0$ in Table \ref{tableFit1PRD} are compatible with those of \cite{Murgia2016,Li2016,Costa2017}, but in our case we were able to attain much higher accuracy, allowing us to claim significant hints of physics beyond the $\CC$CDM (see below).

It is worth mentioning that one can extend these considerations in terms of the so-called generalized second law  of thermodynamics (GSLT), in which the effect of the area entropy associated to the horizon is also included together with the matter entropy contained inside the horizon\,\cite{GSLT1,GSLT2}. For a detailed discussion of the GSLT in the context of this kind of models and related ones, see\,\cite{MimosoPavon2013}.

Let us finally mention that there are other model variants in which matter decays into vacuum\,\cite{Rafael2017,Zhang2017} under special conditions. We have not analyzed these variants here since they involve \textit{ad hoc} massive neutrinos. In these cases, in order to preserve the thermodynamic standards, the chemical potential of the massive neutrino species should be significant, thereby implying a large neutrino-antineutrino asymmetry associated to a net lepton number of the Universe. As shown in our analysis, we can perfectly dispense with this kind of exotic frameworks since we can substantially improve the $\CC$CDM fit under perfectly reasonable conditions for the DVM's which are in full accordance with the GSLT.

\section{Main numerical results}\label{sect:numerical results}

For the statistical analysis, we define the joint likelihood function as the product of the likelihoods for all the data sets. Correspondingly, for Gaussian errors the total $\chi^2$ to be minimized reads:
\be\label{chi2s}
\chi^2_{tot}=\chi^2_{SNIa}+\chi^2_{BAO}+\chi^2_{H}+\chi^2_{LSS}+\chi^2_{CMB}\,.
\ee
Each one of these terms is defined in the standard way and they include the corresponding covariance matrices (see Appendix \ref{chap:App5} for a detailed explanation).

Table \ref{tableFit1PRD} and Figs.\,\ref{fig:PRD1}-\ref{fig:fsigma8PRD} contain the essentials of our analysis. In particular, Fig.\,\ref{fig:MatrixPlot} displays in a nutshell our main results in all possible planes of the fitting parameter space. Among the other figures, let us stand out here Figs.\,\ref{fig:PRD7}-\ref{RecRVMtriad}, which reveal the very clue to the final results. We will further comment on them in the next sections.
We observe from Figs.\,\ref{fig:PRD1}-\ref{fig:MatrixPlot} that the vacuum parameters, $\nu$ and $\nu_{dm}$, are neatly projected non null and positive for the RVM and the Q$_{dm}$. Remarkably enough, the significance of this dynamical vacuum effect attains, in our analysis, up to about $\sim 4\sigma$ c.l. after marginalizing over the remaining parameters.

%%%%%%%%%%%%%%%%%%%%%%%%%%%%%%%%%%%%%%%%%%%%%%%%%%%%%%%%%%%%%%%%%%%%%%%%%%%%%%%%%%%%%%%%%%%%

\begin{figure}[t!]
\begin{center}
\label{EoSplots}
\includegraphics[width=3.6in]{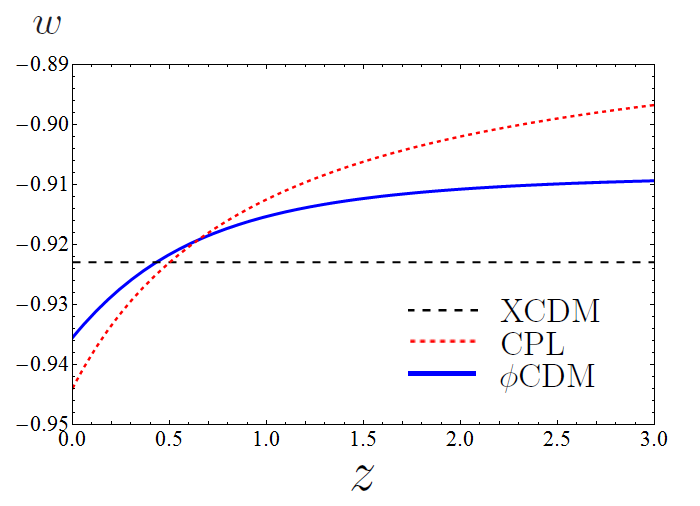}
\caption[EoS parameter $w=w(z)$ for the XCDM and CPL parametrizations and the $\phi$CDM PR model]{{\scriptsize The EoS $w=w(z)$ for the XCDM and CPL parametrizations, as compared to that of a typical $\phi$CDM model. The first two are obtained using the central values of the fitting parameters in Table \ref{tableFit1PRD}. For the XCDM the EoS is of course constant, whereas for the CPL presents some evolution with the redshift, and in both cases the quintessence region $w\gtrsim-1$ is singled out. The value $w_0=-0.923 \pm 0.023$ for the XCDM in Table \ref{tableFit1PRD} points indeed to effective quintessence at $3.35\sigma$ c.l. As for the EoS of the $\phi$CDM model, we have used the original Peebles \& Ratra potential (see Sect.\,\ref{sec:phiCDM}  for details) and fitted it to the same cosmological data (cf. Table \ref{tableFitPhiCDM} in that section). The current value of the EoS can be determined and reads $\omega_{\phi}(z=0)= - 0.936\pm0.019$, which favors once more the quintessence region at $3.37\sigma$ c.l.  These results are compatible with the fitted values for the DVM's in Table \ref{tableFit1PRD}, which signal an effective quintessence behavior ($\nu_i>0$) of vacuum dynamics. For the RVM, $\nu=0.00158 \pm 0.00042$, and thus the significance is at $3.76\sigma$ c.l. In all cases the signal of dynamical DE is favored at a remarkable confidence level.} \label{fig:PRD5}}
\end{center}
\end{figure}

%%%%%%%%%%%%%%%%%%%%%%%%%%%%%%%%%%%%%%%%%%%%%%%%%%%%%%%%%%%%%%%%%%%%%%%%%%%%%%%%%%%%%%%%%%%%%%%

\subsection{Fitting the data with the XCDM and CPL parametrizations}\label{sect:XCDMandCPLnumerical}

Here we further elaborate on the results we have found on exploring the possible time evolution of the DE in terms of the well-known XCDM and CPL parametrizations (introduced in Sect.\,\ref{sect:XCDMandCPL}), in  which the DE is mimicked through the density $\rho_X(a)=\rho_{X0}a^{-3(1+w)}$ associated to some generic (unspecified) entity X, which acts as an ersatz for the  $\CC$ term. Being $\rho_{X0}$ the current energy density value of X, its value must be identified with the measured $\rho_{\CC 0}$. For the XCDM,  $w=w_0$ is the (constant) equation of state (EoS) parameter for X, whereas for the CPL there is also a dynamical component introduced by $w_1$, see Eq.\,(\ref{eq:CPL}). The XCDM trivially boils down to the rigid $\CC$-term for $w_0=-1$, but by leaving $w_0$ free it proves a useful approach to roughly mimic a (non-interactive) DE scalar field with constant EoS. Similarly for the CPL, in which some EoS evolution is allowed; we have, however, in this case two free parameters, $w_0$ and $w_1$, to be determined by the data. The corresponding fitting results for the XCDM parametrization is included in all our tables, along with those for the DVM's and the $\CC$CDM. For the main Table \ref{tableFit1PRD}, and also for Table \ref{tableFit2PRD}, we also include the CPL fitting results. For example, reading off Table \ref{tableFit1PRD} we can see that the best-fit value for $w_0$ in the XCDM is
\begin{equation}\label{eq:woRVM}
w_0=-0.923\pm0.023\,.
\end{equation}
Noteworthy, the obtained EoS value is \emph{not} compatible with a rigid $\CC$-term. It actually departs from $-1$ by precisely $3.35\sigma$. In Fig.\, \ref{fig:PRD3} we depict the contour plot for the XCDM in the $(\Omega_m,w_0)$ plane. Practically all of the $3\sigma$ region lies above the horizontal line at $w_0=-1$. Subsequent marginalization over $\Omega_m$ renders the result (\ref{eq:woRVM}). Concerning the CPL, we can see from Table \ref{tableFit1PRD} that the errors on the fitting parameters are larger, specially on $w_1$. This is not surprising, and was expected given the fact that the CPL has one more parameter than the XCDM. The plot of the CPL EoS function (\ref{eq:CPL}), $w=w(z)$, using the central values of the fitting parameters $w_0$ and $w_1$ in Table \ref{tableFit1PRD}, is shown in Fig. \ref{fig:PRD5}, together with the horizontal line (\ref{eq:woRVM}) for the XCDM. Both the XCDM and the CPL point to the quintessence region ($w\gtrsim-1$) as the preferred one. To ease future comparison with a more realistic quintessence model, in Fig.\, \ref{fig:PRD5} we have also included the evolving EoS, $w=w_\phi(z)$, of a particular scalar field model ($\phi$CDM) with the Peebles \& Ratra potential for $\phi$. This model will be studied in more detail in Sect.\,\ref{sec:phiCDM}.

The following observation is in order. For the first time, even the simplest XCDM parametrization has been able to capture nontrivial signs of dynamical DE and in the form of effective quintessence ($w_0\gtrsim -1$) at more than $3\sigma$ c.l. This is unprecedented, namely in all the analyses in the literature that we are aware of previous to the current one. Obviously, given the significance of this fact, it is highly convenient to compare it with well-known previous fitting studies of the XCDM parametrization, such as the ones performed by the Planck and BOSS collaborations a couple of years ago. The Planck 2015 value for the EoS parameter of the XCDM reads $w_0 = -1.019^{+0.075}_{-0.080}$  \cite{Planck2015} and the BOSS one is $w_0 = -0.97\pm 0.05$\,\cite{Aubourg2015}. These results are perfectly compatible with our own fitting value for $w_0$ given in (\ref{eq:woRVM}), but in stark contrast to it their errors are big enough as to be also fully compatible with the $\CC$CDM value $w_0=-1$. This is not too surprising if we take into account that none of these analyses included LSS data in their fits, as explicitly recognized in the text of their papers. Furthermore, at the time when these analyses appeared they could not have used the important LSS and BAO results from \cite{GilMarin2} -- which we have, of course, incorporated as part of our current data set -- not even the previous ones from\,\cite{GilMarin1}. In Sect.\,\ref{sect:OtherDataSets}, specifically in Table \ref{tableFit6PRD} there, we discuss the result of our fit if we would use precisely the data by Planck 2015\,\cite{Planck2015,PlanckDE2015}.

In the absence of the modern LSS data we would find a similar situation. In fact, as our Table \ref{tableFit4PRD} clearly shows, the removal of the LSS data set in our fit induces a significant increase in the magnitude of the central value of the EoS parameter, as well as the corresponding error. This happens because the higher is $|w|$ the higher is the structure formation power predicted by the XCDM, and therefore the closer is such prediction with that of the $\CC$CDM (which is seen to predict too much power as compared to the data, see Fig. \ref{fig:fsigma8PRD}). Under these conditions our analysis renders $w = -0.992\pm 0.040$, which is definitely closer to the central values obtained by Planck and BOSS teams. In addition, this result is now fully compatible with the $\CC$CDM, as in the Planck 2015 and BOSS cases, and all of them are unfavored by the LSS observations. This is consistent with the fact that both information criteria, $\Delta$AIC and $\Delta$BIC, become now slightly negative in Table \ref{tableFit4PRD}, which reconfirms that if the LSS data are not used the $\CC$CDM performance is comparable or even better than the dynamical DE models. So in order to fit the observed values of $f\sigma_8$, which are generally lower than the predicted ones by the $\CC$CDM, $|w|$ should decrease. This is just what happens for the XCDM; and at the same time it also amounts to an increase of $\nu_i$ for the DVM's when the LSS data are restored in our analysis (in combination with the other data, particularly with BAO and CMB data). It is apparent from Fig.\,\ref{fig:fsigma8PRD} that the $f(z)\sigma_8(z)$ curves for these models are then shifted below and hence adapt significantly better to the data points. Correspondingly, the quality of the fits increases dramatically for the XCDM, RVM and Q$_{dm}$, and this is also borne out by the large and positive values of $\Delta$AIC and $\Delta$BIC (as can be checked in Table \ref{tableFit1PRD}).

The above discussion explains why our analysis of the observations through the XCDM is sensitive to the dynamics of the DE, whereas the previous results in the literature are not. It also shows that the size of the effect found with such a parametrization of the DE essentially concurs with the strength of the dynamical vacuum signature found for the RVM and Q$_{dm}$ using exactly the same data. This was not obvious {\it a priori}, since for the DVM's there is an interaction between vacuum and matter that triggers an anomalous conservation law, whereas for the XCDM we do not have such interaction (meaning that matter is conserved in them, thereby following the standard decay laws for relativistic and nonrelativistic components). The interaction, when occurs, is proportional to $\nu_i$ and thus is small. This probably explains why the XCDM can succeed in nailing down the dynamical nature of the DE. It goes without saying that not all dynamical vacuum models describe the data with the same efficiency. The Q$_{\CC}$ turns out to improve the behavior of the $\Lambda$CDM, but only in a very soft way and, as it is explained below, and applying Occam's razor arguments, we can say that the achieved improvement is not statistically significant.

A detailed comparison is made among models similar (but different) from those addressed here in Chapters \ref{chap:Atype}, \ref{chap:Gtype} and \ref{chap:AandGRevisited}. In the XCDM case the departure from the $\CC$CDM takes the fashion of ``effective quintessence'', whereas for the DVM's it appears as genuine vacuum dynamics. In all cases except in the Q$_{\CC}$, we find compelling signs of DE physics beyond the $\CC$CDM (cf. Table \ref{tableFit1PRD}), and this is a most important result of our work.

\subsection{On the purported tension between the measured values of the Hubble parameter}\label{subsect:Htension}

Recently, a significant tension between the local and nonlocal measurements of the Hubble parameter $H_0$ obtained from different types of observations has generated some perplexity in the literature. There is a kind of bipolarization in two groups of values.  Most conspicuously, Riess {\it et al.} reported  $H_0 =
73.24\pm 1.74$  Km/s/Mpc\,\cite{RiessH0}, based on Hubble Space Telescope data. This value is $\sim 3.3\sigma$ higher than the Planck result $H_0=67.51 \pm 0.64$ Km/s/Mpc\,\cite{Planck2015} inferred from the CMB.  For other measurements and discussions, see e.g. \cite{WMAP9}, \cite{Aubourg2015}, \cite{RiessH02011}, \cite{ChenRatra2011}, \cite{Freedman2012}, and \cite{ACTSievers}. This situation has stimulated a number of discussions and possible solutions in the literature, see e.g. \,\cite{Melchiorri2016,VerdeRiessH0,Melchiorri2017} and references therein. 

As indicated before, owing to this lack of consensus on the experimental value of $H_0=100 h$ Km/s/Mpc, in this work we have fitted $h$ using an uninformative flat prior. This should be fairer in these circumstances, rather than making a subjective choice among the different measurements of $h$ put forward in the aforementioned papers.  One can see from Table \ref{tableFit1PRD} that our fitted value of the local Hubble parameter within the $\CC$CDM is $H_0=69.2\pm 0.4$ Km/s/Mpc. Compared to $H_0 =
73.24\pm 1.74$  Km/s/Mpc from Riess {\it et al.} it is only $2.3\sigma$, i.e one standard deviation less than the discrepancy with the Planck 2015 measurement. Obviously this result helps to alleviate the discrepancy, which is now not so severe.
Let us recall that in our fit we use not only the CMB data from Planck 2015 but a rich variety of observational sources, which were not considered (and some of them not even available) in the fitting analysis performed by Planck in 2015. We are mainly referring to the wealth of BAO+LSS data that we have used in our own analysis, see items DS2, DS3 and DS5 of Sect. \ref{sect:DataSets}. Our fitting result for the $\CC$CDM is by the way perfectly in agreement with the WMAP+ACT+SPT+BAO result $H_0=69.3\pm 0.7$ Km/s/Mpc\,\cite{RiessH0} and is in between the bipolarized local and nonlocal values mentioned above.

It is not obvious to us that the purported tension is suggesting some systematic uncertainties in CMB measurements, as has been claimed. Our $\CC$CDM  fit involving the full set of SNIa+BAO+$H(z)$+LSS+CMB data described in  Sect. \ref{sect:DataSets} is actually in good agreement with Planck and WMAP, and it suggests that once a richer set of observations (specially on LSS) are included the tension loses substantially.  As far as our favorite dynamical vacuum model RVM is concerned, which gives a significantly better fit than the $\CC$CDM, we can see from Table \ref{tableFit1PRD} that we tend to favor lower values of the local Hubble parameter:  $H_0=67.7\pm 0.5$ Km/s/Mpc. This has been the main trend of the cosmological observations for years, see the summary of \cite{ChenRatra2011}, the more recent analysis \cite{ChenKumarRatra2016}, and Chapter \ref{chap:H0tension} for further details and references.

\subsection{Comparing the competing vacuum models through Akaike and Bayesian information criteria}\label{subsect:AICandBIC}

We may judge the fit quality obtained for the different vacuum models in this work from a different perspective. Although the $\chi^2_{\rm min}$ value of the overall fits for the main DVM's (RVM and $Q_{dm}$) and XCDM appear to be definitely smaller than the $\CC$CDM one, it proves extremely useful to reassess the degree of success of each competing model by invoking the time-honored Akaike and Bayesian information criteria, as in previous chapters.

Table \ref{tableFit1PRD} reveals conspicuously that the $\CC$CDM appears very strongly disfavored (according to the above statistical standards) when confronted to some of the dynamical DE models. More precisely, for the RVM and the Q$_{dm}$ the $\Delta$AIC and $\Delta$BIC are in the range $12-13$ and $9-11$, respectively. These results are fully consistent and outstanding; and since both $\Delta$AIC and $\Delta$BIC are above $9-10$ the verdict of the information criteria is conclusive. For instance, the Bayes factor, i.e. $B\equiv e^{\Delta{\rm BIC}/2}$, in favor of the RVM relative to the $\CC$CDM is larger than $B=e^{5.3}\sim 200$. The situation with model $Q_{dm}$ is comparable.
%%%%%%%%%%%%%%%%%%%%%%%%%%%%%%%%%%%%%%%%%%%%%%%%%%%%%%%%%%%%%%%%%%%%%%%%%%%%%%%%%%%%%%%%%%%%%%
%%%%%%%%%%%%%%%%%%%%%%%%%%%%%%%%%%%%%%%%%%%%%%%%%%%%%%%%%%%%%%%%%%%%%%%%%%%%%%%%%%%%%%%%%%%%%%%
\begin{table}
\begin{center}
\begin{scriptsize}
\resizebox{1\textwidth}{!}{
\begin{tabular}{| c | c |c | c | c | c | c | c | c | c |}
\multicolumn{1}{c}{Model} &  \multicolumn{1}{c}{$h$} &  \multicolumn{1}{c}{$\omega_b= \Omega_b h^2$} & \multicolumn{1}{c}{{\small$n_s$}}  &  \multicolumn{1}{c}{$\Omega_m$}&  \multicolumn{1}{c}{{\small$\nu_i$}}  & \multicolumn{1}{c}{$w$} &
\multicolumn{1}{c}{$\chi^2_{\rm min}/dof$} & \multicolumn{1}{c}{$\Delta{\rm AIC}$} & \multicolumn{1}{c}{$\Delta{\rm BIC}$}\vspace{0.5mm}
\\\hline
{\small $\CC$CDM} & $0.679\pm 0.005$ & $0.02241\pm 0.00017$ &$0.968\pm 0.005$& $0.291\pm 0.005$ & - & -1 &  67.86/83 & - & -\\
\hline
XCDM  &  $0.674\pm 0.007$& $0.02241\pm 0.00017 $&$0.968\pm0.005$& $0.298\pm 0.009$& - & $-0.960\pm0.038$ &  66.79/82 & -1.18 & -3.40 \\
\hline
RVM  & $0.679\pm 0.008$& $0.02241\pm 0.00017$&$0.968\pm 0.005$& $0.296\pm 0.015$ & $0.00061\pm 0.00158$ & -1 &  67.71/82 & -2.10 & -4.32 \\
\hline
$Q_{dm}$ &  $0.677\pm 0.008$& $0.02241\pm 0.00017 $&$0.968\pm0.005$& $0.296\pm 0.015 $ & $0.00086\pm 0.00228 $ & -1 &  67.71/82  & -2.10 & -4.32 \\
\hline
$Q_\CC$  &  $0.679\pm 0.005$& $0.02241\pm 0.00017 $&$0.968\pm0.005$& $0.297\pm 0.013$ & $0.00463\pm 0.00922$ & -1 &  67.59/82 & -1.98 & -4.20 \\
\hline
 \end{tabular}
}
\end{scriptsize}
\end{center}
\caption[Fitting results obtained after the removal of the CMB data]{{\scriptsize Same as in Table \ref{tableFit1PRD}, but removing both the $R$-shift parameter and the acoustic length $l_a$ from our fitting analysis.}\label{tableFit3PRD}}
\end{table}
\begin{table}
\begin{center}
\begin{scriptsize}
\resizebox{1\textwidth}{!}{
\begin{tabular}{| c | c |c | c | c | c | c | c | c | c |}
\multicolumn{1}{c}{Model} &  \multicolumn{1}{c}{$h$} &  \multicolumn{1}{c}{$\omega_b= \Omega_b h^2$} & \multicolumn{1}{c}{{\small$n_s$}}  &  \multicolumn{1}{c}{$\Omega_m$}&  \multicolumn{1}{c}{{\small$\nu_i$}}  & \multicolumn{1}{c}{$w$} &
\multicolumn{1}{c}{$\chi^2_{\rm min}/dof$} & \multicolumn{1}{c}{$\Delta{\rm AIC}$} & \multicolumn{1}{c}{$\Delta{\rm BIC}$}\vspace{0.5mm}
\\\hline
{\small $\CC$CDM} & $0.685\pm 0.004$ & $0.02243\pm 0.00014$ &$0.969\pm 0.004$& $0.304\pm 0.005$ & - & -1 &  61.45/72 & - & -\\
\hline
XCDM  &  $0.684\pm 0.009$& $0.02244\pm 0.00015 $&$0.969\pm0.004$& $0.305\pm 0.007$& - & $-0.992\pm0.040$ &  61.41/71 & -2.25 & -4.29 \\
\hline
RVM  & $0.684\pm 0.008$& $0.02242\pm 0.00016$&$0.969\pm 0.005$& $0.304\pm 0.005$ & $0.00014\pm 0.00103$ & -1 &  61.43/71 & -2.27 & -4.31 \\
\hline
$Q_{dm}$ &  $0.685\pm 0.007$& $0.02242\pm 0.00016 $&$0.969\pm0.005$& $0.304\pm 0.005 $ & $0.00019\pm 0.00126 $ & -1 &  61.43/71  & -2.27 & -4.31 \\
\hline
$Q_\CC$  &  $0.686\pm 0.004$& $0.02240\pm 0.00017 $&$0.968\pm0.005$& $0.304\pm 0.005$ & $0.00090\pm 0.00330$ & -1 &  61.37/71 & -2.21 & -4.25 \\
\hline
 \end{tabular}
 }
\end{scriptsize}
\end{center}
\caption[Fitting results obtained after the removal of the LSS data]{{\scriptsize Same as in Table \ref{tableFit1PRD}, but removing the LSS data set from our fitting analysis.}\label{tableFit4PRD}}
\end{table}
%%%%%%%%%%%%%%%%%%%%%%%%%%%%%%%%%%%%%%%%%%%%%%%%%%%%%%%%%%%%%%%%%%%%%%%%

\newpage
It goes without saying that not all dynamical vacuum models describe the data with the same efficiency. In the case of the Q$_{\CC}$ model the improvement is so mild that Occam's razor criterion (``among equally competing models describing the same observations, choose the simplest one'') would probably not recommend its choice. However, since the other dynamical DE models are able to describe the current observations significantly better than the $\CC$CDM, and not just alike,  Occam's razor should definitely bet in their favor. The AIC and BIC criteria can be thought of as a modern quantitative formulation of Occam's razor, in which the presence of extra parameters in a given model is conveniently penalized so as to achieve a fairer comparison with the model having less parameters.

Thus, e.g. model $Q_\Lambda$ despite being also better than the traditional $\Lambda$-picture, the values of $\Delta$AIC and $\Delta$BIC are sitting in the much more moderate range $1-3$ (cf. Table \ref{tableFit1PRD}),
and hence that model is certainly not comparable in efficiency to the main DVM's. Model $Q_\CC$ is also left behind in comparison to the output of the simple XCDM and CPL parametrizations. The corresponding XCDM values of $\Delta$AIC and $\Delta$BIC are in the range $6-9$ (reconfirming the ability of the XCDM to substantially improve the $\CC$CDM fit). Similarly for the CPL parametrization, where $\Delta$AIC and $\Delta$BIC stay in the approximate range $2-6$. This also confirms a superior performance of the CPL versus the $\CC$CDM and the $Q_\CC$ model in their different abilities for describing the data. The lesser quality fit of the CPL with respect to the XCDM, though, is obviously caused by the presence of an extra fitting parameter in the first, as previously pointed out.

It is remarkable the amount of evidence on dynamical DE that can be presently gathered with the simple XCDM parametrization. As formerly noted, to the best of our knowledge it is unprecedented in the literature. It is even more remarkable, if we realize that it equals the level of evidence that we will be able to pick up from the analysis of a full-fledged quintessence model with a given potential, see Sect.\,\ref{sec:phiCDM}.
Nonetheless the difference of roughly $4$ (positive) points in the value of $\Delta$AIC and $\Delta$BIC in favor of the main DVM's with respect to the XCDM is considered significant from the point of view of the information criteria. Therefore, the simple XCDM approach lags behind the main dynamical vacuum models under consideration.
The RVM and Q$_{dm}$ stand out here as superior competing candidates as compared to the XCDM (and in fact compared to all the other models under scrutiny in this work) in their capacity to fit the overall cosmological data.

%%%%%%%%%%%%%%%%%%%%%%%%%%%%%%%%%%%%%%%%%%%%%%%%%%%%%%%%%%%%%%%%%%%%%%%%%%%%%
\begin{table}
\begin{center}
\begin{scriptsize}
\resizebox{1\textwidth}{!}{
\begin{tabular}{| c | c |c | c | c | c | c | c | c | c |}
\multicolumn{1}{c}{Model} &  \multicolumn{1}{c}{$h$} &  \multicolumn{1}{c}{$\omega_b= \Omega_b h^2$} & \multicolumn{1}{c}{{\small$n_s$}}  &  \multicolumn{1}{c}{$\Omega_m$}&  \multicolumn{1}{c}{{\small$\nu_i$}}  & \multicolumn{1}{c}{$w$} &
\multicolumn{1}{c}{$\chi^2_{\rm min}/dof$} & \multicolumn{1}{c}{$\Delta{\rm AIC}$} & \multicolumn{1}{c}{$\Delta{\rm BIC}$}\vspace{0.5mm}
\\\hline
{\small $\CC$CDM} & $0.698\pm 0.005$ & $0.02263\pm 0.00015$ &$0.977\pm 0.004$& $0.288\pm 0.006$ & - & -1 &  61.25/74 & - & -\\
\hline
XCDM  &  $0.681\pm 0.015$& $0.02262\pm 0.00015 $&$0.976\pm0.004$& $0.303\pm 0.014$& - & $-0.949\pm0.042$ &  59.72/73 & -0.75 & -2.83 \\
\hline
RVM  & $0.683\pm 0.007$& $0.02242\pm 0.00016$&$0.969\pm 0.005$& $0.296\pm 0.007$ & $0.00138\pm 0.00049$ & -1 &  52.95/73 & 6.02 & 3.94\\
\hline
$Q_{dm}$ &  $0.685\pm 0.007$& $0.02241\pm 0.00016 $&$0.968\pm0.005$& $0.295\pm 0.007 $ & $0.00192\pm 0.00069 $ & -1 &  52.91/73  & 6.06 & 3.98 \\
\hline
$Q_\CC$  &  $0.701\pm 0.005$& $0.02240\pm 0.00017 $&$0.968\pm0.005$& $0.287\pm 0.006$ & $0.00733\pm 0.00257$ & -1 &  53.36/73 & 5.61 & 3.53 \\
\hline
 \end{tabular}
}
\end{scriptsize}
\end{center}
\caption[Fitting results obtained after the removal of the BAO data]{{\scriptsize Same as in Table \ref{tableFit1PRD}, but removing the BAO data set from our fitting analysis.}\label{tableFitX1}}
\end{table}
%%%%%%%%%%%%%%%%%%%%%%%%%%%%%%%%%%%%%%%%%%%%%%%%%%%%%%%%%%%%%%%%%%%%%%%%%%%%%%%%%%%

\section{Discussion}\label{sect:discussion}

 \subsection{Anomalous matter conservation law}

As we have discussed in Section \ref{sect:DVMs}, for the DVM's there is an interaction between vacuum and matter. Such interaction is, of course, small because the fitted values of $\nu_i$  are small, see Table \ref{tableFit1PRD}. The obtained values are in the ballpark of $\nu_i\sim {\cal O}(10^{-3})$ and therefore this is also the order of magnitude associated to the anomalous conservation law of matter. For example, for the nonrelativistic component in the RVM we have
\begin{equation}
\rho_m(a)=\rho_{dm}(a)+\rho_b(a)=\rho_{m0}a^{-3(1-\nu)}\,.
\end{equation}
Here $\rho_{m0}=\rho_{dm0}+\rho_{b0}$, and use has been made of the conservation law of baryons, Eq.\,(\ref{eq:BaryonsRadiation}), as well as of the DM density in the RVM,  Eq.\,\eqref{eq:rhoRVM}.
As previously mentioned, the possible anomalous behavior of matter conservation has been exploited in devoted works such as \cite{FritzschSola2012,FritzschSola2015,FritzschSolaNunes2017}. These are essentially based on the RVM as a possible explanation for the various hints on the time variation of the fundamental constants, such as coupling constants and particle masses, frequently considered in the literature. See e.g. \cite{Uzan2011,Chiba2011} and \cite{Sola2015editor} for reviews.

%%%%%%%%%%%%%%%%%%%%%%%%%%%%%%%%%%%%%%%%%%%%%%%%%%%%%%%%%%%%%%%%%%%%%%%%%%%%%%%%%%%%%%%%%%%%%%%

\begin{table}[t!]
\begin{center}
\begin{scriptsize}
\resizebox{1\textwidth}{!}{
\begin{tabular}{| c | c |c | c | c | c | c | c | c | c |}
\multicolumn{1}{c}{Model} &  \multicolumn{1}{c}{$h$} &  \multicolumn{1}{c}{$\omega_b= \Omega_b h^2$} & \multicolumn{1}{c}{{\small$n_s$}}  &  \multicolumn{1}{c}{$\Omega_m$}&  \multicolumn{1}{c}{{\small$\nu_i$}}  & \multicolumn{1}{c}{$w$} &
\multicolumn{1}{c}{$\chi^2_{\rm min}/dof$} & \multicolumn{1}{c}{$\Delta{\rm AIC}$} & \multicolumn{1}{c}{$\Delta{\rm BIC}$}\vspace{0.5mm}
\\\hline
{\small $\CC$CDM} & $0.694\pm 0.005$ & $0.02265\pm 0.00022$ &$0.976\pm 0.004$& $0.293\pm 0.007$ & - & -1 &  38.98/39 & - & -\\
\hline
XCDM  &  $0.684\pm 0.010$& $0.02272\pm 0.00023 $&$0.977\pm0.005$& $0.299\pm 0.009$& - & $-0.961\pm0.033$ &  37.61/38 & -1.20 & -2.39 \\
\hline
RVM  & $0.685\pm 0.009$& $0.02252\pm 0.00024$&$0.971\pm 0.006$& $0.297\pm 0.008$ & $0.00080\pm 0.00062$ & -1 &  37.29/38 & -0.88 & -2.07 \\
\hline
$Q_{dm}$ &  $0.686\pm 0.008$& $0.02251\pm 0.00025 $&$0.971\pm0.006$& $0.297\pm 0.008 $ & $0.00108\pm 0.00088 $ & -1 &  37.43/38  & -1.02 & -2.21\\
\hline
$Q_\CC$  &  $0.694\pm 0.006$& $0.02258\pm 0.00029 $&$0.974\pm0.007$& $0.293\pm 0.007$ & $0.00167\pm 0.00471$ & -1 &  38.86/38 & -2.45 & -3.64 \\
\hline
 \end{tabular}
}
\end{scriptsize}
\end{center}
\caption[Fitting results obtained using the same data set as the Planck Collaboration\,\cite{PlanckDE2015}]{{\scriptsize As in Table \ref{tableFit1PRD}, but using the same data set as the Planck Collaboration\,\cite{PlanckDE2015}. See the text for further details.}
\label{tableFit6PRD}}
\end{table}

%%%%%%%%%%%%%%%%%%%%%%%%%%%%%%%%%%%%%%%%%%%%%%%%%%%%%%%%%%%%%%%%%%%%%%%%%%%%%%%%%%%%

Let us note that the time variation of the mass density can be interpreted either as an anomalous change in the number density of particles during the expansion or as a change in their mass values\,\cite{FritzschSola2012}. It is interesting that the current observational limits for such time variation are compatible with the fitted values we have found here\,\cite{FritzschSolaNunes2017}. The possibility that the baryon masses can also change slowly with the cosmic evolution has also been tested, and could be connected with the possible time variation of the fine structure constant\,\cite{FritzschSolaNunes2017}, but the limits are rather strict, much more than the possible time variation of the DM masses. This is the reason why we have assumed from the beginning that the exchange between matter and vacuum energy is strictly confined to the DM sector. If there is some exchange with baryons, it must be so small that it cannot influence significantly the main cosmological considerations that are under study here.

The potential time variation of the fundamental constants is a very active field of research and is therefore nowadays of high interest, see e.g.\,\cite{Sola2015editor}. As we can see, there are different phenomenological possibilities that can be used to test the RVM and other DVM's from various points of view. A positive measurement of that kind of effects could be interpreted as strong support to the `micro and macro connection' hypothesis, viz. the dynamical feedback between the evolution of the cosmological parameters and the time variation of the fundamental constants of the microscopic world\,\cite{FritzschSola2012}.

\subsection{Testing the impact of the different data sets in our main analysis and comparing with Planck 2015, BOSS and DES}\label{sect:OtherDataSets}

The current analysis follows the track of the one carried out in Chapter \ref{chap:Gtype} and is also firmly aligned with the one performed in Chapter \ref{chap:AandGRevisited}. Although the models analyzed in the aforementioned chapters have some differences with the ones treated here, the outcome of the analysis points to the very same direction. Some DVM's and the XCDM do fit considerably better the available data than the $\Lambda$CDM. But we want to emphasize some important aspects of the analysis carried out in this chapter as compared to the other analysis:

\begin{itemize}

\item We have used a large and fully updated set of cosmological  SNIa+BAO+$H(z)$+LSS+ CMB observations. To our knowledge, this is one of the most complete and consistent data sets used in the literature. The current data set has been well tested in the previous two chapters, but differs from the one used in these analyses in that we have now used the updated  BAO and LSS analysis from Ref.\,\cite{GilMarin2}, see more details in Sect.\,\ref{sect:GilMarin2OLD}.

\item The selected string SNIa+BAO+$H(z)$+LSS+CMB of data has been obtained from independent analyses in the literature, see the detailed description of the data sets DS1-DS6) in Sect.\,\ref{sect:DataSets} and references therein.

\item We have taken into account all the known covariance matrices in the total $\chi^2$ function \eqref{chi2s}, which means that we have taken into account all the known correlations among the data. See Appendix \ref{chap:App5} for a detailed explanation.  Not all data sets existing in the literature are fully consistent, sometimes they are affected from important correlations that have not been taken into account. Recall in particular our discussion on correlated data sets in items DS3, DS4 and DS5 of Sect.\,\ref{sect:DataSets}, including the comparison of our fitting results in Tables \ref{tableFit1PRD} and \ref{tableFit2PRD}.

\item We have removed from our analysis all data that are obviously correlated. As an example, we have avoided to use Hubble parameter data extracted from BAO measurements, see DS4. Similarly, we have avoided to perform double counting using ``different'' data releases that are fully or partially overlapping; in particular, we have carefully averted using subsets of data that are part of a bigger set and then treat them all as independent data sets. From our data set list in Sect.\,\ref{sect:DataSets} and the attached explanations there the reader can check that we have thoroughly complied with these important requirements.

\end{itemize}

Altogether, this explains the accuracy obtained in the current fitted values of the vacuum parameters $\nu_i$ (and the EoS parameter $w_0$ of the XCDM),
%In particular, we should stress that for the present analysis we are using a very complete and restrictive BAO data set. Here we are using a total of 11 BAO points. %(none of them based on the $A(z)$ estimator, see DS2-DS3) in Sect. \ref{sect:DataSets}. As explained, these include the recent results from\,\cite{GilMarin2} on BAO and LSS, with the known correlation matrices, which altogether narrow down the allowed parameter space in a very efficient way.
and the significant reduction in the error bars with respect to the ones we had previously obtained in the first part of this thesis.

%%%%%%%%%%%%%%%%%%%%%%%%%%%%%%%%%%%%%%%%%%%%%%%%%%%%%%%%%%%%%%%%%%%%%%%%%%%%%%%%
\begin{table}
\begin{center}
\resizebox{1\textwidth}{!}{
\begin{tabular}{| c | c |c | c | c | c | c | c | c | c |}
\multicolumn{1}{c}{Model} &  \multicolumn{1}{c}{$h$} &  \multicolumn{1}{c}{$\omega_b= \Omega_b h^2$} & \multicolumn{1}{c}{{\small$n_s$}}  &  \multicolumn{1}{c}{$\Omega_m$}&  \multicolumn{1}{c}{{\small$\nu_i$}}  & \multicolumn{1}{c}{$w$} &
\multicolumn{1}{c}{$\chi^2_{\rm min}/dof$} & \multicolumn{1}{c}{$\Delta{\rm AIC}$} & \multicolumn{1}{c}{$\Delta{\rm BIC}$}\vspace{0.5mm}
\\\hline
{\small $\CC$CDM} & $0.691\pm 0.004$ & $0.02251\pm 0.00014$ &$0.973\pm 0.004$& $0.297\pm 0.005$ & - & -1 &  79.67/88 & - & -\\
\hline
XCDM  &  $0.676\pm 0.008$& $0.02257\pm 0.00014 $&$0.974\pm0.004$& $0.309\pm 0.008$& - & $-0.945\pm0.029$ &  76.14/87 & 1.29 & -0.99 \\
\hline
RVM  & $0.680\pm 0.006$& $0.02235\pm 0.00015$&$0.966\pm 0.005$& $0.302\pm 0.006$ & $0.00120\pm 0.00047$ & -1 &  72.74/87 & 4.69 & 2.41 \\
\hline
$Q_{dm}$ &  $0.680\pm 0.006$& $0.02235\pm 0.00015 $&$0.966\pm0.005$& $0.302\pm 0.006 $ & $0.00162\pm 0.00065 $ & -1 &  73.11/87  & 4.32 & 2.04 \\
\hline
$Q_\CC$  &  $0.691\pm 0.004$& $0.02234\pm 0.00017 $&$0.967\pm0.005$& $0.298\pm 0.005$ & $0.00502\pm 0.00261$ & -1 &  76.01/87 & 1.42 & -0.86 \\
\hline
 \end{tabular}
}
\end{center}
\caption[Fitting results obtained using the BOSS data from \cite{Alam2016} instead of \cite{GilMarin2}]{{\scriptsize Same as Table \ref{tableFit1PRD}, but using the BOSS data from \cite{Alam2016} instead of \cite{GilMarin2}. In contradistinction to the latter, the former does not include the bispectrum effects in their results. See the discussion.}\label{tableFitX5}}
\end{table}

%%%%%%%%%%%%%%%%%%%%%%%%%%%%%%%%%%%%%%%%%%%%%%%%%%%%%%%%%%%%%%%%%%%%%%%%%%%%%%%%%%%%%%%%%%%%%%

We have actually performed a practical test in order to check what would be the impact on the fitting quality of our main analysis (i.e. the one presented in Table \ref{tableFit1PRD}) if we would remove some of the data points included in the current work. For instance, if we remove from our analysis crucial ingredients of the CMB data, such as the acoustic length $l_a$  and at the same time replace the current BAO data points by those of \cite{Blake11}, relying on the BAO $A(z)$-estimator. Notice that the CMB part is now left essentially with the $R$-shift parameter only.  This is precisely the old situation that we considered in Chapter \ref{chap:Gtype} with other models, and we obtain indeed results consistent with those presented there: namely, the error bars' size become $\sim 4-5$ times larger than the current ones, i.e. of order $\mathcal{O}(10^{-3})$. We have also checked what would be the effect on our fit if we would remove both the data on the shift parameter and on the acoustic length;  or if we would remove only the data points on LSS. In the first case it means to essentially dispense with the CMB data, and in the second case to ignore the information on structure formation. The results are presented in Tables \ref{tableFit3PRD} and \ref{tableFit4PRD}, respectively. We observe that in both cases the vacuum parameters are compatible with zero at one $\sigma$, and the  $\Delta$AIC and $\Delta$BIC values become $\sim 2-4$ points negative. This means that under these conditions the $\CC$CDM does better than the DVM's. The obvious conclusion is that the full CMB and LSS data are individually very important for the quality of the fit and that without a single one of these data sources the dynamical DE signal gets completely lost. 

We would also like to make some comments on the recent study \cite{Heavens2017}, in which the authors find no evidence for extensions to the standard cosmological model, in particular no evidence of dynamical DE. They find that in the case of evolving DE the Bayes factor $B$ with respect to the $\CC$CDM (cf. Sect.\,\ref{subsect:AICandBIC}) satisfies $\ln B\simeq -3.2$. The negative sign implies negative evidence. We should say that this result is actually compatible with ours. The reason is that these authors did not use LSS data at all. As explained above, no evidence can be found of dynamical DE if the LSS data are missing.
For example, if we look at our Table \ref{tableFit4PRD} (corresponding to the situation when the LSS data are \emph{not} included), we find that the simple XCDM parametrization of the dynamical DE renders $\Delta$BIC$=-4.29$, what implies $\ln B\simeq -2.15$. This negative result agrees with the mentioned one from\,\cite{Heavens2017} since differences of one unit are not significant for the AIC and BIC information criteria. 

It is only when the triplet CMB+BAO+LSS is used that the dynamical DE signal appears with crisp clarity at $3-4\,\sigma$ c.l. depending on the model. In the case of the XCDM (which should be the most comparable one with the generic evolving DE case studied in\,\cite{Heavens2017}) it appears with $3.35\sigma$ c.l. (as emphasized in the main text of our paper), and this translates into \emph{positive} Bayesian evidence of $\ln B=\Delta$BIC/2$=+3.15$ (cf. Table \ref{tableFit1PRD}). This result is 6.3 units higher than the one indicated by these authors, and hence it is highly significant from the point of view of the information criteria. 

Therefore, in our understanding, the reason why the authors of \cite{Heavens2017} did not find evidence for dynamical DE is that they did not use LSS data, apart from the lensing data used by the Planck collaboration. In fact, this is also the reason why the recent one-year results by DES also missed this
signal \cite{DES2017}: they do not use direct data on LSS structure formation despite they recognize that smaller values of $\sigma_8(0)$ than those predicted by the $\CC$CDM are necessary to solve the tension existing between the concordance model and the LSS observations. They focus on the
observable $S_8 = 0.783^{+0.021}_{-0.023}$ as the only tracer for structure formation. Employing that value in our analysis and removing all LSS data from
our fit (i.e. replacing all our structure formation data with the $S_8$-parameter as the only tracer for structure formation) we have checked explicitly that the Bayesian evidence becomes $ln\,B < 0$, therefore in agreement with Ref. \cite{Heavens2017}, and hence a result blind to dynamical DE. This is of course in agreement with the result reported by the DES itself, thus showing that the framework of our analysis is consistent with the good compatibility found between Planck and DES, and therefore no sign of physics beyond the $\Lambda$CDM can be reported by any of them in such conditions.

%%%%%%%%%%%%%%%%%%%%%%%%%%%%%%%%%%%%%%%%%%%%%%%%%%%%%%%%%%%%%%%%%%%%%%%%%%%%%%%%%%%%%%%%%%%%%%%%
%
\begin{figure}
\centering
\includegraphics[angle=0,width=1.03\linewidth]{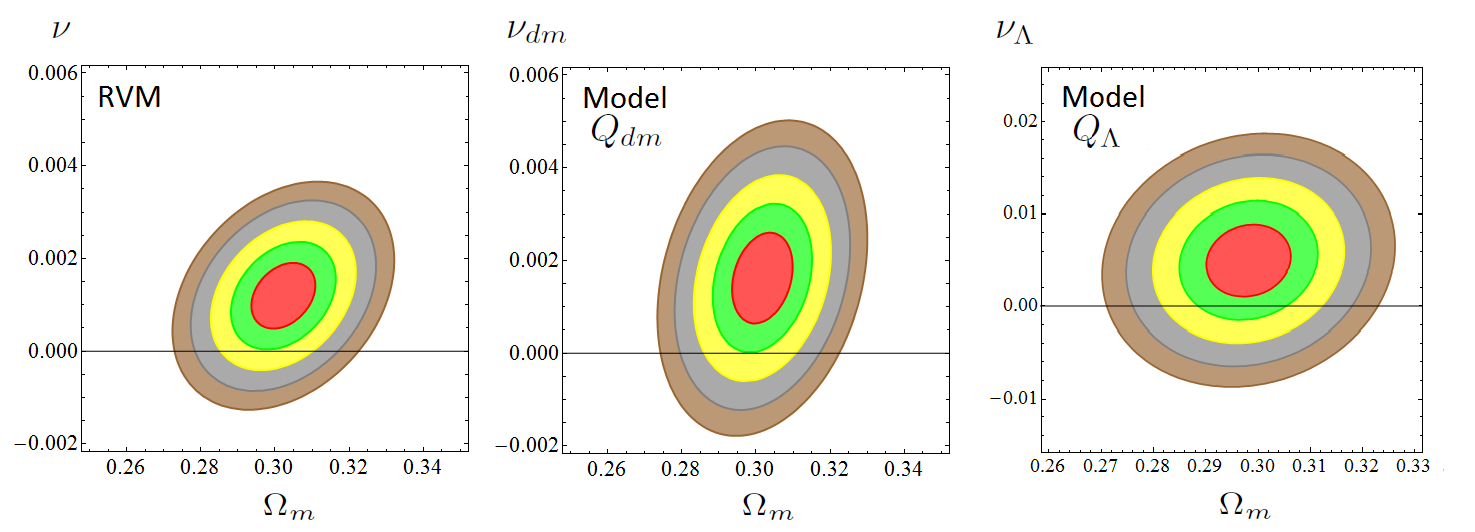}
\caption[Cl's for the DVM's obtained using the BOSS data from \cite{Alam2016} instead of \cite{GilMarin2}]{{\scriptsize As in Fig. \ref{fig:PRD1}, but using the BOSS data of \cite{Alam2016}, which does not include the bispectrum, in contrast to the BOSS data of \cite{GilMarin2}, which does include it. The fitting results associated to this figure are provided in Table \ref{tableFitX5}}.\label{fig:NoBispectrum}}
\end{figure}
%

%%%%%%%%%%%%%%%%%%%%%%%%%%%%%%%%%%%%%%%%%%%%%%%%%%%%%%%%%%%%%%%%%%%%%%%%%%%%%%%%%%%%%%%%%%%%%%

With the same testing spirit we have also refitted the models excluding the BAO data only, and the result is shown in Table \ref{tableFitX1}. It is noticeable in this case that the dynamical DE signal, despite it gets weakened, it still survives at $\sim 2.8\sigma$ for the main DVM's (RVM and $Q_{dm}$), with both information criteria $\Delta$AIC and $\Delta$BIC being $4-6$ and positive. This tells us that the bulk of the signal is probably encoded in the CMB and LSS data, and that the BAO data helps to make it crisper.  However, this requires further confirmation since we have to separate the rest of the data sources. We will do it in Sect.\,\ref{subsect:deconstruction}.

We conclude this section by answering in detail why the dynamical DE signal that we are glimpsing here escaped undetected from the fitting analyses of Planck 2015. The answer can be obtained by repeating our fitting procedure and restricting ourselves to the much more limited data sets used by the Planck 2015 collaboration, specifically in the papers \cite{Planck2015} and \cite{PlanckDE2015}. In contrast to \cite{Planck2015}, where no LSS data were used at all, in the case of \cite{PlanckDE2015} they used only some BAO and LSS data, but their fit is rather limited in scope. Specifically, they used only 4 BAO data points, 1 AP (Alcock-Paczynski parameter) data point, and one single LSS point, namely the value of $f(z)\sigma_8(z)$ at $z=0.57$-- see details in that paper. Using this same data we obtain the fitting results presented in our Table \ref{tableFit6PRD}. They are perfectly compatible with the fitting results mentioned in Sect.\,\ref{sect:XCDMandCPLnumerical} obtained by Planck 2015 and BOSS\,\cite{Aubourg2015}, i.e. none of them carries evidence of dynamical DE, with only the data used by these collaborations two-three years ago.

%%%%%%%%%%%%%%%%%%%%%%%%%%%%%%%%%%%%%%%%%%%%%%%%%%%%%%%%%%%%%%%%%%%%%%%%%%%%%%%%%%%%%%%%
%
\begin{figure}
\centering
\includegraphics[angle=0,width=1.03\linewidth]{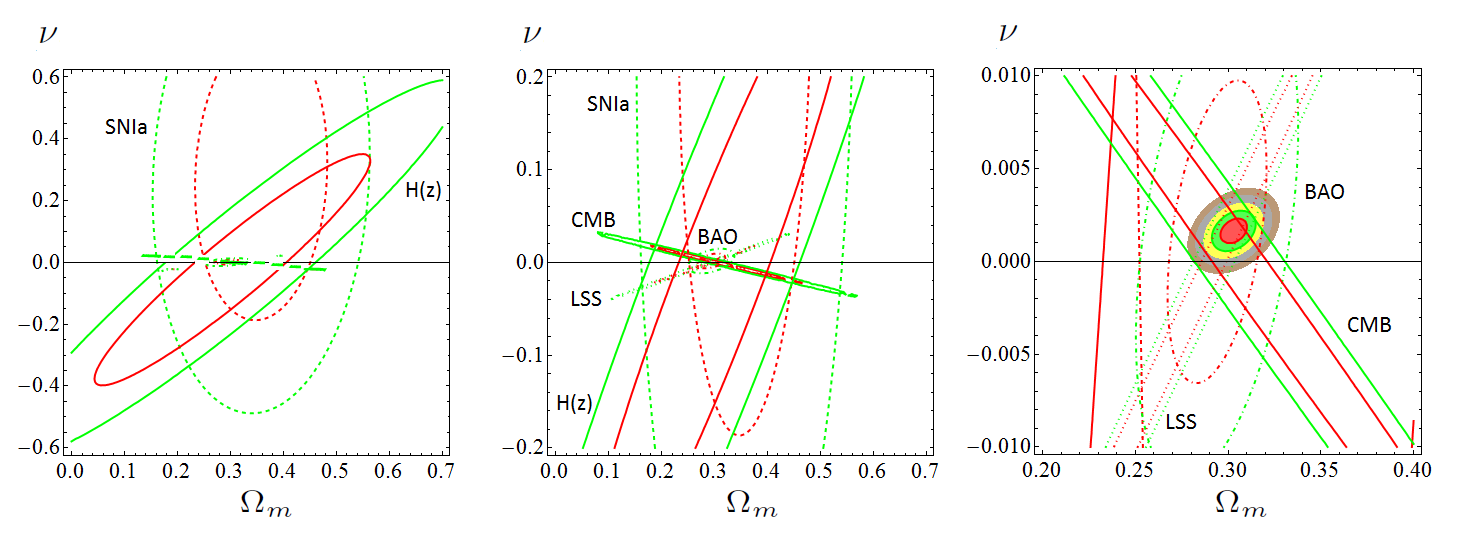}
\caption[Reconstruction of the contour lines for the RVM]{{\scriptsize Reconstruction of the contour lines for the RVM, from the partial contour plots of the different SNIa+BAO+$H(z)$+LSS+CMB data sources. The $1\sigma$ and $2\sigma$ contours are shown in all cases. For the reconstructed final contour lines we also plot the $3\sigma$, $4\sigma$ and $5\sigma$ regions.}\label{fig:PRD7}}
\end{figure}
%
%%%%%%%%%%%%%%%%%%%%%%%%%%%%%%%%%%%%%%%%%%%%%%%%%%%%%%%%%%%%%%%%%%%%%%%%%%%%%%%%%%%%%%%

In contradistinction to them, in our full analysis presented in Table \ref{tableFit1PRD} we used 11 BAO and 13 LSS data points, some of them available only from the recent literature and of high precision\,\cite{GilMarin2}. From Table \ref{tableFit6PRD} it is apparent that with only the data used in \cite{PlanckDE2015} the fitting results for the RVM are poor enough and cannot still detect clear traces of the vacuum dynamics. In particular, the vacuum parameters are compatible with zero at one $\sigma$ and the values of $\Delta$AIC and $\Delta$BIC values in that table are moderately negative, showing that the DVM's do not fit better the data than the $\Lambda$CDM model with only such limited data set. The result is reasonably compatible with what we saw in Tables \ref{tableFit3PRD}-\ref{tableFitX1} since the CMB data and a limited part of the BAO and LSS data are still there, but the Planck 2015 results cannot yet reach the threshold of visibility of the signal. In fact, not even the XCDM parametrization is capable of detecting any trace of dynamical DE with that limited data set, as the effective EoS is compatible with $w_0=-1$ at roughly $1\sigma$ ($w_0=-0.961\pm 0.033$). Compare with the situation obtained with our full data set, see Eq.\,(\ref{eq:woRVM}), where $w_0$ was pinned down to lie above $-1$ at $\gtrsim3.3\sigma$ c.l.

This should explain why the features that we are reporting here have remained hitherto unnoticed in the literature, except in our recent papers\,\cite{ApJnostre,ApJLnostre,MPLAnostre} on which chapters \ref{chap:Gtype}-\ref{chap:MPLAbased} are based, and also by the very recent analysis\,\cite{GongBoZhao2017}. These authors have been able to gather a significant $3.5\sigma$ effect on dynamical DE, presumably in a model-independent way.  The result is well along the lines of the present work and the aforementioned predecessors, where we have been able to collect evidence, in some case at a slightly stronger level, using different models and parametrizations.

\subsection{Results with and without including the bispectrum}

The following question is now in order: why is it important to include the BOSS data from \cite{GilMarin2}? Recently, the cosmological results from the final galaxy clustering data set of BOSS (SDSS-III) were made public\,\cite{Alam2016}. This analysis is even fresher than the one presented in \cite{GilMarin2}, so again, which is the reason to stick to the particular data analysis from\,\cite{GilMarin2} rather than using the more recent one presented in \cite{Alam2016}?  Both actually emerge from the same DR12 galaxy sample of BOSS, but the treatment of the data in each case is different. What is the difference? The important answer to this question is provided explicitly by the authors of \cite{Alam2016} themselves, and we just quote it: {\it one can attribute the improvement in \cite{GilMarin2} when compared to our measurements to the use of the bispectrum, which has not been used in our analysis \cite{Alam2016}}. It is somehow natural that this is so, since the bispectrum is a higher-order statistics, which involves the three-point correlation function instead of the two-point correlation function (the inverse Fourier transform of the power spectrum). It represents the next-to-leading contribution in the analysis of cosmic fluctuations in perturbation theory.

In what follows we check explicitly the implications of using or not using the bispectrum in our analysis.
The results presented in \cite{GilMarin2} are grounded on the RSD measurements of the power spectrum combined with the bispectrum, and the BAO post-reconstruction analysis of the power spectrum. In contrast, the authors of \cite{Alam2016} do not make use of the bispectrum.  Nonetheless, its use is of great importance since it leads to a significant decrease of the error bars of the data inferred from the corresponding
study; and this, of course, has a direct bearing on our own
results, for the smaller the error bars of the data, the smaller are the error bars of the cosmological parameters (in particular those sensitive to the dynamical DE) obtained from our fitting analysis.

While the spectrum  $P({\bf k})$ is connected with the two-point correlator of the density field $D({\bf k})$ in Fourier space, namely $\langle D({\bf k})\,D({\bf k}')\rangle=\delta({\bf k}-{\bf k}') P({\bf k})$, in which $\delta$ is a Dirac delta, the bispectrum $B({\bf k}_1,{\bf k}_2,{\bf k}_3)$ is formally connected with the three-point correlator
\begin{equation}\label{eq:bispectrum}
\langle D({\bf k}_1)\,D({\bf k}_2)\, D({\bf k}_3)\rangle= \delta({\bf k}_1+{\bf k}_2+{\bf k}_3)B({\bf k}_1,{\bf k}_2,{\bf k}_3)\,,
\end{equation}
where the Dirac $\delta$ selects the triangular configurations.
The bispectrum has been described in many places in the literature, see e.g.\cite{Peebles1993,BookAmendolaTsujikawa,LiddleLyth2} and references therein. Now,  while the above definition is the formal one, operationally (namely at the practical level of galaxy counting) a bispectrum estimator $\langle F_3({\bf k}_1)F_3({\bf k}_2) F_3({\bf k}_3)\rangle$ can be defined from the angle-average of closed triangles defined by the $\bf k$-modes, ${\bf k}_1,\,{\bf k}_2,\,{\bf k}_3$, where $F_3({\bf q})$ is the Fourier transform of an appropriately defined weighted field of density fluctuations, namely one formulated in terms of the number density of galaxies\,\cite{GilMarin2}. It can be conveniently written as
\begin{equation}
 \langle F_3({\bf k}_1)F_3({\bf k}_2) F_3({\bf k}_3)\rangle=\frac{k_f^3}{V_{123}}\int d^3{\bf r}\, \mathcal{D}_{\mathcal{S}_1}({\bf r}) \mathcal{D}_{\mathcal{S}_2}({\bf r}) \mathcal{D}_{\mathcal{S}_3}({\bf r})\,,
 \label{eq:bis2}
\end{equation}
i.e. through an expression involving a separate product of Fourier integrals
\begin{equation}
 \mathcal{D}_{\mathcal{S}_j}({\bf r})\equiv \int_{\mathcal{S}_j} d{\bf q}_j\, F_3({\bf q}_j)e^{i{\bf q}_j\cdot{\bf r}}\,.
\end{equation}
Here $k_f$ is the fundamental frequency, $k_f=2\pi/L_{\rm box}$, $L_{\rm box}$ the size of the box in which the galaxies are embedded and
\begin{equation}
 V_{123}\equiv\int_{\mathcal{S}_1} d{\bf q}_1\, \int_{\mathcal{S}_2} d{\bf q}_2\, \int_{\mathcal{S}_3} d{\bf q}_3\, \delta({\bf q}_1+{\bf q}_2+{\bf q}_3),
\end{equation}
is the number of fundamental triangles inside the shell defined by $\mathcal{S}_1$, $\mathcal{S}_2$ and $\mathcal{S}_3$, with $\mathcal{S}_i$ the region of the $k$-modes contained in a $k$-bin, $\Delta k$, around $k_i$. The Dirac $\delta$ insures that only closed triangles are included -- see\,\cite{GilMarin2} for more details.

%%%%%%%%%%%%%%%%%%%%%%%%%%%%%%%%%%%%%%%%%%%%%%%%%%%%%%%%%%%%%%%%%%%%%%%%%%%%%%%%%%%%%%%
%
\begin{figure}
\centering
\includegraphics[angle=0,width=0.55\linewidth]{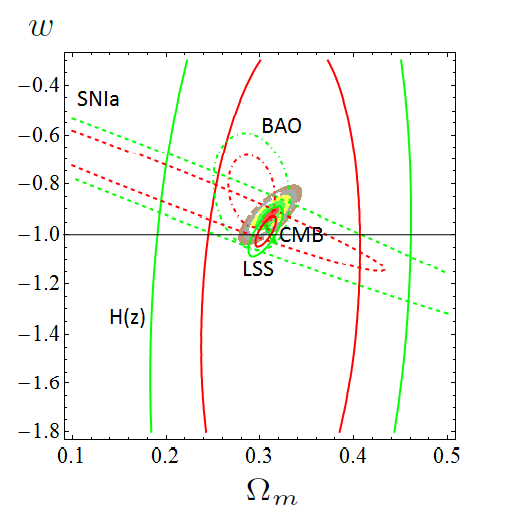}
\caption[Reconstruction of the contour lines for the XCDM]{{\scriptsize As in the previous figure, reconstruction of the contour lines for the XCDM from the partial contour plots of the different SNIa+BAO+$H(z)$+LSS+CMB data sources.}\label{fig:PRD8}}
\end{figure}
%%%%%%%%%%%%%%%%%%%%%%%%%%%%%%%%%%%%%%%%%%%%%%%%%%%%%%%%%%%%%%%%%%%%%%%%%%%%%%%%%%%%%%%%%%%

%%%%%%%%%%%%%%%%%%%%%%%%%%%%%%%%%%%%%%%%%%%%%%%%%%%%%%%%%%%%%%%%%%%%%%%%%%%%%%%%%%%%%%%

The physical importance of including the bispectrum cannot be overemphasized. The bispectrum furnishes important complementary information that goes beyond the spectrum. If fluctuations in the structure formation were strictly Gaussian, their full statistical description would be contained in the two-point correlation function, or equivalently the power spectrum estimator $\langle F_2({\bf k}_1)F_2({\bf k}_2)\rangle$. In such case the formal bispectrum defined above would identically vanish. Therefore, its inclusion is essential to be sensitive to possible higher order effects associated to non-Gaussianities in the distribution of galaxies. Even if one starts from Gaussian initial conditions, gravity makes fluctuations evolve non-Gaussian. Therefore, such deviations with respect to a normal distribution may be due both to the evolution of gravitational instabilities that are amplified from the initial perturbations, or even from some intrinsic non-Gaussianity of the primordial spectrum.

In order to study the effect of the bispectrum on our analysis of dynamical DE in a quantitative way we have completely refitted our models using now the data from \cite{Alam2016} concerning BAO and LSS, leaving of course the remaining data exactly as before. The numerical results corresponding to this new setup are collected in Table \ref{tableFitX5}. We may now compare the results of Table \ref{tableFit1PRD} (in which the bispectrum was included) with those produced in Table \ref{tableFitX5} (where the bispectrum is missing). In addition, we may compare the contour plots of Fig.\,\ref{fig:PRD1}\, with those of  Fig.\,\ref{fig:NoBispectrum}, associated to Table \ref{tableFitX5}.

The upshot of such comparison is quite enlightening, to wit: the inclusion of the bispectrum is very relevant, as it enhances quite significantly the signal of dynamical DE. The bispectrum turns out to be a key element to produce the $3.76\sigma$ signal (resp. $3.60\sigma$) in favor of the RVM (resp.  Q$_{dm}$). Without the inclusion of the bispectrum, the signal gets reduced down to $2.55\sigma$ (resp. $2.49\sigma$). A similar decrease is found for the XCDM (viz. from $3.35\sigma\to 1.90\sigma$) and the Q$_{\CC}$ ($2.38\sigma\to 1.92\sigma$).

We could of course invert the argument in a positive way, and say that even without including the bispectrum the dynamical DE signal remains fair enough -- it reaches near or above $2.5\sigma$ for the favorite models -- specially for the RVM model, i.e. the one best motivated from the QFT point of view. It is reassuring to learn, after this very practical and enlightening exercise of numerical Cosmology, that the quantitative signs of dynamical vacuum energy are sufficiently robust in our data as to be fairly detectable with the power spectrum alone, only to be reinforced with the help of the bispectrum.

The main practical conclusion that we can draw from this section is quite remarkable: the potential of the bispectrum for being sensitive to DE effects is perhaps more important than it was suspected until now. As it turns out, its more conventional application as a tool to estimate the bias between the observed galaxy spectrum and the underlying matter power spectrum may now be significantly enlarged, for the bispectrum (as the leading higher-order correction to the power spectrum) could finally reveal itself as an excellent tracer of dynamical DE effects, and ultimately of the fine structure of the DE.

%%%%%%%%%%%%%%%%%%%%%%%%%%%%%%%%%%%%%%%%%%%%%%%%%%%%%%%%%%%%%%%%%%%%%%%%%%%%%%%%%%%
%%%%%%%%%%%%%%%%%%%%%%%%%%%%%%%%%%%%%%%%%%%%%%%%%%%%%%%%%%%%%%%%%%%%%%%%%%%%%%%%%%%%
\begin{figure}[t!]
\centering
\includegraphics[angle=0,width=0.85\linewidth]{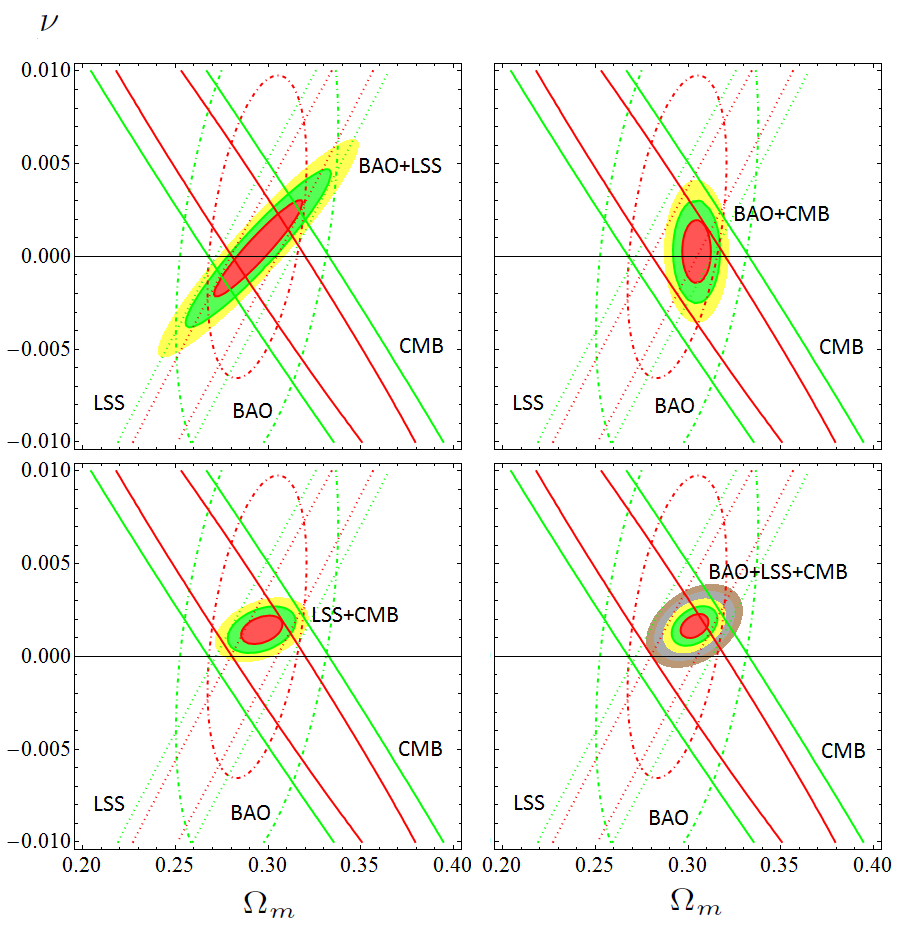}
\caption[Same as in Fig.\,\ref{fig:PRD7}, but considering the effect of only the BAO, LSS and CMB]{{\scriptsize As in Fig.\,\ref{fig:PRD7}, but considering the effect of only the BAO, LSS and CMB in all the possible combinations: BAO+LSS, BAO+CMB, LSS+CMB and BAO+LSS+CMB. As discussed in the text, it is only when such triad of observables is combined that we can see a clear $\sim 4\sigma$ effect, which is comparable to intersecting the whole set of SNIa+BAO+$H(z)$+LSS+CMB data.}\label{RecRVMtriad}}
\end{figure}

%%%%%%%%%%%%%%%%%%%%%%%%%%%%%%%%%%%%%%%%%%%%%%%%%%%%%%%%%%%%%%%%%%%%%%%%%%%%%%%%%%

%%%%%%%%%%%%%%%%%%%%%%%%%%%%%%%%%%%%%%%%%%%%%%%%%%%%%%%%%%%%%%%%%%%%%%%%%%%%%%%%%%%
%%%%%%%%%%%%%%%%%%%%%%%%%%%%%%%%%%%%%%%%%%%%%%%%%%%%%%%%%%%%%%%%%%%%%%%%%%%%%%%%%%%

\subsection{Deconstruction and reconstruction of the final contour plots}\label{subsect:deconstruction}

We complete our analysis by displaying in a graphical way the deconstructed contributions from the different data sets to our final contour plots in Fig.\,\ref{fig:PRD1} for the RVM, although we could have done the same for any of the models under consideration. The result is depicted in Fig.\,\ref{fig:PRD7}, where we can assess the detailed deconstruction of the final contours in  terms of the partial contours from the different SNIa+BAO+$H(z)$+LSS+CMB data sources. A similar deconstruction plot for the XCDM is shown in Fig.\,\ref{fig:PRD8}.

The deconstruction plot for the RVM case is dealt with in Fig.\,\ref{fig:PRD7}, through a series of three plots made at different magnifications. In the third plot of the sequence we can easily appraise that the BAO+LSS+CMB data subset plays a fundamental role in narrowing down the final physical region of the $(\Omega_m,\nu)$ parameter space, in which all the remaining parameters have been marginalized over. This deconstruction process also explains in very obvious visual terms why the conclusions that we are presenting here hinge to a large extent on some particularly sensitive components of the data. While CMB obviously is a high precision component in the fit, we demonstrate in our study (both numerically and graphically) that the maximum power of the fit is achieved when it is combined with the wealth of BAO and LSS data points currently available.

To gauge the importance of the particular BAO+LSS+CMB combination more deeply, in Fig.\,\ref{RecRVMtriad} we try to reconstruct the final RVM plot in Fig.\,\ref{fig:PRD1} (left) from only these three data sources. First we consider the overlapping regions obtained when we cross the data pairs BAO+LSS, BAO+CMB, LSS+CMB and finally the trio BAO+LSS+CMB (in all cases excluding the SNIa and $H(z)$ data). One can see that neither the BAO+LSS nor the BAO+CMB crossings yield a definite sign for $\nu$, despite the obtained regions are very narrow and small as compared to the scale used in Fig.\,\ref{fig:PRD7} (in which all the data are used). This can also be confirmed numerically from Tables \ref{tableFit3PRD} and \ref{tableFit4PRD}, where the removal of either the CMB data or the LSS data renders rather poor fits with negative values of $\Delta$AIC and $\Delta$BIC. Remarkably, it is the LSS+CMB combination the one that carries a well-defined, positive, sign for $\nu$, as it is seen from the lower-left plot in Fig.\,\ref{RecRVMtriad} and also reconfirmed numerically in Table \ref{tableFitX1}, where $\Delta$AIC and $\Delta$BIC are now both positive and above $6$ for the main DVM's (RVM and $Q_{dm}$). Curiously, the XCDM still gives a bad fit when BAO is not included.  Also noticeable is the fact that, in the lower-left plot in Fig.\,\ref{RecRVMtriad}, the significance of $\nu>0$ (and hence that of the dynamical vacuum signature) is ``only'' at $2.8\sigma$ c.l.  It is only when we next intersect the pair LSS+CMB with the BAO data that the signal rockets from $2.8\sigma$ c.l to $3.8\sigma$ c.l., the final contours being now those shown in the lower-right plot of Fig.\,\ref{RecRVMtriad}. Thus, the crossing with BAO further drags the intersection region upwards in that plane and intensifies the signal of dynamical vacuum by one full additional $\sigma$.  At the same time the XCDM also improves dramatically since the combined BAO+LSS+CMB data produces essentially the same output as in Table \ref{tableFit1PRD} with all the data.

%%%%%%%%%%%%%%%%%%%%%%%%%%%%%%%%%%%%%%%%%%%%%%%%%%%%%%%%%%%%%%%%%%%%%%%%%%%%
%%%%%%%%%%%%%%%%%%%%%%%%%%%%%%%%%%%%%%%%%%%%%%%%%%%%%%%%%%%%%%%%%%%%%%%%%%%%%
%%%%%%%%%%%%%%%%%%%%%%%%%%%%%%%%%%%%%%%%%%%%%%%%%%%%%%%%%%%%%%%%%%%%%%%%%%%%%
\begin{table}[t]
\begin{center}
\resizebox{1\textwidth}{!}{
\begin{tabular}{| c | c |c | c | c | c | c | c | c | c |}
\multicolumn{1}{c}{Model} &  \multicolumn{1}{c}{$h$} &  \multicolumn{1}{c}{$\omega_b= \Omega_b h^2$} & \multicolumn{1}{c}{{\small$n_s$}}  &  \multicolumn{1}{c}{$\Omega_m$}&  \multicolumn{1}{c}{{\small$\nu_i$}}  & \multicolumn{1}{c}{$w$} &
\multicolumn{1}{c}{$\chi^2_{\rm min}/dof$} & \multicolumn{1}{c}{$\Delta{\rm AIC}$} & \multicolumn{1}{c}{$\Delta{\rm BIC}$}\vspace{0.5mm}
\\\hline
{\small $\CC$CDM} & $0.692\pm 0.004$ & $0.02252\pm 0.00013$ &$0.974\pm 0.004$& $0.296\pm 0.004$ & - & -1 &  83.14/85 & - & -\\
\hline
XCDM  &  $0.671\pm 0.007$& $0.02262\pm 0.00014 $&$0.976\pm0.004$& $0.312\pm 0.007$& - & $-0.922\pm0.023$ &  71.98/84 & 8.92 & 6.66 \\
\hline
RVM  & $0.677\pm 0.005$& $0.02231\pm 0.00014$&$0.965\pm 0.004$& $0.303\pm 0.005$ & $0.00159\pm 0.00042$ & -1 &  67.72/84 & 13.18 & 10.92 \\
\hline
$Q_{dm}$  &  $0.678\pm 0.005$& $0.02229\pm 0.00015 $&$0.965\pm0.004$& $0.303\pm 0.005 $ & $0.00218\pm 0.00059 $ & -1 &  68.47/84  & 12.43 & 10.17 \\
\hline
$Q_\CC$  &  $0.691\pm 0.004$& $0.02230\pm 0.00016 $&$0.966\pm0.005$& $0.298\pm 0.005$ & $0.00602\pm 0.00253$ & -1 &  77.46/84 & 3.45 & 1.18 \\
\hline
 \end{tabular}
 }
\end{center}
\caption[Fitting results obtained upon replacing the BAO data points of \cite{Delubac2015} with those from \cite{FontRibera2014} and \cite{Bautista2017}]{{\scriptsize As in Table \ref{tableFit1PRD}, but replacing the BAO data of \cite{Delubac2015} with the BAO data from \cite{FontRibera2014} and \cite{Bautista2017}.}\label{tableFit8PRD}}
\end{table}
%

%%%%%%%%%%%%%%%%%%%%%%%%%%%%%%%%%%%%%%%%%%%%%%%%%%%%%%%%%%%%%%%%%%%%%%%%%%%%%

%%%%%%%%%%%%%%%%%%%%%%%%%%%%%%%%%%%%%%%%%%%%%%%%%%%%%%%%%%%%%%%%%%%%%%%%%%%%%
\begin{table}[t]
\begin{center}
\resizebox{1\textwidth}{!}{
\begin{tabular}{| c | c |c | c | c | c | c | c | c | c |}
\multicolumn{1}{c}{Model} &  \multicolumn{1}{c}{$h$} &  \multicolumn{1}{c}{$\omega_b= \Omega_b h^2$} & \multicolumn{1}{c}{{\small$n_s$}}  &  \multicolumn{1}{c}{$\Omega_m$}&  \multicolumn{1}{c}{{\small$\nu_i$}}  & \multicolumn{1}{c}{$w$} &
\multicolumn{1}{c}{$\chi^2_{\rm min}/dof$} & \multicolumn{1}{c}{$\Delta{\rm AIC}$} & \multicolumn{1}{c}{$\Delta{\rm BIC}$}\vspace{0.5mm}
\\\hline
{\small $\CC$CDM} & $0.693\pm 0.003$ & $0.02255\pm 0.00013$ &$0.976\pm 0.003$& $0.294\pm 0.004$ & - & -1 &  90.44/84 & - & -\\
\hline
XCDM  &  $0.670\pm 0.007$& $0.02264\pm 0.00014 $&$0.977\pm0.004$& $0.312\pm 0.007$& - & $-0.916\pm0.021$ &  74.91/83 & 13.23 & 11.04 \\
\hline
RVM  & $0.676\pm 0.005$& $0.02231\pm 0.00014$&$0.965\pm 0.004$& $0.303\pm 0.005$ & $0.00165\pm 0.00038$ & -1 &  70.32/83 & 17.82 & 15.63 \\
\hline
$Q_{dm}$  &  $0.677\pm 0.005$& $0.02229\pm 0.00015 $&$0.964\pm0.004$& $0.303\pm 0.005 $ & $0.00228\pm 0.00054 $ & -1 &  71.19/83  & 16.95 & 14.76 \\
\hline
$Q_\CC$  &  $0.692\pm 0.004$& $0.02229\pm 0.00016 $&$0.966\pm0.005$& $0.297\pm 0.004$ & $0.00671\pm 0.00246$ & -1 &  83.08/83 & 5.06 & 2.87 \\
\hline
 \end{tabular}
}
\end{center}
\caption[Fitting results obtained upon using the original LSS data of \cite{GilMarin2OLD}, instead of that from the revised version \cite{GilMarin2}]{{\scriptsize As in Table \ref{tableFit1PRD}, but making use of the original LSS data of \cite{GilMarin2OLD}, instead of that from the revised version \cite{GilMarin2} (cf. Tables 5 of these two references). In the latter, the uncertainty of the $f(z=0.32)\sigma_8(z=0.32)$ (resp. $f(z=0.57)\sigma_8(z=0.57)$) increases  $\sim 8\%$ (resp. $\sim 26\%$) with respect to the former. See the discussion in the text.}\label{tableFit9PRD}}
\end{table}

For the RVM case, therefore, we have checked that the final BAO+LSS+CMB plot in Fig.\,\ref{RecRVMtriad} is essentially the same as the original RVM plot in Fig.\,\ref{fig:PRD1} (the leftmost one). In other words, the final RVM plot can essentially be reconstructed with only the three data sources BAO+LSS+CMB.

\subsection{Testing the influence of alternative data sets}\label{sect:GilMarin2OLD}

We deem interesting to study the impact of the new BAO measurement extracted from the analysis of the auto-correlation function of the Ly$\alpha$ flux-transmission field using the SDSS-DR12 \cite{Bautista2017}. It is the released version of the BAO auto-correlation data provided in \cite{Delubac2015}, in which the authors made use of the SDSS-DR11 sample. Apart from an increase of the 15\% in the volume sample, in \cite{Bautista2017} they apply some technical improvements in the methodology over the study of \cite{Delubac2015}. After all, the differences in the BAO measurements are fully consistent with those found in \cite{Delubac2015}, so we do not expect that the change in the BAO data set should induce significant differences in our fitting results. But of course putting it to the test is the best way to verify it. The output for this test can be read off Table \ref{tableFit8PRD}, where we can see how the results shown in Table \ref{tableFit1PRD} are modified when we replace the BAO data from \cite{Delubac2015} with those from the very recent analysis of\,\cite{Bautista2017}. But before continuing our comparison of the fitting results in the main analysis shown in Table \ref{tableFit1PRD} with the ones in the new Table \ref{tableFit8PRD}, let us make some technical comments concerning the implementation of this change in our data set. In Table \ref{tableFit1PRD} we used the BAO auto-correlation data \cite{Delubac2015} and the cross-correlation data \cite{FontRibera2014} obtained from the SDSS-DR11. In fact, we used the ``combined'' (auto+cross-correlation) LyaF data at the effective redshift $z = 2.34$ (see data set DS3 for details). In Table \ref{tableFit8PRD}, however, we use 2 data points for the anisotropic BAO estimators, $D_A/r_s(z_d)$ and  $D_H/r_s(z_d)$ at z = 2.36 from  \cite{FontRibera2014} together with the best-determined combination $D_H^{0.3}D_M^{0.7}/r_s(r_d)$ from \cite{Bautista2017} at $z =2.33$, assuming no correlation between the two analyses.  Notice that although the data from \cite{FontRibera2014} and \cite{Bautista2017} are obtained from overlapping SDSS samples, and they correspond to very close redshifts ($z=2.36$ and $z=2.33$, respectively), the cross covariance between the auto and cross-correlation measurements proves to be small for the  SDSS-DR11 sample, as explicitly stated in \cite{Delubac2015,Aubourg2015}. Thus one may judiciously assume that it is also the case for the SDSS-DR12 sample.
Furthermore, we find reasonable to proceed in this way for the following two reasons: i) the analogous of the cross-correlation analysis of the DR12 sample is not available, and therefore the analysis of the  ``combined'' data for the SDSS-DR12 sample is not possible; ii) we do not have the correlation coefficient between the  $D_A/r_s(z_d)$ and  $D_H/r_s(z_d)$ data points of \cite{Bautista2017}, so we have considered more appropriate to follow these authors in using the aforementioned combination $D_H^{0.3}D_M^{0.7}/r_s(r_d)$, in which the powers $0.3$ and $0.7$ have been optimized in order to minimize the variance of the product  (see \cite{Bautista2017} for details). We have also checked that if in the above implemented change we would only use the data from \cite{Bautista2017}, but not that from \cite{FontRibera2014}, the results stand essentially the same. For example, for the RVM we find $\nu=0.00160\pm 0.00042$ and ($\Delta$AIC,$\Delta$BIC)$= (13.33,11.10)$, which favour slightly more the RVM against the $\CC$CDM in comparison to the result in Table \ref{tableFit8PRD}.

Notwithstanding such an alternative and still consistent use of some of the BAO data points, we have opted from the very beginning for presenting our main results as they appear in our Table \ref{tableFit1PRD}, which is based on the aforementioned ``combined'' LyaF DR11 data. In this way we fully incorporate the small correlations between the auto and cross-correlation data, which is the most advisable option. Moreover, as we have mentioned before, the changes in the auto-correlation measurements extracted from the DR12 with respect to the DR11 ones are not significant, so no big difference should be expected between Tables \ref{tableFit1PRD} and \ref{tableFit8PRD}. What we have done here is to check it explicitly. The fitting results from Table \ref{tableFit8PRD} appear to favor all the dynamical vacuum models slightly more than those in Table \ref{tableFit1PRD}, so the latter (our main table) is actually a bit more conservative. This feature also applies to  the XCDM and CPL parametrizations. The differences, however, are not very significant in any of the cases. If we quote them here for, say, the RVM (resp. $Q_{dm}$), we find that the confidence level of the dynamical DE signature has increased now from $3.76\sigma$ (resp. $3.60\sigma$) to $3.78\sigma$ (resp. $3.69\sigma$); and the respective differences in the values of the Akaike and Bayesian information criteria with respect to the $\CC$CDM, ($\Delta$AIC,$\Delta$BIC), have also correspondingly increased, to wit: from $(12.91,10.67)$ to $(13.18,10.92)$ in the RVM case, and from $(12.13,9.89)$ to $(12.43,10.17)$ in the $Q_{dm}$ case. The fact that the differences with respect to our main table are small, and the confirmation that the new results remain well anchored to the high significance levels already attained in Table \ref{tableFit1PRD}, already speaks up both of the robustness of the current analysis and of the significance of the reported results.

We find also pertinent to comment next on the differences between the current results and the ones obtained by using the BAO+LSS data points of \cite{GilMarin2OLD} instead of those from \cite{GilMarin2}. The former were used in Chapters \ref{chap:AandGRevisited} and \ref{chap:MPLAbased} of this dissertation. In Table \ref{tableFit9PRD} we exhibit the fitting outputs obtained with the data points of \cite{GilMarin2OLD}. The differences with respect to the updated ones, i.e. those recorded in the current Table \ref{tableFit1PRD}, are basically caused by the change in the uncertainties of the LSS data presented in \cite{GilMarin2OLD} as compared to the more recent ones in \cite{GilMarin2}. While the central values of the LSS observations quoted in \cite{GilMarin2} are the same as those in the previous version of their work, the errors in the new version are slightly larger than those in the older. It is not clear to us the reason for it, we were unable to trace a justification for that change in \cite{GilMarin2}. We nevertheless find a valuable exercise to check the impact of such difference on our results, as this can be used to test the reaction of our analysis to a change in the errors on such significant data sources.

In Ref.\,\cite{GilMarin2}, the uncertainties affecting the  two LSS data points $f(z_i)\sigma_8(z_i)$ at $z_i=0.32$ and $z_i=0.57$ increased about $\sim 8\%$ and $\sim 26\%$, respectively, with respect to the old version\,\cite{GilMarin2OLD}. The change is not such a trifle, at least {\it a priori}, as it induces a non-negligible increase of the uncertainties for the various cosmological parameters. For example, the comparison of the results for the RVM in Tables \ref{tableFit1PRD} and \ref{tableFit9PRD} tells us that by using the LSS data from \cite{GilMarin2} instead of that from \cite{GilMarin2OLD} the $\nu$ parameter departs from $0$ at $3.76\sigma$ c.l. whereas in the latter case the departure is at $4.34\sigma$ c.l. In both cases, a $\sim 4\sigma$ c.l. signal seems to be secured, but of course there is some difference, which we have been able to quantify. A similar conclusion applies to the other models under study, as we have checked.

At the end of the day the main conclusion is rewarding: in spite of some changes in the uncertainties of the BOSS LSS+BAO data, the signature on dynamical DE stands upright. We devote the following subsection to check if such a signature can also be grasped to some extent by using one of the most well-known quintessence models in the market: the Peebles \& Ratra model\,\cite{PeeblesRatra88b}.

\subsection{Dynamical dark energy and the $\phi$CDM}\label{sec:phiCDM}

A natural question that can be formulated is whether the traditional class of $\phi$CDM models, which have a well-defined local Lagrangian description and in which the DE is described in terms of a scalar field $\phi$ with some standard form for its potential $V(\phi)$, are also capable of capturing clear signs of dynamical DE using the same set of cosmological observations used for fitting the DVM's. In particular we study the Peebles \& Ratra model, amply discussed in the previous chapter. Now we want to redo the fit of this model in order to compare the new results with those obtained for the $\Lambda$CDM, the various DVM's and DE parametrizations, by using the main data set presented in Sect. \ref{sect:DataSets}. The new fitting results for the PR model are presented in Table \ref{tableFitPhiCDM}. These are most conveniently expressed by means of the following $5$-dimensional fitting vector (see Chapter \ref{chap:MPLAbased} for details):
\begin{equation}\label{eq:vfittingPhiCDM}
\vec{p}_{\phi{\rm CDM}}=(\omega_m,\omega_b,n_s,\alpha,\bar{\kappa})\,,
\end{equation}
in which we have traded the original parameter  $\kappa$ in the PR potential for $\bar{\kappa}\equiv \kappa\,M_P^2/\varsigma^2$. Recall the previously defined parameter $\varsigma\equiv 1\, {\rm Km/s/Mpc}=2.1332\times10^{-44} GeV$ (in natural units).
Note that the fitting vector (\ref{eq:vfittingPhiCDM}) has also $5$ parameters (one more than the $\CC$CDM) as for any of the DVM's, see Eq.\,(\ref{eq:fittingvector}), and the XCDM. Therefore the number of parameters is equalized for all these models, except for the CPL parametrization, which has one more parameter and as a result the fitting errors are larger.

%%%%%%%%%%%%%%%%%%%%%%%%%%%%%%%%%%%%%%%%%%%%%%%%%%%%%%%%%%%%%%%%%%%%%%%%%%%%%%%%%%%%%%%%%%%%

\begin{table}
\begin{center}
\resizebox{1\textwidth}{!}{
\begin{tabular}{| c | c |c | c | c | c | c | c | c | c|c|}
\multicolumn{1}{c}{Model} &  \multicolumn{1}{c}{$\omega_m=\Omega_m h^2$} &  \multicolumn{1}{c}{$\omega_b=\Omega_b h^2$} & \multicolumn{1}{c}{{\small$n_s$}}  &  \multicolumn{1}{c}{$\alpha$} &  \multicolumn{1}{c}{$\bar{\kappa}$}&  \multicolumn{1}{c}{$\chi^2_{\rm min}/dof$} & \multicolumn{1}{c}{$\Delta{\rm AIC}$} & \multicolumn{1}{c}{$\Delta{\rm BIC}$}\vspace{0.5mm}
\\\hline
$\phi$CDM  &  $0.1405\pm 0.0008$& $0.02263\pm 0.00014 $&$0.976\pm 0.004$& $0.202\pm 0.065$  & $(32.7\pm1.2)\times 10^{3}$ &  74.08/84 & 8.55 & 6.31 \\
\hline
 \end{tabular}
 }
\caption[Fitting results for the Peebles \& Ratra $\phi$CDM model]{{\scriptsize  The best-fit values for the parameter fitting vector (\ref{eq:vfittingPhiCDM}) of the $\phi$CDM model with Peebles \& Ratra potential (\ref{eq:PRpotential}). We use the same cosmological data set as in Table \ref{tableFit1PRD}. The specific $\phi$CDM fitting parameters are $\bar{\kappa}$ and $\alpha$. The number of independent parameters is $5$, see Eq.\,(\ref{eq:vfittingPhiCDM})-- one more than in the $\CC$CDM, as in the DVM's and XCDM. Using the best-fit values of this table and the overall covariance matrix derived from our fit, we obtain:  $h=0.671\pm 0.006$ and $\Omega_m=0.312\pm 0.006$, which allows direct comparison with Table \ref{tableFit1PRD}. We find $\gtrsim 3\sigma$ evidence in favor of $\alpha>0$. In terms of the EoS of $\phi$ at present, the DE behavior appears quintessence-like at $\sim 3.3\sigma$: $w_\phi(z=0)= -0.936\pm 0.019$, see Eq.\,(\ref{eq:wphinow})}.\label{tableFitPhiCDM}}
\end{center}
\end{table}

%%%%%%%%%%%%%%%%%%%%%%%%%%%%%%%%%%%%%%%%%%%%%%%%%%%%%%%%%%%%%%%%%%%%%%%%%%%%%%%%%%%%%%%%

The motivation for the PR potential is well described in the original paper \cite{PeeblesRatra88b}. First of all let us recall that it admits tracker solutions of the field equations for $\alpha>0$, which is of course very convenient. We can see from Table \ref{tableFitPhiCDM} that we have determined $\alpha$ as being indeed positive at more than $3\sigma$ c.l.

Let us also recall that both in the radiation-dominated (RD) and matter-dominated (MD) epochs the equation of state (EoS) of the scalar field remains stationary since the Hubble function in these epochs is dominated by the usual behavior $\rho\propto a^{-n}$ of the matter component, with $n=4$ and $n=3$ in the respective periods. As a result it is possible to find power-law solutions  $\phi\propto t^p$ ($t$ being the cosmic time) of the Klein-Gordon equation with PR potential (\ref{eq:PRpotential}) in the FLRW metric, $\ddot{\phi}+3H\dot{\phi}+dV/d\phi=0$.
 This allows to establish the initial conditions for the integration of such equation \cite{MPLAnostre}. Finally, with the help of (\ref{eq:rhophi}) we can derive after some calculations a very compact form for the EoS (see Chapter \ref{chap:MPLAbased}):
\begin{equation}
w_\phi=\frac{p_{\phi}}{\rho_{\phi}}=-1+\frac{\alpha n}{3(2+\alpha)}\,.
\end{equation}
This stationary form, valid only in the pure RD and MD regimes, was expected and it perfectly adapts to the well-known result that holds for general tracking solutions\,\cite{ZlatevWangSteinhardt99a,ZlatevWangSteinhardt99b}. The more complicated behavior of the EoS,  namely the time-evolving $w_\phi(a)$ that interpolates between the RD and MD epochs and accounts for the transition from the MD to the DE epoch, obviously requires a numerical integration of the Klein-Gordon equation with the PR potential. We have performed such integration following the method of Chapter \ref{chap:MPLAbased} and using the fitting values of Table \ref{tableFitPhiCDM}. The plot of $w_{\phi}(z)$ in terms of the redshift near our time is shown in Fig.\,\ref{fig:PRD5}, together with the (constant) EoS value of the XCDM parametrization and the evolving one of the CPL parametrization, see Sect.\,\ref{sect:XCDMandCPL}. The curves in Fig. \ref{fig:PRD5} show that the quintessence-like behavior is sustained until the present epoch. The numerically computed EoS value reads:
\begin{equation}\label{eq:wphinow}
w_\phi(z=0)= -0.936\pm 0.019\,.
\end{equation}
This result deviates from $-1$ by $3.37\sigma$ and therefore  points to quintessence behavior at such confidence level.  Comparing with the EoS of the XCDM parametrization in Table \ref{tableFit1PRD}, whose departure from $-1$ is at $3.35\sigma$ c.l. we can see that the agreement with the $\phi$CDM prediction is fairly good: the current EoS values in each case are very close and both predict quintessence-like behavior at $\sim 3.3\sigma$ c.l.

In the present case the Hubble function has to be computed after first numerically solving the Klein-Gordon equation in the FLRW metric. Thus, in contrast to the $\CC$CDM and the other models, for the $\phi$CDM it proves more convenient to fit the parameter $\omega_m=\Omega_m h^2$ rather than $\Omega_m$  and $h$ separately.  Recall that $3H^2=8\pi\,G(\rho_{\phi}+\rho_m)$, where $\rho_{\phi}$ is given in Eq.\,(\ref{eq:rhophi}) and $\rho_m=\rho_{c 0}\Omega_m a^{-3}=(3\times 10^4/8\pi G)\varsigma^2\,\omega_m\,a^{-3}$ is the conserved matter component. In the matter-dominated epoch, one can show that
\begin{equation}\label{eq:barH2}
\bar{H}^2(a)=\frac{\bar{\kappa}\,\phi^{-\alpha}(a)+1.2\times 10^5\,\omega_m\,a^{-3}}{12-a^2\phi^{\prime 2}(a)}\,,
\end{equation}
where we have defined $\bar{H}=H/\varsigma$, and used $\dot{\phi}=a\,H\,\phi^{\prime}(a)$. It is clear that Eq.\,(\ref{eq:barH2}) is a numerical expression since $\phi(a)$ and $\phi'(a)$ are known only after solving by numerical methods the Klein-Gordon equation in the FLRW metric with Peebles \& Ratra potential. In terms of the scale factor variable, the Klein-Gordon equation takes on the form
\begin{equation}\label{eq:KGa}
\phi^{\prime\prime}+\phi^\prime\left(\frac{\bar{H}^\prime}{\bar{H}}+\frac{4}{a}\right)-\frac{\alpha}{2}\frac{\bar{\kappa}\phi^{-(\alpha+1)}}{(a\bar{H})^2}=0\,.
\end{equation}
The initial conditions for $\phi(a)$ and $\phi'(a)$  are fixed as explained in Chapter \ref{chap:MPLAbased}.
Finally, we can determine $h\equiv\bar{H}(a=1)/100$  and $\Omega_m=\omega_m/h^2$. Using the posterior distribution of the analysis in Table \ref{tableFitPhiCDM} we find (as quoted in the table caption):
\begin{equation}\label{eq:handOmegavalues}
h=0.671\pm 0.006\,, \ \ \ \ \  \Omega_m=\frac{\omega_m}{h^2}=0.312\pm 0.006\,.
\end{equation}
They are very close to those for the XCDM in Table \ref{tableFit1PRD}.

The likelihood contours for the PR model in the $(\Omega_m,\alpha)$-plane are depicted in Fig.\,\ref{fig:PRD10}. They clearly point to a nonvanishing and positive value of $\alpha$ at $\sim 3\sigma$ c.l.  Together with the EoS value (\ref{eq:wphinow}), they consistently signal quintessence behavior at more than $3\sigma$ c.l. The numerical details are furnished in Table \ref{tableFitPhiCDM}.  These results, together with the previous chapter (based on a slightly different data set), are unprecedented in the literature. Other studies had previously considered the PR-potential in the light of older or more limited cosmological data sets, see e.g. \cite{Ratra2014,Samushia2014,FarooqRatra2013}, but they did not find comparable signs of dynamical DE as the ones presented here.

If we compare the above results with the DVM's we see that they are compatible and all of them pointing towards the same direction. For the DVM's, being vacuum models, the EoS is of course $-1$ (as emphasized from the very beginning), but since the vacuum is dynamical the quintessence or phantom-like effective behavior manifest here through the sign of the $\nu_i$ parameters. Take e.g. the RVM, for which $\nu=0.00158 \pm 0.00042$. This value is positive at $3.76\sigma$ c.l. and therefore the vacuum energy is larger in the past ($a\to 0$) and decreases towards the future (increasing $a$), see Eq.\,(\ref{eq:rhoVRVM}). It means that the RVM behaves effectively as quintessence.  The same occurs with models $Q_{dm}$ and $Q_{\CC}$ with different levels of significance.  We point out, however, that in contrast to true quintessence the vacuum density of the RVM evolves from a large value in the past (always well below that of the matter density), but it does not asymptote to zero in the remote future. The limiting value for $a\to\infty$  is
\begin{equation}\label{eq:rLRVMasymptote}
\rL\to  \rLo - \frac{\nu\,\rho_{m0}}{1-\nu}\,.
\end{equation}
A similar result holds for model $Q_{dm}$. In contrast, for model $Q_{\CC}$ we have $\rL\to 0$ in the future (for $\nu_\CC>0$). All these results are evident from Eqs.\,(\ref{eq:rhoVRVM})-(\ref{eq:rhoVQL}).

%%%%%%%%%%%%%%%%%%%%%%%%%%%%%%%%%%%%%%%%%%%%%%%%%%%%%%%%%%%%%%%%%%%%%%%%%%%%%%%%%%%%%%%%

\begin{figure}[t!]
\begin{center}
\label{Parella1}
\includegraphics[angle=0,width=0.5\linewidth]{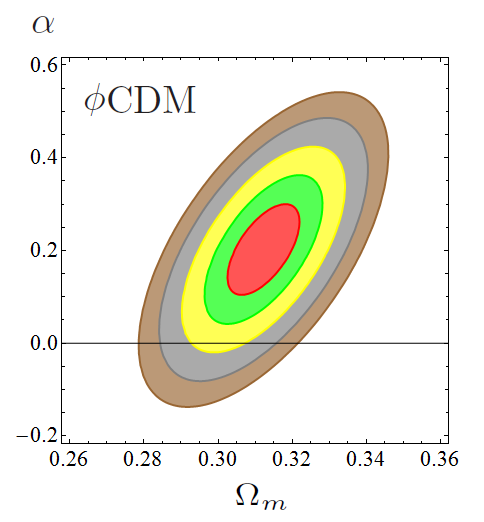}
\caption[CL's for the $\phi$CDM model in the $(\Omega_m,\alpha)$-plane]{{\scriptsize Likelihood contours for the $\phi$CDM model in the $(\Omega_m,\alpha)$-plane after marginalizing over the remaining parameters (cf. Table \ref{tableFitPhiCDM}). The various contours correspond to 1$\sigma$, 2$\sigma$, 3$\sigma$, 4$\sigma$ and 5$\sigma$ c.l. The central value $(0.312,0.202)$ is  $\sim3\sigma$ away from the $\CC$CDM result ($\alpha=0$). Compare with the XCDM contour plots in Fig.\,\ref{fig:PRD3}. Upon marginalizing also with respect to $\Omega_m$, both for the $\phi$CDM and XCDM we find evidence of dynamical DE at $>3\sigma$ c.l. Confronting also with the DVM's contours, specially with the RVM one in Fig. \ref{fig:PRD1}, we find that all of them signal dynamical DE with quintessence-like behavior. For the RVM, however, the level of evidence is closer to $\sim 4\sigma$.} \label{fig:PRD10}}
\end{center}
\end{figure}

%%%%%%%%%%%%%%%%%%%%%%%%%%%%%%%%%%%%%%%%%%%%%%%%%%%%%%%%%%%%%%%%%%%%%%%%%%%%%%%%%%%%%%%%

All in all the signs of dynamical DE perceived with the use of different models and parametrizations  are quite evident. But we should make clear once more that they are only at reach when the crucial triplet of BAO+LSS+CMB data are included in the fitting analysis of these models. Let us stay alert to the evolution of the future cosmological observations. Other possibilities qualitatively very different (e.g. scale invariant theories\,\cite{Maeder2017}) are, in principle, also compatible with observations, but need more phenomenological test. In such context the traditional CC appears as related to the properties of scale invariance of the empty
space and as a result a new dynamical contribution appears that mimics the DE and replaces the role of the rigid $\Lambda$ in Einstein's equations. Or, in a very different vein, dynamical vacuum energy could be the result of a nonperturbative approach to strong field
interactions relying on quasiparticles, which results in a
variable vacuum energy that depends on the state of the
system\,\cite{Trachenko2017}.

Thus, in a rather multifarious way, the dynamical DE germ that seems to be imprinted in the data appears at play and can be unveiled using a variety of independent frameworks. We are, of course, not yet claiming  incontestable evidence of dynamical vacuum energy or, in general, of dynamical DE, but the improvement of the overall fit to the cosmological data under this hypothesis starts to be rampant. We are eager to hear of new observational data, see for instance\,\cite{GongBoZhao2017}. At the moment, the best fit of the data implying dynamical DE signature is fulfilled by the RVM, and it reaches $\sim 4\sigma$ c.l. See the analysis presented in Chapter \ref{chap:H0tension} (based on the paper \cite{PLBnostre}), where we consider small departures of the equation of state parameters of the DVM's in order to check whether they are capable of improving the description of the cosmological data or not, together with their role on the $H_0$ tension issue.

%%%%%%%%%%%%%%%%%%%%%%%%%%%%%%%%%%%%%%%%%%%%%%%%
%%%%%%%%%%%%%%%%%%%%%%%%%%%%%%%%%%%%%%%%%%%%%%%%
%%%%%%%%%%%%%%%%%%%%%%%%%%%%%%%%%%%%%%%%%%%%%%%%
\section{Conclusions}\label{sect:conclusions}

To conclude, in this chapter we have presented a comprehensive study on the possibility that the global cosmological observations can be better described in terms of vacuum models equipped with a dynamical component that evolves with the cosmic expansion. This should be considered a natural possibility in the context of quantum field theory (QFT) in a curved background. It means that we aimed at testing cosmological physics beyond the standard or concordance $\CC$CDM, built upon a rigid cosmological constant. We have focused on three dynamical vacuum models (DVM's): the running vacuum model (RVM) along with two more phenomenological models, denoted $Q_{dm}$ and $Q_\CC$-- see equations (\ref{eq:QRVM})-(\ref{eq:PhenModelQL}). At the same time, we have compared the performance of these models with the general XCDM and CPL parametrizations as well as with specific scalar field models ($\phi$CDM), such as the original Peebles \& Ratra model. We have fitted all these models and parametrizations to the same set of cosmological data based on the observables SNIa+BAO+$H(z)$+LSS+CMB. Needless to say, we have fitted the $\CC$CDM to the same data too. The remarkable outcome of this investigation is that in all the considered cases we find an improvement of the description of the cosmological data in comparison to the $\CC$CDM. Our conclusion is that there are significant signs of dynamical DE in the current data, which we have been able to capture at different intensities with the various analyzed models and parametrizations.

The model that renders the best fit to the overall SNIa+BAO+$H(z)$+LSS+CMB data is the running vacuum model (RVM), which emerges as a serious alternative candidate for the description of the current state of the Universe in accelerated expansion. The RVM is the only DVM among those that we have analyzed here that has a close connection with the possible quantum effects on the effective action of QFT in curved spacetime, see \cite{SolaReview2013,SolGom2015} and references therein. There were previous phenomenological studies that hinted in different degrees at the possibility that the RVM could fit the data similarly as the $\CC$CDM, see e.g. the earlier papers \cite{BPS09,Grande2011,BasPolarSola12,BasSola14a}, as well as the more recent ones by \cite{JCAPnostre1,MNRASnostre,JCAPnostre2} (see, alternatively, Chapters \ref{chap:Atype} and \ref{chap:DynamicalDE}), including the closely related series\,\cite{ApJLnostre,ApJnostre,MPLAnostre} (or Chapters \ref{chap:Gtype}-\ref{chap:MPLAbased}).

Previous analyses by other authors on the more phenomenological DVM's, i.e. models $Q_{dm}$ and $Q_\CC$, are available in the literature, see e.g. \cite{Salvatelli2014,Murgia2016,Li2016}. We have discussed the comparison of our results with theirs and have pointed out some differences with \cite{Salvatelli2014} concerning model $Q_\CC$. We find that this model is far less competitive as compared to the RVM and $Q_{dm}$.

To our knowledge there is no devoted work comparable in scope to the one presented here. The significantly enhanced level of evidence on dynamical DE achieved with the DVM's, the XCDM and CPL parametrizations, as well with specific scalar field models ($\phi$CDM), is unprecedented in the literature, all the more if we take into account the diversified amount of data used. Our study employed for the first time the largest updated SNIa+BAO+$H(z)$+LSS+CMB data set of cosmological observations currently available, which has been submitted to innumerable consistency checks.  The ``deconstruction analysis'' of the contour plots in Sect.\,\ref{subsect:deconstruction} has revealed with pristine clarity which are the most decisive data ingredients responsible for the dynamical vacuum signal. We have identified that the BAO+LSS+CMB components play a momentuous role in the overall fit, as they are responsible for the main effects uncovered here. The impact of the SNIa and $H(z)$ observables appears to be more moderate.  While the SNIa data were of course essential for the detection of a nonvanishing value of the cosmological term $\CC$, these data do not seem to have sufficient sensitivity (at present) for the next-to-leading step, which is to unveil the possible dynamics of $\CC$. The sensitivity for that seems to be reserved for the LSS, BAO and CMB data.

The intersection of the LSS and CMB data is most sensitive to the dynamical vacuum signature, and the simultaneous concurrence of the BAO part enhances even more the effect. Recently the BAO+LSS components have been enriched by more accurate contributions, which have helped to  augment the signs of vacuum dynamics. As we have proven here, in their absence the signal would be blurred or invisible, as it was until very recently.  In addition, we have demonstrated that the inclusion of the leading higher-order correlator data in the spectral analysis, namely the bispectrum data, is instrumental for distilling the maximum strength of the dynamical DE signal.

At the end of the day it has been possible to improve the significance of the hints of dynamical vacuum energy, which were first harvested at a confidence level of ``only'' $\gtrsim2\sigma$ in Chapter \ref{chap:Gtype}, reaching now up to about $\sim 4\sigma$ here. Our results are consistent with those found in Chapter \ref{chap:AandGRevisited} for other type of dynamical models. Overall, the signature of vacuum dynamics seems to be considerably supported by the current cosmological observations. Interestingly enough, we have shown that it is also attainable through a well-known example of $\phi$CDM model endowed with realistic ingredients, the Peebles \& Ratra (PR) model, which we have revisited in Chapter \ref{chap:MPLAbased}. With the updated data we find now that it still renders a dynamical signature comparable to the XCDM parametrization. The upshot is that either with the XCDM or the PR model we are able to exclude the absence of vacuum dynamics ($\CC$CDM) at $>3.3\sigma$ c.l.  Remarkably, this result can be surpassed with the main DVM's (RVM and $Q_{dm}$), with which we are able to find  a significance of up to $3.8\sigma$ and $3.6\sigma$, respectively. Taking, however, into account that the details of the fit depend on small variations caused by the use of different data sources and computational codes, we can assert that the DE dynamics for the main DVM's is secured at a confidence level in between
$3.6-4.3\sigma$ throughout our analysis. These results are, in addition, backed up with compelling $\Delta$AIC and $\Delta$BIC values of the Akaike and Bayesian information criteria, both being firmly anchored in the high ranges $12-18$ and $10-15$ respectively, thereby ensuring very strong evidence (according to the conventional usage of these criteria\,\cite{Burnham}) of the claimed dynamical effects.

We have also found that the signs of dynamical DE correspond to an effective quintessence behavior, in which the vacuum energy decreases with the expansion. Whether or not the ultimate reason  for such signal stems from the properties of the quantum vacuum or from some particular quintessence model, it is difficult to say at this point.
Once more we can only say that, quantitatively, the best description is granted in terms of the RVM and the $Q_{dm}$, and that the results are consistent with the traces of dynamical DE that can be found with the help of the XCDM and CPL parametrizations, as well as with a true quintessence model\,\cite{PeeblesRatra88b}. In all five cases we find unmistakable signs of DE physics beyond the $\CC$CDM.

In our work we have also clarified why previous fitting analyses based e.g. on the simple XCDM parametrization, such as the ones by the Planck 2015\, \cite{Planck2015} and BOSS collaborations\,\cite{Aubourg2015}, did entirely miss any hint of a dynamical vacuum signature. Basically, the reason stems from not using a sufficiently rich sample of the most crucial data, namely BAO and LSS, some of which were unavailable a few years ago, and could not be subsequently combined with the CMB data. The situation has now changed, as we have shown, because a wide span of sensitive data is presently at disposal. It is timely to mention at this point that a significant level of evidence  on dynamical DE at $\sim 3.5\sigma$ c.l. has recently been reported from non-parametric studies of the observational data on the DE, which aim at a model-independent result\,\cite{GongBoZhao2017}. The findings of their analysis are compatible with the ones we have disclosed here and in our recent papers. Finally, it may also be opportune to note that according to a forecast of the Dark Energy Spectropic Instrument (DESI) survey\,\cite{DESI} the dynamics of the dark energy could eventually be detected at $7\sigma$ c.l.\,\cite{GongBoZhao2017b}. It is encouraging to see that we are on the way to have at disposal the necessary technological instruments that will enable to test the dynamics of the DE with a high level of precision, certainly sufficient for unraveling the kind of effects that we are reporting here, if they are finally reconfirmed.

Needless to say, statistical evidence conventionally starts at $5\sigma$ c.l. and we will have to wait for updated observations to see if such level of significance can be achieved in the future. In the meantime, the possible dynamical character of the cosmic vacuum, as suggested by the present study, is pretty high.  It should be welcome since a dynamical vacuum gives more hope for an eventual solution of the old cosmological constant problem, perhaps the toughest problem of fundamental physics.

Let us close this work by noting that if the results presented here would be consolidated in future investigations, the longstanding  rigid status -- in fact hundred years hitherto! --  of the ``cosmological constant'' in Einstein's equations, in its traditionally accepted role for the optimal description of the cosmological data,  would be seriously disputable.

%%%%%%%%%%%%%%%%%%%%%%%%%%%%%%%%%%%%%%%%%%%%%%%%%%%%%%%%%%%%%%%%%
%%%%%%%%%%%%%%%%%%%%%%%%%%%%%%%%%%%%%%%%%%%%%%%%%%%%%%%%%%%%%%%%%
%%%%%%%%%%%%%%%%%%%%%%%%%%%%%%%%%%%%%%%%%%%%%%%%%%%%%%%%%%%%%%%%%

\section{Main bibliography of the chapter}

This chapter is based on the contents of the letter \cite{PRLnostre} and the dedicated study \cite{PRDnostre},
\vskip 0.5cm
\noindent
{\it Dynamical Vacuum against a rigid Cosmological Constant}\newline
J. Solà, J. de Cruz Pérez, and A. Gómez-Valent\newline
Submitted for publication in Phys. Rev. Lett. ; arXiv:1606.00450
\vskip 0.5cm
\noindent
{\it Towards the firsts compelling signs of vacuum dynamics in modern cosmological observations}\newline
J. Solà, J. de Cruz Pérez, and A. Gómez-Valent\newline
Submitted for publication in Phys. Rev. D. ; arXiv:1703.08218

\chapter[The $H_0$ tension in light of vacuum dynamics in the Universe]{The $H_0$ tension in light of vacuum dynamics in the Universe}
\label{chap:H0tension}

Despite the outstanding achievements of modern cosmology, the classical dispute on the precise value of $H_0$, which is the first ever parameter of modern cosmology and one of the prime parameters in the field, still goes on and on after over half a century of measurements. Recently the dispute came to the spotlight with renewed strength owing to the significant tension (at $>3\sigma$ c.l.) between the latest Planck determination obtained from the CMB anisotropies and the local (distance ladder) measurement from the Hubble Space Telescope (HST), based on Cepheids.
In this work, we investigate the impact of the running vacuum model (RVM) and related models on such a controversy. For the RVM, the vacuum energy density $\rL$ carries a mild dependence on the cosmic expansion rate, i.e. $\rL(H)$, which allows to strongly ameliorate the fit quality to the overall SNIa+BAO+$H(z)$+LSS+CMB cosmological observations as compared to the concordance $\CC$CDM model. % Here we show that the main DVMs can surpass the $\CC$CDM fit performance %even after including the local HST measurement of $H_0$.
By letting the RVM to deviate from the vacuum option, the equation of state $w=-1$ continues to be favored by the overall fit. %The kind of cosmic vacuum that is most favored, however, is \emph{not} the traditional cosmological constant but a mildly evolving one with the Hubble rate, i.e. $\rL=\rL(H(t))$. The large scale structure (LSS) formation data play a momentous role in discriminating between the two options.
Vacuum dynamics also predicts the following: i) the CMB range of values for $H_0$ is significantly more favored than the local ones, and ii) smaller values for $\sigma_8(0)$. As a result, a much better account for the LSS structure formation data is achieved as compared to the $\CC$CDM, which is based on a rigid (i.e. non-dynamical) $\CC$ term.

\section{A little of history}\label{intro}

The most celebrated fact of modern observational Cosmology is that the universe is in accelerated expansion\,\cite{SNIaRiess,SNIaPerl}. At the same time, the most paradoxical reality check is that we do not honestly understand the primary cause for such an acceleration.  The simplest picture is to assume that it is caused by a strict cosmological term, $\CC$, in Einstein's equations, but its fundamental origin is unknown\,\cite{WeinbergReview1989}. Together with the assumption of the existence of dark matter (DM) and the spatial flatness of the Friedmann-Lema\^{\i}tre-Robertson-Walker metric (viz. the metric that expresses the homogeneity and isotropy inherent to the cosmological principle), we are led to the ``concordance'' $\CC$CDM model, i.e. the standard model of Cosmology\,\cite{Peebles1993}. The model is consistent with a large body of observations, and in particular with the high precision data from the cosmic microwave background anisotropies\,\cite{Planck2015}.
Many alternative explanations of the cosmic acceleration beyond a  $\CC$-term are possible (including quintessence and the like, see Sect. \ref{subsec:AlternativesLCDM} and references therein).

The current situation with Cosmology is reminiscent of the prediction by the famous astronomer A. Sandage in the sixties, who asserted that the main task of future observational Cosmology would be the search for two parameters: the Hubble constant $H_0$ and the deceleration parameter $q_0$\,\cite{Sandage1961}. The first of them is the most important distance (and time) scale in Cosmology prior to any other cosmological quantity. Sandage's last published value with Tammann (in 2010) is $62.3$ km/s/Mpc\,\cite{TammannSandage2010} -- slightly revised in \,\cite{TammannReindl2013a,TammannReindl2013b} as $H_0 = 64.1 \pm
2.4$ km/s/Mpc. There is currently a significant tension between CMB measurements of $H_0$\,\cite{Planck2015,Planck2016} -- not far away from this value -- and local determinations emphasizing a higher range above $70$ km/s/Mpc\,\cite{RiessH0,RiessH02011}. As for $q_0$, its measurement is tantamount to determining $\CC$ in the context of the concordance model.
On fundamental grounds, however, understanding the value of $\CC$ is not just a matter of observation; in truth and in fact, it embodies one of the most important and unsolved conundrums of theoretical physics and Cosmology: the cosmological constant problem, see Chapter \ref{chap:Introduction} and references therein.

Concerning the prime parameter $H_0$, the tension among the different measurements is inherent to its long and tortuous history. Let us only recall that after Baade's revision (by a factor of one half\,\cite{Baade1944}) of the exceedingly large value $\sim 500$ km/s/Mpc originally estimated by Hubble (which implied a universe of barely two billion years old only), the Hubble parameter was subsequently lowered to $75$ km/s/Mpc and finally to $H_0=55\pm 5$ km/s/Mpc, where it remained for 20 years (until 1995), mainly under the influence of Sandage's devoted observations\,\cite{Tammann1996}. Shortly after that period the first measurements of the nonvanishing, positive, value of $\CC$ appeared\,\cite{SNIaRiess,SNIaPerl} and the typical range for $H_0$ moved upwards to $\sim 65$ km/s/Mpc. In the meantime, many different observational values of $H_0$ have piled up in the literature using different methods (see e.g. the median statistical analysis of $>550$ measurements considered in \cite{ChenRatra2011,BethapudiDesai2017}). As mentioned above, two kinds of \emph{precision} (few percent level) measurements of $H_0$ have generated considerable perplexity in the recent literature, specifically between the latest Planck values ($H^{\rm Planck}_0$) obtained from the CMB anisotropies, and the local HST measurement (based on distance estimates from Cepheids). The latter, obtained by Riess {\it et al.}\,\cite{RiessH0}, is $H_0 = 73.24\pm 1.74$\, km/s/Mpc and will be denoted $H_0^{\rm Riess}$. It can be compared with the CMB value  $H_0 = 67.51\pm 0.64$ km/s/Mpc, as extracted from Planck 2015 TT,TE,EE+lowP+lensing data\,\cite{Planck2015}, or with
$H_0 = 66.93 \pm 0.62$ km/s/Mpc, based on Planck 2015 TT,TE,EE+SIMlow data\,\cite{Planck2016}. In both cases there is a tension above $3\sigma$ c.l. (viz. $3.1\sigma$ and $3.4\sigma$, respectively) with respect to the local measurement. This situation, and in general a certain level of tension with some independent observations
in intermediate cosmological scales, has stimulated a number of discussions and possible solutions in the literature, see e.g.\,\cite{Melchiorri2016,VerdeRiessH0,Shafieloo2017,Cardona2017,Melchiorri2017,Melchiorri2017b,Zhang2017b,Wang2017,Feeney2017}.

We wish to reexamine here the $H^{\rm Riess}_0-H^{\rm Planck}_0$ tension, but not as an isolated conflict between two particular sources of observations, but rather in light of the overall fit to the cosmological data SNIa+BAO+$H(z)$+LSS+CMB. As we have demonstrated in previous chapters, by letting the cosmological vacuum energy density to slowly evolve with the expansion rate, $\rL=\rL(H)$, the global fit can be improved with respect to the $\CC$CDM at a confidence level of $3-4\sigma$.
We devote this chapter to show that the dynamical vacuum models (DVM's) can still give a better fit to the overall data, even if the local HST measurement of the Hubble parameter is taken into account. {However we find that our best-fit values of $H_0$ are much closer to the value extracted from CMB measurements \cite{Planck2015,Planck2016}}. Our analysis also corroborates that the large scale structure formation data are crucial in distinguishing the rigid vacuum option from the dynamical one.

%%%%%%%%%%%%%%%%%%%%%%%%%%%%%%%%%%%%%%%%%%%%%%%%%%%%%%%%%%%%%%%%%
%%%%%%%%%%%%%%%%%%%%%%%%%%%%%%%%%%%%%%%%%%%%%%%%%%%%%%%%%%%%%%%%%
%%%%%%%%%%%%%%%%%%%%%%%%%%%%%%%%%%%%%%%%%%%%%%%%%%%%%%%%%%%%%%%%% 

\begin{table*}
\begin{center}
\begin{scriptsize}
\resizebox{1\textwidth}{!}{
\begin{tabular}{ |c|c|c|c|c|c|c|c|c|c|}
\multicolumn{1}{c}{Model} &  \multicolumn{1}{c}{$H_0$(km/s/Mpc)} &  \multicolumn{1}{c}{$\omega_b$} & \multicolumn{1}{c}{{\small$n_s$}}  &  \multicolumn{1}{c}{$\Omega_m^0$} &\multicolumn{1}{c}{$\nu_i$} &\multicolumn{1}{c}{$w$} &\multicolumn{1}{c}{$\chi^2_{\rm min}/dof$} & \multicolumn{1}{c}{$\Delta{\rm AIC}$} & \multicolumn{1}{c}{$\Delta{\rm BIC}$}\vspace{0.5mm}
\\\hline
$\Lambda$CDM  & $68.83\pm 0.34$ & $0.02243\pm 0.00013$ &$0.973\pm 0.004$& $0.298\pm 0.004$ & - & -1  & 84.40/85 & - & - \\
\hline
XCDM  & $67.16\pm 0.67$& $0.02251\pm0.00013 $&$0.975\pm0.004$& $0.311\pm0.006$ & - &$-0.936\pm{0.023}$  & 76.80/84 & 5.35 & 3.11 \\
\hline
RVM  & $67.45\pm 0.48$& $0.02224\pm0.00014 $&$0.964\pm0.004$& $0.304\pm0.005$ &$0.00158\pm 0.00041 $ & -1  & 68.67/84 & 13.48 & 11.24 \\
\hline
$Q_{dm}$  & $67.53\pm 0.47$& $0.02222\pm0.00014 $&$0.964\pm0.004$& $0.304\pm0.005$ &$0.00218\pm 0.00058 $&-1  & 69.13/84 & 13.02 &10.78 \\
\hline
$Q_\Lambda$  & $68.84\pm 0.34$& $0.02220\pm0.00015 $&$0.964\pm0.005$& $0.299\pm0.004$ &$0.00673\pm 0.00236 $& -1  &  76.30/84 & 5.85 & 3.61\\
\hline
$w$RVM  & $67.08\pm 0.69$& $0.02228\pm0.00016 $&$0.966\pm0.005$& $0.307\pm0.007$ &$0.00140\pm 0.00048 $ & $-0.979\pm0.028$ & 68.15/83 & 11.70 & 7.27 \\
\hline
$w{Q_{dm}}$  & $67.04\pm 0.69$& $0.02228\pm0.00016 $&$0.966\pm0.005$& $0.308\pm0.007$ &$0.00189\pm 0.00066 $& $-0.973\pm 0.027$ & 68.22/83 & 11.63 & 7.20\\
\hline
$w{Q_\Lambda}$  & $67.11\pm 0.68$& $0.02227\pm0.00016 $&$0.965\pm0.005$& $0.313\pm0.006$ &$0.00708\pm 0.00241 $& $-0.933\pm0.022$ &   68.24/83 & 11.61 & 7.18\\
\hline
\end{tabular}}
\caption[Fitting results obtained without using the prior for $H_0^{\rm Riess}$]{{\scriptsize Best-fit values for the $\CC$CDM, XCDM, the three dynamical vacuum models (DVM's) and the three dynamical quasi-vacuum models ($w$DVM's), including their statistical significance ($\chi^2$-test and Akaike and Bayesian information criteria AIC and BIC).
For detailed description of the data and a full list of references, see Chapters \ref{chap:AandGRevisited}-\ref{chap:PRDbased}. The quoted number of degrees of freedom ($dof$) is equal to the number of data points minus the number of independent fitting parameters ($4$ for the $\CC$CDM, 5 for the XCDM and the DVM's, and 6 for the $w$DVM's). For the CMB data we have used the marginalized mean values and {covariance matrix} for the parameters of the compressed likelihood for Planck 2015 TT,TE,EE + lowP+ lensing data from \cite{WangDai2016}, instead of the Planck 2015 TT,TE,EE + lowP data provided in \cite{Huang}, which were used in Chapter \ref{chap:PRDbased}. Each best-fit value and the associated uncertainties have been obtained by marginalizing over the remaining parameters.}\label{tableH0Fit1}}
\end{scriptsize}
\end{center}
\end{table*}

%\newpage

\section{Dynamical vacuum models and beyond}
Let us consider a generic cosmological framework described by the spatially flat FLRW metric, in which matter is exchanging energy with a dynamical DE medium with a phenomenological EoS $p_{\CC}=w\rho_{\CC}$, where $w=-1+\epsilon$ (with $|\epsilon|\ll1$). Such medium is therefore of quasi-vacuum type, and for $w=-1$ (i.e. $\epsilon=0$) we precisely recover the genuine vacuum case. Owing, however, to the exchange of energy with matter, $\rL=\rL(\zeta)$ is in all cases a {\it dynamical} function that depends on a cosmic variable $\zeta=\zeta(t)$.  We will identify the nature of $\zeta(t)$ later on, but its presence clearly indicates that $\rL$ is no longer associated to a strictly rigid cosmological constant as in the $\CC$CDM. The Friedmann and acceleration equations read, however, formally identical to the standard case:
\begin{eqnarray}
&&3H^2=8\pi\,G\,(\rho_m+\rho_r+\rho_\Lambda(\zeta))\label{eq:FriedmannEqH0}\\
&&3H^2+2\dot{H}=-8\pi\,G\,(p_r + p_\Lambda(\zeta))\,.\label{eq:PressureEqH0}
\end{eqnarray}
Here $H=\dot{a}/a$ is the Hubble function, $a(t)$ the scale factor as a function of the cosmic time, $\rho_r$ is the energy density of the radiation component (with pressure $p_r=\rho_r/3$), and $\rho_m=\rho_b+\rho_{dm}$ involves the contributions from baryons and cold DM. The local conservation law associated to the above equations reads:
\begin{equation}\label{eq:GeneralCL}
\dot{\rho}_r + 4H\rho_r + \dot{\rho}_m + 3H\rho_m = Q\,,
\end{equation}
where
\begin{equation}\label{eq:Source}
Q = -\dot{\rho}_\CC - 3H(1+w)\rho_\CC\,.
\end{equation}
For $w=-1$ the last equation boils down to just $Q= -\dot{\rho}_\CC$, which is nonvanishing on account of $\rL(t)=\rL(\zeta(t))$.
%%%%%%%%%%%%%%%%%%%%%%%%%%%%%%%%%%%%%%%%%%%%%%%%%%%%%%%%%%%%%%%%%%%%%%%%%%%%%%%%%%%%%%%%%%%%%
\begin{table*}
\begin{center}
\begin{scriptsize}
\resizebox{1\textwidth}{!}{
\begin{tabular}{ |c|c|c|c|c|c|c|c|c|c|}
\multicolumn{1}{c}{Model} &  \multicolumn{1}{c}{$H_0$(km/s/Mpc)} &  \multicolumn{1}{c}{$\omega_b$} & \multicolumn{1}{c}{{\small$n_s$}}  &  \multicolumn{1}{c}{$\Omega_m^0$} &\multicolumn{1}{c}{$\nu_i$} &\multicolumn{1}{c}{$w$} &\multicolumn{1}{c}{$\chi^2_{\rm min}/dof$} & \multicolumn{1}{c}{$\Delta{\rm AIC}$} & \multicolumn{1}{c}{$\Delta{\rm BIC}$}\vspace{0.5mm}
\\\hline
$\Lambda$CDM  & $68.99\pm 0.33$ & $0.02247\pm 0.00013$ &$0.974\pm 0.003$& $0.296\pm 0.004$ & - & -1  & 90.59/86 & - & - \\
\hline
XCDM  & $67.98\pm 0.64$& $0.02252\pm0.00013 $&$0.975\pm0.004$& $0.304\pm0.006$ & - &$-0.960\pm{0.023}$  & 87.38/85 & 0.97 & -1.29 \\
\hline
RVM  & $67.86\pm 0.47$& $0.02232\pm0.00014 $&$0.967\pm0.004$& $0.300\pm0.004$ &$0.00133\pm 0.00040 $ & -1  & 78.96/85 & 9.39 & 7.13 \\
\hline
$Q_{dm}$  & $67.92\pm 0.46$& $0.02230\pm0.00014 $&$0.966\pm0.004$& $0.300\pm0.004$ &$0.00185\pm 0.00057 $&-1  & 79.17/85 & 9.18 & 6.92 \\
\hline
$Q_\Lambda$  & $69.00\pm 0.34$& ${0.02224}\pm0.00016 $&$0.965\pm0.005$& $0.297\pm0.004$ &$0.00669\pm 0.00234 $& -1  &  82.48/85 & 5.87 & 3.61\\
\hline
$w$RVM  & $67.95\pm 0.66$& $0.02230\pm0.00015 $&$0.966\pm0.005$& $0.300\pm0.006$ &$0.00138\pm 0.00048 $ & $-1.005\pm0.028$ & 78.93/84 & 7.11 & 2.66 \\
\hline
$w{Q_{dm}}$  & $67.90\pm 0.66$& $0.02230\pm0.00016 $&$0.966\pm0.005$& $0.300\pm0.006$ &$0.00184\pm 0.00066 $& $-0.999\pm 0.028$ & 79.17/84 & 6.88 & 2.42\\
\hline
$w{Q_\Lambda}$  & $67.94\pm 0.65$& $0.02227\pm0.00016 $&$0.966\pm0.005$& $0.306\pm0.006$ &$0.00689\pm 0.00237 $& $-0.958\pm0.022$ &   78.98/84 & 7.07 & 2.61\\
\hline
\end{tabular}}

\caption[Fitting results obtained by including the prior for $H_0^{\rm Riess}$]{{\scriptsize The same as Table \ref{tableH0Fit1} but adding the $H_0^{\rm Riess}$ local measurement from Riess {\it et al.}\, \cite{RiessH0}.}\label{tableH0Fit2}}
\end{scriptsize}
\end{center}
\end{table*}
%%%%%%%%%%%%%%%%%%%%%%%%%%%%%%%%%%%%%%%%%%%%%%%%%%%%%%%%%%%%%%%%%%%%%%%%%%%%%%%%%%%%%%%%%%%%
The simplest case is, of course, that of the concordance model, in which $\rL=\rho_{\Lambda 0}=$const and $w=-1$, so that $Q=0$ trivially. However, for $w\neq -1$ we can also have $Q=0$ in a nontrivial situation, which follows from solving Eq.\,(\ref{eq:Source}). It corresponds to the XCDM parametrization\,\cite{XCDM}, in which the DE density is dynamical and self-conserved. It is easily found in terms of the scale factor:
\begin{equation}\label{eq:rhoXCDM}
\rL^{XCDM}(a)=\rho_{\Lambda 0}\,a^{-3(1+w)}=\rho_{\Lambda 0}\,a^{-3\epsilon}\,,
\end{equation}
where $\rho_{\Lambda 0}$ is the current value.
From (\ref{eq:GeneralCL}) it then follows that the total matter component is also conserved. After equality it leads to separate conservation of cold matter and radiation.   In general, $Q$ can be a nonvanishing interaction source allowing energy exchange between matter and the quasi-vacuum medium under consideration; $Q$ can either be given by hand (e.g. through an {\it ad hoc} ansatz), or can be suggested by some specific theoretical framework. In any case the interaction source must satisfy $0<|Q|\ll\dot{\rho}_m$ since we do not wish to depart too much from the concordance model. Despite matter is exchanging energy with the vacuum or quasi-vacuum medium, we shall assume that radiation and baryons are separately self-conserved, i.e. $\dot{\rho}_r + 4H\rho_r =0$ and $\dot{\rho}_b + 3H\rho_b =0$, so that their energy densities evolve in the standard way: $\rho_r(a)=\rho_{r0}\,a^{-4}$ and $\rho_b(a) = \rho_{b0}\,a^{-3}$. The dynamics of $\rL$ can therefore be  associated to the exchange of energy exclusively with the DM (through the nonvanishing source $Q$) and/or with the possibility that the DE medium is not exactly the vacuum, $w\neq -1$, but close to it $|\epsilon|\ll 1$. Under these conditions, the coupled system of conservation equations (\ref{eq:GeneralCL})-(\ref{eq:Source}) reduces to
\begin{eqnarray}
&&\dot{\rho}_{dm}+3H\rho_{dm}=Q\label{eq:Qequations1}\\
&&\dot{\rho}_\CC + 3H\epsilon\rho_\CC=-Q\,.\label{eq:Qequations2}
\end{eqnarray}

%%%%%%%%%%%%%%%%%%%%%%%%%%%%%%%%%%%%%%%%%%%%%%%%%%%%%%%%%%%%%%%%%%%%%%%%%%%%%%%%%%%%%%%%%%%%%
\begin{figure*}
\begin{center}
\label{FigLSS1}
\includegraphics[width=5.7in]{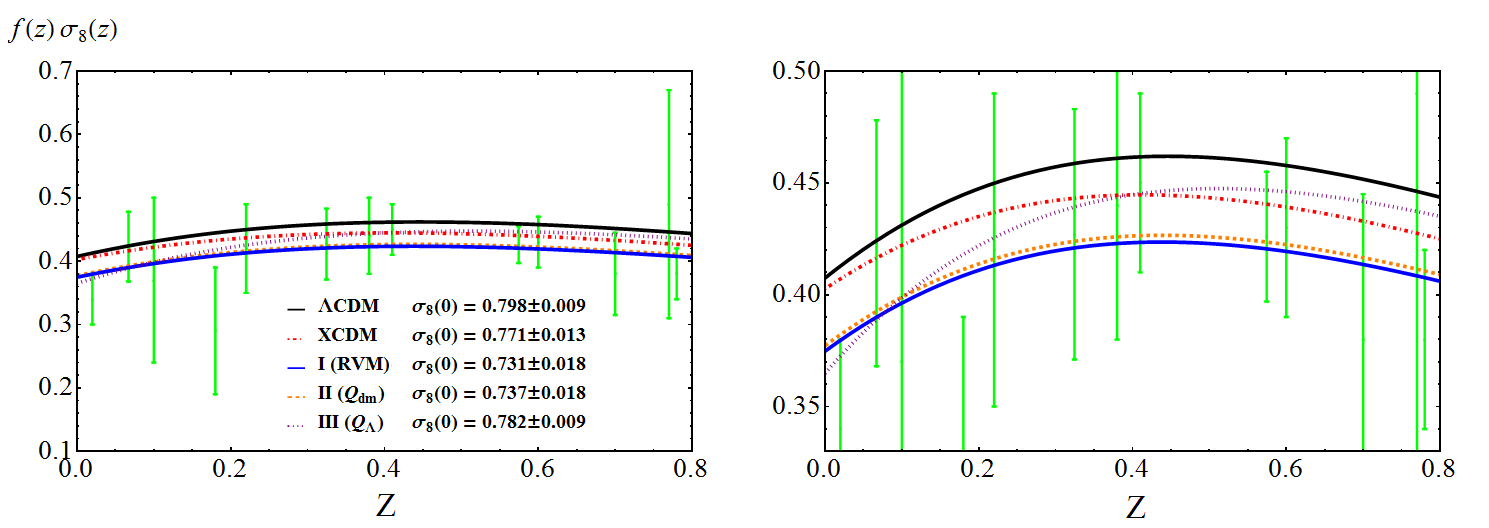}\ \ %\includegraphics[width=3in, height=2.55in]{Fig1bZoomFig1}
\caption[$f(z)\sigma_8(z)$ curves using the best-fit values of Table \ref{tableH0Fit1}]{\label{H01}{\scriptsize {\bf Left}: The LSS  structure formation data ($f(z)\sigma_8(z)$) versus the predicted curves by Models I, II and III, see equations (\ref{eq:QforModelRVM})-(\ref{eq:QforModelQL}) for the case $w=-1$, i.e. the dynamical vacuum models (DVM's), using the best-fit values in Table \ref{tableH0Fit1}. The XCDM curve is also shown. The values of $\sigma_8(0)$ that we obtain for the models are also indicated. {\bf Right}: Zoomed window of the plot on the left, which allows to better distinguish the various models.}}
\end{center}
\end{figure*} 
%%%%%%%%%%%%%%%%%%%%%%%%%%%%%%%%%%%%%%%%%%%%%%%%%%%%%%%%%%%%%%%%%%%%%%%%%%%%%%%%%%%%%%%%%%%%%

\noindent In the following we shall for definiteness focus our study of the dynamical vacuum (and quasi-vacuum) models to the three interactive sources:
\begin{eqnarray}\label{eq:QforModelRVM}
&&{\rm Model\ I\ \ }(w{\rm RVM}):\ Q=\nu\,H(3\rho_{m}+4\rho_r)\label{eq:QforModelQdm}\\
&&{\rm Model\ II\ \ }(wQ_{dm}):\ Q_{dm}=3\nu_{dm}H\rho_{dm}\\
&&{\rm Model\ III\ \ }(wQ_{\CC}):\ Q_{\CC}=3\nu_{\CC}H\rho_{\CC}\,.\label{eq:QforModelQL}
\end{eqnarray}
Here $\nu_i=\nu, \nu_{dm},\nu_{\CC}$ are small dimensionless constants, $|\nu_i|\ll1$, which are determined from the overall fit to the data, see e.g. Tables \ref{tableH0Fit1} and \ref{tableH0Fit2}.
The ordinal number names I,II and III will be used for short, but the three model names are preceded by $w$  to recall that, in the general case, the equation of state (EoS) is near the vacuum one (that is, $w=-1+\epsilon$). These dynamical quasi-vacuum models  are also denoted as $w$DVM's. In the particular case $w=-1$ (i.e. $\epsilon=0$) we recover the dynamical vacuum models (DVM's), which were previously studied in detail in Chapter \ref{chap:PRDbased}, and in this case the names of the models will not be preceded by $w$.

 In all of the above DVM's, the cosmic variable $\zeta$ can be taken to be the scale factor, $\zeta=a(t)$, since they are all analytically solvable in terms of it, as we shall see in a moment. For $w=-1$ model I is just the running vacuum model (RVM) studied in the previous chapter. It is special in that the interaction source indicated in (\ref{eq:QforModelRVM}) is not {\it ad hoc} but follows from an expression for the dynamical vacuum energy density, $\rL(\zeta)$, in which
$\zeta$ is not just the scale factor but the full Hubble rate: $\zeta=H(a)$.  The explicit RVM form reads
\begin{equation}\label{eq:RVMvacuumdadensity8}
\rho_\CC(H) = \frac{3}{8\pi{G}}\left(c_{0} + \nu{H^2}\right)\,.
\end{equation}
\newline
The additive constant $c_0=H_0^2\left(\OLo-\nu\right)$ is fixed from the condition $\rL(H_0)=\rLo$, with $\Omega^{0}_\Lambda=1-\Omega^{0}_m$. Combining the Friedmann and acceleration equations (\ref{eq:FriedmannEqH0})-(\ref{eq:PressureEqH0}), we find $\dot{H}=-(4\pi G/3)\left(3\rho_m+4\rho_r+3\epsilon\rho_\CC\right)$,
and upon differentiating (\ref{eq:RVMvacuumdadensity8}) with respect to the cosmic time we are led to
%\begin{equation}\label{eq:dotrhoCC}
$\dot{\rho}_\CC=-\nu\,H\left(3\rho_m+4\rho_r+3\epsilon\rho_\CC\right)$.
%\end{equation}
Thus, for $\epsilon=0$ (vacuum case) we indeed find $\dot{\rho}_\CC=-Q$ for $Q$ as in (\ref{eq:QforModelRVM}). However, for the quasi-vacuum case ($0<|\epsilon|\ll1$) Eq.\,(\ref{eq:Qequations2}) does not hold if $\rL(H)$ adopts the form (\ref{eq:RVMvacuumdadensity8}). This RVM form is in fact specific to the pure vacuum EoS, and it can be motivated in QFT in curved spacetime though a renormalization group equation for $\rL(H)$. In it, $\nu$ plays the role of the $\beta$-function coefficient for the running of $\rL$ with the Hubble rate, see Sect. \ref{subsec:RVMintro} for details. Thus, we naturally expect $|\nu|\ll1$ in QFT, see \cite{SolaReview2013,Fossil07}. In contrast, models II and III are purely phenomenological models instead, in which the interaction source $Q$ is introduced by hand, see e.g. Refs.\,\cite{Salvatelli2014,Murgia2016,Li2016,Melchiorri2017b} and references therein.

%%%%%%%%%%%%%%%%%%%%%%%%%%%%%%%%%%%%%%%%%%%%%%%%%%%%%%%%%%%%%%%%%%%%%%%%%%%%%%%
The energy densities for the $w$DVM's can be computed straightforwardly. For simplicity, we shall quote here the leading parts only. The exact formulas containing the radiation terms are more cumbersome. In the numerical analysis we have included the full expressions. For the matter densities, we find:
\begin{eqnarray}
\rho^{\rm I}_{dm}(a) &=& \rho_{dm0}\,a^{-3(1-\nu)} + \rho_{b0}\left(a^{-3(1-\nu)} - a^{-3}\right) \nonumber \\
\rho^{\rm II}_{dm}(a) &=& \rho_{dm0}\,a^{-3(1-\nu_{dm})}
\label{eq:rhoms} \\
\rho^{\rm III}_{dm}(a) &=&\rho_{dm0}\,a^{-3} + \frac{\nu_\CC}{\nu_\CC + w}\rLo\left(a^{-3}-a^{-3(\epsilon + \nu_\CC)}\right)\nonumber\,,
\end{eqnarray}
and for the quasi-vacuum DE densities:
\begin{eqnarray}\label{eq:rhowLdensities}
\rho^{\rm I}_\CC(a) &=& \rho_{\Lambda 0}{a^{-3\epsilon}} - \frac{\nu\,\rho_{m0}}{\nu + w}\left(a^{-3(1-\nu)}- a^{-3\epsilon}\right) \nonumber\\
\rho^{\rm II}_\CC(a)&=& \rho_{\Lambda 0}{a^{-3\epsilon}} - \frac{\nu_{dm}\,\rho_{dm0}}{\nu_{dm} + w}\,\left(a^{-3(1-\nu_{dm})}- a^{-3\epsilon}\right) \\
\rho^{\rm III}_\CC(a) &=&\rho_{\Lambda 0}\,{a^{-3(\epsilon + \nu_\CC)}}\,.\nonumber
\end{eqnarray}
Two specific dimensionless  parameters enter each formula, $\nu_{i}=(\nu,\nu_{dm},\nu_\CC)$  and $w=-1+\epsilon$. They are part of the fitting vector of free parameters for each model, as explained in detail in the caption of Table \ref{tableH0Fit1}. For $\nu_{i}\to 0$ the models become noninteractive and they all reduce to the XCDM model case (\ref{eq:rhoXCDM}). For $w=-1$ we recover the DVM's results previously studied in Chapter \ref{chap:PRDbased}. Let us also note that for $\nu_{i}>0$ the vacuum decays into DM (which is thermodynamically favorable, as discussed in Chapter \ref{chap:PRDbased}) whereas for $\nu_{i}<0$ is the other way around. Furthermore, when $w$ enters the fit, the effective behavior of the $w$DVM's is quintessence-like for $w>-1$ (i.e. $\epsilon>0$) and phantom-like for $w<-1$ ($\epsilon<0$).

Given the energy densities (\ref{eq:rhoms}) and (\ref{eq:rhowLdensities}), the Hubble function immediately follows. For example, for Model I:
\begin{equation}\label{eq:HubbewRVM}
H^2(a) = H_0^2\left[a^{-3\epsilon} + \frac{w}{w+\nu}\Omega^{0}_m\left(a^{-3(1-\nu)}- a^{-3\epsilon}\right)\right]\,.
\end{equation}
Similar formulas can be obtained for Models II and III. For $w=-1$ they all reduce to the DVM forms found in the previous chapter. And of course they all ultimately boil down to the $\CC$CDM form in the limit $w\to-1$ {\it and}\ $\nu_i\to0$.

%%%%%%%%%%%%%%%%%%%%%%%%%%%%%%%%%%%%%%%%%%%%%%%%%%%%%%%%%%%%%%%%%%%%%%%%%%%%%%%%%%%%%%%%%%%%%

\begin{table*}
\begin{center}
\begin{scriptsize}
\resizebox{1\textwidth}{!}{
\begin{tabular}{ |c|c|c|c|c|c|c|c|c|c|}
\multicolumn{1}{c}{Model} &  \multicolumn{1}{c}{$H_0$(km/s/Mpc)} &  \multicolumn{1}{c}{$\omega_b$} & \multicolumn{1}{c}{{\small$n_s$}}  &  \multicolumn{1}{c}{$\Omega_m^0$} &\multicolumn{1}{c}{$\nu_i$} &\multicolumn{1}{c}{$w$} &\multicolumn{1}{c}{$\chi^2_{\rm min}/dof$} & \multicolumn{1}{c}{$\Delta{\rm AIC}$} & \multicolumn{1}{c}{$\Delta{\rm BIC}$}\vspace{0.5mm}
\\\hline
$\Lambda$CDM  & $68.23\pm 0.38$ & $0.02234\pm 0.00013$ &$0.968\pm 0.004$& $0.306\pm 0.005$ & - & -1  & 13.85/11 & - & - \\
\hline
RVM  & $67.70\pm 0.69$& $0.02227\pm0.00016 $&$0.965\pm0.005$& $0.306\pm0.005$ &$0.0010\pm 0.0010 $ & -1  & 13.02/10 & -3.84 & -1.88 \\
\hline
$Q_\Lambda$  & $68.34\pm 0.40$& $0.02226\pm0.00016 $&$0.965\pm0.005$& $0.305\pm0.005$ &$0.0030\pm 0.0030 $& -1  &  12.91/10 & -3.73 & -1.77 \\
\hline
$w$RVM  & $66.34\pm 2.30$& $0.02228\pm0.00016 $&$0.966\pm0.005$& $0.313\pm0.012$ &$0.0017\pm 0.0016 $ & $-0.956\pm0.071$ & 12.65/9 & -9.30 & -4.22 \\
\hline
$w{Q_\Lambda}$  & $66.71\pm 1.77$& $0.02226\pm0.00016 $&$0.965\pm0.005$& $0.317\pm0.014$ &$0.0070\pm 0.0054 $& $-0.921\pm0.082$ &   12.06/9 & -8.71 & -3.63\\
\hline
$\Lambda$CDM* & $68.46\pm 0.37$ & $0.02239\pm 0.00013$ &$0.969\pm 0.004$& $0.303\pm 0.005$ & - & -1 & 21.76/12 & - & - \\
\hline
RVM*  & $68.48\pm 0.67$& $0.02240\pm0.00015 $&$0.969\pm0.005$& $0.303\pm0.005$ &$0.0000\pm 0.0010 $ & -1  & 21.76/11 & -4.36 & -2.77 \\
\hline
$Q_\Lambda$*  & $68.34\pm 0.39$& $0.02224\pm0.00016 $&$0.966\pm0.005$& $0.302\pm0.005$ &$0.0034\pm 0.0030 $& -1  &  20.45/11 & -3.05 & -1.46 \\
\hline
Ia ($w$RVM*)  & $70.95\pm 1.46$& $0.02231\pm0.00016 $&$0.967\pm0.005$& $0.290\pm0.008$ &$-0.0008\pm 0.0010 $ & $-1.094\pm0.050$ & 18.03/10 & -5.97 & -1.82 \\
\hline
IIIa ($w{Q_\Lambda}$*)  & $70.27\pm 1.33$& $0.02228\pm0.00016 $&$0.966\pm0.005$& $0.291\pm0.010$ &$-0.0006\pm 0.0042 $& $-1.086\pm0.065$ &   18.64/10 & -6.58  & -2.43 \\
\hline
\end{tabular}}

\caption[Fitting results obtained by excluding the LSS data]{{\scriptsize Best-fit values for the $\CC$CDM and models RVM, Q$_{\Lambda}$, $w$RVM and $w$Q$_\Lambda$  by making use of the CMB+BAO data only. In contrast to Tables \ref{tableH0Fit1}-\ref{tableH0Fit2}, we fully exclude now the LSS data (see Chapter \ref{chap:PRDbased} for the complete reference list) to test its effect. The starred/non-starred cases correspond respectively to adding or not the local value $H_0^{\rm Riess}$ from \cite{RiessH0} as data point in the fit. The AIC and BIC differences of the starred models are computed with respect to the $\Lambda$CDM*. We can see that under these conditions models
tend to have $\Delta$AIC, $\Delta$BIC< 0, including the last two starred scenarios, which are capable to significantly approach $H_0^{\rm Riess}$.}\label{tableH0Fit3}}
\end{scriptsize}
\end{center}
\end{table*}

\section{Structure formation: the role of the LSS data}
The analysis of structure formation plays a crucial role in comparing the various models. For the $\CC$CDM and XCDM  we use the standard perturbations equation\,\cite{Peebles1993}
\begin{equation}\label{diffeqLCDM}
\ddot{\delta}_m+2H\,\dot{\delta}_m-4\pi
G\rmr\,\delta_m=0\,,
\end{equation}
with, however, the Hubble function corresponding to each one of these models.
For the $w$DVM's, a step further is needed: the perturbation equation not only involves the modified Hubble function but the equation itself becomes modified. Trading the cosmic time for the scale factor and extending the analysis of previous chapters and Appendix \ref{sec:LPmatter-vacuum} for the case  $w\neq -1$ ($\epsilon\neq 0$), we find
\begin{equation}\label{diffeqDH0}
\delta^{\prime\prime}_m + \frac{A(a)}{a}\delta_{m}^{\prime} + \frac{B(a)}{a^2}\delta_m = 0  \,,
\end{equation}
where the prime denotes differentiation with respect to the scale factor,
and the functions $A(a)$ and $B(a)$ read
\begin{eqnarray}
&&A(a) = 3 + \frac{aH^{'}}{H} + \frac{\Psi}{H} - 3r\epsilon\label{eq:Afunction}\\
&&B(a) = -\frac{4\pi{G}\rho_m }{H^2} + 2\frac{\Psi}{H} + \frac{a\Psi^{'}}{H} -15r\epsilon- 9\epsilon^{2}r^{2} +3\epsilon(1+r)\frac{\Psi}{H} -3r\epsilon\frac{aH^{'}}{H}\,. \label{eq:Bfunction}
\end{eqnarray}
Here $r \equiv \rho_\Lambda/\rho_m$ and  $\Psi\equiv -\dot{\rho}_{\Lambda}/{\rmr}$. {For $\nu_i=0$ we have $\Psi=3Hr\epsilon$}, and calculations show that (\ref{diffeqDH0})  can be brought back to the common one for  XCDM and $\CC$CDM, Eq.\,(\ref{diffeqLCDM}).

To solve the above perturbation equations we have to fix the initial conditions on $\delta_m$ and ${\delta}'_m$ for each model at high redshift, namely when non-relativistic matter dominates over radiation and DE.
Functions (\ref{eq:Afunction}) and (\ref{eq:Bfunction}) are then approximately constant and Eq.\,(\ref{diffeqDH0}) admits power-law solutions $\delta_m(a) = a^{s}$. The values of $s$ for each model turn out to be:
\begin{eqnarray}\label{eq:svalues}
s^{\rm I} &=& 1 + \frac{3}{5}\nu\left(\frac{1}{w} -4\right) + \mathcal{O}(\nu^2) \nonumber \\
s^{\rm II} &=& 1 -\frac{3}{5}\nu_{dm}\left(  1 + 3\frac{\Omega^{0}_{dm}}{\Omega^{0}_{m}} -\frac{1}{w}   \right) + \mathcal{O}({\nu_{dm}}^{2}) \\
s^{\rm III} &=&1\nonumber\,.
\end{eqnarray}
Once more we can check that for $w=-1$ all of the above equations (\ref{diffeqDH0})-(\ref{eq:svalues}) return the DVM results of Chapter \ref{chap:PRDbased}.

The analysis of the linear LSS regime is usually implemented with the help of the weighted linear growth $f(z)\sigma_8(z)$, where $f(z)=d\ln{\delta_m}/d\ln{a}$ is the growth factor and $\sigma_8(z)$ is the rms mass fluctuation on $R_8=8\,h^{-1}$ Mpc scales. It is computed as follows (see e.g. \cite{ApJnostre,PRDnostre}):
\begin{equation}
\begin{small}\sigma_{\rm 8}(z)=\sigma_{8, \CC}
\frac{\delta_m(z)}{\delta^{\CC}_{m}(0)}
\sqrt{\frac{\int_{0}^{\infty} k^{n_s+2} T^{2}(\vec{p},k)
W^2(kR_{8}) dk} {\int_{0}^{\infty} k^{n_{s,\CC}+2} T^{2}(\vec{p}_\Lambda,k) W^2(kR_{8,\Lambda}) dk}}\,,\label{s88generalH0}
\end{small}\end{equation}
where $W$ is a top-hat smoothing function and $T(\vec{p},k)$ the transfer function. The fitting parameters for each model are contained in $\vec{p}$.
Following the mentioned references, we have defined as fiducial model the $\CC$CDM at fixed parameter values from the Planck 2015 TT,TE,EE+lowP+lensing data\,\cite{Planck2015}. These fiducial values are collected in  $\vec{p}_\CC$.
In Figs. \ref{H01}-\ref{fig:H02} we display  $f(z)\sigma_8(z)$ for the various models using the fitted values of Tables \ref{tableH0Fit1}-\ref{tableH0Fit3}. We remark that our BAO and LSS data include the bispectrum data points from Ref.\,\cite{GilMarin2}
--see Chapter \ref{chap:PRDbased} for a full-fledged explanation of our data sets.  In the next section, we discuss our results for the various models and assess their ability to improve the $\CC$CDM fit as well as their impact on the $H_0$ tension.

%%%%%%%%%%%%%%%%%%%%%%%%%%%%%%%%%%%%%%%%%%%%%%%%%%%%%%%%%%%%%%%%%%%%%%%%%%%%%%%

\begin{figure}
\begin{center}
\label{FigLSS2}
\includegraphics[width=3.3in]{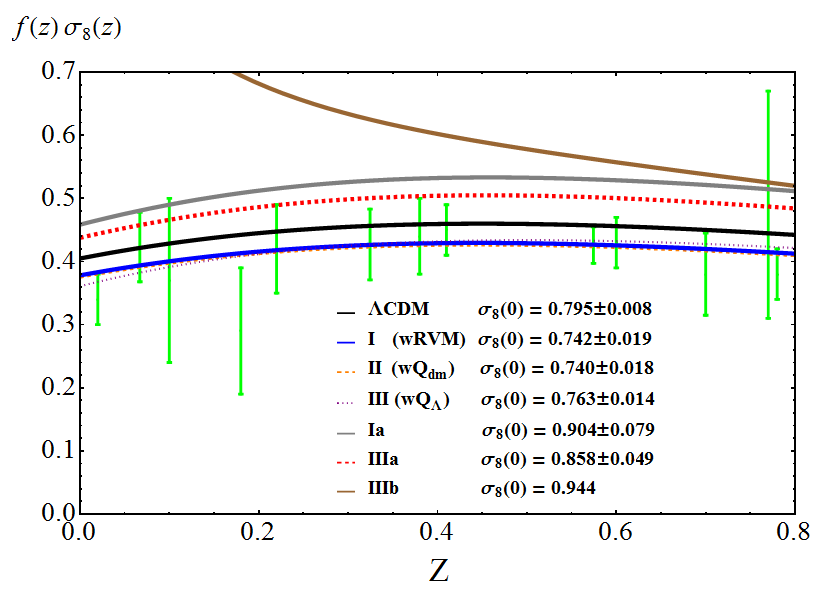}
\caption[$f(z)\sigma_8(z)$ curves using the best-fit values of Tables \ref{tableH0Fit2} and \ref{tableH0Fit3}]{\label{fig:H02}{\scriptsize The LSS  structure formation data ($f(z)\sigma_8(z)$) and the theoretical predictions for models (\ref{eq:QforModelRVM})-(\ref{eq:QforModelQL}), using the best-fit values in Tables \ref{tableH0Fit2} and \ref{tableH0Fit3}. The curves for the cases Ia, IIIa correspond to special scenarios for Models I and III where the agreement with the  Riess {\it et al.} local value $H_0^{\rm Riess}$\,\cite{RiessH0} is better (cf. Table \ref{tableH0Fit3}). The price, however, is that the concordance with the LSS data is now spoiled. Case IIIb is our theoretical prediction for the scenario proposed in \cite{Melchiorri2017b}, aimed at optimally relaxing the tension with $H_0^{\rm Riess}$. Unfortunately, the last three scenarios lead to phantom-like DE and are in serious disagreement with the LSS data.
}}
\end{center}
\end{figure}

%%%%%%%%%%%%%%%%%%%%%%%%%%%%%%%%%%%%%%%%%%%%%%%%%%%%%%%%%%%%%%%%%%%%%%%%%%%%%%%%%%%%

\section{Discussion}
Following Chapter \ref{chap:PRDbased} (see also Appendix \ref{chap:App5}) the statistical analysis of the various models is performed in terms of a joint likelihood function, which is the product of the likelihoods for each data source
and includes the corresponding covariance matrices.
As indicated in the caption of Table \ref{tableH0Fit1},  the $\CC$CDM has $4$ parameters, whereas the XCDM and the DVM's have $5$, and finally any $w$DVM has $6$. Thus, for a fairer comparison of the various nonstandard models with the concordance $\CC$CDM we have to invoke efficient criteria in which the presence of extra parameters in a given model is conveniently penalized, so as to achieve a balanced comparison with the model having less parameters. We do that again through the use of the  Akaike information criterion (AIC) and the Bayesian information criterion (BIC), see the previous chapters for details. Take, for instance, Tables \ref{tableH0Fit1} and \ref{tableH0Fit2}, where in all cases the less favored model is the $\CC$CDM (thus with larger AIC and BIC).
For $\Delta$AIC and $\Delta$BIC in the range $6-10$ one speaks of ``strong evidence'' against the $\CC$CDM, and hence in favor of the nonstandard models being considered. This is typically the situation for the RVM and $Q_{dm}$ vacuum models in Table \ref{tableH0Fit2} and for the three $w$DVM's in Table \ref{tableH0Fit1}.
Neither the  XCDM nor the $Q_\CC$ vacuum model attain the ``strong evidence'' threshold in Tables \ref{tableH0Fit1} or \ref{tableH0Fit2}. The XCDM parametrization, which is used as a baseline for comparison of the dynamical DE models, is nevertheless capable of detecting significant signs of dynamical DE, mainly in Table \ref{tableH0Fit1} (in which $H_0^{\rm Riess}$ is excluded), but not so in Table \ref{tableH0Fit2} (where $H_0^{\rm Riess}$ is included).  In contrast, model $Q_\CC$ does not change much from Table \ref{tableH0Fit1} to Table \ref{tableH0Fit2}.

%%%%%%%%%%%%%%%%%%%%%%%%%%%%%%%%%%%%%%%%%%%%%%%%%%%%%%%%%%%%%%%%%%%%%%%%%%%%%%%

\begin{figure}
\begin{center}
\label{contours}
\includegraphics[width=5in]{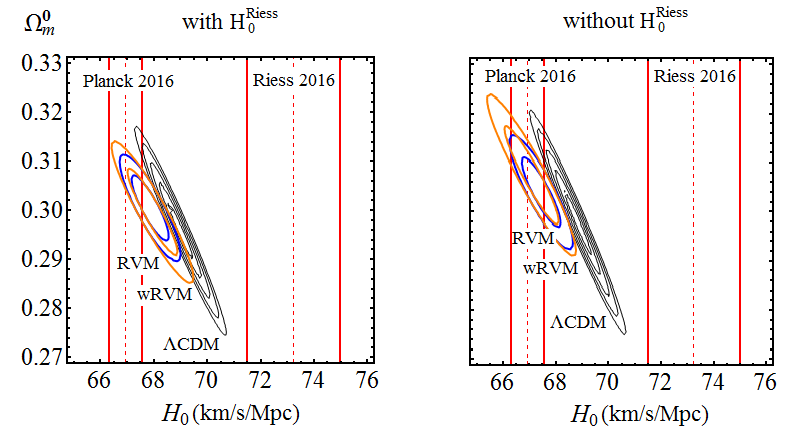}
\caption[CL's for the RVM, the $w$RVM and the $\CC$CDM in the $(H_0,\Omo)$ plane]{\label{H03}{\scriptsize Contour plots for RVM (blue) and $w$RVM (orange) up to $2\sigma$, and for the $\CC$CDM (black) up to $5\sigma$ in the $(H_0,\Omo)$-plane. {Shown are the two relevant cases under study: the plot on the left is for when the local $H_0$ value of Riess {\it et al.}\,\cite{RiessH0} is included in the fit (cf. Table \ref{tableH0Fit2}), and the plot on the right is for when that local value is {\it not} included (cf. Table \ref{tableH0Fit1}). Any attempt at reaching the $H_0^{\rm Riess}$ neighborhood  enforces to pick too small values  $\Omo<0.27$ through extended contours that go beyond  $5\sigma$ c.l.} We also observe that the two ($w$)RVM's are much more compatible (already at $1\sigma$) with the $H_0^{\rm Planck}$ range than the $\CC$CDM. The latter, instead, requires $5\sigma$ contours to reach the $H^{\rm Planck}_0$ $1\sigma$ region when $H_0^{\rm Riess}$ is included in the fit, and $4\sigma$ contours when it is {\it not} included. Thus, remarkably, in both cases when the full data string SNIa+BAO+$H(z)$+LSS+CMB enters the fit the $\CC$CDM has difficulties to overlap also with the $H_0^{\rm Planck}$ range at $1\sigma$, in contrast to the RVM and $w$RVM.
}}
\end{center}
\end{figure} 

%%%%%%%%%%%%%%%%%%%%%%%%%%%%%%%%%%%%%%%%%%%%%%%%%%%%%%%%%%%%%%%%%%%%%%%%%%%%%%%

In actual fact, the vacuum model III ($Q_\CC$) tends to remain always fairly close to the $\CC$CDM. Its dynamics is weaker than that of the main DVM's (RVM and $Q_{dm}$). Being $|\nu_i|\ll 1$ for all the DVM's, the evolution of its vacuum energy density is approximately logarithmic: $\rL^{\rm III}\sim\rLo(1-3\nu_{\CC}\,\ln{a})$, as it follows from  (\ref{eq:rhowLdensities}) with $\epsilon=0$. Thus, it is significantly milder in comparison to that of the main DVM's, for which $\rL^{\rm I, II}\sim \rLo\left[1+(\Omega^{0}_m/\Omega^{0}_\Lambda)\nu_i (a^{-3}-1)\right]$. The performance of $Q_\CC$ can only be slightly better than that of $\CC$CDM, a fact that may have not been noted in previous studies -- see \,\cite{Salvatelli2014,Melchiorri2016,Murgia2016,Li2016,Melchiorri2017b} and references therein.

According to the same jargon, when $\Delta$AIC and $\Delta$BIC are both above 10 one speaks of ``very strong evidence'' against the unfavored model ($\CC$CDM), wherefore in favor of the alternative ones. It is certainly the case of the RVM and $Q_{dm}$ models in Table \ref{tableH0Fit1}, which are singled out as being much better than the $\CC$CDM in their ability to describe the overall observations. From Table \ref{tableH0Fit1} we can see that the best-fit values of $\nu_i$ for these models are secured at a confidence level of $\sim 3.8\sigma$. These two models are indeed the most conspicuous ones in our entire analysis, and remain strongly favored even if $H_0^{\rm Riess}$\,\cite{RiessH0} is included (cf. Table \ref{tableH0Fit2}). In the last case, the best-fit values of $\nu_i$ for the two models are still supported at a fairly large c.l. ($\sim 3.2\sigma$). This shows that the overall fit to the data in terms of dynamical vacuum is a real option since the fit quality is not exceedingly perturbed in the presence of the data point $H_0^{\rm Riess}$. However, the optimal situation is really attained in the absence of that point, not only because the fit quality is then higher but also because that point remains out of the fit range whenever the large scale structure formation data (LSS) are included. For this reason we tend to treat that input as an outlier. In the following, we will argue that a truly consistent picture with all the data is only possible for $H_0$ in the vicinity of  $H_0^{\rm Planck}$ rather than in that of $H_0^{\rm Riess}$.

The conclusion is that the $H_0^{\rm Riess}$-$H_0^{\rm Planck}$  tension cannot be relaxed without unduly forcing the overall fit, which is highly sensitive to the LSS data. It goes without saying that one cannot have a prediction that matches both $H_0$ regions at the same time, so at some point new observations (or the discovery of some systematic error in one of the experiments) will help to consolidate one of the two ranges of values and exclude definitely the other. At present no favorable fit can be obtained from the $\CC$CDM that is compatible with any of the two $H_0$ ranges. This is transparent from Figs. \ref{H03} and \ref{H04}, in which the $\CC$CDM remains always in between the two regions. However, our work shows that a solution (with minimum cost) is possible in terms of vacuum dynamics. Such solution, which inevitably puts aside the $H_0^{\rm Riess}$ range, is however compatible with all the remaining data and tends to favor the Planck range of $H_0$ values. The DVM's can indeed provide an excellent fit to the overall cosmological observations and be fully compatible with both the $H_0^{\rm Planck}$ value and at the same time with the needed low values of the $\sigma_8(0)$ parameter, these low values of $\sigma_8(0)$ being crucial to fit the structure formation data. Such strategy is only possible in the presence of vacuum dynamics, whilst it is impossible with a rigid $\CC$-term, i.e. is not available to the $\CC$CDM.

In Fig. \ref{H01} we confront the various models with the LSS data when the local measurement $H_0^{\rm Riess}$ is not included in our fit. The differences can be better appraised in the plot on the right, where we observe that the RVM and $Q_{dm}$ curves stay significantly lower than the $\CC$CDM one (hence matching better the data than the $\CC$CDM), whereas those of XCDM and $Q_\CC$ remain in between.

Concerning the $w$DVM's, namely the quasi-vacuum models in which an extra parameter is at play (the EoS parameter $w$), we observe a significant difference as compared to the DVM's (with vacuum EoS): they {\it all} provide a similarly good fit quality, clearly superior to that of the $\CC$CDM (cf. Tables \ref{tableH0Fit1} and \ref{tableH0Fit2}) but below that of the main DVM's (RVM and $Q_{dm}$), whose performance is outstanding.

In Table \ref{tableH0Fit3}, in an attempt to draw our fit nearer and nearer to $H_0^{\rm Riess}$\,\cite{RiessH0}, we test the effect of ignoring the LSS formation data, thus granting more freedom to the fit parameter space. We perform this test using the $\CC$CDM and models $(w)$RVM and $(w)Q_\CC$ (i.e. models I and III and testing both the vacuum and quasi-vacuum options), and  we fit them to the CMB+BAO data only. We can see that the fit values for $H_0$ increase in all starred scenarios (i.e. those involving the $H_0^{\rm Riess}$ data point in the fit), and specially for the cases Ia and IIIa in Table \ref{tableH0Fit3}. Nonetheless, these lead to $\nu_i<0$ and $w<-1$ (and hence imply phantom-like DE); and, what is worse, the agreement with the LSS data is ruined (cf. Fig. \ref{fig:H02}) since the corresponding curves are shifted too high (beyond the $\CC$CDM one).  In the same figure we superimpose one more scenario, called IIIb, corresponding to a rather acute phantom behavior ($w=-1.184\pm0.064$). The latter was recently explored in \cite{Melchiorri2017b} so as to maximally relax the $H_0$ tension -- see also\,\cite{Melchiorri2016}. Unfortunately, we find that the associated LSS curve is completely strayed since it fails to minimally describe the $f\sigma_8$ data points. 

%%%%%%%%%%%%%%%%%%%%%%%%%%%%%%%%%%%%%%%%%%%%%%%%%%%%%%%%%%%%%%%%%%%%%%%%%%%%%%%

\begin{figure}
\begin{center}
\label{contours}
\includegraphics[width=5in]{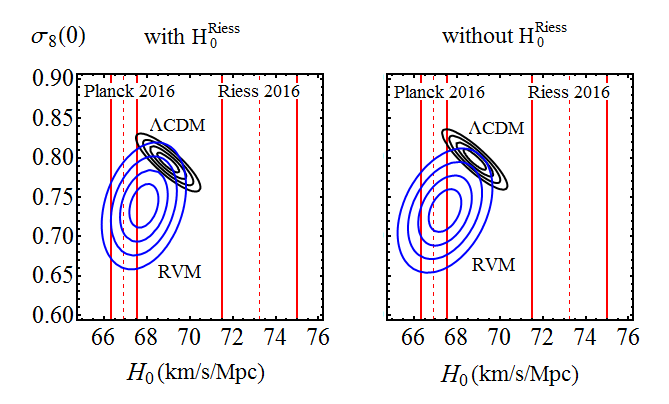}
\caption[CL's for the RVM and the $\CC$CDM in the $(H_0,\sigma_8(0))$ plane]{\label{H04}{\scriptsize Contour lines for the $\CC$CDM (black) and RVM (blue) up to $4\sigma$ in the $(H_0,\sigma_8(0))$-plane. {As in Fig. 3, we present in the {\it left plot} the case when the local $H_0$ value of Riess {\it et al.}\,\cite{RiessH0} is included in the fit (cf. Table \ref{tableH0Fit2}), whereas in the {\it right plot} the case when that local value is {\it not} included (cf. Table \ref{tableH0Fit1}). Again, any attempt at reaching the $H_0^{\rm Riess}$ neighborhood  enforces to extend the contours beyond the $5\sigma$ c.l.}, which would lead to a too low value of $\Omega_m^{0}$ in both cases (cf. Fig. \ref{H03}) and, in addition, would result in a too large value of $\sigma_8(0)$ for the RVM. Notice that $H_0$ and $\sigma_8(0)$ are positively correlated in the RVM (i.e. $H_0$ decreases when $\sigma_8(0)$ decreases), whilst they are anticorrelated in the $\CC$CDM ($H_0$ increases when $\sigma_8(0)$ decreases, and vice versa). It is precisely this opposite correlation feature with respect to the $\CC$CDM what allows the RVM to improve the LSS fit in the region where both $H_0$ and $\sigma_8(0)$ are smaller than the respective $\CC$CDM values (cf. Fig. \ref{H01}). This explains why the Planck range for $H_0$ is clearly preferred by the RVM, as it allows a much better description of the LSS data.}}
\end{center}
\end{figure}
%%%%%%%%%%%%%%%%%%%%%%%%%%%%%%%%%%%%%%%%%%%%%%%%%%%%%%%%%%%%%%%%%%%%%%%%%%%%%%%

In Fig. \ref{H03} we demonstrate in a very visual way that, in the context of the overall observations (i.e.  SNIa+BAO+$H(z)$+LSS+CMB), whether including or not including the data point $H_0^{\rm Riess}$ (cf. Tables \ref{tableH0Fit1} and \ref{tableH0Fit2}), it becomes impossible to getting closer to  the local measurement $H_0^{\rm Riess}$ unless we go beyond the $5\sigma$ contours and end up with a too low value $\Omega_m^0<0.27$. These results are aligned with those of \cite{Zhai2017}, in which the authors are also unable to accommodate the $H_0^{\rm Riess}$ value when a string of SNIa+BAO+$H(z)$+LSS+CMB data (similar but {\it not} equal to the one used by us) is taken into account. Moreover, we observe in Fig. \ref{H03} not only that both the RVM and $w$RVM remain much closer to  $H^{\rm Planck}_0$ than to $H_0^{\rm Riess}$, but also that they are overlapping with the $H^{\rm Planck}_0$  range much better than the $\CC$CDM does. The latter is seen to have serious difficulties in reaching the Planck range unless we use the most external regions of the elongated contours shown in Fig. \ref{H03}.

Many other works in the literature have studied the existing $H_0$ tension. For instance, in \cite{Wang2017} the authors find a value $H_0 = 69.13\pm 2.34$ km/s/Mpc assuming the $\Lambda$CDM model. This result almost coincides with the central values of $H_0$ that we present in Tables \ref{tableH0Fit1} and \ref{tableH0Fit2} for the $\Lambda$CDM. It is also very close to the arithmetic mean of $H^{\rm Planck}_0$ and $H_0^{\rm Riess}$. This fact together with its large uncertainty allows to strongly reduce the aforementioned tension. Nevertheless it is important to notice that the value of \cite{Wang2017} has been obtained using only BAO data, and this explains the large uncertainty that they find. In contrast, we have considered a more complete data set, which has allowed us to better constrain this parameter and to see that, in fact, when a larger data set involving LSS is used in the fitting analysis, the resulting value of the Hubble parameter for the $\Lambda$CDM is also incompatible with the Planck best-fit value at a $4-5\sigma$ c.l. Thus, the $\Lambda$CDM seems to be in conflict not only with the HST estimation of $H_0$, but also with the Planck one! 

Finally, in Figs. \ref{H04} and \ref{H05} we consider the contour plots (up to $4\sigma$ and $3\sigma$, respectively) in the $(H_0,\sigma_8(0))$-plane for different situations. Specifically, in the case of Fig. \ref{H04} the plots on the left and on the right are in exact correspondence with the situations previously presented in the left and right plots of Fig. \ref{H03}, respectively\footnote{The $H_0^{\rm Planck}$ band indicated in Figs. \ref{H03}-\ref{H05} is that of \cite{Planck2016}, which has no significant differences with that of \cite{Planck2015}.}. As expected, the contours in the left plot of Fig. \ref{H04} are slightly shifted (``attracted'') to the right (i.e. towards the $H_0^{\rm Riess}$ region) as compared to those in the right plot because in the former $H_0^{\rm Riess}$ was included as a data point in the fit, whereas $H_0^{\rm Riess}$ was not included in the latter. Therefore, in the last case the contours for the RVM are more centered in the $H_0^{\rm Planck}$ region and at the same time centered at relatively low values of $\sigma_8(0)\simeq0.73-0.74$, which are precisely those needed for a perfect matching with the experimental data points on structure formation (cf. Fig. \ref{H01}). On the other hand, in the case of Fig. \ref{H05} the contour lines correspond to the fitting sets Ia, IIIa of Table \ref{tableH0Fit3} (in which BAO and CMB data, but \emph{no} LSS formation data, are involved). As can be seen, the contour lines in Fig. \ref{H05} can attain the Riess 2016 region for $H_0$, but they are centered at rather high values ($\sim 0.9$) of the parameter $\sigma_8(0)$. These are clearly higher than the needed values $\sigma_8(0)\simeq 0.73-0.74$. This fact demonstrates once more that such option leads to a bad description of the structure formation data.
The isolated point in Fig. \ref{H05} is even worst: it corresponds to the aforementioned theoretical prediction for the scenario IIIb proposed in \cite{Melchiorri2017b}, in which the $H_0^{\rm Riess}$ region can be clearly attained but at the price of a serious disagreement with the LSS data. Here we can see, with pristine clarity, that such isolated point, despite it comfortably reaches the $H_0^{\rm Riess}$ region, it attains a value of  $\sigma_8(0)$ near $1$, thence completely strayed from the observations. This is, of course, the reason  why the upper curve in Fig. \ref{fig:H02} fails to describe essentially all points of the $f(z)\sigma_8(z)$ observable. So, as it turns, it is impossible to reach the $H_0^{\rm Riess}$ region without paying a high price, no matter what strategy is concocted to approach it in parameter space.

%%%%%%%%%%%%%%%%%%%%%%%%%%%%%%%%%%%%%%%%%%%%%%%%%%%%%%%%%%%%%%%%%%%%%%%%%%%%%%%
\begin{figure*}[t!]
\begin{center}
\label{contours}
\includegraphics[width=3in]{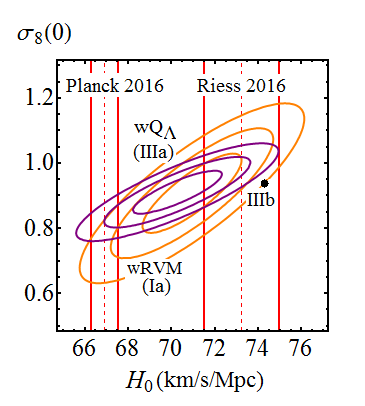}
\caption[CL's for the Ia and IIIa in the $(H_0,\sigma_8(0))$ plane]{\label{H05}{\scriptsize Contour lines for the models $w$RVM (Ia) and $w$Q$_\CC$ (IIIa) up to $3\sigma$ in the $(H_0,\sigma_8(0))$-plane, depicted in orange and purple, respectively, together with the isolated point (in black) extracted from the analysis of Ref. \cite{Melchiorri2017b}, which we call IIIb. The cases Ia, IIIa and IIIb correspond to special scenarios with $w\neq -1$ for Models I and III in which the value $H_0^{\rm Riess}$  is included as a data point and then a suitable strategy is followed to optimize the fit agreement with such value. The strategy consists to exploit the freedom in $w$ and remove the LSS data from the fit analysis. The plot clearly shows that some agreement is indeed possible, but only if  $w$ takes on values in the phantom region ($w<-1$) (see text) and at the expense of an anomalous (too large) value of the parameter $\sigma_8(0)$, what seriously spoils the concordance with the LSS data, as can be seen in Fig. \ref{fig:H02}.
}}
\end{center}
\end{figure*}

%%%%%%%%%%%%%%%%%%%%%%%%%%%%%%%%%%%%%%%%%%%%%%%%%%%%%%%%%%%%%%%%%%%%%%%%%%%%%%%

As indicated, we must still remain open to the possibility that the $H^{\rm Planck}_0$ and/or $H^{\rm Riess}_0$ measurements are affected by some kind of (unknown) systematic errors, although some of these possibilities may be on the way of  being ruled out by recent works. For instance, in \cite{Aylor2017} the authors study the systematic errors in Planck's data by comparing them with the South Pole Telescope data. Their conclusion is that there is no evidence of systematic errors in Planck's results. If confirmed, the class of the $(w)$RVMs studied here would offer a viable solution to both the $H_0$ and $\sigma_8(0)$ existing tensions in the data, which are both unaccountable within the $\CC$CDM. Another interesting result is the ``blinded''  determination of $H_0$  from \cite{Zhang2017b}, based on a reanalysis of the SNIa and Cepheid variables data from the older work by Riess et al. \cite{RiessH02011}. These authors find $H_0 = 72.5\pm 3.2$ km/s/Mpc, which should be compared with $H_0 = 73.8\pm 2.4$ km/s/Mpc\,\cite{RiessH02011}. Obviously, the tension with $H^{\rm Planck}_0$ diminished since the central value decreased and in addition the uncertainty has grown by $\sim 33\%$. We should now wait for a similar reanalysis to be made on the original sample used in \cite{RiessH0}, i.e. the one supporting the value  $H^{\rm Riess}_0$, as planned in \cite{Zhang2017b}. In  \cite{Addison2017}  they show that by combining the latest BAO results with WMAP, Atacama Cosmology Telescope (ACT), or South Pole Telescope (SPT) CMB data produces values of $H_0$  that are $2.4-3.1\sigma$ lower than the distance ladder, independent of Planck. These authors conclude from their analysis  that it is not possible to explain the $H_0$  disagreement solely with a systematic error specific to the Planck data}. Let us mention other works, see e.g. \cite{Cardona2017,Feeney2017}, in which a value closer to $H_0^{\rm Riess}$ is found and the tension is not so severely loosened; or the work\,\cite{FollinKnox2017}, which excludes systematic bias or uncertainty in the Cepheid calibration step of the distance ladder measurement by\,\cite{RiessH0}. We also recall the aforementioned recent study \cite{Lin2017}, where the authors run a new (dis)cordance test to compare the constraints on $H_0$ from different methods and conclude that the local
measurement is an outlier compared to the others, what would favor a systematics-based explanation.

The birth of gravitational-wave multi-messenger astronomy is a now a fact. The detection of the gravitational-wave and the electromagnetic counterpart produced by the merger of the binary neutron-star system GW170817 \cite{GWevent} has allowed to use this event as a standard siren and measure $H_0$ \cite{GWH0}. The value reported in \cite{GWH0} is $H_0=70.0^{+12.0}_{-8.0}\,{\rm km/s/Mpc}$, which lies between the Planck and the HST values. Although this constraint is still very poor and is unable to discriminate between the two values in tension, it is worth to remark that it has been obtained without using any form of cosmic ``distance ladder'', and therefore is free of the systematic errors that could affect the more standard astronomical determinations of cosmological distances. Despite the limited constraining power of this measurement, it is important to stress that gravitational-wave multi-messenger astronomy will be able to reach the few-percent accuracy (as in \cite{RiessH0}) when more events of this sort are available ($20-30$ additional observations of standard sirens will be enough to reach this level of accuracy).

Finally, let us also mention the very recent work \cite{BasilakosH0}, in which the following value for $H_0$ is obtained with an independent approach, $H_0=71.0\pm 3.5\,{\rm km/s/Mpc}$. It also lies between the Planck and the HST values, and is actually compatible with them at $1\sigma$.

Quite obviously, the search for a final solution to the $H_0$ tension is still work in progress.

\section{Conclusions}

The present updated analysis of the cosmological data SNIa+BAO+$H(z)$+LSS+CMB disfavors the hypothesis $\CC=$const. as compared to the dynamical vacuum models (DVMs). This is consistent with our most recent studies\,\cite{ApJLnostre,ApJnostre,MPLAnostre,PRDnostre}. Our results suggest a dynamical DE effect near $3\sigma$ within the standard XCDM parametrization and near $4\sigma$ for the best DVMs.  Here we have extended these studies in order to encompass the class of quasi-vacuum models ($w$DVMs), where the equation of state parameter $w$ is near (but not exactly equal) to $-1$. The new degree of freedom $w$ can then be used to try to further improve the overall fit to the data. But it can also be used to check if values of $w$ different from $-1$ can relax the existing tension between the two sets of  measurement of the $H_0$ parameter, namely those based:  i) on the CMB measurements by the Planck collaboration\,\cite{Planck2015,Planck2016}, and ii) on the local measurement (distance ladder method) using Cepheid variables\,\cite{RiessH0}.

Our study shows that the RVM with $w=-1$ remains as the preferred DVM for the optimal fit of the data. At the same time it favors the CMB measurements of $H_0$ over the local measurement.  Remarkably, we find that not only the CMB and BAO data, but also the LSS formation data (i.e. the known data on $f(z)\sigma_8(z)$ at different redshifts), are essential to support the CMB measurements of $H_0$ over the local one. We have checked that if the LSS data are not considered (while the BAO and CMB are kept), then there is a unique chance to try to accommodate the local measurement of $H_0$, but only at the expense of a phantom-like behavior (i.e. for $w<-1$). In this region of the parameter space, however, we find that the agreement with the LSS formation data is manifestly lost, what suggests that the $w<-1$ option is ruled out. There is no other window in the parameter space where to accommodate the local $H_0$ value in our fit.  In contrast, when the LSS formation data are restored, the fit quality to the overall SNIa+BAO+$H(z)$+LSS+CMB observations improves dramatically and definitely favors the Planck range for $H_0$  as well as smaller values for $\sigma_8(0)$ as compared to the $\CC$CDM.

In short, our work suggests that signs of dynamical vacuum energy are encoded in the current cosmological observations. They  appear to be more in accordance with the lower values of $H_0$ obtained from the Planck (CMB) measurements than with the higher range of $H_0$ values obtained  from the present local (distance ladder) measurements, and provide smaller values of $\sigma_8(0)$ that are in better agreement with structure formation data as compared to the $\CC$CDM. We hope that with new and more accurate observations, as well as with more detailed analyses, it will be possible to assess the final impact of vacuum dynamics on the possible solution of the current tensions in the $\CC$CDM.

%%%%%%%%%%%%%%%%%%%%%%%%%%%%%%%%%%%%%%%%%%%%%%%%%%%%%%%%%%%%%%%%%
%%%%%%%%%%%%%%%%%%%%%%%%%%%%%%%%%%%%%%%%%%%%%%%%%%%%%%%%%%%%%%%%%
%%%%%%%%%%%%%%%%%%%%%%%%%%%%%%%%%%%%%%%%%%%%%%%%%%%%%%%%%%%%%%%%%

\section{Main bibliography of the chapter}

This chapter is based on the contents of the letter-type paper \cite{PLBnostre} and the review \cite{IJMPA2nostre},
\vskip 0.5cm
\noindent
{\it The $H_0$ tension in light of vacuum dynamics in the Universe}\newline
J. Sol\`{a}, A. G\'omez-Valent, and J. de Cruz P\'erez\newline
Phys. Lett. B{\bf 774}, 317 (2017) ; arXiv:1705.06723
\vskip 0.5cm
\noindent
{\it Vacuum dynamics in the Universe versus a rigid $\Lambda$=const.}\newline
J. Sol\`a, A. G\'omez-Valent, and J. de Cruz P\'erez\newline
Int. J. Mod. Phys. A{\bf32}, 1730014 (2017) ; arXiv:1709.07451

\thispagestyle{empty}
\newpage
\null

\pagestyle{fancy}
\fancyhf{}
%\fancyhead[LO,RE]{\thepage}
\fancyhead[CO]{Summary, final conclusions and future prospects}
\fancyhead[CE]{Summary, final conclusions and future prospects}
\cfoot{\thepage}

\chapter*{Summary, final conclusions and future prospects}
\label{chap:conclusions}

\addcontentsline{toc}{chapter}{Summary, final conclusions and future prospects}

In this thesis I have analyzed in detail various dynamical vacuum and dark energy (DE) models and parametrizations. I have especially focused my attention in the so-called running vacuum class, which is strongly motivated in QFT in curved spacetime. In these models the variation of the cosmological term takes the following form, $\delta\rho_\Lambda(H,\dot{H})= C_H H^2+C_{\dot{H}}\dot{H}$, which respects the general covariance of the effective action. This variation can be linked to the dynamical evolution of the Newtonian coupling and/or to anomalous dilution laws for the matter-radiation energy densities. The linkage is carried out through the Bianchi identity, which automatically leads to the corresponding conservation equations. 

I have also studied modifications of these models, as those that include a pure linear term in $H$. They are lesser motivated from a theoretical point of view, since this linear term does not respect the aforementioned general covariance, as it has been amply discussed in the main body of the thesis. They could account, though, for several phenomenological effects, as those induced by bulk viscosity.

Moreover, I have explored other dynamical vacuum models (DVM's) which are mainly of  phenomenological nature, and DE variants of the running vacuum models (RVM's) with an equation of state (EoS) parameter different from $-1$. I have studied the viability of the various models according to the most updated observational data sets at the background and linear perturbation regimes. The latter is of utmost importance, since the structure formation data play a crucial role in the fitting analysis and encrypt part of the observed tension between the standard $\Lambda$CDM model and the experimental data.

The main outcomes of this dissertation are now highlighted:

\begin{itemize}
\item The cosmological constant appearing in the expression of the vacuum energy density in the RVM's is crucial to ensure a good phenomenological behavior of these models. Without this term, they are unable to fit properly the data, since either fail at the background level, e.g showing their incapability of generating a transition from a decelerated to an accelerated Universe, or are unable to fit the structure formation data. Thus, the phenomenological studies carried out in this thesis have served to demonstrate the incapability of those models without the aforesaid constant term in the expression of $\rho_\Lambda$. These pathologies cannot be cured neither by those DE models without constant term in $\rho_{DE}$ which are inside the $\mathcal{D}$-class.

\item In the first part of this thesis we have seen that the RVM's with a well-defined $\Lambda$CDM limit are perfectly viable. In the second part, thanks to a refinement of the observational database and to an improvement of the statistical analysis employed in the study of the various models, we have been able to detect a strong signal in favor of the vacuum dynamics which reaches the $\sim 4\sigma$ c.l. This result is unprecedented in the literature.

\item We have also analyzed different parametrizations of the DE, as the XCDM and CPL. We have found that the DE dynamics can be also traced using these parametrizations. The same happens with more sophisticated scalar field models, with a local action, as for instance the original Peebles \& Ratra (PR) model. The statistical confidence level in these cases is not as high as in the RVM's, but it is also quite remarkable, being in both cases around the $3\sigma$.

\item In all cases, we find that the current observational data favor the decrease of the vacuum/DE density throughout the cosmic history. This means that according with the data, the amount of DE density grows in the past, and this hinders the large scale structure (LSS) formation in the Universe with respect to the $\Lambda$CDM case. Of course, this is caused by the repulsive gravitational properties of these cosmic components, which are clearly enhanced in these models. These effects allow to solve in some cases and alleviate in others the existing tension that is found between the concordance model and the LSS data. 

\item We have seen that these positive signals in favor of non-null DE dynamics can only be detected by using a sophisticated enough set of observational data. We have understood that the role played by the Hubble function and SNIa data is less significant. The SNIa data were crucial to detect almost twenty years ago with a high statistical confidence level the positive acceleration of the Universe, but they are too weak to take the lead in the detection of the DE dynamics. The triad of observables that turns out to be really important is given by the string BAO+CMB+LSS. If we do not consider LSS or CMB, the signal is totally lost, whilst if we do not include the BAO data set, then the signal diminishes, but it is still quite high, being around the $3\sigma$ c.l. for the RVM's. In practice, working with the trio BAO+CMB+LSS is almost indistinguishable from working with the enlarged data set BAO+CMB+LSS+H(z)+SNIa. The fitting results in both cases yield inappreciable differences.

\item The latter item explains why large collaborations as, for instance, Planck, BOSS or DES, have been unable to detect such signals in favor of dynamical DE. This is mainly due to the fact that they do not make use of a rich enough data set. They use few data on BAO and LSS, and exclude some data points from their analyses that prove to be crucial to detect the signal of DE dynamics reported in this thesis.

\item Very recently Heavens {\it et al.} have claimed in \cite{Heavens2017} that there is no evidence for extensions to the standard cosmological model. In contrast, our results are in full consonance with the recent work by Gong-Bo Zhao {\it et al.} \cite{GongBoZhao2017}, see also the glossing of it in \cite{NVnoteValentino}. Both, \cite{Heavens2017} and \cite{GongBoZhao2017}, appeared in the arXiv well after our letter-type paper \cite{ApJLnostre} and even after our devoted works \cite{ApJnostre,MPLAnostre,PRLnostre} proposing the existence of a significant dynamical DE signal in the observational data. Our works are actually previous to any one existing in the literature
making these strong claims using both general parametrizations, such as XCDM and CPL, as well
as specific models such as the RVM's. The case of \cite{Salvatelli2014} is only one of the models that we consider in Chapters \ref{chap:PRDbased} and \ref{chap:H0tension}, $Q_\Lambda$, which is actually the less favored one in our fits to the dynamical vacuum models. Their results are cited, discussed and disputed in Chapter \ref{chap:PRDbased}.

The paper by Gong-Bo Zhao {\it et al.} and our work both reach similar conclusions and we use sources of data involving the main ingredients which prove most sensitive for pinning down this effect, namely CMB+BAO+LSS. This is in contradistinction to the more limited data sources used by the mentioned work by Heavens {\it et al.}, who mainly concentrate on CMB and lensing data and do not find any effect. We interpret that this is the reason for our good resonance with the aforementioned results by Gong-Bo Zhao {\it et al.} and the difference with the other studies. Specifically, in the case of Gong-Bo Zhao {\it et al.} a signal of dynamical DE at $3.5\sigma$ c.l. is found from their fit, whereas we find a peak evidence of $3.8\sigma$ within the RVM and $\sim 3\sigma$ with the general XCDM parametrization (cf. Table \ref{tableH0Fit1}), so very similar to these authors. It is also very important to remark that the results from Heavens {\it et al.} are far from opposite to the ones found by us, since if we stick to the same limited data set used by these authors we do also lose the signal of DE dynamics and obtain absolutely compatible results. The aforesaid positive signal can be only obtained if we include a complete enough data set, with all the crucial ingredients of the string BAO+CMB+LSS.

\item In the last chapter we have seen that the DVM's are more favored than the $w$DVM's, i.e. the generalization of the DVM's studied in Chapter \ref{chap:PRDbased} with a DE component with a constant EoS different from $-1$. We have also checked that if we take into account the LSS data, the $H_0$ value obtained with the $\Lambda$CDM model does not only come in conflict with the estimation of $H_0$ provided by Riess {\it et al.} ($\sim 73$ km/s/Mpc), but also is dragged away 4-5$\sigma$ from Planck's value ($\sim 67$ km/s/Mpc). This is something that has not been previously noted in the literature. We have checked that in the context of some DVM's the Planck $H_0$ value is in full agreement with those that are preferred by other observables we have dealt with in this thesis, as e.g. BAO and LSS. Thus, we have demonstrated that if Planck's data do not suffer from systematic errors (as suggested by some recent works in the literature) the DVM's tend to solve the existing $H_0$ tension in favor of the Planck measurement.

\item Another important point it is worth mentioning is that we have checked the potential power of the Press \& Schechter formalism to test different models, and also to distinguish those that are very similar at the background and linear perturbation levels. Very little variations of the vacuum parameter $\nu$ give rise to big differences in the predicted number of collapsed structures in a given range of masses. It would be convenient to apply this formalism in a future so as to see which of the models that are able to explain the background and linear structure formation better than the $\Lambda$CDM, are also able to explain well the nonlinear structure formation processes. It would also be interesting to see whether we can further constrain some of the studied models in this thesis using this powerful tool, or even make new predictions on the number of collapsed structures that are more in agreement with experimental observations. If so, it would further reinforce the DVM's from the phenomenological point of view.
\end{itemize}

These results are very encouraging and reveal the possible variability of the DE density. Despite this, it is obvious that the ultimate confirmation of this feature must come hand in hand with more data. Exciting times for Cosmology are augured. The Dark Energy Survey (DES), the Extended Baryon Oscillation Spectroscopic Survey (eBOSS), the Dark Energy Spectroscopic Instrument (DESI), the Large Synoptic Survey Telescope (LSST), the Euclid space telescope, the Wide-Field Infrared Survey Telescope (WFIRST), or the Laser Interferometer Space Antenna (LISA), are some examples of projects that are already being or will be run during the next years\footnote{For a list of references, and a concise and nice review of these observational facilities, see \cite{CosmoProjFuture}. For the possible constraints on DE coming from LISA, see \cite{LISA1,LISA2}.}. They will explore a volume greater than $10^3$ millions of cubic light-years. Some of them will be able to collect e.g. LSS data with less than 1\% uncertainties, in redshift ranges between 0 and 3, roughly covering three fourths of the Universe's age. 

In this thesis the firsts compelling signs of DE dynamics are presented. The chapters of the second part report on important (statistically significant) evidences in favor of this DE feature, which are especially conspicuous for the RVM's. Whether these strong indications in favor of the variability of the DE density are or not something real, it will depend on the refinement and amount of the future observational data. In the meanwhile, the cosmological information provided up to now by the different surveys and large collaborations lead to an overwhelming signal in favor of the DE dynamics, which can only be detected if a complete enough data set including CMB+BAO+LSS is taken into account.

In the future, it will be of utmost importance not to lose track of the upcoming observational works. It will be interesting to see e.g. how the inclusion of the bispectrum information affects the new BAO data, and its impact on our fitting results. From the theoretical and phenomenological side, and in view of the high evidence found in this thesis in favor of a dynamical DE component, it would be very convenient to study in deeper detail the constraints coming from the CMB. More concretely, we should carefully check whether the good behavior of the models under study is kept intact when the complete set of data points of the CMB temperature and E-polarization anisotropies (together with the TE cross-correlation ones) are included in the analysis. It will be crucial to see which is the impact of the various DE signatures in these spectra, such as those triggered by the integrated Sachs-Wolfe effect or the CMB lensing. In order to perform this study the complete set of Boltzmann equations must be solved numerically with the use of publicly available codes as CMBFAST \cite{CMBFAST}, CAMB \cite{CAMB}, CMBEASY \cite{CMBEASY}, or CLASS \cite{CLASS}. These programs also allow to study the effect of other important cosmological aspects in a very detailed way, as e.g. the effect of massive and extra neutrinos. This is work to do in a near future.

Many issues concerning the nature of the DE must be still understood. There is still a lot of work to do from both, the theoretical and observational sides. Let us see what the future holds for Cosmology. Whatever the surprises it brings to us, they will probably open new windows, and hopefully new landscapes will appear in front of us. I guess we will be entertained during the next years. Let us be patient and work hard. The answers of many unsolved questions, wherever they are, are surely waiting for us!

\vskip 1.0cm
\begin{flushright}
Barcelona, September 2017
\end{flushright}

\thispagestyle{empty}
\null
\newpage

\pagestyle{fancy}
\fancyhf{}
%\fancyhead[LO,RE]{\thepage}
\fancyhead[CO]{\nouppercase{\leftmark}}
\fancyhead[CE]{\nouppercase{\rightmark}}
\cfoot{\thepage}

\appendix

\chapter[ZPE in flat spacetime]{ZPE in flat spacetime. Two alternative regularization schemes, and renormalization}
\label{ch:appZPE}

In this appendix I explicitly compute the regularized and renormalized ZPE of a free real scalar field of mass $m$ in flat spacetime by making use of two different regularization techniques: cutoff regularization and dimensional regularization. The exclusive dedication of one appendix to this subject turns out to be really useful for two different reasons: i) because I will refer to the results derived in this appendix at several points of the main body of the thesis; ii) because despite being a very basic calculation, it is a good example of how to regularize the ``typical'' integrals involved in the computation of ZPE's. This step is preliminary to the subsequent renormalization of the resulting expressions. In addition, these results, which are obtained in flat spacetime, are compared with those obtained in a curved background in Sect. \ref{subsec:RVMintro}.

The formal ``bare'' expressions for the ZPE density and pressure in the case under study read, respectively \footnote{I explicitly keep the $\hbar$ in these expressions in order to show in a manifest way that they are of pure quantum nature.},
\be\label{eq:DensityFlatScalar}
\rho_{\rm bare}=\hbar\int\frac{d^3k}{(2\pi)^3}\frac{\sqrt{k^2+m^2}}{2}=\frac{\hbar}{4\pi^2}\int dk\,k^2\sqrt{k^2+m^2}\,,
\ee
\be\label{eq:PressureFlatScalar}
p_{\rm bare}=\frac{\hbar}{6}\int\frac{d^3k}{(2\pi)^3}\frac{k^2}{\sqrt{k^2+m^2}}=\frac{\hbar}{12\pi^2}\int dk\,\frac{k^4}{\sqrt{k^2+m^2}}\,,
\ee
where $k\equiv|\vec{k}|$. Obviously, these integrals are ultraviolet (UV) divergent, since the leading contribution of large momenta scales in both cases like $\sim k^4$. Let us discuss now the two chosen alternative ways to regularize them.

\section{Momentum cutoff regularization} 

In this regularization scheme one simply uses a momentum cutoff $\Lambda_c$ in order to keep the integrals finite:
\be\label{eq:rho1}
\rho_{\rm reg}=\frac{\hbar}{4\pi^2}\int_{0}^{\Lambda_c} dk\,k^2\sqrt{k^2+m^2}\,,
\ee
\be\label{eq:p1}
p_{\rm reg}=\frac{\hbar}{12\pi^2}\int_{0}^{\Lambda_c} dk\,\frac{k^4}{\sqrt{k^2+m^2}}\,,
\ee
Of course, these expressions still diverge in the limit $\Lambda_c\to\infty$, but now it is possible to solve the integral and clearly identify which are the problematic divergent terms that should be absorbed by the counterterms in the subsequent renormalization step. Using the following change of variables in \eqref{eq:rho1},
\be
k^2+\frac{m^2}{2}=-\frac{1}{2t}\left(\frac{m^4}{4}+t^2\right)\,,
\ee
one gets 
\be
\rho_{reg}=\frac{\hbar}{8\pi^2}\left[\frac{t^2}{8}-\frac{m^8}{128t^2}-\frac{m^4}{8}\ln{t}\right]_{-m^2/2}^{-\Lambda_c\sqrt{\Lambda_c^2+m^2}-\Lambda_c^2-m^2/2}\,.
\ee
If we consider a momentum cutoff much larger than the mass of the scalar particle, i.e. $\Lambda_c\gg m$, and Taylor-expand the above expression, the following regularized result is obtained,
\be
\rho_{reg}=\frac{\hbar\Lambda_c^4}{16\pi^2}\left[1+\frac{m^2}{\Lambda_c^2}+\frac{m^4}{8\Lambda_c^4}+\frac{m^4}{4\Lambda_c^4}\ln{\left(\frac{m^2}{4\Lambda_c^2}\right)}+...\right]\,,
\ee 
where [...] contains the infinite number of terms that vanish in the limit $\Lambda_c\to\infty$ and, therefore, are not important at all. The momentum cutoff regularization of the bare pressure is straightforward. I just integrate \eqref{eq:p1} by parts. The result reads,
\be\label{eq:NoVacuumEoS}
p_{\rm reg}=\frac{\hbar\Lambda_c^3}{12\pi^2}\sqrt{\Lambda_c^2+m^2}-\rho_{\rm reg}\,.
\ee
Notice that this result does not respect the EoS of the vacuum, i.e. $p=-\rho$. As it is argued in \cite{Akhmedov2002}, this is due to the fact that momentum regularization violates the relativistic Lorentz invariance of the vacuum in a manifest way. The regularized result depends explicitly on the momentum cutoff $\Lambda_c$, and this quantity is not invariant under boost transformations. But this argument is not sufficient to conclude that this regularization scheme is not valid \cite{Maggiore2010}, since \eqref{eq:NoVacuumEoS} is only the (intermediate) regularized expression, not the (final) renormalized one! Renormalization could ultimately cure this (maybe apparent) problem by choosing the appropriate counterterms.  

%%%%%%%%%%%%%%%%%%%
%%%%%%%%%%%%%%%%%%%
%%%%%%%%%%%%%%%%%%%

\section{Dimensional regularization} 
\label{sec:DimReg}

In this case, we regularize the integrals by changing the original dimensionality of the momentum space \cite{GtHooft1973,Bollini1972}, i.e. by assuming that the new number of dimensions is $d\ne 3$,
\be
\rho_{reg}=\hbar\mu^{3-d}\int\frac{d^dk}{(2\pi)^d}\frac{\sqrt{k^2+m^2}}{2}\,.
\ee
The energy scale $\mu$ is the characteristic 't Hooft mass unit of dimensional regularization, which has been introduced in order to keep $\rho_{reg}$ with dimensions of $M^4$. Taking into account that $d^dk=k^{d-1}d\Omega_ddk$, where $\Omega_d$ is the solid angle in $d$ dimensions, we find
\be
\rho_{reg}=\hbar\frac{\Omega_d\mu^{3-d}}{2(2\pi)^d}\int_{0}^{\infty} dk\, k^{d-1} \sqrt{k^2+m^2}\,.
\ee
Now, it is convenient to perform the change of variables $y=\frac{m^2}{k^2+m^2}$,
\be
\rho_{reg}=\hbar\frac{\mu^{3-d}m^{d+1}\Omega_d}{4(2\pi)^d}\underbrace{\int_{0}^{1} dy\, y^{-\frac{d+3}{2}} (1-y)^{\frac{d}{2}-1}}_{B\left(-\frac{d+1}{2},\frac{d}{2}\right)}=\hbar\frac{\mu^{3-d}m^{d+1}}{4(2\pi)^d}\frac{2\pi^{d/2}}{\Gamma\left(\frac{d}{2}\right)}\frac{\Gamma\left(\frac{d}{2}\right)\Gamma\left(-\frac{d+1}{2}\right)}{\Gamma\left(-\frac{1}{2}\right)}\,,
\ee
where $B(\alpha,\beta)=\frac{\Gamma(\alpha)\Gamma(\beta)}{\Gamma(\alpha+\beta)}$ is the beta function, and I have made use of the relation $\Omega_d=\frac{2\pi^{d/2}}{\Gamma(d/2)}$ in the second equality,
\be\label{eq:AppUse0}
\rho_{reg}=\hbar\frac{\mu^{3-d}m^{d+1}}{4}\frac{2\pi^{d/2}}{(2\pi)^d}\frac{\Gamma\left(-\frac{d+1}{2}\right)}{-2\sqrt{\pi}}\,.
\ee
Here, I have used $\Gamma\left(-\frac{1}{2}\right)=-2\sqrt{\pi}$. In general, though, one has to keep in mind the general relation between Gamma functions, $\Gamma(n+1)=n\Gamma(n)$. The latter, together with the redefinition $d\equiv 3-2\epsilon$ allows us to write the Gamma function in the numerator as follows, 
\be\label{eq:AppUse1}
\Gamma\left(-\frac{d+1}{2}\right)=\frac{\Gamma(\epsilon)}{2-3\epsilon+\epsilon^2}\,.
\ee
Notice that we will be interested in performing the limit $\epsilon\to 0$ after renormalization in order to retrieve the original situation with $d=3$ dimensions. This allows us to treat $\epsilon$ as a little perturbation around $d=3$. Let us compute $\Gamma(\epsilon)$ using this fact, by Taylor-expanding the Weierstrass' definition of the Gamma function around $\epsilon=0$,
\be\label{eq:AppUse2}
\frac{1}{\Gamma(\epsilon)}=\epsilon e^{\gamma \epsilon}\prod_{n=1}^{\infty}\left(1+\frac{\epsilon}{n}\right)e^{-\epsilon/n}=\epsilon+\gamma\epsilon^2+\mathcal{O}(\epsilon^3)\,,
\ee
where $\gamma\simeq 0.5772$ is the Euler-Mascheroni constant. Plugging \eqref{eq:AppUse1} and \eqref{eq:AppUse2} in \eqref{eq:AppUse0}, and Taylor-expanding also the remaining terms we finally obtain,
\be\label{eq:regZPEdim}
\rho_{reg}=-\frac{\hbar m^4}{64\pi^2}\left[\frac{1}{\epsilon}+\frac{3}{2}-\gamma+\ln\left(\frac{4\pi\mu^2}{m^2}\right)\right]+...\,,
\ee
with [...] including all the terms that vanish when $\epsilon\to 0$. To obtain the regularized pressure one has to proceed exactly in the same way, performing the same change of variables as before, using again the definition of the beta function, etc. This is the resulting expression,  
\be
p_{reg}=\frac{\hbar\mu^{3-d}m^{d+1}}{2^{2+d}\pi^{\frac{d+1}{2}}}\Gamma\left(-\frac{d+1}{2}\right)=-\rho_{reg}\,.
\ee
In this case, the vacuum EoS is preserved because dimensional regularization respects the relativistic symmetries. 

%%%%%%%%%%%%%%%%%%%
%%%%%%%%%%%%%%%%%%%
%%%%%%%%%%%%%%%%%%%

\section{Minimal subtraction ($MS$) and modified minimal subtraction ($\overline{MS}$) schemes}
\label{sec:MSappendix}

Upon the regularization of the ZPE density, we have been able to split the starting highly divergent integral in a finite plus an explicit divergent part. Renormalization is the procedure that is used to get rid of the latter in order to match the theoretical prediction to the physical (maybe directly measured) value of the quantity under study. It basically consists in splitting the bare quantities appearing in the Lagrangian in order to write them as the sum of a finite (renormalized) quantity plus an (infinite) counterterm, which is in charge of absorbing the divergent contributions appearing in the regularized expressions. 

In this case, the bare CC does the job of renormalizing the ZPE. It can be split as follows,
\be
\rho_{\Lambda,b}=\frac{\Lambda_b}{8\pi G_b}\equiv \rho_\Lambda+\delta\rho_{\Lambda}\,,
\ee
Note that there is some ambiguity in the absorption of the aforementioned infinities, since the counterterm is arbitrary up to a constant. This arbitrariness is fixed by the appropriate choice of the renormalization condition. Let us focus our attention in the regularized formula \eqref{eq:regZPEdim}. In the $MS$ scheme one just subtracts the pole, 
\be
\rho_{\Lambda,{\rm eff}}=\rho_\Lambda(\mu)+ \delta\rho_{\Lambda} +\rho_{reg}(\mu)=\rho_\Lambda(\mu)-\frac{\hbar m^4}{64\pi^2}\left[\frac{3}{2}-\gamma+\ln\left(\frac{4\pi\mu^2}{m^2}\right)\right]\,,
\ee
whereas in the $\overline{MS}$ (``$MS$-bar'') scheme the counterterm also absorbs some constant terms,
\be\label{eq:renZPE}
\rho_{\Lambda,{\rm eff}}=\rho_\Lambda(\mu)-\frac{\hbar m^4}{64\pi^2}\ln\left(\frac{4\pi\mu^2}{m^2}\right)\,.
\ee
In both cases we have performed the physical limit $\epsilon\to 0$ after the removal of the pole. Notice that the energy density $\rho_\Lambda$ automatically acquires a dependence on the arbitrary energy scale $\mu$. This is due the fact that the physical (measured) value of $\rho_{\Lambda,{\rm eff}}$ cannot depend on our choice of $\mu$. This leads us to determine the following $\beta$-function for $\rho_\Lambda$ \footnote{This is a very well-known result, see e.g. \cite{BrownQFT}.},
\be\label{eq:betaFunc}
\beta_\Lambda^{(1)}\equiv\mu\frac{d\rho_\Lambda(\mu)}{d\mu}=\frac{\hbar m^4}{32\pi^2}\,.
\ee
Now, the resulting expressions are perfectly finite. But still, after renormalization $\rho_{\Lambda,{\rm eff}}\propto m^4$ and this is very problematic. In fact, it is the central core of the CC problem, since it produces a value which is many orders of magnitude larger than the measured one. The only way to fulfill Einstein's equations in Minkowski spacetime is to demand that $\rho_{\Lambda,{\rm eff}}^{\rm Mink}=0$, which is tantamount to imposing the non-gravitating feature of vacuum in flat spacetime. This appealing renormalization condition is, at least {\it a priori}, reasonable, and fixes uniquely the counterterm in the $\overline{MS}$ scheme. In curved spacetime, though, vacuum energy can produce sizable effects. They could be due to a sort of gravitational Casimir effect, in which the effective vacuum energy in curved background would just be the difference between the vacuum energy in flat and curved spacetimes. For further details and a extense discussion of this idea I refer the reader to Sect. \ref{sec:MoreVacuumQFT}.

\medskip
Before concluding this appendix, I would like to remark that although momentum cutoff regularization explicitly violates Lorentz invariance, it is also possible to retrieve the generalization of \eqref{eq:renZPE} using this regularization scheme (or even the Paulli-Villars one, see \cite{Koksma2011}), 
\be\label{eq:renZPE2}
\rho_{\Lambda,{\rm eff}}(\mu)=\rho_\Lambda(\mu)-\frac{\hbar}{64\pi^2}\sum_{n}(-1)^{S_n}g_n m_n^4\ln\left(\frac{4\pi\mu^2}{m_n^2}\right)\,,
\ee
where the degeneracy factor $g_n$ includes a spin factor $g_n=2S_n+1$ for massive particles, whereas $g_n=2$ for massless particles. It also includes an extra factor of $2$ when particle and antiparticle are distinct, and an additional factor of $3$ due to color. The index $n$ labels all the particle species. \eqref{eq:renZPE2} is obtained by demanding the fulfillment of the Pauli's three polynomial-in-mass constraints \cite{Pauli1951}. The latter is totally equivalent to force the total zero-point energy-momentum tensor to be Lorentz invariant. See \cite{Visser2016} for a detailed and modern presentation of this idea.

\thispagestyle{empty}
\null
\newpage

\chapter[ZPE in curved spacetime]{ZPE in curved spacetime. A canonical quantization approach}
\label{ch:appZPEcurved}

In this appendix I compute the renormalized ZPE of a non-minimally coupled  and real scalar field in an expanding FLRW Universe by using the canonical quantization approach and the $\overline{MS}$ renormalization scheme. The classical action of such scalar field has been provided in \eqref{eq:SmatterCurved}, but I write it here again just for convenience,
\be\label{eq:SmatterCurvedApp}
S[g_{\mu\nu},\phi]=\frac{1}{2}\int d^4x\,\sqrt{-g}\left(\partial_\mu\phi\partial^\mu\phi-m^2\phi^2-\xi R\phi^2\right)\,.
\ee
The equation of motion for $\phi$ is obtained as usually, by demanding the solution to extremize the action, i.e. $\delta S=0$,
\be
\frac{1}{\sqrt{-g}}\partial_\mu\left(\sqrt{-g}g^{\mu\nu}\partial_\nu\phi\right)+\phi(m^2+\xi R)=0\,.
\ee
Let us particularize now this equation for a FLRW metric, using conformal coordinates, i.e. $g_{\mu\nu}=a^2\eta_{\mu\nu}$,
\be
\phi^{\pp}-\nabla^2\phi+2\mathcal{H}\phi^\p+a^2\phi(m^2+\xi R)=0\,,
\ee
where the prime denotes a derivative w.r.t. the conformal time and $\mathcal{H}=a^\p/a$. By applying the change of variable $\chi=a\phi$ this equation can be written in a more convenient way. Using
\begin{align*}
\phi^\prime&=\frac{\chi^\p}{a}-\frac{a^\p\chi}{a^2}\,,\\
\phi^{\pp}&=\frac{\chi^{\pp}}{a}-\frac{2\chi^\p a^\p}{a^2}-\frac{a^{\pp}\chi}{a^2}+\frac{2(a^\p)^2\chi}{a^3}\,,
\end{align*}
we finally obtain,
\be\label{eq:chiEq}
\chi^{\pp}-\nabla^2\chi+\chi\left(a^2m^2-\frac{a^{\pp}}{a}+a^2\xi R\right)=0\,.
\ee
This equation is the same that we encounter for a real scalar field in flat spacetime. The only difference is that in the case under study we have an effective mass that evolves with the expansion according to
\be
m_{eff}^2(\eta)\equiv a^2m^2-\frac{a^{\pp}}{a}+a^2\xi R\,.
\ee
Taking into account that $a^{\pp}/a=a^2R/6$ (see the non-perturbed part of Eq. \eqref{eq:RicciScalarApp}), the effective mass can be expressed in a more compact way,
\be\label{eq:meff}
m_{eff}^2(\eta)= a^2\left[m^2+R\left(\xi-\frac{1}{6}\right)\right]\,.
\ee
Let us now derive the classical solution of Eq. \eqref{eq:chiEq}. It is highly convenient to Fourier expand it. Thus, by inserting 
\be
\chi(\eta,\vec{x})=\int\frac{d^3k}{(2\pi)^{3/2}}\chi_{\vec{k}}(\eta) e^{i\vec{k}\cdot\vec{x}}
\ee
in Eq. \eqref{eq:chiEq} we find the equation that must be fulfilled by the Fourier modes $\chi_{\vec{k}}(\eta)$,
\be\label{eq:chiEq2}
\chi^{\pp}_{\vec{k}}(\eta)+\omega_k^2(\eta)\chi_{\vec{k}}(\eta)=0\,,
\ee
where $\omega_k(\eta)=\sqrt{k^2+m^2_{eff}(\eta)}$. The general solution of this equation can be written as follows,
\be
\chi_{\vec{k}}(\eta)=\frac{1}{\sqrt{2}}\left[A_{\vec{k}}v^*_{\vec{k}}(\eta)+B_{\vec{k}}v_{\vec{k}}(\eta)\right]\,,
\ee
where the factor $1/\sqrt{2}$ has been introduced for convenience, and $A_{\vec{k}}$ and $B_{\vec{k}}$ are integration constants. Notice that $v_{\vec{k}}=v_{-\vec{k}}=v_k$ because the frequency $\omega_k$ is isotropic. In addition, due to the fact that $\chi(\eta,\vec{x})$ is a real scalar field, the complex Fourier modes must satisfy the condition 
\be
\chi^*_{\vec{k}}(\eta)=\chi_{-\vec{k}}(\eta)\,.
\ee 
Using this relation (which directly leads to $A_{\vec{k}}=B^*_{-\vec{k}}$) it is easy to find the classical solution,
\be\label{eq:classsolution}
\chi(\eta,\vec{x})=\frac{1}{\sqrt{2}}\int\frac{d^3k}{(2\pi)^{3/2}}[A_{\vec{k}}v^*_k(\eta)e^{i\vec{k}\cdot\vec{x}}+A^{*}_{\vec{k}}v_k(\eta)e^{-i\vec{k}\cdot\vec{x}}]\,.
\ee
Now we can proceed to quantize the theory with the canonical quantization procedure, by promoting the integration constants to operators and imposing the standard equal-time commutation relations on the field operator $\hat{\chi}$ and its canonically conjugate momentum $\hat{\pi}=\hat{\chi}^\p$,
\be\label{eq:comm1}
[\hat{\chi}(\eta,\vec{x}),\hat{\pi}(\eta,\vec{y})]=i\delta(\vec{x}-\vec{y})\,,
\ee
\be\label{eq:comm2}
[\hat{\chi}(\eta,\vec{x}),\hat{\chi}(\eta,\vec{y})]=[\hat{\pi}(\eta,\vec{x}),\hat{\pi}(\eta,\vec{y})]=0\,.
\ee
We also need a time-independent normalization condition for the Fourier modes. It proves useful to normalize the them by imposing the Wronskian $W(v_k^*,v_k)=2$ (one can check that $W^\p(v_k^*,v_k)=0$). This is equivalent to impose
\be\label{eq:norm}
Im(v_k^\p v_k^*)=\frac{v_k^\p v_k^*-v_k v_k^{*\p}}{2i}=1\,.
\ee 
Inserting \eqref{eq:classsolution} in the field commutation relations \eqref{eq:comm1}-\eqref{eq:comm2}, and using the normalization condition \eqref{eq:norm}, we obtain the usual commutation relations for the annihilation and creation operators $\hat{A}_{\vec{k}}$ and $\hat{A}^*_{\vec{k}}$. If we make use of the more conventional notation, $\hat{a}^-_{\vec{k}}\equiv \hat{A}_{\vec{k}}$ and $\hat{a}^+_{\vec{k}}\equiv \hat{A}^*_{\vec{k}}$, they read,
\be
[\hat{a}^{-}_{\vec{k}},\hat{a}^+_{\vec{k}^\p}]=\delta(\vec{k}-\vec{k}^\p)\qquad [\hat{a}^{-}_{\vec{k}},\hat{a}^-_{\vec{k}^\p}]=[\hat{a}^{+}_{\vec{k}},\hat{a}^+_{\vec{k}^\p}]=0\,.
\ee
The operators $a^\pm_{\vec{k}}$ can be used to construct the basis of quantum states in the Hilbert space. Notice, though, that these operators only acquire their meaning once we choose the Fourier modes $v_k(\eta)$. It turns out that the general functions
\be
u_k(\eta)=\alpha_k v_k(\eta)+\beta_k v^*_k(\eta)
\ee 
also satisfy Eq. \eqref{eq:chiEq2}, with $\alpha_k$ and $\beta_k$ being time-independent complex coefficients satisfying 
\be                                                  
|\alpha_k|^2-|\beta_k|^2=1\,.
\ee
Obviously, our definition of the vacuum state will depend on our choice of the coefficients $\alpha_k$ (or, equivalently, $\beta_k$). If we use the modes $u_k(\eta)$ instead of $v_k(\eta)$, then we can write the field operator as
\be\label{eq:classsolution2}
\chi(\eta,\vec{x})=\frac{1}{\sqrt{2}}\int\frac{d^3k}{(2\pi)^{3/2}}[\hat{b}^-_{\vec{k}}u^*_k(\eta)e^{i\vec{k}\cdot\vec{x}}+\hat{b}^{+}_{\vec{k}}u_k(\eta)e^{-i\vec{k}\cdot\vec{x}}]\,.
\ee
The matching between \eqref{eq:classsolution} and \eqref{eq:classsolution2} can only be done if 
\be
\hat{b}^{-}_{\vec{k}}=\alpha_k\hat{a}^{-}_{\vec{k}}-\beta_k\hat{a}^{+}_{-\vec{k}}\qquad 
\hat{b}^{+}_{\vec{k}}=\alpha_k^*\hat{a}^{+}_{\vec{k}}-\beta_k^*\hat{a}^{-}_{-\vec{k}}\,.
\ee
These are the so-called Bogolyubov transformations. Recall that the annihilation operators are defined such that 
\be
\hat{a}^{-}_{\vec{k}}|_{(a)}0>=0\qquad \hat{b}^{-}_{\vec{k}}|_{(b)}0>=0
\ee 
where the vacuum states $|_{(a)}0>$ and $|_{(b)}0>$ are, in principle, different. On the other hand, the creation operators fulfill,
\be
|_{(a)}m_{\vec{k}_1}n_{\vec{k}_2}...>=\frac{1}{\sqrt{m!n!...}}\left[(\hat{a}^+_{\vec{k}_1})^m(\hat{a}^+_{\vec{k}_2})^n...\right]|_{(a)}0>\,,
\ee
\be
|_{(b)}m_{\vec{k}_1}n_{\vec{k}_2}...>=\frac{1}{\sqrt{m!n!...}}\left[(\hat{b}^+_{\vec{k}_1})^m(\hat{b}^+_{\vec{k}_2})^n...\right]|_{(b)}0>\,.
\ee
It is easy to check that if $\beta_k\ne 0$ the ``b-vacuum'' contains ``a-particles''. More concretely, in the ``b-vacuum'' one finds the following total mean density of particles,
\be                             
n=\int d^3k\,|\beta_k|^2\,.
\ee
In the curved spacetime case we are dealing with, the Hamiltonian operator
\be
\hat{H}(\eta)=\frac{1}{2}\int d^3x [\hat{\pi}^2+(\vec{\nabla}\hat{\chi})^2+m^2_{eff}(\eta)\hat{\chi}^2]
\ee
depends explicitly on time and, therefore, their eigenstates do not remain constant. In particular, we can only define an instantaneous vacuum (lowest-energy) state. Let us search for the associated energy density of this state at a $\eta_0$. We find,
\be\label{eq:energyDensityLow}
\rho(\eta_0)=\frac{1}{4}\int \frac{d^3k}{(2\pi)^3}(|v_k^\p(\eta_0)|^2+\omega_k^2(\eta_0)|v_k(\eta_0)|^2)\,.
\ee
Using this result we can look for the mode function $v_k$ that minimizes the energy density. It must also fulfill the normalization condition \eqref{eq:norm}. It is useful to define $v_k\equiv r_k e^{if_k}$. Then, a direct calculation yields, 
\be
v_k(\eta_0)=\frac{e^{if_k(\eta_0)}}{\sqrt{\omega_k(\eta_0)}}\qquad v^\p_k(\eta_0)= i\omega_k(\eta_0)v_k(\eta_0)\,.
\ee
The phase factors $f_k(\eta_0)$ are not determined, but they do not play any role in the computation of the lowest-energy density. By plugging the last relations in \eqref{eq:energyDensityLow} we finally find,
\be\label{eq:densityInsta}
\rho(\eta_0)=\frac{1}{2}\int \frac{d^3k}{(2\pi)^3}\,\omega_k(\eta_0)\,.
\ee
This is the same expression that is found in flat spacetime. The only change that must be implemented to recover the last formula starting form the Minkowskian one is $m\leftrightarrow m_{eff}$, where $m_{eff}$ is given by \eqref{eq:meff}. Of course, \eqref{eq:densityInsta} is the energy density associated to the instantaneous lowest-energy state. It is interesting to study how this energy density evolves with time. The expectation value of the Hamiltonian at $\eta_1$ in the vacuum state $|_{\eta_0}0>$ reads,
\be
<_{\eta_0}0|\hat{H}(\eta_1)|_{\eta_0}0>=\delta^3(0)\int \frac{d^3k}{(2\pi)^3}\,\omega_k(\eta_1)\left[\frac{1}{2}+|\beta_k|^2\right]
\ee
Thus, the energy density is given simply by
\be\label{eq:densityInsta2}
\rho(\eta_1)=\int \frac{d^3k}{(2\pi)^3}\,\omega_k(\eta_1)\left[\frac{1}{2}+|\beta_k|^2\right]\,.
\ee
Here it has been made use of the fact that the operators $a^\pm_{\vec{k}}(\eta_0)$ and $a^\pm_{\vec{k}}(\eta_1)$ defining the instantaneous vacuum states $|_{\eta_0}0>$ and $|_{\eta_1}0>$ are related by the Bogoliubov transformation presented before. Notice that the results \eqref{eq:densityInsta} and \eqref{eq:densityInsta2} have basically the same structure, but \eqref{eq:densityInsta2} contains an extra factor that accounts for the creation of particles in the expanding Universe. Nevertheless we could also interpret that the vacuum energy density is just given by 
\be\label{eq:densityInsta3}
\rho(\eta)=\frac{1}{2}\int \frac{d^3k}{(2\pi)^3}\,\omega_k(\eta)\,,
\ee
and that the term accounting for the creation of particles must not be strictly attached to the vacuum energy density, but to an anomalous matter conservation law. This is what one could naively think, which is more aligned with the formalism used throughout this thesis (see e.g. Chapter \ref{chap:Atype}). Thus, we must regularize and renormalize \eqref{eq:densityInsta3}. But this is very easy to do, since we can take advantage of the results of Appendix \ref{ch:appZPE}. We can take the result \eqref{eq:renZPE} and simply apply the change $m^2\leftrightarrow m^2_{eff}$,
\begin{eqnarray}
\sqrt{-g}\left(\frac{R+2\Lambda_{{\rm eff}}}{16\pi G_{{\rm eff}}}\right)+\sqrt{-g}\,\alpha_{4,{\rm eff}}R^2 &=  \sqrt{-g}&\left(\frac{R}{16\pi G(\mu)}+\rho_{\Lambda}(\mu)+\alpha_{4}(\mu)R^2\right)-\nonumber\\
&&-\frac{\hbar (m^2_{eff})^2}{64\pi^2}\ln\left(\frac{4\pi\mu^2}{m_{eff}^2}\right)\,.
\end{eqnarray}
Taking into account that $\sqrt{-g}=a^{4}$ in the FLRW metric we can rewrite this equation as follows,
\begin{eqnarray}
\frac{R+2\Lambda_{{\rm eff}}}{16\pi G_{{\rm eff}}}+\alpha_{4,{\rm eff}}R^2&=&\left(\frac{R}{16\pi G(\mu)}+\rho_{\Lambda}(\mu)+\alpha_{4}(\mu)R^2\right)-\nonumber\\&&-\frac{\hbar}{64\pi^2}\left[m^2+R\left(\xi-\frac{1}{6}\right)\right]^2\ln\left(\frac{4\pi\mu^2}{m_{eff}^2}\right)\,.
\end{eqnarray}
From this expression we can easily extract the $\beta$-functions of interest in the low-energy Universe. 
\be
\beta_\Lambda\equiv \frac{d\rho_\Lambda}{d\ln\mu}=\frac{\hbar m^4}{32\pi^2}\,,
\ee   
\be
\beta_{G^{-1}}\equiv \frac{d}{d\ln\mu}\left(\frac{1}{16\pi G}\right)=\frac{\hbar m^2}{16\pi^2}\left(\frac{1}{6}-\xi\right)\,.
\ee
These are the same $\beta$-functions that are obtained using the path-integral formalism and the adiabatic expansion of the matter field propagator (see Sect. \ref{subsec:RVMintro} and Ref. \cite{BirrellDavies,ParkerToms}). The method employed in this appendix also provides the correct $\beta$-function associated to the parameter $\alpha_4$, which parametrizes the higher derivative effects of the $R^2$-term,
\be                                                                           \beta_{\alpha_4}=\frac{d\alpha_4}{d\ln\mu}=\frac{\hbar}{32\pi^2}\left(\frac{1}{6}-\xi\right)^2\,.
\ee
\thispagestyle{empty}
\null
\newpage

\chapter[Cosmological perturbations in the DVM's]{Cosmological perturbations in the Dynamical Vacuum Models}
\label{ch:appPert}

Before the decoupling of photons and baryons, at $z_{\rm dec}\approx 1090$, these two components (and also electrons) were tightly coupled. In that time, the radiation pressure inhibited the growth of gravitational instabilities in the baryonic sector. The ``sound'' waves generated in the photo-baryon plasma during that epoch gave rise to the well-known baryon acoustic oscillations (BAO's). After the decoupling time, baryons released from the radiation (and its pressure), and gravitational collapse became effective. Dark matter, which had not been susceptible to the aforementioned interactions with photons, had formed some gravitational potential wells before the decoupling, and now baryons could fall in them. In that period of matter-dominated expansion, the first large-scale structures in the Universe formed and grew up to our time. 

In this appendix I derive the equations that govern the evolution of small matter perturbations in the Dynamical Vacuum Models (DVM's). The formalism that will be used is not new at all. It was introduced by E. Lifshitz in \cite{Lifshitz1946}, more than seventy years ago, when he applied GR to study the cosmological perturbations for closed and open Universes in presence of matter. A special effort to present and explain these calculations with a certain degree of detail is not useless at all, since as it is exhaustively remarked in the main body of this dissertation, measurements of large-scale structure are nowadays providing crucial information that allows us to constrain in an outstanding way the various models under study. Thus, this point deserves its own space in this thesis.

\section{DVM's with interaction between matter and vacuum, and $G={\rm const}$}
\label{sec:LPmatter-vacuum}

In this first section I derive the same result in two different gauges: the Newtonian gauge and the synchronous one. This will be useful so as to take a first contact with these two alternative approaches, and also to show the robustness of the equation that governs the growth of matter perturbations in the linear regime.

\subsection{Conformal Newtonian gauge}

The perturbed (flat) FLRW metric expressed in comoving coordinates reads,

$$\bar{g}_{\mu\nu}=a^2\eta_{\mu\nu}\qquad \delta g_{\mu\nu}=a^2 h_{\mu\nu}$$
\be
ds^2=g_{\mu\nu}dx^\mu dx^\nu=(\bar{g}_{\mu\nu}+\delta g_{\mu\nu})dx^\mu dx^\nu=a^{2}(\eta_{\mu\nu}+h_{\mu\nu})dx^\mu dx^\nu\,,
\ee
where $\bar{g}_{\mu\nu}$ coincides, of course, with the background FLRW metric \eqref{eq:FLRW} after setting $k=0$ \footnote{Recall that $\eta_{\mu\nu}$ is Minkowski's metric and that we use the sign convention $(+,-,-,-)$ for it.}. The perturbation of the metric tensor, $\delta g_{\mu\nu}$, must be symmetric in order to preserve the symmetry of $g_{\mu\nu}$. Therefore, $h_{\mu\nu}$ has $10$ degrees of freedom (dof), but only $6$ are really physical, meaning that we can freely fix $4$ of them due to the diffeomorphism invariance of GR. But before going on with the calculations, let me open a brief parenthesis to explain how we can decompose $h_{\mu\nu}$ in a very special way so as to be able to classify the various perturbations in three different types that do not affect each other. This decomposition must encapsulate the original number of dof, $10$. 

\begin{itemize}
\item $h_{00}\equiv 2\Phi$ ; The original dof of $h_{00}$ is transferred to the scalar function $\Phi$.

\item $h_{0i}\equiv\omega^{\parallel}_i+\omega^{\perp}_i$ ; This is the Helmholtz's decomposition of a vector in a longitudinal part $\vec{\omega}^{\parallel}$ and a transverse part $\vec{\omega}^{\perp}$, which satisfy 
\be\vec{\nabla}\times\vec{\omega}^\parallel=\vec{0}\qquad \vec{\nabla}\cdot\vec{\omega}^\perp=0\,.
\ee
The vector $\vec{\omega}^\perp$ encapsulates $2$ of the $3$ original dof carried by the vector $h_{0i}$. In contrast, $\vec{\omega}^\parallel$ is irrotational, so we can express it as $\vec{\omega}^\parallel=\vec{\nabla}w$, where $w$ is a potential function. Therefore, $\vec{\omega}^\parallel$ carries (through $\omega$) the remaining dof of $h_{0i}$.

\item $h_{ij}\equiv h^\parallel_{ij}+h^\perp_{ij}+\tilde{h}_{ij}$, where the vector $\partial_i h^\parallel_{ij}$ is longitudinal, i.e. $\epsilon_{ijk}\partial_j\partial_l h^\parallel_{lk}=0$, and $\partial_i h^\perp_{ij}$ is transverse, i.e. $\partial_i\partial_jh^\perp_{ij}=0$. The longitudinal part can be written in terms of a potential function $B$,
\be
h^{\parallel}_{ij}=\left(\partial_i\partial_j-\frac{1}{3}\delta_{ij}\nabla^2\right)B\,.
\ee
Notice that $h^\parallel_{ij}$ is traceless and only carries $1$ dof. The transverse part can be expressed as
\be
h^\perp_{ij}=\partial_iW_j^T+\partial_jW^T_i,
\ee
with $\vec{W}^T$ being a transverse vector. The tensor $h^{\perp}_{ij}$ is also traceless by construction, i.e. $h_{ii}^\perp=2\vec{\nabla}\cdot\vec{W}^T=0$. Thus, $h_{ij}^\perp$ (or, equivalently, $\vec{W}^T$) carries two dof. So the remaining piece of $h_{ij}$, i.e. $\tilde{h}_{ij}$, must contain $3$ dof. It is split in two pieces,
\be
\tilde{h}_{ij}=h_{ij}^{TT}-2\Psi\delta_{ij}\,.
\ee
Obviously, $\Psi$ carries $1$ dof and $h^{TT}_{ij}$ is chosen to be traceless. Moreover, the latter is symmetric (just because the other parts of $h_{ij}$ defined above are also symmetric and $h_{ij}$ is symmetric too) and transverse, i.e. $\partial_i h^{TT}_{ij}=0$, which means that it carries the remaining $2$ dof.
 
\end{itemize} 

This is the so-called scalar-vector-tensor (SVT) decomposition\footnote{See the seminal work \cite{Lifshitz1946}, and the more modern exposition provided in the books \cite{BookAmendolaTsujikawa,BookGorbunovRubakov} and the comprehensive paper \cite{MaBertschinger1995}.}, since the various perturbations are classified in one of these three categories depending on how they transform under rotations in momentum space (around the wave vector $\vec{k}$). Scalar perturbations are those that remain unmodified under such transformations, $\Phi$, $\Psi$, $\omega$, and $B$; the vector ones are encoded in $\vec{w}^\perp$ and $\vec{W}^T$; and, finally, tensor perturbations in $h^{TT}_{ij}$. But why is it useful to decompose the perturbed metric in this way? Well, because by doing this one can split the perturbations in Einstein's equations in three different types that do not talk to each other, meaning that the equations that govern their evolution are not coupled. This allows us to study the three types of perturbations independently. 

We are interested in studying the growth of scalar matter perturbations in the linear regime, so we must worry about $\Phi$, $\Psi$, $\omega$, and $B$. Let us now fix the gauge by choosing a concrete system of spacetime coordinates. In the Newtonian gauge, 
\be\label{eq:NewtonGauge}
h_{0i}=0\qquad B=0\,.
\ee
Thus, after the decomposition and the gauge fixing we effectively deal with the following metric in the analysis of scalar perturbations,
\be
ds^2=a^2\left[(1+2\Phi)d\eta^2-(1+2\Psi)d\vec{x}^2\right]\,.
\ee
It is quite practical to write the metric elements in both, the covariant and contravariant forms,

$$g_{00}=a^2(1+2\Phi)\qquad g_{ij}=-a^2(1+2\Psi)\delta_{ij}\,,$$
\be
g^{00}=\frac{1}{a^2}(1-2\Phi)\qquad g^{ij}=-\frac{\delta^{ij}}{a^2}(1-2\Psi)\,.
\ee
Using these expressions together with \eqref{eq:ChrystoffelSymbols}, \eqref{eq:RiemannTensor}, \eqref{eq:RicciTensor}, \eqref{eq:RicciScalar} and \eqref{eq:EinsteinOrFielEq}, respectively, one finds the following geometrical quantities up to corrections of second order in perturbations:
\vskip 0.3cm 
Christoffel symbols

$$\Gamma^{0}_{00}=\mathcal{H}+\Phi^\prime\qquad\Gamma^{0}{}_{i0}=\Gamma^{i}{}_{00}=\partial_i\Phi\qquad \Gamma^{0}{}_{ij}=\delta_{ij}\left[\mathcal{H}(1+2\Psi-2\Phi)+\Psi^\prime\right]$$
\be\label{eq:ChristoffelNewton}
\Gamma^{i}{}_{jk}=\delta_{ij}\partial_k\Psi+\delta_{ik}\partial_j\Psi-\delta_{jk}\partial_i\Psi\qquad\Gamma^{i}{}_{0j}=\delta^{i}{}_{j}(\mathcal{H}+\Psi^\prime)\,.
\ee

Ricci tensor

\bea             
R_{00}&=&3\mH^\prime+3\Psi^{\pp}+3\mH(\Psi^\p-\Phi^\p)-\nabla^2\Phi\,,\\
R_{i0}&=&-2\mH\partial_i\Phi+2\partial_i\Psi^\p\,,\nonumber\\
R_{ij}&=&\partial_i\partial_j(\Phi+\Psi)+\delta_{ij}\nabla^2\Psi-\delta_{ij}[(2\mH^2+\mH^\p)(1+2\Psi-2\Phi)+\Psi^{\pp}+5\mH \Psi^\p-\Phi^\p\mH]\,.\nonumber
\eea
\newpage
Ricci scalar

\bea\label{eq:RicciScalarApp}
a^2 R=6(1-2\Phi)(\mH^\p+\mH^2)-2\nabla^2(\Phi+2\Psi)+6\psi^{\pp}+6\mH(3\Psi^\p-\Phi^\p)\,.
\eea

Einstein's tensor
\bea
G_{00}&=&-3\mH^2+2\nabla^2\Psi-6\mH\Psi^\p\,,\nonumber\\
G_{i0}&=&2\partial_i\Psi^\p-2\mH\partial_i\Phi\,,\\
G_{ij}&=&\partial_i\partial_j(\Phi+\Psi)+\delta_{ij}(\mH^2+2\mH^\p)(1+2\Psi-2\Phi)+\nonumber\\
&&\qquad\qquad\qquad\qquad\qquad+\delta_{ij}[2\Psi^{\pp}-\nabla^2(\Phi+\Psi)+2\mH(2\Psi^\p-\Phi^\p)]\nonumber\,,
\eea
where $\mathcal{H}\equiv\frac{1}{a}\frac{da}{d\eta}=aH$, $\nabla^2$ is the spatial Laplace operator in comoving coordinates, and the prime denotes a derivative with respect to the conformal time. Let us now compute the perturbed $4$-velocity $u^\mu=\bar{u}^\mu+\delta u^\mu$ entering the formula of the perturbed energy-momentum tensor. Obviously, the unperturbed velocity is simply $\bar{u}^\mu=(\frac{1}{a},\vec{0})$, since in the background cosmology a test particle remains comoving to the expansion of the Universe. Of course, every fluid has its own $4$-velocity. The perturbed time component (up to second order perturbations) is obtained in a simple way,
\be
u^\mu u_\mu=1\longrightarrow \delta u^0=-\frac{\Phi}{a}\,,
\ee
and the physical $3$-velocity is defined as $v^i\equiv au^i=a\delta u^i$. Thus, we have,
\be
u^\mu=\frac{1}{a}(1-\Phi,v^j)\qquad u_{\mu}=a(1+\Phi,-v^j)\,.
\ee
We can also decompose $\vec{v}$ {\it a la} Hemholtz, i.e. as a sum of a longitudinal plus a transverse part. The former can be expressed as $\vec{v}^\parallel=\vec{\nabla}v$, and the latter does not play any role in the analysis of the scalar perturbations, simply because it is of vector type. Thus the velocity only enters these calculations through the velocity potential $v$. 

Now we have all the needed ingredients for obtaining the equations that govern the matter scalar perturbations. We just have to plug all the above computed expressions into Einstein's field equations \eqref{eq:EinsteinModFielEq} and the covariant conservation equations $\nabla^\mu T_{\mu\nu}=0$, making also use of \eqref{eq:EMTPF}. I only show the perturbed equations, since the non-perturbed ones are presented in \ref{subsec:BasicsLCDM} \footnote{Eqs. \eqref{eq:FriedmannLCDM}, \eqref{eq:PressureLCDM}, \eqref{eq:acceEq} and \eqref{eq:consBianchi} are not only valid for the $\Lambda$CDM model, but also for a general type of DVM, being \eqref{eq:consBianchi} only valid in the concrete case $G$=const, the one we are studying in this section.}. I express the densities and pressures of the various fluids under consideration as a sum of a background part plus a perturbed one, i.e. $\rho=\bar{\rho}+\delta\rho$ and $p=\bar{p}+\delta p$.

\begin{itemize}
\item $G_{00}=8\pi G T_{00}$
\be\label{eq:Poisson0}
3\mH^2\Phi+3\mH\Phi^\p-\nabla^2\Phi=-4\pi Ga^2\sum_{n=1}^{N}\delta\rho_n\,.
\ee
\item $G_{0i}=8\pi G T_{0i}$
\be
\mH\Phi+\Phi^\p=-4\pi Ga^2\sum_{n=1}^{N}(\bar{p}_n+\bar{\rho}_n)v_n\,.
\ee
\item $G_{ij}=8\pi G T_{ij}$
\be
\Phi^{\pp}+3\mH\Phi^\p+\Phi(\mH^2+2\mH^\p)=4\pi G a^2\sum_{n=1}^{N}\delta p_n\,.
\ee
\item $\nabla^\mu T_{\mu 0}=0$ 
\be\label{eq:continuity0}
\sum_{n=1}^{N}\left[\delta\rho_n^\p+(\bar{p}_n+\bar{\rho}_n)(\nabla^2v_n-3\Phi^\p)+3\mH(\delta\rho_n+\delta p_n)\right]=0\,.
\ee
\item $\nabla^\mu T_{\mu i}=0$
\be\label{eq:euler0}
\sum_{n=1}^{N}\left[(\bar{p}_n+\bar{\rho}_n)^\p v_n+(\bar{p}_n+\bar{\rho}_n)(4\mH v_n+v_n^\p+\Phi)+\delta p_n\right]=0\,.
\ee
\end{itemize}
In all these equations I have already used $\Psi=-\Phi$, which is also obtained from the ij component of Einstein's field equations. It is important to remark that only $3$ out of these $5$ equations are independent. It turns out to be very useful to express the various perturbed quantities in momentum space, e.g.
\be
\Phi(\eta,\vec{x})=\int \Phi_{\vec{k}}(\eta) e^{i\,\vec{k}\cdot\vec{x}}\,d^3k\,,
\ee
since as we are working at leading order in perturbation theory, the different modes do not mix in the perturbed equations, so they are automatically decoupled. Here the $\vec{k}$'s are comoving wave vectors. From now on I will only consider physical modes much shorter than the horizon, i.e. those modes that at present satisfy $\lambda\ll 3000h^{-1}$ Mpc. Moreover, I will study matter perturbations growing in a matter and DE dominated Universe. In this context, we can neglect the effect of radiation at the background and perturbed levels. In this thesis I analyze DVM's in which the vacuum interacts with baryons+DM or only with DM. Both cases are different, in principle, in the sense that we cannot expect {\it a priori} that the aforementioned interaction affects in the same way the growth of the density perturbations of DM and baryons. The point here is that we are not interested in the growth of baryons and DM separately, but in the total matter density perturbations. This is what the LSS observables used in this dissertation are sensitive to. For example, the Redshift Space Distortions (RSD's) are caused by the total amount of matter, not only by a particular type. Bearing this in mind, we can obtain an equation for the total nonrelativistic matter energy density perturbations that is valid for both scenarios. From \eqref{eq:Poisson0}, \eqref{eq:continuity0} and \eqref{eq:euler0}, we obtain the Poisson, continuity and Euler equations, respectively:

\begin{align}
k^2\Phi&=-4\pi G a^2(\delta\rho_{m}+\delta\rho_\Lambda)\,,\label{eq:PoissonNewton}\\
      0&=\delta\rho_m^\p+3\mathcal{H}\delta\rho_m-k^2v_m\bar{\rho}_m+\delta\rho_\Lambda^\p\,,\label{eq:ContinuityNewton}\\
       0&=\frac{d}{d\eta}\left(\bar{\rho}_mv_m\right)+4\mathcal{H}\bar{\rho}_mv_m+\bar{\rho}_m\Phi-\delta\rho_\Lambda\,,\label{eq:EulerNewton}
\end{align}
where $\delta\rho_m=\delta\rho_b+\delta\rho_{dm}$, and $v_m=(v_{dm}\bar{\rho}_{dm}+v_b\bar{\rho}_b)/\bar{\rho}_m$ is the total matter velocity potential, obtained upon weighting the contribution of the two matter components. Let us assume that $\delta\rho_m\gg\delta\rho_\Lambda$ and $\delta\rho_m^\p\gg\delta\rho_\Lambda^\p$. This is a very natural assumption, since in the $\Lambda$CDM model $\delta\rho_\Lambda=0$, so we expect that if we have a little deviation from the concordance model the perturbation of the vacuum energy density, despite being different from zero, it will still be much less important than the matter energy density perturbations\footnote{See Sect. \ref{sect:DEorNot} for a related study on the effect of DE perturbations at subhorizon scales in the context of the $\mathcal{D}$-class of dynamical DE models.}. Thus, \eqref{eq:PoissonNewton} and \eqref{eq:ContinuityNewton} can be rewritten in a simpler way, as follows:
\begin{align}
k^2\phi &= -4\pi G a^2\bar{\rho}_m\delta_m\,,\label{eq:PoissonNewton2}\\
\delta^\p_m+\psi\delta_m&=k^2v_m\,,\label{eq:ContinuityNewton2}
\end{align}
where $\delta_m=\delta\rho_m/\bar{\rho}_m$ is the matter density contrast and $\psi=-\bar{\rho}^\p_\Lambda/\bar{\rho}_m$. The continuity equation \eqref{eq:ContinuityNewton2} is modified with an extra term, i.e. $\psi\delta_m$, with respect to the $\Lambda$CDM case because now there is an injection/extraction of energy in the matter sector caused by the decay/increase of the vacuum energy density. On the other hand, by using the background continuity equation,
\begin{equation}
\bar{\rho}^\p_\Lambda+\bar{\rho}^\p_m+3\mathcal{H}\bar{\rho}_m=0\,,
\end{equation}
one can write \eqref{eq:EulerNewton} in a more standard way:
\begin{equation}\label{eq:EulerNewton2}
v^\p_m+\mathcal{H}v_m+\Phi+\psi v_m-\frac{\delta\rho_\Lambda}{\bar{\rho}_m}= 0\,.
\end{equation}
Notice that, in principle, we cannot remove the $\delta\rho_\Lambda$ term because it could be, for instance, of the same order of $\psi v_m$. We can only throw away the $\delta\rho_\Lambda$ terms when they are directly compared with a $\delta\rho_m$ term. Otherwise, we could be removing relevant contributions or, at least, contributions as important as the remaining ones. The first three terms of the last expression correspond to those appearing in the $\Lambda$CDM Euler equation. In fact, they can be obtained in a simple nonrelativistic way, just by perturbing the Newtonian gravitational law, i.e. $\frac{d\vec{v}}{dt}=-\vec{\nabla}\phi$. The two extra terms of \eqref{eq:EulerNewton2} could be interpreted as the change of the matter velocity that is induced by the matter-vacuum interaction. We should ask ourselves if it is possible that this interaction modifies the velocity of the matter particles. The loss of energy of the vacuum sector can only happen in two different ways: by vacuum decay through the generation of particle pairs or because of an increase of the particles' masses. Let us analyze the first case. In order to do that let us firstly define the velocity of the matter component as the discrete sum of the particles' peculiar velocities. We associate the velocity of the matter fluid in a given location of space $\vec{x}$ to the averaged value of the individual velocities in a small volume of the fluid centered in $\vec{x}$:
\begin{equation}
\vec{v}_m(\vec{x})\equiv\frac{1}{N}\sum_{j=1}^{N}\left[\vec{v}_{j,\phi}+\vec{v}_{j,int}\right]\,,
\end{equation}
where $N$ is the number of particles contained in such volume, $\vec{v}_{j,\phi}$ is the peculiar velocity of the j$_{th}$ particle induced by the gravitational potential and $\vec{v}_{j,int}$ is the peculiar velocity of the j$_{th}$ particle due to the vacuum decay. Notice that the latter is just $\vec{0}$ if the particle has not been produced by vacuum decay. Moreover, it is obvious that in each decay two particles (let us call them A and B) will be produced, one with velocity $\vec{v}_{A,int}$ and another one with velocity $\vec{v}_{B,int}=-\vec{v}_{A,int}$. Each decay product has a companion particle with the same velocity in modulus and direction, but opposite way. This is a direct consequence of the momentum conservation if we assume that vacuum has no intrinsic velocity, i.e. $\vec{v}_\Lambda=\vec{0}$. Thus, if no decay product leaves the differential volume where we are performing the average, the sum of $\vec{v}_{j,int}$ is exactly $\vec{0}$. We do not expect it to be the case, though. Some companion particles leave the volume where they had been created in order to enter another one. But assuming that the decay process is homogeneous inside a region larger than the differential volume (at least, in the immediate surrounding volumes), one can see that the excess of particles created by the decay in the neighborhood of the wall of one side of the volume is compensated by the particles created in the other side. As these two sets of particles have, on average, contrary velocities, they cancel out. At each instant of time the same reasoning can be applied, so we find,
\begin{equation}
\vec{v}_m(\vec{x})=\frac{1}{N}\sum_{j=1}^{N}\vec{v}_{j,\phi}\,.
\end{equation}
This is telling us that vacuum decay cannot modify the peculiar velocity of the matter fluid. The latter only seems to be sensible to the gravitational potential generated by the matter inhomogeneities. But can an increase of the particles' masses give rise to a modification of $\vec{v}_m$? The answer is no, since this is totally forbidden by the equivalence principle. Thus, if we assume that the equivalence principle is also fulfilled by the dark matter particles we must also accept that the variation of the particles' masses does not affect their acceleration if the only external force generating this acceleration is the gravitational one. This reasoning leads us to impose an extra relation in order to ensure the correct physical behavior of the RVM's at the linear perturbation level,
\begin{equation}\label{eq:ExtraRelation}
\delta\rho_\Lambda=\psi v_m\bar{\rho}_m\,,
\end{equation}
so as to the usual ($\Lambda$CDM) Euler equation to be fulfilled:
\begin{equation}\label{eq:EulerNewton3}
v^\p_m+\mathcal{H}v_m+\phi=0\,.
\end{equation}
By using \eqref{eq:ContinuityNewton2} in \eqref{eq:ExtraRelation} one can check that $\frac{\delta\rho_\Lambda}{\bar{\rho}_m}\propto\nu_i\left(\frac{H}{k}\right)^2$, where $\nu_i$ is the DVM parameter (see e.g. Chapter \ref{chap:PRDbased}). Therefore, the vacuum energy density perturbations are very suppressed at scales deeply inside the horizon ($k\gg \mathcal{H}$). In fact, the most recent phenomenological studies on the RVM's show that $\nu$ is expected to be of order $\mathcal{O}(10^{-3})$ (cf. e.g. Table \ref{tableFit1PRD}) and this, of course, also helps to suppress even more the value of $\delta\rho_\Lambda$. Thus, \eqref{eq:ExtraRelation} is totally consistent with our initial assumption, i.e. $\delta\rho_m\gg\delta\rho_\Lambda$.  Moreover, it is quite reassuring to recover the $\Lambda$CDM result, i.e. $\delta\rho_\Lambda=0$, when we set $\nu=0$. Equation \eqref{eq:ExtraRelation} can also be written in terms of the physical velocity of matter and the gradient of the vacuum perturbations, 
\begin{equation}
\vec{\nabla}\delta\rho_\Lambda=-\bar{\rho}^\p_\Lambda\vec{v}_m\,.
\end{equation}
Combining \eqref{eq:PoissonNewton2}, \eqref{eq:ContinuityNewton2} and \eqref{eq:EulerNewton3}, the following equation for the matter density contrast is obtained,
\begin{equation}\label{eq:DensityContrastEq}
\delta_m^{\prime\prime}+\delta_m^\prime\left(\psi+\mH\right)+\delta_m\left(-4\pi G a^2\bar{\rho}_m+\psi^\p+\psi\mH\right)=0\,.
\end{equation}
As expected, if there is no matter-vacuum interaction, i.e. $\psi=0$, we retrieve the equation of the standard $\Lambda$CDM model. To solve this second-order differential equation we need to set two initial conditions, at a time at which the expansion is strongly matter-dominated, i.e. $\delta_m(z_i)$ and $\delta_m^\p(z_i)$ at, let us say, $z_i\sim 100$. These initial conditions are model-dependent and are provided in the main body of the thesis for each one of the models under study.

The same equation \eqref{eq:DensityContrastEq} can be obtained by using a source 4-vector $Q_\mu=Q_p u_\mu$, with $Q_p$ being the perturbed source function $Q_p=Q+\delta Q$, and $Q=-\bar{\rho}^\p_\Lambda/a$ the background one. The use of this interaction vector ensures the automatic fulfillment of \eqref{eq:ExtraRelation} and the usual Euler equation \eqref{eq:EulerNewton3}. Let us see this more in detail. Due to the Bianchi identity we find,
\begin{equation}
\nabla^\mu (T^{\rm m}_{\mu\nu}+T^{\rm \Lambda}_{\mu\nu})=0\,,
\end{equation}
with $T^{\rm m}_{\mu\nu}$ and $T^{\rm \Lambda}_{\mu\nu}$ being the matter and vacuum energy-momentum tensors, respectively. We can split this equation in two parts by means of $Q_\mu$,
\begin{equation}\label{eq:splitQ}
\nabla^\mu T^{\rm m}_{\mu\nu}\equiv Q_\nu\qquad\nabla^\mu T^{\rm \Lambda}_{\mu\nu}\equiv -Q_\nu\,.
\end{equation}
The perturbed source vector yields,
\begin{equation}
\delta Q_\mu=\delta Q\bar{u}_\mu+Q\delta u_\mu=(a\delta Q+a\Phi Q,-aQ\vec{v}_m)\,.
\end{equation}
Let us introduce these relations in \eqref{eq:splitQ} and take the perturbed part of the resulting expression. By choosing the spatial component and defining $\delta Q_i\equiv \partial_i\delta V$, where $\delta V=-aQv_m=\bar{\rho}^\p_\Lambda v_m$, we find,
\begin{equation}
v^\p_m+\mathcal{H}v_m+\Phi=\frac{-\delta V+\bar{\rho}^\p_\Lambda v_m}{\bar{\rho}_m}=0\,,
\end{equation} 
so the usual Euler equation is retrieved, as promised. If we choose the $0$ component of the Bianchi identity, instead of the spatial one, then we find
\begin{equation}
\delta^\p_m-k^2v_m+\psi \delta_m=\frac{\delta Q_0}{\bar{\rho}_m}=\frac{a}{\bar{\rho}_m}(\delta Q+\phi Q)\,,
\end{equation}
or, equivalently,
\begin{equation}
\delta^\p_m-k^2v_m+\psi \delta_m=-\left(\frac{\delta\rho^\p_\Lambda+\phi \bar{\rho}^\p_\Lambda}{\bar{\rho}_m}\right)\,.
\end{equation}
The first term on the {\it r.h.s.} can be clearly neglected when compared with the first term of the {\it l.h.s.} in order to be consistent with our initial assumption, $\delta\rho_m\gg\delta\rho_\Lambda$ and $\delta\rho^\p_m\gg\delta\rho^\p_\Lambda$. The second too, but for a different reason. Notice that because of the Poisson equation \eqref{eq:PoissonNewton2} $\phi\propto G \delta\rho_m/k^2$. Thus, at low scales $|\psi \delta_m|\gg |\phi \bar{\rho}^\p_\Lambda/\bar{\rho}_m|$ and this directly allows us to neglect the second term of the {\it r.h.s.} too. The continuity equation \eqref{eq:ContinuityNewton2} is then recovered. Therefore, the source $4$-vector $Q_\mu$ defined above is capable of describing the vacuum-matter interaction in a consistent and physical way, since it does not modify the standard Euler equation. As it has been explained before, the modification of the latter would imply a breach of the equivalence principle or/and a violation of the momentum conservation, which is completely intolerable. In addition, we have seen that the source 4-vector $Q_\mu$ generates the correct system of linear perturbations equations which directly lead to \eqref{eq:DensityContrastEq}.    

%%%%%%%%%%%%%%%%%%%%%%%%%%%%%%%%%%%%%%%%%%%%%%%%%%%%%%%%%%%%%%%%%%%%%%%%%
%%%%%%%%%%%%%%%%%%%%%%%%%%%%%%%%%%%%%%%%%%%%%%%%%%%%%%%%%%%%%%%%%%%%%%%%%
%%%%%%%%%%%%%%%%%%%%%%%%%%%%%%%%%%%%%%%%%%%%%%%%%%%%%%%%%%%%%%%%%%%%%%%%%

\subsection{Synchronous gauge}

In the synchronous gauge, we choose
\be
\omega=0\qquad \Phi=0\,.
\ee
In this case we work with cosmic time, instead of conformal time. The perturbed metric can be written as follows,
\be
ds^2=dt^2-(a^2\delta_{ij}+\chi_{ij})dx^idx^j\,.
\ee 
where $\chi_{ij}\equiv\delta g_{ij}$. Thus, the covariant and contravariant elements of the metric tensor read, respectively,

$$g_{0i}=0\qquad g_{00}=1\qquad g_{ij}=-a^2\delta_{ij}+\chi_{ij}\,,$$
\be
g^{0i}=0\qquad g^{00}=1\qquad g^{ij}=-\frac{\delta^{ij}}{a^2}+\chi^{ij}\,. 
\ee
It is convenient to obtain some relations involving $\chi_{ij}$, $\chi^{ij}$, and their trace. They are used in the calculation of the geometrical quantities that enter the field equations. Neglecting the second-order perturbations we find,
\be
g_{i\mu}g^{\mu j}=\delta^{j}{}_{i}\longrightarrow \chi^{ij}=-\frac{\chi_{ij}}{a^4}\,,
\ee
\be                                     
\chi\equiv \chi_{\mu\nu} g^{\mu\nu} =-\frac{\chi_{ii}}{a^2}\,,
\ee
or, equivalently, 
\be
\chi_{ii}=-a^2\chi\qquad \chi^{ii}=\frac{\chi}{a^2}\,.
\ee
The Christoffel symbols read,

$$\Gamma^{0}_{00}=\Gamma^{i}_{00}=\Gamma^{0}{}_{i0}=0 \qquad \Gamma^{0}{}_{ij}=a^2H\delta_{ij}-\frac{\chi_{ij,0}}{2}$$
\be\label{eq:ChristoffelSyn}
\Gamma^{i}{}_{jk}=-\frac{1}{2a^2}(\chi_{ij,k}+\chi_{ik,j}-\chi_{jk,i}) \qquad\Gamma^{i}{}_{0j}=H\delta^{i}{}_{j}-\frac{\chi_{ij,0}}{2a^2}-a^2H\chi^{ij}\,.
\ee
The Einstein field equations can be written as 
\be\label{eq:inter}
R_{\mu\nu}=8\pi G \left[T_{\mu\nu}-\frac{g_{\mu\nu}}{2}(T_{00}+T_{ij}g^{ij})\right],
\ee
just by using the relation $R=-8\pi G T_{\mu\nu}g^{\mu\nu}$, which has been obtained from the trace of the starting equation \eqref{eq:EinsteinOrFielEq}. The $00$ component of \eqref{eq:inter} is simply given by
\be\label{eq:R00}
R_{00}=4\pi G (T_{00}-T_{ij}g^{ij})\,.
\ee
Let us compute the perturbed equations in a more general case than the one analyzed with the conformal Newtonian gauge in the last subsection, by also considering perturbations on the gravitational coupling $G$. In this way, we will be able to obtain the equations for G-type models just as a particular case of our results. Upon plugging 
\be
R_{00}=3(\dot{H}+H^2)+\frac{\ddot{\chi}}{2}+H\dot{\chi}
\ee
in \eqref{eq:R00} and doing $G=\bar{G}+\delta G$ one obtains at zeroth order in perturbations the acceleration equation \eqref{eq:acceEq}, and at first order,
\be\label{eq:syn1}
\frac{\dot{\hat{\chi}}}{2}+H\hat{\chi}=4\pi\sum_{n=1}^{N}\left[\bar{G}(\delta\rho_n+3\delta p_n)+\delta G(\bar{\rho}_n+3\bar{p}_n)\right]\,,
\ee
where $\hat{\chi}\equiv -\dot{\chi}$. The Bianchi identity in this case translates to 
\be\label{eq:BianchiGen}
\nabla^\mu (GT_{\mu\nu})=0\,,
\ee
since now $G$ is not a rigid constant and, therefore, $\nabla^\mu G\ne 0$. By perturbing the time component of this equation we find at leading order the following conservation equation
\be
\sum_{n=1}^{N}\left[\dot{\bar{G}}\bar{\rho}_n+\bar{G}\{\dot{\bar{\rho}}_n+3H\bar{\rho}_n(1+\omega_n)\}\right]=0\,,
\ee
which is the direct generalization of \eqref{eq:consBianchi}. The resulting linear perturbation equation reads,
\bea
0&=&\sum_{n=1}^{N} \bar{G}\left[\delta\dot{\rho}_n+3H\delta\rho_n(1+\omega_n)+\bar{\rho}_n(1+\omega_n)\left(\theta_n-\frac{\hat{\chi}}{2}\right)\right]+\label{eq:syn2}\\
&&\qquad\qquad\qquad\qquad+\dot{\bar{G}}\delta\rho_n+\delta\dot{G}\bar{\rho}_n+\delta G [\dot{\bar{\rho}}_n+3H\bar{\rho}_n(1+\omega_n)]\nonumber\,,
\eea
where $\theta_n=\nabla_\mu\delta u^\mu_n$, with $\delta u^\mu_n=(0,\vec{\nabla}v_n)/a$. The spatial component of \eqref{eq:BianchiGen} gives us the following equation in momentum space,
\be\label{eq:syn3}
0=\sum_{n=1}^{N} (1+\omega_n)\left[\dot{\bar{G}}\bar{\rho}_n\theta_n+\bar{G}(\dot{\bar{\rho}}_n\theta_n+\bar{\rho}_n\dot{\theta}_n+5H\bar{\rho}_n\theta_n)\right]-\frac{k^2}{a^2}\left[p_n\delta G+\bar{G}\delta p_n\right]\,.
\ee
Equations \eqref{eq:syn1}, \eqref{eq:syn2} and \eqref{eq:syn3} make up our coupled system. Let us now particularize these expressions for those models with a constant Newtonian coupling. At the end of the day, we would have to recover equation \eqref{eq:DensityContrastEq}. In this case the system is given by
\begin{equation}\label{eq:syn1v2}
\dot{\hat{\chi}}+2H\hat{\chi}=8\pi G\sum_{n=1}^{N}(\delta\rho_n+3\delta p_n)\,,
\end{equation}
\begin{equation}\label{eq:syn2v2}
\sum_{n=1}^{N}\dot{\delta\rho_n}+(\bar{\rho}_n+\bar{p}_n)\left(\theta_n-\frac{\hat{\chi}}{2}\right)+3H(\delta\rho_n+\delta p_n)=0\,,
\end{equation}
\begin{equation}\label{eq:syn3v2}
\sum_{n=1}^{N} \dot{\theta}_n(\bar{\rho}_n+\bar{p}_n)+\theta_n\left[\dot{\bar{\rho}}_n+\dot{\bar{p}}_n+5H(\bar{\rho}_n+\bar{p}_n)\right]-\frac{k^2}{a^2}\delta p_n=0\,,
\end{equation}
As in the Newtonian gauge, we assume that the vacuum has no peculiar velocity and that $\delta\rho_m\gg\delta\rho_\Lambda$ and $\delta\dot{\rho}_m\gg\delta\dot{\rho}_\Lambda$. Therefore, we find,

\be\label{eq:syn1a}
\dot{\hat{\chi}}+2H\hat{\chi}=8\pi G \delta\rho_m
\ee
\be\label{eq:syn2a}
\delta\dot{\rho}_m+\bar{\rho}_m\left(\theta_m-\frac{\hat{\chi}}{2}\right)+3H\delta\rho_m=0
\ee
\be\label{eq:syn3a}
\bar{\rho}_m\dot{\theta}_m+(\dot{\bar{\rho}}_m+5H\bar{\rho}_m)\theta_m+\frac{k^2}{a^2}\delta\rho_\Lambda=0\,.
\ee
The last equation can be rewritten as follows,
\begin{equation}\label{eq:thetaEq}
\dot{\theta}_m+\theta_m\left(\bar{\psi}+2H\right)=-\frac{k^2}{a^2}\frac{\delta\rho_\Lambda}{\bar{\rho}_m}\,.
\end{equation}
In coordinate space,
$$
\theta_m=\nabla_\mu\delta u^\mu=\nabla_\mu(g^{\mu\nu}\delta u_\nu)=g^{ij}\partial_j \delta u_i\,,
$$
\begin{equation}
\theta_m=\frac{-\vec{\nabla}\cdot(-a\vec{v}_m^L)}{a^2}=\frac{\nabla^2v_m}{a}\,,
\end{equation}
up to corrections of second order, and in momentum space,
\begin{equation}
\theta_m=-\frac{k^2}{a}v_m\,.
\end{equation}
Substituting the latter expression in \eqref{eq:thetaEq} we find,
\begin{equation}
\dot{v}_m+Hv_m+v_m\bar{\psi}=\frac{1}{a}\frac{\delta\rho_\Lambda}{\bar{\rho}_m}\,.
\end{equation}
This is the momentum conservation equation for the matter particles in the synchronous gauge. As in the Newtonian gauge, we do not want this equation to be modified with respect to the $\Lambda$CDM one ($\dot{v}_m+Hv_m=0$, $\delta\rho_\Lambda=0$), since we do not expect that the vacuum-matter interaction has any effect on the velocity of the matter fluid. Thus, we impose $\delta\rho_\Lambda=\psi v_m\bar{\rho}_m$, which is exactly the same relation that we have obtained in the Newtonian gauge. But we must still fix the residual gauge freedom, since it is not completely fixed by the choice $\omega=\Phi=0$. In order to do that we take $\theta_m=0$ \cite{MaBertschinger1995} and, therefore, $\delta\rho_\Lambda=0$. Thus, combining equations \eqref{eq:syn1a} and \eqref{eq:syn2a} we are automatically led again to \eqref{eq:DensityContrastEq}, so we find that at scales deeply inside the horizon, the differential equation that controls the evolution of the matter density perturbations is the same as the one derived with the Newtonian gauge.

\section{DVM's with $\dot{\Lambda}\ne 0$ and $\dot{G}\ne 0$, and self-conserved matter}
\label{sec:AppenGpert}

In the context of the G-type models, matter is a self-conserved fluid and $\rho_\Lambda$ varies at the expense of a variation of $G$. We can make use of \eqref{eq:syn1}, \eqref{eq:syn2} and \eqref{eq:syn3} in order to derive the equations that control the growth of matter perturbations. The resulting system reads,
\begin{equation}\label{eq:Gpert1}
\dot{\hat{\chi}}+2H\hat{\chi}=8\pi \left[\bar{G}(\delta\rho_m-2\delta\rho_\Lambda)+\delta G(\bar{\rho}_m-2\bar{\rho}_\Lambda)\right]\,,
\end{equation}
\begin{equation}\label{eq:Gpert2}
\bar{\rho}_m\left(\theta_m-\frac{\hat{\chi}}{2}\right)+3H\delta\rho_m+\delta\dot{\rho}_m=0\,,
\end{equation}
\begin{equation}\label{eq:Gpert3}
\dot{\bar{\rho}}_\Lambda\delta G+\delta \dot{G}(\rho_m+\rho_\Lambda)+\dot{\bar{G}}(\delta\rho_m+\delta\rho_\Lambda)+\bar{G}\delta\dot{\rho}_\Lambda=0\,,
\end{equation}
\be\label{eq:Gpert4}                 
\dot{\bar{\rho}}_m\theta_m+\bar{\rho}_m\dot{\theta}_m+5H\bar{\rho}_m\theta=0\,,
\ee
\be\label{eq:Gpert5}
\dot{\bar{G}}\bar{\rho}_m\theta_m+\frac{k^2}{a^2}\left(\bar{\rho}_\Lambda\delta G+\bar{G}\delta\rho_\Lambda\right)=0\,.
\ee
This is a system with $5$ equations and $5$ unknown functions: $\delta\rho_m$, $\delta\rho_\Lambda$, $\theta_m$, $\hat{\chi}$, and $\delta G$. Using \eqref{eq:Gpert4} and $\dot{\bar{\rho}}_m+3H\bar{\rho}_m=0$ one obtains:
\be
\theta_m(a)=\theta_0a^{-2}\,.
\ee   
It is a decaying mode that plays no role in the current time, specially at scales deeply inside the horizon. By means of \eqref{eq:Gpert5} we find,
\be
\frac{\delta\rho_\Lambda}{\bar{\rho}_\Lambda}=-\frac{\delta G}{\bar{G}}\,.
\ee
Introducing this result in \eqref{eq:Gpert3} we obtain,
\be
\delta_m=\frac{\delta\rho_m}{\bar{\rho}_m}=-\frac{\delta\dot{G}}{\dot{\bar{G}}}\,.
\ee
Now, from \eqref{eq:Gpert2} one gets another important relation, 
\be
\hat{\chi}=2\dot{\delta}_m\,.
\ee
Introducing these results in \eqref{eq:Gpert1} and after some algebraic operations, one finally gets the equation for the matter density contrast,
\be
\delta^{\pp\p}_m+\frac{\delta_m^{\pp}}{2a}(16-9\Omega_m)+\frac{3\delta_m}{2a^2}(8-a\Omega_m^\p+3\Omega_m^2-11\Omega_m)=0\,,
\ee
where here the prime does not denote a derivative with respect to the conformal time, but with respect to the scale factor in this case.

\chapter{Collapse density threshold $\delta_c$ for type-A and B RVM's}
\label{ch:appCollapse}

In this appendix we present the necessary formulas to compute the
linearly extrapolated density threshold above which structures
collapse, i.e, $\delta_{c}$, for the type-A and type-B dynamical
vacuum models. We follow the standard methods available
in the literature (see e.g. \cite{Pace10}, \cite{Abramo1} and \cite{Abramo2}, and
references therein).  Details of the procedure were also amply
provided in Ref. \cite{Grande2011} where it was applied to G-type dynamical vacuum models (see the aforementioned reference and e.g. Chapter \ref{chap:Gtype}). We will therefore not repeat these details,
but just the initial setup and the final results for the nonlinear
perturbation equations corresponding to the models under
consideration. The linearized part of these equations reduces, of
course, to the linear perturbations equation \eqref{diffeqDa}. Recall that in order
to derive the nonlinear equations it is convenient to start from
the Newtonian formalism for the cosmological fluid, and use  the
continuity, Euler and Poisson equations in the matter dominated
epoch, see Appendix \ref{ch:appPert} for details:
\begin{eqnarray}
  \frac{\partial\rho_m}{\partial t}+\nabla_{\vec{r}}\cdot(\rho_m\vec{v})= 0
\label{eqn:cnpert}\;,\\
  \frac{\partial\vec{v}}{\partial
    t}+(\vec{v}\cdot\nabla_{\vec{r}})\,\vec{v}+
  \nabla_{\vec{r}}\,\Phi=0\label{eqn:enpert}\;,\\
  \nabla^2\Phi=4\pi G\sum_i\rho_i(1+3\omega_i)\;,\label{eqn:pnpert}
\end{eqnarray}
where $\vec v$ is the total velocity of the co-moving observer in
three-space, $\Phi$ is the Newtonian gravitational potential,
$\vec{r}$ is the physical coordinate, and  $\omega_i=p_i/\rho_i$ is
the EoS parameter for each component. Introducing comoving
coordinates $\vec{x}=\vec{r}/a$  the perturbations are defined in
the following way:
\begin{eqnarray}
\rho_i(\vec{x},t) & = & \bar{\rho}_i(t)+\delta\rho_i(\vec{x},t)=\bar{\rho}_i(t)(1+\delta_i(\vec{x},t))\;, \label{eqn:rpert} \\
\Phi(\vec{x},t) & = & \Phi_0(\vec{x},t)+\phi(\vec{x},t)\;, \label{eqn:fpert}\\
 \vec{v}(\vec{x},t) & = & a(t)[H(t)\vec{x}+\vec{u}(\vec{x},t)] \label{eqn:vpert}\;.\label{eqn:gpert}
\end{eqnarray}
Here $\vec{u}(\vec{x},t)$ is the comoving peculiar velocity.  Next
we have to insert Eqs.~(\ref{eqn:rpert})--(\ref{eqn:gpert})
into Eqs.~(\ref{eqn:cnpert})--(\ref{eqn:pnpert}) and use the
definition of the gradient with respect to co-moving coordinates.
Notice that $\vec{\nabla}\delta_m=0$, which holds for the spherical
collapse of a top-hat distribution. Following the same systematics
as described in  \cite{Grande2011} we arrive at the nonlinear
perturbations equations for the models under consideration.

\section{Type-A models}

In this case the corresponding fully nonlinear evolution equation
reads as follows:

$$a^2H^2\delta^{\prime\prime}_m+aH\delta_m^\prime\left[3H+\Psi-\frac{\rho_m}{2H}\right]+\delta_m\left[2H\Psi+aH\Psi^\prime-\frac{\rho_m}{2}(1+\delta_m)\right]-$$
\begin{equation}\label{eq:SDE}
-\left[\frac{4a^2H^2\delta_m^{\prime
2}+5aH\Psi\delta_m\delta_m^\prime+\Psi^2\delta_m^2}{3(1+\delta_m)}\right]=0\,,
\end{equation}

%%%%%%%%%%%%%%%%%%%%%%% FIGURE  2 %%%%%%%%%%%%%%%%%%%%%%%%%%%%%%%%%%%%%

\begin{figure}[t!]
\begin{center}
{\includegraphics[scale=0.50]{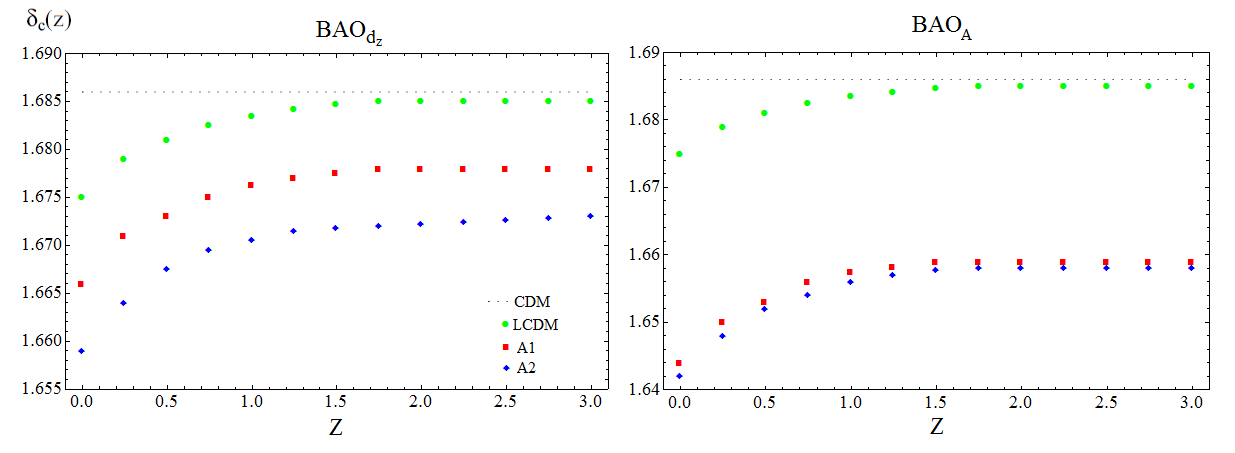}}
\end{center}
\caption[$\delta_c(z)$ for type-A models obtained from Tables \ref{tableFitBAOdz} and \ref{tableFitBAOA}]{\scriptsize{Computation of the collapse density threshold function $\delta_c(z)$ using the best fit values to SNIa+CMB+BAO$_{dz}$ data (left plot) and to SNIa+CMB+BAO$_{A}$ data (right plot) of Tables \ref{tableFitBAOdz} and \ref{tableFitBAOA}. In both plots we include the constant CDM value $\delta_c=\frac{3}{20}(12\pi)^{2/3}\simeq 1.686$ (horizontal dotted line) as well as the $\CC$CDM curve (solid points, in green). The $\delta_c(z)$ curves for the vacuum models A1 and A2 are represented with squares (in red) and with  diamonds (in blue), respectively. The corresponding values at $z=0$ define $\delta_c$ for each model, and are indicated in the last column of Tables \ref{TableModelsBAOdz} and \ref{TableModelsBAOA}.} \label{deltac BAOdz}}
\end{figure}

%%%%%%%%%%%%%%%%%%%%%%%%%%%%%%%%%%%%%%%%%%%%%%%%%%%%%%%%%%%%%%%%%%%%%%%%

\noindent where the primes denote derivatives with respect to the
scale factor and
\begin{equation}
\Psi(a)=-\frac{\dot{\rho}_\Lambda}{\rho_m}=3H(a)(1-\xi).
\end{equation}
\noindent The formulas for the non-relativistic matter energy
density $\rmr$ and the Hubble function $H$ can be found in Sect.
\ref{subsec:solvingA1A2}.  The numerical solution of the above nonlinear equation is used to compute $\delta_c(z)$ for models A1 and A2 in Fig.\,\ref{deltac BAOdz}, see Sect.\,\ref{sec:numericaldeltac} for details.

\section{Type-B models}

For this type of models the nonlinear equation for the perturbations
can be obtained with some extra effort since on this occasion the
calculations cannot be performed analytically in terms of the scale
factor. We write the final result using the variable $y$, which has
been defined in \eqref{eq:defx}. We find:
$$\frac{9}{16}H_0^2\mathcal{F}^2(1-y)^2\delta_m^{\prime\prime}+\frac{3}{4}H_0\mathcal{F}\delta_m^\prime(1-y^2)\left[2H+\Psi-\frac{3}{2}yH_0\mathcal{F}\right]+$$
$$+\left[2H\Psi+\frac{3}{4}H_0\mathcal{F}(1-y^2)\Psi^\prime-\frac{\rho_m}{2}(1+\delta_m)\right]\delta_m-\frac{\Psi^2\delta_m^2}{3(1+\delta_m)}-$$
\begin{equation}\label{eq:SDE2}
-\left[\frac{4\left(\frac{3}{4}H_0\mathcal{F}(1-y^2)\delta_m^\prime\right)^2+
\frac{15}{4}\Psi H_0\mathcal{F}(1-y^2)\delta_m\delta^\prime_m}{3(1+\delta_m)}\right]=0\,,
\end{equation}
where the primes indicate on this occasion derivatives with respect to $y$ -- the notation should not be confusing with the previous use of primes since we make explicit the argument. The expressions for $\Psi(y)$, $\rho_m(y)$ and $H(y)$ for the type-B models can be found in Sect. \ref{sec:perturbationsTypeB}, and the formula for the constant $\mathcal{F}$ is given by (\ref{defmathF}). The numerical solution of the above nonlinear equation is used to compute $\delta_c(z)$ for models B1 and B2 in Fig.\,\ref{deltac BAOA}, cf. Sect. \ref{sec:numericaldeltac}.

The corresponding nonlinear
equation for type-C1 models is a particular case of
Eq.\,(\ref{eq:SDE2}) and is obtained as indicated in
Sect.\,\ref{subsec:solvingC1C2}. In particular, for $\epsilon=\OLo $
and $\nu=0$ (hence $\mathcal{F}=\OLo$) we obtain the corresponding
equation for the pure linear model $\rL\propto H$, which we have
ruled out. We shall not consider the computation of the number counts for these models here.

\section{Numerical procedure to determine $\delta_c$}
\label{sec:numericaldeltac}

%%%%%%%%%%%%%%%%%%%%%%%% FIGURE  2 %%%%%%%%%%%%%%%%%%%%%%%%%%%%%%%%%%%%%

\begin{figure}[t!]
\begin{center}
{\includegraphics[scale=0.50]{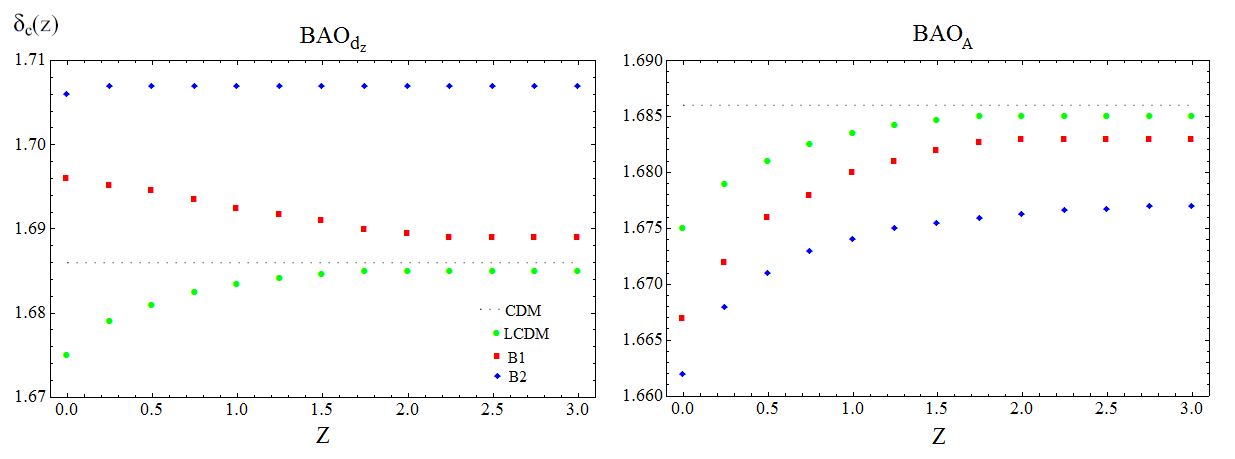}}
\end{center}
\caption[$\delta_c(z)$ for type-B models obtained from Tables \ref{tableFitBAOdz} and \ref{tableFitBAOA}]{\scriptsize{Computation of the collapse density threshold function $\delta_c(z)$} for the B1 and B2 vacuum models obtained from the fitting values of Tables \ref{tableFitBAOdz} and \ref{tableFitBAOA}. The rest of the notation is as in Fig.\,\ref{deltac BAOdz}. The values of $\delta_c(z)$ at $z=0$ are indicated in the last column of Tables \ref{TableModelsBAOdz} and \ref{TableModelsBAOA}. \label{deltac BAOA}}
\end{figure}

%%%%%%%%%%%%%%%%%%%%%%%%%%%%%%%%%%%%%%%%%%%%%%%%%%%%%%%%%%%%%%%%%%%%%%%%

Next we follow the prescriptions of \cite{Pace10}, which was also
described in detail in
\cite{Grande2011}. We compute $\delta_c(z_f)$  by numerically
integrating the above nonlinear equations between $z_i$ and $z_f$
(where the initial redshift $z_i$ is sufficiently large, for
instance $10^{6}$). The aim is to find the initial value
$\delta_m(z_i)$ for which the collapse takes place at $z=z_f$, i.e.
such that $\delta_m(z_f)$ is very large, say $10^5$ or $10^9$ (the
result does not change significantly). Second, we use the previously
determined value of $\delta_m(z_i)$ together with a small value of  $\delta'_m(z_i)$. In fact, we know it is zero for a sphere, so we may take $\delta'_m(z_i)\sim 10^{-6}-10^{-4}$\,\cite{Pace10}. These are then used as the initial conditions for solving the corresponding linear perturbations equations. The value
of $\delta_m(z_f)$ obtained in the second step of this procedure
defines $\delta_c(z_f)$, and the value of this quantity at $z_f=0$
defines $\delta_c\equiv\delta_c(z_f=0)$.  Notice that the linear
equations are simply obtained from  (\ref{eq:SDE}) and (\ref{eq:SDE2}) upon neglecting all terms of ${\cal O}(\delta_m^2,\delta_m'^2,\delta_m\delta_m')$. The equations obtained in this way are, of course,
the ones already presented in Sect.\,\ref{sec:perturbations} for
both types of models A and B.  The values of $\delta_c$ obtained by
this method for each model are displayed in the last column of Tables \ref{TableModelsBAOdz} and \ref{TableModelsBAOA}
(cf. Sect. \ref{subsec:HaloMasFunction}). The numerical solutions $\delta_c(z)$ for each model are displayed in Figs.\,\ref{deltac BAOdz} and \ref{deltac BAOA}.
\newpage

\chapter{Cosmological observables and statistical analysis}
\label{chap:App5}

In this Appendix we explicitly define and discuss the various observables used in the phenomenological analysis of Chapters \ref{chap:PRDbased} and \ref{chap:H0tension}. In fact, the information presented in this appendix is also valid for the analyses of Chapters \ref{chap:AandGRevisited} and \ref{chap:MPLAbased}, with two slight differences: i) in these two chapters we used the data of Ref. \cite{GilMarin2OLD} instead of Ref. \cite{GilMarin2}; and ii) In Chapter \ref{chap:AandGRevisited} we did not marginalize the SNIa M parameter \cite{BetouleJLA} (see below for details). In addition, we also collect some basic (essentially technical) information about the fitting procedure that has been followed, as e.g. the correlation matrices entering the complete $\chi^2$ function \eqref{chi2s} or the precise values (with the corresponding uncertainties) of the BAO and CMB data sets. These aspects of the analysis were omitted in the main body of the thesis.

%%%%%%%%%%%%%%%%%%%%%%%%%%%%%%%%%
%%%%%%%%%%%%%%%%%%%%%%%%%%%%%%%%%

Each of the $\chi^2_s$ functions in the total $\chi^2$ to be minimized can be written as follows,
\be
\chi^2_s(\vec{p})=[\vec{x}_s(\vec{p})-\vec{d}_s]^{T}C_s^{-1}[\vec{x}_s(\vec{p})-\vec{d}_s]\,,
\ee
where the data set vector $\vec{d}_s$ runs over all the data sets DS1-DS6 described in detail in Sect. \ref{sect:DataSets}. They are labeled here in a compact way as $s$=(SNIa, BAO, LSS, BAO$\&$LSS, H, CMB). In the particular case of DS3, part of the data are from \cite{GilMarin2} and in this case BAO and LSS are correlated. We have denoted this situation with  BAO$\&$LSS: it reflects the contribution from the combined BAO+LSS covariance matrices for the LOWZ and CMASS data samples from that reference, see \eqref{LOWZmatrix} and \eqref{CMASSmatrix} below. Similarly $\vec{x}_s$ are the vectors containing the theoretically computed values from the different models; and $C_s$ is the covariance matrix for each data set $s$, which can be constructed from the correlation matrix $\rho_s$ and the $1\sigma$ uncertainties according to $C_{s,ij}=\rho_{s,ij}\sigma_{s,i}\sigma_{s,j}$.

%%%%%%%%%%%%%%%%%%%%%%%%%%%%%%%%%
%%%%%%%%%%%%%%%%%%%%%%%%%%%%%%%%%

\section{SNIa}
For the SNIa sector, we have used the binned distance modulus fitted to the JLA sample shown in Table F.1 of \cite{BetouleJLA}, together with the $31\times31$ covariance matrix presented in Table F.2 of the same paper. The fitted quantity is the distance modulus, defined as
\be
{\cal \mu}(z,\vec{p})=5\log{d_{L}(z,\vec{p})}+M\,,
\ee
where $M$ is a free normalization parameter (see \cite{BetouleJLA} for details) and $d_{L}(z,\vec{p})$ is the luminosity distance \eqref{eq:LumDist}. Parameter $M$ is treated as a nuisance parameter and, therefore, it can be integrated out through the corresponding marginalization of the SNIa likelihood. It can be done analytically, so this also helps to improve the computational efficiency by reducing in one dimension the fitting parameter space. After this marginalization, the resulting effective $\chi^2$-function for this data set reads,
\begin{equation}\label{chi2SNIa}
\begin{small}
\chi^2_{SNIa}(\vec{p})=\vec{y}(\vec{p})^{T}J\vec{y}(\vec{p})-\frac{[\sum_{i,j=1}^{31} J_{ij}y_{i}(\vec{p})]^2}{\sum_{i,j=1}^{31} J_{ij}}\,,
\end{small}
\end{equation}
where $J\equiv C_{SNIa}^{-1}$ and $y_i(\vec{p})\equiv 5\log{d_{L}(z_i,\vec{p})}-{\cal \mu}_{\rm obs,i}$. \eqref{chi2SNIa} is the direct generalization of formula (13.16) of \cite{BookAmendolaTsujikawa}, since we are now considering a non-diagonal covariance matrix, instead of a diagonal one. Let us prove \eqref{chi2SNIa}. The original SNIa likelihood can be written as follows,
\begin{equation}
\mathcal{L}_{SNIa}(\vec{p},M)=\mathcal{L}_0\,e^{-\frac{1}{2}[y_i(\vec{p})+M]J_{ij}[y_j(\vec{p})+M]}\,,
\end{equation}
where we have assumed the Einstein summation convention in order to write the result more concisely, and $\mathcal{L}_0$ is the normalization constant. Let us now marginalize this distribution over the parameter $M$,
\begin{equation}
\tilde{\mathcal{L}}(\vec{p})\equiv \int_{-\infty}^{+\infty}\mathcal{L}_{SNIa}(\vec{p},M)\,dM\,.
\end{equation}
This can be expressed as
\begin{equation}\label{demo0}
\tilde{\mathcal{L}}(\vec{p})={\mathcal{L}}_0 e^{-\frac{1}{2}B_0(\vec{p})}F(\vec{p})\,,
\end{equation}
with
\begin{equation}
F(\vec{p})\equiv \int_{-\infty}^{+\infty}e^{-\frac{B_1}{2}M^2-MB_2(\vec{p})}\,dM\,,
\end{equation}
and
\begin{equation}
\begin{array}{rcl}
B_0 (\vec{p}) &\equiv &  \sum_{i,j=1}^{31}y_{i}(\vec{p})J_{ij}y_{j}(\vec{p})\,,\\
B_1 & \equiv & \sum_{i,j=1}^{31}J_{ij}\,, \\
B_2(\vec{p}) & \equiv & \sum_{i,j=1}^{31}J_{ij}y_{i}(\vec{p})\,.
\end{array}
\end{equation}
The problem is basically reduced to compute the function $F(\vec{p})$ appearing in \eqref{demo0}. Let us perform the derivative of the latter with respect to $M$ and the i{th}-component of the fitting vector $\vec{p}$,
\begin{equation}\label{demo1}
\frac{\partial F}{\partial M}=0=-\int_{-\infty}^{+\infty}\left[MB_1+B_2({\vec{p}})\right]e^{-\frac{B_1}{2}M^2-MB_2(\vec{p})}\,dM\,,
\end{equation}
\begin{equation}\label{demo2}
\frac{\partial F}{\partial p_i}= -\frac{\partial B_2(\vec{p})}{\partial p_i}\int_{-\infty}^{+\infty}M\,e^{-\frac{B_1}{2}M^2-MB_2(\vec{p})}\,dM\,.
\end{equation}
Combining these two expressions, we obtain a differential equation for $F(\vec{p})$,
$$\frac{1}{F(\vec{p})}\frac{\partial F}{\partial p_i}=\frac{\partial B_2}{\partial p_i}\frac{B_2(\vec{p})}{B_1}\,,$$
whose solution is very simple:
\begin{equation}
F(\vec{p})\propto e^{\frac{B_2^2(\vec{p})}{2B_1}}\,.
\end{equation}
The integration constant is irrelevant since it can be absorbed by the normalization factor $\mathcal{L}_0$ in \eqref{demo0}, giving rise to a new one, $\tilde{\mathcal{L}}_0$. Thus, the marginalized likelihood reads,
\begin{equation}
\tilde{\mathcal{L}}(\vec{p})=\tilde{\mathcal{L}}_0\,e^{-\frac{\chi^2_{SNIa}(\vec{p})}{2}}\,,
\end{equation}
with $\chi^2_{SNIa}(\vec{p})$ given by \eqref{chi2SNIa}.

%%%%%%%%%%%%%%%%%%%%%%%%%%%%%%%%%
%%%%%%%%%%%%%%%%%%%%%%%%%%%%%%%%%

\section{BAO and LSS}
Galaxy surveys measure angular and redshift distributions of galaxies. Using these data, they are able to obtain the matter power spectrum $P(\vec{k},z)$ in redshift space upon modeling the bias factor and transforming these angles and redshifts into distances.
However, the distance of two sources along the line of sight, being
given by the difference of redshift $\Delta z$, depends on the cosmological model. Similarly, the distance of two sources in the direction perpendicular to the line of sight is given by the angular separation, $\Delta\theta$, and also depends on the cosmological model. Therefore, it is convenient to first define a fiducial model to which one refers the predictions of the new models being studied. Usually such model is the $\CC$CDM with appropriately chosen values of the free parameters. We have already used a fiducial model in our analysis, cf. sect. \ref{sect:FiducialModel}, but different papers may use slightly different fiducial parameters and one has to take this feature carefully into account.
If that is not enough, one has to disentangle the effect of the redshift space distortions that are due to the peculiar velocities of galaxies falling into the gravitational potential wells. The comoving wavevector $\vec{k}$ is usually decomposed in a parallel to the line of sight component, $k_{\parallel}$, and another one which is perpendicular to it, $k_{\perp}$. These two directions determine the two-dimensional (2D) space in which the distribution of galaxies is studied. The analysis of the BAO signal in the 2D power spectrum carries precious information about the angular and longitudinal size of the BAO standard ruler at the measured redshift $z$, namely

\be\label{eq:BAO1}
\Delta\theta_s (z)=\frac{a r_s(a_{d})}{D_A(a)}=\frac{r_s(z_{d})}{(1+z)D_A(z)}\,,
\ee
where $a=1/(1+z)$ is the scale factor, and
\be\label{eq:BAO2}
\Delta z_s (z)= \frac{r_s(z_{d})H(z)}{c}\,,
\ee
$r_s(z_d)$ being the sound horizon at the redshift of the drag epoch (see below), with $D_A$ the proper diameter angular distance,
\begin{equation}\label{SomDist}
D_A(z,\vec{p}) \equiv \frac{c}{(1+z)}\int^{z}_{0}\frac{dz'}{H(z')}\,.
\end{equation}
In principle, galaxy surveys are able to extract anisotropic BAO information from their analyses. That is to say, they can afford constraints on the quantities $r_s(z_d)/D_A(z)$ and $H(z)r_s(z_d)$ through the perpendicular and parallel dilation scale factors \cite{GilMarin2,Delubac2015,Aubourg2015},
\begin{eqnarray}\label{DilaScales}
\alpha_\perp &\equiv & \frac{r^{fid}_s(z_d)/D^{fid}_A(z)}{r_s(z_d)/D_A(z)}=\frac{D_A(z)r^{fid}_s(z_d)}{D^{fid}_A(z)r_s(z_d)}\nonumber\\ \alpha_\parallel &\equiv& \frac{H^{fid}(z)r^{fid}_s(z_d)}{H(z)r_s(z_d)}\,,
\end{eqnarray}
where $r^{fid}_s(z_d)$, $D^{fid}_A(z)$ and $H^{fid}(z)$ are obtained in the fiducial cosmology that the survey has used to convert redshift and angles into distances. The quantity $r_s(z_d)/D_A(z)$ is the sound horizon at the drag epoch measured in units of the angular diameter distance at redshift $z$, and similarly $r^{fid}_s(z_d)/D^{fid}_A(z)$ is the corresponding value in the fiducial model. These ratios are just the comoving angular sizes (\ref{eq:BAO1}), i.e. the angular sizes divided by the scale factor.  Similarly, $H(z)r_s(z_d)$ and $H^{fid}(z)r^{fid}_s(z_d)$ are, respectively, the sound horizon in units of the Hubble horizon in the given model and in the fiducial model.

Sometimes the surveys are not equipped with enough BAO data to obtain \eqref{eq:BAO1} and \eqref{eq:BAO2} separately. In this case they limit themselves to compute the 
volume-averaged spectrum in a conveniently defined volume $V$ and obtain a constraint which is usually encapsulated in the isotropic BAO estimator $r_s(z_d)/D_V(z)$ (sometimes denoted $d_z$\,\cite{Blake11}). Here $D_V(z)$ is an effective distance or dilation scale\,\cite{Eisenstein2005} obtained from the cubic root of the volume $V$ (whose value is defined from the square of the transverse dilation scale times the radial dilation scale):
\begin{equation}\label{eq:DV}
D_V(z,\vec{p}) = \left[zD_H(z,\vec{p})D^{2}_M(z,\vec{p})\right]^{1/3}\,.
\end{equation}
In this expression, $D_M=(1+z) D_A$ is the comoving angular diameter distance (playing the role of the transverse dilation scale at redshift $z$)
and $D_H$ is the Hubble radius
\begin{equation}\label{SomDist2}
D_H(z,\vec{p}) \equiv \frac{c}{H(z)},
\end{equation}
with $zD_H$ acting as the radial dilation scale at redshift $z$. Therefore, the distilled BAO estimator $d_z$  measures the sound horizon distance at the drag epoch in units of the effective dilation scale \eqref{eq:DV}.

Again a comparison with a fiducial value is necessary. The isotropic BAO parameter relating the fiducial model value of $D_V$ with the actual value of a given model is defined in \cite{Beutler2011} as
\begin{equation}\label{CombDilaScales0}
\alpha\equiv\frac{D_V(z)}{D^{fid}_V(z)}\,.
\end{equation}
In other cases $\alpha$ is defined in terms of $d_z$,
\begin{equation}\label{CombDilaScales}
\alpha\equiv \frac{r^{fid}_s(z_d)/D^{fid}_V(z)}{r_s(z_d)/D_V(z)}= \frac{D_V(z)r^{fid}_s(z_d)}{D^{fid}_V(z)r_s(z_d)}\,,
\end{equation}
as in \cite{Ross,Kazin2014}. These isotropic BAO estimators (\ref{CombDilaScales0} and \ref{CombDilaScales}) are akin to the Alcock-Paczynski (AP) test \cite{AP1979}, in which the ratio of the observed radial/redshift to the angular size at different redshifts, $\Delta z_s/\Delta\theta_s$, is used to obtain cosmological constraints on the product $H(z)D_M(z)$ and hence of $H(z)D_A(z)$. Both for the AP test and the isotropic BAO measurement, the value of the Hubble parameter $H(z)$ cannot be disentangled from $D_A(z)$, there is a degeneracy. This is in contradistinction to the situation with the anisotropic BAO parameters (\ref{DilaScales}), in which a measurement of both $\alpha_\perp$ and $\alpha_\parallel$ permits to extract the individual values of  $H(z)$ and $D_A(z)$.

%%%%%%%%%%%%%%%%%%%%%%%%%%%%%%%%%%%%%%%%%%%%%%%%%%%%%%%%%%%%%%%%%%%%%%%%%%%
%
\begin{table}[t!]
\begin{center}
\begin{tabular}{| c | c | c |}
\multicolumn{1}{c}{Model} &  \multicolumn{1}{c}{$r_s(z_d)_{ff}$} & \multicolumn{1}{c}{$r_s(z_d)_{{\rm Boltz}}$}
\\\hline
$\CC${CDM} & $150.85\pm0.24$ & $147.32\pm0.22$
\\\hline
XCDM & $151.02\pm0.23$ & $147.48\pm0.22$
\\\hline
RVM & $153.34\pm0.74$ & $148.25\pm0.33$
\\\hline
$Q_{dm}$ & $153.34\pm0.72$ & $148.25\pm0.34$
\\\hline
$Q_{\CC}$ & $151.39\pm0.32$ & $147.32\pm0.21$
\\\hline
\end{tabular}
\caption[Values of the sound horizon distance at the drag epoch for different models under study in Chapter \ref{chap:PRDbased}, and using two different fitted formulas]{{\scriptsize The obtained values of the sound horizon distance at the drag epoch for the different models under study in Chapter \ref{chap:PRDbased}, $r_s(z_d)$ (in Mpc), using \eqref{comdist}-\eqref{zdrag} in the first column and \eqref{rsBoltz} in the second.
We show the central values and associated uncertainties of $r_s(z_d)$ for each model, which have been obtained using the posterior distributions of the analysis in Table \ref{tableFit1PRD}.}\label{tableSound}}
\end{center}
\end{table}
%
%%%%%%%%%%%%%%%%%%%%%%%%%%%%%%%%%%%%%%%%%%%%%%%%%%%%%%%%%%%%%%%%%%%%%%%%%%%

%%%%%%%%%%%%%%%%%%%%%%%%%%%%%%%%%%%%%%%%%%%%%%%%%%%%%%%%%%%%%%%%%%%%%%%%%%%
\begin{table}
\begin{center}
\resizebox{1\textwidth}{!}{
\begin{tabular}{| c | c |c | c | c | c | c | c | c | c | c|}
\multicolumn{1}{c}{Model} &  \multicolumn{1}{c}{$h$} &  \multicolumn{1}{c}{$\omega_b= \Omega_b h^2$} & \multicolumn{1}{c}{{\small$n_s$}}  &  \multicolumn{1}{c}{$\Omega_m$}&  \multicolumn{1}{c}{{\small$\nu_i$}}  & \multicolumn{1}{c}{$\omega_0$} & \multicolumn{1}{c}{$\omega_1$} &
\multicolumn{1}{c}{$\chi^2_{\rm min}/dof$} & \multicolumn{1}{c}{$\Delta{\rm AIC}$} & \multicolumn{1}{c}{$\Delta{\rm BIC}$}\vspace{0.5mm}
\\\hline
{\small $\CC$CDM} & $0.692\pm 0.004$ & $0.02253\pm 0.00013$ &$0.974\pm 0.004$& $0.296\pm 0.004$ & - & -1 & - & 84.83/85 & - & -\\
\hline
XCDM  &  $0.672\pm 0.007$& $0.02262\pm 0.00014 $&$0.976\pm0.004$& $0.311\pm 0.007$& - & $-0.923\pm0.023$ & - & 74.15/84 & 8.54 & 6.19  \\
\hline
CPL  &  $0.674\pm 0.009$& $0.02263\pm 0.00014 $&$0.976\pm0.004$& $0.310\pm 0.009$& - & $-0.946\pm0.085$ & $0.070\pm0.250$ & 74.08/83 & 6.31 & 1.77 \\
\hline
RVM  & $0.682\pm 0.004$& $0.02240\pm 0.00014$&$0.968\pm 0.004$& $0.296\pm 0.004$ & $0.00152\pm 0.00038$ & -1 & - & 66.95/84 & 15.74 & 13.39 \\
\hline
$Q_{dm}$ &  $0.683\pm 0.004$& $0.02238\pm 0.00014 $&$0.967\pm0.004$& $0.296\pm 0.004 $ & $0.00217\pm 0.00055 $ & -1 & - &  66.99/84  & 15.70 & 13.35 \\
\hline
$Q_\CC$  &  $0.693\pm 0.004$& $0.02226\pm 0.00016 $&$0.964\pm0.005$& $0.296\pm 0.004$ & $0.00785\pm 0.00256$ & -1 & - &  75.56/84 & 7.13 & 4.78 \\
\hline \end{tabular}
 }
\end{center}
\caption[Fitting results for the various models under study in Chapter \ref{chap:PRDbased} using \eqref{rsBoltz}, instead of \eqref{comdist}-\eqref{zdrag}, in the computation of the comoving sound horizon at the drag epoch]{{\scriptsize Fitting results for the various models under study in Chapter \ref{chap:PRDbased} using \eqref{rsBoltz}, instead of \eqref{comdist}-\eqref{zdrag}, in the computation of the comoving sound horizon at the drag epoch. It is noteworthy that in the RVM case we find $\nu>0$ at exactly $4\sigma$ c.l.}\label{tableFitXIII}}
\end{table}
%
%%%%%%%%%%%%%%%%%%%%%%%%%%%%%%%%%%%%%%%%%%%%%%%%%%%%%%%%%%%%%%%%%%%%%%%%%%%

In our fitting analysis, we use data of both types of BAO estimators: isotropic and anisotropic ones, see the data set items DS2 and DS3 in Sect. \ref{sect:DataSets}. The anisotropic BAO data contains more information than the isotropic one because the aforementioned degeneracy between $H(z)$ and $D_A(z)$ has been broken. Thus, anisotropic BAO is richer and yields stronger cosmological constraints \cite{ShojiJeongKomatsu2009,Anderson2013}.

To compute the theoretical predictions for the above BAO estimators, we need some extra formulas. For instance, the sound horizon at redshift $z$, i.e. $r_s(z)$, is given by the expression
\begin{equation}\label{comdist}
r_s(z) = \int^{\infty}_{z}\frac{c_s(z^\prime)dz^\prime}{H(z^\prime)}\,,
\end{equation}
where
\begin{equation}\label{soundSpeed}
c_s(z) = \frac{c}{\sqrt{3(1+\mathcal{R}(z)})}\,
\end{equation}
is the sound speed in the photo-baryon plasma and $\mathcal{R}(z)=\frac{\delta\rho_b(z)}{\delta\rho_\gamma(z)}$. For all the  models studied in Chapter \ref{chap:PRDbased} the energy densities for radiation and baryons evolve in the standard way. Thus, this function takes the usual form $\mathcal{R}(z) = (3\Omega_b/4\Omega_\gamma)/(1+z)$, being $\Omega_\gamma$ the photon density parameter. The redshift at the drag epoch is $z_d\sim \mathcal{O}(10^{3})$. Below such redshift value the baryon perturbations effectively decouple from the photon ones and start to grow with dark matter perturbations. The precise value of $z_d$ depends in a complicated way on the different cosmological parameters. The fitted formula for determining it is given in \cite{EisensteinHu98}:
\begin{align}\label{zdrag}
&z_d = \frac{1291\,\omega_m^{0.251}}{1 + 0.659\,{\omega_m}^{0.828}}\left[1 + \beta_1{\omega_b}^{\beta_2}\right]\nonumber\\
&\beta_1 = 0.313\,{\omega_m}^{-0.419}\left[1 + 0.607\,{\omega_m}^{0.674}\right]\\
&\beta_2 = 0.238\,{\omega_m}^{0.223}\nonumber\,.
\end{align}
Let us point out that many experimental groups do not make use of these fitted formulas (denoted with a subindex ff below), but of the complete set of Boltzmann (Boltz) equations, which must be solved numerically. Both approaches can give rise to differences that are around $2-3\%$. Consequently, if the observational values are computed with a Boltzmann code, then we need to correct them in order to perform the comparison with our theoretical predictions, which make use of the above expressions. In these cases we follow the same procedure applied in \cite{Kazin2014} based on re-scaling the observational data by $f_{r}\equiv r_s(z_d)^{fid}_{ff}/r_s(z_d)^{fid}_{{\rm Boltz}}$. The two quantities involved in this ratio are computed using the same fiducial $\Lambda$CDM cosmology chosen by the observational teams, but while the first is obtained with the fitted formula \eqref{zdrag} together with \eqref{comdist} and \eqref{soundSpeed}, the second is obtained with a Boltzmann code. One can also compute the latter using the existing approximated formulas for the sound horizon at $z_d$ that are obtained by fitting the data extracted from e.g. the CAMB code \cite{CAMB}, since they are very accurate, with errors less than the 0.1\% \cite{Anderson2013}:

\begin{equation}\label{rsBoltz}
r_s(z_d)_{\rm Boltz}=\frac{55.234(1+\Omega_\nu h^2)^{-0.3794}\,{\rm Mpc}}{(\Omega_{dm}h^2+\Omega_b h^2)^{0.2538}(\Omega_b h^2)^{0.1278}}\,,
\end{equation}
where $\Omega_\nu h^2$ is the neutrino reduced density parameter and depends, of course, on the effective number of neutrino species and the photon CMB temperature, which is taken from \cite{Fixsen2009}, see also Sect. \ref{sect:Background}.

We point out  that the aforementioned re-scaling does not change the values of the dilation scale factors \eqref{DilaScales}, \eqref{CombDilaScales0} and \eqref{CombDilaScales} furnished by the experimental teams. If they, for instance, deliver their results through the product $\alpha_\perp(D^{fid}_A/r_s(z_d)^{fid}_{{\rm Boltz}})$, then we just multiply it  by $f_r^{-1}$ in order to keep our fitting procedure consistent. Note that we are also using the fitting formulas \eqref{zlastscattPRD} and \eqref{zlastscatt1PRD} in the computation of the decoupling redshift $z_*$ (see below, Sect. \ref{sec:CMBobsApp}). For the same reason, we re-scale $(H^{fid}r_s(z_d)^{fid}_{{\rm Boltz}})/\alpha_\parallel$ on multiplying it by $f_r$. But we stress once more that the dilation scale factors, which are the fundamental outputs extracted from the analysis of the BAO signal, are not modified by this re-scaling.

We have studied the changes induced in our fitting results when we use \eqref{rsBoltz} instead of \eqref{comdist}-\eqref{zdrag}. We show the corresponding values of the sound horizon at the drag epoch, $r_s(z_d)$, in Table \ref{tableSound}, and the corresponding new fitting results for the various models in Table \ref{tableFitXIII}. As it can be seen, they are close to the original ones shown in Table \ref{tableFit1PRD}. The statistical significance of the new results is exactly $4\sigma$ for the RVM and very near to it ($3.94\sigma$) also for the $Q_{dm}$. The $\Delta$AIC and $\Delta$BIC values of these models are above $15$, which denote very strong evidence in favor of these models as compared to the $\CC$CDM. The central fit values did not undergo significant changes with respect to Table \ref{tableFit1PRD}, which again reflects the robustness of our results and conclusions.

We use data from the 6dFGS survey \cite{Beutler2011}, $D_V(z=0.106)=(456\pm 27){\rm Mpc}$. Also from the SDSS DR7 MGS one \cite{Ross}, $D_V(z=0.15)\left(\frac{r_s(z_d)^{fid}}{r_s(z_d)}\right) = (664\pm25){\rm Mpc}$. For the data of the WiggleZ Dark Energy survey \cite{Kazin2014},
\begin{equation}
\begin{array}{lcl}
D_V(0.44)\left(\frac{r_s(z_d)^{fid}}{r_s(z_d)}\right) & = & (1716.4\pm83.1)\,{\rm Mpc}\\
D_V(0.60)\left(\frac{r_s(z_d)^{fid}}{r_s(z_d)}\right) & = & (2220.8\pm100.6)\,{\rm Mpc}\\
D_V(0.73)\left(\frac{r_s(z_d)^{fid}}{r_s(z_d)}\right) & = & (2516.1\pm86.1)\,{\rm Mpc}\,.
\end{array}
\end{equation}
We have taken also into account the correlations between the different measurements through the corresponding inverse covariance matrix,
\begin{equation}
C^{-1}_{\rm BAO/WZ} =
b\begin{small}
 \begin{pmatrix}
  2.17898878 & -1.11633321 & 0.46982851  \\
  -1.11633321 & 1.70712004 & -0.71847155 \\
  0.46982851  & -0.71847155  & 1.65283175
 \end{pmatrix}\,,
\end{small}
\end{equation}
with $b=10^{-4}\,{\rm Mpc}^{-2}$.
\newline
\newline
As explained in the data set item DS3) of Chapter \ref{chap:PRDbased}, we also use BOSS LSS and anisotropic BAO data, more concretely from the LOWZ ($z=0.32$) and CMASS ($z=0.57$) samples\,\cite{GilMarin2}. For each of these redshifts the authors provide three data points, $f(z)\sigma_8(z)$, $D_A(z)/r_s(z_d)$ and $H(z)r_s(z_d)$, which are extracted by a thorough study of RSD measurements of the power spectrum combined with the bispectrum, and the BAO post-reconstruction analysis of the power spectrum. For the LOWZ sample, we have
\begin{equation}
\begin{array}{lcrcl}
f\sigma_8 & = & 0.42660 &\pm& 0.05627 \\
Hr_s(z_d)\cdot 10^{-3} & = & f_r\,(11.549 &\pm& 0.385)\,{\rm km/s}\\
D_A/r_s(z_d) & = & f_r^{-1}(6.5986 &\pm&0.1337)\,,
\end{array}
\end{equation}
and for the CMASS sample,
\begin{equation}
\begin{array}{lcrcl}
f\sigma_8 & = & 0.42613 &\pm& 0.02907 \\
Hr_s(z_d)\cdot 10^{-3} & = & f_r\,(14.021 &\pm& 0.225)\,{\rm km/s}\\
D_A/r_s(z_d) & = & f_r^{-1}(9.3869 &\pm&0.1030)\,,
\end{array}
\end{equation}
where $f_r=(151.79 {\rm Mpc}/148.11 {\rm Mpc})$ is the sound horizon re-scaling introduced above. We need to apply such rescaling because the authors of \cite{GilMarin2} compute $r_s(z_d)^{fid}$ with the Boltzmann code, whereas we make use of the fitting formula \eqref{zdrag}. The corresponding covariance matrices read, respectively,
\begin{equation}\label{LOWZmatrix}
C_{\rm LOWZ} =10^{-3}
 \begin{pmatrix}
  3.1667 & 14.726f_r & 5.0871f_r^{-1}  \\
  14.726f_r & 148.099f_r^2 & 28.929 \\
  5.0871f_r^{-1}  & 28.929  & 17.883f_r^{-2}
 \end{pmatrix}\,,
\end{equation}
and
\begin{equation}\label{CMASSmatrix}
C_{\rm CMASS} =10^{-3}
 \begin{pmatrix}
  0.84506 & 4.3722f_r & 2.0151f_r^{-1}  \\
  4.3722f_r & 50.717f_r^2 & 13.827 \\
  2.0151f_r^{-1} & 13.827  & 10.613f_r^{-2}
 \end{pmatrix}\,.
\end{equation}
We also use the data from the combined LyaF analysis at $z=2.34$ presented in \cite{Delubac2015},
\begin{equation}
\begin{array}{lcrcl}
D_A/r_s(z_d) & = & 10.93^{+0.35}_{-0.34}\cdot f_r^{-1} \\
D_H/r_s(z_d) & = & 9.15^{+0.20}_{-0.21}\cdot f_r^{-1}\,,
\end{array}
\end{equation}
with $f_r=(153.4 {\rm Mpc}/149.7 {\rm Mpc})$ and their correlation coefficient, $\rho_{12}=-0.48$ \cite{Aubourg2015}.

%%%%%%%%%%%%%%%%%%%%%%%%%%%%%%%%%
%%%%%%%%%%%%%%%%%%%%%%%%%%%%%%%%%
\section{H(z)}
The covariance matrix for the $H(z)$ data is diagonal and therefore, $C_{H,ij}=\sigma_{H,i}^2\delta_{ij}$, (see DS4 in Chapter \ref{chap:PRDbased} and Table \ref{compilationH}). The possibility of having non zero off-diagonal terms in the correlation matrix is not excluded. But these coefficients are not found in the literature, and as we have explained in the Discussion section of Chapter \ref{chap:PRDbased}, the $H(z)$ data set plays a completely secondary role in comparison to the BAO, CMB and LSS sets. The impact of introducing these (unknown) correlations would be negligible. Thus, the corresponding $\chi^2$-function adopts the following simple form,
\begin{equation}\label{chi2H}
\chi^{2}_{\rm H}(\vec{p})=\sum_{i=1}^{30} \left[ \frac{ H(z_{i},\vec{p})-H_{\rm obs}(z_{i})}
{\sigma_{H,i}} \right]^{2}\,.
\end{equation}
The theoretical values for the DVM's are computed with equations \eqref{HRVM}-\eqref{HQL}, and the ones for the XCDM and CPL parametrizations with \eqref{eq:HXCDM} and \eqref{Hzzzquint}, respectively. The observed points and their uncertainties are presented in Table \ref{compilationH}.

%%%%%%%%%%%%%%%%%%%%%%%%%%%%%%%%%
%%%%%%%%%%%%%%%%%%%%%%%%%%%%%%%%%

\section{CMB}
\label{sec:CMBobsApp}

As explained in DS6) of Chapter \ref{chap:PRDbased}, we have made use of the CMB data and correlation matrix of the Planck 2015 TT,TE,EE+lowP analysis presented in \cite{Huang}. The data points are the following:
\begin{equation}
\begin{array}{lcrcl}
R & = & 1.7448 &\pm& 0.0054 \\
l_a  & = & 301.460 &\pm& 0.094\\
 \omega_b & = & 0.02240 &\pm& 0.00017\\
n_s &=& 0.9680 &\pm& 0.0051\,,
\end{array}
\end{equation}
and the corresponding correlation matrix:
\begin{equation}
\rho_{cmb} =
 \begin{pmatrix}
  1 & 0.53 & -0.73 & -0.80 \\
  0.53 & 1 & -0.42 & -0.43 \\
  -0.73  & -0.42  & 1 & 0.59  \\
  -0.80 & -0.43 & 0.59 & 1
 \end{pmatrix}\,.
\end{equation}
The theoretical expression for the $R$ ``shift parameter'' is given by
\begin{equation}\label{shiftparameter}
R(\vec{p})=\sqrt{\Omega_{m}}\int_{0}^{z_{*}} \frac{dz}{E(z)}\,,
\end{equation}
where $z_*$ is the redshift at decoupling. Its precise value depends weakly on the parameters, and it is obtained from the fitting formula\,\cite{HuSugiyama}:
\begin{equation}\label{zlastscattPRD}
z_{*}=1048\,\left(1+0.00124\,\omega_b^{-0.738}\right)\left(1+g_1 \omega_m^{g_2}\right)\,,
\end{equation}
with
\begin{align}\label{zlastscatt1PRD}
&g_1 = \frac{0.0783\,\omega_b^{-0.238}}{1+39.5\,\omega_b^{0.763}}\nonumber\\
&g_2 = \frac{0.560}{1+21.1\omega_b^{1.81}}\,.
\end{align}
The formula for the acoustic length $\ell_a$ is:
\begin{equation}
\ell_a(\vec{p})=\pi(1+z_*)\frac{D_A(z_*)}{r_s(z_*)}\,,
\end{equation}
 in which the angular diameter distance $D_A$ is given in \eqref{SomDist} and the expression for the sound horizon in \eqref{comdist}.

%%%%%%%%%%%%%%%%%%%%%
%%%%%%%%%%%%%%%%%%%%%
%%%%%%%%%%%%%%%%%%%%%

\section[Some comments on the fitting procedure and the Fisher matrix
formalism]{Some comments on the computational fitting procedure and Fisher matrix
formalism}

In a fitting parameter space of dimension greater than or equal to 4 it is quite unpractical to use the simple gridded minimization procedure that was applied in the first part of this dissertation to search for the minimum of the $\chi^2$ functions. The gridded minimization method has two important drawbacks. First of all, the grid must be built in such a way that the minimum of the function to be minimized is contained in it. So, in principle, in order to use this method one has to know something about the location of the minimum one is looking for. At least, one must have a rough estimate of the range in which the minimum lies. Obviously, the better is the hint about such location, the faster is the method. The second disadvantage is that one has to select a step for each one of the fitting parameters. One has to make sure that this step is small enough to ensure a good determination of the minimum, but on the other hand one has to take care, since the smaller the step the slower is the gridded minimization procedure. These facts make the method usually non-viable. And the problems worsen even more when we work with higher dimensional fitting vectors, because then the expense in computational time is completely unaffordable. Thus, if we want to explore larger parameter spaces, we are forced to look for a more efficient method.

In the second part of this thesis, in which we have used 4, 5 and even 6 dimensional parameter spaces, we have opted to use a quite simple alternative, but much more efficient in terms of computational time. It is the so-called Newton-Raphson method. The idea is the following. Let us consider an N-dimensional fitting vector $\vec{p}$. We want to minimize the $\chi^2$ function with respect to each of the elements of the aforementioned vector. We can approximate the $\chi^2$ function around a given point $\vec{p}_0$ as follows,
\be\label{eq:chi2Taylor}
\chi^2(\vec{p})=\chi^2(\vec{p}_0)+\frac{\partial\chi^2}{\partial p_i}\Bigr|_{\vec{p}_0}(\vec{p}-\vec{p}_0)_{i}+\frac{1}{2}\frac{\partial^2\chi^2}{\partial p_i\partial p_j}\Bigr|_{\vec{p}_0}(\vec{p}-\vec{p}_0)_{i}(\vec{p}-\vec{p}_0)_{j}\,.
\ee
This is simply the Taylor expansion of the $\chi^2$ function, which is only a good approximation in the vicinity of $\vec{p}_0$. Notice that we are neglecting the cubic and higher power terms. The Newton-Raphson method is an iterative one. One estimates the location of the minimum of the $\chi^2$ by directly using \eqref{eq:chi2Taylor}. If we denote $\vec{p}_1$ the estimation of the location of the minimum we find, 
\be
\vec{p}_1=\vec{p}_0-\left(\frac{\partial^2\chi^2}{\partial p^2}\Bigr|_{\vec{p}_0}\right)^{-1}\vec{\nabla}\chi^2\Bigr|_{\vec{p}_0}\,,
\ee
where $\left(\frac{\partial^2\chi^2}{\partial p^2}\Bigr|_{\vec{p}_0}\right)^{-1}$ is the inverse of the Hessian matrix of the $\chi^2$ function at $\vec{p}_0$. We can now repeat the procedure exchanging $\vec{p}_0\to\vec{p}_1$, and later on $\vec{p}_1\to\vec{p}_2\to...\to\vec{p}_k$ until the convergence to the real minimum is achieved. This is a quite fast method. Certainly, much more than the gridded minimization procedure used in Part I. Here we also have to select the correct step for each of the involved fitting variables in order to compute the numerical derivatives, so a previous study of the fitting program is needed to make a correct choice. Nevertheless, once we determine the optimum size of these steps, the method is quite agile and the convergence fast. When the Hessian is unavailable or too expensive to compute at each iteration, a quasi-Newton method can also be used, but in our case this has not been necessary.

Another very interesting point is that for the same price, this method, which allows us to determine the minimum of the $\chi^2$ function, can be combined with the Fisher matrix formalism in order to obtain a practical (but, in general, approximated) form of the posterior distribution of the fitting parameters. The idea of the Fisher formalism is to approximate the full (exact) likelihood with a multivariate Gaussian distribution in parameter space,
\be\label{eq:FisherLike}
\mathcal{L}(\vec{p})\approx \frac{1}{(2\pi)^{N/2}\sqrt{|F|}}\,e^{-\frac{1}{2}(p_i-\hat{p}_i)F_{ij}(p_j-\hat{p}_j)}\,,
\ee
where $\hat{p}_i$ are the best-fit values of the fitting variables (those that minimize the $\chi^2$ function or, equivalently, maximize the likelihood), and $F$ is the so-called Fisher matrix. Assuming Gaussian errors in the experimental data, the exact likelihood that governs the distribution of the fitting vector elements reads,
\be                                                          
\mathcal{L}(\vec{p})= \mathcal{L}_0\,e^{-\frac{\chi^2(\vec{p})}{2}}\,.
\ee
Notice that the exponent can be in this case a quite involved object (in comparison with the exponent of \eqref{eq:FisherLike}). If we expand in Taylor series the $\chi^2$ function around $\vec{p}=\hat{\vec{p}}$ (where $\vec{\nabla}\chi^2\Bigr|_{\hat{\vec{p}}}=\vec{0}$) using \eqref{eq:chi2Taylor}, i.e. neglecting higher order corrections, we find,
\be
F_{ij}=\frac{1}{2}\frac{\partial^2\chi^2}{\partial p_i\partial p_j}\Bigr|_{\hat{\vec{p}}}\,.
\ee
The {\it r.h.s.} of this expression is an ingredient that we already have from the Newton-Raphson minimization procedure, since we have previously computed it in order to find the best-fit parameters. Thus, with this method we automatically obtain the Fisher matrix. And the inverse of the Fisher matrix is the approximated covariance matrix, which do not only contain the information of the approximate uncertainties of each fitting parameter, but also the approximate existing correlations among them. 

In the studies of Chapters \ref{chap:AandGRevisited}-\ref{chap:H0tension} we have explicitly checked that the posterior distributions are purely Gaussian\footnote{This check has been carried out making use of the Metropolis-Hastings Monte Carlo algorithm \cite{Metropolis,Hastings}, in order to sample the exact distribution and compare it with the normally distributed one obtained with the Fisher matrix formalism. For further details see Sect. \ref{sec:discussionApJ}, together with the useful references \cite{AlanHeavens,Trotta}}. This is not obvious at all because experimental data distributed in a pure Gaussian way do not lead (in general) to Gaussian distributions in the fitting parameters. Thus, in the analyses carried out in the second part of the thesis the Fisher matrices contain all the (exact) statistical information concerning the fitting parameters, since now they are not just working as first approximations, but as exact objects. Therefore, from these Fisher matrices we can obtain uncertainties, correlation coefficients, derive the corresponding confidence intervals in a reduced parameter space, etc. But in order to extract all this information we have to marginalize \eqref{eq:FisherLike}. The marginalization of a multivariate normal distribution turns out to be immediate. One only has to eliminate from the inverse Fisher matrix the columns and rows associated to those variables one wants to get rid of. This is something very easy to prove. I start defining the marginalized likelihood over the variables ($p_{r+1}$,...$p_N$), with $N$ being the dimension of the original fitting vector,
\be                                                               
\mathcal{L}_m(p_1,...,p_r)=\int...\int dp_{r+1}...dp_N\,\mathcal{L}(\vec{p})\,.
\ee
Now let us use Greek subscripts to denote those fitting variables which are {\it not} marginalized out, and derive the above expression with respect to one of these variables,
\be
\frac{\partial \mathcal{L}_m}{\partial p_\alpha}=-\int...\int dp_{r+1}...dp_N\,\mathcal{L}(\vec{p})F_{\alpha j}(p_j-\hat{p}_j)\,,
\ee
where the Latin subscript refers to any variable (marginalized or not), i.e. $j\in [1,...,N]$. Now we can multiply both sides of the last expression by the reduced matrix $(F^{-1})_{i\alpha}$,
\be
(F^{-1})_{i\alpha}\frac{\partial \mathcal{L}_m}{\partial p_\alpha}=-\int...\int dp_{r+1}...dp_N\,\mathcal{L}(\vec{p})(p_i-\hat{p}_i)\,.
\ee
If we select $i=\beta$, we obtain
\be
(F^{-1})_{\beta\alpha}\frac{\partial \mathcal{L}_m}{\partial p_\alpha}=-(p_\beta-\hat{p}_\beta)\mathcal{L}_m\,.
\ee
Notice that $(F^{-1})_{\beta\alpha}$ is obtained by inverting the original Fisher matrix $F$ and deleting the columns and rows of the variables that have been integrated out. It is convenient to define such matrix as $(F^{-1})_{\beta\alpha}\equiv (M^{-1})_{\beta\alpha}$ in order not to confuse the reader (notice that if we invert $(M^{-1})_{\beta\alpha}$ we do not obtain the original N-dimensional matrix $F$, but a different $r$-dimensional matrix $M$!). 

If we derive the last expression with respect to $p_\gamma$ and rearrange terms we obtain,
\be
(M^{-1})_{\beta\alpha}\frac{\partial^2\ln \mathcal{L}_m}{\partial p_\alpha\partial p_\gamma}=-\delta_{\gamma\beta}\,.
\ee
By multiplying both sides by $M_{\beta\delta}$ we finally get
\be
\frac{\partial^2\ln \mathcal{L}_m}{\partial p_\delta\partial p_\gamma}=-M_{\gamma\delta}\,.
\ee
This expression is telling us that $M$ is the Fisher matrix of the marginalized likelihood $\mathcal{L}_m$. Thus, the prove is completed. The marginalization of a multivariate Gaussian distribution is quite simple and can be performed without need of numerical brute force, which allows us to save an important amount of computational time and power when the Fisher formalism can be used. This is quite remarkable. If the Fisher matrix formalism was not good enough to describe the real shape of the exact likelihood at point located not very far away from the maximum, i.e. if the posterior distribution was highly non-Gaussian in the problem under study, one would have to make use of Markov-Chain Monte Carlo methods to obtain the correct uncertainties, contour lines, etc. associated to the various fitting parameters. In addition, one would have to carry out the numerical marginalization before getting the desired results, which is also quite time-consuming. 

In those cases in which non-gaussianities are important, one can also apply intermediate procedures which allow to obtain an approximate analytical expression for the likelihood which can be considerably lighter (easier to evaluate) than the starting one, by taking into account higher order corrections to the $\chi^2$ function, i.e. extra terms in the Taylor expansion \eqref{eq:chi2Taylor} used in the Fisher matrix formalism. This helps to improve the efficiency of the Monte Carlo algorithm, with the cost of deviating from the exact distribution in the final result. The aforementioned generalizations could seem straightforward from a mathematical point of view, since in principle one only has to introduce more derivatives in the expansion of the $\chi^2$ function, but in practice it is not so easy. One has to take due account of certain technical aspects. For instance, the construction of a positive definite and normalizable distribution is not guaranteed by just adding third (or higher) order derivatives in the $\chi^2$ expansion. The implementation of these corrections must be carried out in a very specific way as in the DALI (Derivative Approximation for Likelihoods) method \cite{SellentinQuartinAmendola,Sellentin2015}. These methods allow a fast evaluation of the posterior distribution, which consequently makes possible a faster running of the Monte Carlo method. But still, in order to obtain e.g. uncertainties or confidence contours, one has to marginalize the distribution, and this marginalization cannot be performed as fast as in the Fisher formalism. Of course, the Fisher matrix formalism is faster, but in general it is only a first approximation! In our case, though, we have checked that the posterior distribution is perfectly described by a multivariate Gaussian, which of course helps us to reduce a lot the computational time without sacrificing at all the precision at which the various statistical quantities (and plots) of interest are obtained. 

The Fisher matrix formalism is also used in many other applications in observational Cosmology, e.g. in Fisher forecasts. We do not discuss them here. For more details see Chapter 13 of \cite{BookAmendolaTsujikawa}.

As we have seen before, if we have the Fisher matrix we can automatically marginalize out as many fitting parameters as we want. In particular, we can obtain the one and two-dimensional marginalized distributions, which of course are also Gaussian. This allows us to compute confidence intervals for individual variables, and also likelihood contour lines in two-dimensional planes. This is very easy to do. Taking into account that the distribution of 
\be
\Delta\chi^2(\vec{p})\equiv \chi^2(\vec{p})-\chi^2_{\rm min} 
\ee  
is governed by the following probability density,
\be
dP(\Delta\chi^2,\nu)=\frac{2^{-\nu/2}e^{-\Delta\chi^2/2}(\Delta\chi^2)^{-1+\frac{\nu}{2}}}{\Gamma(\nu/2)}\,d\Delta\chi^2\,,
\ee
where $\nu$ is in this case the number of parameters entering the fit, we can compute the $1\sigma$, $2\sigma$, etc. confidence intervals in the reduced parameter space that we want. The probability to find a value of $\Delta\chi^2$ between $0$ and $\Delta\chi^2_l$ is given by
\be
P(\Delta\chi^2\leq\Delta\chi^2_l,\nu)=\int_{0}^{\Delta\chi^2_l}\frac{2^{-\nu/2}e^{-\Delta\chi^2/2}(\Delta\chi^2)^{-1+\frac{\nu}{2}}}{\Gamma(\nu/2)}\,d\Delta\chi^2\,.
\ee
%
%%%%%%%%%%%%%%%%%%%%%%%%%%%%%%%%%%%%%%%%%%%%%%%%%%%%%%%%%%%%%%%%%%%%%%%%%%%
%
\begin{table}[t!]
\centering
\begin{tabular}{cc|c|c|}
\cline{3-4}
                       &  & \multicolumn{2}{c|}{$\Delta\chi^2_l$} \\ \cline{2-4} 
\multicolumn{1}{c|}{}  & $P(\Delta\chi^2\leq\Delta\chi^2_l,\nu)$ &       1D    & 2D           \\ \hline
\multicolumn{1}{|c|}{$1\sigma$} & $68.3\%$ &     1      &     2.30      \\ \hline
\multicolumn{1}{|c|}{$2\sigma$} & $95.45\%$ &      4     &     6.18      \\ \hline
\multicolumn{1}{|c|}{$3\sigma$} & $99.73\%$ &      9     &      11.81     \\ \hline
\multicolumn{1}{|c|}{$4\sigma$} & $99.9937\%$ &    16       &    19.33       \\ \hline
\multicolumn{1}{|c|}{$5\sigma$} & $99.9999\%$ &     25      &     27.63      \\ \hline
\end{tabular}
\caption[Values $\Delta\chi^2_l$ that set the border between the different $1-5\sigma$ confidence regions for both, the one and two-dimensional Gaussian cases]{{\scriptsize Values of $\Delta\chi^2_l$ that set the border between $1\sigma$, $2\sigma$, $3\sigma$, $4\sigma$, and $5\sigma$ confidence regions for both, the one and two-dimensional multivariate Gaussian distributions.}\label{statisticsGauss}}
\end{table}
%
%%%%%%%%%%%%%%%%%%%%%%%%%%%%%%%%%%%%%%%%%%%%%%%%%%%%%%%%%%%%%%%%%%%%%%%%%%%

\noindent We are basically interested in determining the uncertainties of individual variables and confidence regions in 2D spaces. In Table \ref{statisticsGauss} we show some practical information. Concretely, the values of $\Delta\chi^2_l$ that set the border between the $1-5\sigma$ confidence regions for both, the one and two-dimensional Gaussian cases. The corresponding values of the parameters delimiting the aforementioned regions can be directly inferred by solving the following equation,
\be\label{eq:prelast}
M_{\alpha\beta}(p_\alpha-\hat{p}_\alpha)(p_\beta-\hat{p}_\beta)=\Delta\chi^2_l\,,
\ee
where $M$ is the Fisher matrix associated to the marginalized likelihood. The values of $\Delta\chi^2_l$ can be found in Table \ref{statisticsGauss} for the one and two-dimensional spaces. In the 1D case, the result is obvious,
\be\label{eq:last}
p_{n\sigma}=\hat{p}\pm nM^{-1/2}\,,
\ee 
where, of course, $M^{-1/2}$ is the $1\sigma$ uncertainty of the fitting parameter $p$. The expression for \eqref{eq:last} could not be different, since the $n\sigma$ regions are defined precisely from the 1D Gaussian distribution. Thus, the last relation is fulfilled just by definition. In the 2D case, the confidence regions are delimited by elliptical contours, as it is clear from \eqref{eq:prelast}.

\thispagestyle{empty}
\null
\newpage
\thispagestyle{empty}
\null
\newpage

\backmatter
\pagestyle{fancy}
\fancyhf{}
%\fancyhead[LO,RE]{\thepage}
\fancyhead[CO]{Resum de la tesi}
\fancyhead[CE]{Resum de la tesi}
\cfoot{\thepage}
\chapter[Resum de la tesi]{Energia de buit en Teoria Quàntica de Camps i Cosmologia (Resum de la tesi)}
\label{chap:ResumCatala}

Sabem que el nostre Univers es troba en expansió des de fa aproximadament 90 anys, quan el treball observacional d'Edwin Hubble sobre la relació distància-velocitat de les nebuloses espirals va veure exitosament la llum, confirmant així la predicció teòrica de G. Lema\^itre. Aquest fet s'ha reconfirmat de manera continuada des d'aleshores. Per exemple, gràcies a la detecció de la radiació còsmica de microones a mitjans dels anys 60, que va suposar la confirmació de l'existència de la teoria del Big Bang. En el 1998, la Humanitat va donar un pas més vers la comprensió de les lleis que regeixen l'Univers com un tot. Les mesures de la relació distància-lluminositat en funció del despla\c{c}ament al roig de la llum provinent de supernoves de tipus Ia portades a terme pels grups observacionals liderats per S. Perlmutter, A. Riess i B.P. Schmidt van permetre entendre que l'Univers no només es troba en expansió, sinó que aquesta és accelerada i positiva. En altres paraules, vam comprendre que el teixit còsmic actualment es dilata cada cop més ràpidament. Així doncs, sabem prou acuradament quina és la dinàmica de l'Univers passat i present, però paradoxalment, desconeixem molt sobre quina és la composició exacta del Cosmos que dóna lloc a la dinàmica observada. Podem dir que coneixem amb alta precisió només un 5\% del contingut d'energia-moment de l'Univers. Aquest 5\% és que el queda enmarcat en el model estàndard de partícules elementals, àmpliament testejat als acceleradors de partícules d'arreu del món, com ara en l'LHC. Sabem, també, que sigui el que sigui el 95\% restant, també conegut com el sector fosc, ha de tenir unes certes característiques molt concretes. Aproximadament un 25\% ha de ser descrit per matèria no relativista i, per tant, ha d'estar caracteritzat per tenir una pressió molt baixa (a efectes pràctics, inexistent). Aquesta component, que rep el nom de matèria fosca, interactúa gravitacionalment amb la resta de components que omplen l'Univers, però de manera molt dèbil (si és que ho fa) a partir d'altres tipus d'interacció. El 70\% restant és el que es coneix com energia fosca, que té propietats de repulsió gravitatòria i és la responsable de l'acceleració positiva amb la que s'expandeix l'Univers en el moment present. L'explicació més senzilla d'aquesta energia ve de la mà de la coneguda constant cosmològica, o $\Lambda$, la qual juga en les equacions de camp d'Einstein un paper anàleg al de l'energia ``pura'' de buit. El model $\Lambda$CDM, que incorpora matèria fosca i $\Lambda$, constitueix el model estàndard cosmològic, on també prenen un paper fonamental l'isotropia i l'homogeneïtat de l'Univers que s'estableixen en el Principi Cosmològic, així com també l'existència del període inflacionari primigeni en la seva versió estesa.

Tot i que a nivell fenomenològic la constant cosmològica és capaç de descriure prou bé una gran varietat de dades, aquesta no està lliure de tensions observacionals ni tampoc de problemes teòrics importants. Existeix una discrepància brutal (de més de 55 ordres de magnitud) entre el valor mesurat de la densitat d'energia associada a $\Lambda$ i la predicció teòrica que s'obté a partir de la Teoria Quàntica de Camps. Aquest és el problema més sever del terme cosmològic, i podem dir que és un dels problemes oberts més importants de la Física Teòrica actual. A més, cal dir que aquest problema és agreujat (encara més si cal) per l'existència de transicions de fase en l'Univers primigeni, així com també per la inestabilitat de l'energia de buit sota correccions radiatives.  

La solució última d'aquests problemes potser no sigui possible amb les eines de les que disposem ara per ara. En aquest sentit, estudis fenomenològics que ajudin a entendre de manera més precisa quina és la natura de $\Lambda$ són molt benvinguts. I una de les preguntes més inmediates que ens podríem plantejar té a veure amb la possible dinàmica d'aquest terme que es suposa constant en el model estàndard. Roman la densitat d'energia fosca constant en el temps? Una dinàmica de l'energia de buit que portés a un creixement de la mateixa en el passat implicaria que en èpoques remotes de la història còsmica $\rho_\Lambda$ va prendre valors del mateix ordre de magnitud que les prediccions teòriques, i això faria veure el problema des d'una òptica bastant diferent. De totes maneres, per què el terme constant que es fa visible en l'Univers recent pren un valor tan petit? Aquesta qüestió que té a veure amb la ``naturalitat'' del valor de $\Lambda$, que està lluny del que són les escales naturals de la Física de partícules, encara quedaria lliure de resposta. Potser s'hauria de pensar aquesta constant com un {\it input} purament fenomenològic, com molts altres paràmetres ``running'' del model estàndard de partícules, i esperar que una eventual ``Teoria del tot'' pogués dirimir sobre aquest punt crucial de la Física Teòrica. El problema al què ens enfrontem és molt profund. Al capdavall, el valor precís d'aquesta constant (així com el valor d'altres constants fonamentals de la Natura), tot i ser tan petit, és el que ha permès que pugui haver vida a l'Univers. Parafrasejant al mestre A. Einstein, ``I want to know God's thoughts, the rest are details'', potser les respostes a totes aquestes preguntes siguin el que més ens apropen als pensaments de Déu, sigui quina sigui la concepció particular que cadascú tingui de Déu.  

En aquesta tesi he analitzat diferents models de buit i d'energia fosca variable. He centrat la meva atenció principalment en els models de ``running'', que estan fortament motivats en el context del grup de renormalització en Teoria Quàntica de Camps en espais corbats, i en els quals la variació de $\Lambda$ està vinculada a una llei de conservació anòmala de la radiació i/o matèria no relativista, o bé a la variació de l'acoblament Newtonià amb el temps. El vincle es duu a terme a través de la identitat de Bianchi, que ens porta directament a la llei  de conservació corresponent. També he analitzat altres models de buit dinàmic amb una motivació teòrica més feble, bàsicament fenomenològica, així com variants dels models de buit dinàmic amb un paràmetre de l'equació d'estat diferent de $-1$. He estudiat la viabilitat dels diferents models segons les dades observacionals més actualitzades, no solament les que involucren les diferents funcions cosmològiques particulars de cada model a nivell de background, sinó també el comportament dels models a nivell de pertorbacions, tant lineals com no lineals. Això últim és important, ja que les dades de formació d'estructura juguen un paper molt rellevant i encripten part de la tensió observada entre el model $\Lambda$CDM i les dades experimentals. 

Els principals resultats obtinguts en els treballs duts a terme durant el període de recerca d'aquesta tesi són:

\begin{itemize}
\item La constant cosmològica en el models de buit dinàmic juga un paper fonamental. Sense aquest terme, els models són incapaços d'ajustar correctament les dades observacionals, ja sigui perquè fallen a l'hora de generar una transició entre un Univers desaccelerat i un Univers amb acceleració positiva, o ja sigui perquè no són capaços d'ajustar alhora les dades de background i les de formació d'estructura. Així doncs, l'anàlisi fenomenològic dels models sense aquesta constant additiva ha demostrat la inviabilitat dels mateixos. Aquestes patologies tampoc es curen en els models sense terme constant que es troben dins del que hem anomenat la $\mathcal{D}$-class, aquells models de DE autoconservada, amb $\omega_{DE}\ne -1$ i amb una estructura de l'energia fosca en funció d'$H$ com la de l'energia de buit en els RVM's. 
\item A la primera part de la tesi també hem vist que els models de ``running'' amb terme additiu constant són perfectament viables. A la segona part, gràcies al refinament de la base de dades observacionals emprada i a l'anàlisi més minuciós portat a terme dels models des del punt de vista estadístic i metodològic, hem estat capa\c{c}os de detectar una senyal a favor de la variabilitat del buit còsmic de $\sim 4\sigma$, un resultat sense precedents a la literatura.
\item També hem analitzat diferents parametritzacions de l'energia fosca, com la XCDM i la CPL. Hem vist que l'esmentada dinàmica de l'energia fosca també es pot tra\c{c}ar amb aquestes parametritzacions, així com amb models més sofisticats de camps escalars, amb un Lagrangià ben definit, com en el cas del model de Pebbles \& Ratra. El nivell de significància estadística d'aquests models i parametritzacions no arriba a ser tan alt com en els models de buit variable, però també és considerable, estant en ambdós casos al voltant de les $3\sigma$.
\item En tots els casos s'obtenen resultats coherents, predint una augment de la densitat d'energia fosca (o de buit) en el passat, que fa que la formació d'estructura d'alguna manera quedi frenada respecte la que es troba en el model cosmològic estàndard, degut, és clar, a les propietats de repulsió gravitatòria d'aquesta component còsmica.
\item També hem vist que aquesta senyal positiva a favor de la dinàmica de l'energia fosca solament pot ser detectada si s'empra un conjunt suficientment complet i refinat de dades observacionals. Hem entés que el rol de la dades de SNIa i $H(z_i)$ és molt secundari. Les dades de SNIa van ser crucials per detectar fa aproximadament vint anys amb un nivell alt de significància estadística l'acceleració positiva de l'Univers, però no són suficientment riques com per prendre el mateix protagonisme en l'estudi de la dinàmica de l'energia fosca. La tríada de conjunts de dades que juga un paper realment cabdal ve donada per la combinació BAO+CMB+LSS. Eliminant LSS o CMB es perd totalment la senyal, mentre que eliminant de l'anàlisi les dades de BAO es disminueix bastant la seva força, sent però encara prou elevada (de gairebé $3\sigma$ en el cas dels RVM's). A efectes pràctics, treballar amb BAO+CMB+LSS és gairebé indistingible de fer-ho amb la base de dades ampliada BAO+CMB+LSS+H(z)+SNIa, ja que els resultats de l'ajust queden pràcticament inalterats.
\item El punt anterior explica per què, per exemple, els equips Planck, BOSS o DES no han estat capa\c{c}os de detectar senyals a favor de la dinàmica de l'energia fosca. Això es deu principalment a que no han fet servir una base de dades suficientment rica, tenint en compte per una banda poques dades de BAO i de formació d'estructura, i per l'altra excloent també dades que són crucials en la detecció de l'esmentada senyal.
\item Molt recentment Heavens {\it et al.} han afirmat en el seu article \cite{Heavens2017} que no hi ha cap evidència a favor d'extensions del model estàndard cosmològic. En canvi, els nostres resultats ressonen molt bé amb els del treball (també recent) de Gong-Bo Zhao {\it et al.} \cite{GongBoZhao2017}, llegiu també l'article divulgatiu sobre el mateix \cite{NVnoteValentino}. Ambdós articles, \cite{Heavens2017} and \cite{GongBoZhao2017}, van aparèixer a l'arXiv després del nostre article curt \cite{ApJLnostre}, inclús després dels nostres treballs més extensos \cite{ApJnostre,MPLAnostre,PRLnostre}, en els quals hem defensat la presència d'una evidència significativa a favor d'energia fosca dinàmica en el si de les dades observacionals. Els nostres articles són, de fet, previs a qualsevol altre fent aquestes contundents afirmacions i emprant tant parametritzacions generals, com l'XCDM o la CPL, com models específics com els RVM's. A l'article \cite{Salvatelli2014} només s'analitza el model $Q_\Lambda$, també estudiat per nosaltres en els capítols \ref{chap:PRDbased} i \ref{chap:H0tension} d'aquesta tesi, on hem posat de manifest les discrepàncies que trobem respecte els resultats d'aquest article. El model $Q_\Lambda$ és, de fet, dels models de buit dinàmic que hem estudiat, el menys efectiu a l'hora d'ajustar les dades observacionals.

Tant el treball de Gong-Bo Zhao {\it et al.} com el nostre arriben a conclusions semblants i en ambdós casos s'empren conjunts de dades que involucren els ingredients que demostren tenir una especial sensibilitat a l'hora de captar l'efecte de la dinàmica de l'energia fosca, CMB+BAO+LSS. En canvi, en el treball de Heavens {\it et al.} s'utilitza una base de dades cosmològiques molt més limitada, que bàsicament es centra en el CMB i les dades de lensing. Nosaltres interpretem que aquesta és precisament la raó per la qual no troben cap efecte. Més concretament, en el cas de Gong-Bo Zhao {\it et al.} troben una senyal d'energia fosca dinàmica de $3.5\sigma$, mentre que nosaltres trobem un pic d'evidència de $3.8\sigma$ amb el RVM, i de $\sim 3\sigma$ amb la parametrització general de l'XCDM (veure la Taula \ref{tableH0Fit1}). Per tant, trobem resultats molt semblants als de \cite{GongBoZhao2017}. També és important remarcar que els resultats de Heavens {\it et al.} estan lluny de ser oposats als nostres, ja que si nosaltres ens cenyim al mateix conjunt limitat de dades que aquests autors també perdem la senyal d'energia fosca dinàmica. Per tant, els nostres resultats són totalment compatibles amb els seus. L'esmentada senyal positiva només pot ser obtinguda si incloem un conjunt suficientment complet de dades, amb tots els ingredients crucials continguts en la combinació BAO+CMB+LSS. 
\item En el darrer capítol hem vist que els models dinàmics de buit estan més afavorits que la seva generalització en forma d'energia fosca amb un paràmetre de l'equació d'estat constant diferent de -1. També hem vist que si es tenen en compte les dades de formació d'estructura, el valor d'$H_0$ que es troba amb el model $\Lambda$CDM no només es troba en conflicte amb la dada d'$H_0$ de Riess {\it et al.} ($\sim 73$ km/s/Mpc), sinó que també s'allunya unes 4-5$\sigma$ del valor de Planck ($\sim 67$ km/s/Mpc), la qual cosa no sembla haver estat notada abans a la literatura. Hem comprovat que en el marc d'alguns models dinàmics de buit el valor de Planck ressona totalment amb els valors preferits per les dades de BAO i LSS. Hem vist que si les dades de Planck estan lliures d'errors sistemàtics els models de buit variable poden donar una explicació viable a favor dels valors baixos d'$H_0$ i $\sigma_8(0)$. 
\item Un altre punt important que hem comprovat és el potencial que té el formalisme de Press \& Schechter per testejar models, i també per distingir aquells que a nivell de background o de pertorbacions lineals són molt similars. Petites variacions del paràmetre de buit $\nu$ donen lloc a grans diferències en el nombre d'estructures predites en un cert rang de masses. En un futur seria interessant poder aplicar aquest formalisme per detectar quins dels models d'energia fosca o de buit que són capaços d'explicar les dades millor que el $\Lambda$CDM a nivell de background o de pertorbacions lineals segueixen mostrant un bon comportament a nivell de pertorbacions no lineals. També per veure si en el procés de formació no lineal d'estructura apareixen diferències que permetin una caracterització més clara dels diferents models de buit dinàmic estudiats.  
\end{itemize}

Aquests resultats són molt encoratjadors i posen de manifest la possible variabilitat de la densitat de l'energia fosca (o del buit). És evident, però, que la confirmació definitiva ha de venir de la mà d'un augment del nombre de dades observacionals, que en un futur podrien ser proporcionades pels projectes DES, eBOSS, DESI, PFS, LSST, Euclid, i WFIRST, els quals exploraran un volum superior a 1000 milions cúbics d'anys-llum. Alguns d'ells seran capaços d'obtenir dades relatives a la formació d'estructura amb incerteses inferiors a l'1\% en redshifts que estaran en el rang $0<z<3$, aproximadament cobrint $3/4$ parts de l'edat de l'Univers present. 

En aquesta tesi es presenten les primeres senyals positives a favor de la dinàmica temporal de la densitat d'energia fosca. Els treballs de la segona part de la tesi mostren les primeres evidències importants (estadísticament significatives) a favor d'aquest fet, sent especialment rellevants en el cas de l'energia de buit dinàmica. L'autor confia que aquesta tesi aporti una miqueta de llum a un dels sectors més foscos de l'Univers. Si aquests primers indicis de pes a favor de la variabilitat de la densitat d'energia fosca són o no una manifestació del comportament real d'aquesta component còsmica, dependrà en gran mesura de la quantitat i qualitat de les dades observacionals futures. Mentrestant, les evidències proporcionades per les dades actuals són aclaparadores i semblen fer-se només visibles amb força quan un conjunt suficientment depurat i complet de dades de CMB+BAO+LSS és tingut en compte. 

Anem a veure què depara el futur. De moment, gaudim del camí.

\vskip 1.0cm
\begin{flushright}
Barcelona, Setembre de 2017
\end{flushright}

\thispagestyle{empty}
\null
\newpage

\pagestyle{fancy}
\fancyhf{}
%\fancyhead[LO,RE]{\thepage}
\fancyhead[CO]{\nouppercase{\leftmark}}
\fancyhead[CE]{\nouppercase{\rightmark}}
\cfoot{\thepage}

\listoffigures
\listoftables

%\addcontentsline{toc}{chapter}{Bibliography}

\end{document}